\documentclass[3p,times]{elsarticle}

\usepackage{ecrc}

\volume{
}

\firstpage{1}

\journalname{Physics reports}

\runauth{G. \R\ et al.}

\jid{pr}

\jnltitlelogo{
Physics reports}

\usepackage{amssymb}

\usepackage[figuresright]{rotating}
\newcommand{\Lu}{\mathrm{S}}
\newcommand{\Rey}{\mathrm{Re}}

\newcommand{\Rb}{\mathrm{Rb}}

\newcommand{\Rm}{\mathrm{Rm}}
\newcommand{\Rmquer}{\overline{\mathrm{Rm}}}
\newcommand{\Omin}{{\it \Omega}_{\mathrm{in}}}
\newcommand{\Omout}{{\it \Omega}_{\mathrm{out}}}
\newcommand{\Rin}{R_\mathrm{in}}
\newcommand{\Rout}{R_\mathrm{out}}

\def\R{R\"udiger}

\newcommand{\alf}{\alpha}
\newcommand{\Pm}{\mathrm{Pm}}
\newcommand{\Ha}{\mathrm{Ha}}
\newcommand{\Mm}{\mathrm{Mm}}
\newcommand{\ord}[1]{$O(#1)$}

\newcommand{\etaT}{\eta_{\rm T}}
\newcommand{\nuT}{\nu_{\rm T}}
\newcommand{\DT}{D_{\rm T}}
\newcommand{\mdiffquer}{\overline{\mathrm{\eta}}}
\def\rot{\mathop{\rm curl}\nolimits}
\def\div{\mathop{\rm div}\nolimits} 
\def\gsim{\lower.4ex\hbox{$\;\buildrel >\over{\scriptstyle\sim}\;$}} 
\def\lsim{\lower.4ex\hbox{$\;\buildrel <\over{\scriptstyle\sim}\;$}} 

\def\beg{\begin{eqnarray}}
\def\ende{\end{eqnarray}}
\renewcommand{\vec}[1]{\mbox{\boldmath $#1$}}
\newcommand{\rin}{r_{\rm in}}
\newcommand{\Om}{{\it \Omega}}
\def\A{Alfv\'en}
\def\urms{u_{\rm rms}}

\def\omdr{\omega_{\rm dr}}
\def\omgr{\omega_{\rm gr}}
\def\omquer{\overline{\omega}}
\def\i{{\rm i}}
\def\p{\kappa}
\def\d{{\rm d}}

\def\ara\&a{ Ann. Rev. Astronomy Astrophysics}

\begin{document}
\begin{frontmatter}

\dochead{DRAFT}

\title{Stability and instability of hydromagnetic Taylor-Couette flows}
\author{G\"unther \R$^{a,*}$,
Marcus Gellert$^a$, Rainer Hollerbach$^b$, Manfred Schultz$^a$, Frank Stefani$^c$}

\cortext[cor1]{Corresponding author.\\ E-mail addresses: gruediger@aip.de, mgellert@aip.de, R.Hollerbach@leeds.ac.uk, mschultz@aip.de,  f.stefani@hzdr.de}

\address{$^a$Leibniz-Institut f\"ur Astrophysik Potsdam (AIP), An der Sternwarte 16, D-14482 Potsdam, Germany\\
$^b$Department of Applied Mathematics, University of Leeds, Leeds, LS2 9JT, United Kingdom\\
$^c$Helmholtz-Zentrum Dresden-Rossendorf, Bautzner Landstr. 400, D-01328  Dresden, Germany}

\begin{abstract}
Decades ago S.~Lundquist, S.~Chandrasekhar,  P.~H.~Roberts and R.~J.~Tayler first posed questions about the stability of  Taylor-Couette flows of conducting material under the influence of large-scale  magnetic fields.  These   and many new questions can now  be answered numerically where the nonlinear simulations even provide the instability-induced  values of several transport coefficients. The cylindrical containers are  axially unbounded and penetrated by  magnetic background fields with axial and/or azimuthal components.
The influence of the magnetic Prandtl number $\Pm$ on the onset of the instabilities is shown to be substantial. The potential flow subject to {\em axial fields} becomes unstable against axisymmetric perturbations for a certain supercritical value    of the  averaged Reynolds number $\Rmquer=\sqrt{\Rey\cdot\Rm}$  (with $\Rey$  the Reynolds number of rotation, $\Rm$ its magnetic Reynolds number). Rotation profiles as flat as the quasi-Keplerian rotation law scale similarly but only for $\Pm\gg1$ while 
for $\Pm\ll1$ the instability instead sets in for supercritical $\Rm$ at an optimal value of the magnetic field.
Among the considered instabilities of {\em azimuthal fields}, those of the Chandrasekhar-type,  where the background field and the background flow  have identical  radial profiles,  are particularly interesting. They are unstable against nonaxisymmetric perturbations if at least one of the diffusivities  is non-zero. For $\Pm\ll 1$ the onset of the instability   scales with $\Rey$  while it scales with   $\Rmquer$  for $\Pm\gg 1$.  
 Even     superrotation can be destabilized by azimuthal and current-free magnetic fields; this recently discovered nonaxisymmetric instability   is   of a double-diffusive character, thus excluding   $\Pm= 1$.  It scales with $\Rey$ for  $\Pm\to 0$ and with $\Rm$ for  $\Pm\to \infty$. 
 
 The presented results allow the construction of  several new  experiments with liquid metals as the conducting fluid.
 Some of them  are described here and their results will be  discussed  together with relevant diversifications of the magnetic instability theory including nonlinear numerical studies of the kinetic and magnetic energies, the azimuthal spectra  and  the influence of the Hall effect.
\end{abstract}

\end{frontmatter}
\newpage
\tableofcontents
\newpage

\section{Introduction}\label{history}
A large variety of astrophysical phenomena
involves the interaction of rotating fluids and 
magnetic fields. An important case in point is 
the magnetorotational
instability, which is commonly considered the main
driver of angular momentum and mass transport 
in accretion disks, with enormous implications for 
cosmic structure formation. Magnetically triggered 
instabilities also influence the rotational 
structure and  chemical composition of stars 
at various stages of 
their evolution, and might even contribute to  
the stellar dynamo mechanism. 
Beyond that, they play a crucial
role in more earthly applications such as 
fusion reactors, silicon crystal growth, 
aluminum reduction cells, and 
liquid metal batteries.

Taylor-Couette flow as the flow between two coaxial
rotating cylinders is one of the most important paradigms
of fluid dynamics, exhibiting a great diversity of unstable
flow regimes when changing the rotation rate of the two 
cylinders. Exposing the (electrically conducting) fluid
to magnetic fields leads to a further
enhancement of flow phenomena which then depend 
on the geometry and the strength of the magnetic 
field as well as on the ratio of viscosity and resistivity 
of the fluid.

This review aims at giving a systematic and comprehensive
overview about the diverse instabilities that occur 
in Taylor-Couette flows
under the influence of axial, azimuthal, and helical magnetic 
fields. Particular emphasis will be placed on the 
recent liquid metal experiments, and their numerical
simulations. Yet, we will also try to 
apply the gained insight for tackling specific 
problems in the original astrophysical motivation.

\subsection{History}
\subsubsection{Hydrodynamics}
We shall set the scene by giving a
historical account of the research on 
(magnetized) Taylor-Couette flows. In doing so, we 
also introduce the most relevant dimensionless
numbers such as the magnetic Prandtl number, the 
hydrodynamic and magnetic Reynolds 
numbers, and the Hartmann number (which in later 
sections might be adapted to the needs of the 
specific problem though).

For  inviscid flows  with an arbitrary    rotation law $\Om=\Om(R)$ the  `Rayleigh condition'
\beg
\frac{1}{R^3}\frac{{\rm d}}{{\rm d}R}(R^2\Om)^2 > 0
\label{Ray2}
\ende
 is sufficient and necessary for stability against axisymmetric perturbations \cite{LR17}. Flows steeper than $1/R^2$ are unstable, but the so-called potential flow $\Om\propto 1/R^2$ is of neutral stability.  It is easy to see that it represents the radial profile  with $\rot\vec{U}=0$  if $\Om$ does not depend on $z$. The specific angular momentum $R^2\Om$ of the potential flow does not depend on radius $R$.
In 1923 G.~I.~Taylor considered the stability of a {\em viscous} flow between two axially unbounded cylinders rotating about the same axis with different frequencies but the same sign \cite{T23}. By use of the narrow-gap approximation he found that the flow can only be stable for rotation frequencies (normalized with the diffusion frequency) below a critical value that can be expressed by a critical Reynolds number whose theoretical value has been confirmed by experiments. This was the start of many theoretical developments towards an increasingly successful theory of hydrodynamic instabilities to understand the experimental findings. 

The standard model for Taylor-Couette flow uses a stationary outer cylinder. If the outer cylinder  rotates, this tends to stabilize the flow, the more so the flatter the rotation profile is. Flows with
\beg
\mu_\Om= \rin^2,
\label{Ray1}
\ende
where
\beg
\mu_\Om=\frac{\Omout}{\Omin}, \ \ \ \ \ \ \ \ \ \ \ \ \ \ \ \ \ \ \ \ \rin=\frac{\Rin}{\Rout},
 \label{mu} 
 \ende
form the limit of neutral hydrodynamical stability as there the Reynolds number 
\beg
{\Rey}=\frac{\Om_{\rm in} R_0^2}{\nu}
\label{Reynolds}
\ende
for instability goes to infinity. Here $\Rin$ and $\Rout$ are the radii of the inner and outer cylinders, $\Omin$ and $\Omout$ are their rotation rates, $\nu$ the microscopic viscosity and $R_0=\sqrt{R_{\rm in}(R_{\rm out}-R_{\rm in}) }$. The condition (\ref{Ray1}) is also called the `Rayleigh limit'  and the  associated flow is the potential flow with $\Om\propto 1/R^2$.


For the often used standard model with stationary outer cylinder, with $\Rout=2 \Rin$ and for no-slip boundary conditions, 
\beg
u_R=u_\phi=u_z=0,
\label{bc1}
\ende
Chandrasekhar \cite{C61} first calculated for this geometry the critical Reynolds number  $\Rey_0=68.2$  characteristic for neutral stability. For the nonaxisymmetric modes with the lowest azimuthal wave numbers $m=1$ and $m=2$ Roberts found $\Rey_0=75$ and $\Rey_0=127$ (see \cite{DO62}). As these numbers exceed Chandrasekhar's value for $m=0$ the Taylor vortices excited for the lowest rotation rate are basically axisymmetric about the $z$-axis.

\subsubsection{With azimuthal fields}
The present article reviews several new  results for modifications of the stability condition (\ref{Ray2}) if the fluid is electrically conducting and in the presence of magnetic fields with  relatively simple geometry. The fields may have only axial components or only azimuthal components or combinations of both. Michael \cite{M54} formulated the question how {azimuthal} background magnetic fields modify the condition (\ref{Ray2}) for stability of ideal fluids (inviscid and perfectly conducting). His criterion
\beg
\frac{1}{R^3}\frac{{\rm d}}{{\rm d}R}(R^2\Om)^2 - \frac{R}{\mu_0 \rho}
\frac{{\rm d}}{{\rm d}R}\left(\frac{B_\phi}{R} \right)^2 > 0
\label{bfcr}
\ende
only ensures stability against axisymmetric perturbations. For $\Om=0$ the requirement for stability is
\cite{V59,V72,T73}
\beg
\frac{{\rm d}}{{\rm d}R}\left(\frac{B_\phi}{R} \right)^2 < 0.
\label{bf}
\ende
It shows that an azimuthal magnetic field in stationary cylinders is unstable against axisymmetric perturbations for positive $n$ if it scales with radius $R$ as $R^{1+n}$. In contrast, the field $B_\phi\propto 1/R$ due to an electric current along the central axis proves to be stable, while the field $B_\phi\propto R$ due to a uniform axial current has only marginal stability.

The  condition (\ref{bfcr}) implies that  combinations of stable flows with stable fields are always stable and that  combinations of unstable flows with unstable fields are always unstable, while the combination of stable and unstable flows and fields leads to stability/instability depending on the relative amplitudes of the effects. Flows with high Mach numbers (ratio of the frequencies of global rotation and \A~rotation) are unstable if the rotation is unstable and stable if the rotation is stable. However, the condition (\ref{bfcr}) is a local one which means that in dependence on the radial profiles $\Om(R)$ and $
B_\phi(R)$ its left-hand side can change in sign between the boundaries and the system is unstable.
This can  in particular be  true if 
$B_\phi(R)$ changes its sign between the cylinders. 


The full magnetohydrodynamic problem  for real fluids with finite values of viscosity and magnetic diffusivity has been formulated by Edmonds \cite{E58} and Gotoh \cite{G62}  for a finite gap between two corotating cylinders of perfectly conducting material. As Michael  did, only axisymmetric perturbations were considered. The equation system was able to provide the critical Reynolds number for marginal stability as a function of the magnetic field  and the prescribed values of $\rin$, $\mu_\Om$ and the magnetic Prandtl number
\beg
\Pm=\frac{\nu}{\eta}
\label{Pmdef}
\ende
as the ratio of the microscopic viscosity and the magnetic diffusivity $\eta=1/\mu_0\sigma$ ($\mu_0$ the vacuum permeability, $\sigma$ the electric conductivity) which we shall  call -- following \cite{B03} -- the resistivity. Characteristically, the liquid metals used in MHD experiments have very small magnetic Prandtl numbers, between $10^{-7}$ and $10^{-5}$. The idea that it might be reasonable to put $\Pm=0$ in the equations (the so-called quasi-static or inductionless approximation) dominated the magnetohydrodynamic theory over several decades \cite{R67,ZT98,YB06}. The equations have been solved numerically  for finite values of $\mu_\Om$ within a narrow gap between the cylinders. Instability only occurred for $\mu_\Om<\rin^2$  which means that the magnetic field only suppressed the centrifugal instability. The magnetic field did not generate any new instability against axisymmetric perturbations, which indeed do not exist.

The stability criterion (\ref{bfcr}) for ideal fluids only holds for axisymmetric perturbations. Indeed, the inclusion of nonaxisymmetric perturbations into the stability theory drastically changes the situation. Tayler considered the problem of stability against nonaxisymmetric perturbations of an electric current within a stationary and  axially unbounded cylinder \cite{T57}. The fluid itself may be a perfect conductor surrounded by vacuum while  the azimuthal field $B_\phi$ is proportional to $R$. A sufficient condition for stability in this case resulted as $m\geq 2$, so that among the nonaxisymmetric modes only the azimuthal wave number $m=1$ can be unstable, excluding the instability of the modes $m> 1$.

A particular version of Tayler's inequality for the azimuthal wave number $m=1$ is
\beg
\frac{{\rm{d}}}{{\rm{d}}R}( R B_\phi^2) \leq 0
\label{tay}
\ende
as the sufficient and necessary condition for stability of a stationary ideal fluid against nonaxisymmetric perturbations \cite{T73}. All uniform and/or outwardly increasing fields are therefore not necessarily stable against perturbations with the mode number $m=1$. This, in particular, is true for the field $B_\phi\propto R$ due to a uniform electric current.
\bigskip

Lundquist \cite{L51} argued that a uniform electric current can be stabilized by application of a uniform axial magnetic field if their energies are of the same order, i.e.~$2 \langle B_z^2\rangle>\langle B_\phi^2\rangle$. The first experiments using mercury as a liquid conductor indeed seem to point in this direction \cite{DL58}.  Roberts \cite{R56} found instability against perturbations with high azimuthal mode numbers $m$ for all ratios of azimuthal to axial field components. In his detailed paper, Tayler \cite{T60} discussed the overall problem of current-driven instability under the influence of a {\em twisted} magnetic field without rotation. The innovation is that here the background field has its own nonvanishing current helicity $\vec{J}\cdot\vec{B}$. Valid only for inviscid fluids, his Fig.~7 demonstrates how positive growth rates of $m=1$ perturbations without axial field are transformed to negative growth rates under the presence of an axial field of the same magnitude. Chandrasekhar showed that a sufficiently strong axial field will always suppress any axisymmetric instability of an
azimuthal field by deriving the stability condition
\beg
{\cal I} B^2_z > \int \frac{\xi_R^2}{R^2} \frac{{\rm d}}{{\rm d}R}(R B_\phi)^2
\ \ {\rm d}R,
\label{chandra}
\ende
where ${\cal I}>0$ and $\xi_R$ is the (purely real) radial eigenfunction. 
The condition  (\ref{chandra}) reduces to
\beg
\frac{{\rm{d}}}{{\rm{d}}R}( R B_\phi)^2< 0
\label{chandracond}
\ende
as a sufficient condition for stability against axisymmetric perturbations \cite{C61}. Howard \& Gupta \cite{HG62} included differential rotation to extend this condition to 
\beg
R\frac{{\rm d}\Om^2}{{\rm d}R} - \frac{1}{\mu_0 \rho R^3}
\frac{{\rm d}}{{\rm d}R}\left( RB_\phi \right)^2 > 0.
\label{gupta}
\ende
That  this condition is violated somewhere between inner and outer cylinder is   {\em necessary} for instability \cite{K92}. For the current-free field $B_\phi\propto 1/R$ only superrotating flows are stable against axisymmetric perturbations. Note that the condition (\ref{chandra}) only applies to axisymmetric perturbations and to ideal fluids. 
Below we shall demonstrate that dissipative super-potential flows which are hydrodynamically stable can easily (i.e.~with moderate Reynolds numbers) be destabilized by helical magnetic fields with current-free azimuthal components. The resulting axisymmetric traveling wave instability has become known as the Helical MagnetoRotational Instability (HMRI).

We also mention  because of its astrophysical relevance a particular result by  Tayler who also discussed the adiabatic ($\nu=\eta=0$) stability of stars with mixed poloidal and toroidal fields \cite{T80}. For poloidal and toroidal field components of the same order he suggested stability of the system but the final answer to this complex  question remained open until now.

Taylor-Couette flows with stationary inner cylinder  have been considered as  the prototype of hydrodynamic stability \citep{W33,SG59}. Now we know, however, that for dissipative fluids with $\Pm\neq 1$ even superrotation may become unstable against nonaxisymmetric perturbations under the influence of weak, strictly toroidal magnetic fields and for moderate Reynolds numbers.   Recently,   for  very large  Reynolds numbers  even the existence of  a linear instability for superrotating nonmagnetic Taylor-Couette flows has been 
reported  \citep{D17}. 
\subsubsection{With axial fields}
The question how purely { axial} fields modify the rotating Taylor-Couette flow of conducting fluids has been addressed by Chandrasekhar in Ref.~\cite{C53}. For axisymmetric perturbations in an axially unbounded cylinder he formulated the complete set of MHD equations, which leads to a 10$^{\rm th}$ order system of  differential equations. After elimination of the pressure by means of the incompressibility condition $\div \vec{u}=0$, six equations remain for the components of $\vec{u}$, and four equations for the two potentials of the field-perturbations $\vec{b}$. Applying the inductionless approximation $\Pm\to 0$ (which is not identical to taking $\nu=0$, see \cite{PG07}) the system is reduced to 8$^{\rm th}$ order.

\begin{figure}[htb]
\centering
\includegraphics[width=6.0cm]{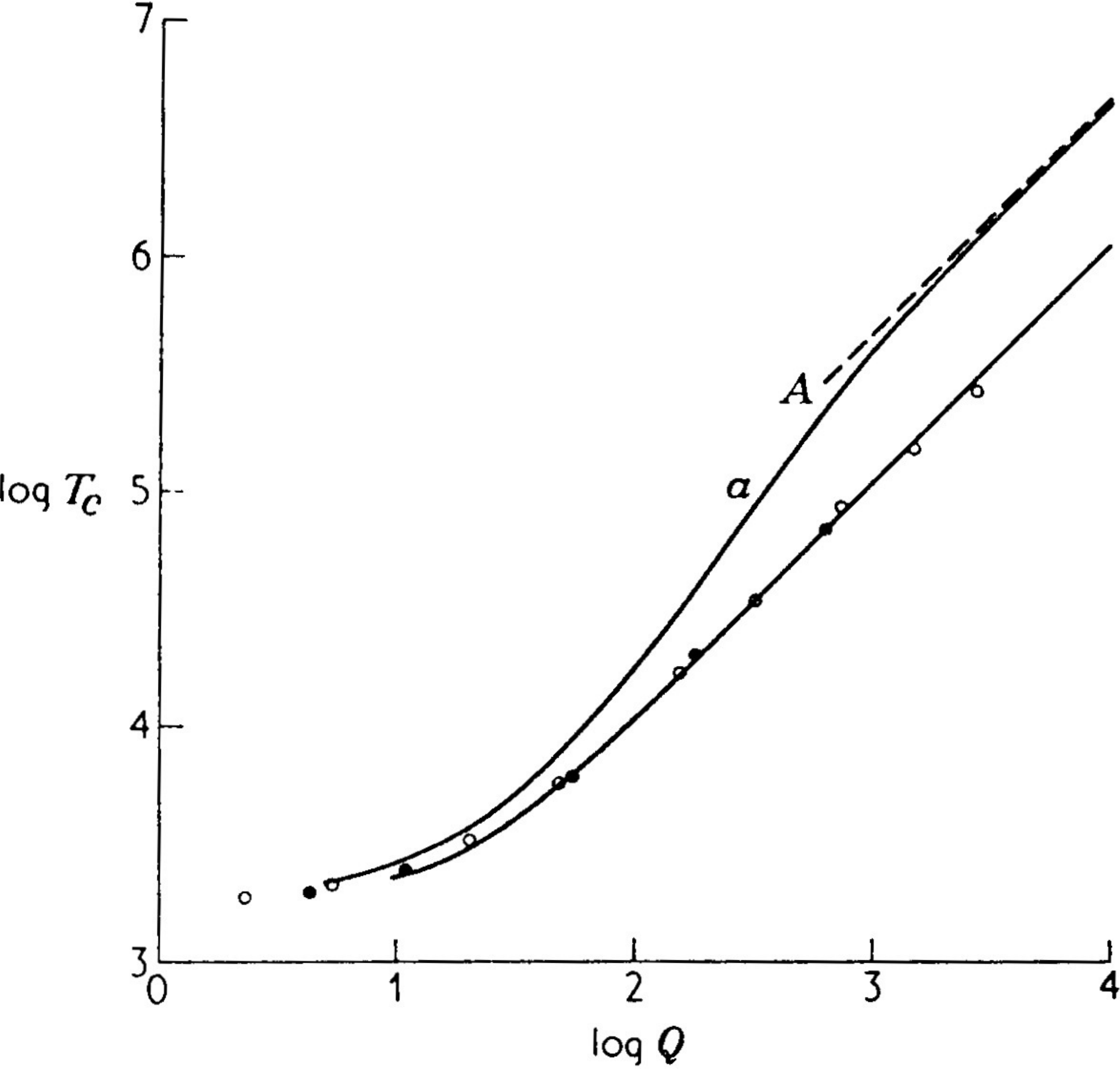}
\caption{
Using the inductionless approximation ($\Pm=0$), Chandrasekhar found for steep rotation laws the magnetic suppression of the Rayleigh instability by a uniform  axial magnetic field, in agreement with the measurements \cite{C61,DO60}. Open circles for $\rin=0.9$, solid circles are for $\rin=0.95$. $Q$ symbolizes the Hartmann number and  $T_{\rm c}$  the critical Reynolds number. Boundary 
conditions: upper line for conducting walls, lower line for insulating walls. The dashed curve marks the asymptotic behavior.  }
\label{donelly}
\end{figure}

The corresponding boundary conditions besides (\ref{bc1}) follow from the general rule of electrodynamics that the normal component $b_R$ of the magnetic field and the tangential component $E_z$ of the electric field are continuous at the transition from the fluid to either cylinder walls. If it is assumed that the cylinders are made from a highly conducting material, then $E_z=b_R=0$ at $R=\Rin$ and $R=\Rout$, resulting in the `Fermi conditions'
\beg
b_R= \frac{{\rm d} b_\phi}{{\rm d}R}+\frac{b_\phi}{R}=0.
\label{bc2}
\ende
For a given magnetic field amplitude, Chandrasekhar then computed the critical Reynolds number for the onset of instability, namely the smallest Reynolds number for all possible axial wave numbers. In all models the onset of the axisymmetric Taylor vortices is suppressed, where the suppression is weaker for the insulating cylinders (Fig.~\ref{donelly}). For these boundary conditions the results perfectly reflect the experimental results of Donnelly \& Ozima \cite{DO60,DO62} obtained with mercury as the conducting fluid, with $\Pm\simeq 10^{-7}$. Both cylinders were made from stainless steel with $\rin=0.95$, where the outer cylinder was stationary. Niblett stressed the importance of insulating boundary conditions in theory and experiments \cite{N58}.

Within the narrow-gap approximation and imposing axisymmetry, Kurzweg solved the 10$^{\rm th}$ order system without any restriction on the magnetic Prandtl number \cite{K63}. For small $\Pm$ the magnetic field suppresses the Taylor instability but for large $\Pm$ and weak fields the instability is enhanced, leading to subcritical Reynolds numbers compared with the nonmagnetic case. 
The magnetic boundary conditions in this work  are somewhat oversimplified, and do not completely match the formulation (\ref{bc2}). Nevertheless, the new step to allow finite values of $\Pm$ was an important one for the following reason. Assume that some unknown instability exists which for small $\Pm$ scales with moderate values of the magnetic Reynolds number 
$\Rm=\Pm\ \Rey$. Then for small $\Pm$ the critical Reynolds numbers yield values that are too large for numerical methods to cope with, since $\Pm\to 0$ and finite $\Rm$ yields $\Rey\to \infty$. The numerical codes for the 8$^{\rm th}$ order system (which always contain $\Rey$ rather than $\Rm$) could never find  instabilities  scaling with $\Rm$ for $\Pm\to 0$. For small $\Pm$ the numerical calculations only lead to enhanced Reynolds numbers $\Rey\simeq 10.4\cdot\Ha$ with the Hartmann number 
\beg
 \Ha=\frac{B_0 R_0}{\sqrt{\mu_0 \rho \nu \eta}},
\label{Hartmann}
\ende
while quite another scaling appears for $\Pm\to\infty$, i.e.~$\Rm\simeq 3.2~\Lu$ with the Lundquist number 
\beg
\Lu=\frac{B_0 R_0}{\sqrt{\mu_0 \rho}\ \eta},
\label{Lund1}
\ende
or $\Lu=\sqrt{\Pm}\Ha$. This scaling leads to a magnetic Mach number $\Mm=\Rm/\Lu\simeq 3.2$, so the instability exists for large magnetic Mach numbers.

Our calculations below for axial fields and $\mu_\Om=0$ confirm the result of Kurzweg that for large $\Pm$ and weak fields the critical Reynolds numbers lie below the hydrodynamic value of 68 valid for $\rin=0.5$, which increases to 185 for the narrow gap with $\rin=0.95$. The latter value describes the wide-gap mode of the viscosimeter of Donnelly. 
Both values $\rin=0.5$ and $\rin=0.95$ are still in  use in MHD laboratories. Obviously, if the field is not too strong it can play a destabilizing role for a Taylor-Couette flow. For the ideal hydromagnetic Taylor-Couette flow this was first discovered by Velikhov \cite{V59,C60}. In the MHD regime the Rayleigh criterion for stability against axisymmetric perturbations, $\mu_\Om>\rin^2$, changes to 
\beg
\frac{{\rm d} \Om}{{\rm d} R}>0
\label{veli}
\ende
i.e.~only flows with superrotation are stable (see  Fig.~1 in \cite{V59}). Velikhov found a growth rate along the Rayleigh line of $2\Om_{\rm in} \rin$. A dispersion relation has been derived for the Fourier frequency $\omega$ which only indicates instability if the \A~velocity $U_{\rm A}=B_0/\sqrt{\mu_0\rho}$ is smaller than the shear $-R^2 {\rm d}\Om/{\rm d}R$. His instability is thus again an instability for large magnetic Mach numbers. We shall show  that for dissipative fluids this new `magnetorotational instability' (MRI) indeed scales for $\Pm\to 0$ with the magnetic Reynolds number
\beg
{\Rm}=\frac{\Om_{\rm in} R_0^2}{\eta},
\label{Reynoldsmag}
\ende
which explains the absence of this mode in the early theories based on the inductionless approximation with $\Pm\to 0$ \cite{GJ02}. For $\Pm\gg 1 $, on the other hand,  the critical $\Rm$ does not remain constant but we shall find it growing with $\sqrt{\Pm}$. 

The most complete theory of the subject at the time was formulated by Roberts \cite{R64}. The MHD equations were written for general magnetic Prandtl number, for a finite gap and with nonaxisymmetric modes included. The formulation of the boundary conditions avoided the Fermi conditions for perfectly conducting cylinders: fluid and walls have different but finite electric conductivities where the conductivity of the cylinders exceed the conductivity of the fluid by a factor of only 1.37. This problem proved much more difficult to solve than the problem with insulating walls. The critical Reynolds numbers (meaning minimal with respect to all wave numbers) have been computed for given magnetic Hartmann number, with the result that the Taylor instability is suppressed by the magnetic field and this happens more effectively for conducting boundaries than for insulating boundaries (his Fig.~2).

Following the experiments of Donnelly \& Ozima, the magnetic Prandtl number used by Roberts was that of mercury ($10^{-7}$), and the outer cylinder was stationary. This was the reason that the standard MRI did not appear in this study. As shown below for the rotation law satisfying Eq.~(\ref{Ray1}), i.e.~$\mu_\Om=0.25$ for $\rin=0.5$, the critical Reynolds number for standard MRI is $\Rey\simeq 66/\sqrt{\Pm}$ (see Section \ref{carlos}). This is a rather small numerical value for, e.g., $\Pm\simeq 1$, indicating a new (magnetorotational) instability, as the Reynolds number for the nonmagnetic system at the Rayleigh limit is infinite. Roberts' code was certainly able to handle the magnetohydrodynamics near the Rayleigh limit for not too small $\Pm$.
\subsection{Outline of the review}

We shall revisit many of the mentioned questions (and preliminary answers) in the following where the stability of cylindrical Taylor-Couette flows under the influence of large-scale magnetic background fields is considered when the fluid between the cylinders is electrically conducting. Present-day and future experiments will always form the focus of the calculations and simulations, as has already been done in the first papers  initiating this special branch of Taylor-Couette research at the beginning of this century \cite{RZ01,RS01,JG01,GJ02,RS02}.

As a warm-up, we start  by considering the 
suppressing effect of axial and azimuthal magnetic 
fields on the instabilities in 
classical Taylor-Couette flows with 
stationary outer cylinder. 
For much flatter rotation laws, at 
and beyond the Rayleigh limit,
we discuss in Section \ref{standardMRI} the important standard
version of the MRI, with a purely axial field being applied.
We will focus here on nonlinear simulations
and on the resulting transport coefficient for 
angular momentum.

Section \ref{AMRI} deals with another magnetic 
field topology,
i.e.~a purely azimuthal field being 
produced by a central axial current that is 
insulated from the fluid. 
After a discussion of the so-called Azimuthal
MRI (AMRI) for potential flow and Keplerian rotation, 
we assess in detail the results of
a liquid metal experiment having shown 
AMRI slightly beyond  the Rayleigh 
limit. A further detailed  discussion is devoted to the 
so-called Super-AMRI, the surprising magnetic 
double-diffusive destabilization 
of flows whose angular frequency is  
steeply {\it increasing} with radius.

A particular aspect of AMRI is discussed in 
Section \ref{Chandra}. Here we reconsider Chandrasekhar's 
theorem that states, for ideal fluids, 
the stability of rotating flows of any radial dependence 
under the influence of an azimuthal magnetic field 
whose corresponding Alfv{\'e}n 
velocity has the same
amplitude and radial dependence as 
the rotation. For three representative cases, i.e.~
potential flow,
Keplerian rotation, and the rigidly-rotating $z$-pinch,
we show that finite diffusivities can even 
destabilize this class of Chandrasekhar-type flows.   

Section \ref{HMRI} is devoted to the combination of axial and 
azimuthal fields which are current-free between the cylinders. Actually, the resulting axisymmetric 
helical MRI
(HMRI) had been found earlier than AMRI, with
which it shares the inductionless character and the
corresponding scaling with the Reynolds and Hartmann 
numbers.
The transition between HMRI and AMRI will
also be described before the results of the
PROMISE experiments are  discussed. 

The additional or 
complementary energy source of axial electrical currents 
within the fluid, briefly mentioned
in Section \ref{Chandra}, will dominate the discussions of 
Sections \ref{TI} -- \ref{Twisted}. In 
Section \ref{TI} we start with the basic case of the 
Tayler instability in a stationary current-carrying 
cylinder, as realized in the 
liquid metal experiment  GATE. 
Rotation will re-enter the scene in 
Section \ref{Tayler}, where the various effects of 
rigid-body rotation and negative or positive shear
flows 
are investigated. The additional complication of
superimposing an axial field to this setting is 
discussed in Section \ref{Twisted}. 

After this comprehensive study of different
combinations of rotation and background 
magnetic fields, 
Sections \ref{Transport} and  \ref{Helicities} are concerned with  
questions of specific 
astrophysical relevance. This applies to the
numerical estimations (in Section \ref{Transport}) 
of the eddy viscosity and the
effective diffusivity which play a key role for
angular momentum and species transport in
accretion disks and stars.
The question of whether magnetic 
instabilities
can lead to helicity and a corresponding $\alpha$ effect, 
which may play an important role 
in nonlinear dynamo concepts such as 
the MRI dynamo or the so-called Tayler-Spruit dynamo,
is dealt with in Section \ref{Helicities}.
In Section \ref{Influence} we assess the special effects
that arise when the Hall effect is taken into account, which 
is particularly important for neutron stars.
The paper concludes with a short summary and 
a discussion of some future developments.

\section{Equations and model}
The general MHD equations for the conducting fluid are 
\begin{equation}
\rho\left(\frac{\partial\vec{U}}{\partial t} + (\vec{U}\cdot\vec{\nabla})\vec{U}\right) = -\nabla P+ \rho \,
\nu\ \Delta\vec{U} + \frac{1}{\mu_0} {\rm curl}\,\vec{B} \times \vec{B}
\label{Navier}
\end{equation} 
and 
 \begin{equation}
\frac{\partial\vec{B}}{\partial t} = {\rm curl}\,(\vec{U}\times\vec{B})+
\eta\ \Delta\vec{B}, 
\label{Ind1}
\end{equation}
where $\vec{U}$ is the fluid flow, $P$ the pressure, and $\vec{B}$ the magnetic field. The solutions must also fulfill the source-free conditions 
 \begin{equation}
 \div \vec{U}=\div \vec{B}=0.
\label{div}
\end{equation} 
The quantity $R_0=\sqrt{R_{\rm in}(R_{\rm out}-R_{\rm in}) }$ is used as the unit of length, $\eta /R_0$ as the unit of the perturbed velocity, $\nu/R_0^2$ as the unit of frequency (inverse time). For both very wide and very narrow gaps it is often reasonable to replace $R_0$ by the gap width $d=\Rout- \Rin$. Note that $\Rout=2 \Rin$ is the only model with $R_0=d$. We also define a characteristic magnetic field amplitude $B_0$ as the unit of the magnetic field fluctuations, $R_0^{-1}$ as the unit of the wave number and $\Om_{\rm in}$ as the unit of $\Om$. The dimensionless numbers of the problem are then the Reynolds number (\ref{Reynolds}), the magnetic Prandtl number (\ref{Pmdef}) and the Hartmann number (\ref{Hartmann}) which is formed with the geometric average of the diffusivities, $\bar\eta=\sqrt{\nu \eta}$. We shall see that in most cases where no hydromagnetic  instability  exists, the magnetic Reynolds number ${\rm Rm}={\rm Pm} \   {\rm Re}$ and the Lundquist number ${\rm S}= \sqrt{{\rm Pm}} \, {\rm Ha}$ are better representations of the characteristic eigenvalues. There are also exceptions to this rule when the stability/instability of rather steep rotation laws in the presence of toroidal fields is considered. Sometimes it also makes sense to use the averaged Reynolds number
\beg
\Rmquer=\sqrt{\Rey\Rm}= \frac{\Om_{\rm in} R_0^2}{\bar\eta},
\label{rmquer}
\ende
formed with $\bar\eta$ instead of $\eta$ hence $\Mm=\Rmquer/\Ha$. The magnetic Mach number
\beg
\Mm=\frac{\Rm}{\Lu}=\frac{\Rmquer}{\Ha},
\label{omaomain}
\ende
which does not involve any diffusivities, can be considered as a rotation rate normalized with the \A\ frequency $B_0/\sqrt{\mu_0\rho R_0^2}$. The magnetic Mach numbers of astrophysical objects often exceed unity. Galaxies have $\Mm$ between 1 and 10, for the solar tachocline with a magnetic field of 1 kG one obtains $\Mm\simeq 30$, and for typical white dwarfs and neutron stars $\Mm\simeq 1000$. For magnetars with fields of $\sim10^{14}$~G and a rotation period of $\sim$1 s, the magnetic Mach number is $\sim0.1-1$.

In general, $\vec{U}$, $\vec{B}$ and $P$ may be split into mean and fluctuating components $\vec{U}=\bar{ \vec{U}}+\vec{u}$, $\vec{B}=\bar{ \vec{B}}+\vec{b}$ and $P=\bar P+p$. In this work we immediately drop the bars from the variables again, so that the upper-case letters $\vec{U}$, $\vec{B}$ and $P$ represent the large-scale or background quantities. By developing the disturbances $\vec{u}$, $p$ and $\vec{b}$ into normal modes, the solutions of the linearized MHD equations are considered in the form
\beg
\vec{u}=\vec{u}(R) {\rm e}^{{\rm i}(\omega t+kz+ m\phi)},
 \quad \quad
 p=p(R) {\rm e}^{{\rm i}(\omega t+kz+ m\phi)},
 \quad \quad 
\vec{b}= \vec{b}(R){\rm e}^{{\rm i}(\omega t+kz+ m\phi)}
\label{72.1}
\ende
for axially unbounded cylinders. Here $k$ is the axial wave number, $m$ the azimuthal wave number and $\omega$ the complex frequency including growth rate and a possible drift (or oscillation) frequency.

For viscous flows in the absence of any longitudinal pressure gradient the basic form of the radial rotation law in the container is
\begin{equation}
\Om(R) = a_\Om+\frac{b_\Om}{R^2},
\label{1.1}
\end{equation}
where $a_\Om$ and $b_\Om$ are two constants related to the angular velocities $\Om_{\rm in}$ and $\Om_{\rm out}$ with which the inner and outer cylinders rotate (we shall only be interested in positive $\Om_{\rm in}$ and $\Om_{\rm out}$). With $R_{\rm in}$ and $R_{\rm out}$ being the radii of the two cylinders, one obtains the coefficients 
\beg
a_\Om=\frac{\mu_\Om-r_{\rm in}^2}{1-r_{\rm in}^2}
 \Om_{\rm in}, \ \ \ \ \ \ \ \ \ \ \ \ \ \ \ \quad \quad \quad \quad
b_\Om= \frac{1-\mu_\Om}{1-r_{\rm in}^2} \Om_{\rm in} R_{\rm in}^2,
\label{1.2}
\ende
using the definitions (\ref{mu}).

\subsection{Axial field}\label{axfi}
\begin{figure}[htb]
\centering
 \includegraphics[height=8cm]{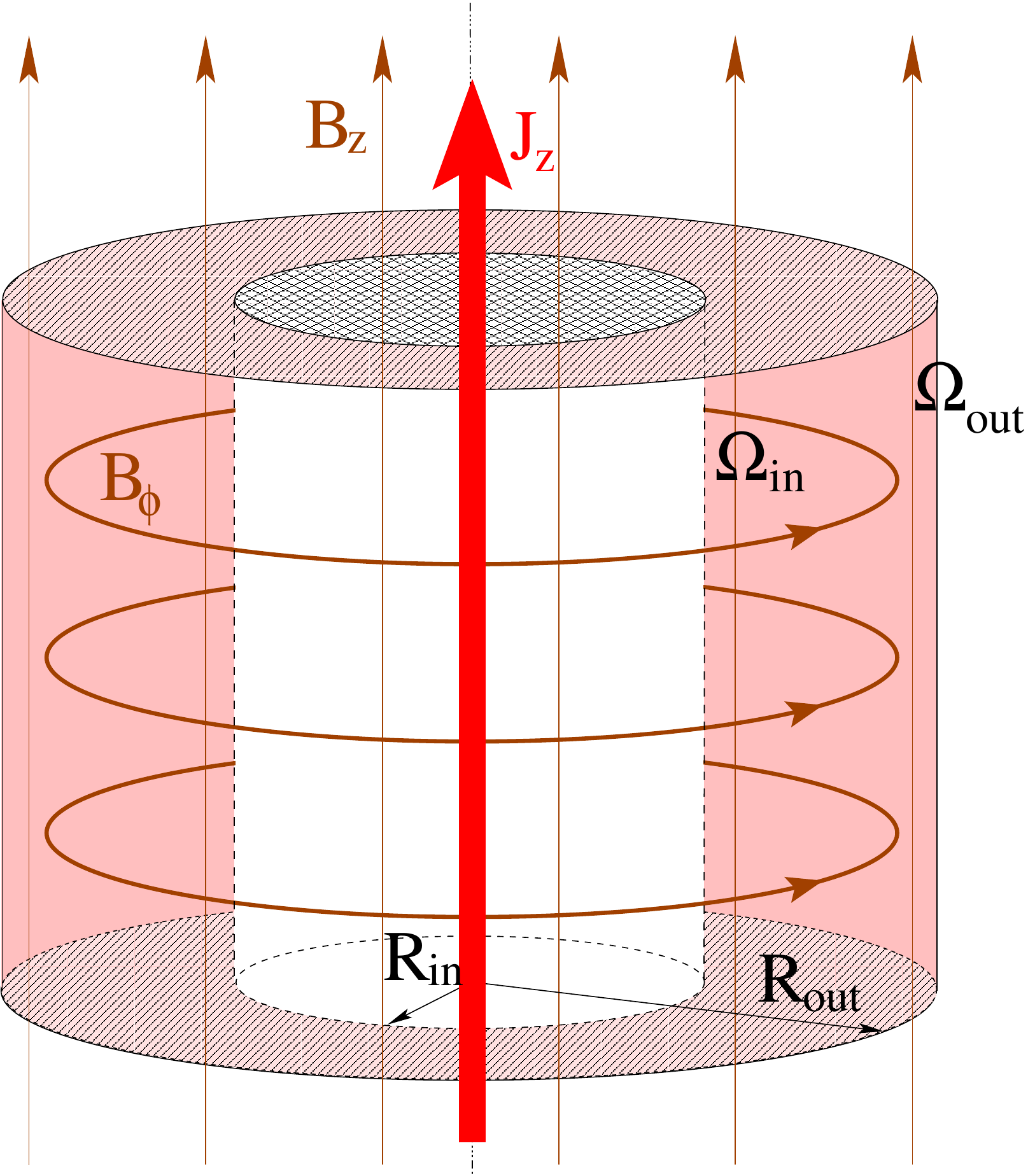}
 \caption{Geometry of hydromagnetic Taylor-Couette flows with uniform axial fields $B_0$ and/or circular azimuthal fields $B_\phi$ due to axial electric  currents inside the outer cylinder. The conducting fluid resides between the two concentric and axially unbounded cylinders with radii $\Rin$ and $\Rout$ rotating with $\Om_{\rm in}$ and $\Om_{\rm out}$  prescribed  by the boundary conditions. The cylinders are made either from perfectly conducting or  insulating material.  The relation $\Rout B_{\rm out} = \Rin B_{\rm in}$ 
 characterizes azimuthal  fields which are current-free between the cylinders.  Endplate effects are only discussed related to existing experiments. The standard container  is  defined by $\Rout=2\Rin$.}
 \label{f0}
\end{figure}

Figure \ref{f0} displays the geometrical setup and repeats the main definitions of the input parameters. The relevant equations follow from Eqs.~(\ref{Navier}) - (\ref{div}) and can be written as a system of ten first order equations. After eliminating $p$ and $b_z$, the linearized equations become
\begin{eqnarray}
\lefteqn{{{\rm d}^2 u_\phi \over {\rm d} R^2} + {1\over R} {{\rm d} u_\phi
\over {\rm d} R} - {u_\phi \over R^2} - \left({m^2 \over R^2} + k^2\right)
u_\phi 
-{\rm i} \left(m {\rm Re}\, {\Om} + \omega\right)
u_\phi +}\nonumber\\
&& \quad\quad + {2{\rm i}m \over R^2} u_R - {\rm Re} {1\over R} 
{{\rm d} \over {\rm d} R}
\left(R^2 \, {\Om}\right) u_R 
 - {m \over k} \left[{1\over R} {{\rm d}^2 u_z \over {\rm d} R^2} + 
{1\over R^2} {{\rm d} u_z \over {\rm d} R} - \left({m^2 \over R^2} + 
k^2\right) {u_z\over R} - {\rm i}\left(m {\rm Re}\, {\Om} + \omega\right) 
{u_z\over R}\right]\nonumber\\
&& \quad\quad
 + {m\over k} {\rm Ha}^2
 \left[{1\over R} {{\rm d} b_R \over 
{\rm d} R} +
{b_R \over R^2}\right] 
 + {{\rm i}\over k} {\rm Ha}^2 \left({m^2\over R^2} + k^2\right)
b_\phi = 0,
\label{p22}
\end{eqnarray}
\begin{eqnarray}
\lefteqn{{{\rm d}^3 u_z \over {\rm d} R^3} + {1\over R} 
{{\rm d}^2 u_z \over {\rm d}
R^2} - {1\over R^2} {{\rm d} u_z \over {\rm d} R} -
 \left({m^2\over R^2} +
k^2\right) {{\rm d} u_z \over {\rm d} R} +
{2m^2 \over R^3} u_z -
 {\rm i}\left( m
{\rm Re}\, {\Om} +\omega\right) {{\rm d} u_z \over
{\rm d} R} -}\nonumber\\ 
&& \quad\quad - {\rm i}m {\rm Re} {{\rm d}\Om \over {\rm d} R} u_z 
 - {\rm Ha}^2 \left[{{\rm d}^2 b_R \over {\rm d} R^2} + {1\over
R} {{\rm d} b_R \over {\rm d} R} - {b_R \over R^2} - k^2 b_R + 
{{\rm i}m \over R}
{{\rm d} b_\phi \over {\rm d} R} - {{\rm i}m\over R^2} b_\phi 
 \right] \nonumber\\
&& \quad\quad-{\rm i}k\left[{{\rm d}^2 u_R \over
 {\rm d} R^2} + {1\over R} {{\rm d} u_R
\over {\rm d} R} - {u_R \over R^2} - 
 \left(k^2 + {m^2\over R^2}\right)
u_R\right] - k \left(m {\rm Re}\, {\Om} + \omega\right)
u_R -
 2 {km \over R^2} u_\phi - 2 {\rm i}k {\rm Re}\, {\Om}
u_\phi = 0.
\label{p23}
\end{eqnarray}

The field perturbations fulfill
\begin{eqnarray}
{{\rm d} ^2 b_R \over {\rm d} R^2}+ {1\over R} {{\rm d} b_R \over 
{\rm d} R}
- {b_R \over R^2}- \left({m^2 \over R^2} + k^2\right) b_R -
 {2{\rm i}m\over R^2}
b_\phi -
 {\rm i} {\rm Pm} \left(m {\rm Re}\, {\Om}
+\omega \right) b_R + {\rm i}k u_R=0
\label{p24}
\end{eqnarray}
and
\begin{eqnarray}
{{\rm d}^2 b_\phi \over {\rm d} R^2} + {1\over R} 
{{\rm d} b_\phi \over
{\rm d} R} -{b_\phi \over R^2} - \left({m^2 \over R^2} + k^2\right) b_\phi
 +{2{\rm i}m \over R^2} b_R 
 -{\rm i} {\rm Pm} \left(m {\rm Re} \,
{\Om}
+ \omega\right) b_\phi + {\rm i}k u_\phi+ 
 {\rm Pm} \ {\rm Re} \ R {{\rm d} \Om
 \over {\rm d} R} b_R = 0
\label{p25}
\end{eqnarray}
\cite{SR02}. The last  term in Eq.~(\ref{p25}) describes the energy input by the induction of the global shear. It vanishes for $\Pm=0$ so that in the inductionless approximation differential rotation cannot be destabilized by uniform axial fields (no MRI, see next section). The hydrodynamic continuity equation 
\beg
{{\rm d} u_R \over {\rm d} R} + {u_R \over R}
 + {{\rm i}m \over R} u_\phi + {\rm i}k u_z = 0,
\label{p26}
\ende
completes the system. The rotation law $\Om=\Om(R)$ in these relations is normalized with $\Om_{\rm in}=\Om(R_{\rm in})$. The vertical component $b_z$ follows from the continuity condition
\beg
{\frac{{\rm d} b_R}{{\rm d} R}
+\frac{b_R}{R}+ \frac{{\rm i}m}{R} b_\phi + {\rm i} k b_z=0.}
\label{p27}
\ende
An appropriate set of ten boundary conditions is needed to solve the system. For the hydrodynamic quantities we always use the no-slip conditions for the velocity 
$
u_R=u_\phi=u_z=0. 
$
Generally, the normal component of the magnetic field and the tangential component of the electric field must be continuous. For perfectly conducting walls the conditions (\ref{bc2}) apply at $R_{\rm in}$ and $R_{\rm out}$. For insulating walls the magnetic field at the boundaries must match the vacuum  field with $\rot {\vec b}=0$, hence 
\begin{equation}
b_R+\frac{{\rm i}b_z}{I_m(kR)} \left(\frac{m}{kR} I_m(kR)+I_{m+1}(kR)\right)=0
\label{72.7}
\end{equation}
for $R=R_{\rm in}$, and 
\begin{equation}
b_R+ \frac{{\rm i}b_z}{K_m(kR)} \left(\frac{m}{kR} K_m(kR)-K_{m+1}(kR)\right)=0
\label{72.8}
\end{equation}
for $R=R_{\rm out}$, where $I_m$ and $K_m$ are the modified Bessel functions. The conditions for the toroidal field  are simply $k R b_\phi =m\, b_z$ at $\Rin$ and $\Rout$. In both cases five conditions exist at each boundary, so that the necessary ten conditions can be formulated. For both sorts of magnetic boundary conditions the resulting eigenvalues are often close together but not always. It is important in such cases to know the influence of  a finite conductivity $\sigma_{\rm cyl}$ of the cylinder material in relation to the conductivity $\sigma_{\rm fluid}$ of the conducting fluid between the cylinders. Note that the electric  conductivity of copper (as the
cylinder material) is only five times higher than the conductivity of 
sodium, hence  this constellation leads to $\hat\sigma\simeq 5$ for the ratio
\beg
\hat\sigma=\frac{\sigma_{\rm cyl}}{\sigma_{\rm fluid}}.
\label{hatsigma}
\ende
 One has to ask whether  this value
 leads to stability maps   close to those 
for perfectly conducting material or not.
As the derivation of these condition is rather cumbersome, only the final results may be given here, i.e.
 \begin{eqnarray}
b_R+\frac{\i k b_z}{\p I_m(\p\Rin)} \left(\frac{m}{\p \Rin} I_m(\p \Rin)+I_{m+1}(\p \Rin)\right) = -\frac{\i m {\hat\sigma}}{\p^2\Rin^2} \left( \frac{\d R b_\phi}{\d R} - \i m b_R \right),
\label{bc17}
\end{eqnarray}
 \begin{eqnarray}
\p b_\phi -\frac{km}{\p\Rin} b_z = \frac{\hat\sigma}{\Rin {I_m(\p\Rin)}} \left(\frac{m}{\p \Rin} I_m(\p \Rin)+I_{m+1}(\p \Rin)\right) 
 \left( \frac{\d R b_\phi}{\d R} - \i m b_R \right)
 \label{bc18}
\end{eqnarray}
for  $R=\Rin$ and 
 \begin{eqnarray}
b_R+\frac{\i k b_z}{\p K_m(\p\Rout)}  \left(\frac{m}{\p \Rout} K_m(\p \Rout)-K_{m+1}(\p \Rout)\right)  = -\frac{\i m {\hat\sigma}}{\p^2\Rout^2} ( \frac{\d R b_\phi}{\d R} - \i m b_R ),
\label{bc12}
\end{eqnarray}
\begin{eqnarray}
\p b_\phi -\frac{km}{\p\Rout} b_z = \frac{\hat\sigma}{\Rout {K_m(\p\Rout)}} \left(\frac{m}{\p \Rout} K_m(\p \Rout)-K_{m+1}(\p \Rout)\right)  \left( \frac{\d R b_\phi}{\d R} - \i m b_R \right)
 \label{bc13}
\end{eqnarray}
for $R=\Rout$.  The modified wave number $\p$ results from the definition
\begin{eqnarray}
\p^2=k^2 + \frac{\i (\omega+m\Om)}{\eta_{\rm cyl}}   
\label{bc3}
\end{eqnarray} 
including the skin effect \cite{R64,RS18}. Because $\Om$ is different at the two boundaries, they each have their own separate value of $\p$. The boundary conditions for perfectly conducting or for insulating cylinder material obviously follow in the limits $\hat\sigma\to \infty$ or $\hat\sigma\to 0$. 
For  axisymmetric perturbations  Eqs.~(\ref{bc18}) and  (\ref{bc13}) for $m=0$ approximately provide
 \begin{eqnarray}
\p b_\phi   \simeq \frac{\hat\sigma}{\Rin} \frac{\d R b_\phi}{\d R},\ \ \ \ \ \ \ \ \ \ \ \ \ \ \ \ \ \ \ \ \    \p b_\phi   \simeq -\frac{\hat\sigma}{\Rout} \frac{\d R b_\phi}{\d R}
 \label{bc14}
\end{eqnarray}
for   the inner  and the outer boundary condition. 

The homogeneous set of linear equations together with the choice of boundary conditions determines the eigenvalue problem for any given value of $\Pm$. The real part ${\Re}(\omega)$ of $\omega$ describes a drift of the pattern depending on the rotational symmetry: the drift is along the $z$-axis for $m=0$ and it is along the azimuth for $m\neq 0$. For a fixed Hartmann number, a fixed Prandtl number and a given axial wave number one finds the eigenvalues ${\Rey}$ and ${\Re}(\omega)$. For a certain axial wave number a minimum of the Reynolds numbers exists, which is the desired critical Reynolds number. 

\subsection{Azimuthal field}\label{azfi}
The radial profile of an azimuthal background field in a dissipative system is
\beg
B_\phi=a_B R+\frac{b_B}{R},
\label{basic}
\ende
where $a_B$ and $b_B$ are defined by the values of the azimuthal magnetic field at the inner ($B_{\rm in}$) and outer ($B_{\rm out}$) boundaries as 
\beg
a_B=\frac{B_{\rm in}}{R _{\rm in}}\frac{\rin
 (\mu_B - \rin)}{1- \rin^2},  \ \ \ \ \ \ \ \ \ \ \ \ \ \ \ \quad \quad \quad \quad
b_B=B_{\rm in}R _{\rm in}\frac{1-\mu_B \rin}
{1-\rin^2}
\label{abB}
\ende
with 
\beg
\mu_B=\frac{B_{\rm{out}}}{B_{\rm{in}}}.
\label{abC}
\ende
The constants $B_{\rm in}$ and $B_{\rm out}$ are defined by the vertical electric currents inside the inner and outer cylinders. For $\mu_B=1/\rin$ we have $b_B=0$ so that the magnetic field is of the form $B_\phi\propto R$, describing a uniform axial current within $R<R_{\rm out}$ (`$z$-pinch'). For $\mu_B=\rin$ we have $a_B=0$ and $B_\phi\propto 1/R$, which is current-free outside $R_{\rm in}$. A field of the form $b_B/R$ is generated by running an axial current only through the inner region $R<R_{\rm{in}}$, whereas a field of the form $a_B R$ is generated by running a uniform axial current through the entire region $R<R_{\rm{out}}$ including the fluid. As the standard choice in this paper will be $\rin=0.5$ one finds $\mu_B=0.5$ for the solution which is current-free between the cylinders and $\mu_B=2$ for the solution with uniform axial electric current between the cylinders. Another important radial profile of the background field which we shall often consider is given by $\mu_B=1$, describing a solution with almost uniform magnetic field between the cylinders. We have $b_B/a_B= \Rin^2/\rin$ in this case.
Expressing the electric currents in Ampere we obtain 
\beg
I_{\rm axis}=5R_{\rm in} B_{\rm in },  \ \ \ \ \ \ \ \ \ \ \ \ \ \ \ \quad \quad \quad \quad I_{\rm fluid}= 5 (R_{\rm out}B_{\rm out}-R_{\rm in} B_{\rm in})
\label{bi}
\ende
with $I_{\rm{axis}}$ the axial current inside the inner cylinder and $I_{\rm{fluid}}$ the axial current through the fluid. Here $R$, $B$ and $I$ are measured in centimeter, Gauss and Ampere. Expressing $I_{\rm{axis}}$ and $I_{\rm{fluid}}$ in terms of the Hartmann number formed with the azimuthal field strength $B_{\rm in}$ at the inner cylinder,
\beg
 \Ha=\frac{B_{\rm in} R_0}{\sqrt{\mu_0 \rho \nu \eta}}, 
\label{Hartmannin}
\ende
so that
\beg
I_{\rm{axis}}=
5 {\rm Ha}\ \sqrt{\frac{r_{\rm in}}{1-r_{\rm in}}} \sqrt{\mu_0\rho\nu\eta},\ \ \ \ \ \ \ \ I_{\rm{fluid}}=\frac{\mu_B-r_{\rm in}}{r_{\rm in}}I_{\rm{axis}}.
\label{Iin}
\ende
Quite similar relations can be formulated by means of  the Lundquist number for azimuthal magnetic fields
\beg
\Lu=\frac{B_{\rm in} R_0}{\sqrt{\mu_0 \rho}\ \eta}.
\label{Lund2}
\ende
For $\mu_B= r_{\rm in}$ we find $I_{\rm{fluid}}=0$ for the solution with $a_B=0$. 
On the other hand, for $ \mu_B=1/\rin$ it is $(1-\rin^2)I_{\rm axis}=\rin^2 I_{\rm fluid} $ hence $I_{\rm axis}=0$  for $\rin=0$ and $I_{\rm axis}=I_{\rm fluid}/3$ for $\rin=0.5$. Note that for $\mu_B<\rin$ the currents $I_{\rm{axis}}$ and $I_{\rm fluid}$ have {opposite} signs. In the present review the Hartmann number (\ref{Hartmannin}) -- formed with the azimuthal field amplitude $B_{\rm in}$ -- will be used in all sections where azimuthal magnetic background field exist. As explained later on, Section \ref{HMRI} forms the only  exception. 

The dimensionless parameters of the instability problem are the same as defined above, but with $B_{\rm in}$ instead of $B_0$ as in (\ref{Hartmann}). The necessary and sufficient condition for ideal flow stability is (\ref{bfcr}). Using (\ref{1.1}) for the angular velocity and (\ref{basic}) for the magnetic field and normalizing with $r=R/R_0$, Eq.~(\ref{bfcr}) takes the form 
\beg
a_\Om^2+\frac{a_\Om b_\Om}{r^2}+\frac{b_B}{(\Mm)^2 r^2}
\left( {a_B }+\frac{b_B}{r^2} \right) >0
\label{abc}
\ende
with $\Mm=\Om_{\rm in}/(B_{\rm{in}}/{\mu_0\rho R_0^2})^{1/2}$ as the magnetic Mach number representing a normalized rotation rate. The angular velocity part of (\ref{abc}) is positive for hydrodynamically stable flows beyond the Rayleigh limit. The magnetic part has a simple structure. It vanishes for $b_B=0$. Hence, magnetic fields $B_\phi\propto R$ have no influence on the axisymmetric mode of the instability for any rotation profiles. On the other hand, the magnetic part in (\ref{abc}) is positive definite for $a_B=0$ so that magnetic fields which are current-free in the fluid ($B_\phi\propto 1/R$) always stabilize any rotation profile.

Beyond these extremes it is always possible that the magnetic influence destabilizes rotation profiles beyond the Rayleigh limit against axisymmetric perturbations. It is also obvious that for negative magnetic parts in (\ref{abc}) (i.e.~$\mu_B>1/\rin$) one always finds values of the magnetic Mach number which are small enough to provide negative values for any $\mu_\Om$. Some sorts of magnetic fields with sufficiently strong currents can thus destabilize any rotation law even against axisymmetric perturbations. This is in particular true in the Rayleigh limit where $a_\Om=0$ so that the nonmagnetic part in (\ref{abc}) vanishes and all fields with $b_B<0$ become unstable, which according to (\ref{abB}) means $\mu_B>1/\rin$.

 The normalized equations with toroidal background fields are
\beg
{\frac{ {\rm d}^2 u_R}{ {\rm d} R^2}
+\frac{1}{R}\frac{{\rm d} u_R}{{\rm d} R}
-\frac{u_R}{R^2} -\left(k^2+\frac{m^2}{R^2}\right) u_R
-2{\textrm{i}}\frac{m}{R}u_\phi-}
{\textrm{i Re}}(\omega+m\Om) u_R
+2{\textrm{Re}}\Om u_\phi
-\frac{ {\rm d} p}{{\rm d} R}
 +{\textrm{i}}\frac{m}{R}{\textrm{Ha}}^2 B_\phi b_R
-2{\textrm{Ha}}^2 \frac{B_\phi}{R} b_\phi=0
\ende
\beg
\frac{ {\rm d}^2 u_\phi}{{\rm d} R^2}
+\frac{1}{R}\frac{{\rm d} u_\phi}{{\rm d} R}
-\frac{u_\phi}{R^2} -\left(k^2+\frac{m^2}{R^2}\right) u_\phi
+2{\textrm{i}}\frac{m}{R}u_R-
 {\textrm{i Re}}(\omega+m\Om) u_\phi
&-&{\textrm{i}}\frac{m}{R}p-\frac{{\textrm{Re}}}{R}\frac{{\rm d}}{{\rm d} R}(R^2 \Om) u_R+\nonumber\\
 &&+\frac{{\textrm{Ha}}^2}{R}\frac{{\rm d}}{{\rm d}R}\left(B_\phi R \right) b_R
+{\textrm{i}}\frac{m}{R}{\textrm{Ha}}^2 B_\phi b_\phi=0,
\label{uphi}
\ende
\beg
{\frac{ {\rm d}^2 u_z}{{\rm d} R^2}
+\frac{1}{R}\frac{{\rm d} u_z}{{\rm d} R}
-\left(k^2+\frac{m^2}{R^2}\right) u_z
-{\textrm{i Re}}(\omega+m\Om) u_z
-{\textrm{i}}\,kp
+{\textrm{i}}\frac{m}{R}{\textrm{Ha}}^2 B_\phi b_z=0}
\ende
and
\beg
{\frac{{\rm d}^2 b_R}{{\rm d} R^2}
+\frac{1}{R}\frac{{\rm d} b_R}{{\rm d} R}
-\frac{b_R}{R^2}-\left(k^2+\frac{m^2}{R^2}\right) b_R
-2{\textrm{i}}\frac{m}{R^2}b_\phi-} {\textrm{i Pm Re}}(\omega+m\Om) b_R
+{\rm i}\frac{m}{R}B_\phi u_R=0,
\label{bR}
\ende
\beg
\frac{ {\rm d}^2 b_\phi}{{\rm d} R^2}
+\frac{1}{R}\frac{{\rm d} b_\phi}{{\rm d} R}
-\frac{b_\phi}{R^2} -\left(k^2+\frac{m^2}{R^2}\right) b_\phi
+2{\textrm{i}}\frac{m}{R^2}b_R-
 {\textrm{i Pm Re}}(\omega+m\Om) b_\phi
&+&{\rm Pm Re} R\frac{{\rm d} \Om}{{\rm d} R}b_R-\nonumber\\
&&-R\frac{{\rm d}}{{\rm d} R}\left(\frac{B_\phi}{R}\right) u_R
+{\rm i}\frac{m}{R}B_\phi u_\phi=0,
\label{bphi}
\ende
with the boundary conditions described above (\cite{RK13}). The system is again supplemented by the incompressibility condition (\ref{p26}). The vertical component $b_z$ follows from (\ref{p27}).

The axial wave number $k$ is again varied until the Reynolds number for a given Hartmann number reaches its minimum. The resulting wave number corresponds to the most unstable mode. Both the background flow and magnetic field are normalized with their values at $R=R_{\rm in}$, hence $\hat\Om=\Om/\Om_{\rm in}, \, \hat B_\phi = B_\phi/B_{\rm in}$ (and the hats are then immediately dropped).

This system has the characteristic symmetry that if $k$ is kept fixed, but $m$ is replaced by $-m$, and simultaneously the eigenvalue ${\rm i}\omega$, the flow $\vec u$ and the field $\vec b$ are transformed to their complex conjugates, then the overall system remains unchanged. This means that $m=\pm 1$ constitute a single solution, with the same drift rate $\Re(\omega)/m$, Reynolds and Hartmann numbers. 
\section{Stationary outer cylinder}\label{Stationary}
According to the Rayleigh criterion the ideal flow is stable whenever the specific angular momentum increases outwards.
It is thus not stable if the outer cylinder is stationary so that (\ref{1.1}) becomes
\begin{equation}
\Om(R) = \frac{\Om_{\rm in}}{1-\rin^2}\left( \frac{\Rin^2}{R^2}-\rin^2\right).
\label{1.4}
\end{equation}
This is the rotation law whose stability characteristics in the presence of either axial or azimuthal magnetic background fields are now discussed.
\begin{figure}[htb]
\centering
 \includegraphics[width=8cm]{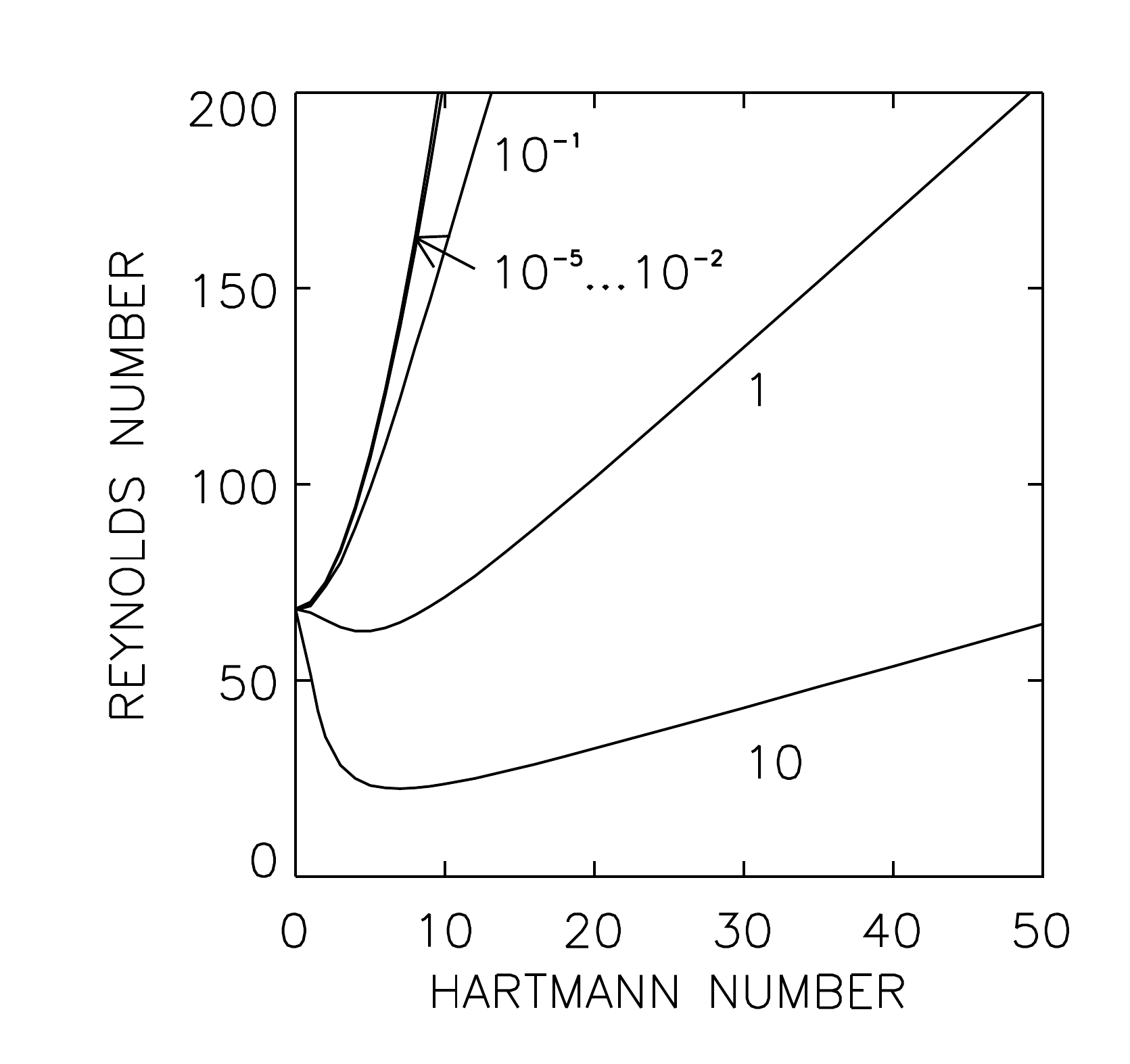}
 \includegraphics[width=8cm]{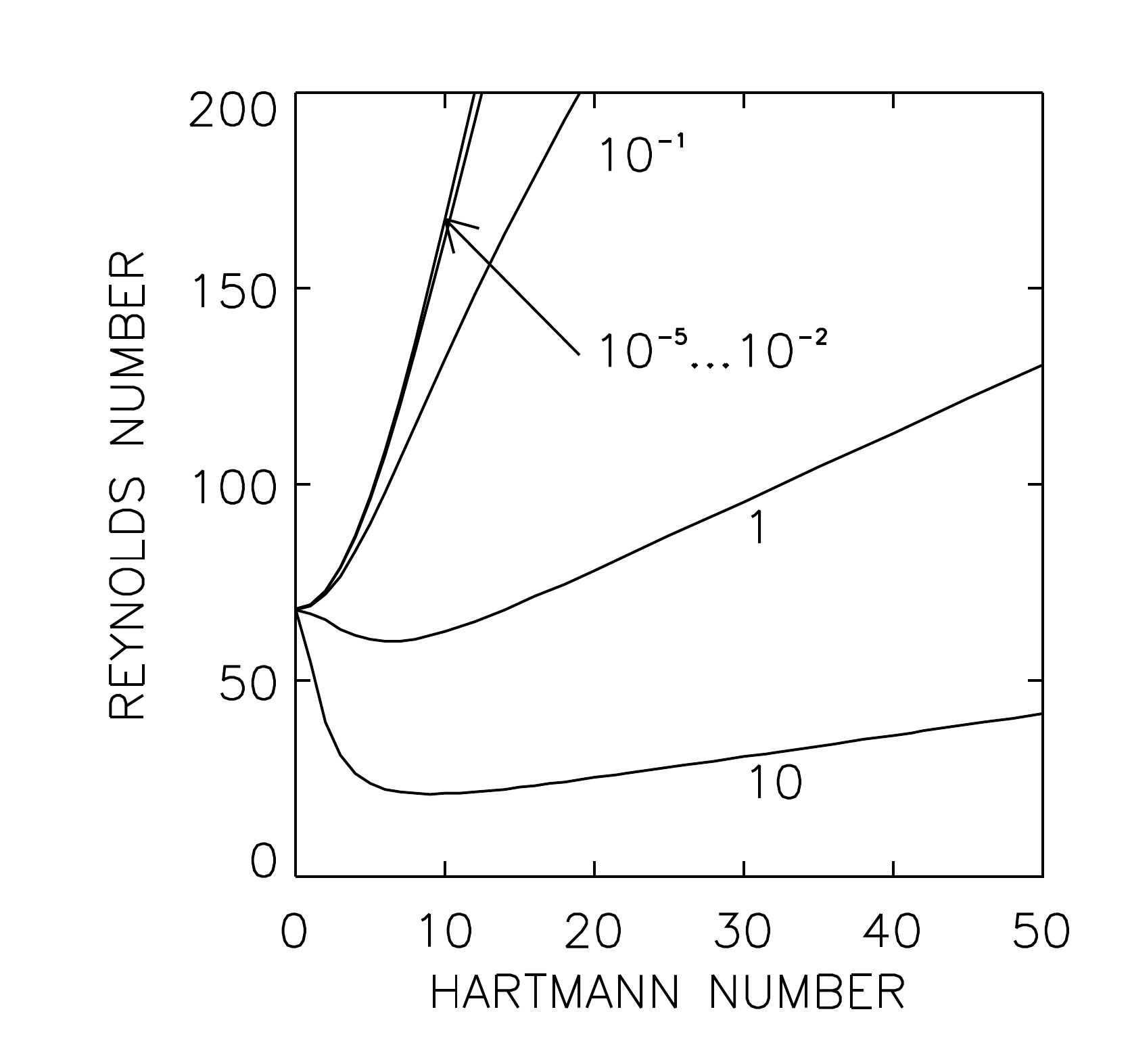}
 \caption{Axial background fields: stability maps for the axisymmetric modes Taylor-Couette flows with stationary outer cylinder for perfectly
conducting (left) or insulating (right) boundary conditions versus the magnetic Prandtl number (marked). $\Rey_0=68$ is the  eigenvalue for marginal stability of the hydrodynamic flow.  Note the existence of  magnetically induced subcritical excitation of instability for large $\Pm$ \cite{K63}. The influence of the two differing boundary conditions is here only weak. $m=0$, $\mu_\Om=0$, $\rin=0.5$, \cite{RS03}.}
 \label{f1}
\end{figure}

\subsection{Axial field}
Figure~\ref{f1} shows the neutral stability of axisymmetric modes for containers with both conducting and insulating walls with stationary outer cylinder and for fluids of various magnetic Prandtl number. These results are merely a generalization of the early findings in Ref.~\cite{R64}, where very similar methods were used to analyze the narrow-gap case $\rin=0.95$ for both types of magnetic boundary conditions. For the small magnetic Prandtl number of mercury the phenomenon of the magnetic stabilization of the centrifugal instability has already been found, which can be observed in Figs.~\ref{f1} presenting the stability maps of the axisymmetric perturbations under the presence of axial fields. The magnetic suppression of the onset of the centrifugal instability is stronger for conducting walls than for insulating walls. $\Rey_0= 68$ is the classical hydrodynamic eigenvalue for $m=0$,  $\mu_\Om=0$ and $\rin=0.5$. Note the strong difference of the bifurcation lines for $\Pm \gsim 1$ and ${\Pm}<1$. For small $\Pm$ the magnetic field always suppresses the instability so that all the given critical Reynolds numbers exceed the value 68. For $\Pm\to 0$ the stability lines no longer differ for different $\Pm$, which may be expressed as a statement that for small $\Pm$ the magnetically suppressed instability scales with $\Ha$ and $\Rey$. On the other hand, for $\Pm \gsim 1$ the resulting Reynolds numbers can be smaller than the nonmagnetic value $\Rey_0 =68$. For small Hartmann numbers (\ref{Hartmann})  the magnetic field, therefore, does {\em not} stabilize the flow. This high-$\Pm$ phenomenon -- which we shall often meet in the following -- becomes more effective for increasing $\Pm$, but in all cases it vanishes for stronger magnetic fields. One can show that the minima which appear for high $\Pm$ scale as $\Rmquer\simeq {\rm const}$, so that $\Rey\propto \Pm^{-1/2}$, leading to $\Om\propto \bar\eta$  for fixed gap width with $ \bar\eta=\sqrt{\nu \, \eta}$. The critical rotation rate of the inner cylinder  only depends on the product of $\nu$ and $\eta$.
\begin{figure}[htb]
\centering
 \includegraphics[width=5.25cm]{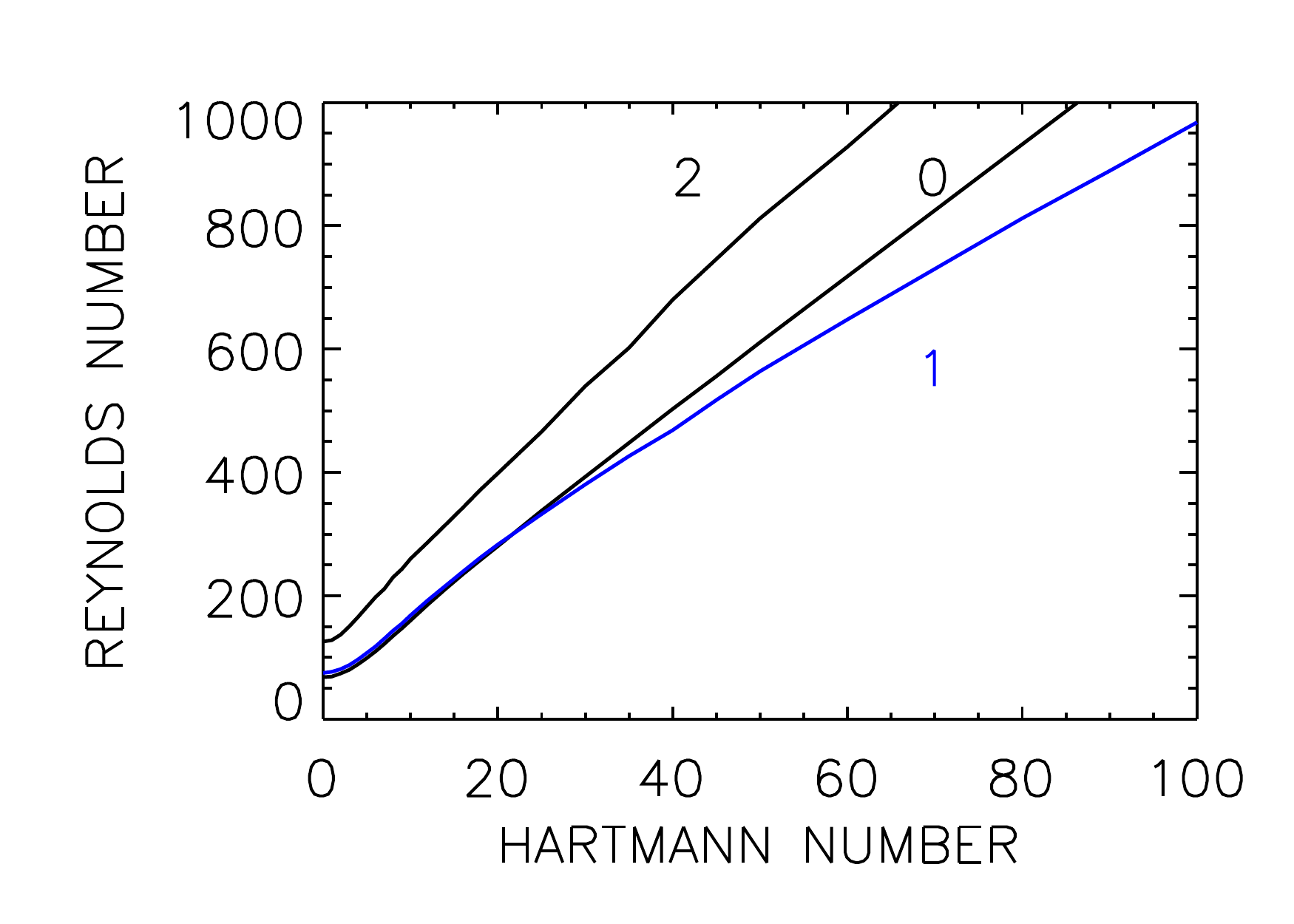}
 \includegraphics[width=5.25cm]{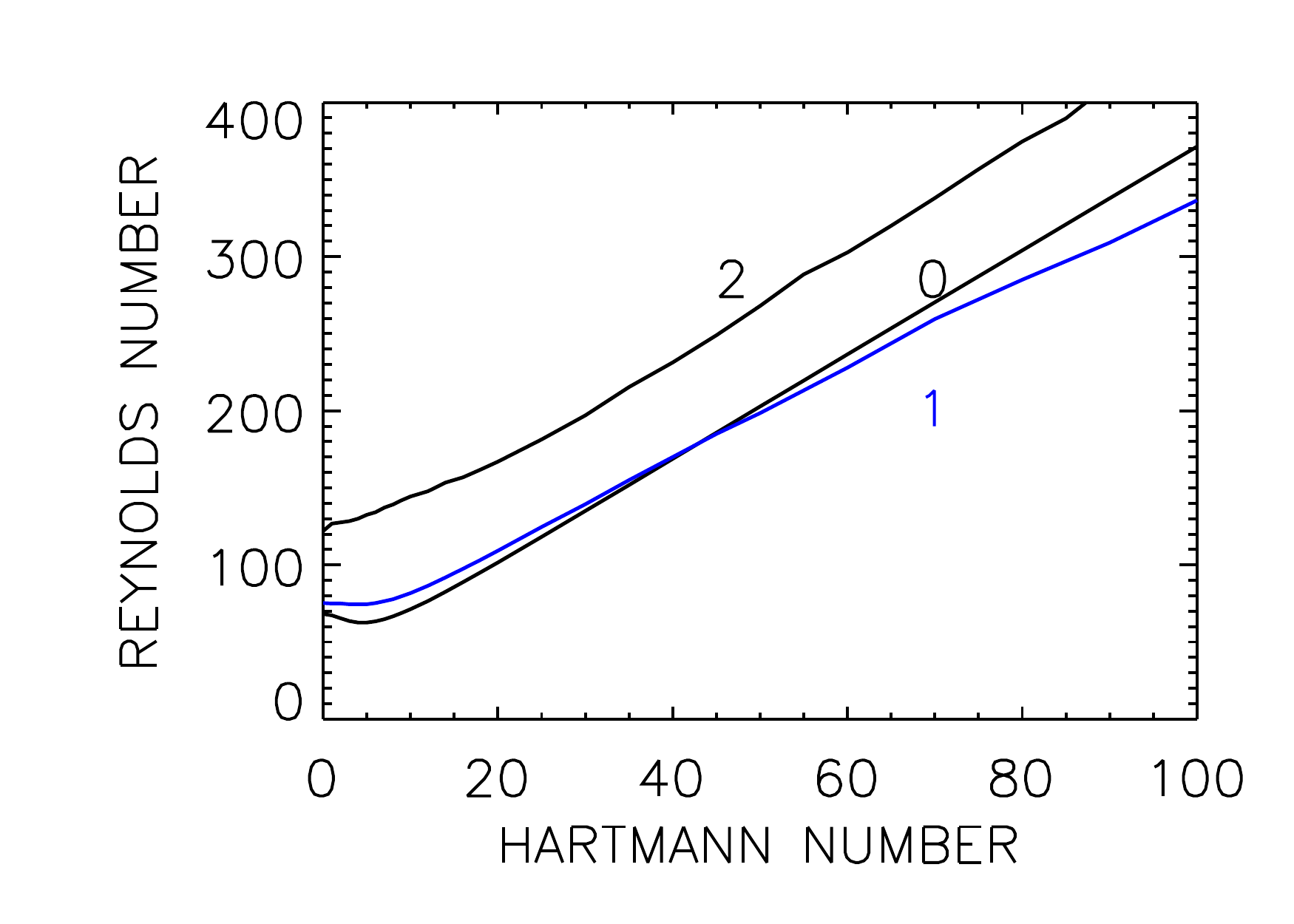}
 \includegraphics[width=5.25cm]{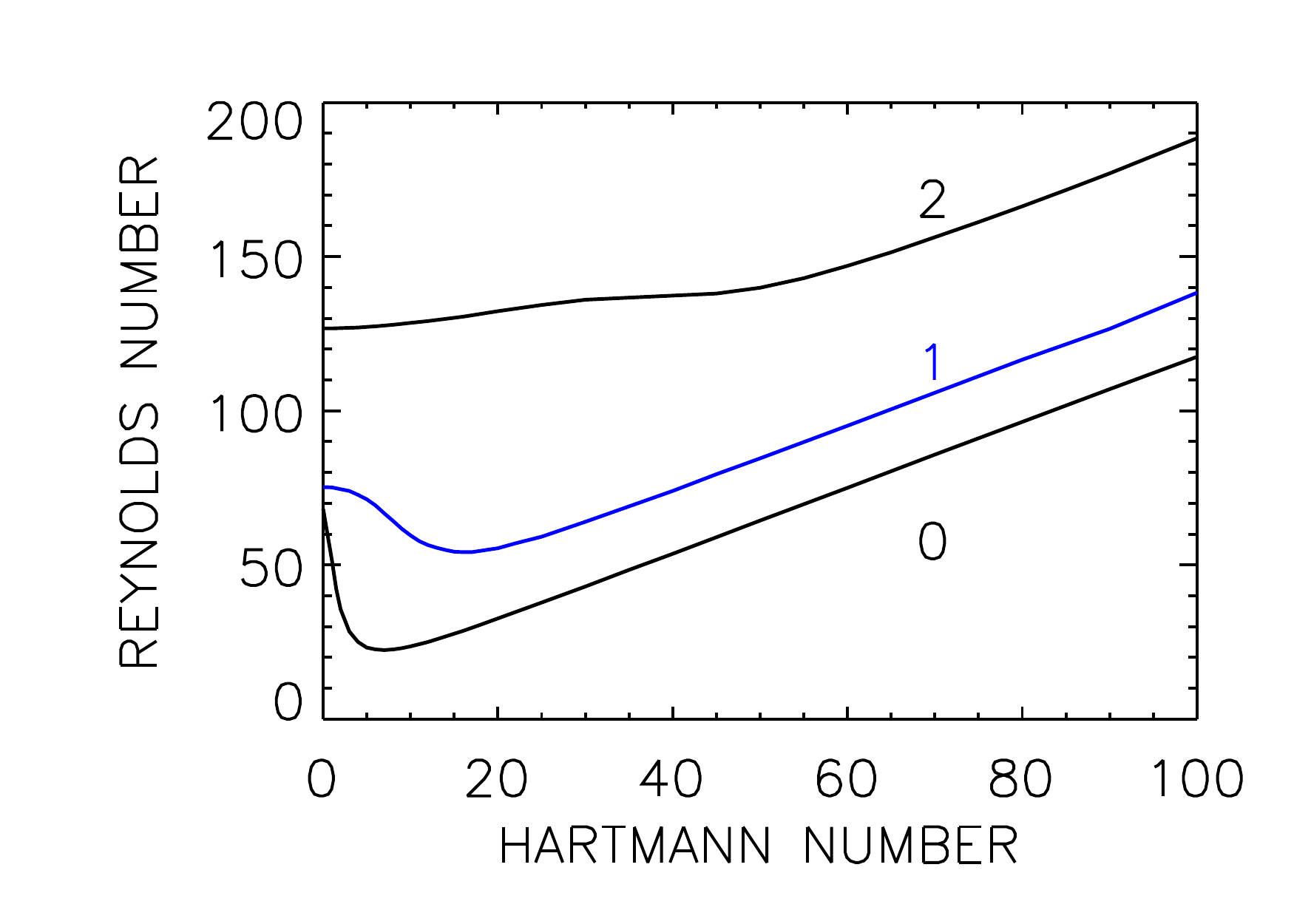}
\caption{Stability maps of the modes $m=0$, $m=1$ (blue lines)  and $m=2$  for perfectly conducting cylinders and axial fields for various $\Pm$. From left to right: $\Pm=0.1$, $\Pm=1$, $\Pm=10$. Observe  the line crossings for $m=0$ and $m=1$ for small magnetic Prandtl number and large Hartmann numbers which lead to   nonaxisymmetric modes as the preferred excitations. $\mu_\Om=0$, $\rin=0.5$.} 
 \label{f2}
\end{figure}

While Fig.~\ref{f1} only provides the bifurcation lines for the axisymmetric modes, Fig.~\ref{f2} demonstrates the excitation conditions of  the nonaxisymmetric modes $m=1$ and $m=2$ for various $\Pm$. The nonmagnetic Rayleigh instability for $m=0$ leads to $\Rey_0=68, 75, 127$ for $m=0,1,2$. Without magnetic fields the axisymmetric mode  always has  the lowest Reynolds number. However, the plots in Fig.~\ref{f2} also show crossings of the instability lines for axisymmetric and nonaxisymmetric modes of the MHD flows with $\Pm\leq 1$. Below we shall demonstrate that this phenomenon also appears for containers with rotating outer cylinder.

So far however, the crossover phenomenon only appeared in calculations using perfectly conducting boundary conditions. In these cases the magnetic suppression of the instability against axisymmetric perturbations is much stronger than for insulating boundary conditions. One can find this phenomenon also by comparison of the data in Fig.~\ref{f1}. The differences of the critical Reynolds numbers of the nonaxisymmetric modes are much smaller than the differences for axisymmetric modes so that crossovers of the lines for insulating boundary conditions cannot happen. The most striking phenomenon is that for insulating cylinders the magnetic suppression of the axisymmetric mode is much weaker than the suppression of the nonaxisymmetric modes so that  $m=0$ is always the mode with the lowest Reynolds number \cite{RS03}.
\bigskip

\subsection{Azimuthal field}\label{azifield}
\begin{figure}[htb]
\centering
 \includegraphics[width=8cm]{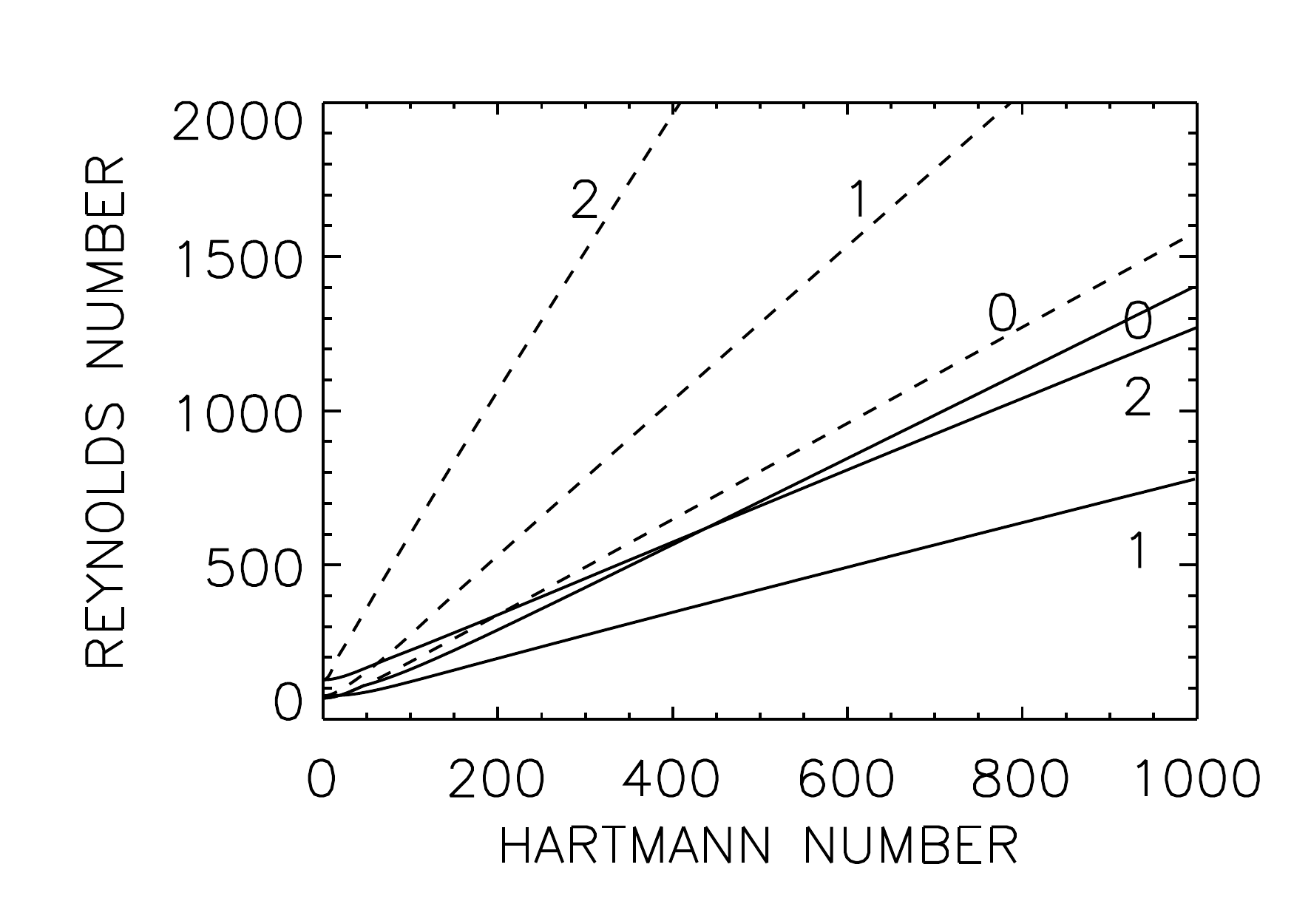}
 \includegraphics[width=8cm]{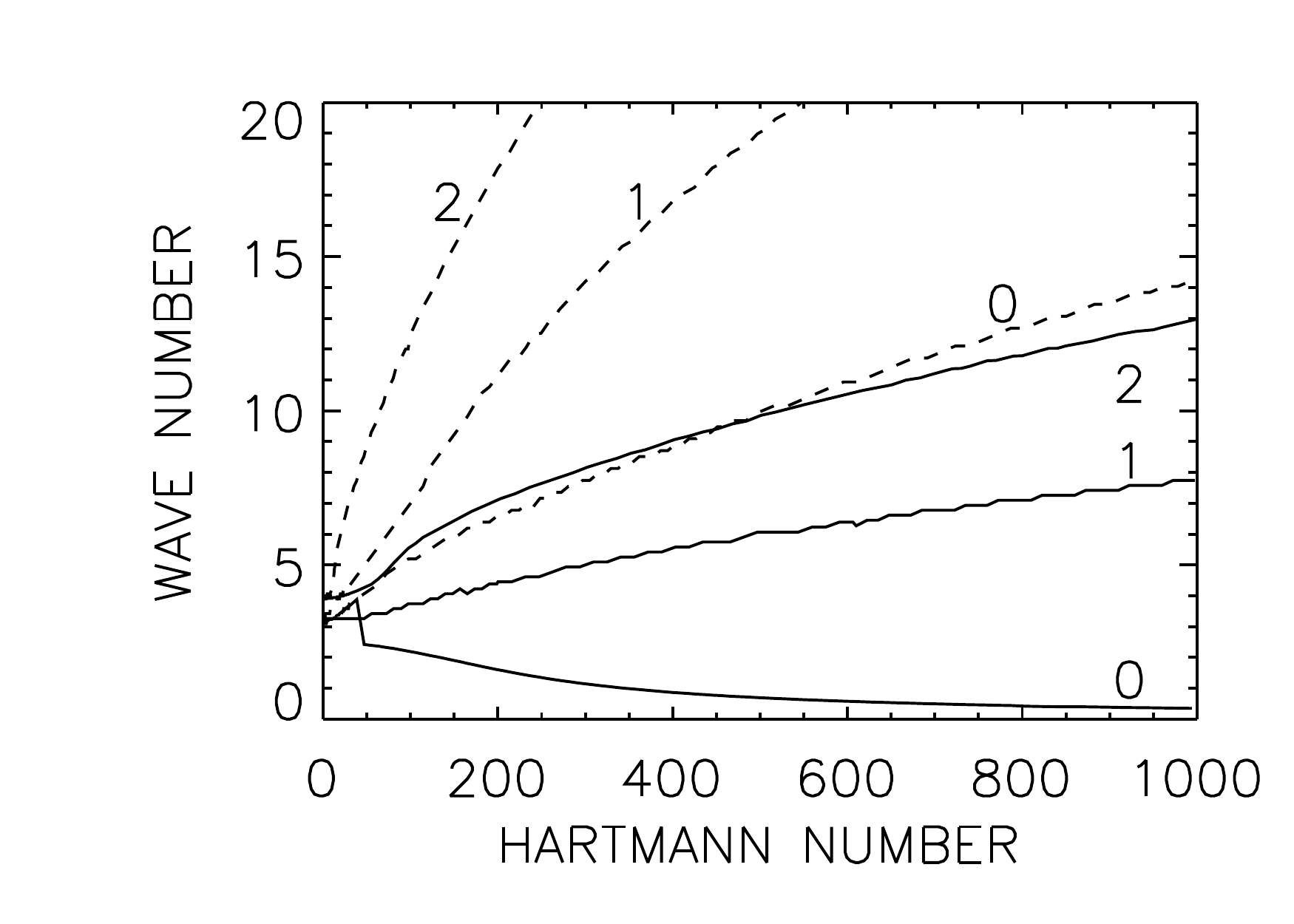}
 \caption{Azimuthal  background fields: critical Reynolds numbers (left) and the corresponding axial wave numbers (right) for the modes with $m=0$, $m=1$ and $m=2$ of the flow with stationary outer cylinder subject to  azimuthal fields which are current-free between the cylinders for $\Pm=1$ (solid) and $\Pm=10^{-5}$ (dashed). $\rin=0.5$, $\mu_\Om=0$, $\mu_B=0.5$. Perfectly conducting cylinders.}
 \label{restamri}
\end{figure}

We next consider Taylor-Couette flows with a stationary outer cylinder under the influence of an azimuthal magnetic field. Ref.\ \cite{E58} showed that current-free toroidal fields ($B_\phi\propto1/R$) suppress the axisymmetric Taylor vortices, at least in the narrow-gap limit, with conducting boundaries, and dissipative fluids. This result holds true even if the narrow-gap approximation is not made.  Allowing electric currents to flow within the fluid though can dramatically change the results.
In the following we shall apply the two extreme azimuthal magnetic fields (with $a_B=0$ and with $b_B=0$) to the rotation profile having a stationary outer cylinder, and find completely different classes of solutions.
In the {\em first case} $B_\phi$ may be assumed as current-free, i.e.~$a_B=0$ or $\mu_B=0.5$ if $\rin=0.5$. Figure \ref{restamri} (left) gives the resulting critical Reynolds numbers as functions of the Hartmann number  (\ref{Hartmannin}) for the modes with $m=0$, $m=1$ and $m=2$. The three corresponding Reynolds numbers $\Rey_0$ for the modes are (again) 68, 75 and 127 for $m=0,1,2$. The above statement for ideal fluids is confirmed that current-free fields always suppress the axisymmetric modes as shown here for $\Pm=1$ and $\Pm=10^{-5}$. The suppression is stronger for smaller magnetic Prandtl numbers.

It is obvious that strong differential rotation  leads to a suppression of the instability, as nonuniform rotation always suppresses nonaxisymmetric modes for sufficiently high electric conductivity. On the other hand, weak differential rotation may support the excitation of nonaxisymmetric modes in contrast to rigid rotation. A Taylor-Couette flow with stationary outer cylinder may easily serve as a model to study such problems. 

From Fig.~\ref{restamri}  we take  that even current-free azimuthal fields suppress nonaxisymmetric modes. The stabilizing action of the field is stronger on nonaxisymmetric rather than on axisymmetric modes. This finding complies with the above mentioned idea that differential rotation strongly amplifies the dissipation of nonaxisymmetric modes. Note that the calculated lines of neutral stability of the mode $m=0$ hardly differ for $\Pm=1$ and $\Pm=10^{-5}$. The eigenvalues along the line of neutral stability of the axisymmetric modes, therefore, appear to scale with $\Rey$ and $\Ha$ for $\Pm\to 0$. In both cases the magnetic field simply suppresses the axisymmetric mode as predicted by Eq.~(\ref{abc}). The results, however, for the nonaxisymmetric modes and for $\Pm=1$ are surprising with respect to the line crossings in the left panel of Fig.~\ref{restamri}. For $\Ha<18$ the lowest Reynolds number for instability is for $m=0$ but for larger values $m=1$ is preferred. For higher values of $\Ha$ even the $m=2$ mode overcomes the axisymmetric solution. The same phenomenon might happen for small $\Pm$ but for much higher Hartmann numbers (not shown). We thus find again crossover effects for the instability of the rotation law with stationary outer cylinder, quite similar to the interaction with axial fields (see Fig.~\ref{f2}). Magnetically influenced Taylor-Couette flows -- if the field is strong enough -- appear to form nonaxisymmetric structures much easier than nonmagnetic flows. We shall see below that the nonaxisymmetry of the instability pattern shown by Fig.~\ref{restamri} (left) proves to be a characteristic property also of all Taylor-Couette flows subject to azimuthal fields formed by stable rotation with no electric current and/or no rotation with electric current. We call phenomena related to the first case the Azimuthal MagnetoRotational Instability (AMRI) and the second case the Tayler Instability (TI).

Another finding concerns the axial wavelengths of the unstable modes. Under the magnetic influence they become shorter and shorter except for $m=0$, $\Pm=1$ and $\Ha\geq 40$. For this curve the axisymmetric Taylor vortex as the mode with the lowest Reynolds number (see \cite{K93}) develops from nearly spherical cells to cells strongly elongated in the axial direction under the influence of the current-free azimuthal magnetic field. The pattern becomes two-dimensional for $\Ha\to\infty$. This surprising effect disappears for higher mode numbers $m$ and for smaller magnetic Prandtl numbers. The real part of the eigenfrequency $\omega$, which  for axisymmetric modes often vanishes,  has here finite values, indicating that the unstable patterns oscillate or migrate in the azimuthal or the axial direction.
\medskip

\begin{figure}[htb]
\centering
 \includegraphics[width=9cm]{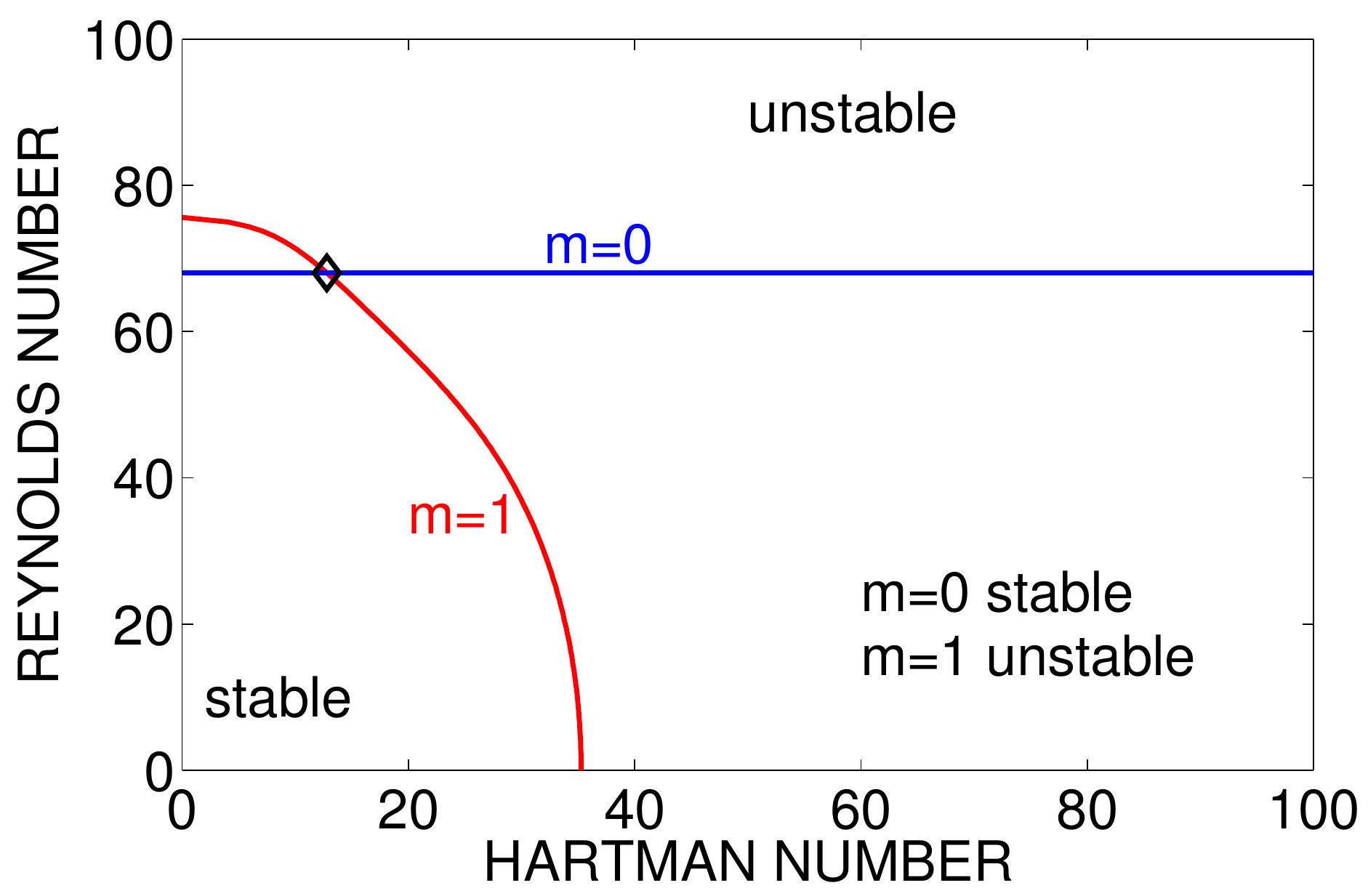}
 \caption{Critical Reynolds numbers as function of the Hartmann number for the modes with $m=0$ (blue) and $m=1$ (red) for uniform axial electric current and stationary outer cylinder. The diamond indicates the eigenvalues where the symmetry of the modes with the lowest Reynolds number changes.  
  $\rin=0.5$, $\mu_\Om=0$, $\mu_B=2$, $\Pm=1$. Perfectly conducting cylinders. Adapted from \cite{GR15}.}
 \label{f9b}
\end{figure}

For the { second case} the combination of differential rotation and a magnetic field due to a uniform axial electric current is considered ($b_B=0$). We shall find a completely different situation with respect to the axisymmetry of the solutions. Figure \ref{f9b}  shows that the axisymmetric mode is not influenced by the magnetic field, in agreement with the consequences of Eq.~(\ref{abc}). However, already for Hartmann numbers of order 10, the $m=1$ mode crosses the line for $m=0$. For stronger fields the most easily excited azimuthal mode is that with $m=1$. For $\Ha_0 = 35.3$ the line for the neutral stability of the mode $m=1$ even crosses the abscissa defined by $\Rey=0$. The uniform axial electric current (the `$z$-pinch') becomes unstable even without any rotation (TI). We shall stress below that the characteristic Hartmann number $\Ha_0$ for $\Rey=0$ never depends on the magnetic Prandtl number $\Pm$. The red line in this plot depend on the magnetic Prandtl number but the Hartmann number $\Ha_0$ for stationary cylinders does not (see Section \ref{TI}).  We shall meet the value $\Ha_0=35.3$ for the stationary $z$-pinch inside perfectly conducting cylinders several times in this paper. The corresponding value for insulating cylinders is $\Ha_0=28.1$.

One can also show that  the growth rates of the $m=0$  modes of the flow field for various magnetic field strengths are identical.  For $m=1$ they become positive  for $\Rey\geq 75$ for weak fields, but for sufficiently  strong fields they are already positive for $\Rey=0$. At the vertical axis ($\Rey=0$) the growth rates increase with increasing $\Ha$ so that for large fields the growth rate scales with the \A~frequency $\Om_{\rm A}$  in perfect agreement with  Fig.~\ref{f9b}.

For information about the instability pattern and the energies which are stored in the various modes and in the flow and field components, one needs a  code solving the nonlinear MHD equations. To this end a spectral element code has been developed from the hydrodynamic code of Fournier et al. \cite{FB04}. It works with an expansion of the solution in azimuthal Fourier modes. A set of meridional problems results, each of which is solved with a Legendre spectral element method as in Ref.~\cite{DF02}. Between 8 and 16 Fourier modes are used. The polynomial order is varied between 10 and 16, with four or five elements in radial 
direction where the largest resolution is used for the smallest magnetic Prandtl numbers. The number of 
elements in axial direction ensures that the spatial resolution is the same as for the radial 
direction.
At the inner and outer walls perfect conducting boundary conditions are applied together with no-slip 
conditions for the flow.  With a semi-implicit approach consisting of second-order backward differentiation and third-order Adams-Bashforth for the nonlinear forcing terms time-stepping is done with second-order accuracy. Periodic conditions in the axial direction are applied to minimize finite size effects. With the aspect ratio $\Gamma=8$  (the height of the numerical domain in units of the gap width) all excitable modes in the analyzed parameter region fit into the system.
\begin{figure}[htb]
\centering
 \includegraphics[width=5cm]{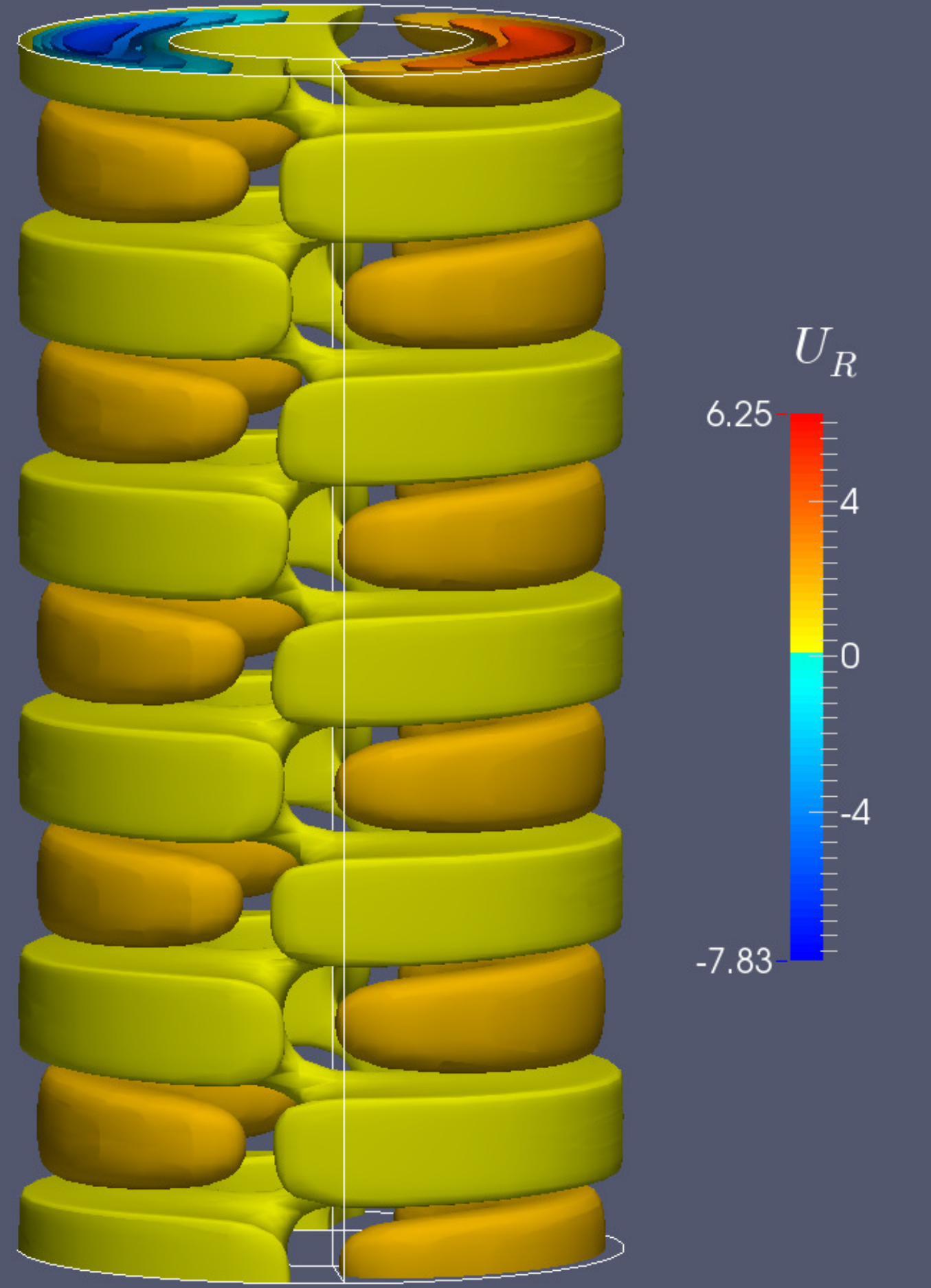}
 \includegraphics[width=5cm]{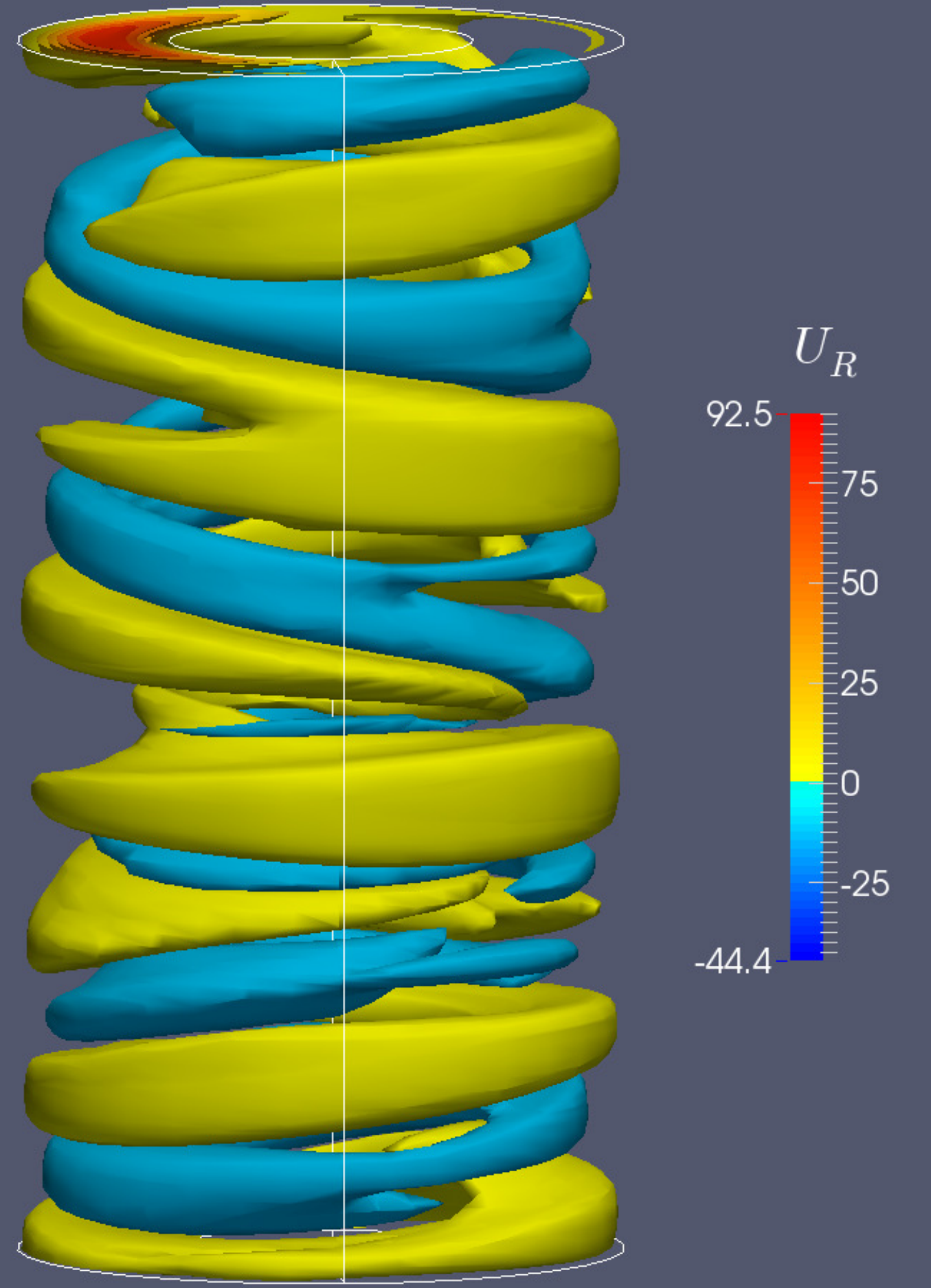}
 \caption{Isolines of the radial flow component measured as Reynolds numbers $u_R d/\nu$ for a $z$-pinch with $\Rey=0$ (left)  for pure Tayler instability and  with $\Rey=350$ (right). The numerical values of the maxima and minima are shown by the color bars (the negative signs in the left panel are hidden by the yellow colors). The TI of the stationary flow produces a single nondrifting   mode with $m=\pm 1$ (left panel). For $\Rey=350$ also the axisymmetric mode is visible. $\rin=0.5$, $\Ha=80$, $\mu_\Om=0$, $\mu_B=2$, $\Pm=1$. Perfectly conducting boundaries.}
 \label{f9c}
\end{figure}

As a first application of this code, the right panel of Fig.~\ref{f9c} shows the patterns of the radial flow component $u_R$ for uniform axial electric current ($\mu_B=2$), rapid rotation ($\Mm= 4.4$) and stationary outer cylinder \cite{GR07}. As expected, the instability is highly nonaxisymmetric. The axisymmetric mode also exists but does not dominate the structure which as a whole drifts in the positive azimuthal direction. Clearly, the pattern with differential rotation is of the mixed-mode type, but without rotation it is formed by a single nondrifting mode $m=1$ (left panel). There is no axisymmetry in the solution as Fig.~\ref{f9b} suggests, and the complete pattern is stationary. 
\begin{figure}[htb]
\centering
 \includegraphics[width=7.4cm]{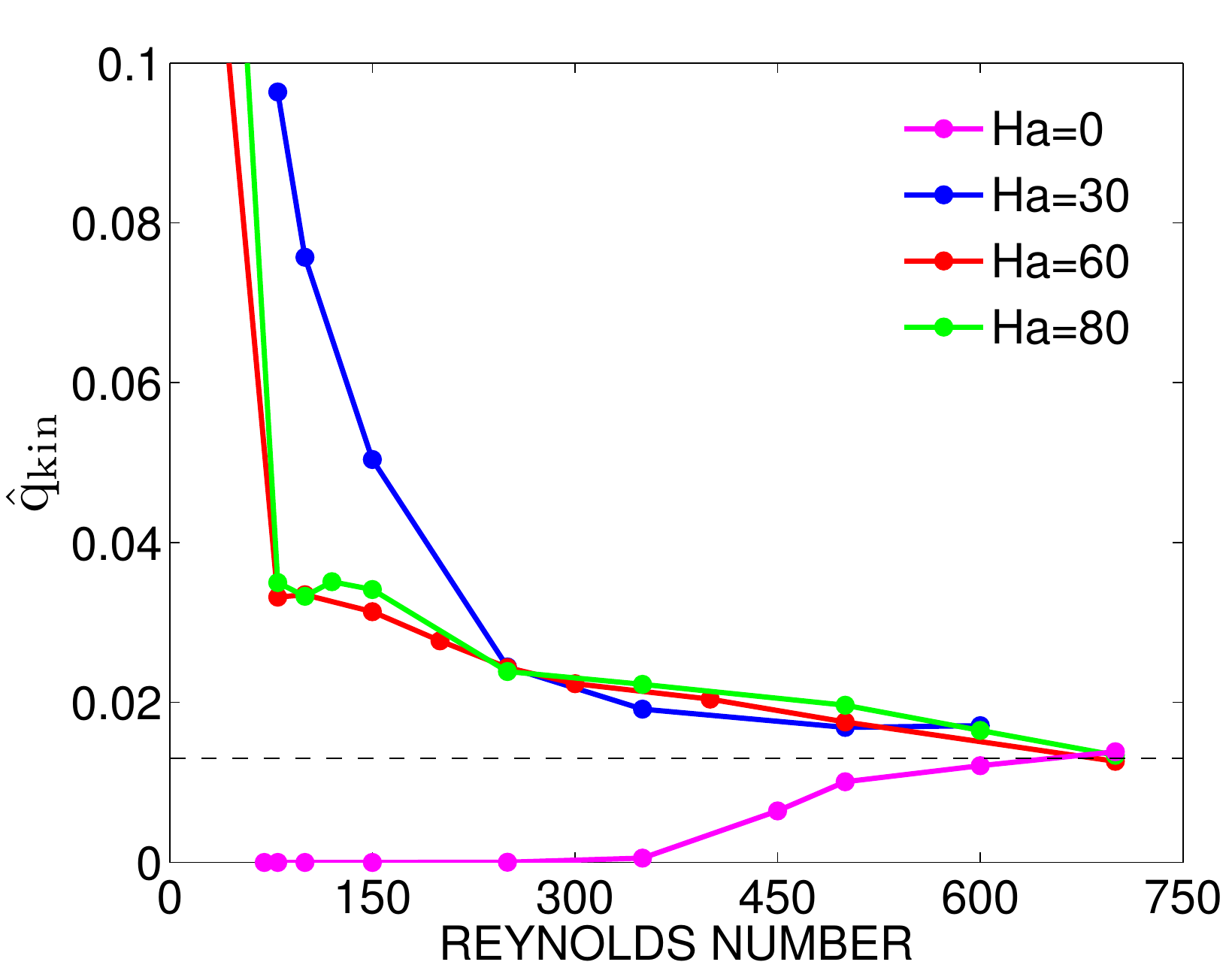}
 \includegraphics[width=8cm]{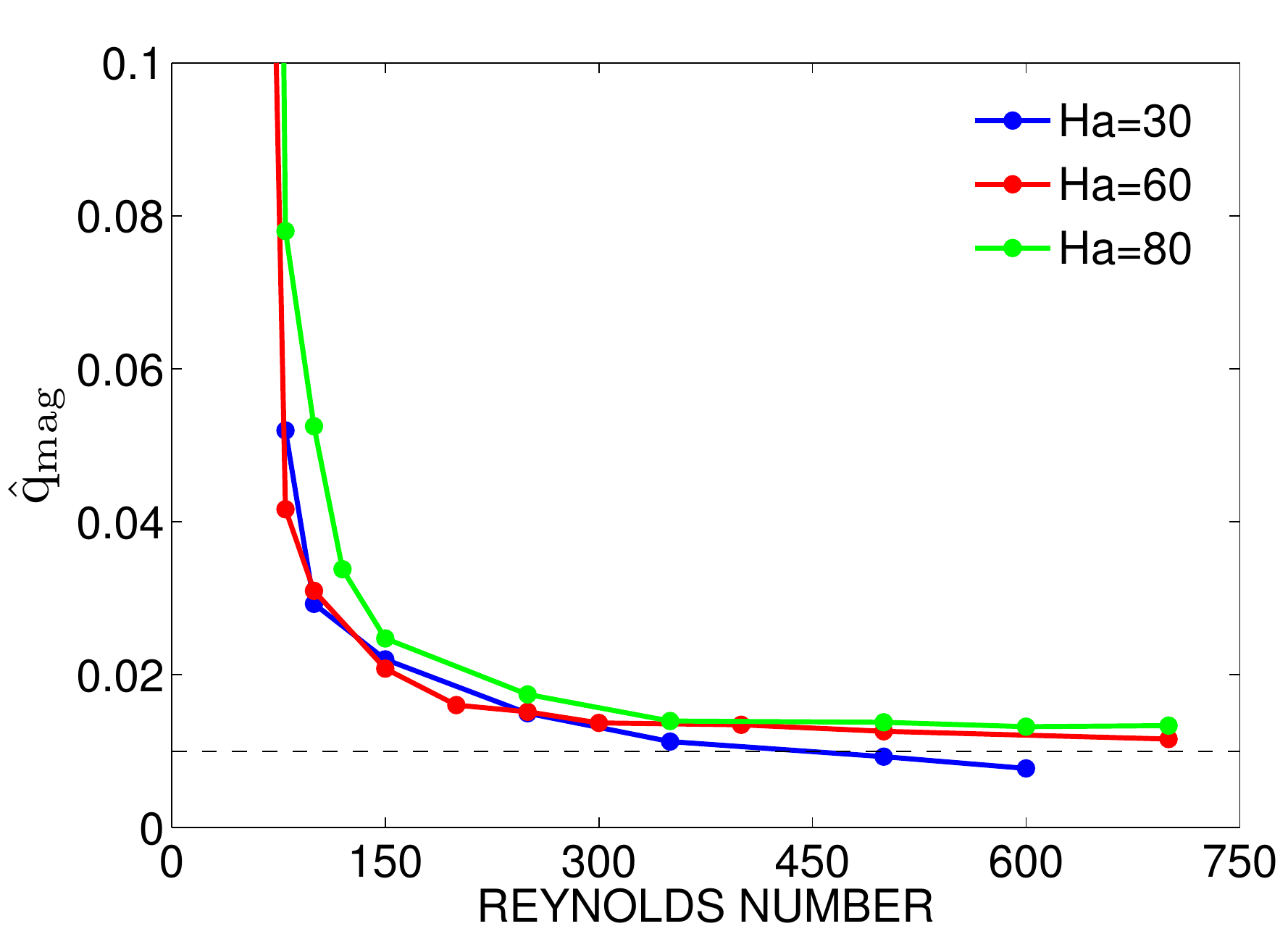}
 \caption{The kinetic (left) and magnetic (right) energy of the nonaxisymmetric modes normalized with the centrifugal energy $\Om_{\rm in}^2 d^2$ for various Hartmann numbers. The pink line in the left plot gives the kinetic energy  of the nonmagnetic flow. $\rin=0.5$, $\mu_\Om=0$, $\mu_B=2$, $\Pm=1$. Perfectly conducting boundaries.}
 \label{f9d}
\end{figure}

For the flow with stationary outer cylinder and uniform axial electric current the kinetic and magnetic energy (normalized with the centrifugal energy $d^2 \Om_{\rm in}^2$) have also been computed. The question is how much centrifugal energy is stored in the nonaxisymmetric modes of flow and field and which sort of energy dominates. We write
\begin{equation}
\label{result}
{\langle \vec{u}^2 \rangle} = {\hat{\rm q}_{\rm kin}} \ \Om_{\rm in}^2 d^2, \ \ \ \ \ \ \ \ \ \ \ \ \ \ \ \ \ \ \ \ \ \ \ \ \frac{\langle \vec{b}^2 \rangle}{\mu_0\rho} = {\hat{\rm q}_{\rm mag}} \ \Om_{\rm in}^2 d^2
\end{equation}
and find the numerical values $\hat{\rm q}_{\rm kin}\simeq 0.015$ and $\hat{\rm q}_{\rm mag}\simeq 0.012$ for very rapid rotation (Fig.~\ref{f9d}, dashed lines). For $\Mm\gg 1$ the coefficients ${\hat{\rm q}_{\rm kin}}$ and ${\hat{\rm q}_{\rm mag}}$ no longer depend on the Reynolds number. The faster the rotation of the inner cylinder, therefore, the more energy is stored in the nonaxisymmetric modes of the instability.  Both energies can thus easily be expressed by the global energy $\Om^2_{\rm in}d^2$. 
 
A very similar formulation can be used for the nonmagnetic Taylor-Couette flow. The pink curve in the left panel of Fig.~\ref{f9d} gives the kinetic energy in the nonaxisymmetric modes of the hydrodynamic Taylor-Couette flow. Clearly, it starts at $\Rey=75$ and grows for faster rotation. Surprisingly, for very large Reynolds numbers the energy approaches (from below) the kinetic energy values of the MHD pattern for rapid rotation. Hence, for magnetized rapid rotators (with $\Pm=1$) the energies in the hydromagnetic modes are continuously reduced by increasing rotation until they both reach just the same value as the hydrodynamic Taylor-Couette flow produces. Figure \ref{f9d} also demonstrates that for $\Pm=1$ the kinetic and magnetic energies are almost in equipartition. We shall later see that the magnetic energy in such simulations only exceeds the kinetic energy for large $\Pm$.

\section{Standard magnetorotational instability (MRI)}\label{standardMRI}
So far we have discussed the stability of the Couette flows (\ref{1.1}), which by themselves can be hydrodynamically unstable. If the fluid is electrically conducting and an axial magnetic field is applied then for small $\Pm$ the critical Reynolds number increases with increasing magnetic field. Chandrasekhar explained the experimental data of Donnelly \& Ozima for narrow gaps and with $\Pm=0$ by a magnetic suppression of the Rayleigh instability (see Fig.~\ref{donelly}, \cite{C61,DO60}).
 
The hydrodynamic Taylor-Couette flow is stable if its angular momentum increases with radius, but according to (\ref{veli}) the hydromagnetic Taylor-Couette flow is only stable if the angular velocity itself increases with radius. This remains true also for nonideal fluids subject to axial magnetic fields. Weak magnetic fields reduce the critical Reynolds number for hydrodynamically unstable flows, and destabilize the otherwise hydrodynamically stable flow for $\rin^2 < \mu_\Om < 1$.

As we shall demonstrate, for small $\Pm$ and given Hartmann number  (\ref{Hartmann}) the Reynolds numbers for neutral stability scale as $1/\Pm$ for hydrodynamically stable flows, so that it is the magnetic Reynolds number $\Rm$ which controls the instability. Because of the high value of the molecular magnetic resistivity $\eta$ for liquid metals (Table \ref{t2}) it is not easy to reach magnetic Reynolds numbers of the required order of 10. This is the reason why the standard MRI has not yet been unambiguously observed experimentally in the laboratory \cite{SISAN,NS10}.
\begin{table}[htb]
\caption{Parameters of the liquid metals as conducting fluids, where $\bar\eta=\sqrt{\nu \, \eta}$. From \cite{C61,NP02}.}
 \begin{center}
\begin{tabular}{@{}llllcr@{}}
\hline
&&&&&\\[-1.5ex]
 & $\rho$ [g/cm$^3$] & $\nu$ [cm$^2$/s] & $\eta$ [cm$^2$/s] &
 $\bar\eta$ [cm$^2$/s] &${\rm Pm}$\\[1ex]
\hline
&&&&&\\[-1.5ex]
mercury & 5.4 & 1.1$\cdot 10^{-3}$ & 7600 & 2.9 & 1.4$\cdot10^{-7}$\\
gallium & 6.0 & 3.2$\cdot 10^{-3}$ & 2060 & 2.6 & 1.5$\cdot 10^{-6}$\\
galinstan (GaInSn) & 6.4 & 3.4$\cdot 10^{-3}$ &2428 & 2.9 & 1.4$\cdot 10^{-6}$ \\
sodium & 0.92 & 7.1$\cdot 10^{-3}$ & 810 & 2.4 & 0.88$\cdot 10^{-5}$\\[1ex]
 \hline
 \end{tabular}
 \label{t2}
 \end{center}
\end{table}
\subsection{Potential flow}\label{carlos}
From all possible Couette flows only those with vanishing $a_\Om$ form an irrotational vortex with $\rot {\vec{U}}=0$. For this flow the specific angular momentum $R^2\Om$ is uniform in the radial direction. The rotation profile with $\mu_\Om =\rin^2$ (hence $\mu_\Om=0.25$ for $\rin=0.5$) is called the Rayleigh limit while the associated flow is called the `potential flow'. It plays an important role in the general theory of Taylor-Couette flows. In pure hydrodynamics, negative values of the radial gradient of $R^2\Om$ are destabilizing and positive values are stabilizing. One might expect that instabilities subject to uniform $R^2\Om$ should be easiest, i.e.~the excitation needs minimal Reynolds numbers.

\begin{figure}[htb]
\centering
 \includegraphics[width=7.0cm]{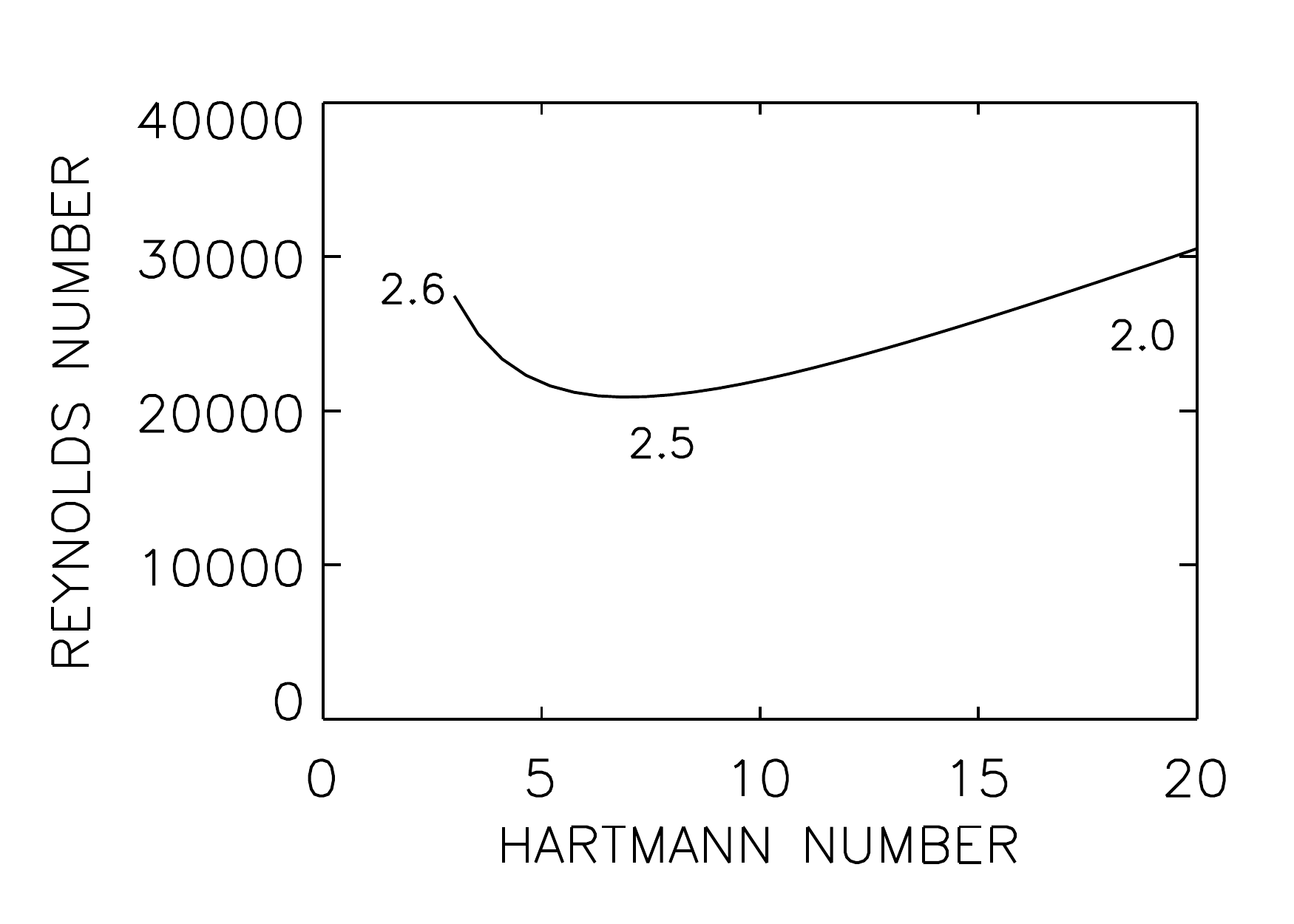}
 \includegraphics[width=7.0cm]{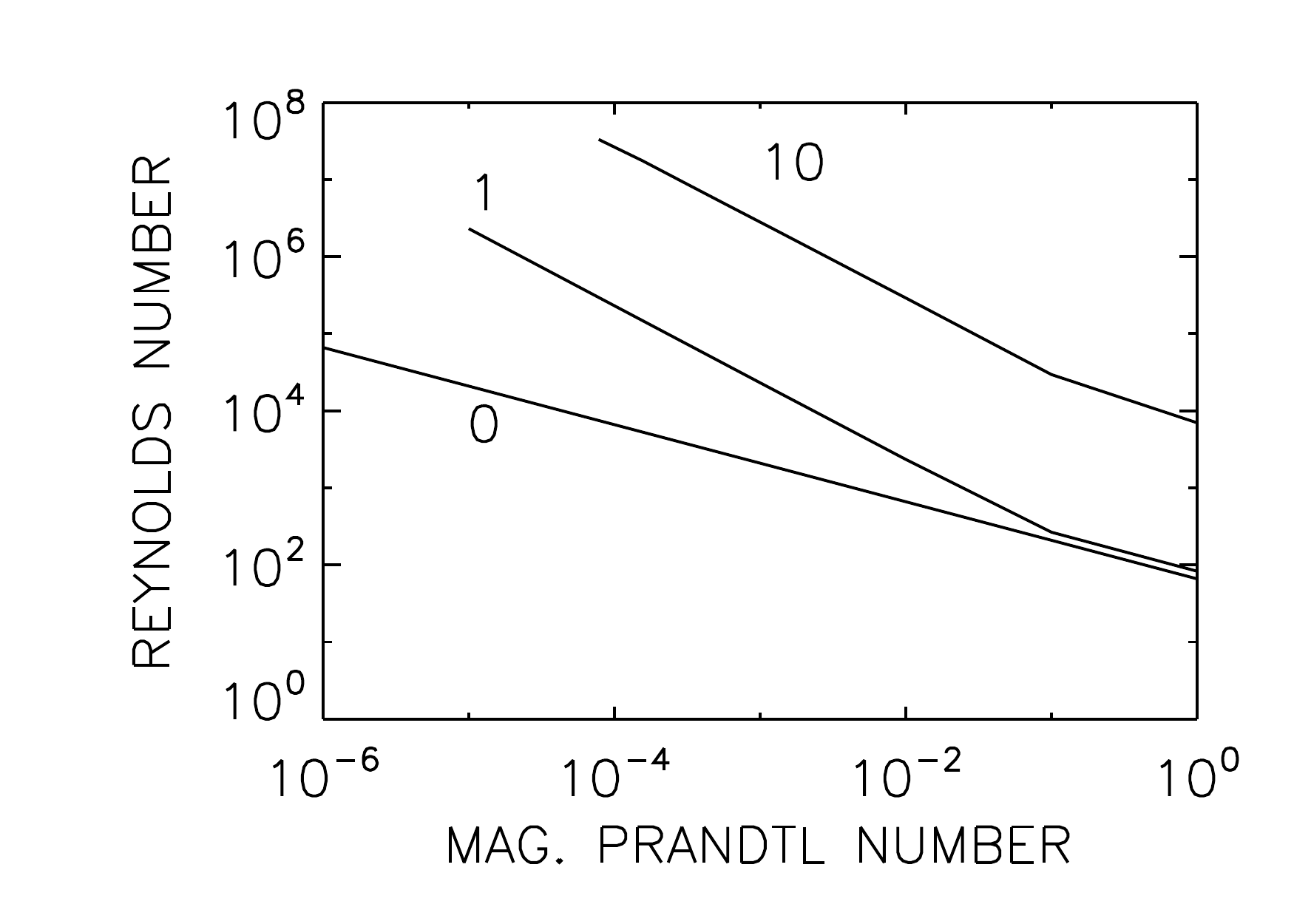}
 \caption{Stability maps for  standard MRI of the potential flow   for $m=0$ with $\Pm=10^{-5}$ (left) and the minimum Reynolds numbers  for $m=0$, $m=1$ and $m=10$ as functions of $\Pm$ (right). The curve in the left panel shows the axial wave numbers normalized with the characteristic scale $R_0$. $\rin=0.5$, $\mu_\Om=0.25$. Perfectly conducting boundaries. Adapted from \cite{RS08}.}
 \label{f10}
\end{figure}

For the potential flow a particular scaling of the solutions of the MHD equations  (\ref{p22}) - (\ref{p27}) with axial background field exists with respect to the magnetic Prandtl number. The quantities $u_R, u_z, b_R$ and $b_z$ scale as ${\rm Pm}^{-1/2}$ while $u_\phi, b_\phi, k$ and $\Ha$ scale as ${\rm Pm}^0$. Then for the axisymmetric modes it follows that  the minimum Reynolds number scales as $1/\sqrt{\Pm}$, so that  $\Rmquer$=const, independent of the boundary conditions \cite{WB02}. One has thus only to solve the equations for $\Pm=1$ and simultaneously knows the solutions for all other $\Pm$. The minimum Reynolds number for $\Pm=1$ is 66 at a Hartmann number ${\Ha}\simeq 7$ \cite{RS03}. Hence, the small minimum value of ${\Rey}=22,248$ for $\Pm=10^{-5}$ (liquid sodium) is needed (together with $\Ha=7$) which seems to be very promising for experiments (Fig.~\ref{f10}, left). Below we shall argue that significant problems with accuracy prevent its realization thus far.

The axial wave numbers marked  in the left panel of  Fig.~\ref{f10}  demonstrate the increase of the axial scales with increasing magnetic field in accordance with  the  magnetic analog  of the Taylor-Proudman theorem. It is indeed known from early experiments \cite{M68,KT71} and  theoretical studies \cite{M67,R74,Ra74,S76} that the correlation lengths in MHD turbulence become longer in the field direction the stronger the applied background field is. On the other hand, if the wave number normalized with the gap width is smaller than $\pi$ (for our standard container) then the axisymmetric vortices in the Taylor-Couette flow  are axially aligned. The given  numbers in the plot predict that the cells become longer and longer for growing $\Ha$.  

The right panel of Fig.~\ref{f10} also demonstrates that the simple scaling of the Reynolds number with $\Pm^{-1/2}$ for the potential flow only exists for the axisymmetric mode. For the modes with $m=0$, $m=1$ and $m=10$ the dependencies of the characteristic Reynolds numbers on the magnetic Prandtl number are plotted. One finds that for $\Pm\to0$ the nonaxisymmetric modes follow a much steeper scaling with $\Pm$ than the axisymmetric mode. For $\Pm$ of order unity, however, the various Reynolds numbers for excitation of $m=0$ and $m=1$ do not differ much, as is also true down to $\Pm\simeq 0.1$.
\begin{figure}[h]
\centering
 \includegraphics[width=4.5cm]{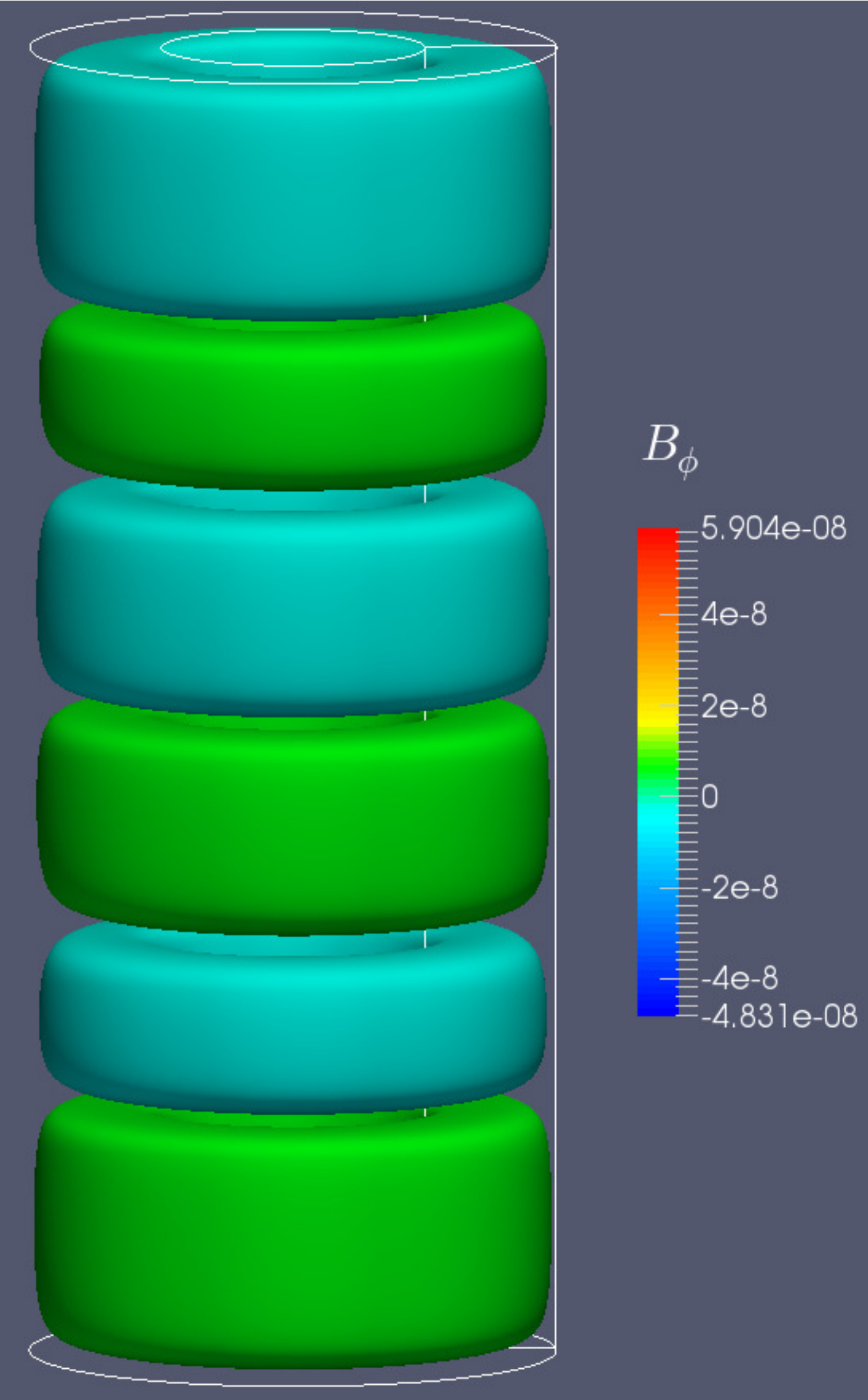}
 \includegraphics[width=4.5cm]{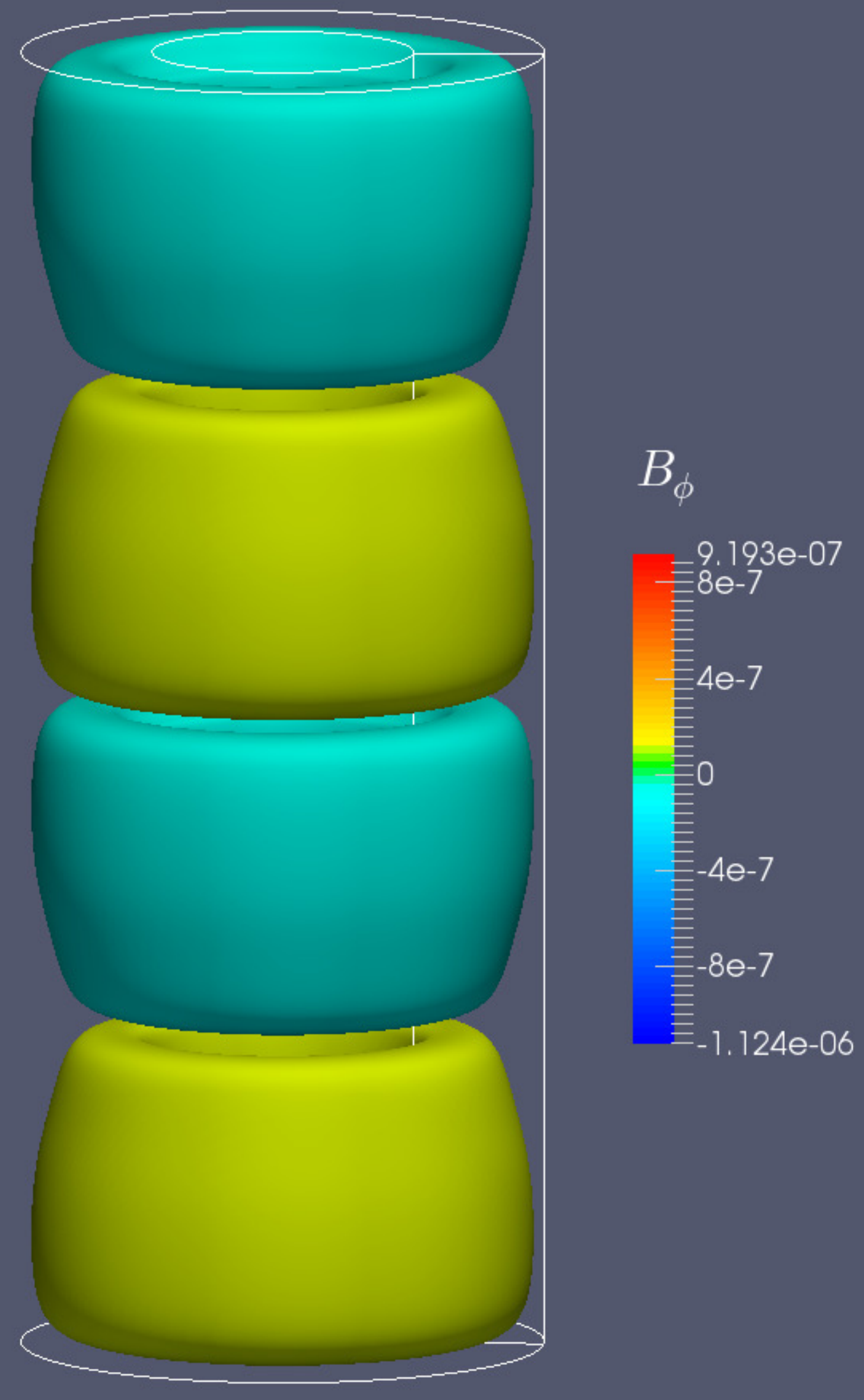}
 \caption{Standard MRI of the potential flow subject to axial background field. Isolines of the azimuthal field  component (normalized with  $B_0$) are shown  for $\Ha=7$ (left)   and  $\Ha=50$ (right). Only the axisymmetric mode   is excited despite the high Reynolds number. $\Gamma=10$, $\rin=0.5$, $\Rey=50,000$, $\mu_\Om=0.25$,  $\Pm=10^{-5}$. Insulating boundaries, see Section \ref{nonsim}.  }
 \label{fpot1}
\end{figure}
This is not true  for $\Pm\ll1$. Figure \ref{fpot1} demonstrates for  two different Hartmann numbers that for the supercritical  Reynolds number  $\Rey=50,000$  only the axisymmetric mode is excited. All modes with $m>0$  decay. The instability pattern even remains axisymmetric for similar examples with $\Rey=10^5$ (not shown). The  models  prove the axisymmetry of standard MRI also for very  small magnetic Prandtl numbers such as $\Pm=10^{-5}$, and this   for not too high Reynolds numbers.  Observe that the resulting normalized magnetic perturbations are only weak compared with the background field. The two given models  with different Hartmann numbers  may also serve  to probe the prediction  that  the axial wavelength   increases with increasing magnetic field. This is indeed the case.   

Only slightly beyond the Rayleigh limit (e.g. for $\mu_\Om=0.255$) and  the other parameters left unchanged,  the numerical simulations no longer yield standard MRI. This is a direct  consequence of different scalings of the solutions for small $\Pm$ for different rotation laws  (see Section \ref{quasikep}). 
\subsection{Quasi-Keplerian flow}\label{quasikep}
A quasi-Keplerian Couette flow may be defined by requiring that the cylinders rotate like planets following the Kepler law $\Om\propto R^{-3/2}$. This becomes $\mu_\Om=\rin^{1.5}=0.35$ for $\rin=0.5$.
\begin{figure}[htb]
\centering
 \includegraphics[width=7.0cm]{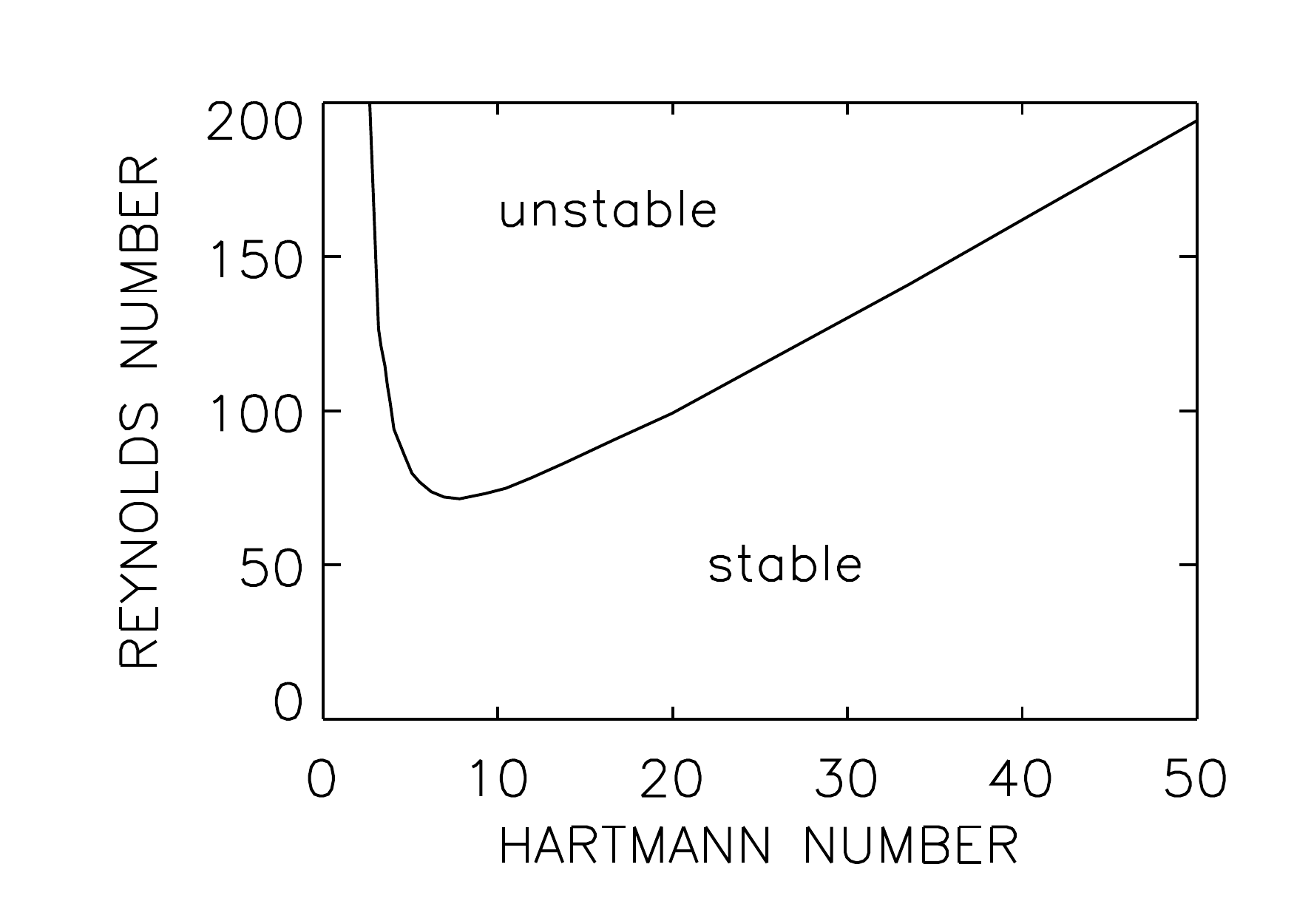}
 \includegraphics[width=7.0cm]{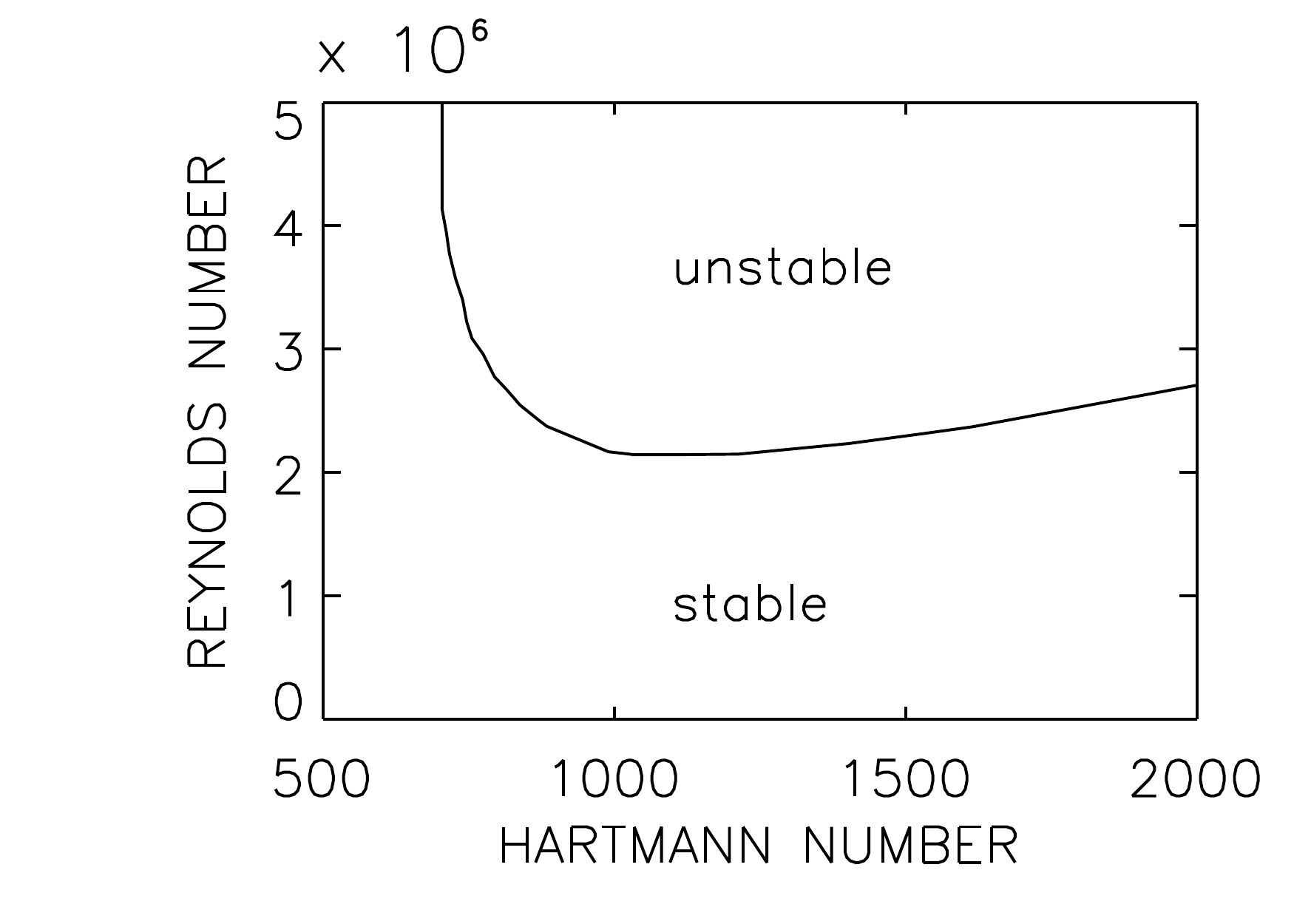}
 \caption{Stability maps of standard MRI for quasi-Keplerian MHD Taylor-Couette flows for $\Pm=1$ (left) and $\Pm=10^{-5}$ (right). The combinations of Reynolds and Hartmann numbers below the curves are  stable. There are strong differences in the ($\Ha/\Rey$) coordinate plane for small and large magnetic Prandtl numbers. $\rin=0.5$, perfectly conducting boundaries.  From \cite{RS03}.}
 \label{f122}
\end{figure}
Figure \ref{f122} shows the eigenvalues of the axisymmetric modes for this flow\footnote{For historical reasons  given for $\mu_\Om=0.33$ instead of $\mu_\Om=0.35$ (quasi-Keplerian rotation).} for the two magnetic Prandtl numbers $\Pm=1$ (left) and $\Pm=10^{-5}$ (right). Compared with Fig.~\ref{f1} the eigenvalues for $\Ha=0$ along the vertical axis disappear to infinity, but the minima for both flows remain almost unchanged. For both magnetic Prandtl numbers the characteristic minima are at very different locations in the ($ \Ha$/$\Rey$) plane. Minimum Reynolds and Hartmann numbers increase for decreasing magnetic Prandtl number. The characteristic Reynolds numbers scale as $\Rey\propto 1/\Pm$ -- much steeper than the $1/\sqrt{\Pm}$ scaling for the potential flow. For the Hartmann number the relation $\Ha\propto 1/\sqrt{\Pm}$ results  -- also steeper than $\Ha\simeq$~const for the potential flow \cite{RS02}. For the quasi-Keplerian flow one finds the simple relations $\Rm\simeq$~const and $\Lu\simeq$~const for $\Pm\to 0$. For small magnetic Prandtl numbers the minima thus have very similar coordinates in the ($\Lu$/$\Rm$) plane. We conclude that for $\Pm\to 0$ the characteristic minima for standard MRI scale with the magnetic Reynolds number $\Rm$ and the Lundquist number $\Lu$. Note that the microscopic viscosity does not play any role in that formulation. This is in contrast to the inductionless approximation for $\Pm=0$, where the remaining eigenvalues are $\Ha$ and $\Rey$ and thus include the microscopic viscosity (see Section \ref{inductionless}). The solutions of the MHD equations for  $\Pm=0$, therefore,  also  scale with  the Hartmann number and the Reynolds  number. As the standard  MRI for finite magnetic Prandtl numbers scales with $\Lu$ and $\Rm$ for small $\Pm$ the limit  for $\Pm\to 0$ yields  $\Ha\to \infty$ and $\Rey\to \infty$, which  can never form a solution of the equations of the inductionless approximation. These equations, therefore, cannot contain any  MRI solution for  uniform axial fields. The solutions only exist for arbitrarily small $\Pm$, but they do not exist for $\Pm=0$  (see \cite{C61}).

It follows that for small $\Pm$ the transition from the potential flow to the non-potential flow might be a dramatic one. For small $\Pm$ a vertical jump along the Rayleigh line from $1/\sqrt{\Pm}$ to $1/\Pm$, i.e.~by a factor of $1/\sqrt{\Pm}$ must exist within a very small interval $ \delta\mu_\Om$. For $\Pm=10^{-5}$ the vertical jump is by more than two orders of magnitudes. For ${\Pm}=1$, on the other hand, the transition from the potential flow to flatter radial flow profiles is much smoother.

Table \ref{MRI} gives the numerical values for the excitation of the standard MRI in quasi-Keplerian flows, for perfectly conducting and insulating boundaries. The critical Reynolds numbers are lower for insulating cylinders, whereas the critical Hartmann numbers are lower for conducting cylinders. The magnetic Mach numbers of the two examples, therefore, differ by a factor of two. One also finds that for all $\Pm\leq1$ the strong-field branches of the lines of neutral stability can be described with $\Mm\simeq 4$. The standard magnetorotational instability, therefore, only works for large Lundquist numbers ($\Lu>1$) {\em and} large magnetic Mach numbers. However, we shall see below that for the nonaxisymmetric modes maximal $\Mm$ exist above which the fluid again becomes stable to these modes. 
\begin{table}
\caption{Coordinates of the absolute minima of the Reynolds numbers  for quasi-Keplerian rotation. $\rin=0.5$, $\mu_\Om=0.33$, $\Pm= 10^{-5}$.}
 \begin{center}
\begin{tabular}{@{}lll@{}}
\hline
&&\\[-1.5ex]
 & perfectly conducting walls & insulating walls\\[1ex]
\hline
&& \\[-1.5ex]
Reynolds number& $2.13 \cdot 10^6$ & $1.42 \cdot 10^6$\\
mag. Reynolds number & 21 & 14\\
Hartmann number & $1100$ & $1400$\\
Lundquist number & 3.47 & 4.42\\
magnetic Mach number & 6.05 & 3.16 \\[1ex]
\hline
\end{tabular}
\label{MRI}
 \end{center}
\end{table}

The MRI is so elementary that its main rules already follow from a simple analysis with a local short-wave approximation of the MHD equations (\ref{p22}) - (\ref{p27}). For disturbances with $k R \gg1$, the differential rotation can be approximated by a plane shear flow \cite{HB91}. For the simplest case of plane-wave disturbances with the  axial wave number $k_z$ and Eq. (\ref{72.1}) one finds the algebraic relation
\begin{eqnarray}
 {\left({\rm i}\omega + \eta k^2\right)^2\left( \left({\rm i}\omega + \nu k^2\right)^2
 + 2\left(2 - q\right)\tilde\Om^2\right)}
 + \Om_{\rm A}^2\left( \Om_{\rm A}^2 - 2 q \tilde\Om^2
 + 2\left({\rm i}\omega + \nu k^2\right)\left({\rm i}\omega +\eta k^2\right)
 \right)
 = 0 
\label{11}
\end{eqnarray}
for the  Fourier frequency  $\omega$, where $\tilde\Om=(k_z/k)\Om$,  the \A~frequency $\Om_\mathrm{A} = kV_\mathrm{A}$ (with the \A\ velocity $V_{\rm A}=B_0/\sqrt{\mu_0\rho}$)  and the local shear $q = -\mathrm{d}\log\Om /\mathrm{d}\log R$. The neutral line between stability and instability defined by $\Im(\omega)=0$ yields 
\begin{equation}
 {\widetilde{\Rm}} = \frac{{\Pm} + {\tilde\Lu}^2}
 {\sqrt{2\left( q{\tilde\Lu}^2 - 2 + q\right)}} .
 \label{12}
\end{equation}
Here, the dimensionless quantities $\tilde\Lu$ and $\widetilde\Rm$ are redefined in terms of the wave number $k$ and the modified rotation rate $\tilde\Om$ ($R_0 \rightarrow k^{-1}$,  $\Om\rightarrow  \tilde\Om$). Equation (\ref{12}) shows that the instability requires sufficiently large $\tilde\Rm$ exceeding
\begin{equation}
 {\widetilde\Rm}_\mathrm{min} = \sqrt{\frac{2}{q} 
 \left( \Pm + \frac{2-q}{q}\right)}.
 \label{13}
\end{equation}
The associated Lundquist number is $\tilde\Lu_{\rm min} = \sqrt{ \mathrm{Pm} +2(2-q)/q}$. For ${\widetilde\Rm}> {\widetilde\Rm}_\mathrm{min}$ the instability only exists for $\tilde\Lu$ between a lower and an upper limit, i.e.~$\tilde\Lu\geq \sqrt{(2-q)/q}$ and $\Mm\geq 1/\sqrt{2q}$. For small $\Pm$ the expression (\ref{12}) loses its dependence on $\Pm$ so that the viscosity disappears from the theory. The instability in this limit is controlled by $\widetilde\Rm$ and $ \tilde\Lu$, which are only formed with the magnetic resistivity $\eta$. For quasi-Keplerian flows ($q=3/2$) one finds $\tilde\Lu\geq 1/\sqrt{3} $ and for the slope of the strong-field branch $\Mm\simeq 1/\sqrt{3}$.  On the other hand, for {large} $\Pm$ the minima of the curves fulfill the conditions $\Rmquer= 2/\sqrt{3}$ and $\Ha=1$.

\begin{figure}[htb]
\centering
 \includegraphics[width=8cm]{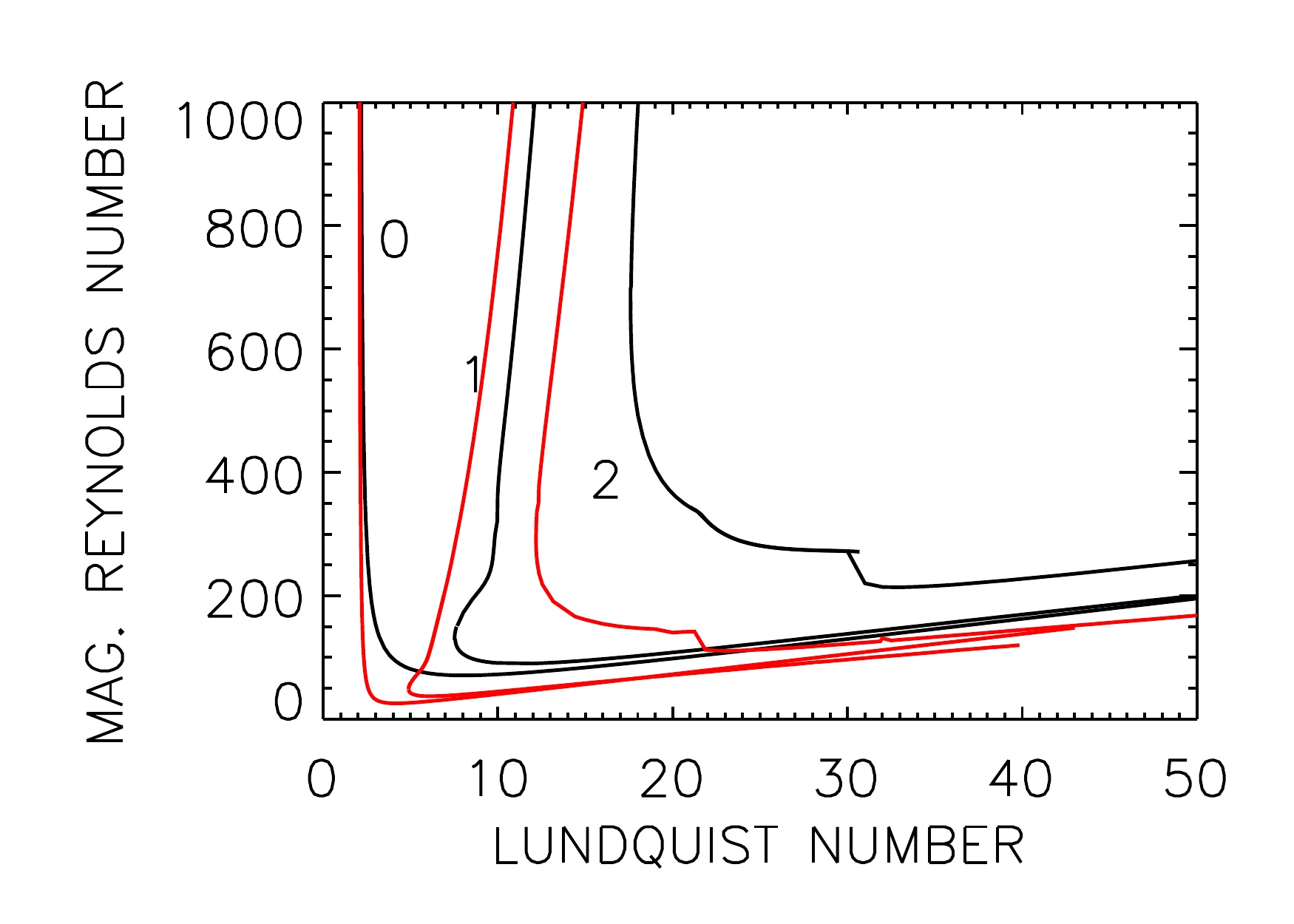}
  \includegraphics[width=8.0cm]{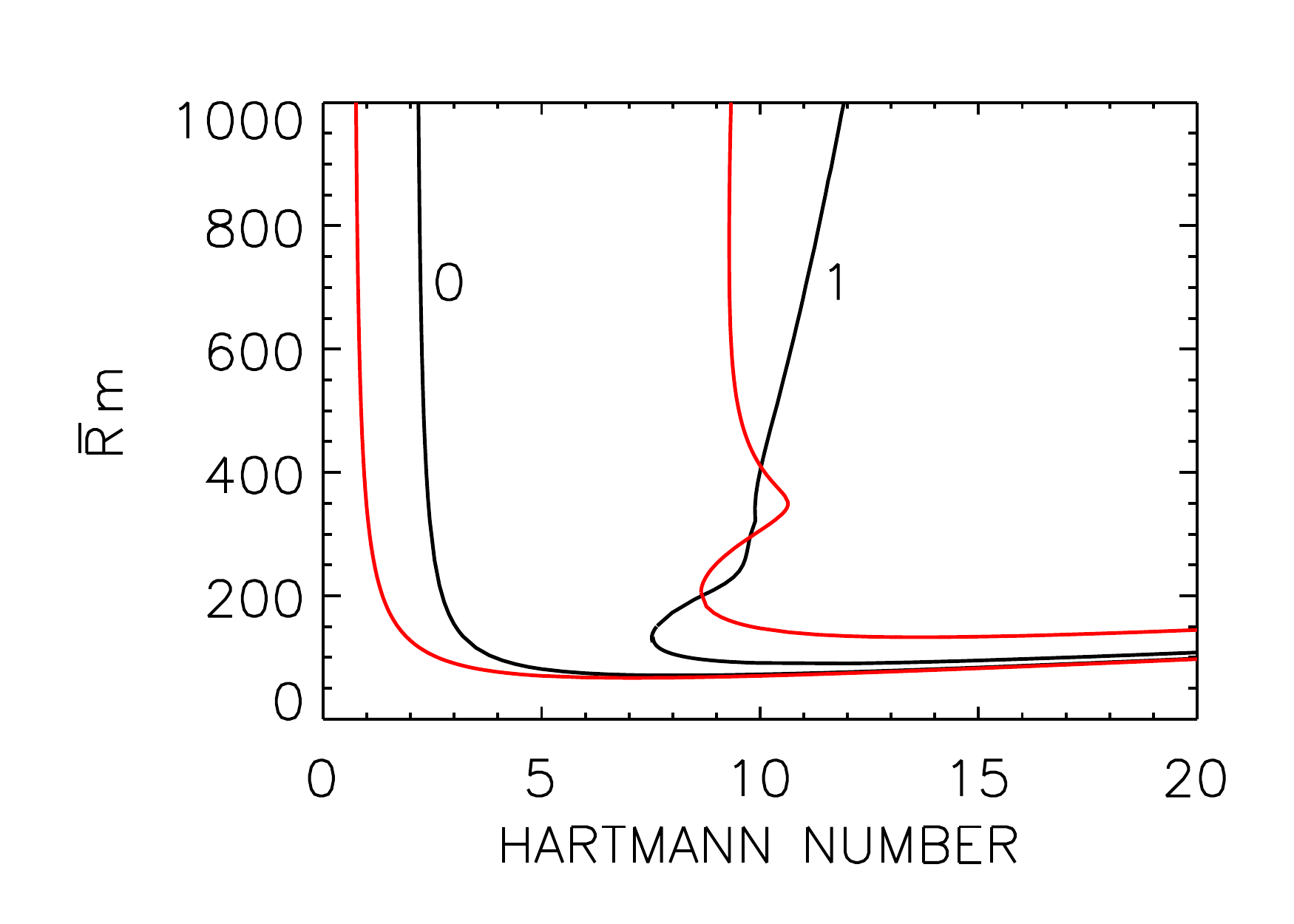}
 \caption{Stability lines for standard MRI with quasi-Keplerian flow  for the  modes $m=0$, $m=1$ and $m=2$ (marked). Dark lines:  $\Pm=1$. Left:  $\Pm=0.01$ (red). Right:  $\Pm=10$ (red). For small $\Pm$ the curves scale with $\Rm$ and $\Lu$, for large $\Pm$ they scale with $\Rmquer$ and $\Ha$. The weak-field branches show opposite slopes for axisymmetric and nonaxisymmetric modes. Nonaxisymmetric modes  decay for too high Reynolds numbers.  $\rin=0.5$,  $\mu_\Om=0.35$.  Perfectly conducting boundaries.}
 \label{f14}
\end{figure}

The stability lines in Figs.~\ref{f14} as the solutions of the  MHD equations    (\ref{p22}) - (\ref{p27}) for quasi-Keplerian rotation indeed show  for small $\Pm$  no  clear dependence on $\Pm$  in the ($\Lu/\Rm$) plane, at least for the axisymmetric mode, true both for conducting (left panel) and for insulating (right panel) boundary conditions. The right branch of the neutral stability lines at sufficiently high $\Rm$ can be characterized by $\Mm=1/\sqrt{3}$. The left branch is controlled by the diffusivities. Its expression for large Rm can be obtained by setting the denominator in (\ref{12}) to zero, hence $\Lu=1/\sqrt{3}$ \cite{KR04}.

In Fig.~\ref{f14} (right) the lines of neutral stability are compared for $\Pm=1$ (dark lines) with those for $\Pm=10$ (red lines) in the ($\Ha/\Rmquer$) plane. The scaling with $\Ha$ and $\Rmquer$ for both axisymmetric and nonaxisymmetric solutions for large $\Pm$ is obvious. This finding remains robust also for larger $\Pm$. The numerical results for the global quasi-Keplerian flow with $\rin=0.5$ confirm these findings for the axisymmetric and nonaxisymmetric modes. 
The graphs   also demonstrate that nonaxisymmetric modes require stronger fields for their excitation than axisymmetric modes. There is, however, an even more interesting difference between the axisymmetric and nonaxisymmetric modes. For $m=0$ and $\Lu \geq 1$  a single critical Reynolds number always exists above which the  MRI is excited for all larger $\Rm$. The nonaxisymmetric modes behave differently. For $\Lu> S_{\rm min}\simeq 1 $ ($\Lu_{\rm min}$ the smallest possible Lundquist number) there are always two critical Reynolds numbers between which the nonaxisymmetric modes can exist. The nonaxisymmetric modes are thus stabilized by too slow and by too fast rotation. If it is too strong, the differential rotation suppresses the nonaxisymmetric parts of the instability pattern. As an estimation one finds that $\Mm \simeq 300$ is the highest possible magnetic Mach number for the excitation of nonaxisymmetric modes. The dependence of this value on  the magnetic Prandtl number is weak. For $m=0$ such an upper limit does not exist.

For small magnetic Prandtl numbers (here $\Pm=0.01$) we again find a crossing phenomenon for strong fields in the neutral-stability curves for $m=0$ and $m=1$ \cite{GR12}. In Fig.~\ref{f14} (left) for perfectly conducting cylinders lines for $m=0$ and $m=1 $ cross for $\Lu\simeq 20$. For weaker fields the mode with the lowest Reynolds number is always axisymmetric, but for stronger fields the Reynolds numbers for $m=1$ are smaller than those for $m=0$. In these cases the MRI sets in as a nonaxisymmetric flow pattern. The nonaxisymmetric structure is lost, however, for too fast rotation when the magnetic Reynolds number reaches the upper value of the marginal stability of the $m=1$ curve. We have found this sort of mode-crossing only for MHD flows with perfectly conducting boundary conditions.

One can show that a solution with a certain positive $k$ is always accompanied by a solution with $-k$ with the same Reynolds number and drift frequency (for given $\rm Ha$ and $m$). As the pitch angle of the resulting spirals is given by $\partial z/\partial \phi= -m/k$, it is clear that the two solutions have opposite pitch angles, so that the solution is always a combination of a left screw and a right screw. In the ideal case the same number of left and right spirals will be excited as there is no reason for a preference. Both the kinetic and magnetic helicities thus vanish on average. As a consequence, standard MRI does not produce any $\alf$ effect (see Section \ref{Helicities}).

The governing equation system is also invariant under the simultaneous transformations $m\to -m$, $k\to -k$ and $\Re{(\omega})\to -\Re{(\omega})$, with $\Re({\omega})$ as the real part of the mode frequency $\omega$. Hence, the drift of both solutions and also the pitch angles, i.e.
\beg
\frac{\partial\phi}{\partial t}=-\frac{\Re{(\omega)}}{m} \ \ \ \ \ \ \ \ \ \ \ \ \ \ \ \ \ \ \ \ \ \frac{\partial z}{\partial \phi}= -\frac{m}{k}, 
\label{pitch}
\ende
are equal so that the solutions are identical. It is thus enough to assume $k>0$.
\begin{figure}[htb]
\centering
 \includegraphics[width=8cm]{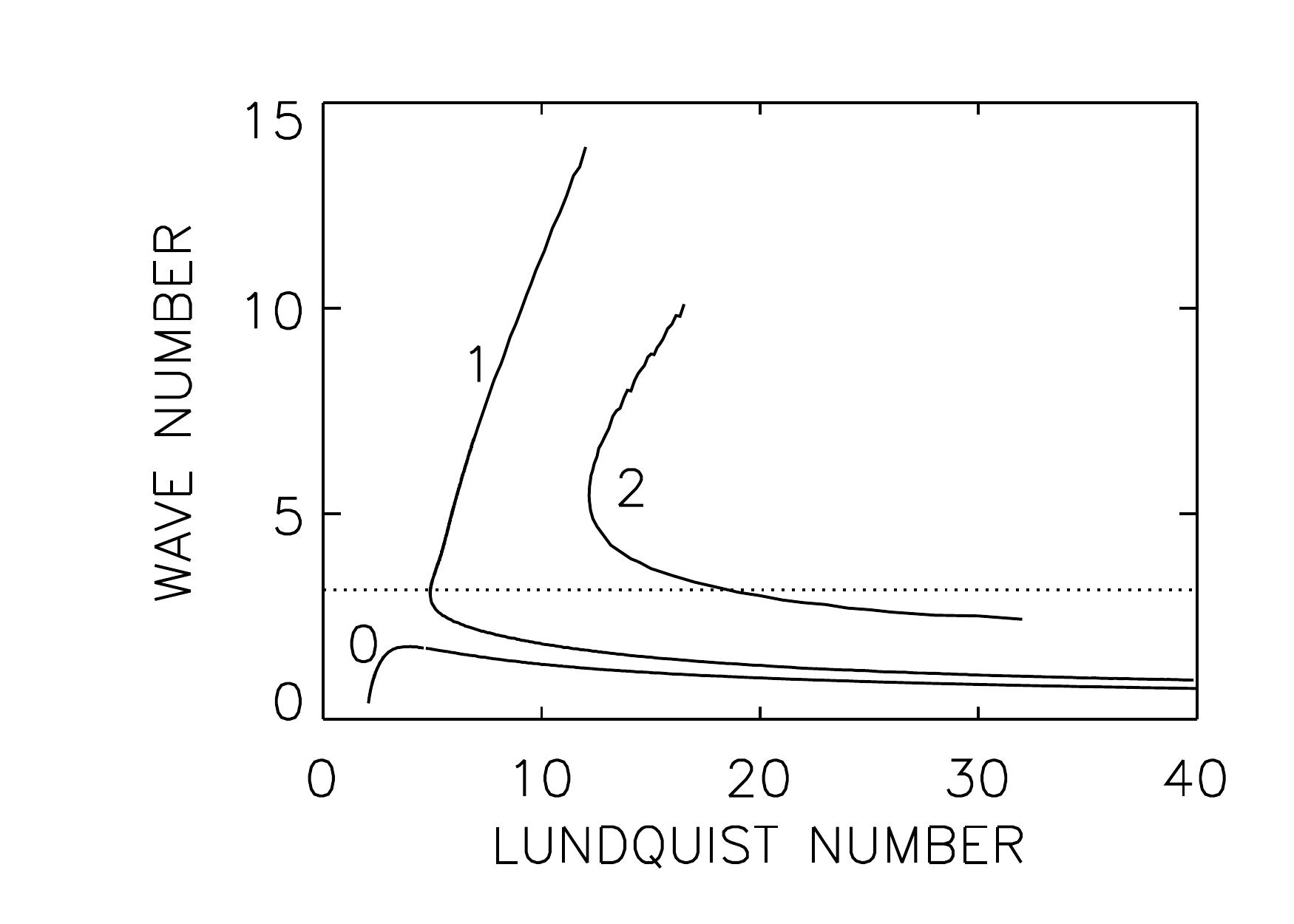}
 \includegraphics[width=8cm]{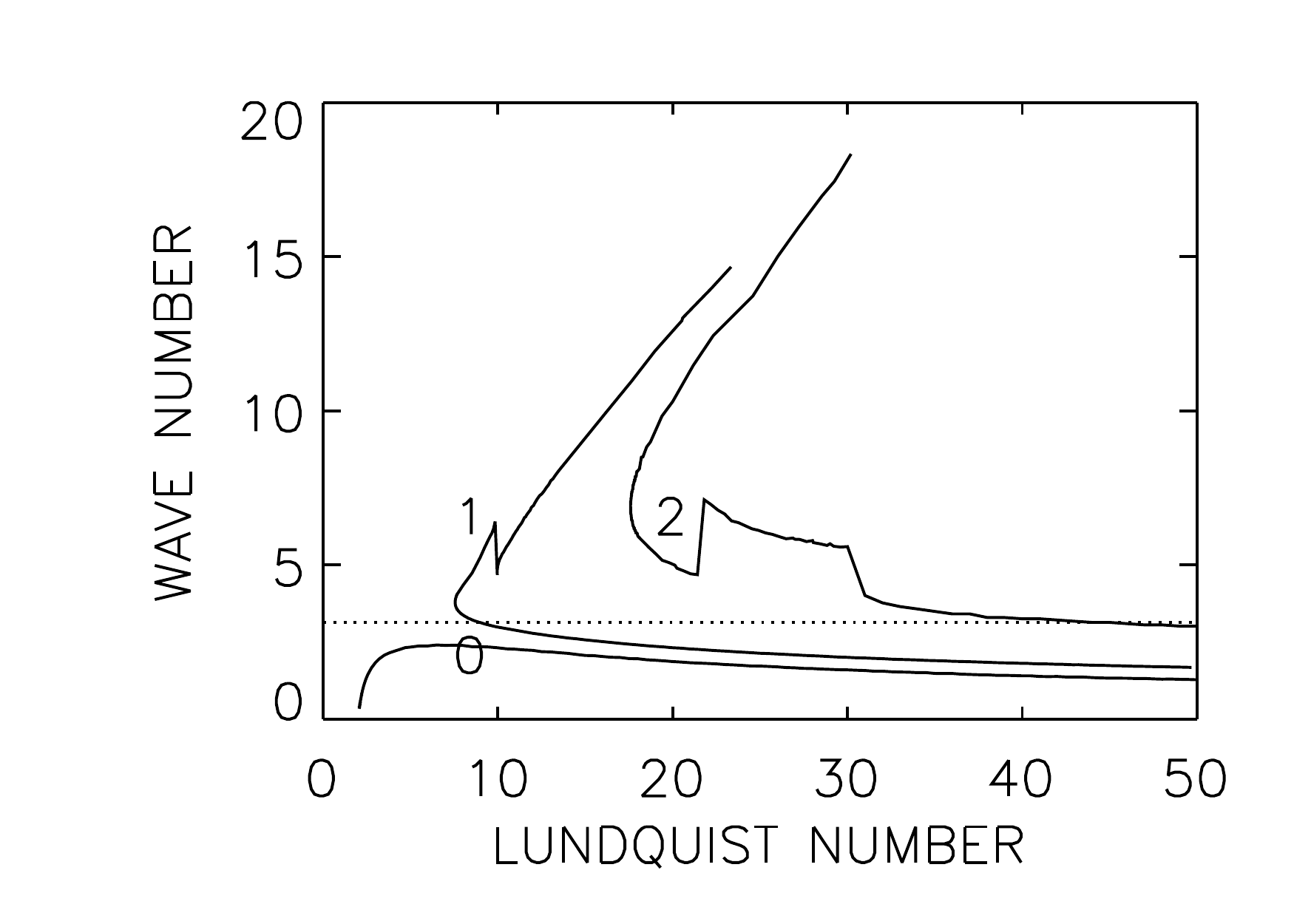}
\caption{Wave numbers along the lines of neutral  stability for the azimuthal modes $m=0$, $m=1$ and $m=2$ (marked by $m$) for standard  MRI with quasi-Keplerian flow. Left: $\Pm=0.01$, right: $\Pm=1$. Dotted lines mark the limit $k=\pi$ for circular cells in the $(R/z)$ plane. The axisymmetric modes and the strong-field solutions of the nonaxisymmetric modes are always prolongated in axial direction. $\rin=0.5$, $\mu_\Om=0.35$.}
 \label{f15}
\end{figure}

The vertical extent $\delta z$ of the cells of the instability pattern, normalized by the gap width $d=R_{\rm out} - R_{\rm in}$ between the cylinders, is given by
\beg
{\delta z \over d} = {\pi \over k}
\sqrt{{\rin \over 1-\rin}}.
\label{delz}
\ende
For $\rin=0.5$ it is simply ${\delta z }/{ d} = {\pi}/{ k}$ so that for $k\simeq \pi$ the cells are almost circular in the ($R/z$) plane, and for $k\gg\pi$ the cells are very flat. Figure \ref{f15} shows that for both values of $\Pm$ the azimuthal rolls of the axisymmetric modes become more and more elongated in the vertical direction. Generally, only the cells of the weak-field branches of the nonaxisymmetric modes are very flat while the other modes possess circular or prolate cells.

The real part $\Re{(\omega)}$ of the frequency $\omega$ of the Fourier mode in units of the rotation rate of the inner cylinder
\beg
\omega_{\rm dr}= \frac{\Re{(\omega)}}{\Om_{\rm in}},
\label{omdr}
\ende
which for $m\neq 0$ describes an azimuthal drift 
\beg
\frac{\dot \phi}{\Omin} =- \frac{\omega_{\rm dr}}{m }
\label{dotfi}
\ende
of the instability pattern, in units of the inner cylinder's rotation rate. For negative $\omega_{\rm dr}$ the pattern migrates in the direction of the global rotation (eastward). Because of these definitions a drift value of $-\mu_\Om$ describes an exact corotation of the flow pattern with the {\em outer} cylinder as we are working in the fixed laboratory system.
 
 \subsection{Nonlinear simulations}\label{nonsim}
Nonlinear numerical simulations reveal the axisymmetric character of the standard MRI for large magnetic Mach numbers. The nonlinear three-dimensional time-stepping problem is solved using the MPI-parallelized code \cite{GW15}, which itself is based on an earlier pipe flow solver by A.P. Willis\footnote{See www.openpipeflow.org.}. The spatial structures in $z$ and $\phi$ are described by the standard Fourier mode approximation, allowing energy spectra in these two directions to be easily constructed. The periodic domain length in the axial direction is chosen as 10 times the gap width, to allow sufficiently large structures in $z$. The resolution varies from $127 \times 64\times 32$ and $511 \times 256 \times 128$, depending on the Reynolds number. In its present form the code only works without endplates in axial direction and only for insulating radial boundaries.
\begin{figure}[htb]
\centering
\includegraphics[width=4cm,height=6cm]{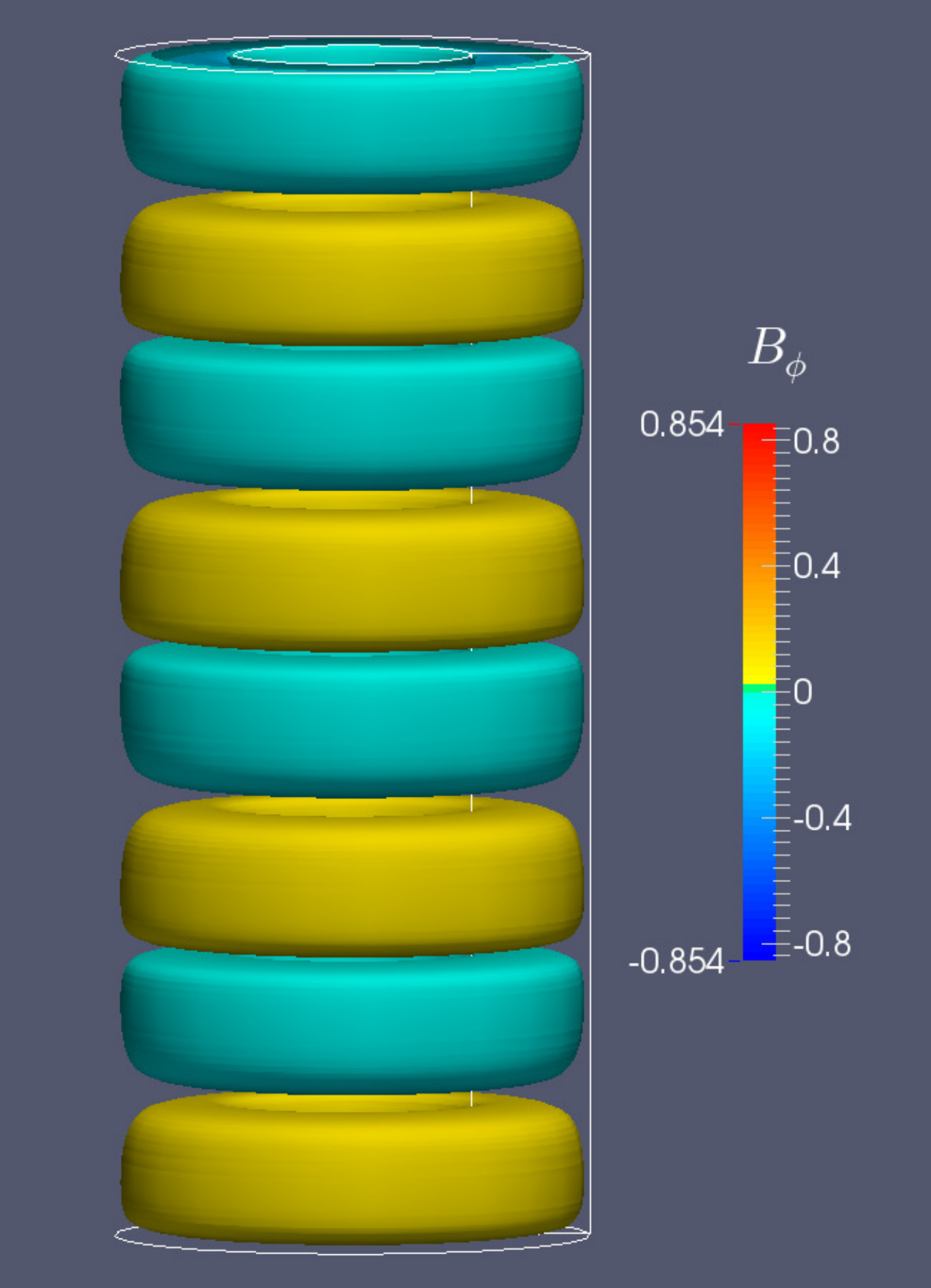}
\includegraphics[width=4cm,height=6cm]{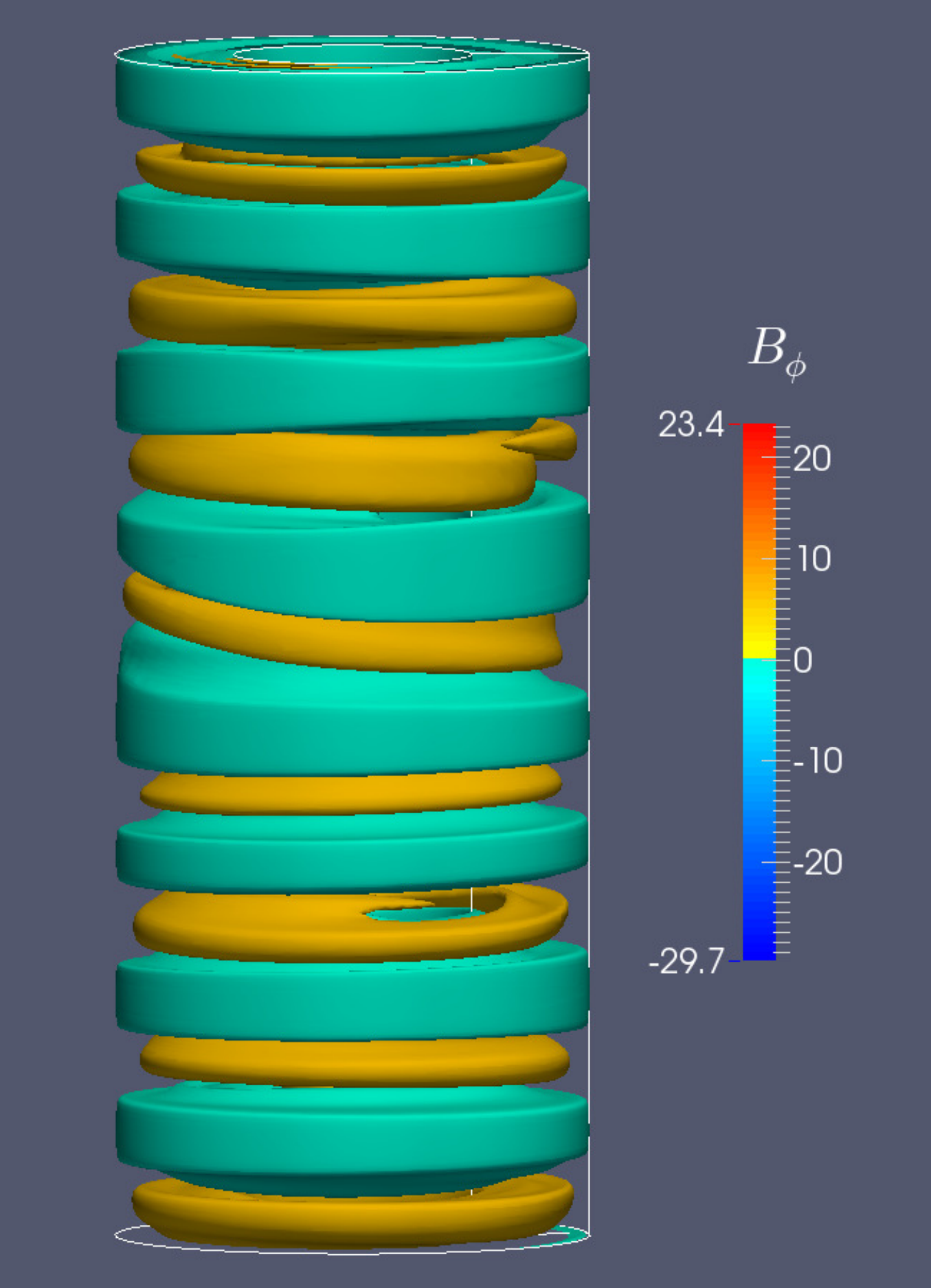}
 \includegraphics[width=4cm,height=6cm]{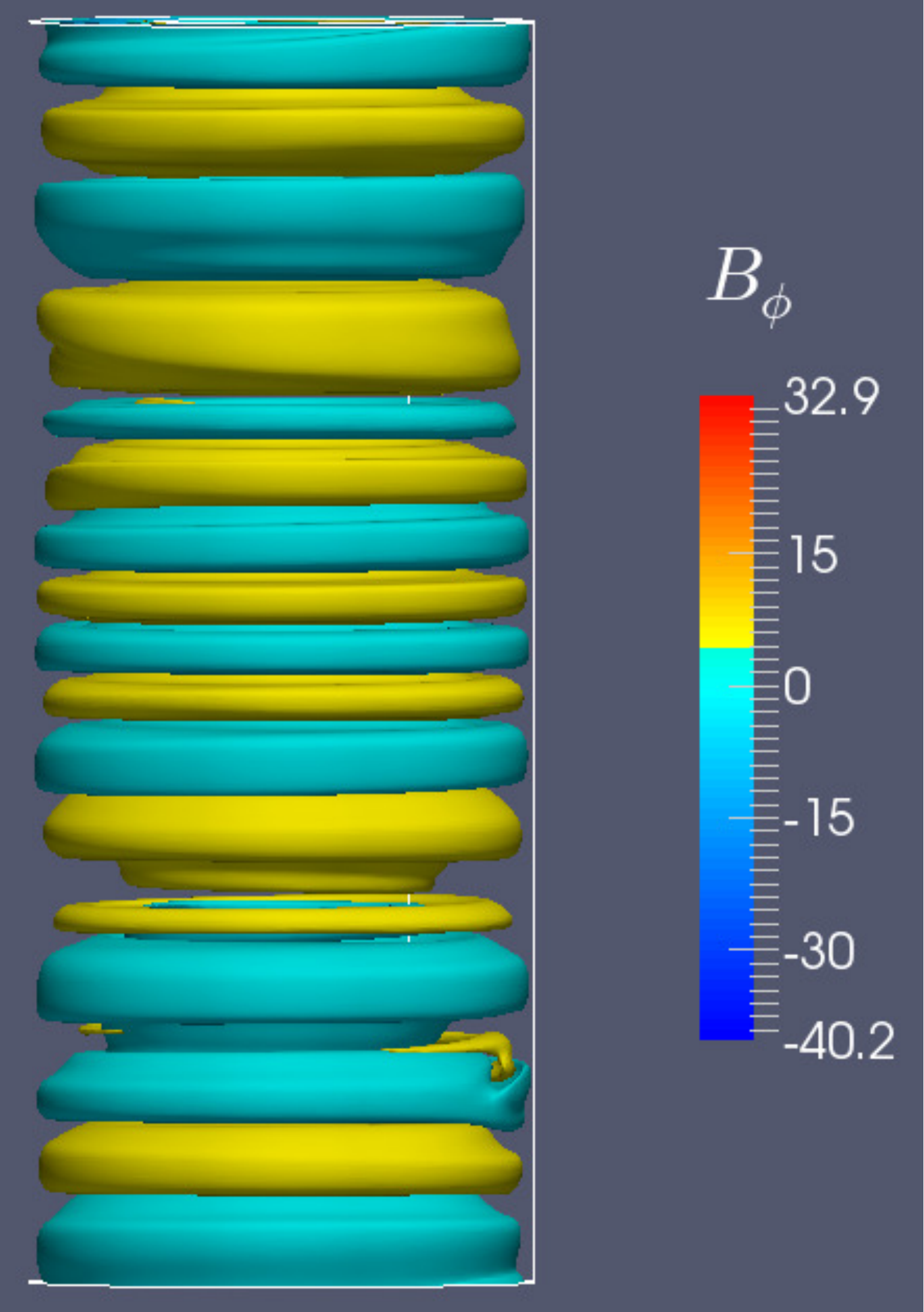}
 \caption{Instability patterns of standard MRI for  quasi-Keplerian rotation for weak background field and increasing magnetic Mach numbers  (from left to right: $\Mm = 6.8, 96.2, 307.7$). The   magnetic Reynolds numbers are 88, 1250 and 4000 (from left to right). The numbers on the color bars yield the amplitude of the local quantity $b_\phi/B_0$. $\rin=0.5$, $\mu_\Om=0.35$,  $\Lu=13$, $\Pm=1$. Insulating boundaries ($b_\phi=0$ at the cylinders).}
 \label{ha13}
\end{figure}

The equations for the quasi-Keplerian flow have been solved in axially unbounded containers with insulating boundary conditions. The right panel of Fig.~\ref{f14} shows the neutral stability curves. For a weak field Fig.~\ref{ha13} shows the isolines of the azimuthal components of the magnetic field for models with increasing Reynolds numbers. At $\Rm=88$ the lowest Reynolds number lies below the instability curve of the nonaxisymmetric $m=1$ mode, so that the exact ringlike geometry of the left plot in Fig.~\ref{ha13} is not a surprise. The cells are nearly circular in the ($R/z$) plane. For faster rotation ($\Rm=1250$, middle) nonaxisymmetric structures occur but remain weak. Nevertheless, the cell structure changes as the cells become more oblate, which cannot be understood by means of the nonmagnetic Taylor-Proudman theorem. This trend is continued for even faster rotation ($\Rm=4000$) where again the axisymmetry of the solution prevails.
\begin{figure}[htb]
\centering
\includegraphics[width=4cm,height=6cm]{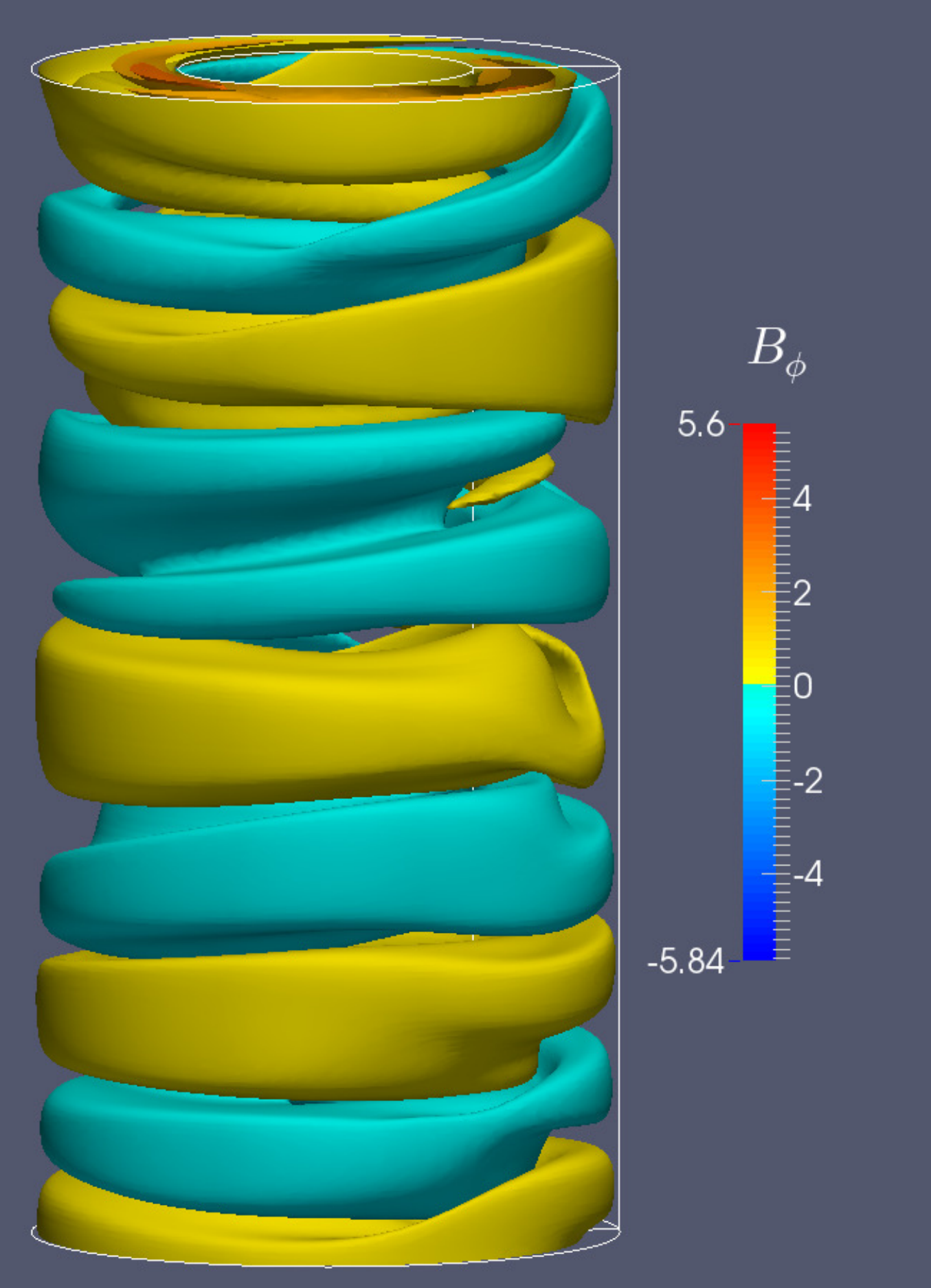}
 \includegraphics[width=4cm,height=6cm]{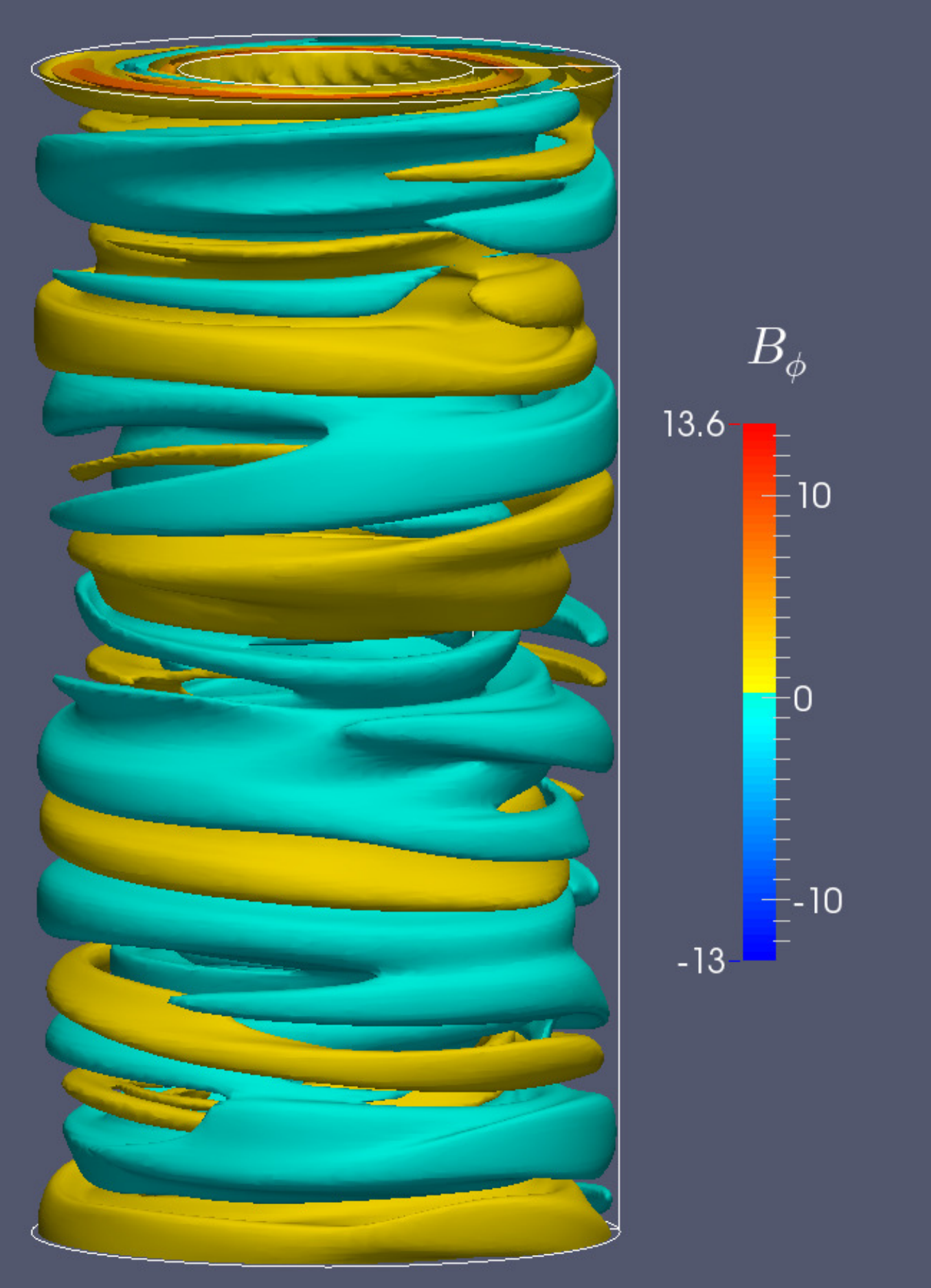}
 \includegraphics[width=4cm,height=6cm]{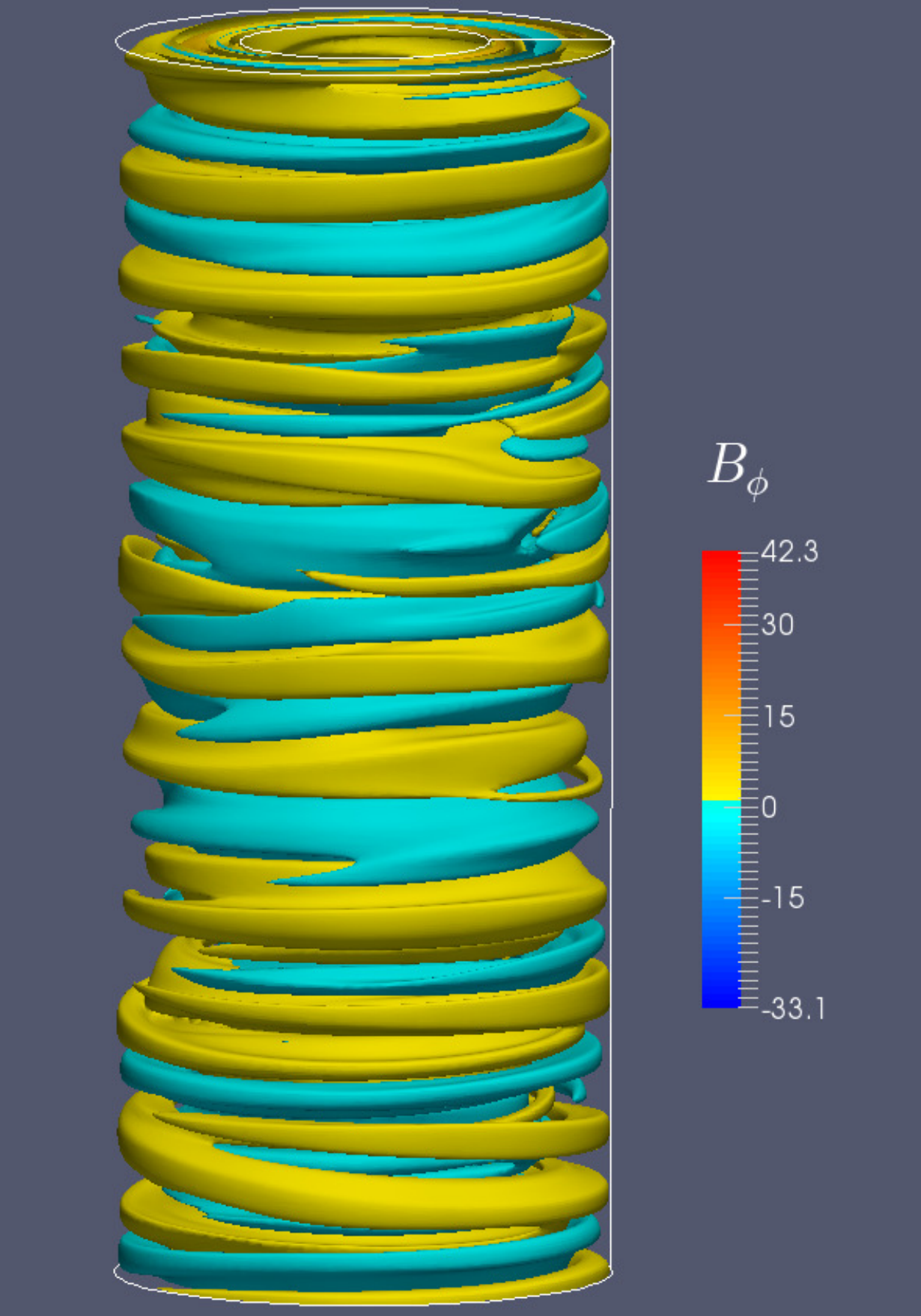}
 \caption{As in Fig.~\ref{ha13} but for $\Lu=30$. Magnetic Reynolds numbers are $\Rm=500, 1500, 4000$ and magnetic Mach numbers are $\Mm = 16.7,
  50, 133$  (from left to right). }
 \label{ha30}
\end{figure}

\begin{figure}[h]
\centering
 \includegraphics[width=4cm,height=6cm]{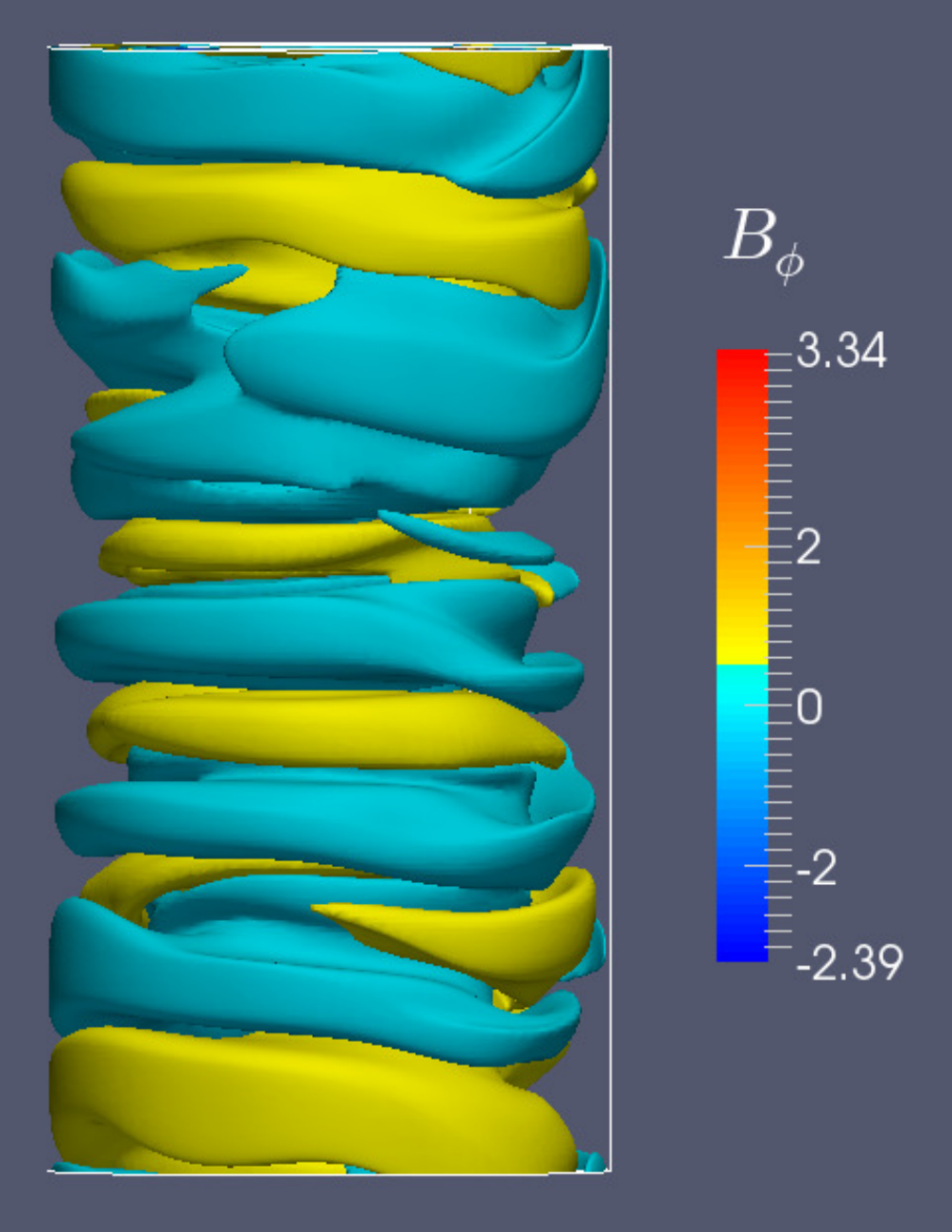}
 \includegraphics[width=4cm,height=6cm]{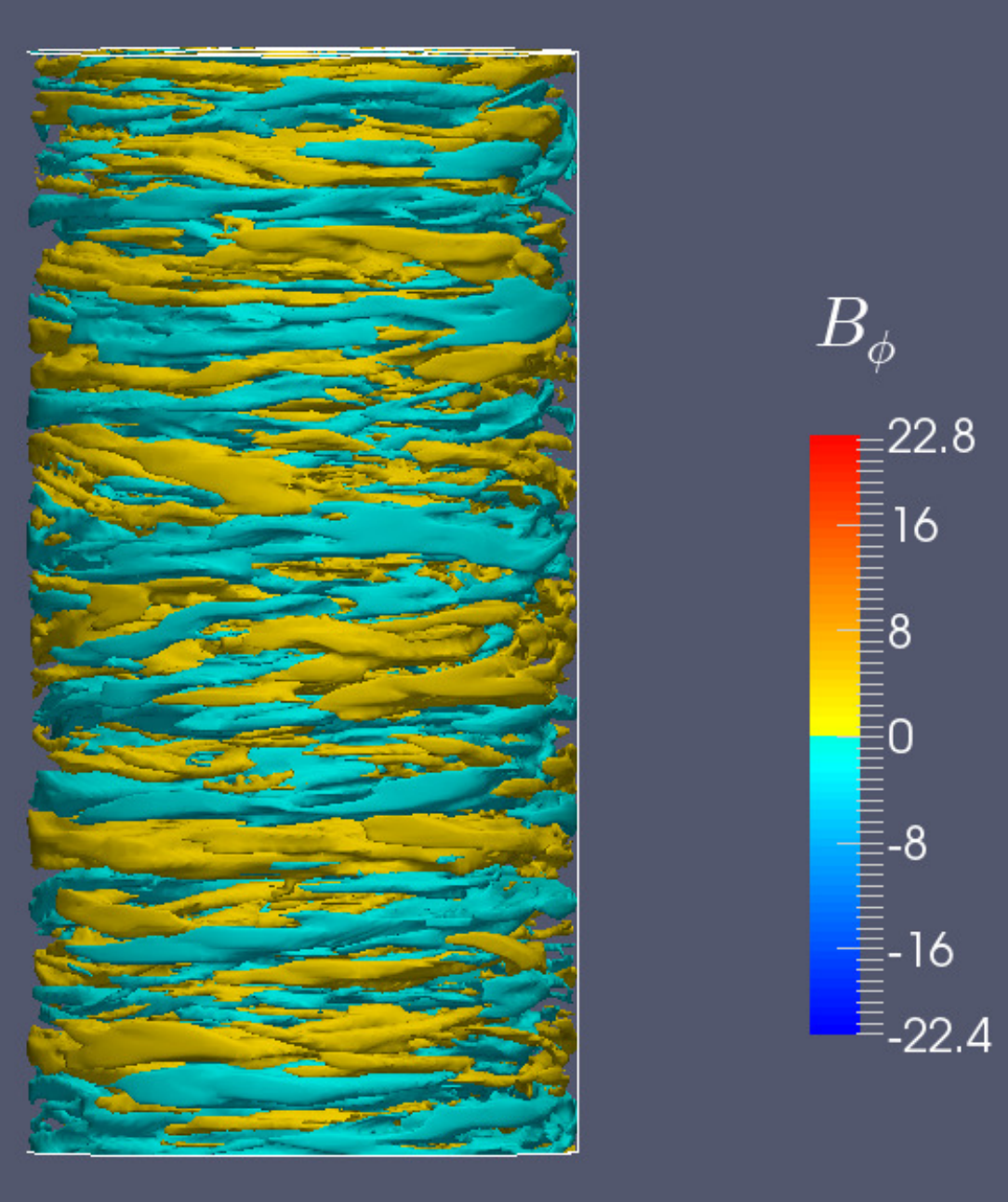}
 \includegraphics[width=4cm,height=6cm]{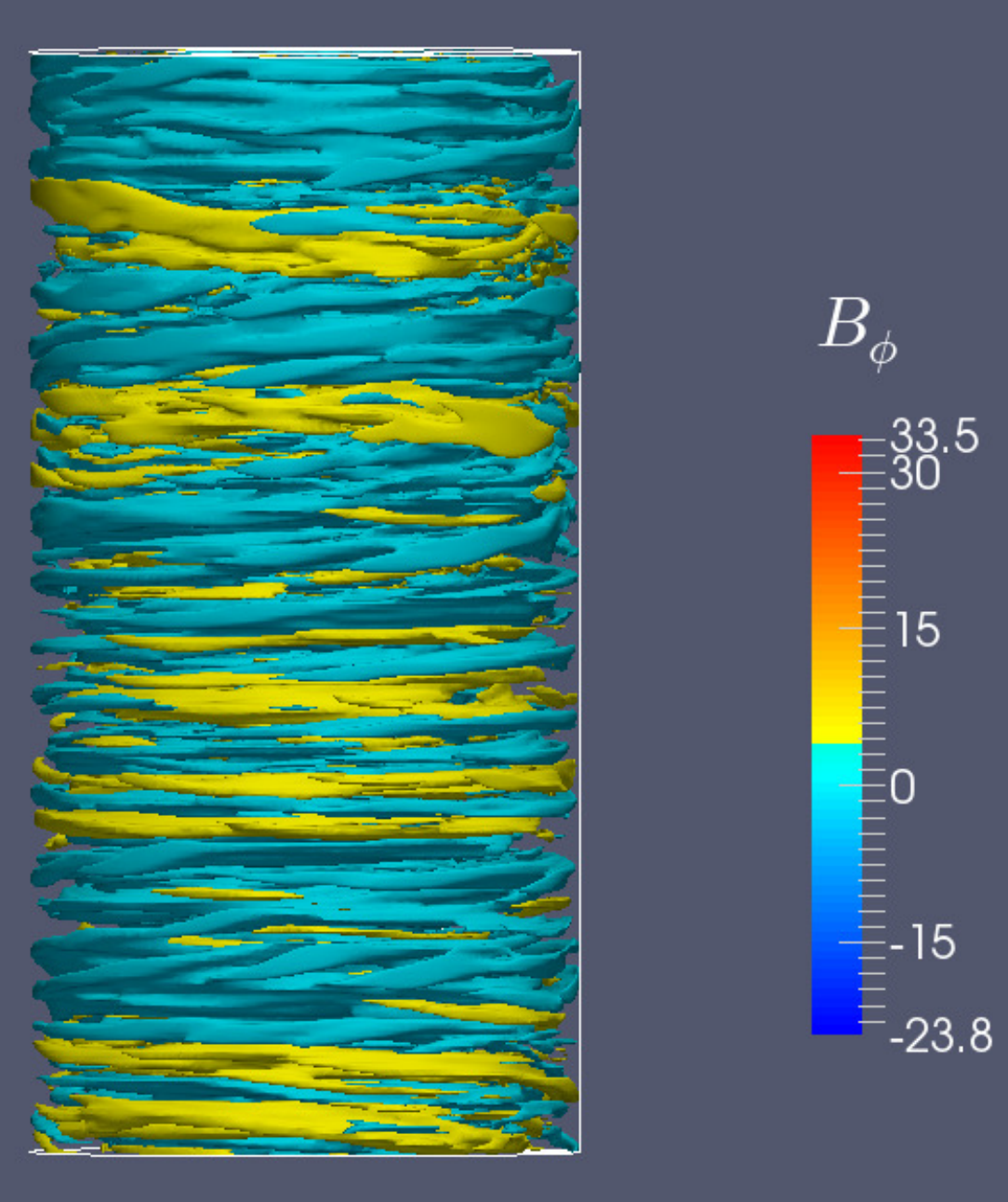}
 \caption{As in Fig.~\ref{ha13} but for $\Lu=100$.   Magnetic Reynolds numbers are $\Rm=1000, 10000, 20000$ and magnetic Mach numbers are $\Mm = 10,
  100, 200 $  (from left to right).
 }
 \label{ha100}
\end{figure}

For stronger fields the nonaxisymmetric modes are obviously excited, but only for not too low and not too high Reynolds numbers. The right panel of Fig.~\ref{ha30} shows that very large Reynolds numbers indeed prevent the excitation of nonaxisymmetric modes. The instability map suggests that for $\Lu=30$ the Reynolds number 4000 lies outside the instability domain for $m=1$ (see Fig.~\ref{f14}, right). For $\Lu=100$ only the model with $\Rm=1000$ (i.e.~magnetic Mach number of order 10) shows a nonaxisymmetric pattern with low $m$, while the models with faster rotation become more and more axisymmetric with increasing axial wave numbers (see Fig.~\ref{ha100}). 

\begin{figure}[htb]
\centering
 \includegraphics[width=8cm]{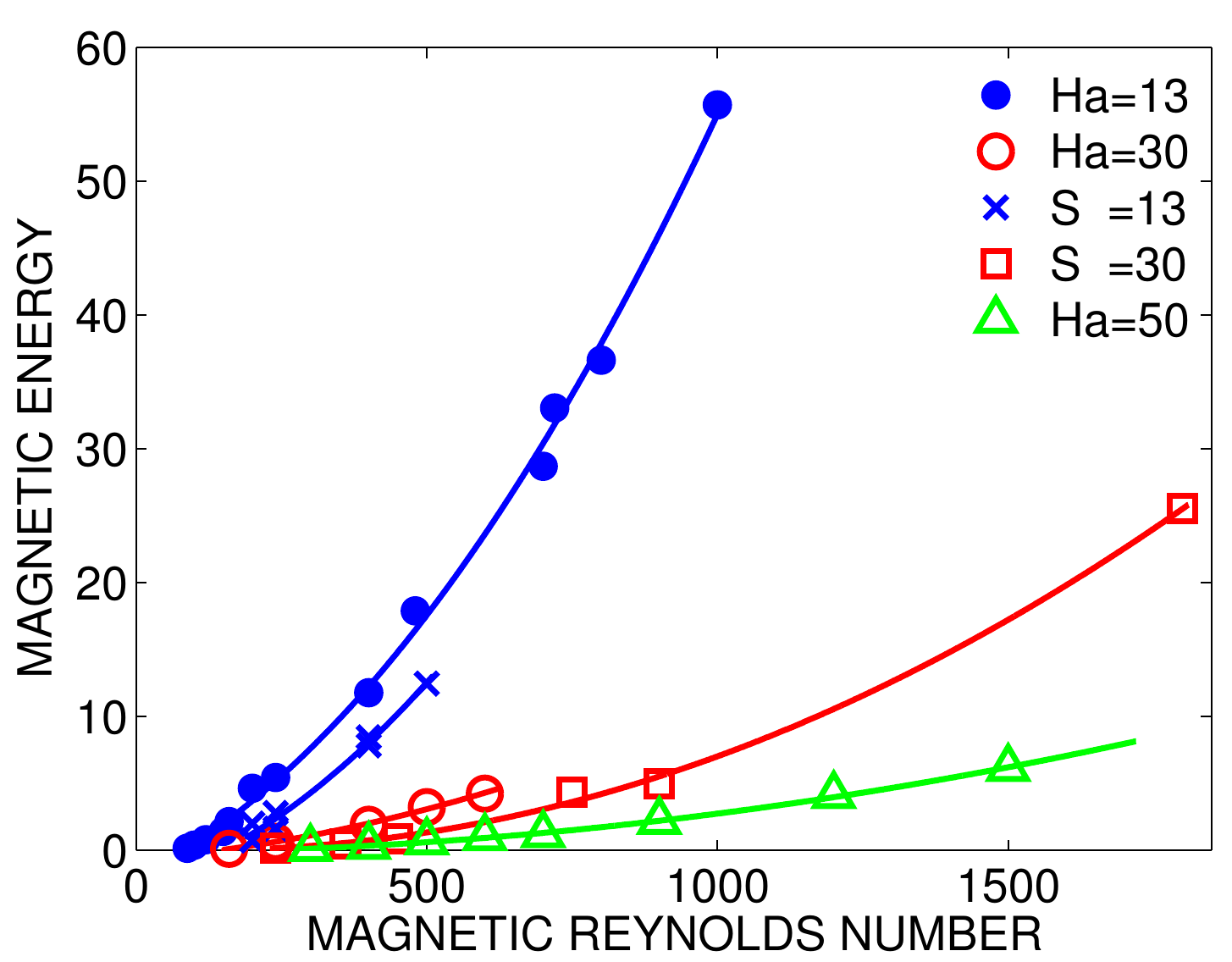}
 \includegraphics[width=8cm]{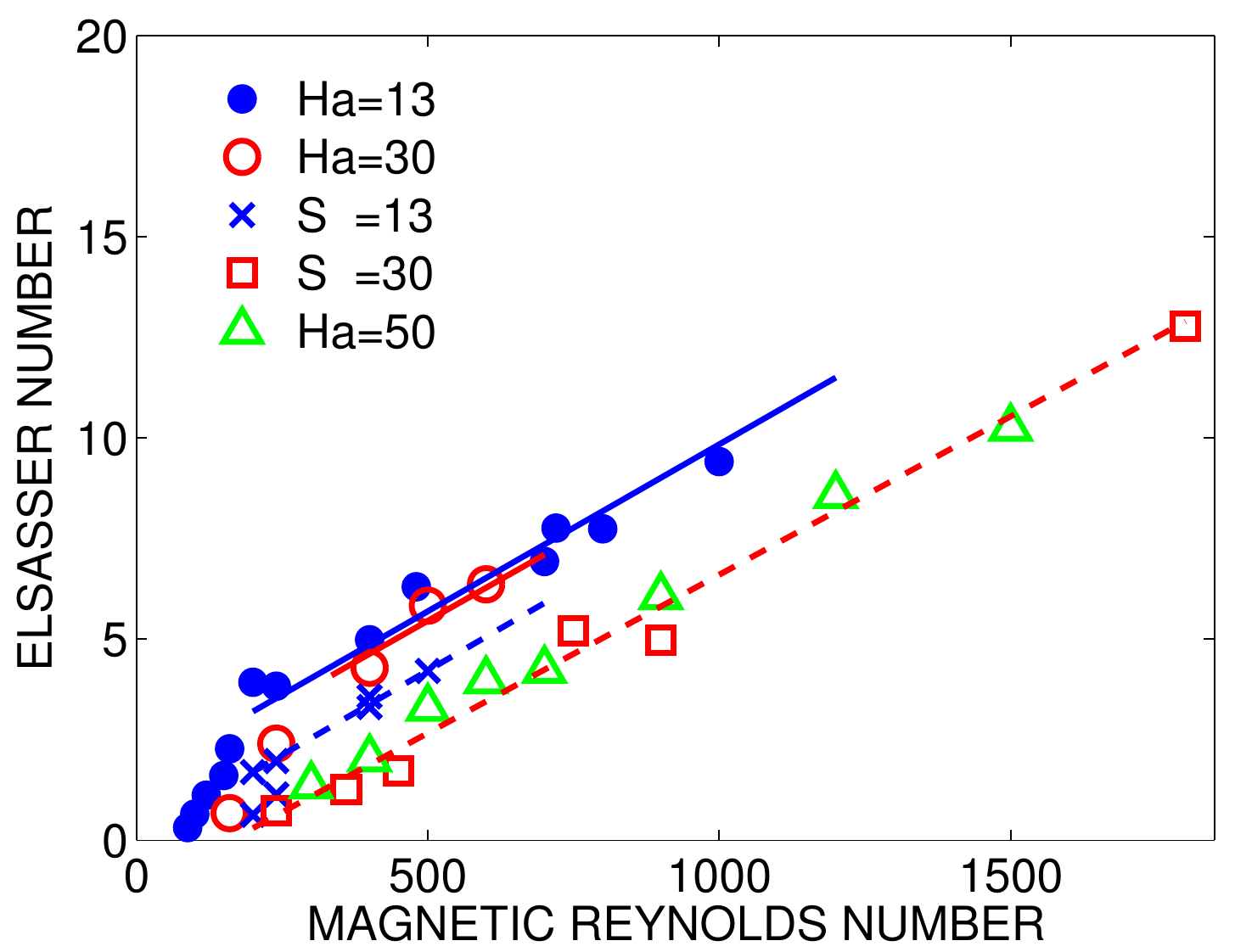}
 \caption{Left: Magnetic energy (\ref{qu}) for standard  MRI averaged over the whole container as  functions of the magnetic Reynolds number and for various magnetic field amplitudes.  Right: The magnetic Elsasser number (\ref{Els1}). The circles and the green triangles correspond to $\Pm=1$ (increasing $\Ha$) while the other symbols belong to $\Ha=50$ and $\Pm=0.1$ (blue crosses) and  $\Pm=0.2$ (red squares). The dependencies on $\Lu$ and $\Pm$ are rather weak. $\rin=0.5$,  $\mu_\Om=0.35$, perfectly conducting cylinders.}
 \label{energ}
\end{figure}

\begin{figure}[h]
\centering
 \includegraphics[width=4cm,height=6cm]{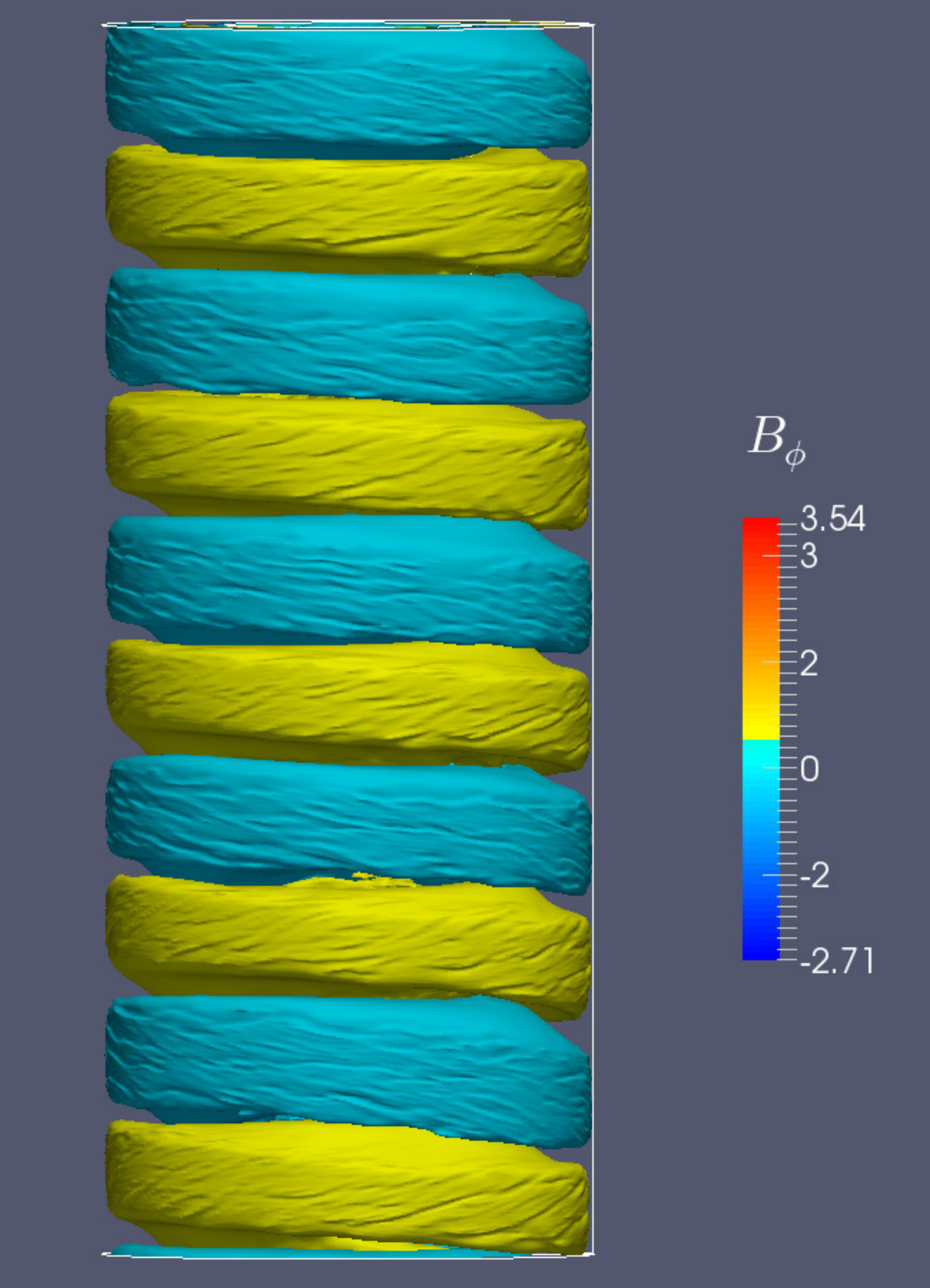}
  \includegraphics[width=4cm,height=6cm]{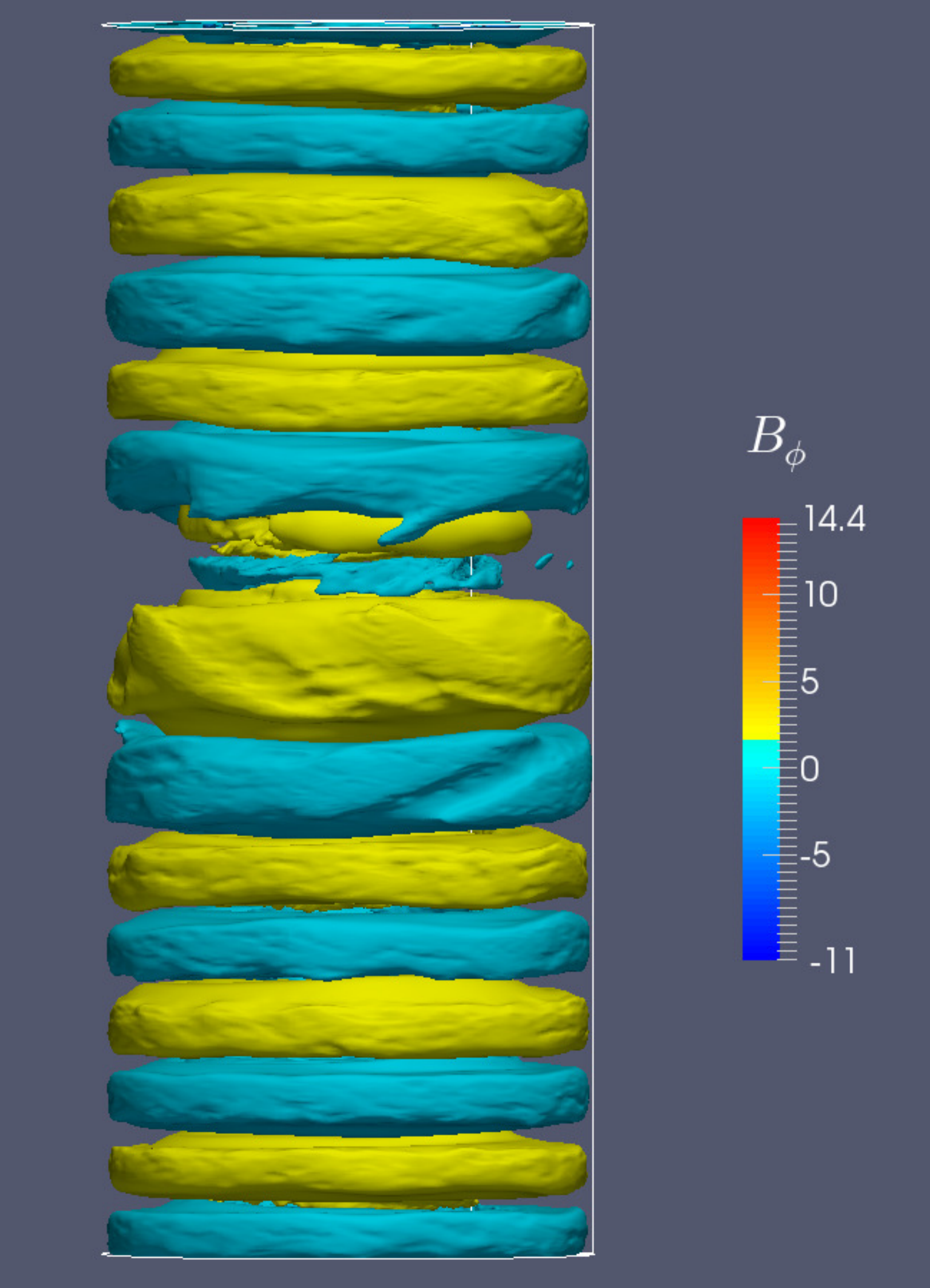}
 \includegraphics[width=4cm,height=6cm]{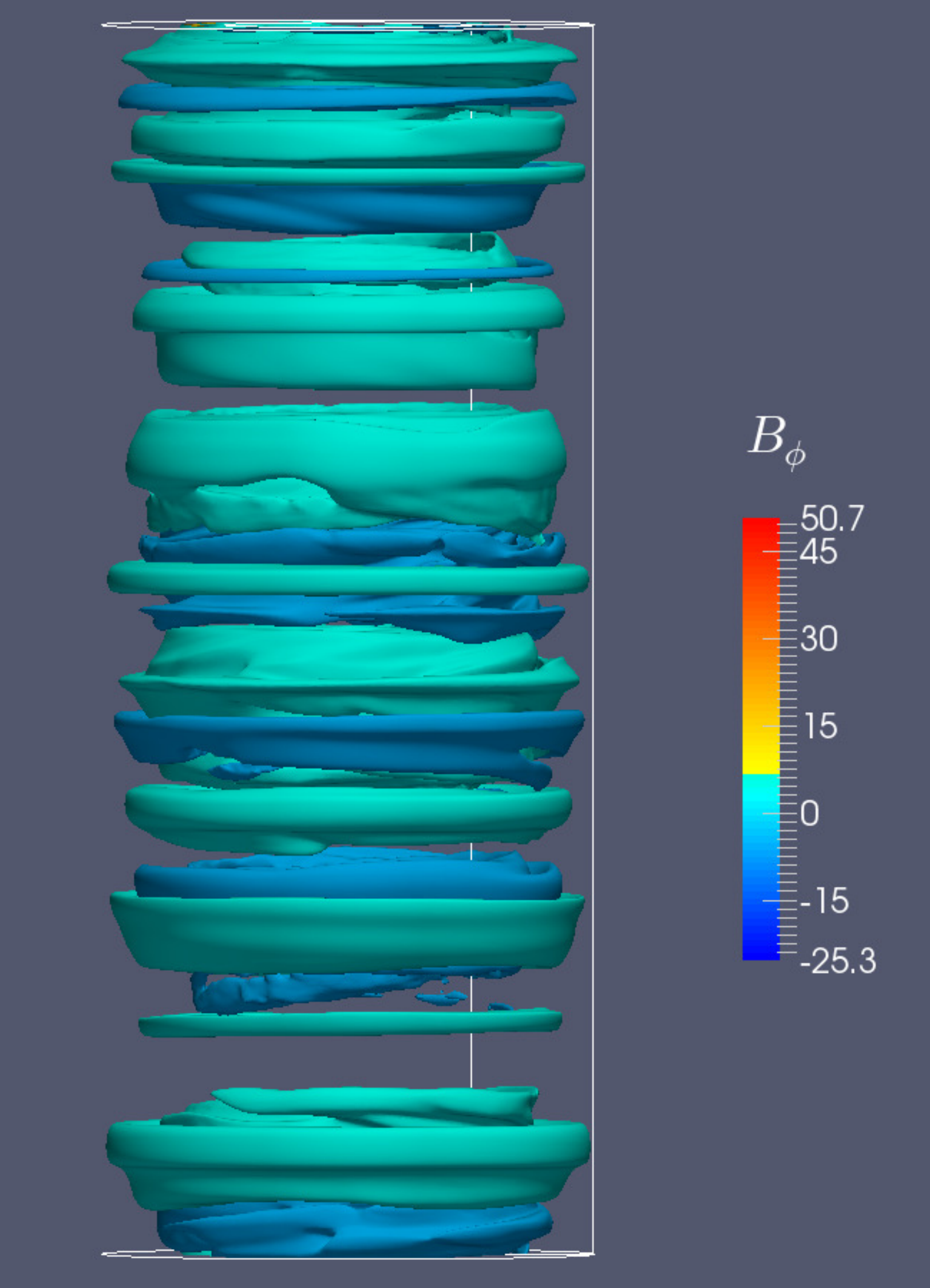}
\caption{As in Fig.~\ref{ha13} but for  $\Pm=0.01$ and $\Lu=13$.   Magnetic Reynolds numbers are $\Rm=400, 1250, 4000$ and magnetic Mach numbers are $\Mm = 30.8, 96.2, 307.7$  (from left to right).
 }
 \label{pm001}
\end{figure}

\begin{figure}[h]
\centering
 \includegraphics[width=4cm,height=6cm]{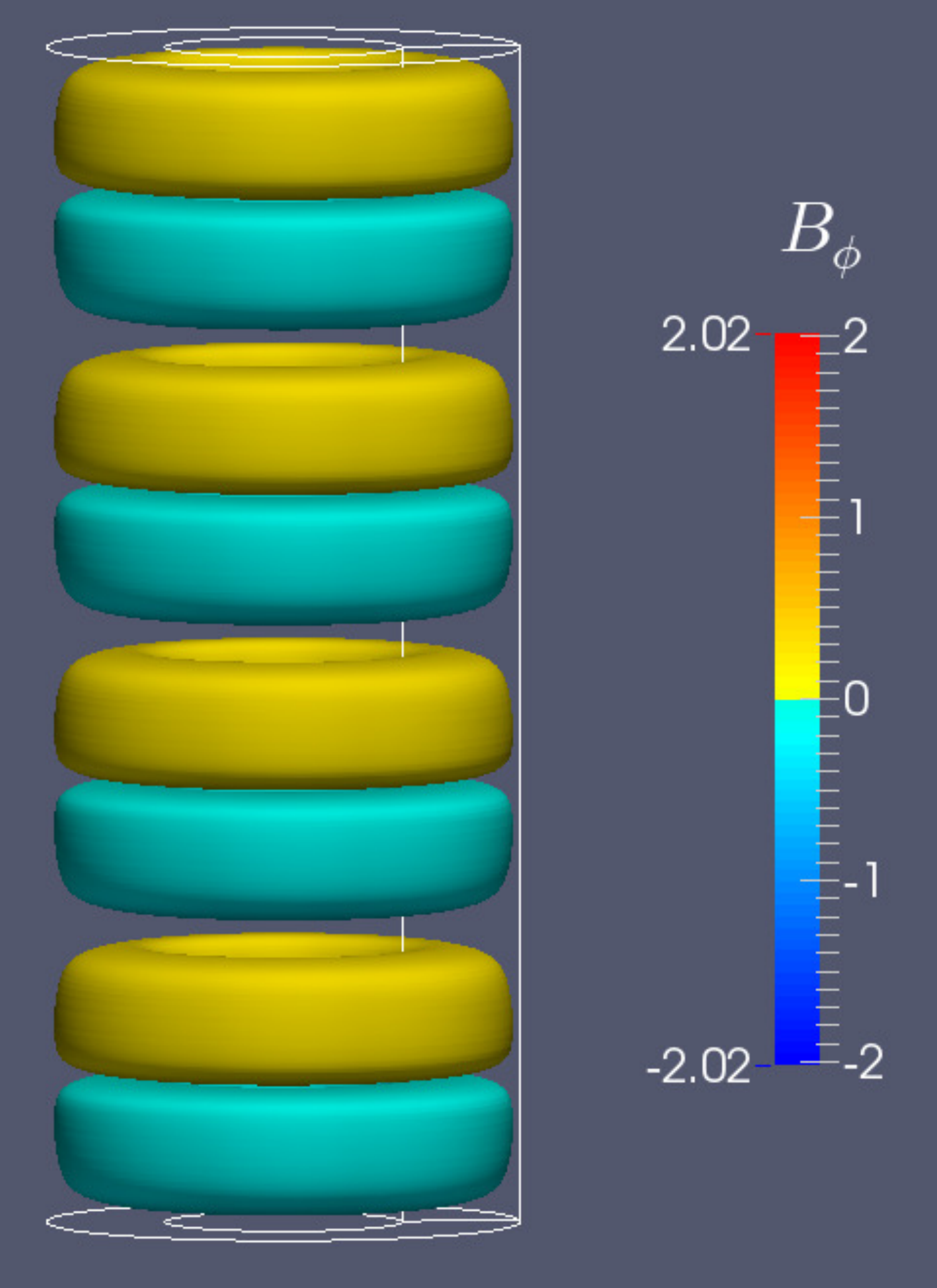}
  \includegraphics[width=4cm,height=6cm]{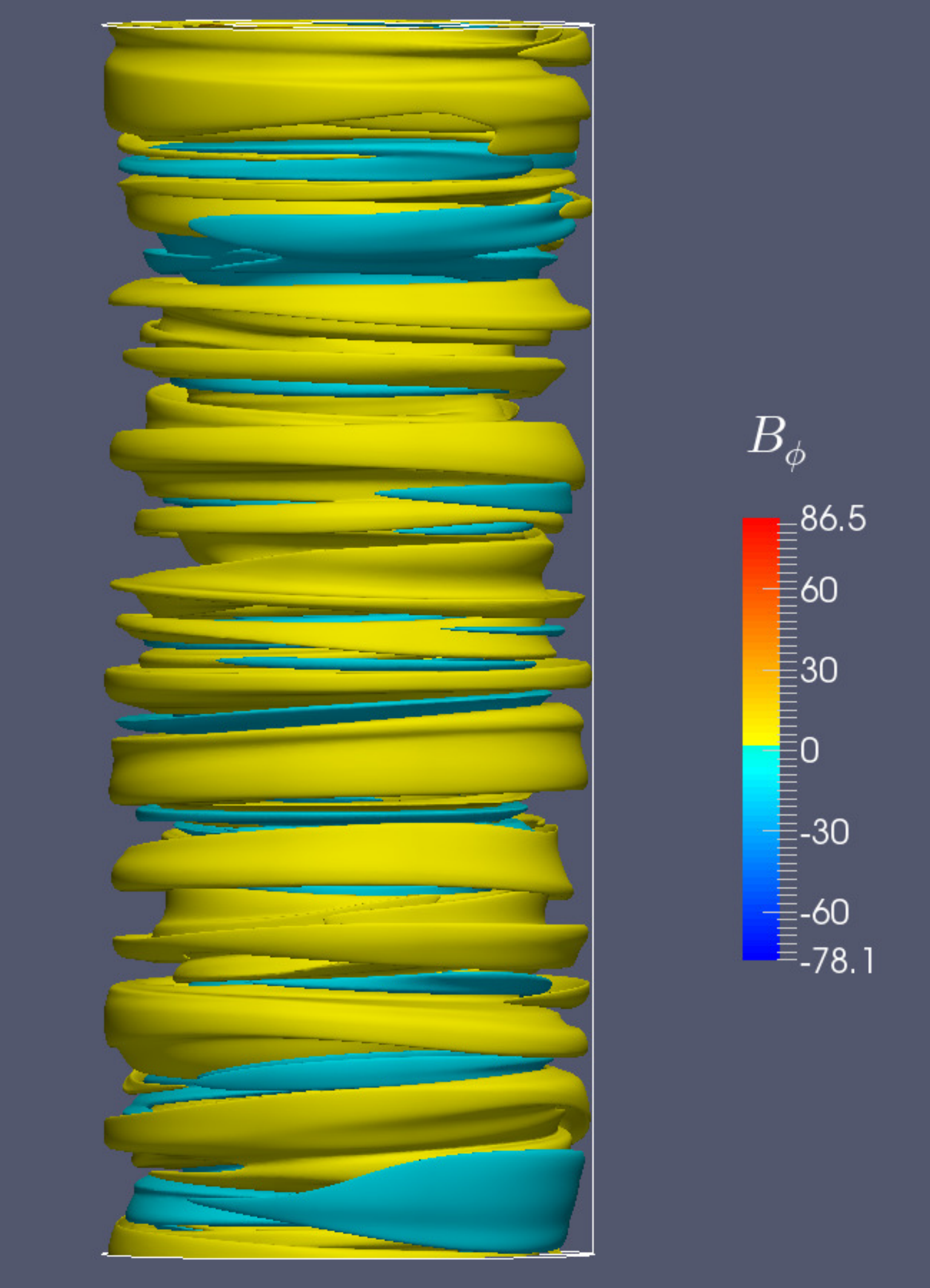}
   \includegraphics[width=4cm,height=6cm]{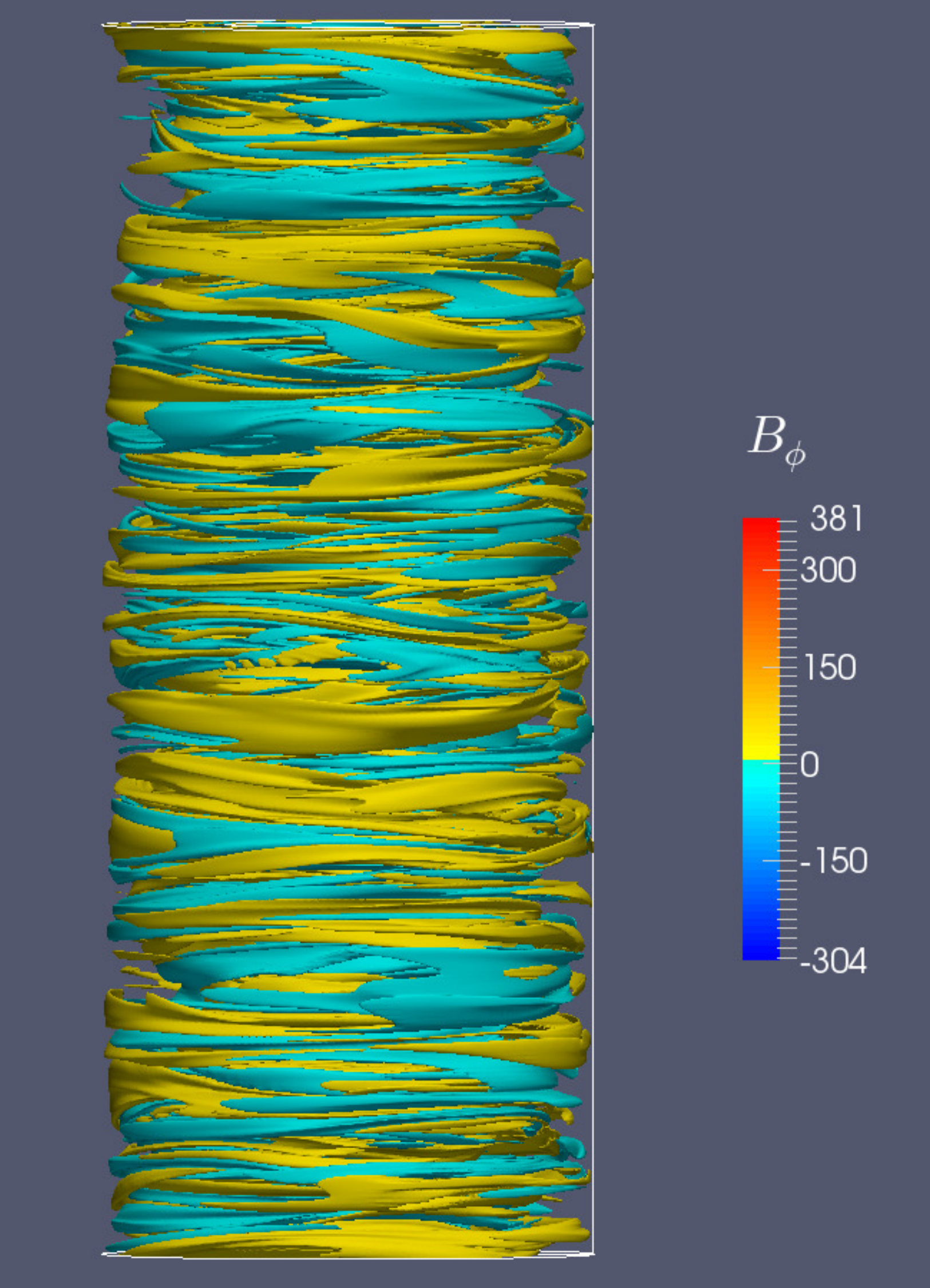}
\caption{As in Fig.~\ref{ha13} but for  $\Pm=100$. $\Ha=13$. It is $\Rmquer=88, 1250, 4000$ for the averaged Reynolds number   and the magnetic Mach numbers are $\Mm = 6.8, 96.2, 307.7$  (from left to right).}
 \label{pm100}
\end{figure}

We also note that for all models with fixed Lundquist number the amplitude of the $b_\phi$-component grows for growing $\Rm$, i.e.~for stronger shear. The magnetic energy averaged over the whole container should be increasingly relevant. For several models with different magnetic Prandtl number $\Pm$ the normalized magnetic energy
\beg
{\rm Q}=\frac{\langle \vec{b}^2\rangle}{B_0^2}
\label{qu}
\ende
is given in Fig.~\ref{energ} in its dependencies on the magnetic Reynolds number and the Hartmann number. The blue (red) curves are for weak (medium) background fields; they only differ by $\Pm$ (the circles and triangles are for $\Pm=1$). One finds $\rm Q\propto \Rm$, the $\Pm$-dependence as rather weak, and an anticorrelation between $\rm Q$ and $\Ha$. There is, however, another clear relation to report. For the magnetic Elsasser number 
\beg
\Lambda= \frac{\langle \vec{b}^2\rangle}{\mu_0\rho\eta\Omin}
\label{Els1}
\ende
one finds from the right panel of Fig.~\ref{energ} the linear relation
$\Lambda \simeq 0.007\ \Rm$ for large $\Rm$ and independent of $\Pm$, 
leading to the simple result $\langle \vec{b}^2\rangle\simeq 0.007\cdot \mu_0\rho R_0^2\Omin^2$ which is identical to 
\beg
{\rm Q}\simeq 0.007\ \Mm^2.
\label{Els11}
\ende
Note that for Kepler disks  $\Mm^2$ equals the plasma-$\beta$ as the ratio of  kinetic  pressure and  magnetic pressure as in  such disks the averaged pressure equals $\rho d^2\Om^2$. The plasma-$\beta$ value of 400 used in Ref.~\cite{FP07a}
corresponds to $\Mm=20$, close to the minimum values used in the simulations  which lead to  Figs.~\ref{ha13} - \ref{pm100}. According to  Eq.~(\ref{Els11}) the resulting  $\rm Q$ will be  expected  as of order unity.

The normalized magnetic energy of the perturbations does  not depend on the microscopic diffusivities. Not even 1\% of the rotation energy of the Taylor-Couette flow exists in the form of stochastic perturbations of the magnetic field. Nevertheless, as the MRI occurs for large magnetic Mach numbers, Eq.~(\ref{Els11}) leads to the conclusion that the energy of the magnetic perturbations may easily exceed the energy of their magnetic background fields. It is unlikely that this finding is changed for much smaller or larger  magnetic Prandtl numbers, as the dependence of the standard MRI on $\Pm$ is basically weak. If a magnetically induced viscosity is defined in a heuristic manner by $\nu_{\rm T}\simeq \langle \vec{b}^2\rangle /\mu_0\rho\Om$ one finds $\nu_{\rm T}\simeq 0.007 \Rin^2\Omin$, which might be relevant for the angular momentum transport in the unbounded differentially rotating container. One can also understand such expressions as a realization of the $\beta$ viscosity concept \cite{DS00,HR01,SJ12}.

Closing this section, the calculations presented in Fig.~\ref{ha13} may be repeated with 
a basically smaller magnetic Prandtl number. The identical models represented in the $(\Lu/\Rm)$ system are numerically repeated for $\Pm=0.01$  rather than $\Pm=1$ (Fig.~\ref{pm001}). The magnetic Mach numbers are thus reduced by a factor of 100; they are now of order unity. The differences between the results in Figs.~\ref{ha13} and \ref{pm001} are surprisingly small, which  demonstrates the basic role of the magnetic Reynolds number (for fixed Lundquist number) for geometry and energy of the MRI perturbations with axial fields. In this sense the role of the magnetic Prandtl number  for excitation  {\em and} formation of the MRI is only small.

Realizations of standard MRI for large magnetic Prandtl number ($\Pm=100$) are given by Fig.~\ref{pm100}.  The values of the averaged  Reynolds number $\Rmquer$ and Hartmann number $\Ha$  correspond to those used in Fig.~\ref{ha13}. Figure \ref{ha13} for $\Pm=1$ and Fig.~\ref{pm100} for $\Pm=100$ with $\Mm=7,100, 300$ provide  the same series of magnetic Mach numbers.  In all cases  the maximum values of $b_\phi/B_0$ grow linearly with growing magnetic Mach numbers so that  the relation (\ref{Els11}) is indeed approached.

\begin{figure}[htb]
\centering
 \includegraphics[width=9.0cm]
 {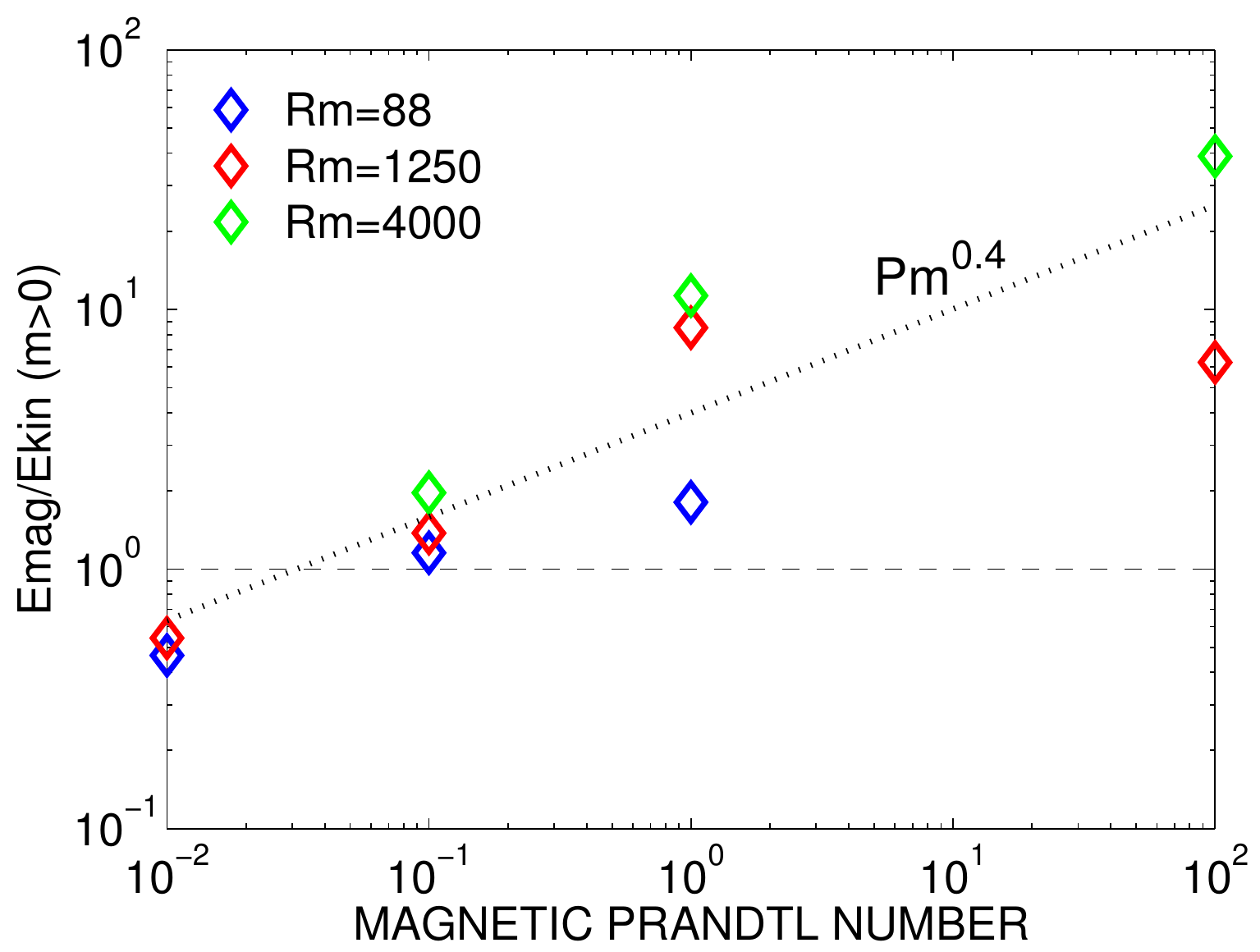}
 \caption{The ratio (\ref{ratio}) of the magnetic and the kinetic energy for the modes with $m>0$ of standard MRI with quasi-Keplerian rotation law and magnetic Prandtl numbers in the interval between 0.01 and 100, \ref{pm001}, \ref{pm100}). $\rin=0.5$,  $\mu_\Om=0.35$, $\Lu=13$.}
 \label{energies}
\end{figure}

The  models with  magnetic Prandtl numbers in the interval between 0.01 and 100 (Figs.~\ref{ha13}, \ref{pm001}, \ref{pm100}) have been used to calculate the   ratio $\varepsilon$ of magnetic to kinetic energy 
\begin{equation}
 \varepsilon=\frac{\Ha^2}{\Pm}\ \frac{\langle \vec{b}^2\rangle}{ \langle \vec{u}^2\rangle}
 \label{ratio}
\end{equation}
averaged over the container for various  magnetic Reynolds numbers. Here $\vec{u}^2$ and $\vec{b}^2$ have the same dimension. The results show only a slight  dependence of the energy ratio on the magnetic Prandtl number $\varepsilon\propto {\Pm}^\kappa$ with $\kappa\sim 0.4$ (Fig.~\ref{energies}). The  larger $\Pm$ the larger the magnetic energy related to the kinetic energy. For small $\Pm$ the fluid becomes less and less magnetized. On the other hand, the magnetic energy  dominates the kinetic energy only for large values of $\Pm$. Written with a more  appropriate normalization one finds for the normalized kinetic energy 
\begin{equation}
 \frac{ \langle \vec{u}^2\rangle}{\Omin^2 R^2_0}=\frac{0.007}{\varepsilon}\simeq 0.007\  \Pm^{-\kappa}. 
 \label{kinen}
\end{equation}
with $\kappa\lsim 0.4$. Again, for $\varepsilon\simeq 1$  the kinetic energy is also  only about 1\% of the rotational energy of the system, but it is much higher for small $\Pm$. The influence of the magnetic Prandtl number on this result is not very strong. It is weaker than the expected coefficient of order unity and it is slightly larger than the $\kappa\simeq 0.2$ which has been derived from numerical shearing-box simulations  \cite{KK11}. For the very small magnetic Prandtl numbers of liquid metals, however,  one expects much smaller $\varepsilon$-values hence the MHD turbulence is only weakly magnetized. Nevertheless, the small exponents $\kappa$ suggest the standard  MRI as rather robust against variations of the magnetic Prandtl number. 

\subsection{Angular momentum transport}\label{AMT}
In Keplerian accretion disks the rotation velocities are supersonic with $\Rin\Omin\gg c_{\rm ac}$ with $c_{\rm ac}$ as the speed of sound, so that in particular the magnetically induced viscosity may adopt high values. The angular momentum transport by MRI thus plays an important role in  theoretical astrophysics and  should thus be considered here  in more detail. The radial angular momentum transport by MHD turbulence can be expressed by the component
\begin{equation}
 T_{R\phi}=\langle u_R u_\phi - \frac{1}{\mu_0 \rho} b_R b_\phi\rangle 
\label{Trfi}
\end{equation}
of the Reynolds and Maxwell stresses. The averaging procedure may be an integration over time and the whole container. Within the Boussinesq approximation one always has $T_{R\phi}\cdot{\rm d}\Om/{\rm d}R<0$, as the angular momentum transport  $T_{R\phi}$ is thought to be opposite to the gradient of $\Om$ \cite{B97}. It is thus convenient to  introduce a scalar factor, the so-called eddy viscosity $\nu_{\rm T}$, by
\begin{equation}
 T_{R\phi}=- \nu_{\rm T} R \frac{{\rm d}\Om}{{\rm d}R}
\label{T1}
\end{equation}
with positive $\nu_{\rm T}$. The sign of these correlations can even be computed with the linear theory.  Here we shall use nonlinear simulations to also compute the amplitude of the eddy viscosity. 

One may introduce dimensionless coefficients $\alpha$ via $T_{R\phi} =\alpha \Om^2 R_0^2$ \cite{SS73}. Hence, the MRI $\alpha$ can be computed with the definition $\alpha_{\rm mri}= {T_{R\phi}}/{\Om^2 R_0^2}$. Note that this definition differs from the one used in astrophysics unless $R_0\simeq H$, which is only fulfilled for thick accretion disks. It follows that
\beg
\frac{\nu_{\rm T}}{\nu}= \frac{\alpha_{\rm mri}}{q}~\Rey,
\label{alf1}
\ende
where the rotation profile $\Om\propto R^{-q}$ has been used ($q=3/2$ for Keplerian rotation).

\begin{figure}[htb]
 \centering
 \includegraphics[width=9cm]{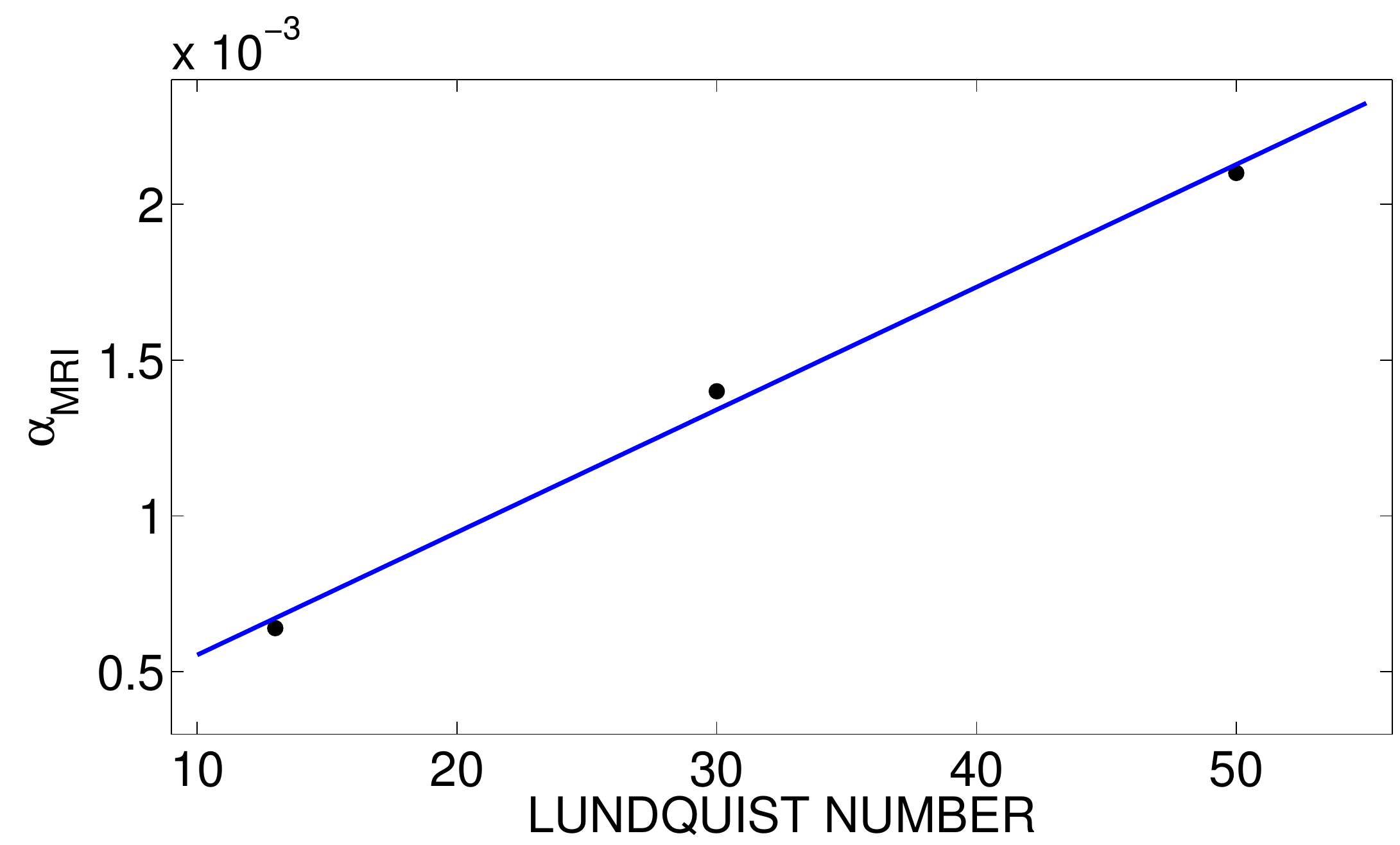}
\caption{$\alpha_{\rm mri}$ versus the Lundquist number  $\Lu$   for quasi-Keplerian rotation. 
Each symbol represents many models with varying $\Rey$ and for $\Pm$ between $\Pm=0.1$ and $\Pm=1$. $\alpha_{\rm mri}$ linearly grows with $\Lu$. $\rin=0.5$, $\mu_\Om=0.35$. Perfectly conducting boundaries.  From \cite{GR12}.}
 \label{alphamri}
\end{figure}

As a first step we compute $\alpha_{\rm mri}$ with $\mu_\Om=0.35$ by averaging only over the azimuth. One finds that the angular momentum transport is positive everywhere, with a rather weak indication of a cell structure. The angular momentum transport shown in Fig.~\ref{alphamri} is again only due to the nonaxisymmetric modes with $m>0$. Only these modes have here been defined as the fluctuations in the definitions of $\vec{u}$ and $\vec{b}$. Let the averaging procedure concern the entire container. Our results for $\alpha_{\rm mri}$ lead to  the linear relation
\begin{equation}
\alpha_{\rm mri}= 5 \cdot 10^{-5}\ {\rm S} 
\label{alf2}
\end{equation}
(Fig.~\ref{alphamri}). The numerical value of  $\alpha_{\rm mri}$ depends linearly on the amplitude of the magnetic field, the size of the disk or torus and the electric conductivity. This relation proves to hold for all Reynolds numbers and magnetic Prandtl numbers. Note that $\alpha_{\rm mri}$ does not vary with the rotation rate and/or the microscopic viscosity. There is thus no dependence of the $\alpha_{\rm mri}$ on the magnetic Prandtl number. It does not, in particular, decrease for  decreasing magnetic Prandtl number  as  suggested by a few  shearing-box simulations \cite{LL07a,MF15}. For the models used in Fig.~\ref{energies}  which all belong to one and the same Lundquist number $\Lu=13$, Eq.~(\ref{alf2}) leads to $\alpha_{\rm mri}= 0.65\cdot 10^{-3} $,  in  accordance with   results of the box simulations in Ref.~\cite{KK11} -- also with respect to the nonexistence of a  $\Pm$-dependence of  $\alpha_{\rm mri}$.

For two examples for $\Pm=1$ (green diamonds in Fig.~\ref{alphamri}) even the outer boundary condition has been changed from perfectly conducting to insulating. The numbers do not show any influence of the boundary conditions on the resulting $\alpha_{\rm mri}$. Equation (\ref{alf2}) also implies that the microscopic viscosity has no essential influence on the angular momentum transport parameter $\alpha_{\rm mri}$ and, moreover, does not influence the eddy viscosity values, see Eq.~(\ref{alf1}).

As an astrophysical application of the compact result (\ref{alf2}), we ask how strong the axial magnetic field must be in order to produce $\alpha_{\rm mri}=1$. For a protoplanetary disk $\eta=10^{15}$cm$^2$/s and $\rho= 10^{-10}$g/cm$^3$ can be assumed \cite{BS05}. Hence $\Lu\simeq 10^3 (B_0/1{\rm G})(R_0/10 {\rm AU})$, so that $B_0=1$~G is needed for $\alpha_{\rm mri}=0.05$. It is obvious that the magnetic field amplitude must not be much smaller than about 1 G in order to get $\alpha_{\rm mri}$ values of the needed  order. The immediate consequence  is that dipolar large-scale stellar fields as the source of the background fields for MRI-induced eddy viscosities must be excluded.

\section{Azimuthal magnetorotational instability (AMRI)}\label{AMRI}
According to Michael's criterion (\ref{bfcr}) hydrodynamically stable flows are also stable under the influence of curl-free azimuthal magnetic fields, i.e.~$B_\phi\propto 1/R$. On the other hand, all rotation laws between two insulating cylinders in the presence of toroidal fields due to an axial current inside the inner cylinder are stable against axisymmetric perturbations \cite{V59,HS06}. The reason is simple: the axisymmetric version of Eq.~(\ref{bR}) fully decouples from the system so that this magnetic component  decays because of missing energy sources. In particular, it cannot generate  induction energy by the differential rotation term in Eq.~(\ref{bphi}).
An axisymmetric magnetorotational instability with purely azimuthal fields is thus  not possible. These results, however, only hold for axisymmetric perturbations so that we have to ask for possible instability of nonaxisymmetric modes, which can indeed  arise \cite{OP96,RHS07}. Because of the absence of large-scale electric currents in the fluid between the cylinders we have called this phenomenon the Azimuthal MagnetoRotational Instability (AMRI). We shall derive in this section the theoretical background of this nonaxisymmetric instability, including its first experimental realization in a laboratory. In the entire section the Hartmann number is defined in accordance with  (\ref{Hartmannin}).

\subsection{Potential flow}\label{potflow}
\begin{figure}[htb]
\centering
 \includegraphics[width=8cm]{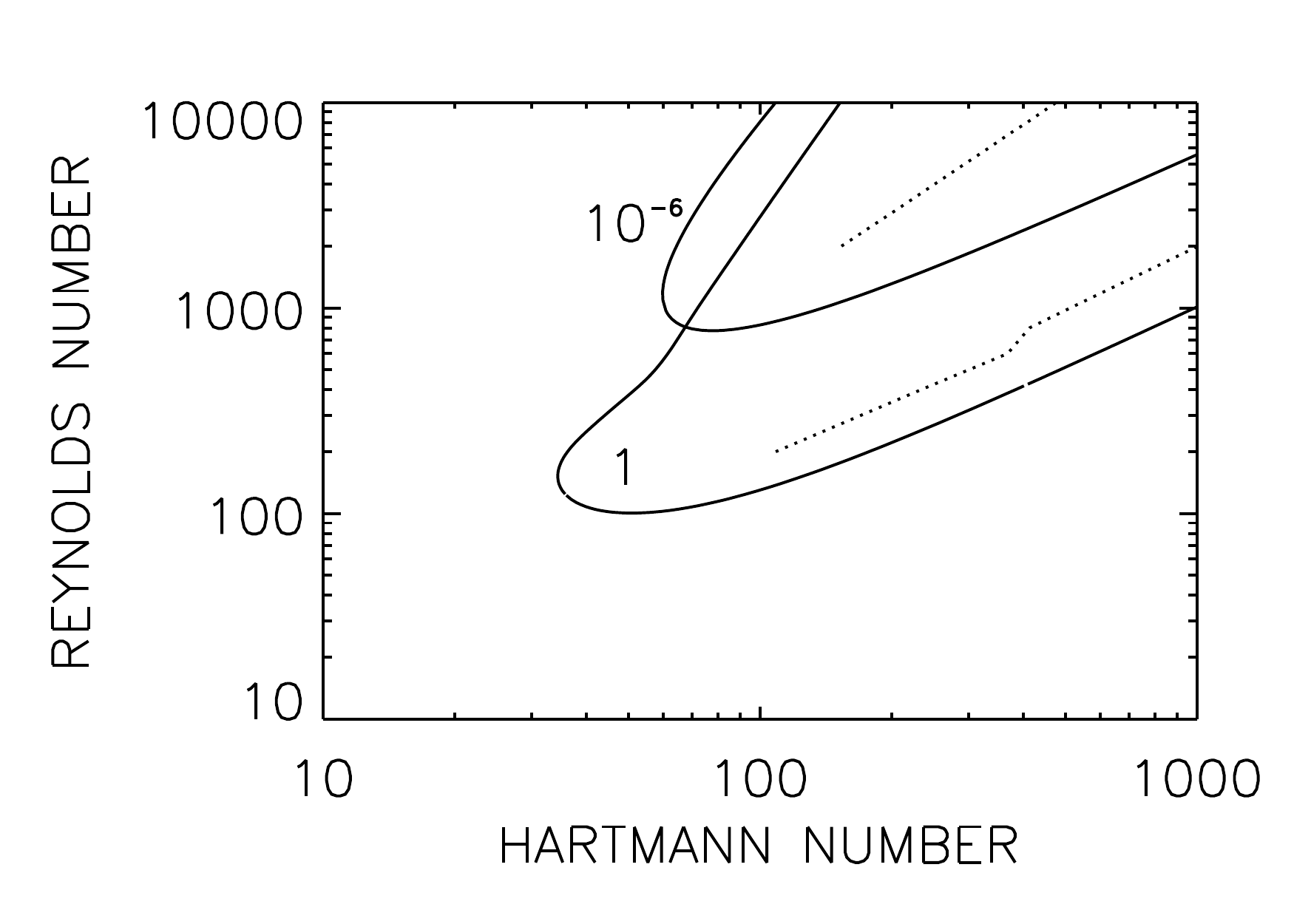}
\includegraphics[width=8cm]{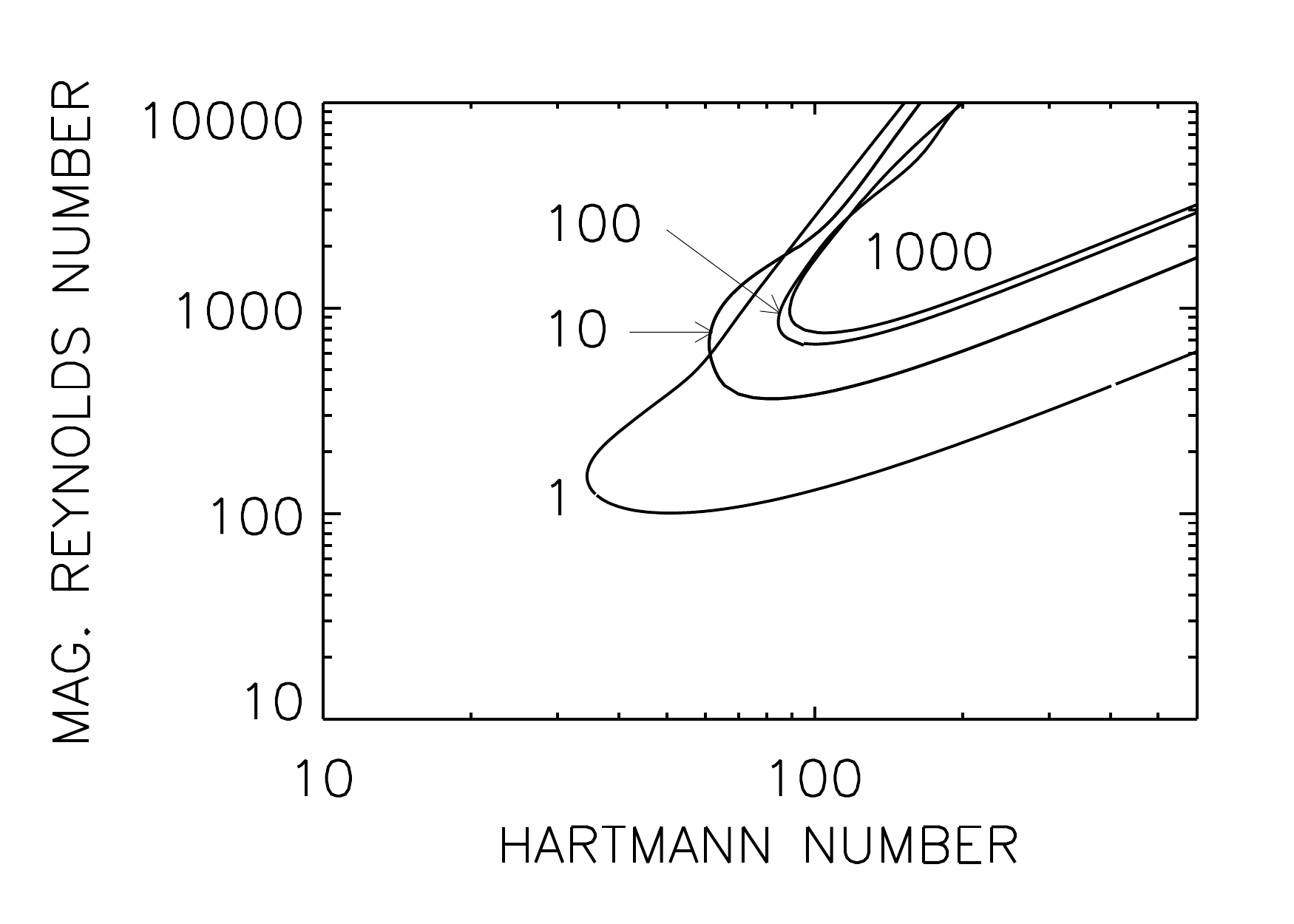}
 \caption{Stability lines of the potential flow for various magnetic Prandtl numbers influenced by current-free azimuthal magnetic fields. Left: all curves for $\Pm<10^{-4}$ are basically identical (for small $\Pm$ the curves scale with $\Ha$ and $\Rey$).  The dotted lines mark the location with maximal growth rates. Right: for large $\Pm$ the potential flow scales with the magnetic Reynolds number $\Rm$. $m=1$,  $\mu_B=\rin=0.5$,  $\mu_\Om=0.25$, perfectly conducting boundaries. From \cite{RS15}.}
 \label{f16}
\end{figure}

For the curl-free magnetic field with $B_\phi\propto 1/R$ (i.e.~$\mu_B=\rin$), Fig.~\ref{f16} shows the lines of marginal stability for the potential flow with $\Om\propto 1/R^2$. Note that precisely this combination fulfills the condition $B_\phi\propto U_\phi$ which corresponds to a very special type of MHD flow (Chandrasekhar-type flows, see Section \ref{Chandra}). One finds that the instability for $m=1$ always exists between a minimum and a maximum Reynolds number. Too slow or too fast rotation enforces stability. The upper branch limits the instability domain by suppressing the nonaxisymmetric instability by too strong shear while the lower branch is defined by the minimum shear energy needed for the instability. The location of the maximum growth rate marked by dots in the left panel of Fig.~\ref{f16} is closer to the lower branch than to the upper branch.

The curves in the $(\Ha/\Rey)$-plane converge for small $\Pm$ and are no longer visible as distinct curves. For increasing magnetic Prandtl number the value of the minimum Reynolds number decreases, and the smallest critical Hartmann number is reached for $\Pm\simeq 0.1$. For very small $\Pm$ the minimum of the instability cone scales with $\Rey$, here with a value of about $\Rey\simeq 800$, while the associated Hartmann number is ten times less. The coordinates of the characteristic minimum of the lines of neutral stability for all $m$ and for small magnetic Prandtl numbers are given in Fig.~\ref{f18}. The curves in this plot demonstrate how for $\Pm\to 0$ the lines in the ($\Ha/\Rey$) plane no longer depend on the value of $\Pm$, that is, the instability scales with $\Rey$ and $\Ha$ for small $\Pm$.

The right panel of Fig.~\ref{f16} demonstrates the scaling of the instability curves for large $\Pm$ in the ($\Ha/\Rm$) plane. The curves converge for $\Pm\to \infty$ for magnetic Reynolds number of about 1000 and with minimal Hartmann number of about 100. 
The use of the average Reynolds number ${\Rmquer}=\sqrt{\Rey\cdot \Rm}$ as the vertical axis leads to additional findings. The dotted line in Fig.~\ref{f16} (right) represents the location of the limit $\Mm=1$. Note that the main part of the cones for $\Pm>1$ lies above the dotted line while it lies below this line for $\Pm<1$. For $\Pm\to 0$ the entire instability domain no longer reaches values with $\Mm>1$. The relevance of AMRI for super-Alfv\'enic astrophysical applications might thus be rather restricted. On the other hand,  for $\Pm\to \infty$ the instability cone never reaches values with $\Mm<1$. 
\begin{figure}[htb]
\centering
\includegraphics[width=8cm]{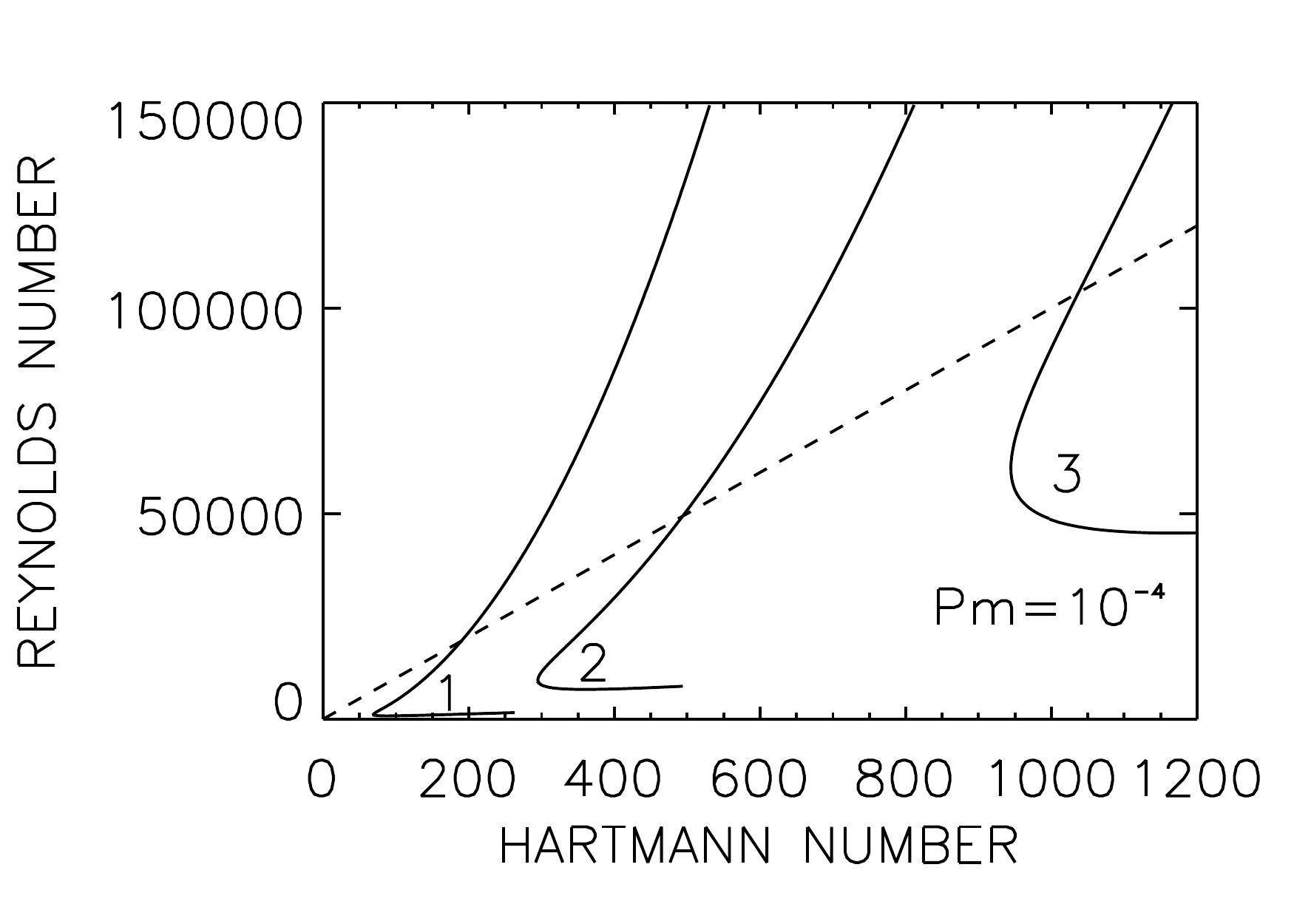}
 \includegraphics[width=8cm]{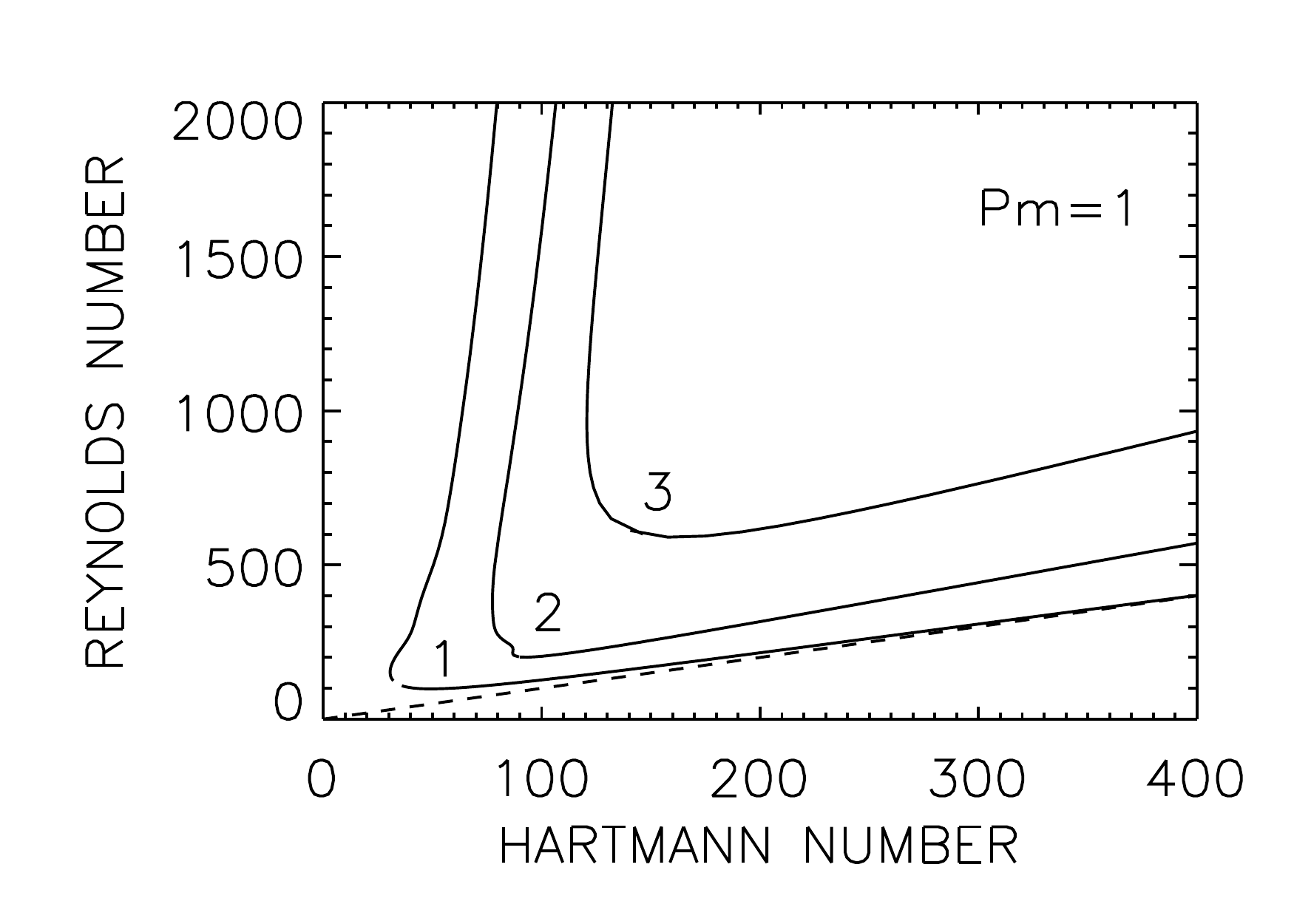}
 \caption{Stability maps for the modes with  $m\geq1$ (marked) for potential flow with $\Pm=10^{-4}$ (left) and $\Pm=1$ (right). All curves for $\Pm<10^{-4}$ are identical. The dotted lines mark $\Mm=1$. For sufficiently small $\Pm$ all curves are located below $\Mm=1$. For fixed $\Ha$ the unstable modes  decay for too slow and  too fast rotation.  $\mu_B=\rin=0.5$,   $\mu_\Om=0.25$. Insulating boundaries.}
 \label{f17}
\end{figure}

The eigenvalues for modes with $m>1$ are given in Fig.~\ref{f17} for two different magnetic Prandtl numbers. These curves also have the characteristic form consisting of lower and upper branches with positive slopes, so again the rotation can be too slow or too fast for instability. For higher $m$ the instability domains are smaller than for lower $m$. For all $m$ the minima of the curves for $\Pm\to 0$ move below the dashed line $\Mm=1$. The absolute minimum values of $\Rey$ and $\Ha$ of all curves are plotted in Fig.~\ref{f18}. It shows the $m=1$ mode as the most unstable mode with the lowest Reynolds and Hartmann numbers. Decreasing $\Pm$ shifts the minimum values to higher values of $\Rey$ and $\Ha$, and this the more the greater $m$ is. For small $\Pm$ the excitation of the higher modes requires much higher Reynolds and Hartmann numbers than for $\Pm=1$. The plots also show that for $\Pm\to 0$ all the considered azimuthal modes scale with $\Rey$ and $\Ha$. Because of $\Mm=\sqrt{\Pm}\Rey/\Ha$ for small $\Pm$ all minima are thus sub-\A{ic}. We shall demonstrate below that these results are typical for the  Chandrasekhar-type  MHD flows. These results do not remarkably depend on the choice of the boundary conditions.
\begin{figure}[htb]
\centering
 \includegraphics[width=8cm]{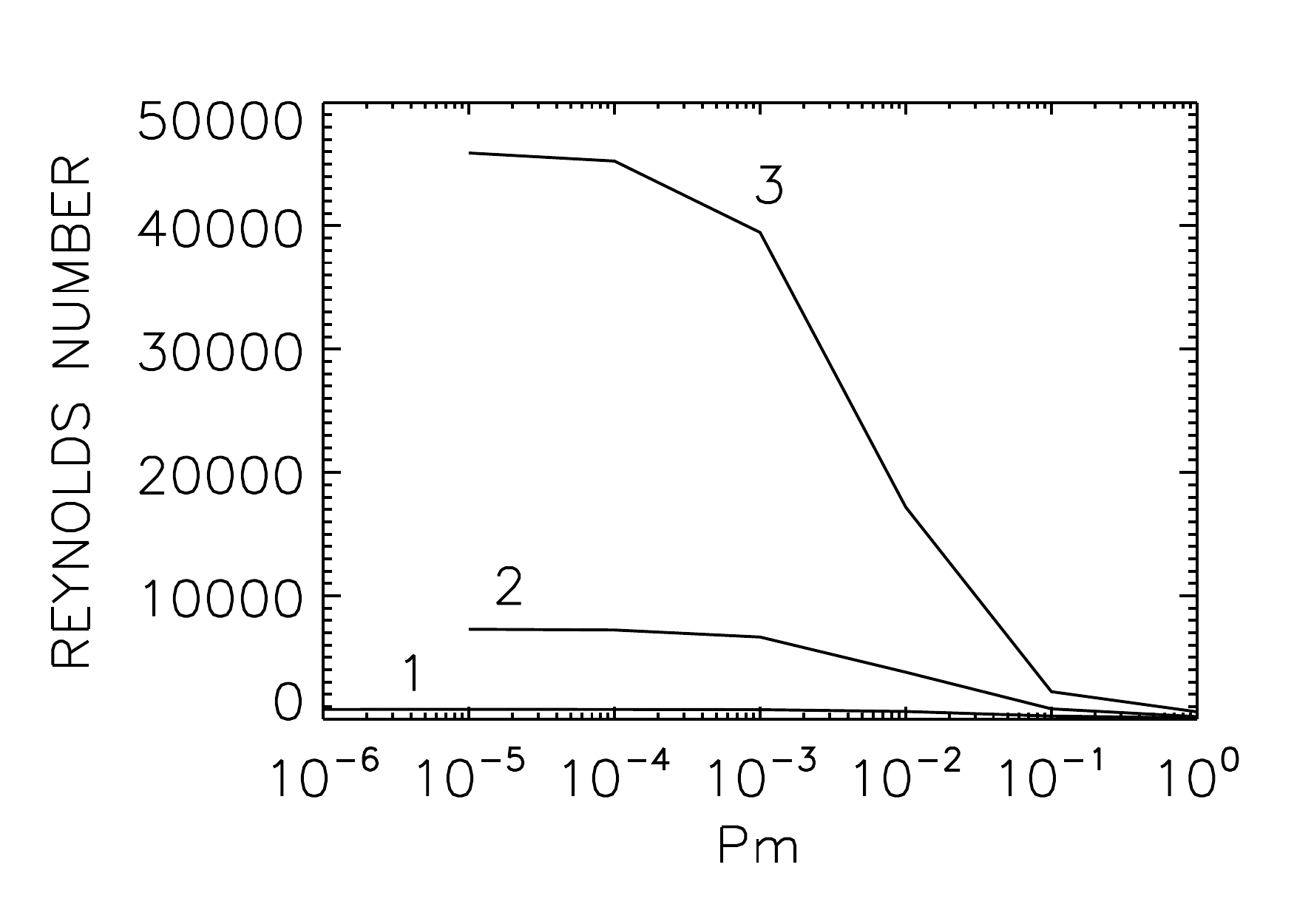}
 \includegraphics[width=8cm]{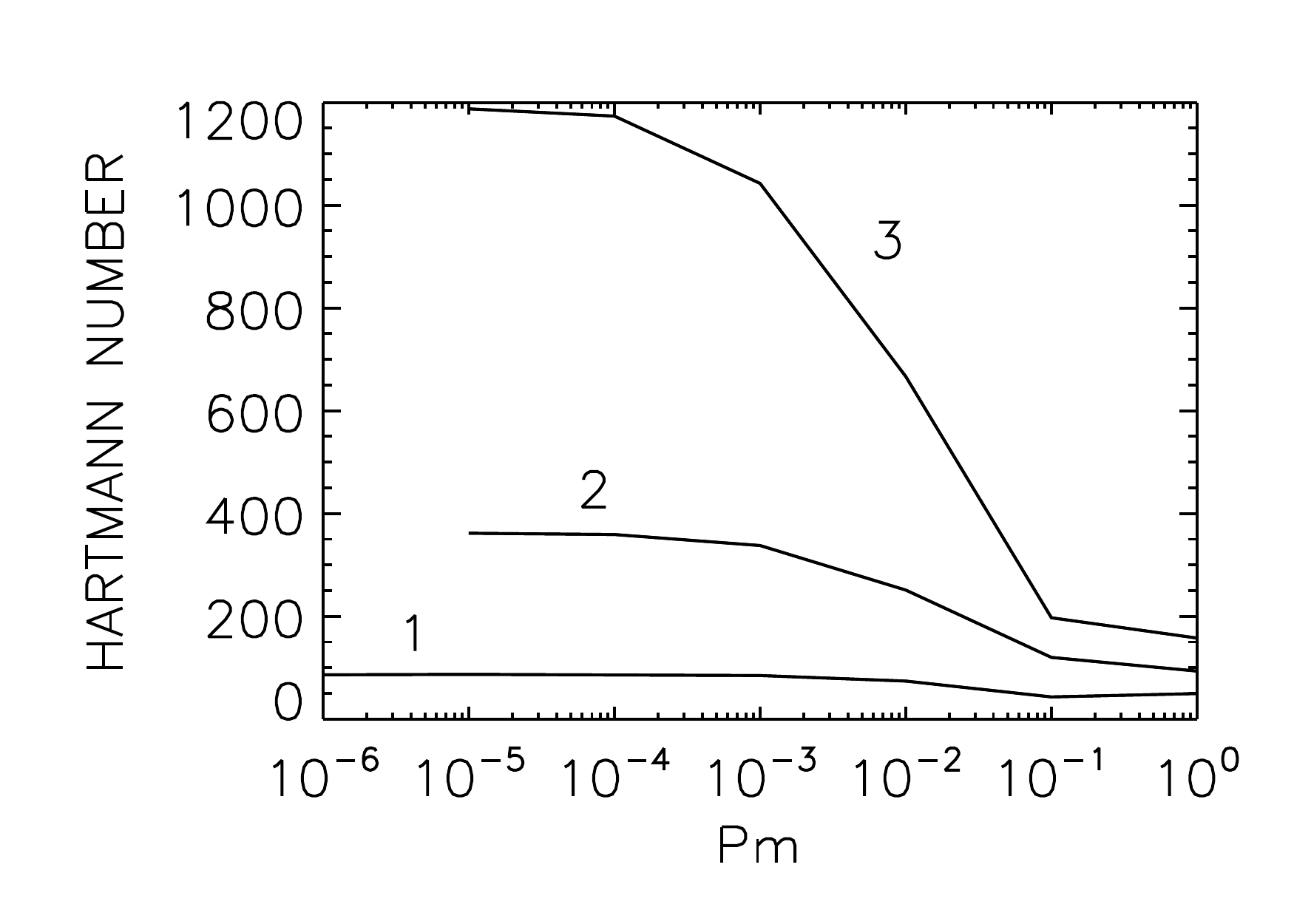}
 \caption{The $\Pm$-dependence of the minimum Reynolds numbers (left) and Hartmann numbers (right)   of the modes  $m=1,2,3$ taken from many models  similar to those used in  Fig.~\ref{f17}.}
 \label{f18}
\end{figure}

\begin{figure}[htb]
\centering
 \includegraphics[width=8.0cm]{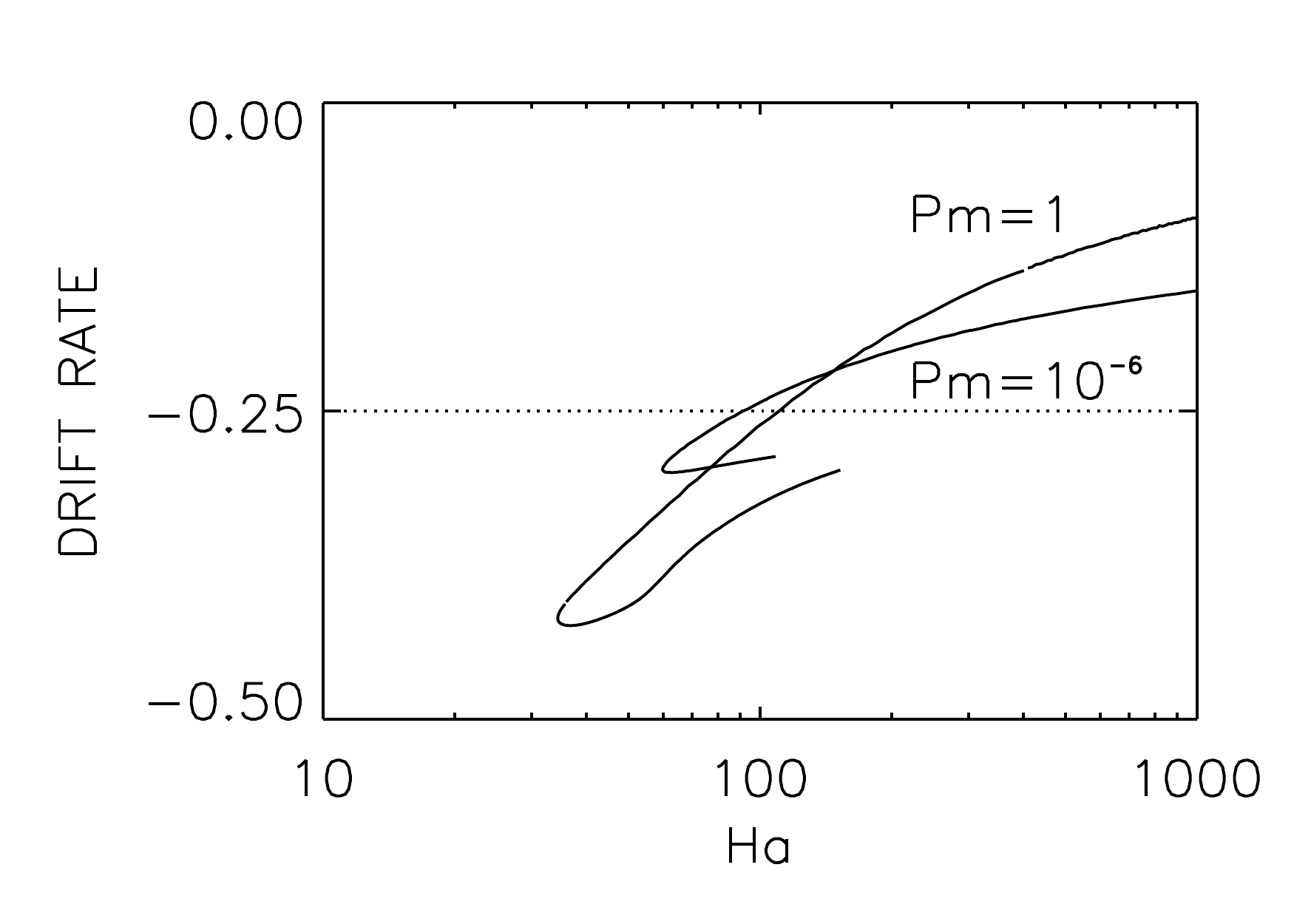}
  \includegraphics[width=8cm]{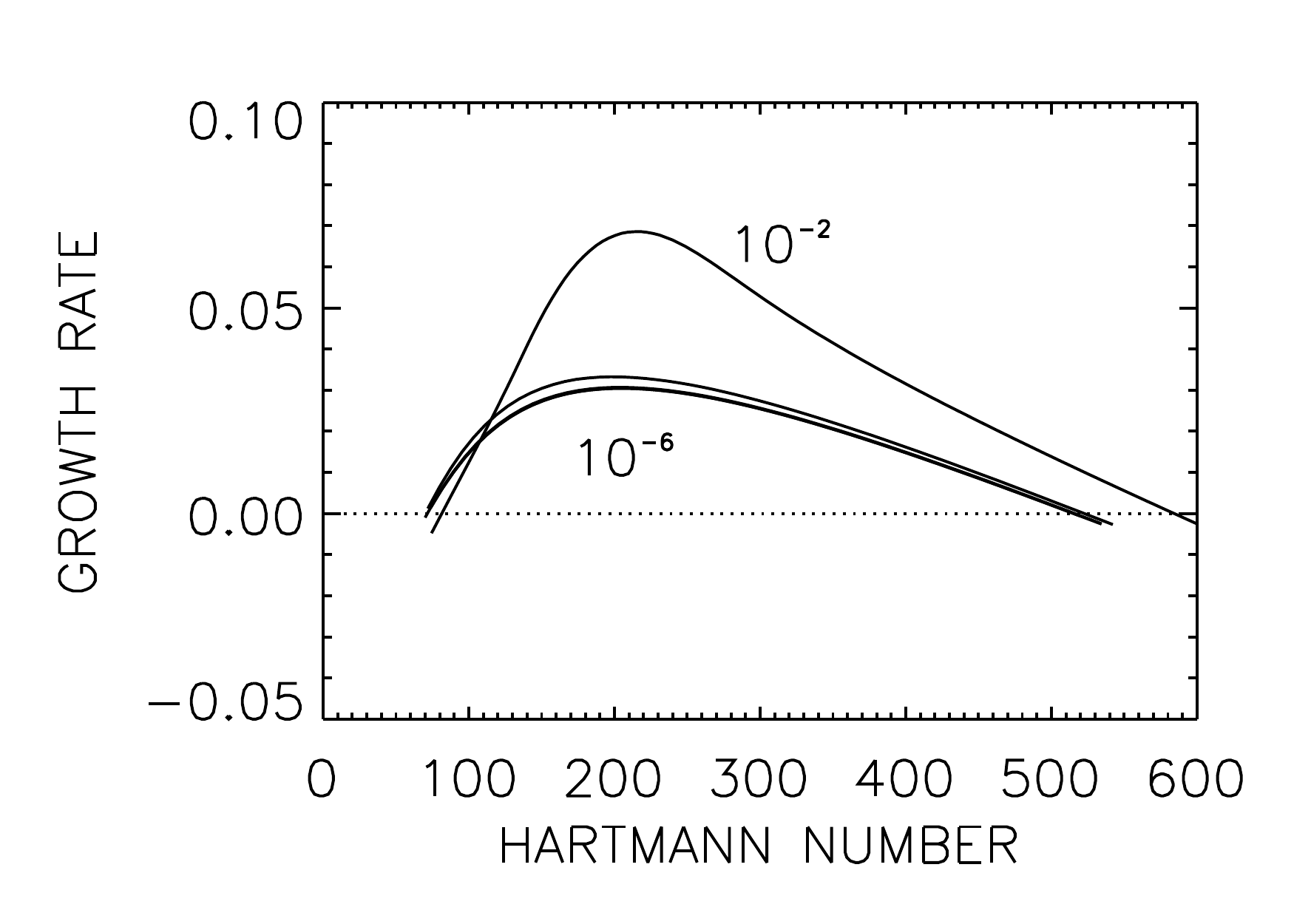}
 \caption{Left: the drift frequency (\ref{omdr}) along the lines of neutral stability are always negative (prograde migration). The dotted line marks the value $-0.25$ indicating exact corotation with the outer cylinder. All curves for $\Pm<10^{-4}$ are identical. 
 Right: growth rates (\ref{growthrate}) along a line of constant Reynolds number ($\Rey=3000$). The lines from top to bottom correspond to $\Pm=10^{-2}-10^{-6}$.   The curves for $\Pm<10^{-4}$ are almost identical.
    $m=1$, $\mu_B=\rin=0.5$, $\mu_\Om=0.25$. Perfectly conducting boundaries.}
 \label{f19}
\end{figure}

Figure \ref{f19} (left) gives an example for the drift rate (\ref{omdr}) along the lines of neutral stability. In this normalization the outer cylinder has a rotation rate of $\mu_\Om$. The real part of the Fourier frequency $\omega$ has the opposite sign as the azimuthal migration of the pattern. Since  always  $\omega_{\rm dr}<0$, the instability pattern drifts in the direction of the basic rotation (prograde migration). A typical value of the drift in units of $\Omin$ for small $\Pm$ is $-0.25$ (marked in the plot), so that for $\mu_\Om=0.25$ the pattern basically corotates with the outer cylinder. For stronger fields the drift is slower. 

We have still to ask how the growth rates behave between the two branches of neutral stability for a given Reynolds number. The growth rate is the negative imaginary part of the eigenfrequency $\omega$. In relation to the experiment described in Section \ref{expamri} we take $\Rey=3000$ for the fixed Reynolds number. The right panel of Fig.~\ref{f19} clearly shows the convergence of the growth rates 
\beg
\omega_{\rm gr}=-\frac{\Im(\omega)}{\Om_{\rm in}}
\label{growthrate}
\ende
for small $\Pm$ so that it makes sense to probe their saturation between the two branches. It is obviously enough for the limit of small $\Pm$ to calculate the growth rates along the upper dotted line in Fig.~\ref{f16}. The growth rates grow for growing Hartmann numbers. The maximum growth rate for very rapid rotation is 0.050 $\Om_{\rm in}$, so that the shortest growth time of AMRI for the potential flow is about 0.9 rotation times of the {\em outer} cylinder, which is just of the order of the growth time for the standard MRI \cite{RG14}.
\begin{figure}[htb]
\centering
 \includegraphics[width=4cm,height=8cm]{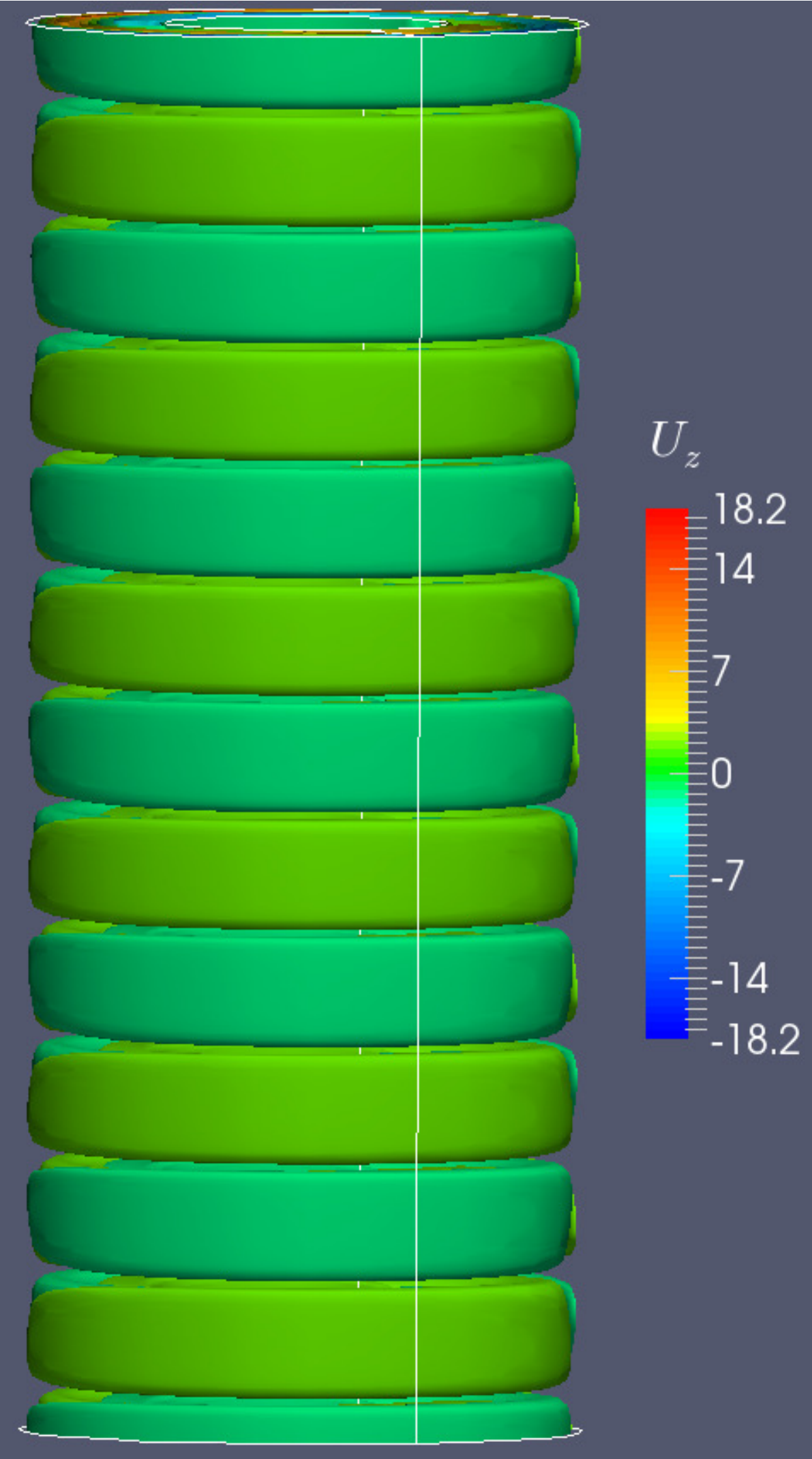}
 \includegraphics[width=4cm,height=8cm]{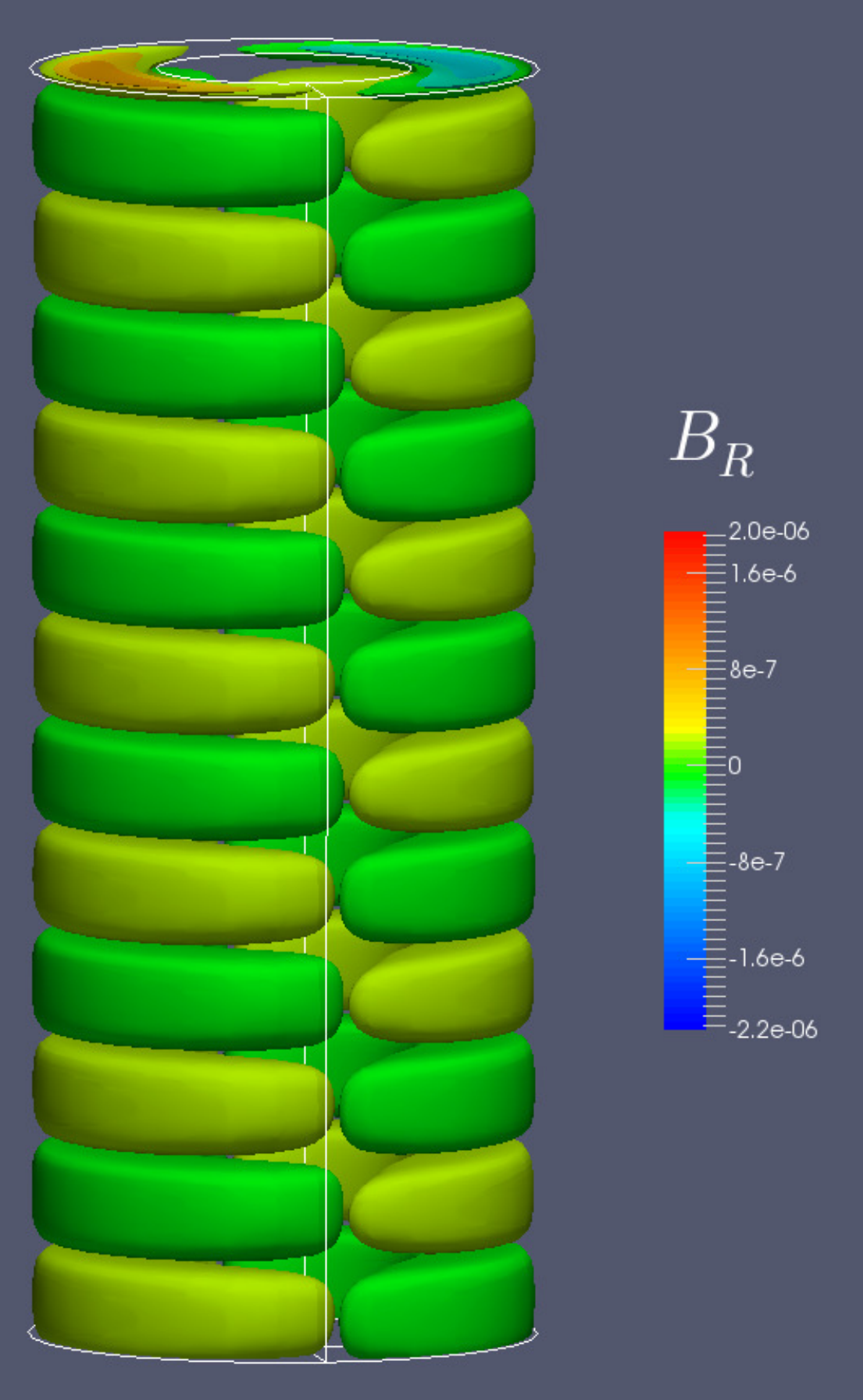}
 \caption{Snapshot of the isolines of axial flow (left) and radial magnetic field (right) patterns for axially unbounded  containers. The flow component $u_z$ is given in form of Reynolds numbers $u_z d/\nu$ and the field component $b_R$ is normalized with $B_{\rm in}$. The modes with $m=\pm 1$ drift in the positive $\phi$ direction with identical  rates. The axial flow pattern is also nonaxisymmetric, but there is a remarkable phase shift  to the radial field. The cells are slightly elongated in axial direction. The energy ratio  (\ref{ratio}) is \ord{10^{-5}}  determined with  the maximal values of $u_z$ and $b_R$. $\Rey=1500$, $\Ha=100$, $\mu_B=\rin=0.5$, $\mu_\Om=0.26$, ${\rm Pm}=10^{-5}$. Insulating boundary conditions.}
 \label{f20}
\end{figure}
\begin{figure}[ht]
\centering
\includegraphics[width=9cm]{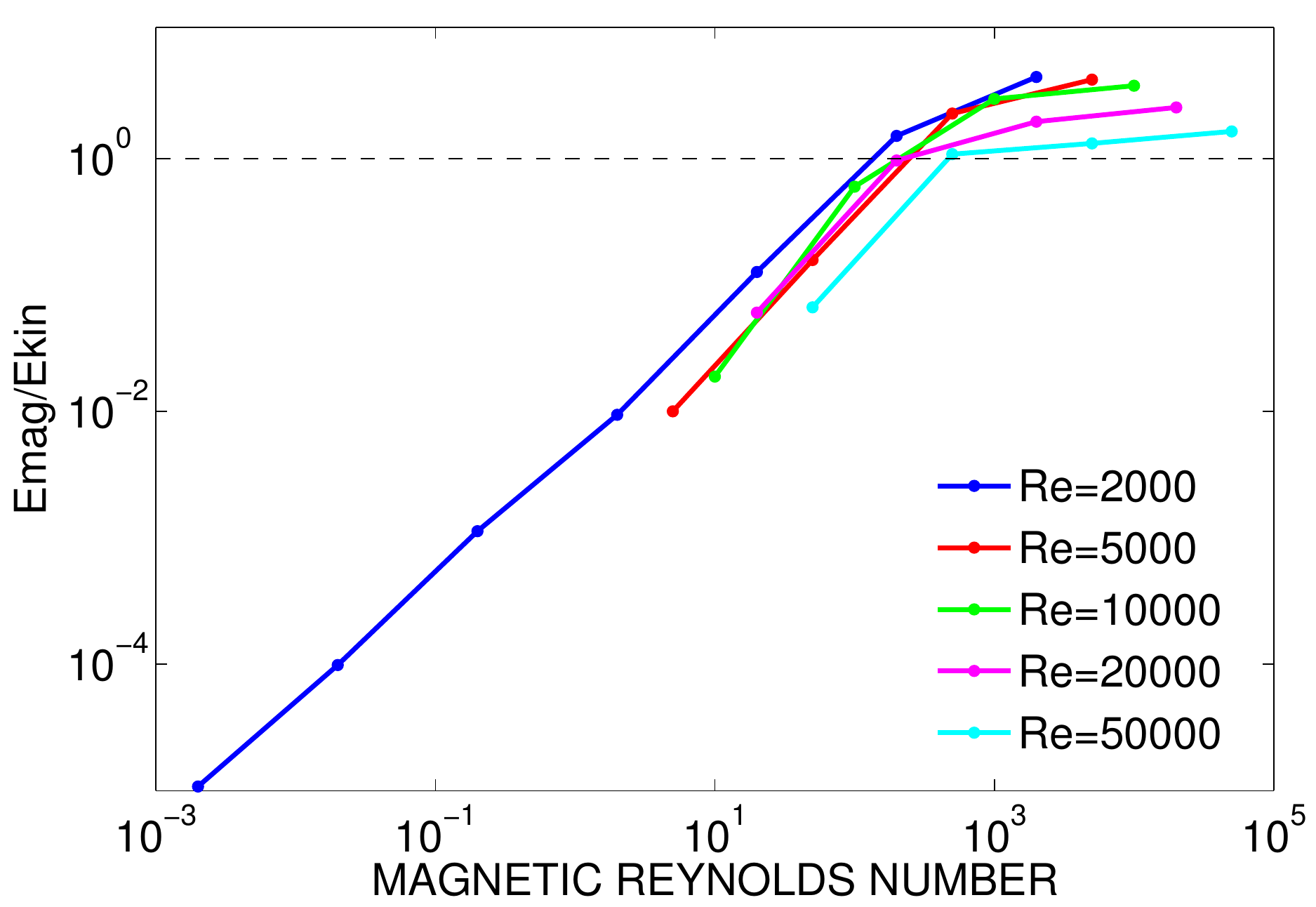}
\caption{Ratio (\ref{ratio}) of magnetic to kinetic energy as  function of $\Rm$.  $\mu_B=\rin=0.5$, $\mu_\Om=0.25$. $\Ha=600$,  insulating boundaries.}
\label{f25d}
\end{figure}

For the small magnetic Prandtl number $\Pm=10^{-5}$ in Fig.~\ref{f20} the isolines of the axial  component of the flow and the radial component of the magnetic field for axially unbounded containers with nearly potential flow are given. The flow is measured in the form of Reynolds numbers $u_R d/\nu$, and the field is normalized with $B_{\rm in}$. The shear parameter $\mu_\Om=0.26$ (as also in Figs.~\ref{f20a} and \ref{f20b}) has been chosen here to correspond to the experiments described in Section \ref{expamri}. For small Reynolds numbers the pattern is nonaxisymmetric with $m=1$. The maximal flow amplitude exceeds the maximal field amplitude by many orders of magnitudes, contrary to the common assumption that kinetic and magnetic energies of magnetohydrodynamic turbulence `ought'  to be equipartitioned. It is thus necessary to study the ratio of both energies in more detail, which leads to a surprising result. The left panel of Fig.~\ref{f25d} shows the ratio (\ref{ratio}) 
for various Reynolds numbers as a function of the magnetic Prandtl number. The Hartmann number is fixed. The result is that for small magnetic Prandtl number ($\Pm\lesssim 10^{-2}$) a relation $\varepsilon\propto\Pm$ seems to hold, which implies that $\eta \langle \vec{b}^2\rangle/\mu_0\rho\simeq \nu \langle {\vec u}^2\rangle$, or equivalently $b_{\rm rms}=$\ord{\sqrt{\Pm} u_{\rm rms}}. This dependence is weaker than that used earlier as $b_{\rm rms} =$\ord{\Pm u_{\rm rms}} for small $\Pm$ \cite{R64}. For Reynolds numbers up to 50,000, and magnetic Prandtl numbers smaller than a value of (say) 0.01, the instability pattern is always dominated by the kinetic fluctuations. The critical $\Pm$, however, depends on the applied Reynolds number; it becomes smaller for increasing $\Rey$. The plot also shows that the influence of the global Reynolds number on this relation is only weak. For faster rotation the ratio (\ref{ratio}) is somewhat larger than for slower rotation. For forced MHD turbulence models a similar behavior for the viscous and Ohmic dissipation has been found \cite{B14}.

In Fig.~\ref{f25d} (right) the same ratio $\varepsilon$ is plotted as it depends on the magnetic Reynolds number $\Rm$, which yields a much clearer scaling of the data. The magnetic energy exceeds the kinetic energy only for $\Rm\gsim 200$. For larger $\Rm$ the energy ratio seems to remain constant. The models with $\Rm<200$ are only weakly magnetized, while for larger $\Rm$ the pattern is magnetically dominated. If this is true, experiments with liquid metals as the fluid between the cylinders will always lead to $\varepsilon\ll 1$ unless the Reynolds number exceeds $10^7$. Working with the maximal values of field and flow given in Fig.~\ref{f20}, we have $\varepsilon=$\ord{10^{-5}} for $\Rm=0.015$, in agreement with the numbers given in Fig.~\ref{f25d} (right) and far away from equipartition. It thus makes sense in related experiments to observe the flow pattern rather than the magnetic pattern.

\subsection{Quasi-Keplerian rotation and beyond}\label{amrikep}
\begin{figure}[h]
\centering
 \includegraphics[width=8cm]{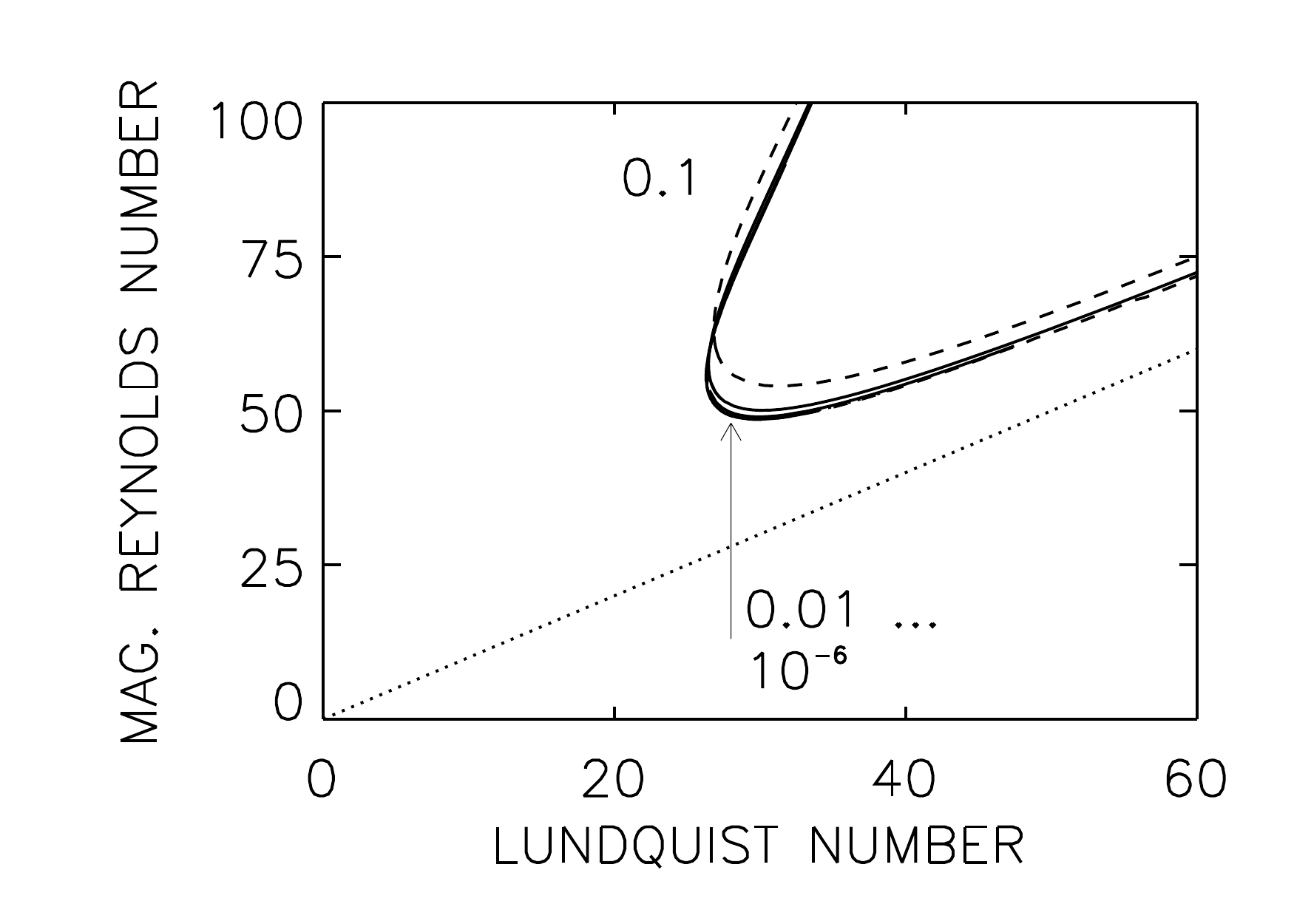}
 \includegraphics[width=8cm]{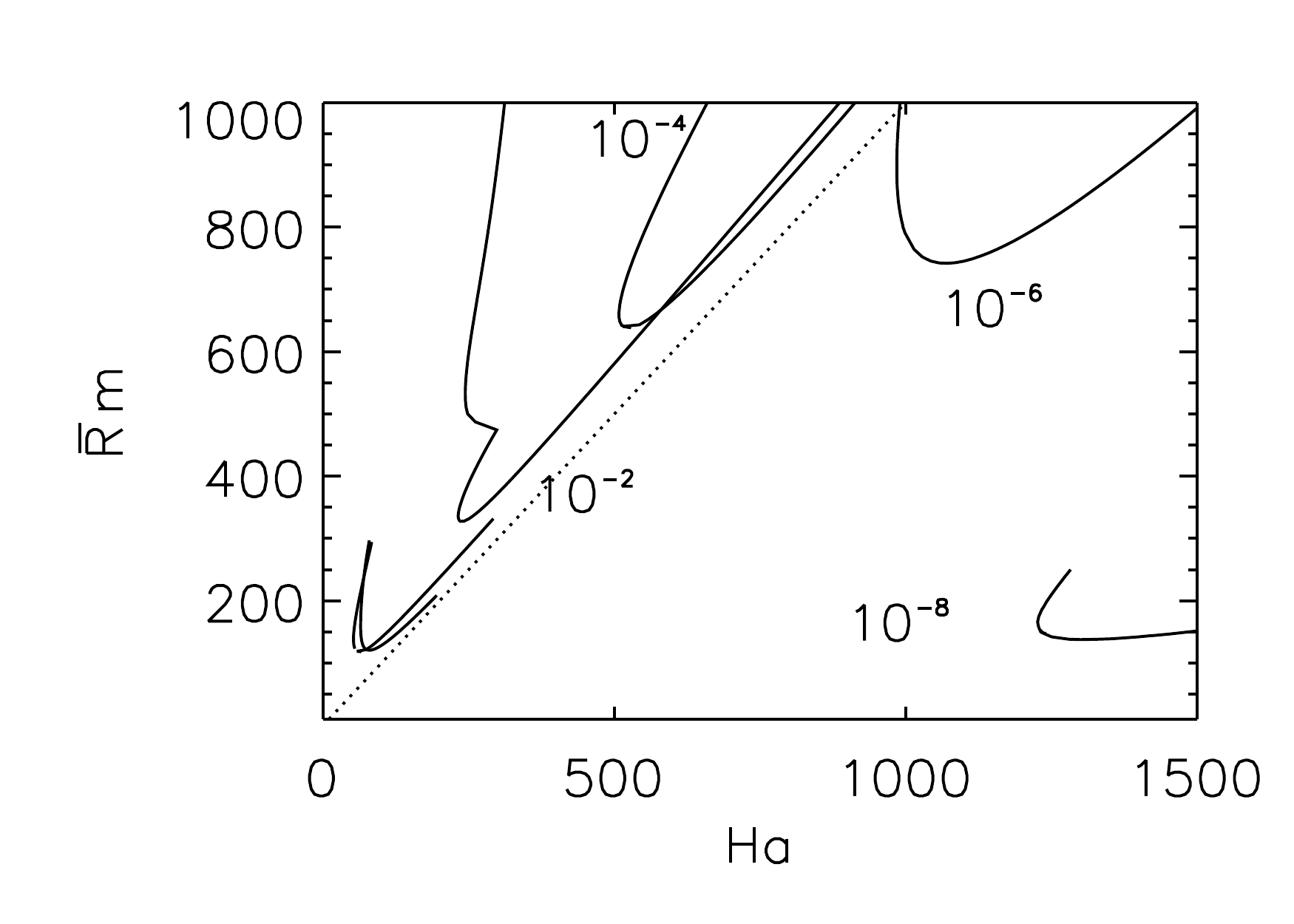}
 \caption{Stability maps of AMRI  for quasi-Keplerian flow. The curves are marked with $\Pm$. For insulating boundaries (left) the curves scale with $\Lu$ and $\Rm$ for $\Pm\to 0$, unlike for perfectly conducting boundaries (right). The dotted lines represent  $\Mm=1$. The instability  flow becomes super-\A{ic} (at least for insulating boundary conditions). $\mu_B=\rin=0.5$. $\mu_\Om=0.35$.}
 \label{f26a}
\end{figure}

If a flatter rotation profile is considered the situation can be different. For quasi-Keplerian rotation ($\mu_\Om=0.35$) the neutral stability curves for the two possible boundary conditions are plotted in Fig.~\ref{f26a}. They show a different scaling behavior for $\Pm\to 0$. For insulating boundaries (left panel) the curves with $\Pm\ll 1$ are almost identical in the ($\Lu/\Rm$) plane. They lie {\em above} the line $\Mm=1$. The instability, therefore, also exists for rapid rotation. In contrast, the potential flow for small $\Pm$ always scales with $\Ha$ and $\Rey$ (see Fig.~\ref{f16}) with severe consequences for the excitation conditions for rapid rotation. As $\Mm=\sqrt{\Pm}\Rey/\Ha$ one always finds $\Mm\propto \sqrt{\Pm}$ for instabilities which scale with $\Ha$ and $\Rey$ for $\Pm\to 0$. Hence, for $\Pm\to 0$ the Mach number also vanishes and the instability only exists for slow rotation.

Surprisingly,  the instability  for quasi-Keplerian flow with perfectly conducting boundaries does not scale with $\Lu$ and $\Rm$ for small $\Pm$. It thus makes sense to use another coordinate system. We have plotted the stability maps for quasi-Keplerian flow and perfectly conducting cylinders in the ($\Ha/\Rmquer$) plane (Fig.~\ref{f26a}, right). Note that simply $\Mm=\Rmquer/\Ha$, hence $\Rmquer=\Ha$ defines the location of  $\Mm=1$. One finds that the curves for large and medium $\Pm$ satisfy $\Mm>1$ (as also for insulating boundaries), while very small $\Pm$ yields $\Mm<1$, similar to the potential flow. It should also be  stressed that for $\Pm=1$ the excitation of the $m=\pm 1$ modes is (slightly) easier for insulating than for perfectly conducting conditions. Nevertheless, substantial differences of the excitation conditions for small $\Pm$ for different boundary conditions are unexpected. The smooth transition from the scaling with $\Ha$ and $\Rey$ for $\mu_\Om=0.25$ to the scaling with $\Lu$ and $\Rm$ for $\mu_\Om=0.35$ and $\Pm\to 0$ for insulating boundary conditions \cite{HT10} is not visible for fluids between conducting cylinders. It is not known whether  this transition only needs much smaller $\Pm$ for different boundary conditions. 

\begin{figure}[htb]
\centering
 \includegraphics[width=8cm]{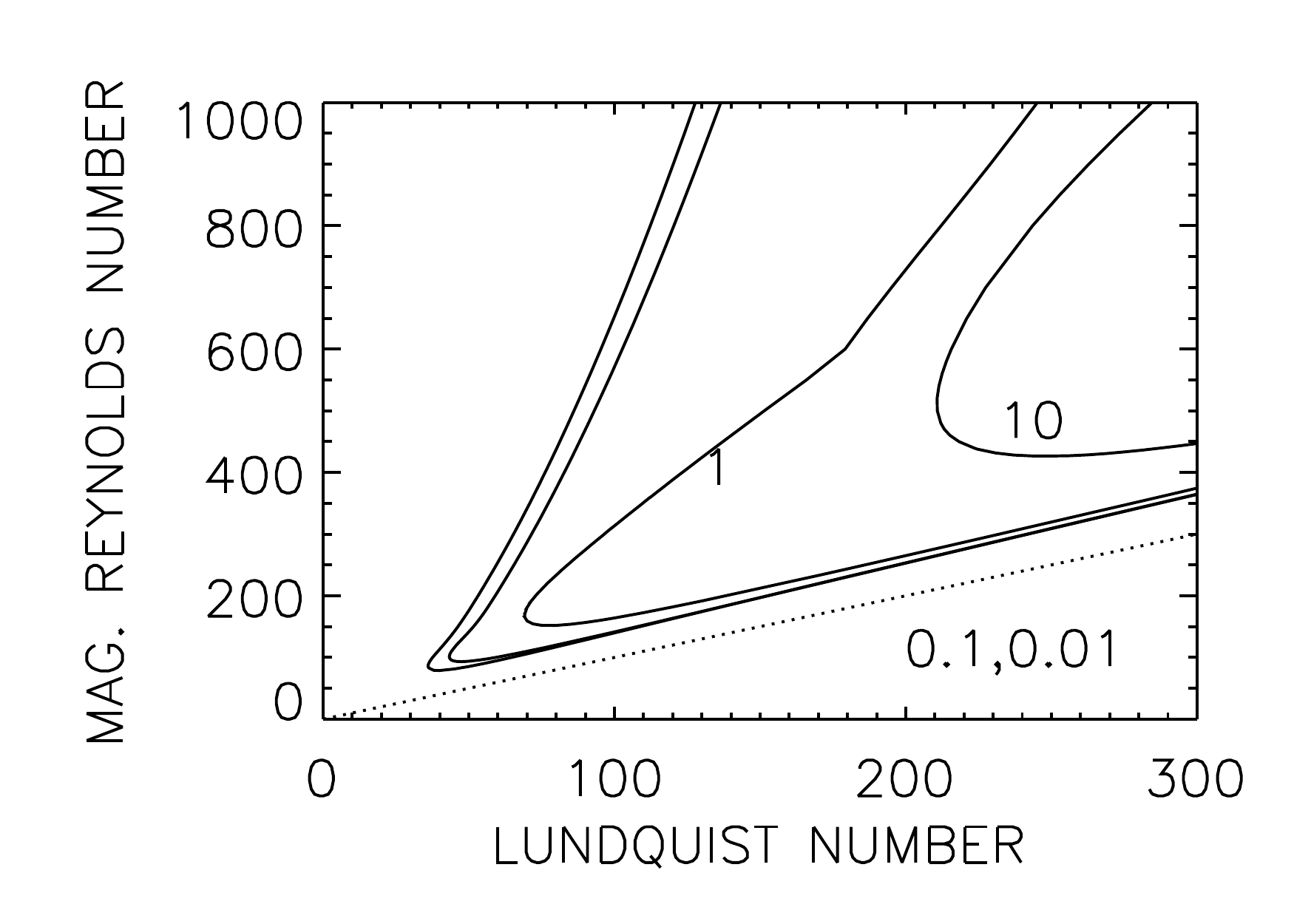}
 \includegraphics[width=8cm]{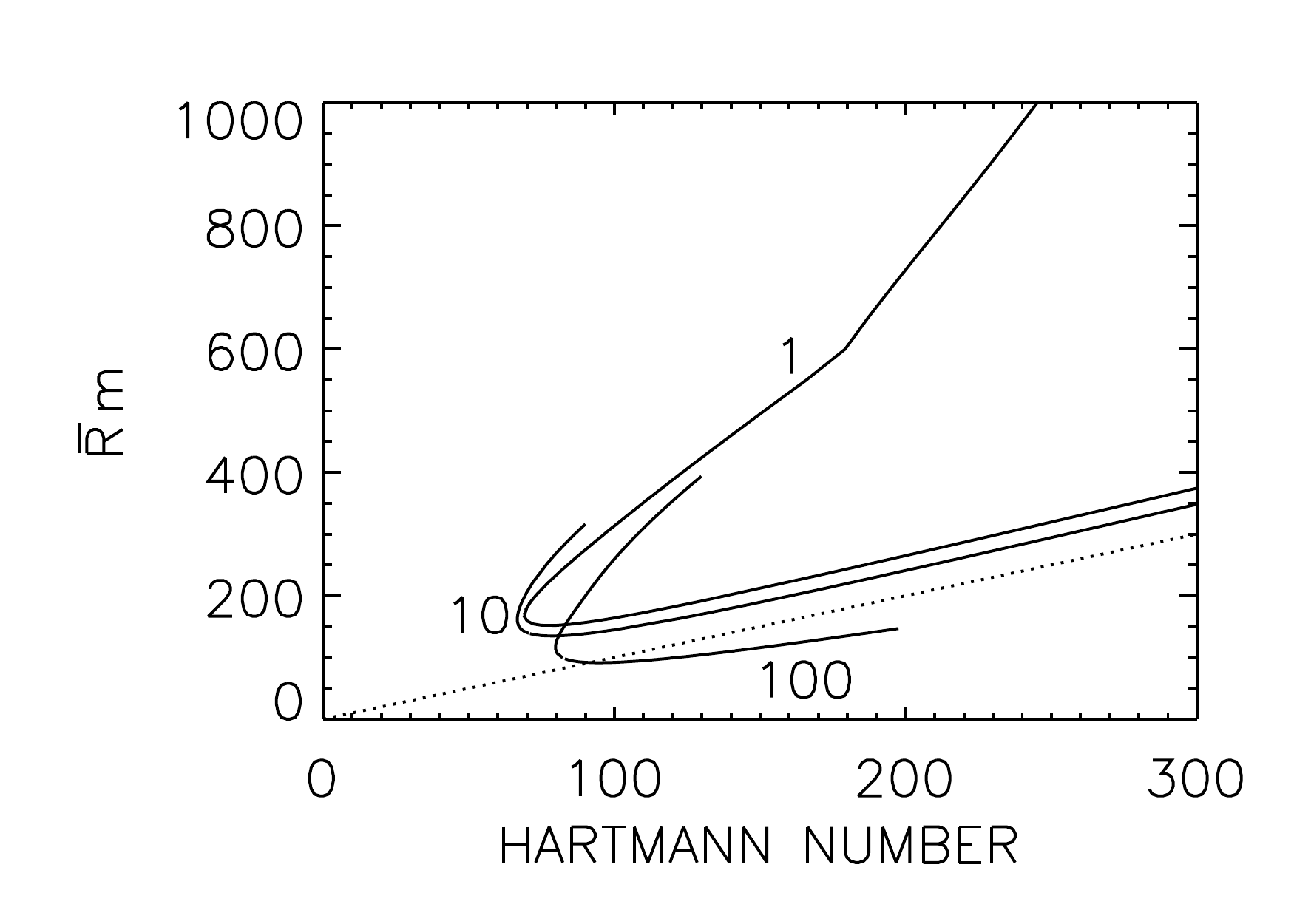}
 \caption{Stability maps for AMRI with quasi-uniform flow of the modes $m=\pm 1$ for  small $\Pm$ (left) and large $\Pm$ (right). The curves scale with $\Lu$ and $\Rm$ for $\Pm\to 0$ and with $\Ha$ and $\Rmquer$ for $\Pm\to \infty$. The dotted lines define $\Mm=1$, the instability is   super-\A{ic} for  small $\Pm$ and   sub-\A{ic} for  large $\Pm$. $\mu_B=\rin=0.5$, $\mu_\Om=0.5$. Perfectly conducting and insulating  boundaries, no differences.}
 \label{f26b}
\end{figure}

The  even flatter rotation profile with $\mu_\Om=0.5$ as the next example   demonstrates that now the same scaling laws for $\Pm\to 0$ exists for both sorts of boundary conditions (Fig.~\ref{f26b}). 
A much weaker influence of the boundary conditions than  for the quasi-Keplerian flow (Fig. \ref{f26a}) appears. The differences between the neutral stability curves for both boundary conditions are very small. The scaling for $\Pm\to 0$ with $\Lu$ and $\Rm$ no longer depends on the boundary conditions; the instability curves for ${\rm Pm}\to 0$ always converge in the ($\Lu/\Rm$) plane. The important difference to the potential flow is that now all curves lie {\em above} the line $\Mm=1$, so that this instability also exists for rapid rotation. For all $\Pm$ the magnetic Mach number $\Mm$ lies between low and high rotation limits but it is always super-\A{ic}, e.g., for ${\Pm}=1$ AMRI exists for 
\begin{equation}
1\lsim \Mm \lsim 3.
\label{ratio1}
\end{equation}
Again, both the magnetic field and the rotation rate can be too weak or too strong for AMRI and again the excitation of the instability for $\Pm=1$ is slightly easier for insulating boundary conditions.  Figure \ref{f26b} also  demonstrates that for $\Pm>1$ the scaling switches from $\Lu$ and $\Rm$ (valid for $\Pm<1$) to $\Ha$ and $\Rmquer$. For both limits the influence of the boundary conditions is very weak. It seems to be clear, however, that the magnetic Mach numbers move from large values for small $\Pm$ to small values for large $\Pm$. This is insofar surprising as the definition of the magnetic Mach number is entirely free of  diffusivities.

\begin{figure}[htb]
\centering
 \includegraphics[width=8cm]{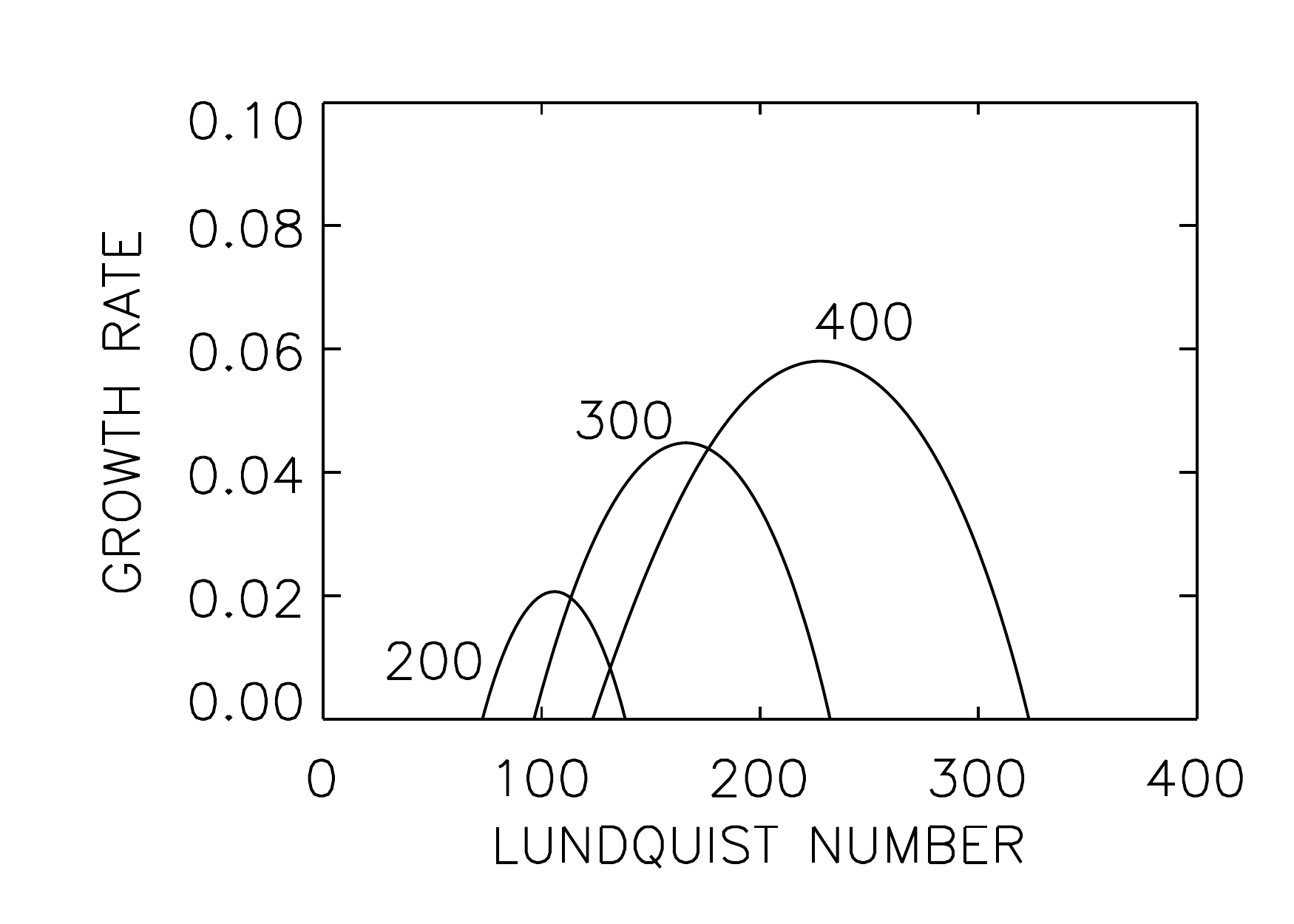}
 \includegraphics[width=8cm]{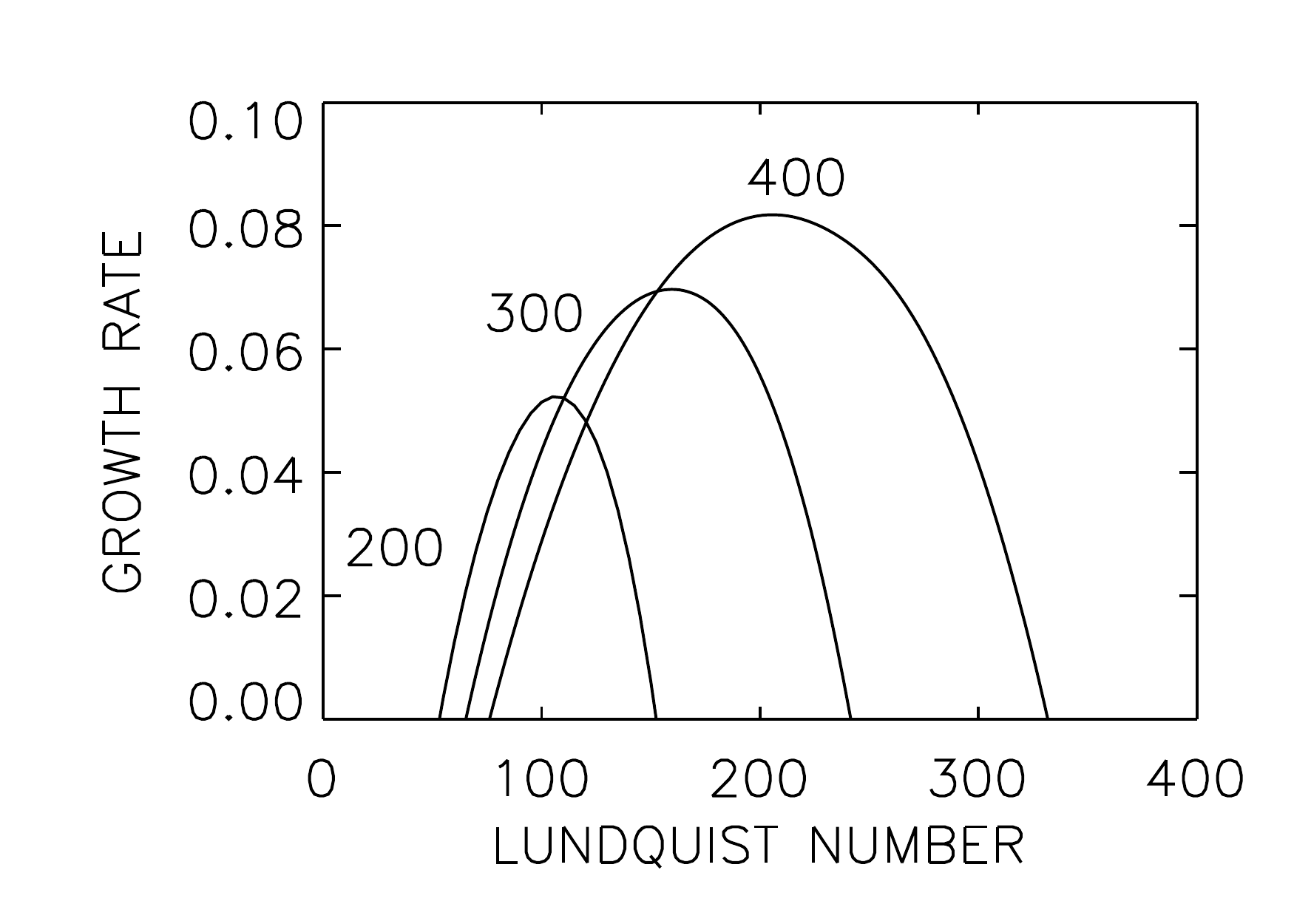}
\caption{Growth rate (\ref{growthrate}) for quasi-uniform flow versus Lundquist number  for various magnetic Reynolds numbers. The curves are marked with $\Rm$. The normalized growth rates  grow linearly with $\Rm$. $m=\pm 1$, ${\rm Pm}=1$ (left), 
${\rm Pm}=0.01$ (right),   $\mu_B=\rin=0.5$, $\mu_\Om=0.5$. Perfectly conducting cylinders.}
\label{f26c}
\end{figure}

For various ${\Rm}$ and ${\Pm}$ the growth rates (\ref{growthrate}) have been calculated between the two limiting values ${\rm S}$ where it vanishes; it is maximal somewhere between the two limits. In Figs.~\ref{f26c} the normalized growth rate is plotted for the parameters ${\rm Rm}$ and ${\rm Pm}$. One finds quasilinear relations
\begin{equation}
\omega_{\rm gr}\simeq
 \epsilon_{\rm gr} \, {\rm Rm} 
\label{omgr}
\end{equation}
with $\epsilon_{\rm gr}$ varying slightly from $1.5\cdot 10^{-4}$ for ${\Pm}=1$ to $2.1 \cdot 10^{-4}$ for ${\Pm}=0.01$ \cite{RG14}. The growth rate slowly increases for smaller $\Pm$ but this effect is weak. The growth time in units of the rotation time is thus $\tau_{\rm gr}/\tau_{\rm rot}\simeq 10^3/{\Rm}$. Of course, the linear relation can only hold for small ${\Rm}$. For $\Rm\gg1$ the growth rate no longer depends on $\Rm$, so then $\omega_{\rm gr}\leq 0.14$. The growth time of the instability for $\mu_\Om=0.5$ can therefore never be shorter than one rotation time. 

The growth rates of the instability pattern and its axial wave number can be used to compute the characteristic Strouhal number\footnote{Often in fluid dynamics the  reciprocal  definition $\rm Sr=1/\rm St$ is called the Strouhal number.}  The 
\begin{equation}
{\rm St} = \frac{u_{\rm rms} }{\ell \omgr}, 
\label{turn}
\end{equation} 
with the axial cell size $\ell=\pi/k$. The growth rates have been calculated from linear models for various Hartmann numbers along the lines of maximal instability such as in Fig.~\ref{f16} (left). The rotation profiles vary in the wide interval between $\mu_\Om=0.25$ and $\mu_\Om=0.5$. For the rms velocity only the axial intensity ${\langle u_z^2\rangle}^{1/2}$ is derived by the nonlinear code described in Section \ref{azifield} for conducting boundary conditions ($\Gamma=8$).
\begin{figure}[htb]
\centering
 \includegraphics[width=9cm]{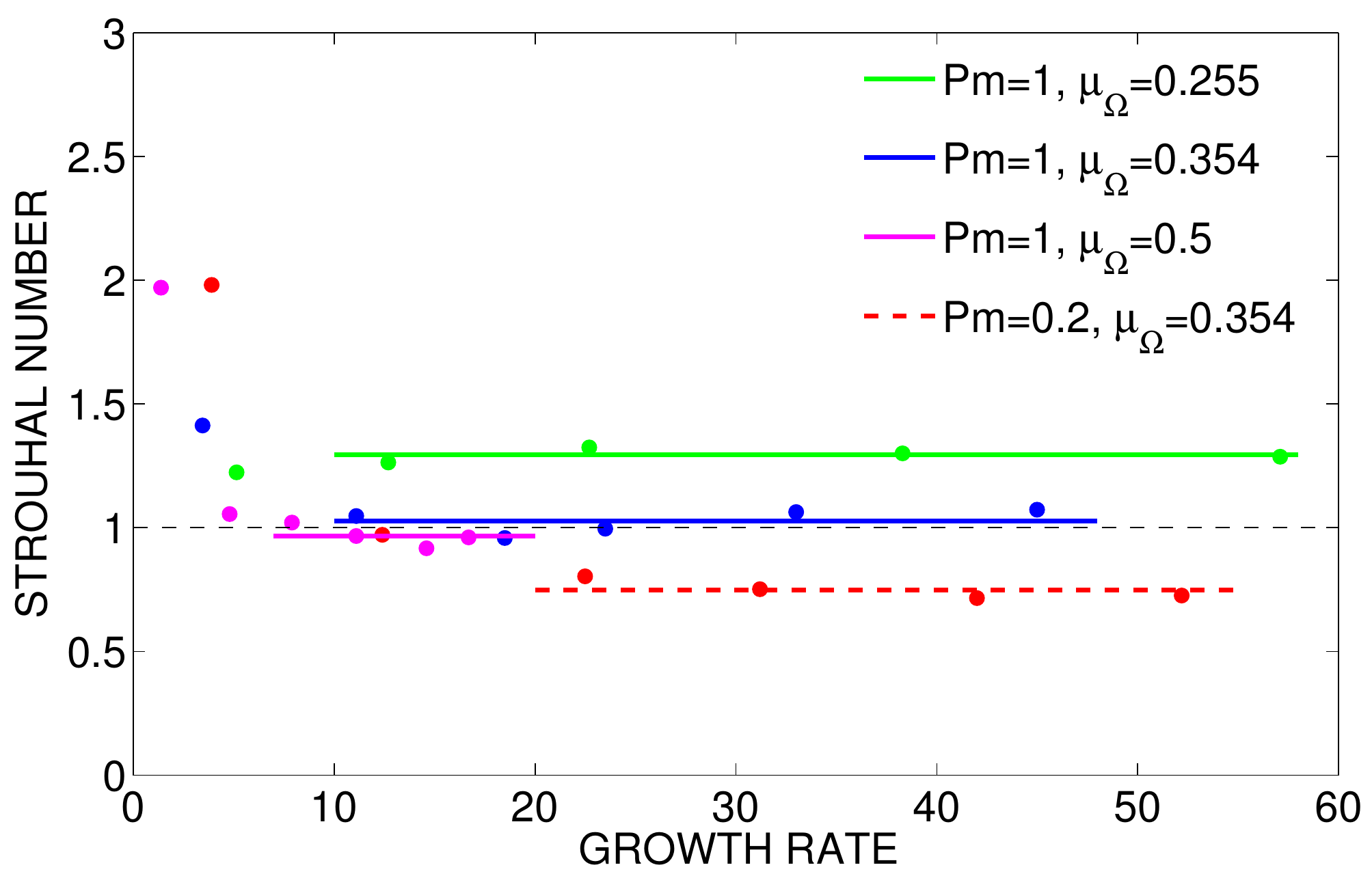}
\caption{Strouhal number  (\ref{turn}) versus growth rate for various  $\Pm$ and  shear parameters $\mu_\Om$ as indicated. The models possess the maximal growth rates. The Strouhal numbers of AMRI are always of order unity.  $\mu_B=\rin=0.5$.  Perfectly conducting boundaries.}
\label{turnover}
\end{figure}

The numerical results  underline the exceptional importance of the Strouhal number. The Strouhal number is almost unity for steep rotation profiles and  magnetic Prandtl numbers of order unity, in confirmation of often-used assumptions. This is certainly not a trivial result for consistent models of MHD flows. The self-consistent models of MHD instability (not driven turbulence!) indeed lead to Strouhal numbers of order unity. It becomes only slightly smaller  for smaller $\Pm$ and for flatter rotation profiles (Fig. \ref{turnover}). Consequently, the nonlinear turbulence intensity $\langle u_z^2\rangle$ can indeed be estimated by means of the characteristic quantities $k$ and $\omega_{\rm gr}$ of the linear theory alone. For dissipation coefficients such as the magnetic resistivity it should be allowed to move from the well-founded relation
\begin{equation}
\eta_{\rm T} \simeq \int_{-\infty}^\infty \langle u_i(t)u_i(t-\tau) \rangle\ {\rm d} \tau
\label{integral}
\end{equation} 
to $\eta_{\rm T}\simeq u_{\rm rms} \ell$ as a good estimation \cite{KR74}. This does not mean, however, that the effective viscosity can be estimated just with these quantities if the effect of the magnetic fluctuations cannot be neglected \cite{VK83}.
\subsection{The AMRI experiment}\label{expamri}
In this section, we will present results of a liquid metal experiment devoted to the investigation of AMRI and the helical version of the magnetorotational instability (to be discussed in Section \ref{HMRI}). The switch of the scalings for fluids with low magnetic Prandtl number from $\Rm$ and $\Lu $ to $\Rey$ and $\Ha$ allows experiments to work with slow rotation and weak fields, provided that the rotation profile is not too far from the potential flow. The most popular candidates for experiments with liquid metals are given in Table \ref{t2}. Generally, they combine the viscosity of water with the electrical conductivity of the solar plasma. The low values of the magnetic Prandtl numbers of liquid metals in comparison to the solar plasma are due to their low viscosities. If liquid metal AMRI experiments are carried out close to the Rayleigh line, only the values of $\nu$ and $\bar\eta=\sqrt{\nu \, \eta}$ are relevant. The magnetic Prandtl numbers vary by two orders of magnitudes, but close to the Rayleigh limit this is not really important. The viscosity varies by a factor of only seven among the metals in Table \ref{t2}, and the averaged diffusivity $\bar\eta=\sqrt{\nu \, \eta}$ is also very similar for all fluids.

While sodium, with its low magnetic diffusivity, is the liquid of choice for experiments that require high values of $\Rm$ (and $\Lu$), such as dynamo experiments and experiments on standard MRI, GaInSn is more convenient for experiment governed by $\Rey$ (and $\Ha$). This has mainly to do with the much milder safety requirements compared to sodium, but also with the fact that GaInSn is liquid at room temperatures. While the latter advantage is shared by mercury, the health risks in dealing with that metal  made it disappear from most liquid metal labs.
\begin{figure}[htb]
\centering
 \includegraphics[width=7.0cm]{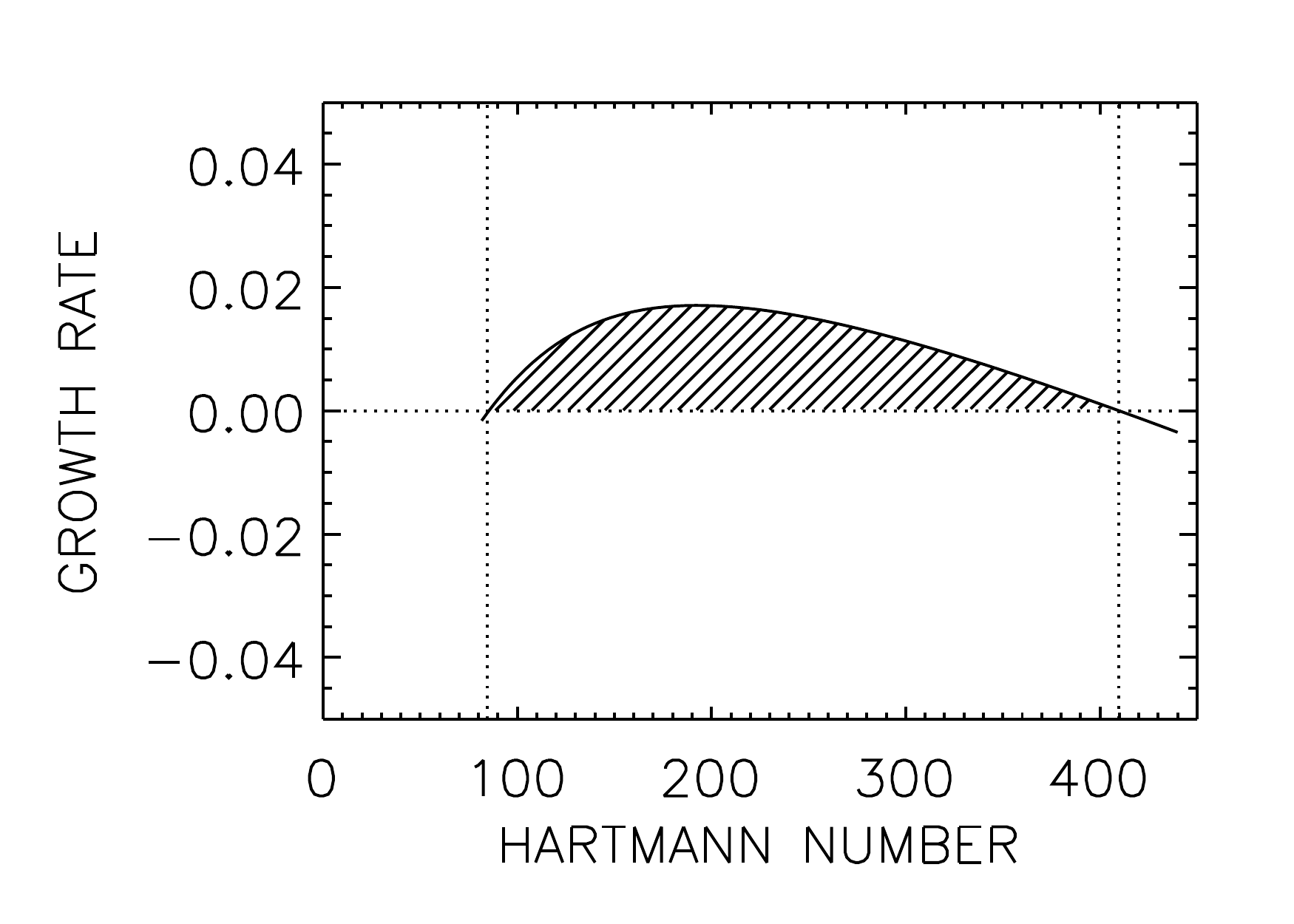}
 \includegraphics[width=7.0cm]{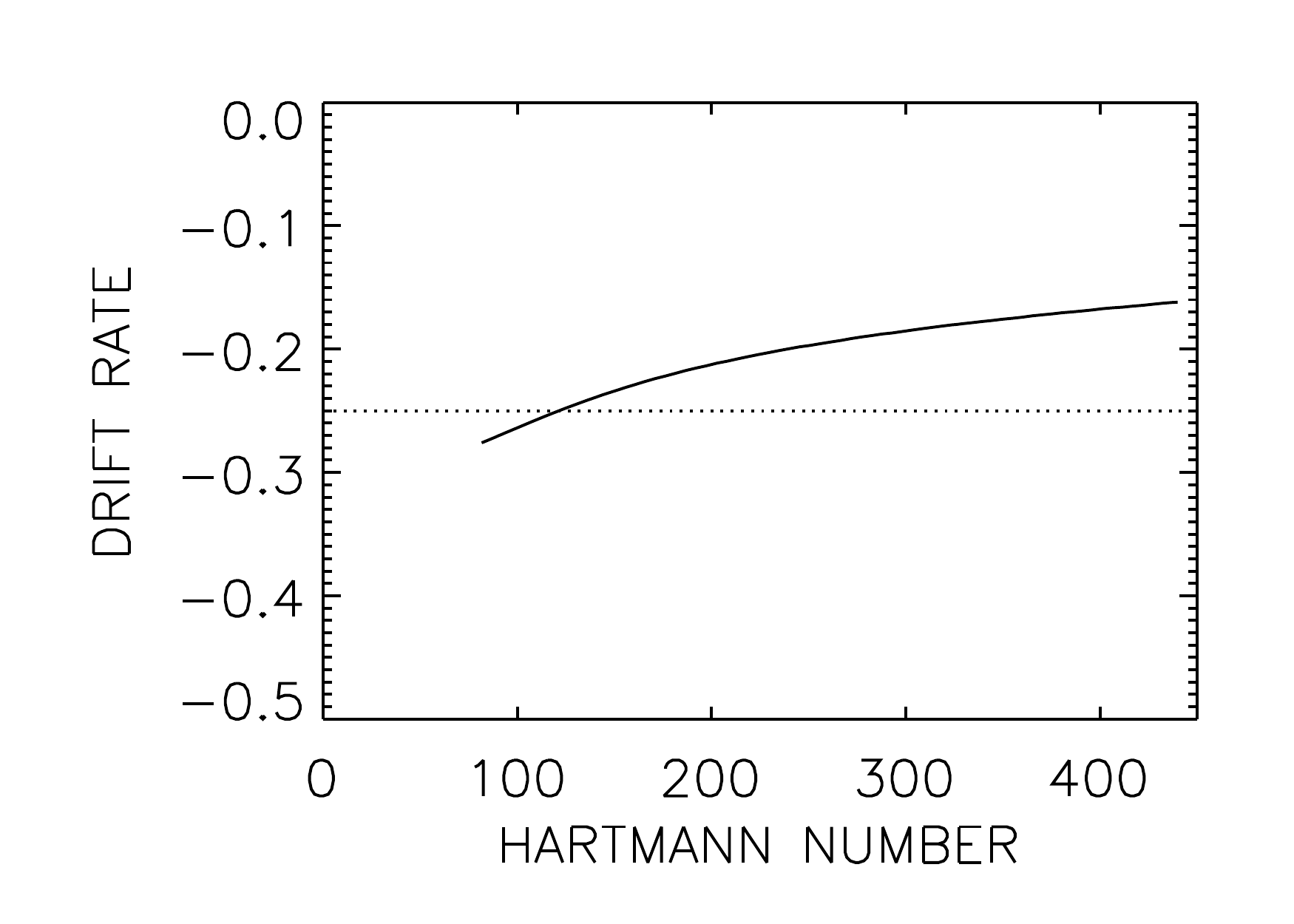}
 \caption{Growth rate $\omgr$ (left) and  drift rate $\omdr$ (right)  as functions of $\Ha$ for quasi-potential flow. The hatched area marks instability between the two limits of neutral stability, the dotted line in the right panel marks corotation with the outer cylinder. The maximal growth time of the instability is about 2 rotation times of the outer cylinder. $m=1$, $\Rey=3000$,  $\mu_B=\rin=0.5$, $\mu_\Om=0.26$, $\Pm=10^{-6}$. Perfectly conducting boundaries, from  \cite{RG14}.}
 \label{f22}
 \end{figure}
 \begin{figure}[h]
\centering
 \includegraphics[width=4cm,height=6cm]{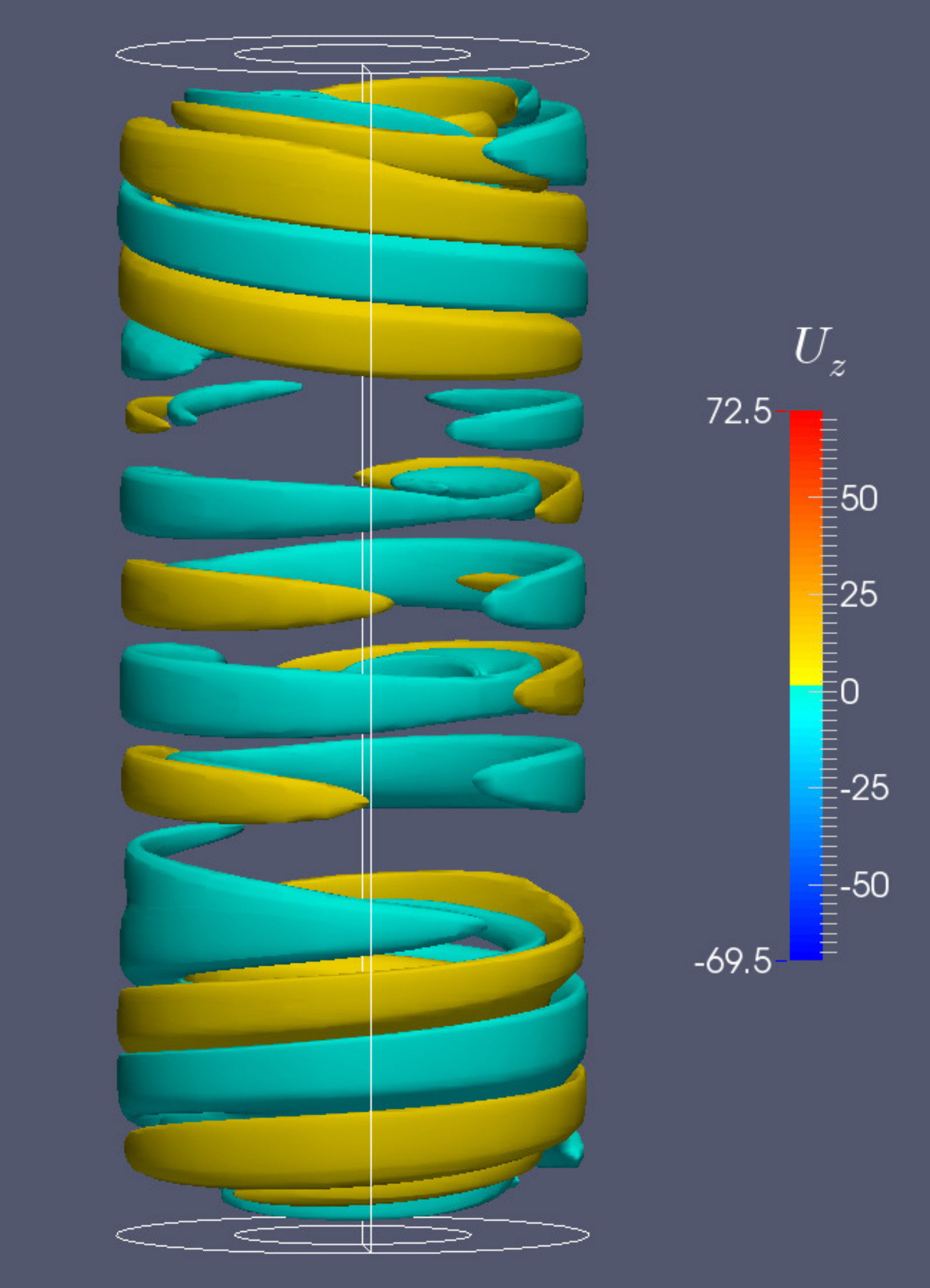}
 \includegraphics[width=4cm,height=6cm]{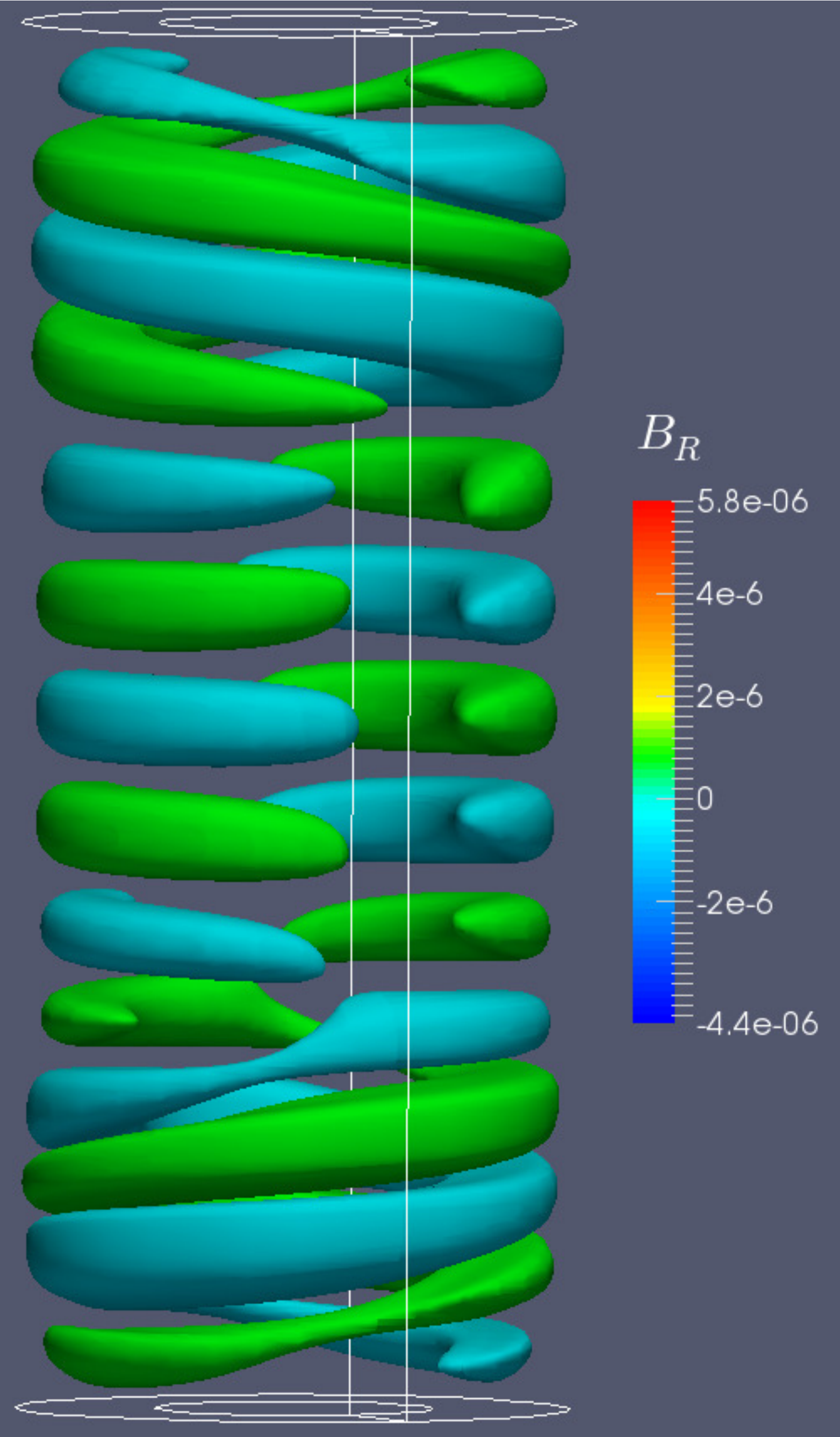}\hfill
 \caption{As in Fig.~\ref{f20} but for an axially bounded container with $\Gamma=10$  and insulating endplates which are split like the endplates of the {\sc Promise} experiment (see Section \ref{promise}). Because of the endplates a slight equatorial antisymmetry occurs and the flow amplitudes enhance near the two lids. The magnetic perturbations are hardly modified.  The power supply is perfectly axisymmetric along the $z$-axis. $\Rey=1500$, $\rm Ha=100$, $\mu_B=\rin=0.5$, $\mu_\Om=0.26$, ${\rm Pm}=10^{-5}$. Perfectly conducting cylinders.}
 \label{f20a}
\end{figure}
\begin{figure}[h]
\centering
\includegraphics[width=4cm,height=6cm]{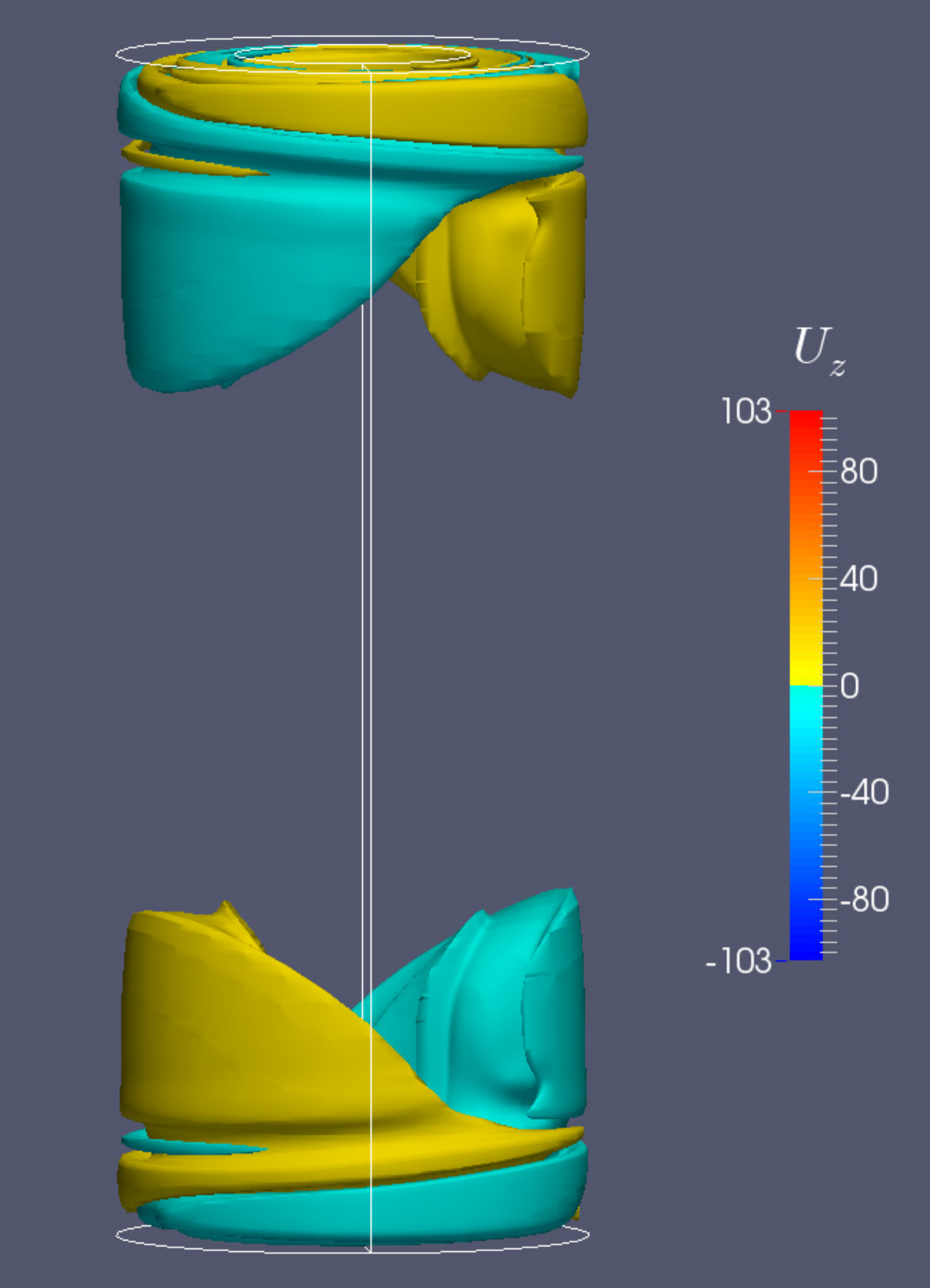}
 \includegraphics[width=4cm,height=6cm]{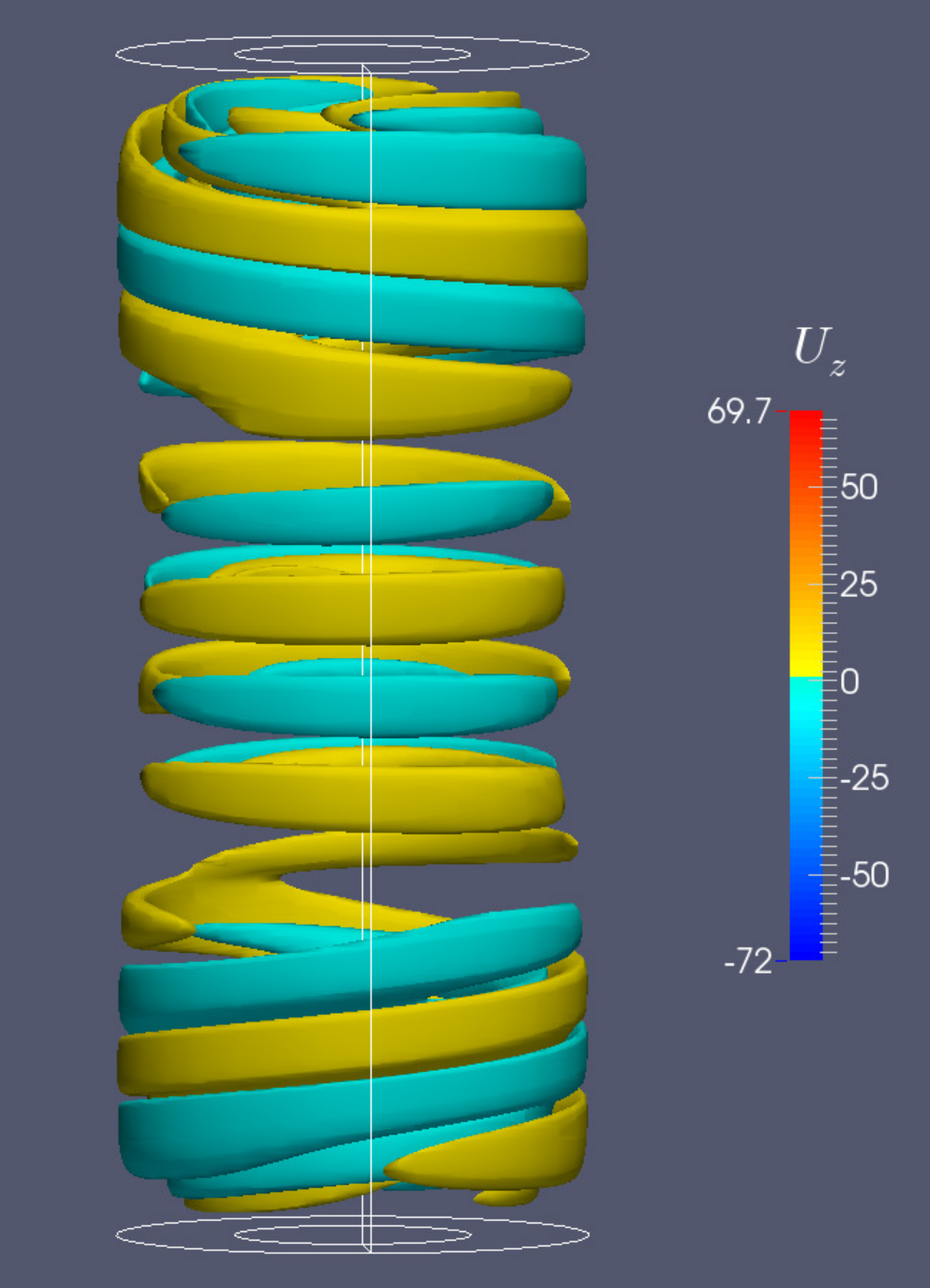}
 \caption{Vertical flow speeds as in Fig.~\ref{f20a} (left) but for a very nonaxisymmetric lead wire system with $m=1$ symmetry (left) and a lead wire system with $m=2$ symmetry (right).}
 \label{f20b}
\end{figure}
As a first guide to the experimentally relevant parameter space, Fig.~\ref{f22} shows the growth rate and the drift rate (both normalized with the rotation rate of the inner cylinder) for a Reynolds number of $\Rey=3000$ and low magnetic Prandtl number. The parameters strongly differ from  the values used in Fig.~\ref{f26c}. The main difference, however, is the different scalings of AMRI for steep ($\mu_\Om=0.26$)  and flat ($\mu_\Om=0.5$) rotation laws \cite{HT10} so that, as a consequence,  (\ref{omgr}) does not hold for the potential flow. The maximum growth rate lies between the two values of neutral stability and takes a value of 0.02, which corresponds to a growth time of about 8 rotation times of the inner cylinder. Compared with MRI, the AMRI also scales with the rotation time but is somewhat slower. The onset of the instability is at $\Ha\simeq 85$, corresponding to an axial electric current of 10.9 kA. It is very characteristic that for much higher Hartmann numbers ($\Ha>400$ in Fig.~\ref{f22}, left) the instability disappears. The right panel of Fig.~\ref{f22} gives the normalized azimuthal drift $\omdr$ of the pattern of the $m=1$ mode. The dotted line represents the relation $\dot \phi=\mu_\Om \Omin=\Omout$ where the pattern corotates with the {\em outer} cylinder. Note that the instability pattern indeed corotates with the outer cylinder for $\Ha\simeq 110$, which is only slightly greater than the lower Hartmann number for neutral stability. The measurements will confirm this prediction.

A serious difficulty to realize AMRI (and all other versions of MRI) in the laboratory are the endplate effects of finite-length devices. Figure \ref{f20a} shows simulations for a data set close to experimental realizations for a height-to-gap ratio $\Gamma=10$ and with very small $\Pm$. The corresponding version for $\Gamma\to \infty$ (i.e., with periodic boundary conditions), as given by Fig.~\ref{f20}, leads to an energy ratio $\varepsilon\simeq 10^{-5}$ which for $\Gamma=10$ is not basically changed. The endplates have two main effect: First, there is some concentration of the energy close to the endplate where the flow intensity is drastically enhanced. Second, we observe a symmetry breaking between left and right handed spirals which appear now preferentially in the lower/upper half of the cylinder (in contrast to the equal distribution as visible in Fig.~\ref{f20}).

\begin{figure}[thb]
\includegraphics[width=0.99\textwidth]{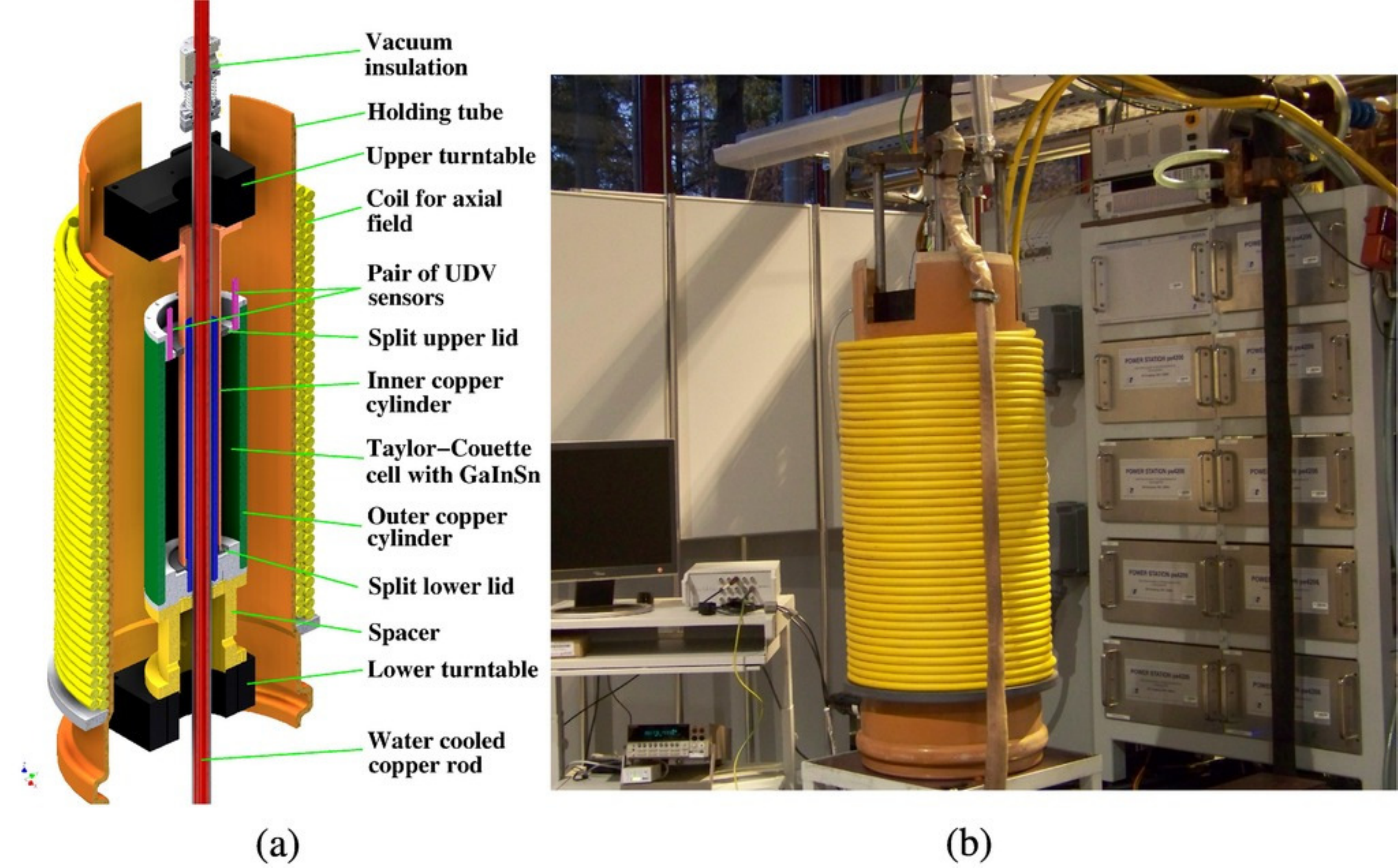}
\caption{The  {\sc Promise} facility. Schematic sketch of the central Taylor-Couette setup and the coil for producing the axial field (left) and   photograph of the installation, with the 20\,kA power supply (right).}
\label{anlageamri}
\end{figure}

The facility {\sc Promise} ({\it P\/}otsdam {\it RO\/}ssendorf {\it M\/}agnetic {\it I\/}n{\it S\/}tability {\it E\/}xperiment) is shown in Fig.~\ref{anlageamri}. Its heart is a cylindrical vessel made of copper. The inner wall extends in radius from 22 to 32 mm; the outer wall extends from 80 to 95 mm. This vessel is filled with the liquid alloy GaInSn whose material parameters are given in Table \ref{t2}.

In the real experiment, however, the electric current is provided by a closed wire system which forms an external  magnetic field which modulates the prescribed axisymmetric azimuthal field by a weak nonaxisymmetric component. One easily finds that this modulation corresponds to an additional $m=1$ component which strongly influences the excitation of the $m=1$ AMRI mode (see Fig.~\ref{f20b}, left). This unfortunate situation can be overcome by a more complicated lead wire system providing an external magnetic modulation with $m\neq 1$. To produce the constellation given in Fig.~\ref{f20b} (right) two lead wires have been used in the same plane (at top and bottom of the container) separated by angles of 180$^\circ$. The results are a good match to the vertical velocity component of the unbounded system (Fig.~\ref{f20}, left).
\begin{figure}[ht]
\centering
\includegraphics[width=9cm]{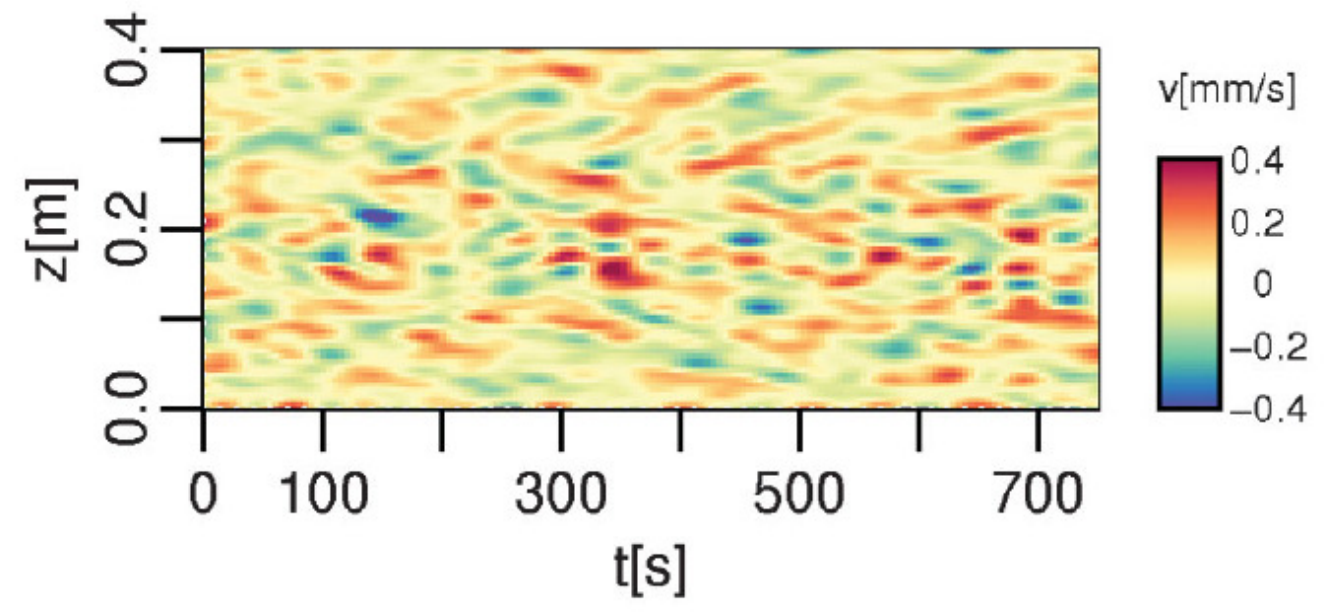}
 \includegraphics[width=6.5cm]{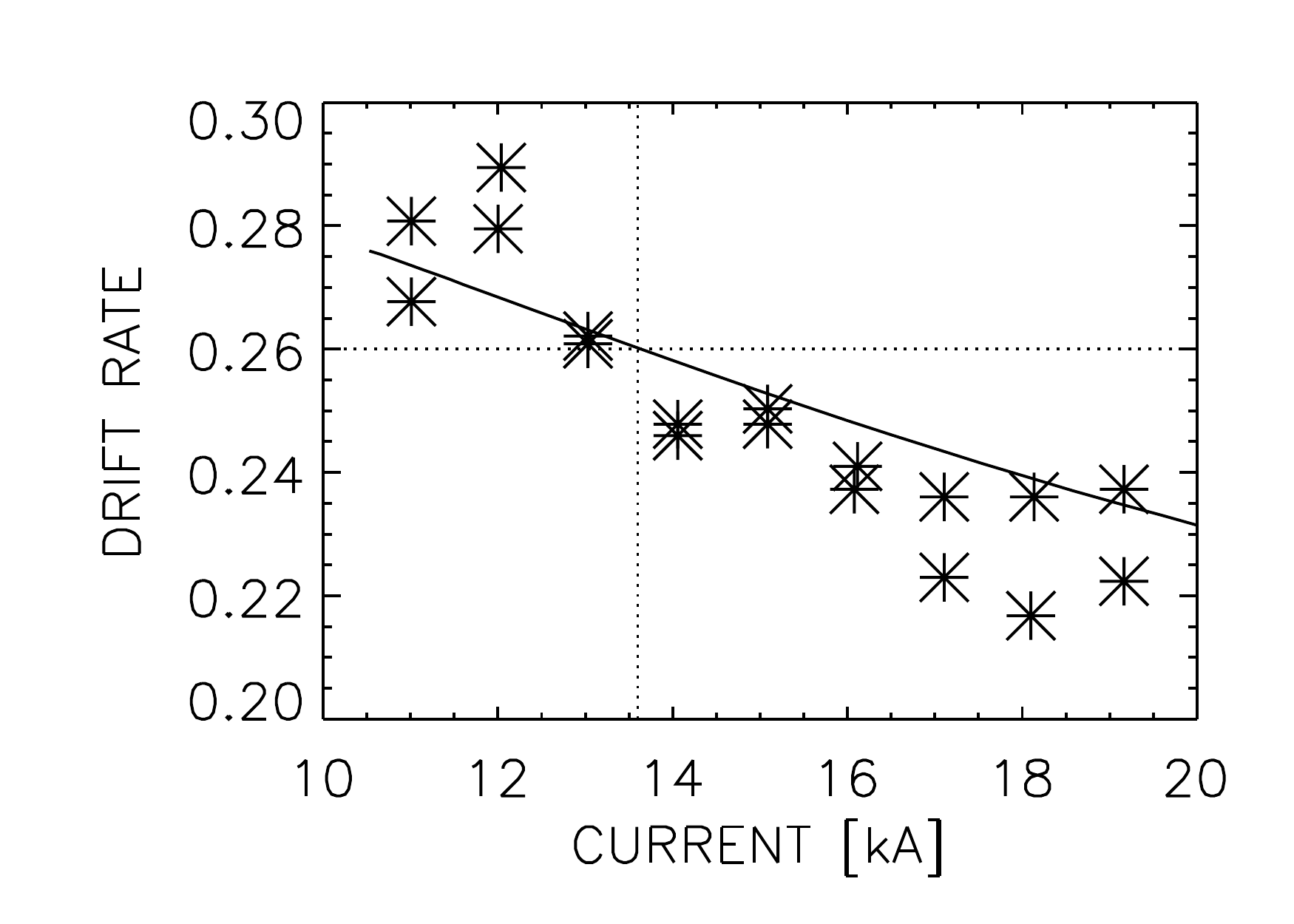}
 \caption{Results of the AMRI experiment for the original asymmetric lead wire system. Left: experimental results for the vertical flow in mm/s. The data also exhibit the (slight) equatorial antisymmetry predicted by the simulations (see Fig.~\ref{f20b}, left). The magnetic symmetry breaking due to 
the one sided wiring of the central current makes the upward and downward 
traveling waves interpenetrate each other in the upper and lower halves 
of the cylinder. $\Rey=1480$. Right: drift frequencies measured in positive azimuthal direction. The solid line corresponds to the line  in Fig.~\ref{f22} (right) but for $\mu_\Om=0.26$. One finds a nearly perfect agreement of the  experiment with the simulations.  $\Rey=2960$, $\mu_B=\rin=0.5$, $\mu_\Om=0.26$. Experimental data from \cite{SS14}.}
\label{frank1}
\end{figure}

The copper vessel is fixed via a spacer on a precision turntable. The outer wall of the vessel thus serves as the outer cylinder of the Taylor-Couette device. The inner cylinder is fixed to an upper turntable, and is immersed into the GaInSn from above. It has a thickness of 4 mm, extending from 36 to 40 mm. The actual Taylor-Couette flow then extends between $R_{\rm in}= 40$ mm and $R_{\rm out}=80$ mm. In the present configuration of the experiment the lower and upper lids are electrically insulating and split at a well defined intermediate radius of 56 mm which had been found in \cite{S07,SR07,SG07,SG08} to minimize the Ekman pumping. The endplates are made of plexiglass which are split into two rings where the inner one is attached to the inner cylinder and the outer one to the outer cylinder.

This represents a major advantage compared to the initial version of {\sc Promise} \cite{SG06,RH06,SG07} in which the upper endplate was a plexiglass lid fixed to the frame while the bottom was simply part of the copper vessel, and hence rotated with the outer cylinder, producing strong Ekman pumping and a clear top/bottom asymmetry with respect to both rotation rates and electrical conductivity. This central module is embedded into a 2 x 39-winding coil for the production of a vertical field (which only becomes relevant when discussing the helical MRI in Section \ref{HMRI}). The axial velocity perturbations are measured by two ultrasonic sensors from Signal Processing SA with a\ working frequency of 4 MHz which are fixed into the outer plastic ring, 12 mm away from the outer copper wall, flush mounted at the interface to the GaInSn (see Fig.~\ref{anlageamri}). Since this outer ring is rotating, it is necessary to transfer the signals into the laboratory frame by the use of a slip ring contact. The advantage of the ultrasound Doppler Velocimetry is that it provides full profiles of the axial velocity $u_z$ along the beam-lines parallel to the axis of rotation.

The azimuthal magnetic field is produced by a water-cooled copper rod going through the center of the setup. In the present configuration the current is supplied by a 20\,kA switching mode power supply. Significant effort was spent on severe problems of electromagnetic interference \cite{SS16}, before the (initially extremely noisy) UDV data could be utilized for characterizing the AMRI. As mentioned above. the central copper rod is connected to the power supply in an asymmetric, one-sided manner.

With the  container data the unit of velocity is $\nu/d\simeq 8.5\cdot 10^{-3}$~mm/s. With the maximal $u_z\simeq 100 $ taken from Fig.~\ref{f20b} (left) the maximal axial velocity which can be expected for the AMRI experiment as 0.85 mm/s. The experimental data have been analyzed in detail resulting in maximal values of 0.4~mm/s which can be considered as a rather good empirical confirmation of the simulations \cite{SS14}. Note that the simulations of cylinders with $\Gamma\to \infty $ provide the lower value of about 20 and also the optimized container with $\Gamma=10$ exhibits only 70 as the relevant quantity. The more perfect the experiment the lower amplitudes of the maximal $u_z$ appear.

The relation $I_{\rm axis}= 5 \Rin B_{\rm in}$ connects the toroidal field amplitude $B_{\rm in}$ at $R_{\rm in}$ with the axial current inside the inner cylinder. $I_{\rm axis}$, $\Rin$ and $B_{\rm in}$ must be measured in A, cm and G. Hence,
\begin{equation}
{\Ha}= \frac{1}{5}\frac{I_{\rm axis}}{\sqrt{\mu_0\rho\nu\eta}}.
\label{Hadef}
\end{equation}
The radial size of the container does not appear in this relation. For the gallium alloy GaInSn the value of the square root in (\ref{Hadef}) is 25.6. The resulting electric current for marginal instability is 10.9 kA, hence $B_{\rm in}=545$~G. With the largest fluctuations of $b_\phi/B_{\rm in}\simeq 6\cdot 10^{-6}$ taken from Fig.~\ref{f20a}, one finds 3~mG as the maximum field fluctuation.

Analyzing more experimental runs we have compiled the dependence of quantities on the applied axial currents. Figure~\ref{f22} (left) shows the theoretical growth rate for the infinite length system. The growth rates under the axisymmetric field condition give a consistent picture with a sharp onset of AMRI at $\rm Ha\simeq 80$ corresponding to current of 10.9~kA.

The left panel of Fig.~\ref{frank1} shows a typical experimental result for  
$\Rey=1480$,  $\mu_{\Om}=0.26$, and $\Ha=124$, which demonstrates 
how the upward and downward traveling waves interpenetrate each other in 
the upper and lower halves of the cylinder. Evidently, the selective 
occurrence of upward and downward traveling waves in the upper and 
lower halves which resulted from the first symmetry breaking in 
{\it axial direction} (due to the endcaps, see Fig.~\ref{f20b}, left), is 
neutralized here by the second symmetry breaking in {\em azimuthal direction} 
(due to the one sided wiring).

Here we focus on the dependencies of the numerically and experimentally determined drift frequencies on the applied current, which proves to be a very robust property of the instability. From Fig.~\ref{f22} (right),the pattern corotates with the outer cylinder for $ \Ha\simeq 120$. In the right panel of Fig.~\ref{frank1} a nearly perfect agreement between theory and experiment can be seen. The theoretically expected enhanced frequency for lower $\Ha$ and a slightly reduced frequency for higher $\Ha$ can also easily be identified in the experimental data.

To summarize this section, the experiments revealed the existence of AMRI close to the Rayleigh line. While the observed and numerically confirmed effects of the double symmetry breaking on the AMRI are interesting in their own right, a new system of wiring of the central current, comprising a `pentagon' of 5 back-wires situated around the experiment, is presently commissioned for further investigations.

\subsection{Eddy viscosity}
The instability-induced angular momentum transport which was calculated in Section \ref{AMT} for MRI under the influence of an axial field will now similarly be computed for AMRI under the influence of a current-free azimuthal field.
\begin{figure}[htb]
\centering
\includegraphics[width=8cm]{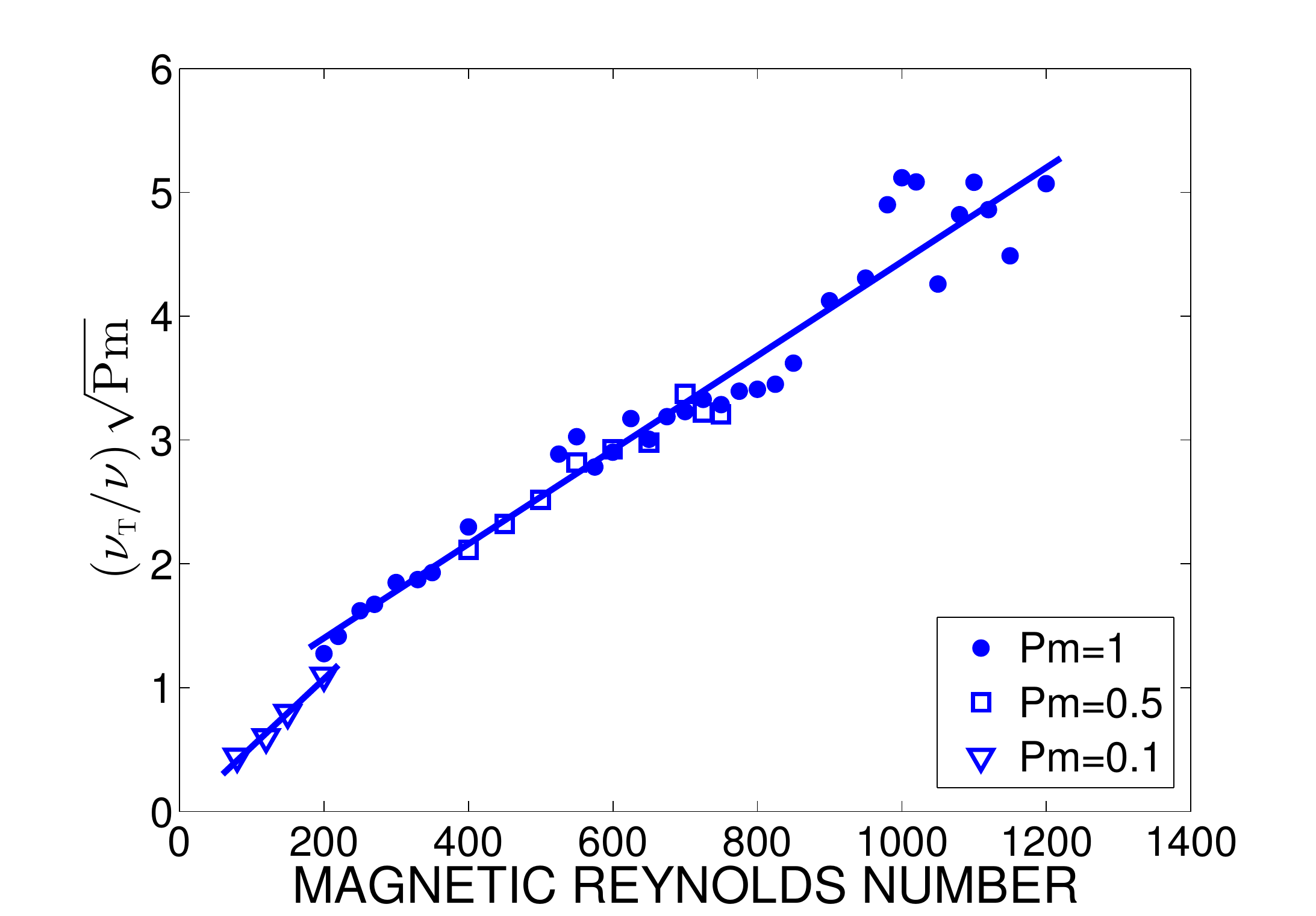}
\includegraphics[width=8cm]{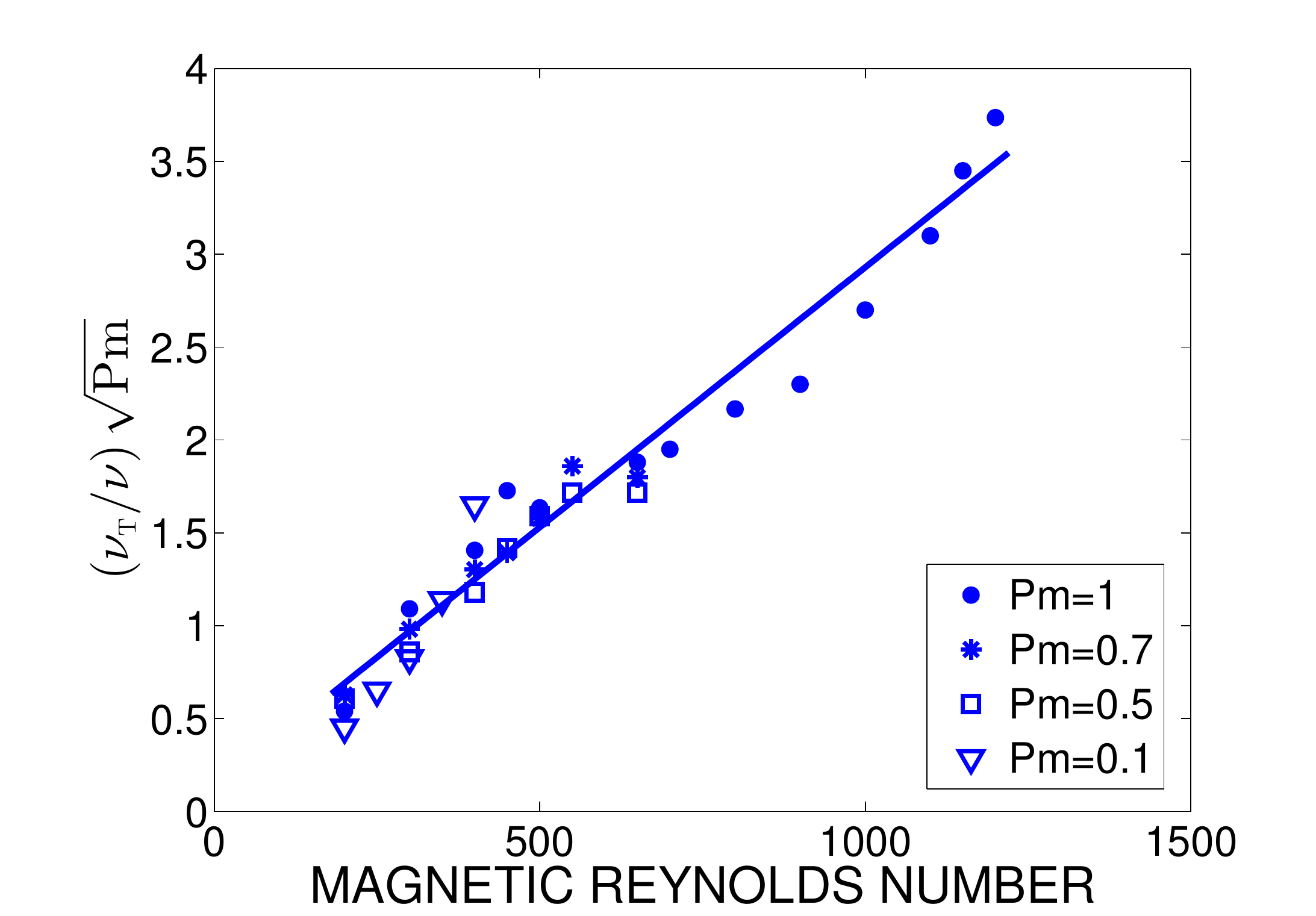}
\caption{Normalized viscosity for $\mu_\Om=0.25$ (left) and $\mu_\Om=0.5$ (right). The maximal values optimized with $\Ha$ are given for fixed $\Rm$. $\Pm=0.1-1$ as indicated. For variation of the rotation law the numerical values seem to vary as $1/\sqrt{\mu_\Om}$. $\mu_B=\rin=0.5$, perfectly conducting boundaries \cite{RG15}.
 }
\label{AMT1}
\end{figure}

We shall simulate AMRI for two different rotation profiles, i.e.~$\Om \propto 1/R^2$ and $\Om \propto 1/R$. The eddy viscosities are numerically computed in the instability cones for fixed Reynolds and Hartmann numbers, with the general result that $\nu_{\rm T}$ peaks at the location of the maximum growth rates (dotted lines in Fig.~\ref{f16}, left). The effective viscosity is calculated by computing the right-hand side of the relation (\ref{Trfi}) within the instability domain in Fig.~\ref{f16} (left). For a given Reynolds number, the Hartmann number is varied until the maximum value of $\nu_{\rm T}$ is found, always close to the line of maximum growth rate. Finally, the maximum viscosity between the inner and outer cylinder is taken. The average procedure in (\ref{Trfi}) concerns only the azimuthal and axial directions.

For various magnetic Reynolds numbers, this procedure yields viscosities which grow linearly for increasing ${\Rm}$. This is true for all rotation profiles between $1/R^2$ and $1/R$, including Keplerian (Fig.~\ref{AMT1}). For the magnetic Reynolds numbers of the order of 10$^3$ we do not find any indication of a saturation.  For ${\Pm} <1$ the resulting viscosity scales as $\nu_{\rm T}/\nu\propto \Rm/\sqrt{\Pm}$, which can also be written as 
\begin{equation}
\frac{\nu_{\rm T}}{\nu}\simeq 5\cdot 10^{-3}  {\Rmquer}
\label{nuT2}
\end{equation}
using the averaged Reynolds number (\ref{rmquer}). Unlike the MRI case, for AMRI we find a (weak) dependence of the viscosity on the magnetic Prandtl number, i.e.~$\nu_{\rm T}\propto \sqrt{\Pm} \Om_{\rm in}R_0^2$. The numerical factor is taken from Fig.~\ref{AMT1}. Note that we always only looked for the maximal values belonging to a given $\Rey$. We can thus assume that at least for $\Rey\lsim 10^3$ the effective viscosity does not exceed the given value. Refs.~\cite{GusevaMHD,GusevaApJ} have suggested though that the effective viscosities can become considerably enhanced once $\Rm\gsim 10^2$, when the turbulence is effectively triggered twice over, once by having $\Rey$ sufficiently large, and again by having $\Rm$ sufficiently large.

With the results in Section \ref{AMT} the eddy viscosities arising from MRI and AMRI can be compared. One finds
\begin{equation}
\frac{\nu_{\rm T, amri}}{\nu_{\rm T, mri}}\simeq \frac{100 } {\Ha_{\rm mri}} .
\label{nuT3}
\end{equation}
Obviously, the effective angular momentum transport for both instabilities also  depends on the  axial magnetic field strength $B_0$. As the AMRI values are maximal values and as $\Ha_{\rm mri}\lsim 100$ in our simulation in Section \ref{AMT}, we find both viscosities to be of the same order. If, however, the relation (\ref{nuT3}) is still valid for stronger fields (which we do not know) then ultimately the angular momentum transport by axial fields would be more effective than that by azimuthal fields.

\subsection{Super-AMRI}\label{Super}
In the following the stability of superrotation is considered,  i.e.~rotation profiles with positive shear ${\rm d}\Om/{\rm d}R>0$. Flows with stationary inner cylinder are the prototype of stable flows in hydrodynamics \cite{T36,SG59}, but see \cite{D17}. The nonlinear behavior is less clear as  Taylor-Couette experiments have shown instability in this regime \cite{W33,RZ99,BS10,PV12}. There are several present-day experiments with Reynolds numbers of order $10^6$ with various gaps between the cylinders and various aspect ratios $\Gamma= H/d$ (with $H$ as the height of the container). The Princeton experiment has the smallest aspect ratio ($\Gamma\simeq 2$) with precisely controlled endplates split into several independently rotating rings \cite{JB06,SJ09,BS10,SJ12,GGJ2012,Wei2016}. Other experiments have considerably greater aspect ratios, and also direct torque measurements, but no split-ring endplates, and hence potentially greater end-effects \cite{PL11,VH11,HV12,B11,MB12}. The measured torques are significantly greater than the results inferred in the Princeton experiment.

It is clear that superrotation cannot be destabilized by the standard magnetorotational instability with axial background fields. A WKB method for inviscid fluids in current-free helical fields has been applied providing two limits of instability in terms of the shear in the rotation law \cite{LG06,KS12,KSF12}. In the same WKB framework, the existence of the upper threshold was also found for purely azimuthal fields \cite{
SK15}. Any upper threshold suggests a magnetic destabilization of superrotating flows for sufficiently strong positive shear. It has already been shown, however, that for {\em rapid} rotation the current-driven instability of toroidal fields may always be stabilized by positive shear \cite{A78}. It thus only remains to probe superrotating flows with slow rotation for instability.

 \begin{figure}[htb]
 \centering
 \includegraphics[width=8cm]{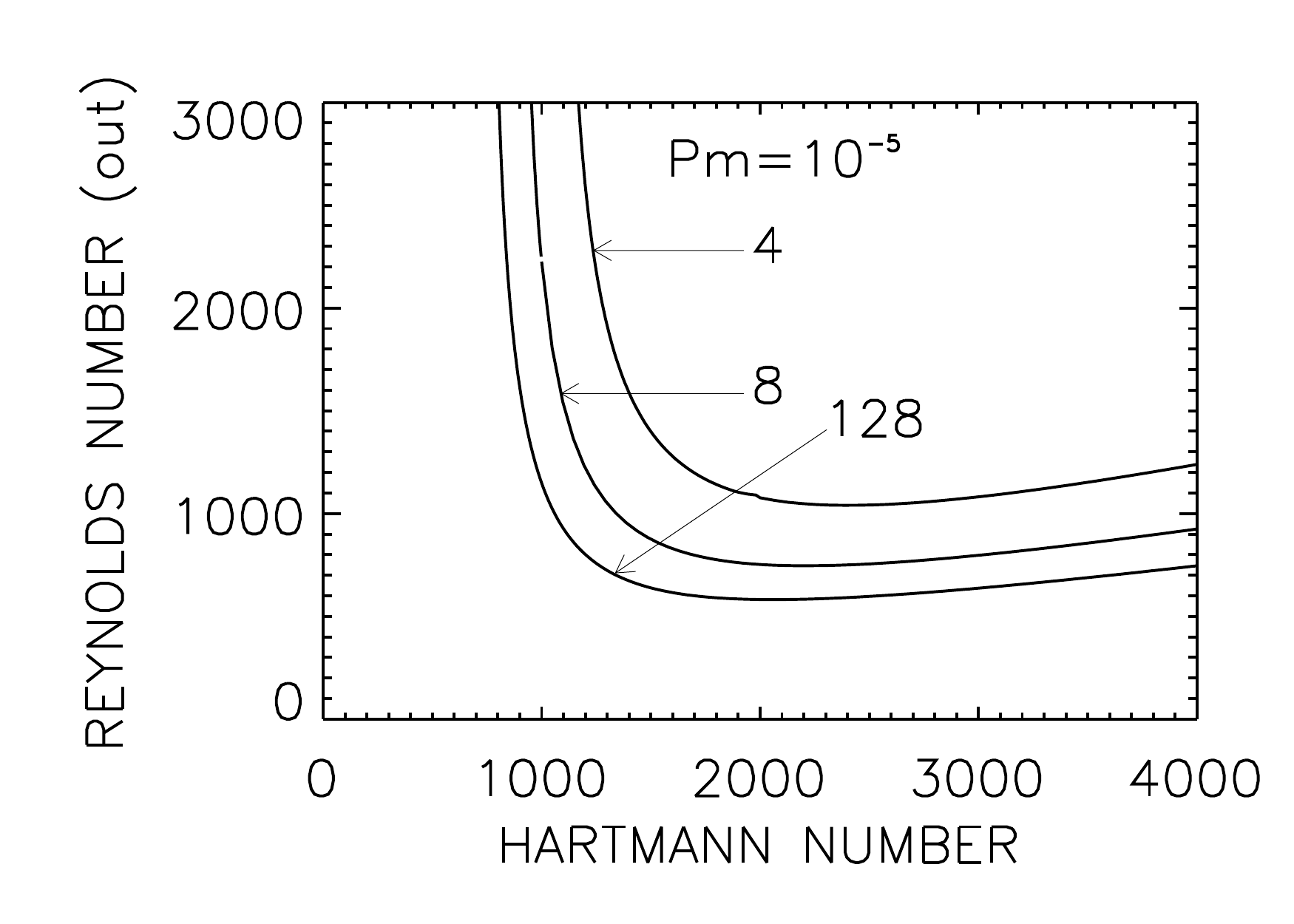}
 \includegraphics[width=8cm]{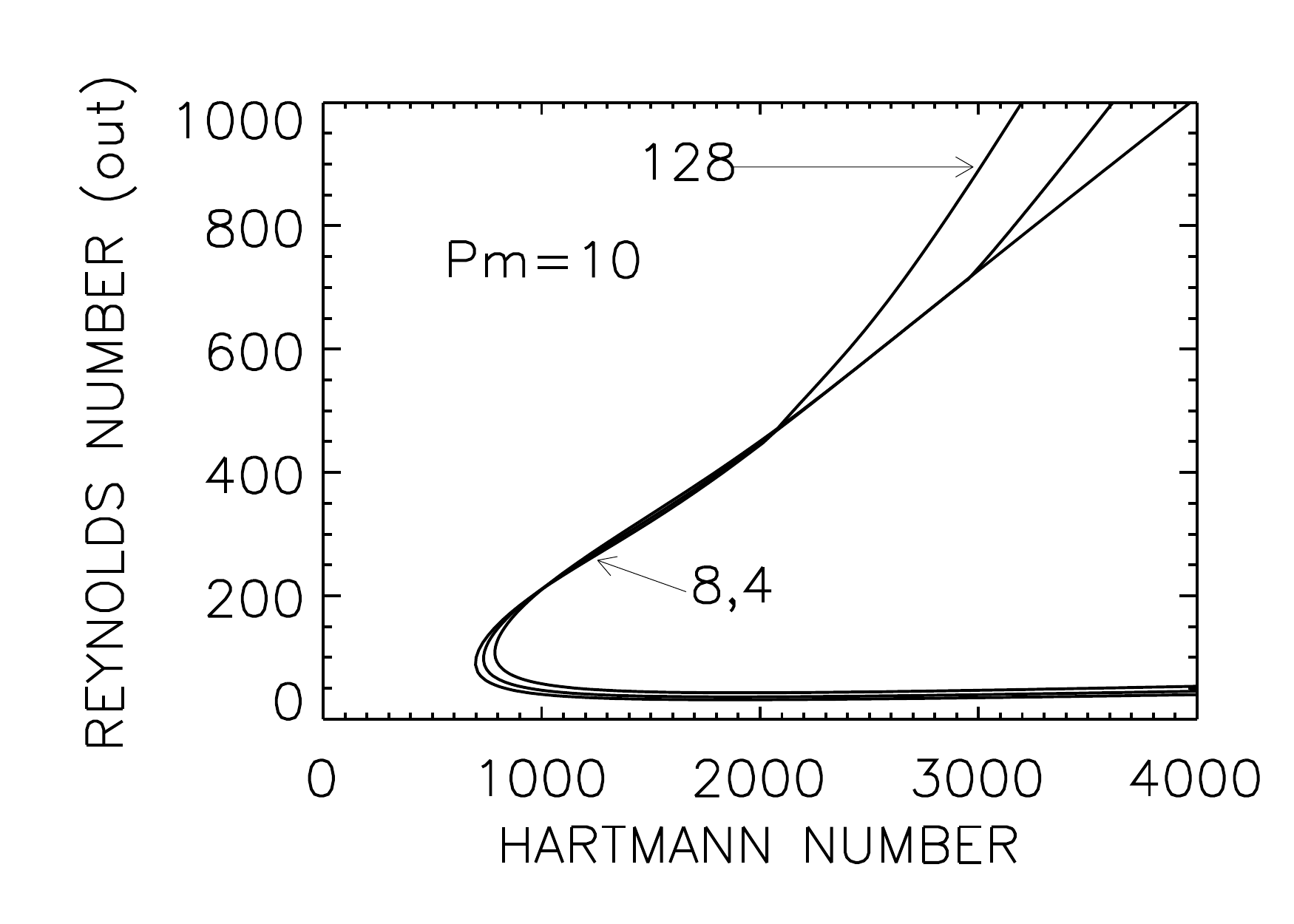}
 \caption{Stability maps  for superrotation and  small $\Pm$ ($\rm Pm=10^{-5}$, left) and large $\Pm$ ($\Pm=10$, right). The lines are marked with their values of $\mu_\Om>1$. Reynolds numbers are formed with the outer rotation rate.  We find the curves converging for large $\mu_\Om$.  A stationary inner cylinder
 can thus be modeled with $\mu_\Om\to \infty$. $\mu_B=\rin=0.95$, perfectly conducting cylinders.}
 \label{samri1}
 \end{figure}

The following models of Taylor-Couette flows in narrow gaps between perfectly conducting cylinders are considered, where the outer cylinder rotates faster than the inner one. It thus makes sense to here modify the definitions of the Hartmann and Reynolds numbers  as 
 \beg
\Ha =\frac{B_{\rm in} d}{\sqrt{\mu_0\rho\nu\eta}},  \ \ \ \ \ \ \ \ \ \ \ \ \ \ \ \quad \quad \quad \quad
 \Rey_{\rm out} =\frac{\Om_{\rm out} d^2}{\nu}
\label{Reout}
\ende
with $d$ as the gap width. The wave numbers $k$ and the eigenfrequencies $\omega$ will also be normalized with $d$ and the rotation rate $\Om_{\rm out}$ of the outer cylinder. Wave numbers of $\pi$, therefore, describe a circular cell geometry in the meridional plane between the cylinders, and a drift value of $\omega_{\rm dr}=-1$ describes corotation with the outer cylinder. Cells with $k<\pi$ are prolate while cells with $k>\pi$ are oblate with respect to the rotation axis. 

\begin{figure}[htb]
 \centering
 \includegraphics[width=10cm]{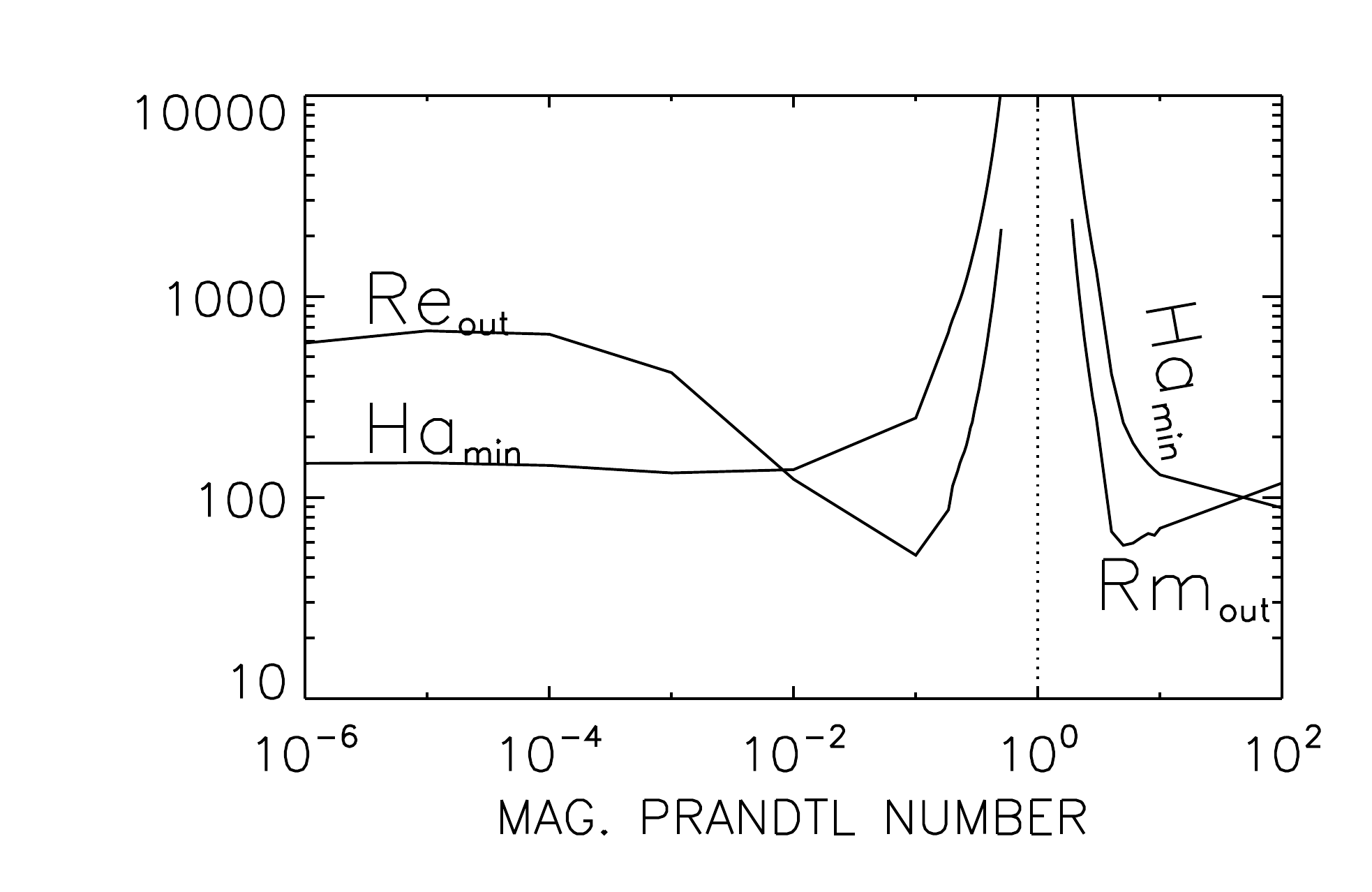}
 \caption{Minimal Hartmann numbers and corresponding $\Rey_{\rm out}$  for superrotating flows with (almost)  stationary   inner cylinder vs. $\Pm$. No solution exists for $\Pm=1$. $\mu_\Om=128$. $m=1$, $\mu_B=\rin=0.90$, perfectly conducting cylinders.}
 \label{ddnew}
 \end{figure}

According to our experience the instability of superrotation in wide gaps requires very high Reynolds numbers. The critical Reynolds number only falls below $10^4$  for $\rin\gsim 0.7$. In terms of future experiments it makes thus sense to restrict ourselves to consider narrow gaps in the present section. The lines in Fig.~\ref{samri1} represent the instability limit for the background field which is current-free in the very narrow gap ($\rin=0.95$) between the cylinders. The curves cannot cross the horizontal axis. The three hydrodynamically stable rotation laws have positive shear with $\mu_\Om=4,8,128$ and are magnetically destabilized in fluids with $\Pm=10^{-5}$ (left panel) and $\Pm=10$ (right panel). The instability curves disappear for $\Pm=1$, demonstrating that the differential rotation is able to deliver the entire energy for the maintenance of the instability patterns only for $\Pm\neq 1$; the magnetic field only acts as a catalyst. Instabilities which only exist for $\nu\neq \eta$ belong to the class of double-diffusive instabilities \cite{A78}. They do not appear for $\Pm$ of order unity. The basic Pm-dependence of the characteristic eigenvalues for a Taylor-Couette flow with almost stationary inner cylinder ($\mu_\Om=128$) is shown by Fig. \ref{ddnew}. The model with $\mu_\Om=128$ already gives an excellent approximation for the rotation profile with stationary inner cylinder.  $\Ha_{\rm min}$ denotes the smallest possible Hartmann number and $\Rey_{\rm out}$ the corresponding Reynolds number after (\ref{Reout}). Both values go to infinity for $\Pm\to 1$.   For $\Pm>1$ the magnetic Reynolds number is given instead of the ordinary Reynolds number for $\Pm<1 $ because of the different scaling behaviors for $\Pm\neq 1$.
 
The numerical results given in Fig.~\ref{samri1} also show that for small $\Pm$ the instability scales with the Reynolds number of the outer cylinder. The frequency of the inner cylinder does not play an important role. Note, however, that the rotation profile with the slowest inner cylinder becomes unstable most easily. The curves converge for $\Omin\to 0$. It is also interesting to see how easily the flow can be destabilized for large $\Pm$. While the Hartmann numbers for small and large $\Pm$ are very similar, the Reynolds numbers  differ strongly. Obviously, for given molecular viscosity the excitation is easier the smaller the magnetic diffusivity.

Edmonds \cite{E58} argued that in narrow gaps the radial profiles of the azimuthal fields between the cylinders are almost uniform with only small influences on the excitation conditions. Test calculations indeed provided instability even for fields with uniform $B_\phi$ for very similar Reynolds numbers and Hartmann numbers. One may assume that for $\Pm\neq 1$ the superrotation becomes unstable under the mere presence of any toroidal field, but for $\Pm=1$ the dissipation processes prevent the excitation of this slow instability. 

\subsubsection{Influence of boundary conditions}\label{AMRIBC}
\begin{figure}[h]
\centering
\includegraphics[width=8cm]{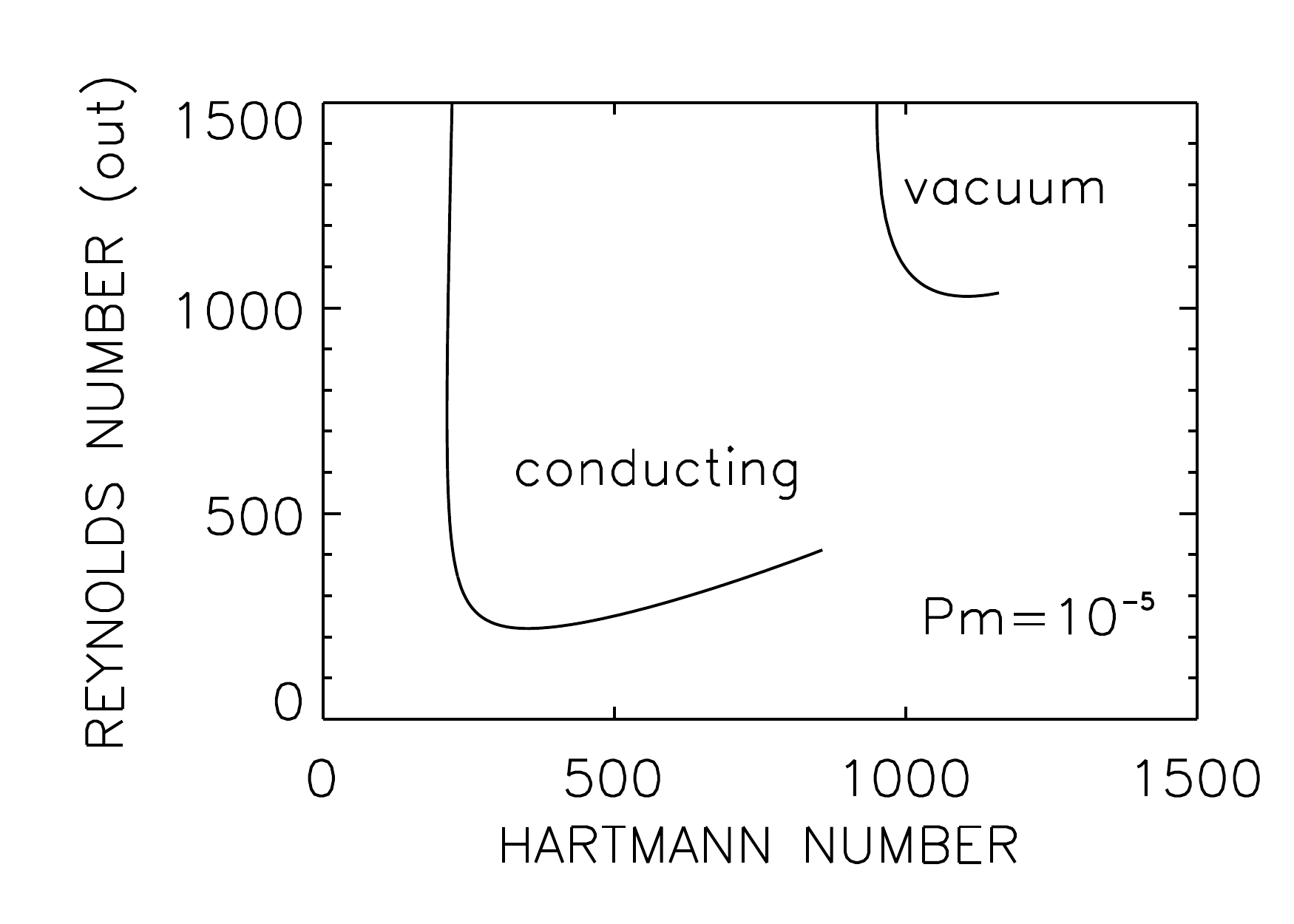} 
\includegraphics[width=8cm]{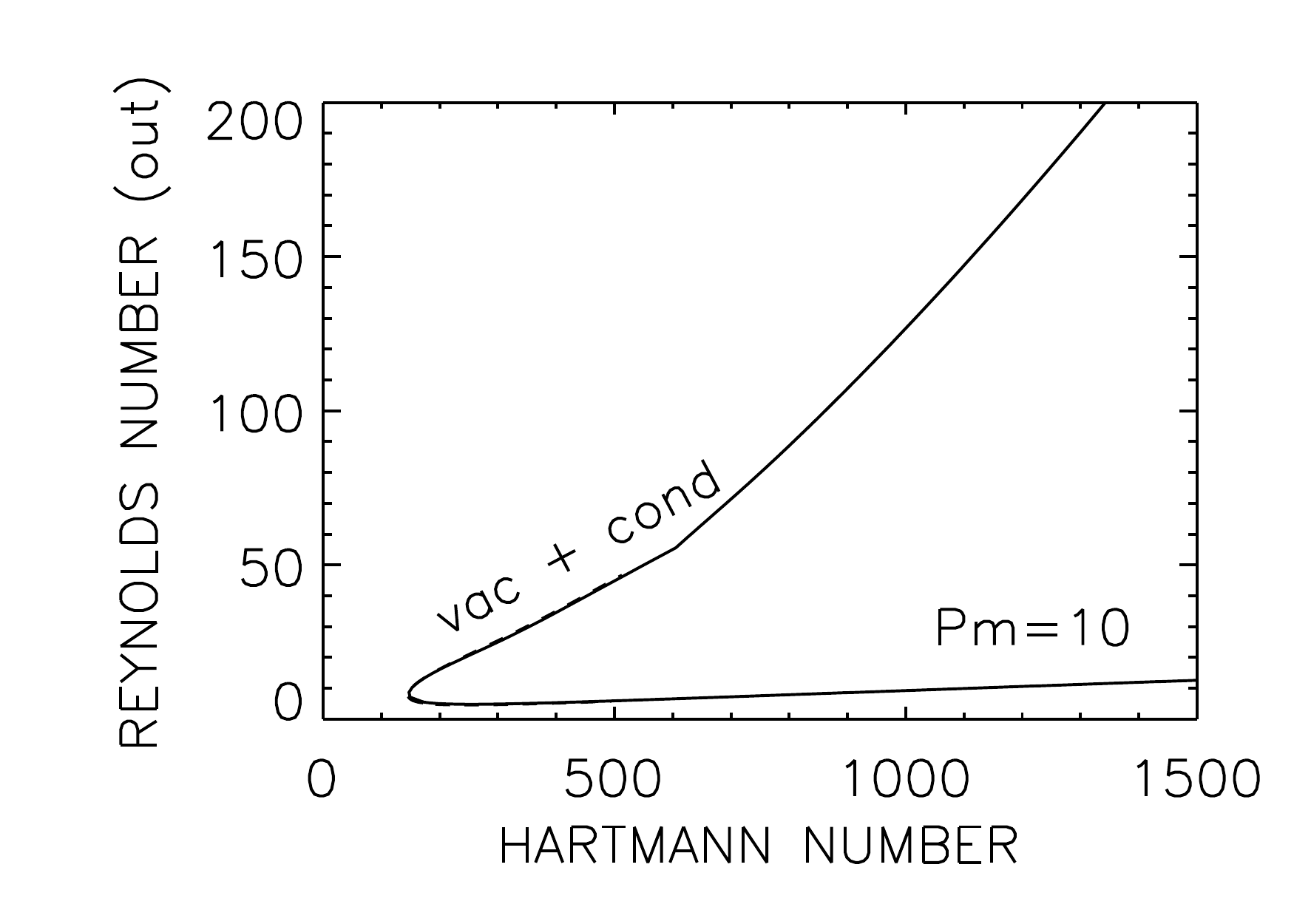} 
\caption{$\Rey_{\rm out}$ versus Hartmann number for superrotation  in a narrow gap with  perfectly conducting and insulating cylinders. Left: $\Pm=10^{-5}$. Right: $\Pm=10$, here the two lines cannot be separated. All solutions are sub-\A{ic} and  do not exist for $\Pm=1$. $m=\pm 1$, $\mu_B=\rin=0.9$. $\mu_\Om=5$.} 
\label{samri2}
\end{figure}

To investigate the influence of the boundary conditions, Fig.~\ref{samri2}  gives the instability map for the rotation law $\mu_\Om=5$ in models with slightly broader gaps ($\rin=0.9$) for the two cases of perfectly conducting and insulating boundary conditions. The magnetic Prandtl number is taken to be small ($\Pm=10^{-5}$, left) and large ($\Pm=10$, right). In the first case for insulating boundaries the superrotation laws are much more stable than for perfectly conducting boundaries. For conducting walls, both the Reynolds and Hartmann numbers are much smaller than they are for the insulating case. This is a striking difference to other magnetic instabilities. For the classical AMRI with negative shear the critical Hartmann numbers for both kinds of boundary conditions only differ slightly. Often, however, insulating boundary conditions lead to an easier excitation of the instability than conducting boundaries do. For large $\Pm$, however, the differences for the two boundary conditions completely disappear as the two curves cannot be distinguished. One finds again  positive slopes of both branches of the lines of neutral instability; only between them  the system is unstable.
Note also that all curves of marginal stability fulfill the condition $\Mm<1$ describing slow rotation.

\begin{figure}[h]
\centering
\includegraphics[width=8cm]{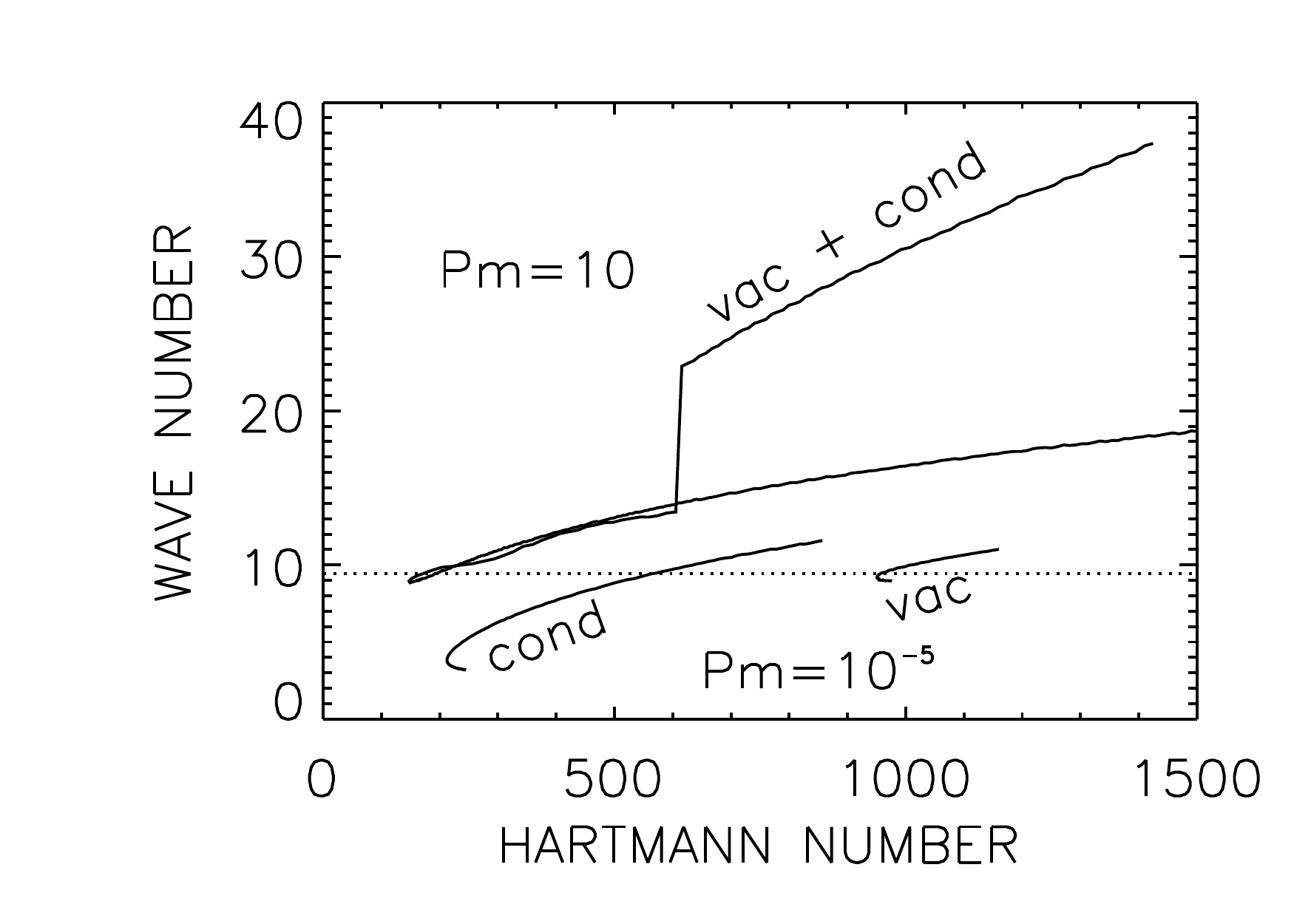} 
\includegraphics[width=8cm]{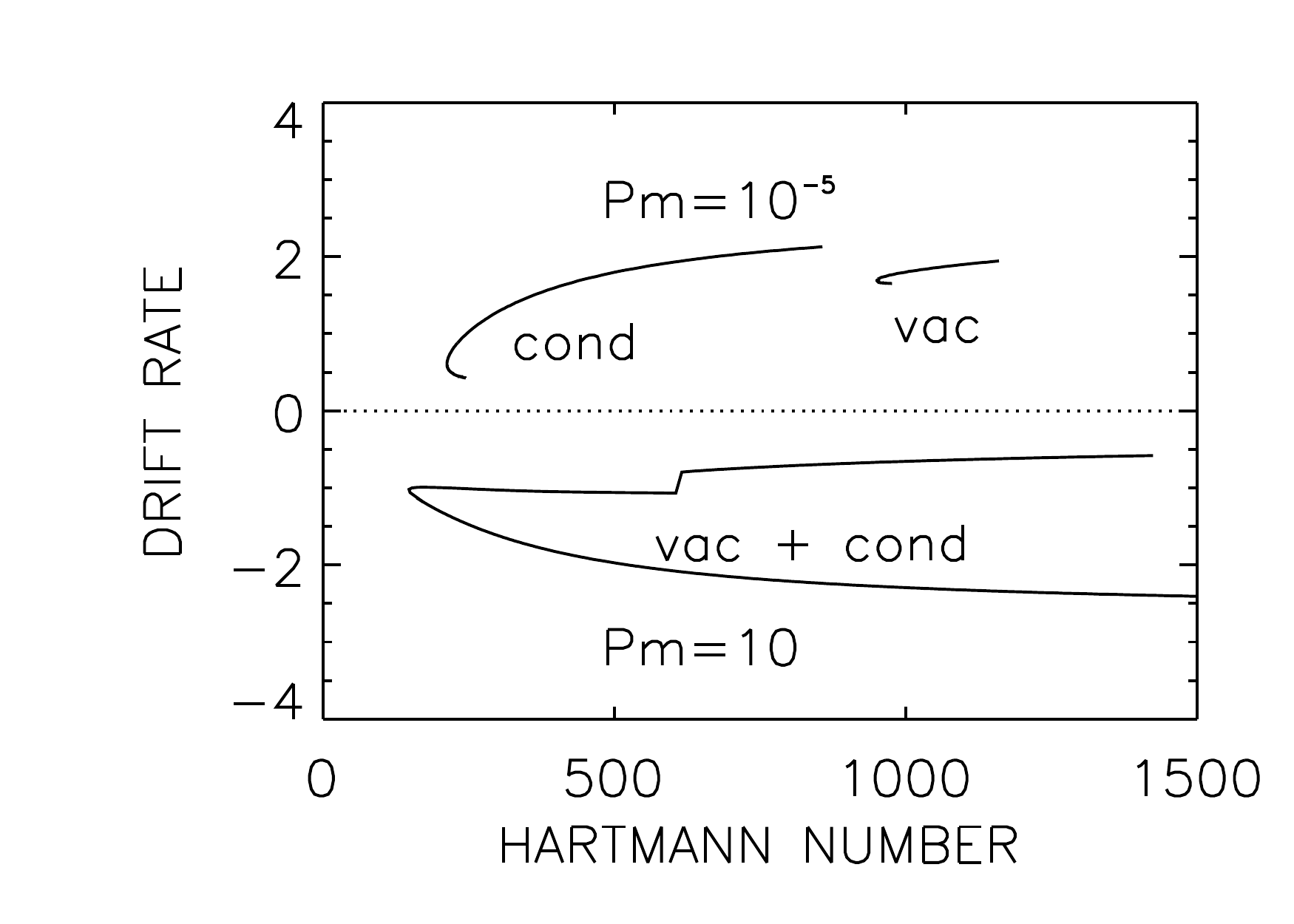} 
\caption{As in Fig.~\ref{samri2} but for axial wave numbers $k$ (left) and drift frequencies $\omega_{\rm dr}$ (right) for $\Pm=10^{-5}$  and $\Pm=10$.
The dotted line ($3\pi$) in the left panel gives the location of  cells circular in the meridional  ($R/z$) plane. The cells for small (large) magnetic Prandtl number are prolate (oblate) in this meridian.  The drifts have opposite signs for small and large $\Pm$.}
\label{samri3} 
\end{figure}

The left panel of Fig.~\ref{samri3} gives the axial wave numbers of the flow pattern  along the branches of neutral stability for small and large $\Pm$. The limit $k=\pi$ for nearly circular  cells in the meridional $(R/z)$ plane is marked by a horizontal dotted line. The cell geometry indeed depends on the magnetic Prandtl number. For small $\Pm$ the axial  wave numbers are smaller than for large $\Pm$, hence the cells are prolate. Along the strong-field branch of the instability cone the wave numbers exceed those at the weak-field branch where the cells are almost circular in the meridional plane. For $\Pm\gg 1$, however, the wave numbers at both branches are much larger so that the cells are always very flat. Note that the influence of the  boundary conditions is only weak; for $\Pm=10$ it vanishes  completely.

The drift rates even  possess a very strong $\Pm$-dependence. They are given in the right panel of Fig.~\ref{samri3} as the real parts $\omega_{\rm dr}$ of the frequency $\omega$ of the Fourier mode of the instability normalized with the rotation rate of the {\em outer} cylinder. From (\ref{dotfi}) the azimuthal migration has the opposite sign of $\omega_{\rm dr}$. For small $\Pm$ we find positive $\omega_{\rm dr}$ hence the instability pattern rotates {\em backwards}. Unlike for the AMRI with negative shear, large $\Pm$ yield negative drift values, and the pattern migrates with the rotation. For the lowest Hartmann number one even finds $\omega_{\rm dr}=-1$ so that in this particular case the pattern corotates with the outer cylinder. Again, for high values of $\Pm$ the influence of the boundary conditions even vanishes.
\medskip

We have seen that 
 for small $\Pm$ the minimal Hartmann numbers  for perfectly conducting cylinders are { much} smaller than those  for insulating cylinders. Almost all theoretical  investigations only worked with these extremal  boundary conditions.  They, however, are far from reality. E.g., the conductivity of copper (as the cylinder material) is only about five times higher than that of liquid sodium (as the  fluid), hence the conductivity ratio 
(\ref{hatsigma})
for this combination approaches the value of  5. The question is thus whether such a small  conductivity ratio of cylinders and fluid still leads to magnetic fields for the onset of instability close to the results for perfectly conducting material or not.
\begin{figure}[h]
\centering
\includegraphics[width=8cm]{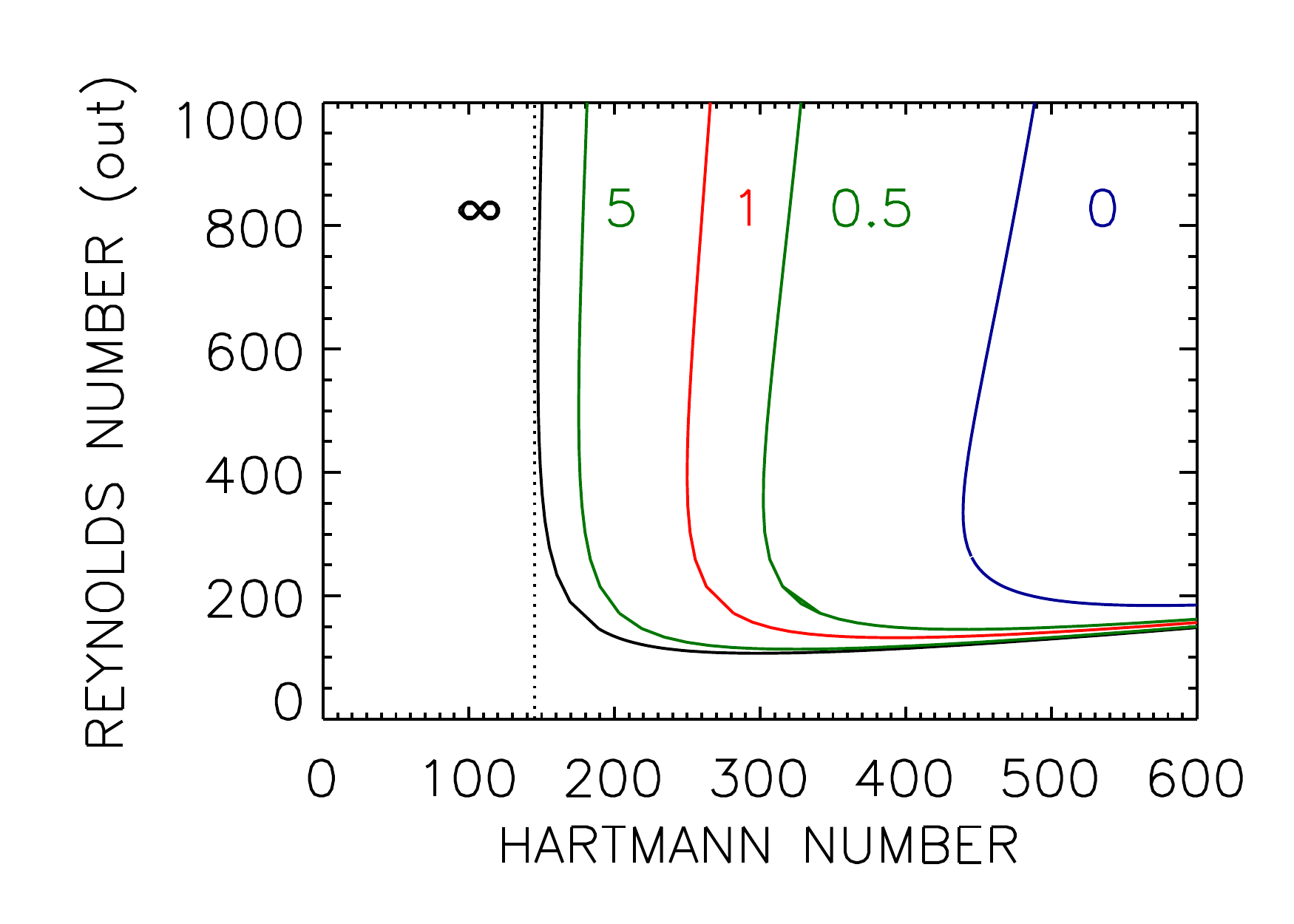} 
\includegraphics[width=8cm]{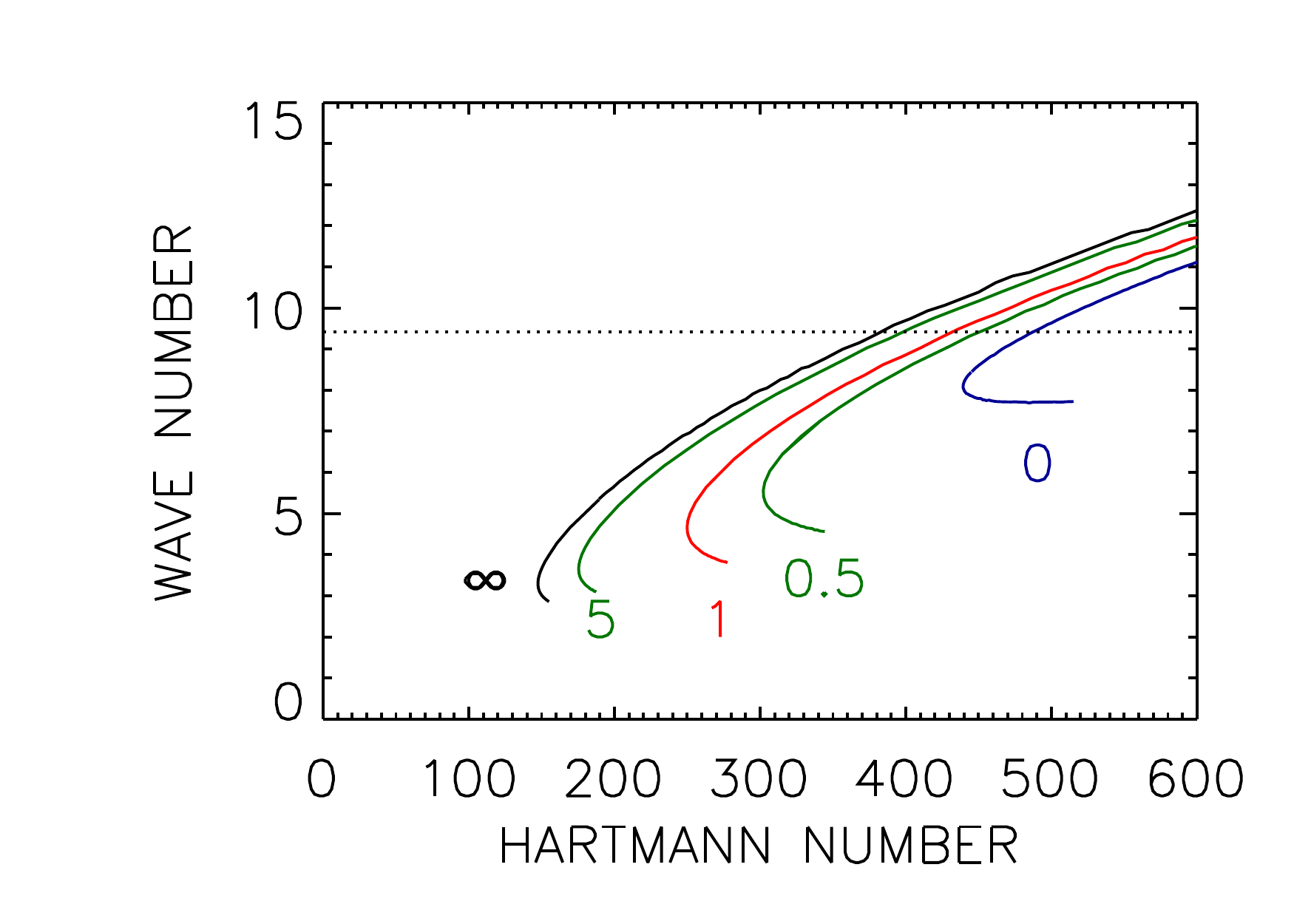} 
\caption{$\Rey_{\rm out}$ (left panel) and normalized wave numbers (right panel) versus Hartmann number for superrotation  in a narrow gap between  cylinders of finite conductivity. The curves are marked with their values of $\hat\sigma$:  $\hat\sigma=0$ (blue), $\hat\sigma=1$ (red), $\hat\sigma=\infty$ (black). The vertical dotted line indicates the global minimum Hartmann number. At the  horizontal dotted line the cells are almost circular in the $(R/z)$ plane. $m=\pm 1$, $\mu_B=\rin=0.9$. $\mu_\Om=128$.  $\Pm=10^{-5}$.}
\label{samri33} 
\end{figure}

The boundary conditions for finite values of the ratio $\hat\sigma$  are given in Section \ref{axfi} following  Eq. (\ref{bc17}). With these conditions  the  left panel of  Fig. \ref{samri33} presents the instability maps for $\mu_\Om=128$,  $\Pm=10^{-5}$ and for various $\hat\sigma$ in the $(\Ha/\Rey)$ plane.  While the critical Reynolds numbers only slightly depend on the cylinder conductivity the critical Hartmann numbers do not. The absolute minimum  of the critical Hartmann number belongs to the  perfectly conducting boundary condition. The (red) line for $\hat\sigma\simeq 1$ approximately lies in the middle of the instability domain defined by  the two extremes for the  cylinder conductivity.  The solutions for (say) $\hat\sigma>5$ are located close to the line for $\hat\sigma=\infty$. On the other hand,   the solutions for  $\hat\sigma<1/5$ are located rather close to the line for $\hat\sigma=0$.  Both the minimum Hartmann number and the associated Reynolds number for $\hat\sigma>5$ thus only differ slightly from the values for perfect conductors. 

The  influence of the boundary conditions on the shape of the instability cells is also  strong.
The wave numbers in Fig. \ref{samri33} must be interpreted using (\ref{delz}) so that for $k R_0\simeq 3 \pi$ (horizontal dotted line) the cells are almost circular in the ($R/z$) plane. Below the horizontal  dotted line the cells  are all oblong with respect to the rotation axis. Contrary to the Hartmann numbers  the wave numbers for $\hat\sigma=1$  are already close to the values for perfectly-conducting cylinders.

\subsubsection{Higher modes}\label{Higher}
\begin{figure}[htb]
\centering
\includegraphics[width=5.25cm]{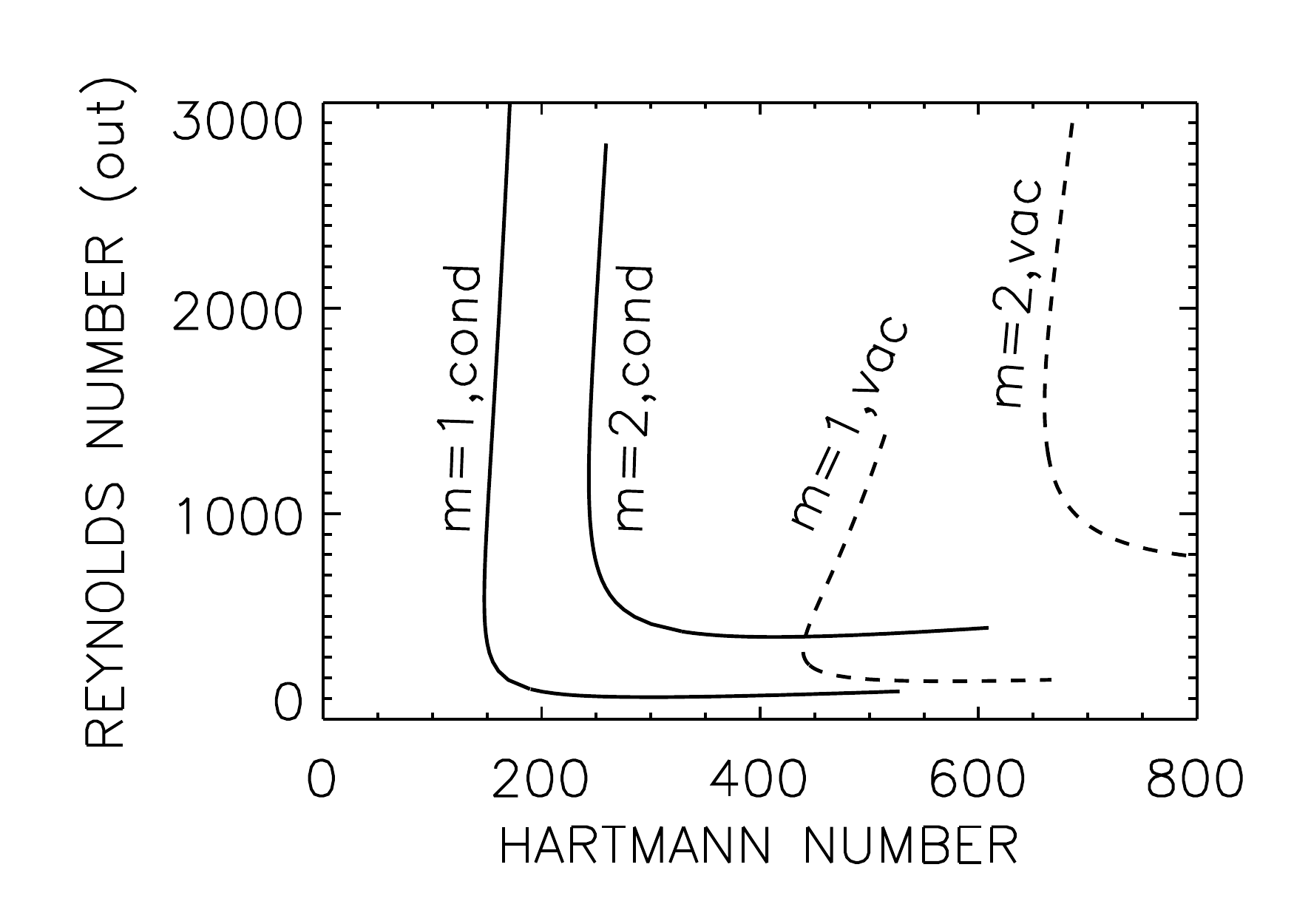} 
\includegraphics[width=5.25cm]{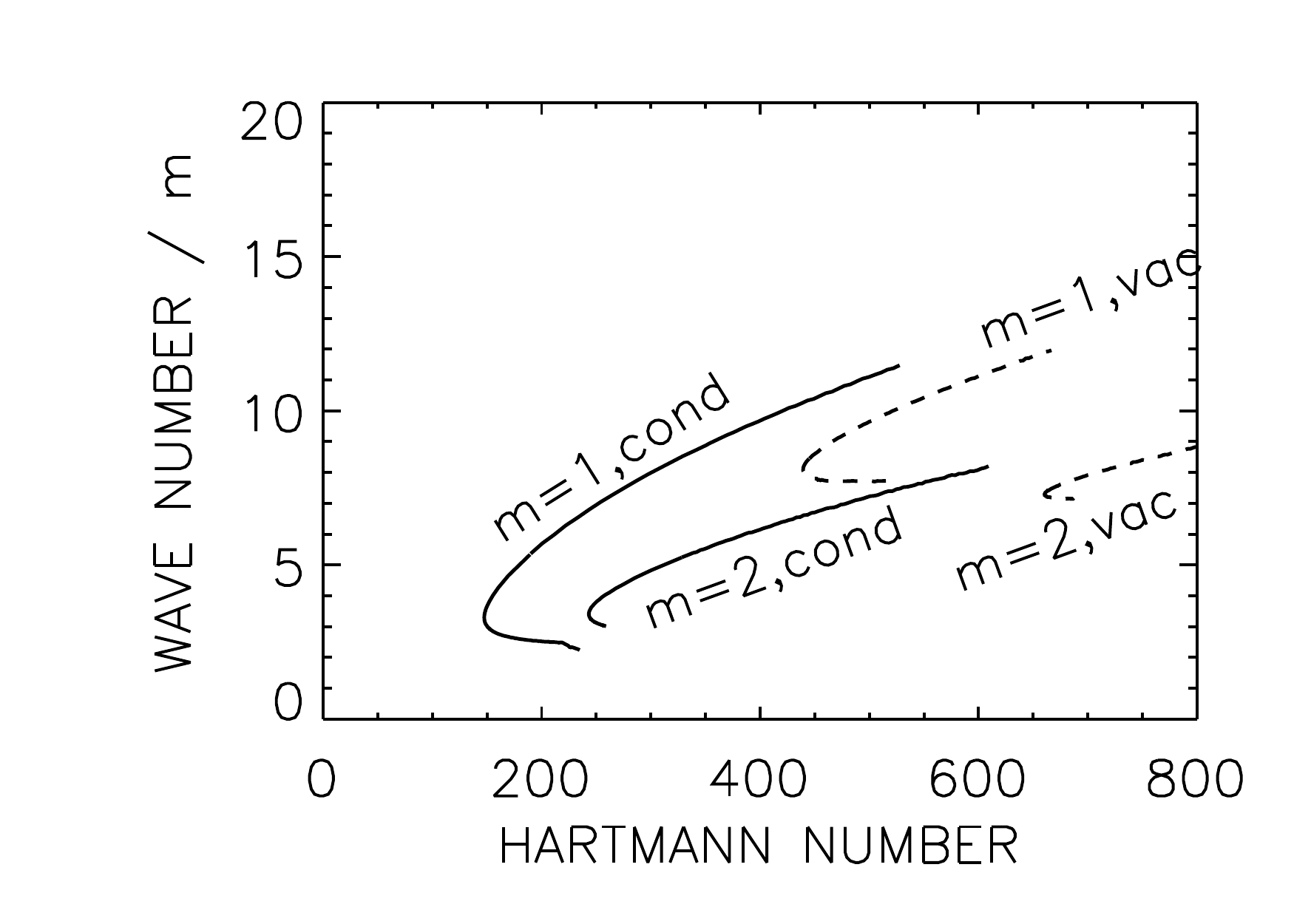} 
\includegraphics[width=5.25cm]{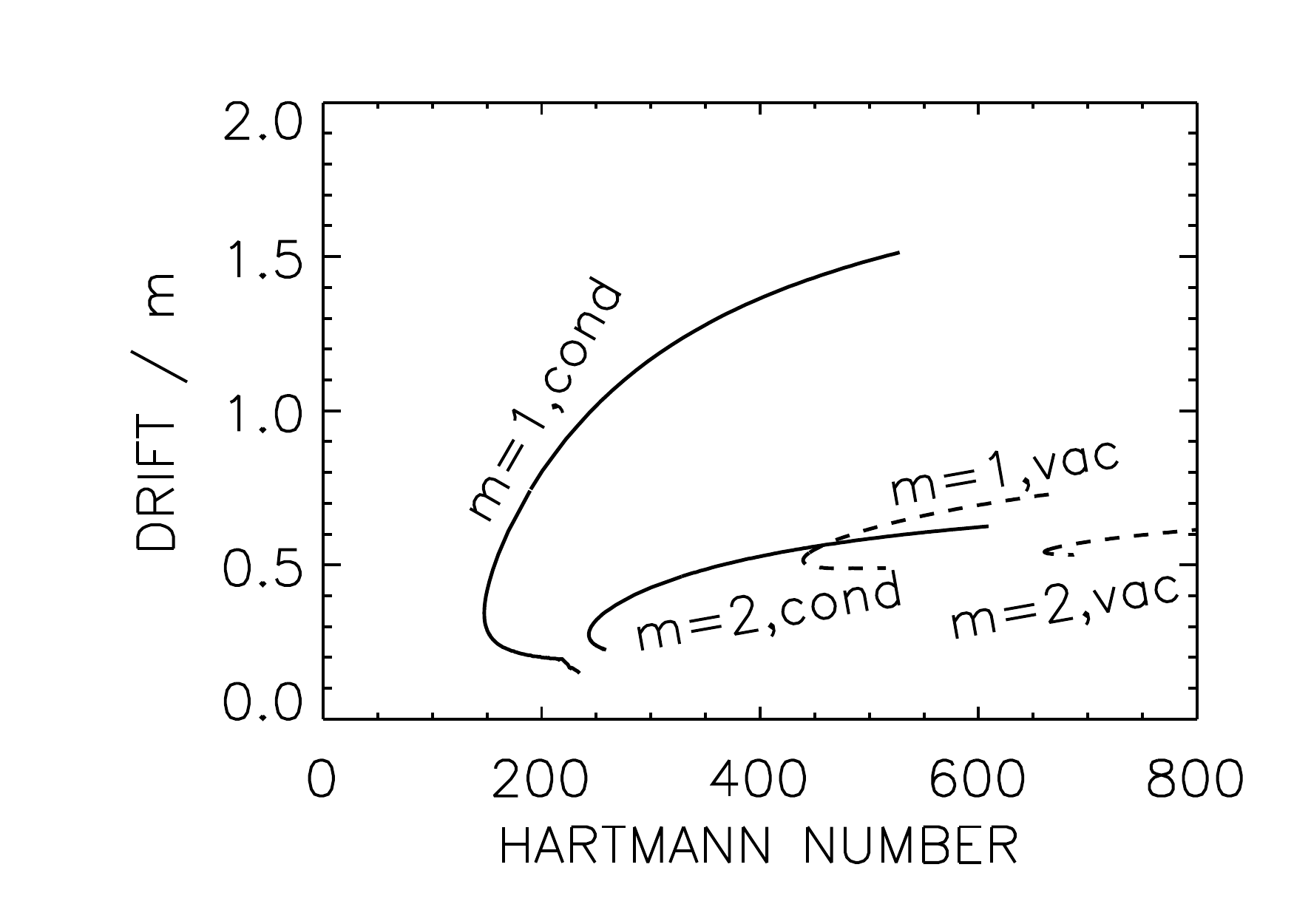} 
\caption{Critical Reynolds numbers (left panel), critical normalized wave numbers  (central panel) and the corresponding drift rates  (right panel) versus  Hartmann number for superrotating flows with  (almost) resting inner cylinder  ($\mu_\Om=128$). The cells prove to be always oblate.  $\mu_B=\rin=0.9$, $\rm Pm=10^{-5}$, perfectly conducting (`cond') and insulating (`vac') cylinders. $m=\pm1,2$.}
\label{samriplotsnew}
\end{figure}

The plots in Fig. \ref{samriplotsnew} show the Reynolds numbers, the critical wave numbers and the corresponding drift rates as function of the Hartmann number defined by  Eq.~(\ref{Reout}) for marginal instability  for a  narrow-gap model with $\rin=0.9$ and a very high value of $\mu_\Om$. The results for such large $\mu_\Om$  are also  representative for $\mu_\Om\to \infty$. The results hardly change for even slower  inner rotation. As the stability lines for the fixed small magnetic Prandtl number ($\Pm=10^{-5}$) are given for the two boundary conditions, i.e.  perfect-conducting  (solid lines)  and insulating  (dotted lines). The lines are calculated for the two azimuthal modes with $m=1$ and $m=2$. From the stability lines of the left panel in Fig. \ref{samriplotsnew} one immediately finds that in both cases  the excitation of the higher modes requires higher values of  Reynolds number and  Hartmann number than the excitation of the lower mode. There is no indication, however, for a different scaling of the critical values with $\Pm$ for the different mode numbers $m$ (as we shall find below  in Figs. \ref{g4} and \ref{h55}). Also for the higher modes  the insulating boundary condition increases the  critical magnetic fields   for the onset of the instability.

As expected the azimuthal drift (\ref{dotfi}) shown in the right panel of Fig.  \ref{samriplotsnew} is almost independent of the mode number $m$. The same is true for the normalized wave number divided by $m$ which, however, strongly depends on the choice of the boundary conditions. Similar to the argumentation for the azimuthal drift rate (\ref{dotfi}) as almost independent of $m$ it is also true that the pitch angle $\partial z/\partial \phi= -m/k$ does not depend on the mode number $m$. It thus  makes  sense to expand the solutions as the Fourier modes $ {\rm exp}({{\rm i}m(\omega t+kz+ \phi)})$.

\subsubsection{Electric currents}\label{Ec}
\begin{figure}[h]
\centering
\includegraphics[width=8cm]{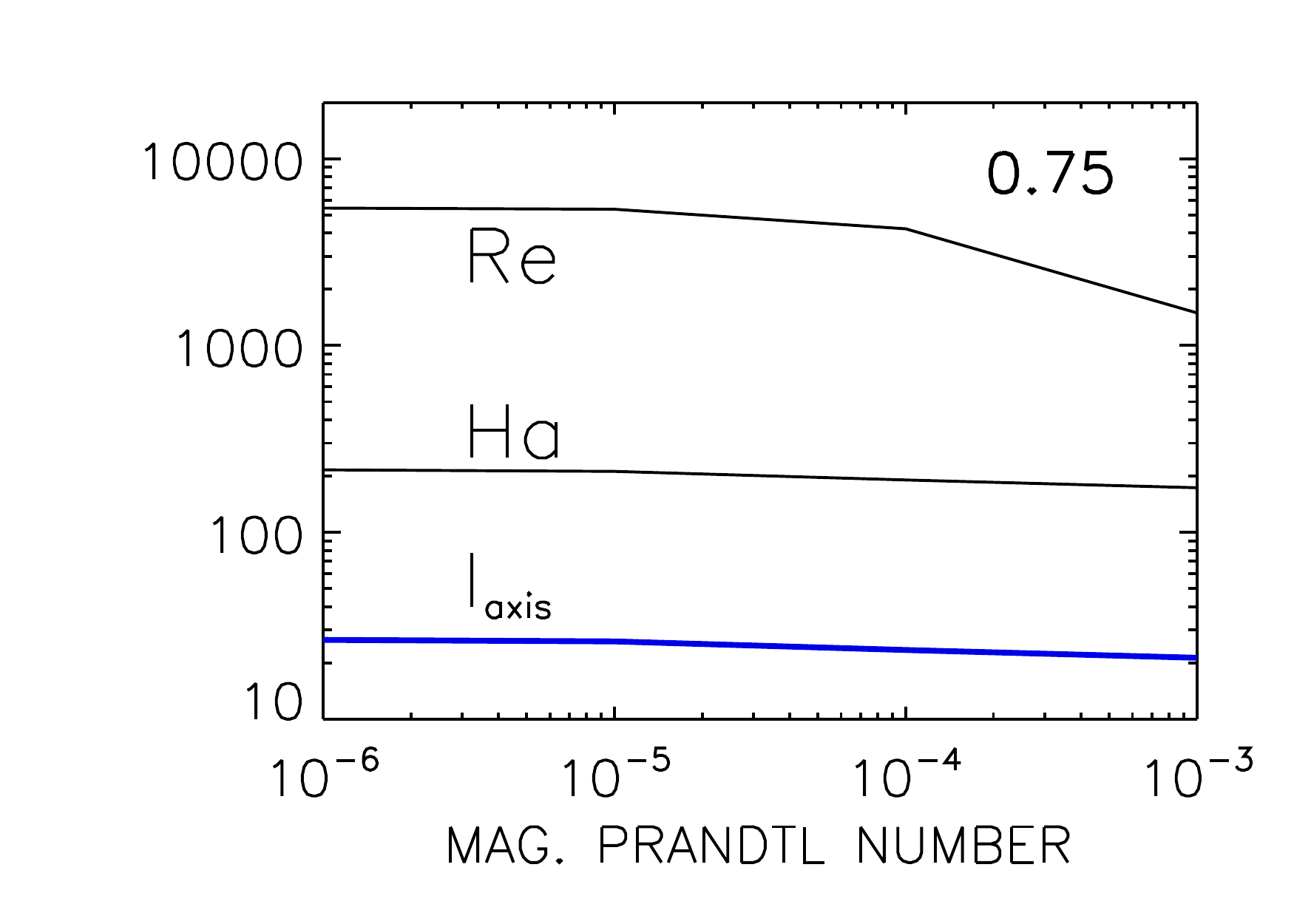}
\includegraphics[width=8cm]{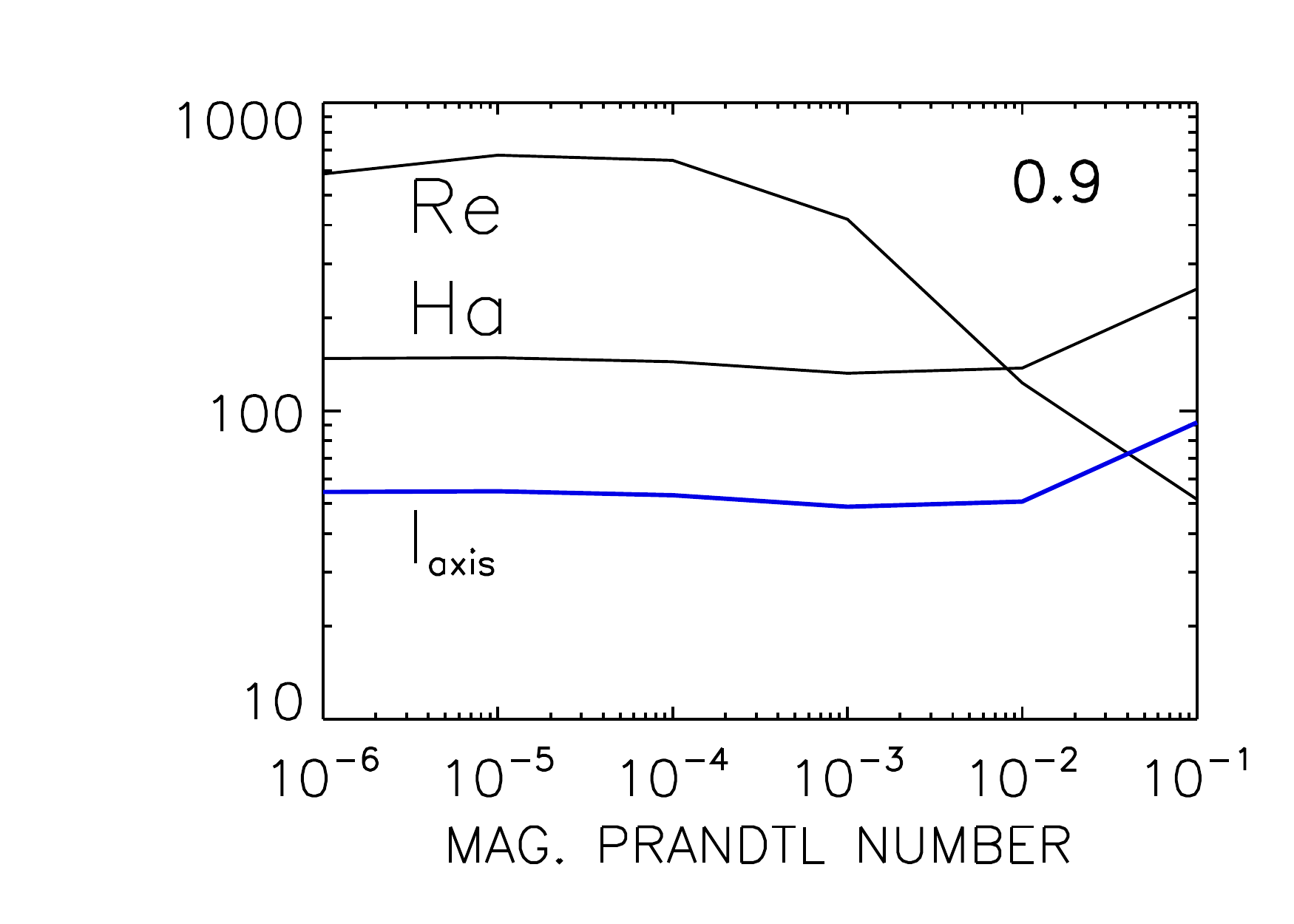}
\caption{Minimal Hartmann numbers, the related outer Reynolds numbers and the necessary axial electric currents (in kA, along the inner rod) for $\rin=0.75$ (left) and $\rin=0.9$ (right) with almost stationary inner cylinder  as functions of the magnetic Prandtl number.  Left: $\mu_B=\rin=0.75$, right: $\mu_B=\rin=0.9$. In all cases the instability appears to scale with $\Ha$ and $\Rey$ for $\Pm\to 0$. The necessary axial electric currents $I_{\rm axis}$ (blue lines) become  weaker (!) for wider gaps. They do not depend on the size of the container. $\mu_\Om=128$. Perfectly conducting boundaries.}
\label{samriplots}
\end{figure}

Here we are also interested in the values of the absolute minimum
of the Hartmann number for neutral stability in order to discuss the possibility of laboratory
experiments. The minimum Hartmann numbers have been defined for  Fig. \ref{ddnew}
and characterizes  the absolutely minimum magnetic field for possible experimental realizations of the super-AMRI phenomena. For two  narrow-gap models with (almost) stationary inner cylinders for the modes with $m=1$ the $\Ha_{\rm min}$ and the related Reynolds numbers  are given in Fig. \ref{samriplots} as functions of the magnetic Prandtl number.

For small $\Pm$ the resulting values of $\Ha$ and $\Rey$ do not depend on $\Pm$, hence the eigenvalues scale  for small $\Pm$ with $\Ha$ and $\Rey$. For large enough  positive shear these solutions thus also exist in the inductionless approximation.  The question is still open  whether an absolute  minimum value of $\mu_{\Om,0}>1$ exists below which only stability occurs. The limit mentioned at the end of Section \ref{QKr}  for $\rin=0.9$ is only  $\mu_{\Om,0}=2.7$ but it concerns  the $m=0$ solutions of the HMRI.  No solution exists for $\Pm\leq 1$ larger than the  values shown in Fig.~\ref{samriplots}; fluids with $\Pm=1$ are always stable. Also solutions for $\Pm\gg 1$ exist (Fig.~\ref{samri2}, right) but the scaling of the eigenvalues for $\Pm\to \infty$ is still unknown.

Note also the rather different Reynolds numbers between the left and the right panel of Fig.~\ref{samriplots} if $\Pm\ll 1$.  For the reduction of the gap width from 0.25 to 0.10 the critical Reynolds number for the onset of instability decreases by a factor of 10.

In order to transform the Hartmann numbers to the generating axial electric currents
 (i.e.~within the inner cylinder) the relation $I_{\rm axis}=5 R_{\rm in} B_{\rm in}$ can be rewritten as
\begin{eqnarray}
I_{\rm axis}= 5 \frac{R_{\rm in}}{d} \ \Ha\ \sqrt{\mu_0\rho\nu\eta},
\label{current} 
 \end{eqnarray}
where $R_{\rm in}\Ha/d$ represents a Hartmann number formed with $R_{\rm in}$ instead with $d$. 
Equation (\ref{current}) gives the minimum current for instability, since the numbers in Fig.~\ref{samriplots} also hold for the Hartmann numbers which are the lowest for marginal instability.

The marginal values are independent of $\Pm$ provided $\Pm\ll 1$. One also finds the minimal Hartmann numbers almost independent of the gap width. The consequence is that the corresponding axial electric current decreases for wider gaps. The minimum current in the calculations is 26 kA for the gap with $\rin=0.75$. For wider gaps the necessary Hartmann number increases strongly, and also the necessary electric current. The linear size of the container does not influence the excitation of the instability. In all cases the critical Reynolds number of the outer cylinder (the inner-one is stationary) is of order $10^3$, which should also be possible to be realized in the laboratory.

\section{Chandrasekhar-type flows}\label{Chandra}
The combination of a magnetic field $B_\phi\propto 1/R$ (current-free for $R> 0$) and a rotation profile $\Om\propto1/R^2$ (the potential flow) constitutes an example of a particular class of MHD flows defined by Chandrasekhar \cite{C56} as
\begin{eqnarray}
 \vec{U}=\vec{U}_{\rm A},
 \label{chandra1}
\end{eqnarray}
or more generally,
\begin{eqnarray}
 \vec{U}=\Mm\ \vec{U}_{\rm A}
 \label{chandra2}
\end{eqnarray}
with the magnetic Mach number $\Mm$ taken as constant here. The radial profiles of $\vec{U}$ and $\vec{U}_{\rm A}=\vec{B}/ \sqrt{\mu_0\rho}$ are required to be identical, but there may be a constant of proportionality between the two \cite{TM87}. As shown by Chandrasekhar, all basic states satisfying (\ref{chandra1}) are stable in the absence of dissipation. In Section \ref{potflow}, however, we found that the potential flow can be destabilized by a toroidal magnetic field with $B_\phi\propto 1/R$ if at least one of the two molecular diffusivities $\nu$ and $\eta$ is non-zero.

Taking $\Om\propto R^{-q}$ and $B_\phi\propto R^{1-q}$ (thereby satisfying Eq.~(\ref{chandra2}) with non-negative $q$) Michael's relation (\ref{bfcr}) yields 
$
 (2-q){\Mm}^2 +q>0
$
as a sufficient condition for stability of the $m=0$ mode. All Chandrasekhar states with $0 \leq q \leq 2$ are thus stable against axisymmetric perturbations, as any state satisfying (\ref{chandra2}) is stable for ideal fluids. It becomes clear that the condition (\ref{bfcr}) is not a necessary condition for stability. The relations $\Om\propto R^{-q}$ and $B_\phi\propto R^{1-q}$ defining the class of Chandrasekhar-type flows which we shall consider lead to 
\beg
\mu_\Om=\rin\mu_B,
\label{chancon}
\ende 
so that for $\rin=0.5$ simply $\mu_B=2 \mu_\Om$. As an example, for the very wide gap with $\rin=0.05$ and for rigid rotation the value is $\mu_B=20$.
 
The Chandrasekhar condition (\ref{chancon}) can also be fulfilled with negative values describing  profiles  $\Om(R)$ and $B_\phi(R)$ changing in sign  somewhere between the boundaries. The cylinders are then counterrotating. Both  the magnetic field and the rotation law with negative $\mu$'s can be unstable against  axisymmetric perturbations ( \cite{G62}, see also Fig.~\ref{ti2}, below). The combination of both unstable profiles leaves stability  for $\Pm=1$  only in a very narrow strip    along the line $\Mm=1$ \cite{Sh07,CKH2015}. 

\subsection{Inductionless approximation}\label{inductionless}
Following  \cite{C61,R67} we transform  Eqs.~(\ref{Navier}) and (\ref{Ind1})  with respect to the inductionless approximation. A linearized and dimensionless  version of these equations reads
\begin{equation}
\Rey\ \Big(   
\frac{\partial \vec{u}}{\partial t} + ({\vec{ U}}\cdot \nabla)\vec{u}+  
({\vec{u}}\cdot \nabla){\vec{ U}}
\Big) =
- \nabla P +  \Delta \vec{u} + 
\Ha^2 (  \rot\vec{b} \times {\vec{ B}}+\rot\vec{ B} \times \vec{b}   )
\label{mom}
\end{equation}
and 
\begin{equation}
\Pm\ \Rey\ \Big( \frac{\partial \vec{b}}{\partial t}- 
{{\rot}}~(\vec{U} \times \vec{b})\Big)= \rot~(\vec{u\times \vec{B}})
+  \Delta\vec{b}.
\label{mhd}
\end{equation}
The   magnetic fields in these equations  are normalized with  
characteristic   $B_0$ of the background field. The mean 
flow $\vec U$ is normalized with a  flow 
amplitude while  the flow perturbations are normalized with $\eta/d$ (with distance  $d$). Reynolds number and Hartmann number are formed with these scales. 
Obviously, the limit $\Pm\to 0$ is only allowed for finite  $\Rey$, hence   the solutions within the inductionless approximation must possess  finite Reynolds number. On the other hand, solutions of the linearized MHD equations which do not scale for small $\Pm$ with $\Rey$ and $\Ha$ cannot possess a solution for $\Pm=0$. We shall see in this section that the entire class of Chandrasekhar-type flows (\ref{chandra1}) possesses marginal instabilities scaling with $\Rey$ and $\Ha$ for small $\Pm$ (at least for the fundamental mode $m=1$) so that  they also exist for $\Pm=0$ -- in great contrast to the eigensolutions of the standard MRI in Section \ref{standardMRI} which do not exist for $\Pm=0$ \cite{GJ02}. Also,  the results of the inductionless approximation basically differ from those of the inviscid approximation. All  eigensolutions which for small $\Pm$ scale with $\Lu$ and $\Rm$
should also fulfill the inviscid  MHD equations with $\nu=0$. 
\subsection{Potential flow}
The potential flow with $q=2$ under the influence of a current-free background field simultaneously  belongs to the classes of Chandrasekhar-type flows and of AMRI. $U_\phi$ and $B_\phi$ are both proportional to $1/R$, hence $\mu_B=2\mu_\Om=0.5$ for $\rin=0.5 $. Stability maps (Figs.~\ref{f16} and \ref{f17}) show that just for this case and for $\Pm\to 0$ the Reynolds and Hartmann numbers  (\ref{Hartmannin}) for neutral stability do not depend on the magnetic Prandtl number. We show here that this particular scaling (which is the basis of the technical realization of several MHD experiments with fluid metals) is characteristic for {\em all} Chandrasekhar-type flows fulfilling the relation  (\ref{chancon}). The potential flow with $\mu_\Om=0.25$ fulfills this condition and therefore scales with $\Rey$ and $\Ha$ for small $\Pm$, while quasi-Keplerian flows $\mu_\Om=0.35$ or quasi-uniform flows ($U_\phi\simeq $~const) with $\mu_\Om=0.5$ together with current-free fields ($\mu_B=0.5$) do not fulfill this condition, resulting in a different scaling for $\Pm\to 0$ as known from Section \ref{AMRI}.
\begin{figure}[htb]
\centering
 \includegraphics[width=8cm]{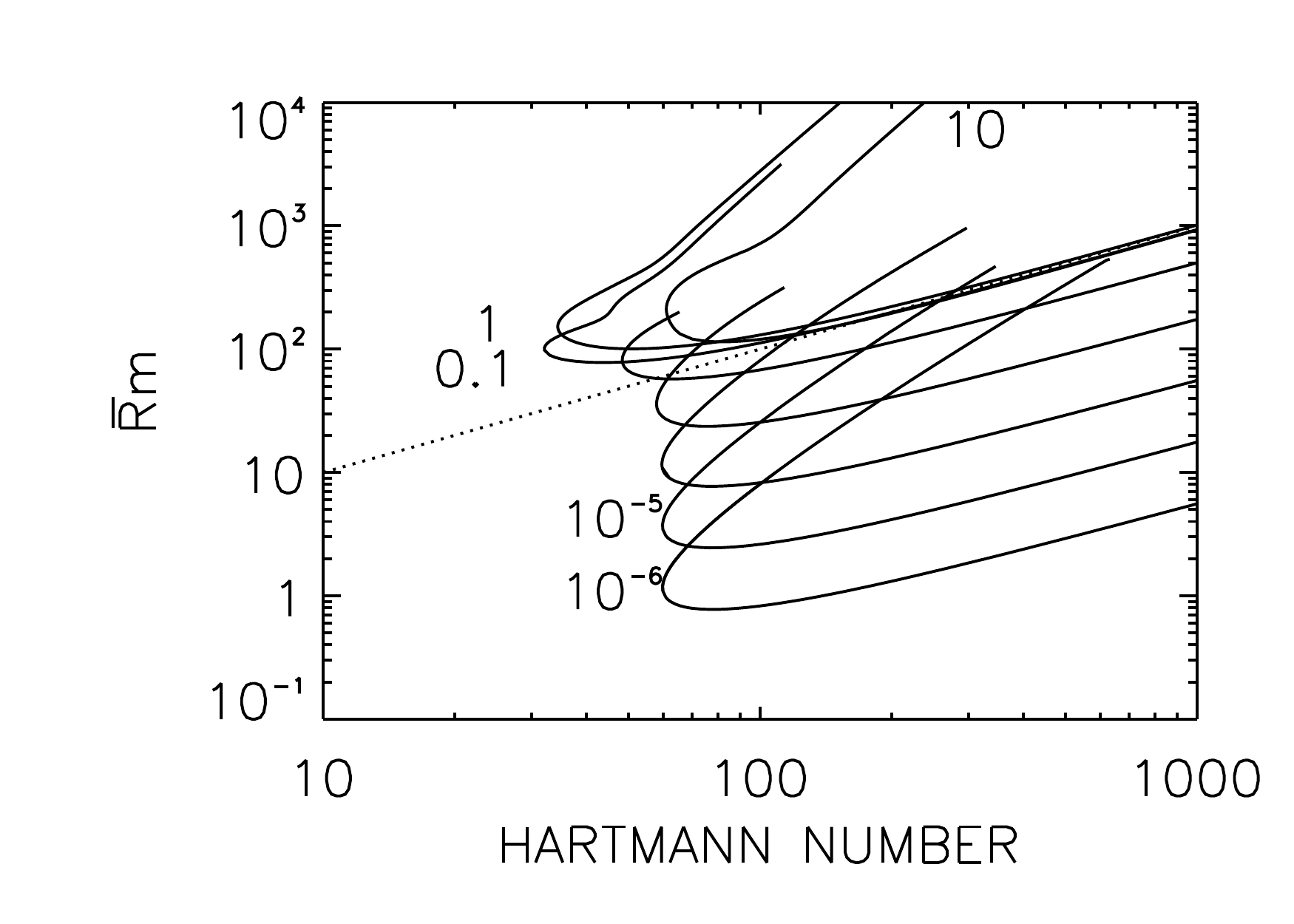}
 \includegraphics[width=8cm]{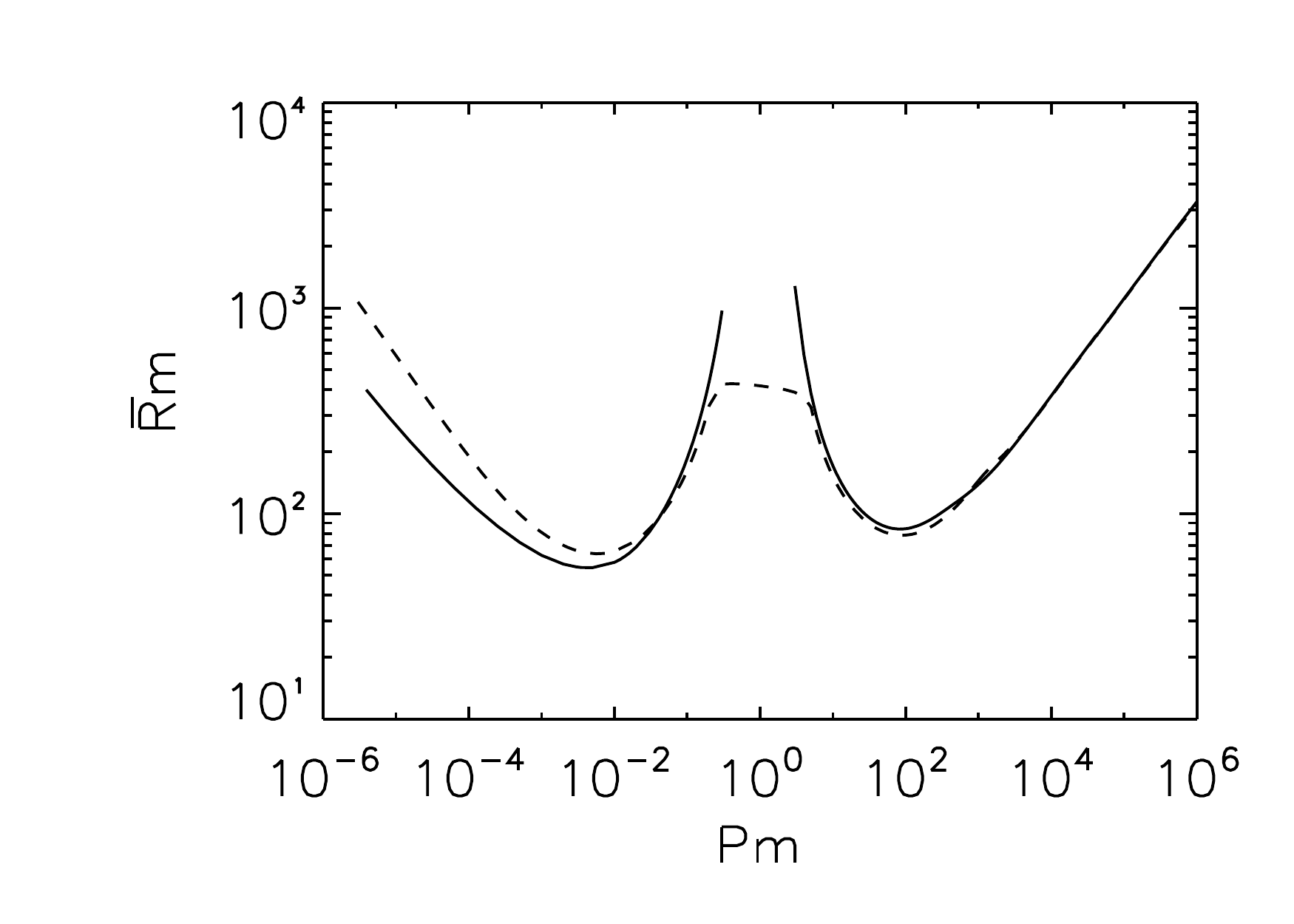}
 \caption{Potential flow of various magnetic Prandtl numbers. Left:  Lines of neutral  stability in the ($\Ha/\Rmquer$) plane.  The dashed line marks $\Mm=1$.             Right:  averaged Reynolds number  $\Rmquer$  of the crossing points  where the parameters for   neutral stability fulfill the Chandrasekhar condition (\ref{chandra2}) with  $\Mm=1$. Solid lines: perfectly conducting boundaries, dashed line: insulating boundaries.  $\mu_B=2\mu_\Om=\rin=0.5$, $m=1$. From \cite{RS15}.}
 \label{g1}
\end{figure}

Consider the dotted line in Fig.~\ref{g1} (left), which represents the location of $\Mm=1$. For different $\Pm$ it crosses the lines of neutral stability at different values of the averaged Reynolds number $\Rmquer$. Following Chandrasekhar such solutions do not exist for ideal media with $\Mm=1$  \cite{C56}. As they only exist for finite values of the diffusivities, the described instability is of {\em diffusive} nature. For small $\Pm$ the numerical values $\Rmquer$ of the crossing points increase for decreasing $\Pm$, which is true for the models with perfectly conducting and insulating cylinders. Both cases lead to very similar results. In the limit $\Pm\to 0$ the Hartmann numbers and $\overline{\Rm}$ of the crossing points scale with $\Pm^{-1/2}$, so that $\Rm$ and $\Lu$ remain finite. One finds for $\Pm\to 0$ values of $\Rm\simeq \Lu\simeq 0.8$ (perfectly conducting cylinders) and $\Rm\simeq\Lu\simeq 2$ (insulating cylinders), for which solutions with $\Mm=1$ exist. The molecular viscosity no longer appears in the theoretical results.

In the limit $\Pm\to \infty$ the opposite is true. Solutions with $\Mm=1$ only exist for finite values of $\Rey=\Ha/\sqrt{\Pm}$. For $\Pm\to \infty$ the averaged Reynolds number $\Rmquer$ grows with $\Pm^{1/2}$, so that the Reynolds number  $\Rey$ remains finite in this limit. The magnetic resistivity completely drops out of the theory. Obviously, the Chandrasekhar theorem of the nonexistence of unstable solutions with $\Mm=1$ fails for potential flows if either of the two molecular diffusivities is non-zero. The suppression of the instability with $\Mm=1$ which appears in the right panel of Fig. \ref{g1} for $\Pm\to 1$  again reflects the original result of Chandrasekhar that this flow is stable for ideal fluids.

In Section \ref{potflow} it was also mentioned that for the potential flow the instability domain for very large $\Pm$ lies above the line $\Mm=1$, while for very small $\Pm$ it lies below this line. In the first case the crossing points belong to the lower branches of the instability cone while in the second case they belong to the upper branches.
Figure \ref{f16} also contains the scaling laws of the lines of neutral instability for the two limits of $\Pm$. For $\Pm\to 0$  the lines converge in the ($\Ha/\Rey$) plane while for $\Pm\to \infty$ they converge in the ($\Ha/\Rm$) plane.  We shall demonstrate  that the $\Rey$-scaling for small $\Pm$ is a general feature of the Chandrasekhar-type flows but the scaling laws for $\Pm\to \infty$ are more diverse. Systems with less but finite (negative) shear will scale with $\Rmquer$  while the system with vanishing shear again changes the scaling law for large $\Pm$ (see Section \ref{pinch}).
\subsection{Quasi-Keplerian flow}\label{keplergalactic}
\begin{figure}[htb]
\centering
 \includegraphics[width=8cm]{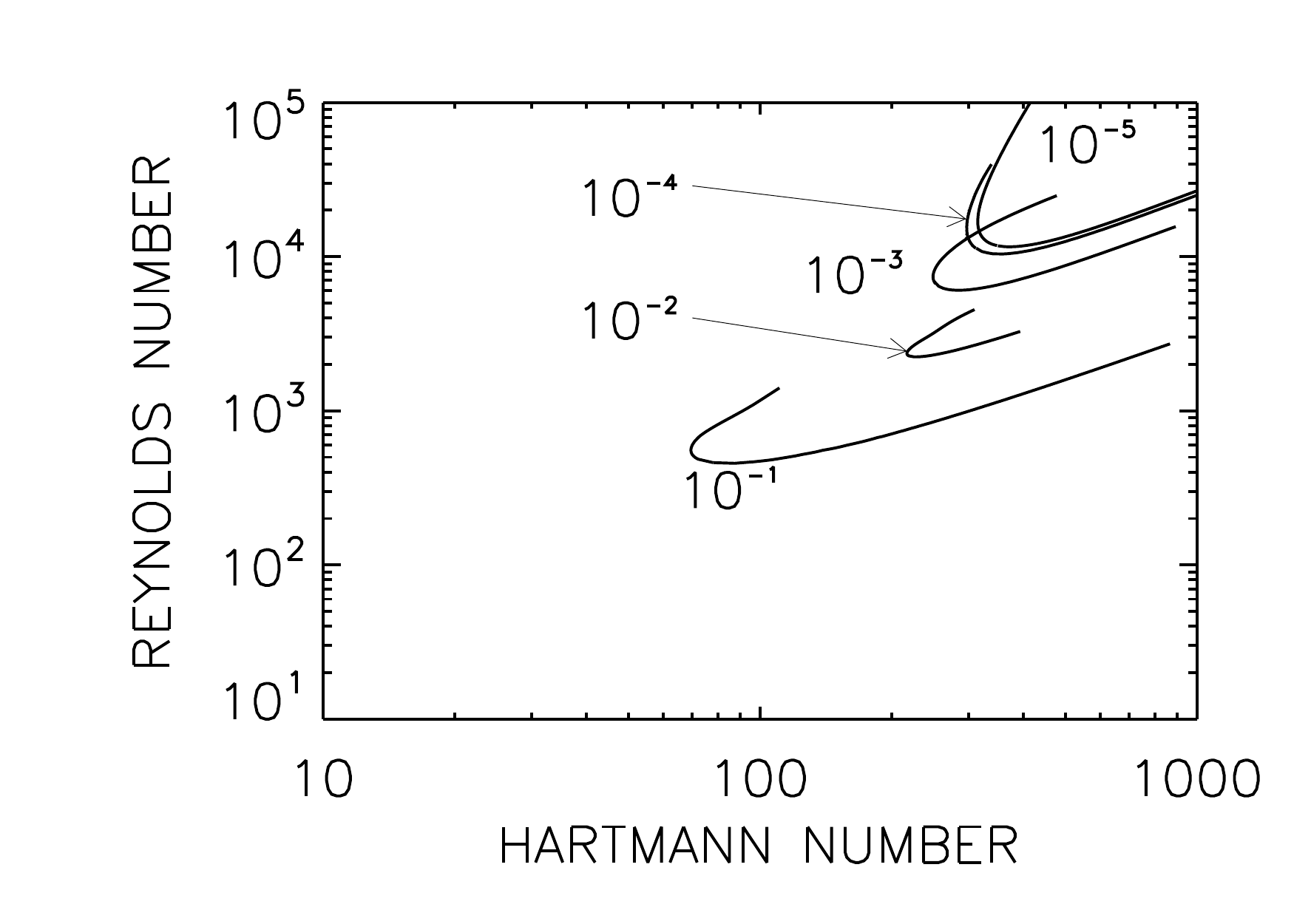}
 \includegraphics[width=8cm]{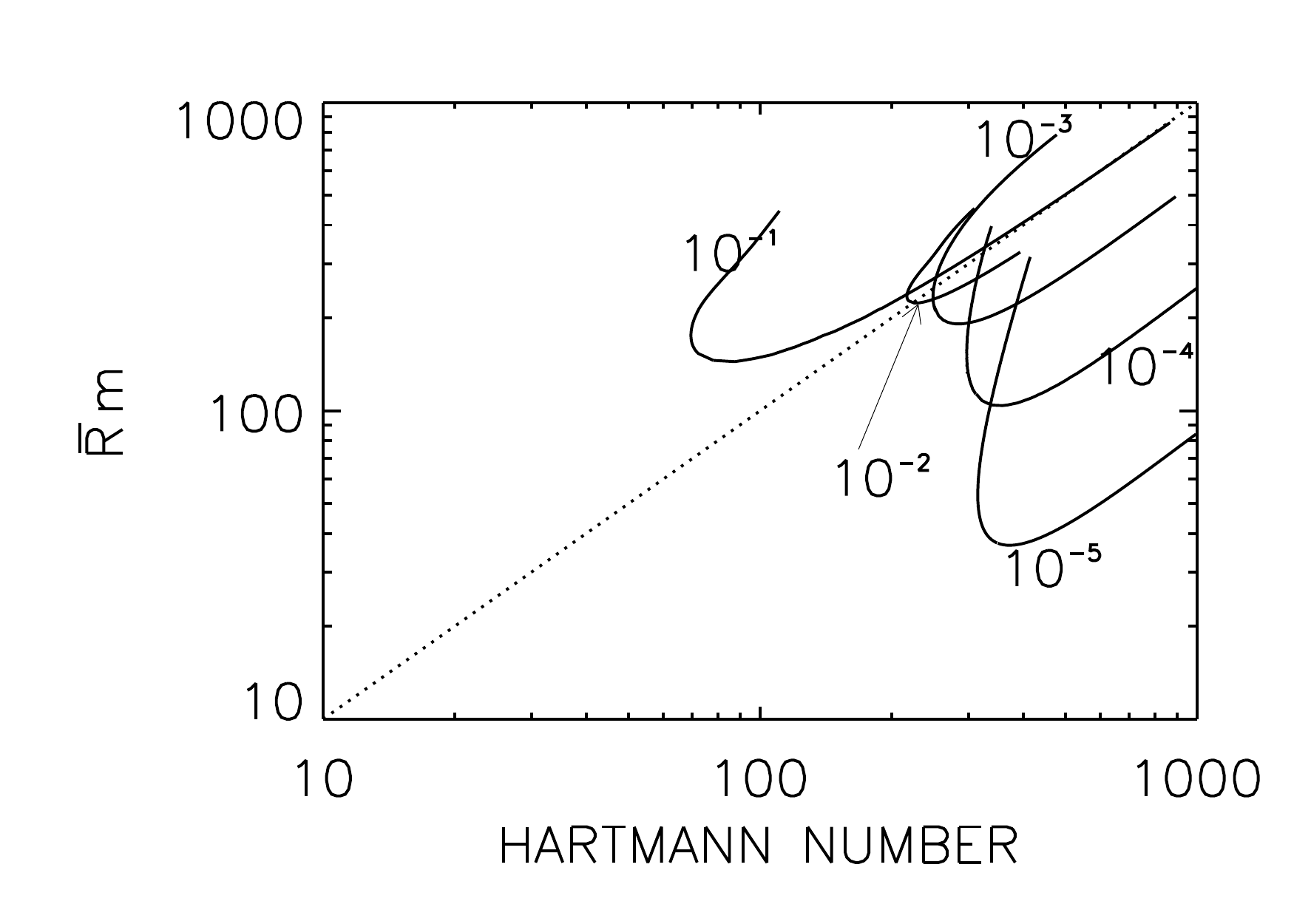}
 \caption{Lines of neutral stability for quasi-Keplerian Chandrasekhar-type flow in two different coordinate systems. In the ($\Ha$/$\Rey$) plane (left) the convergence of the curves for $\Pm\to 0$ is visible. The solutions plotted in the ($\Ha/\Rmquer$) plane show that the curves for $\Pm\to 0$ lie below $\Mm=1$ (dotted line), that is, they are sub-\A{ic} (right). $\rin=0.5$, $\mu_B=2\mu_\Om=0.7$. Insulating boundaries. Adapted from \cite{RS15}.}
 \label{g2}
\end{figure}

For the quasi-Keplerian flow within the Chandrasekhar class, Fig.~\ref{g2} provides quite a similar behavior. The left panel demonstrates that for small $\Pm$ the $m=1$ mode also scales with $\Ha$ and $\Rey$. The minimum critical Hartmann and Reynolds numbers exceed the corresponding values for the potential flow by almost one order of magnitude. The scaling with $\Ha$ and $\Rey$ for small $\Pm$ differs strongly from that of the AMRI combination of quasi-Keplerian rotation ($\mu_\Om=0.35$) with the current-free magnetic field ($\mu_B=0.5$), which is known to scale with $\Rm$ and $\Lu$ (see Section \ref{amrikep}). Here the additional energy source connected with the axial electric current in the fluid determines the scaling rules for small $\Pm$.

\begin{figure}[htb]
\centering
 \includegraphics[width=8cm]{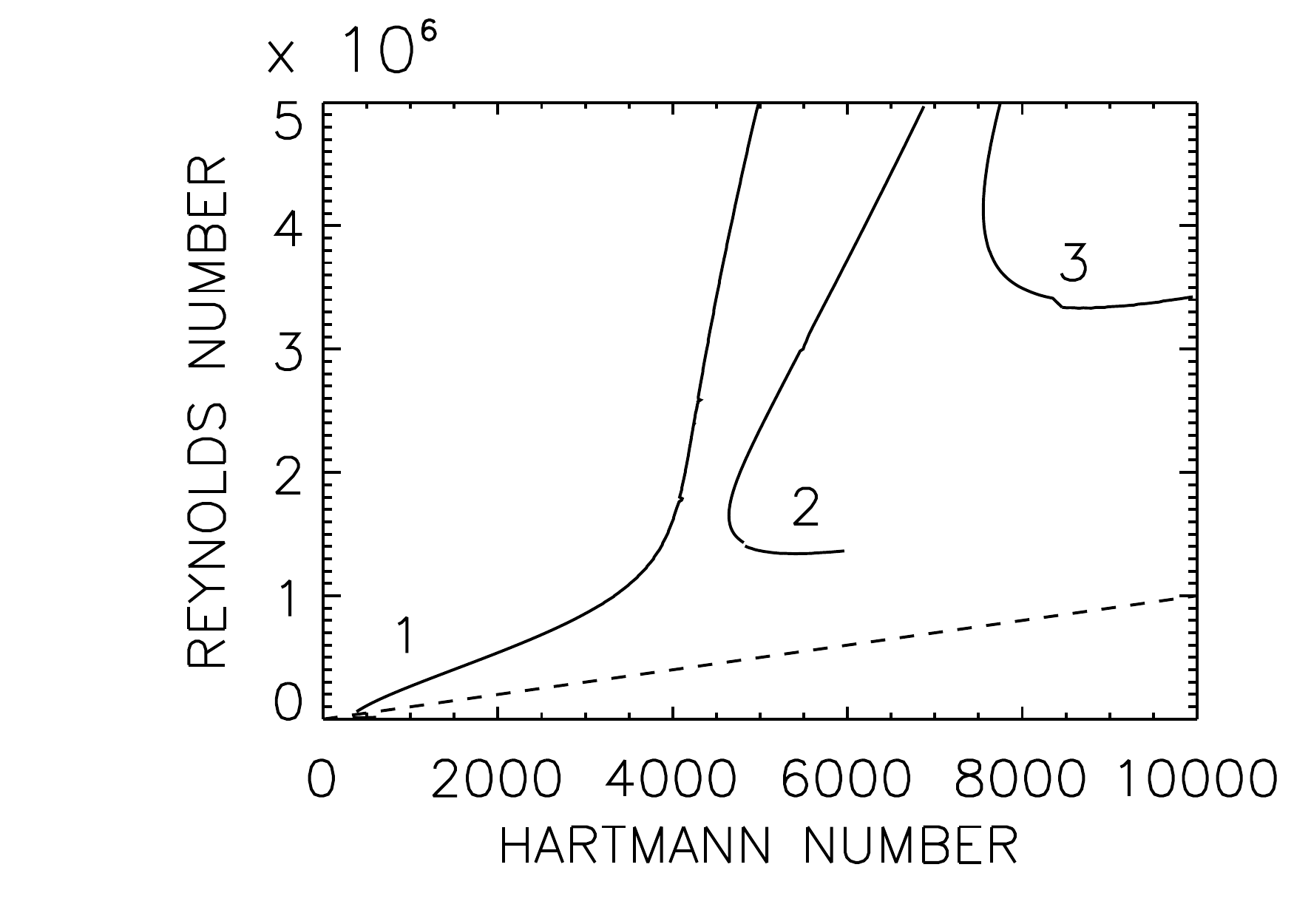}
 \includegraphics[width=8cm]{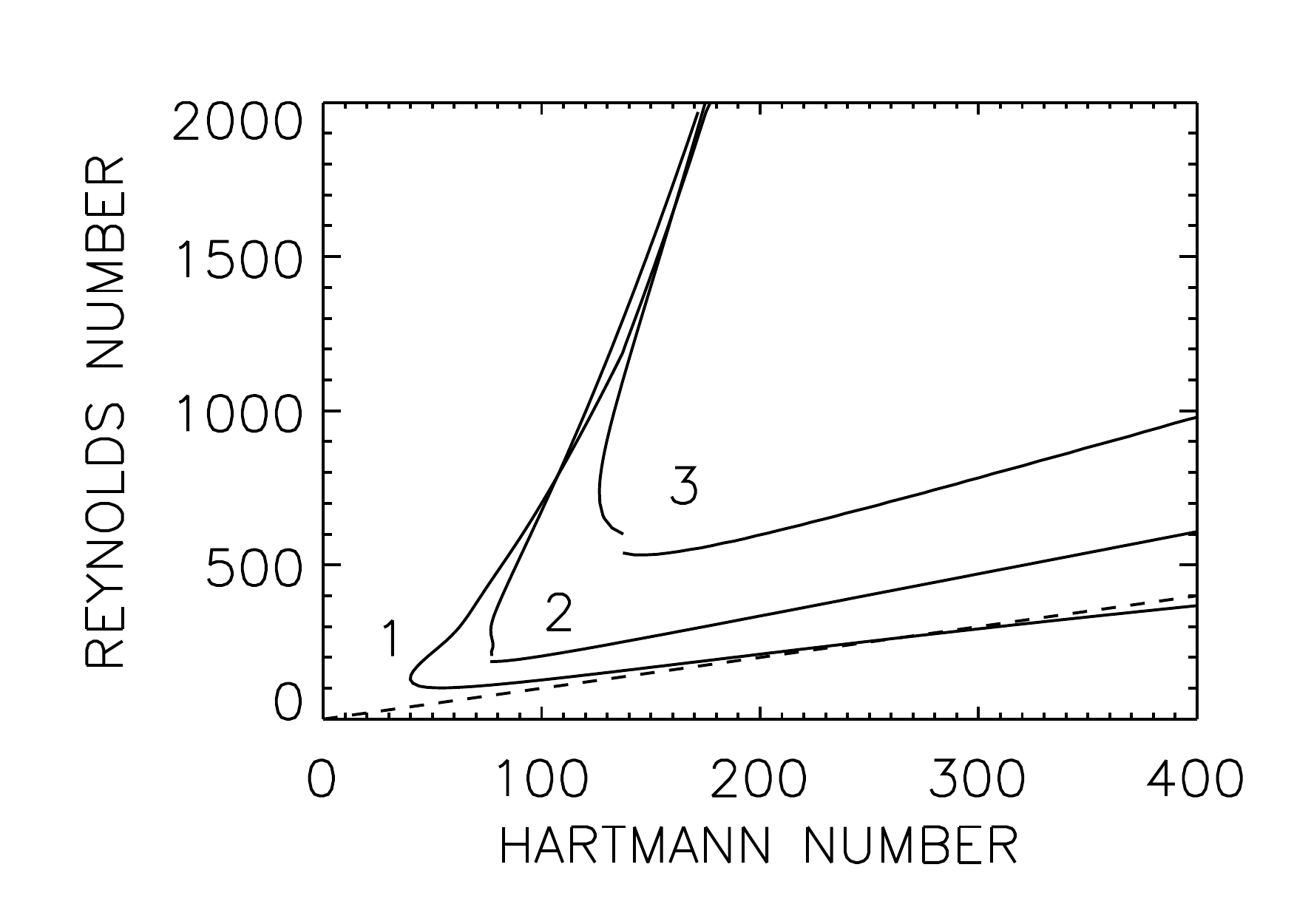}
\caption{Neutral stability curves for quasi-Keplerian Chandrasekhar-type  flow for $\Pm=10^{-4}$  (left)  and $\Pm=1$ (right). The curves are marked with their values of $m$. For small $\Pm$ only the $m=1$ curve lies below the $\Mm=1$ line (dashed). $\rin=0.5$, $\mu_B=2\mu_\Om=0.7$.  Insulating boundaries.}
 \label{g3}
 \end{figure}

\begin{figure}[htb]
\centering
 \includegraphics[width=8cm]{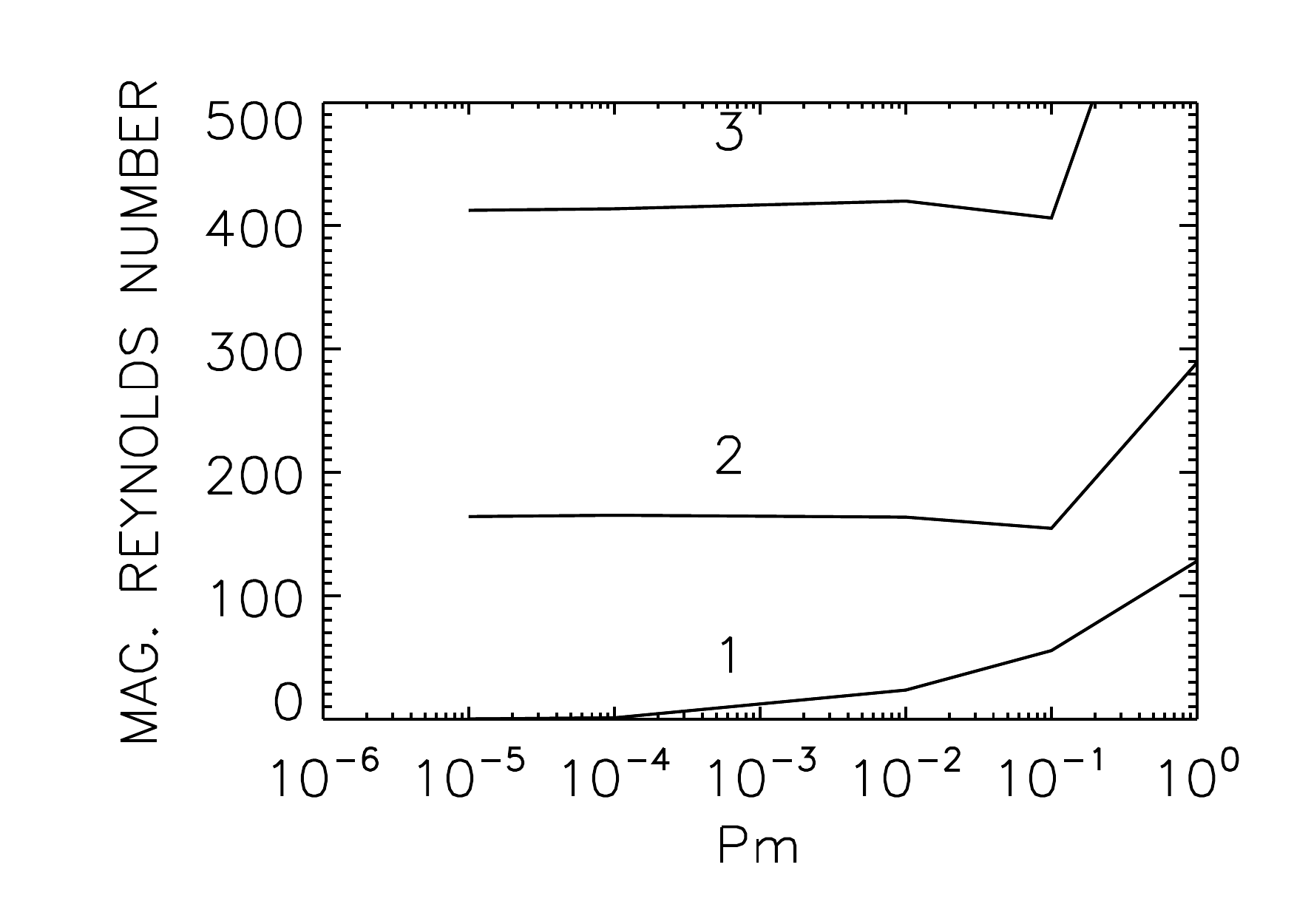}
 \includegraphics[width=8cm]{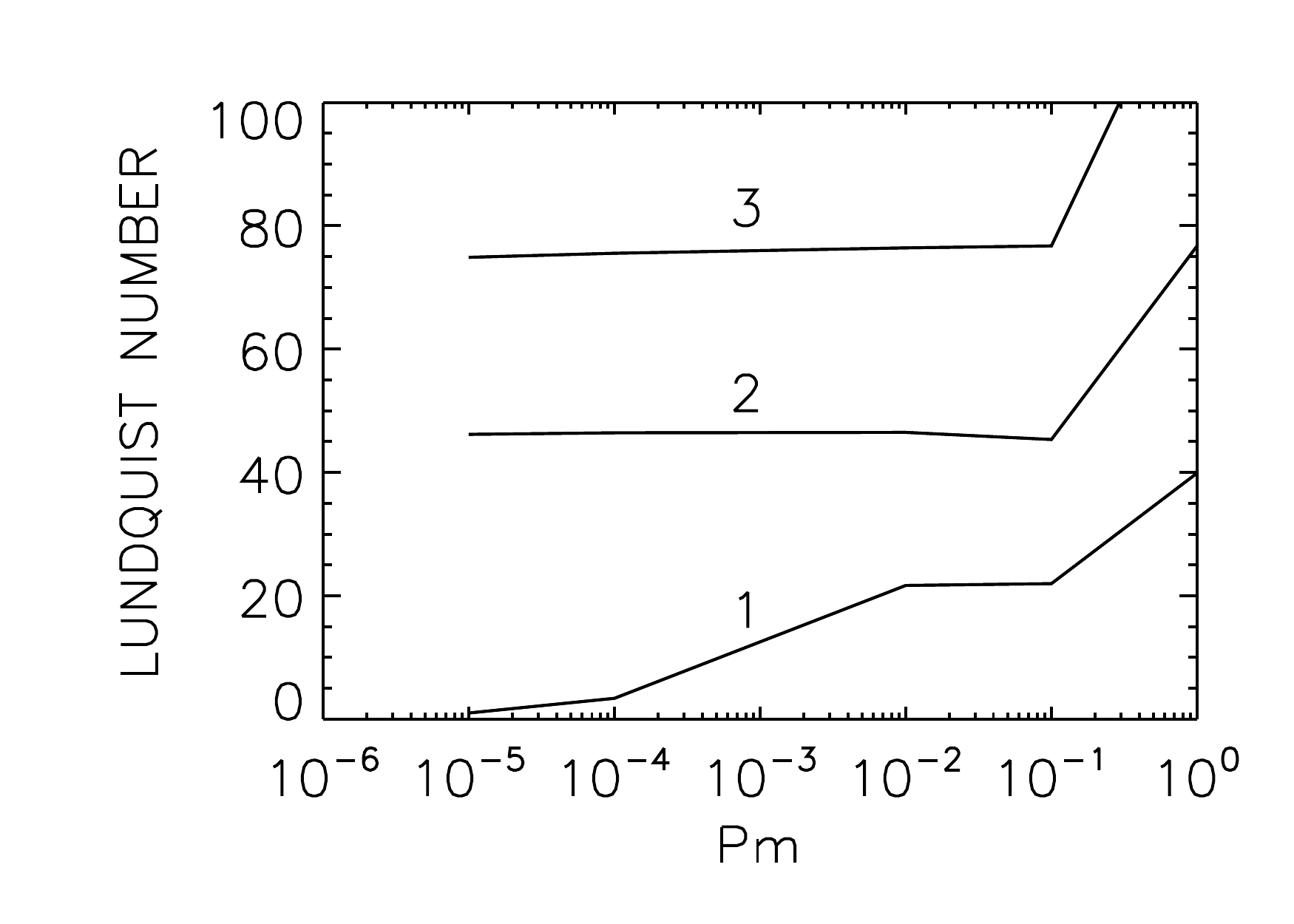}
 \caption{Minimal magnetic Reynolds numbers  (left) and Lundquist numbers  (right) of the neutral stability curves in Fig.~\ref{g3} as functions of $\Pm$. The scaling rules are two-fold:  for  $\Pm\to 0$ the $m>1$ curves scale with $\Rm$ and $\Lu$, unlike for  $m=1$  which scales with $\Rey$ and $\Ha$. The solution with the lowest Reynolds and Hartmann numbers is always $m=1$.  For small $ \Pm$ the instability is super-\A{ic} only for the modes $m>1$.}
 \label{g4}
\end{figure}

A serious consequence of the result is that the instability of the $m= 1$ modes only exists for slow rotation, $\Mm<1$. Including higher azimuthal modes, however, changes the situation. As seen in Figs.~\ref{g3} and \ref{g4}, the critical parameters are $\Rey$ and $\Ha$ only for $m=1$. For $m=2$ and $m=3$, the instabilities scale with $\Rm$ and $\Lu$. As a consequence, these modes should also exist for vanishing viscosity. Figure \ref{g4} shows that for $\Pm\to 0$ the magnetic Mach number $\Mm=\Rm/\Lu$ easily exceeds unity. The new scalings, therefore, generate astrophysical applications of these instabilities, where small $\Pm$ and large $\Mm$ are often associated.

The crossing points of the instability lines with $\Mm=1$ are given as a function of $\Pm$ for quasi-Keplerian flows for both sets of boundary conditions in Ref.~\cite{RS15}. In contrast to the situation for the potential flow there is no clear scaling with $\Rm$ or $\Rmquer$ for small $\Pm$. One finds $\Rm\propto \Pm^{1/3}$. For $\Pm\to 0$ the magnetic Reynolds number does not remain finite. There is thus no solution for $\nu=0$ as exists for the potential flow. 
\subsection{Quasi-uniform flow}
\begin{figure}[htb]
\centering
\includegraphics[width=8cm]{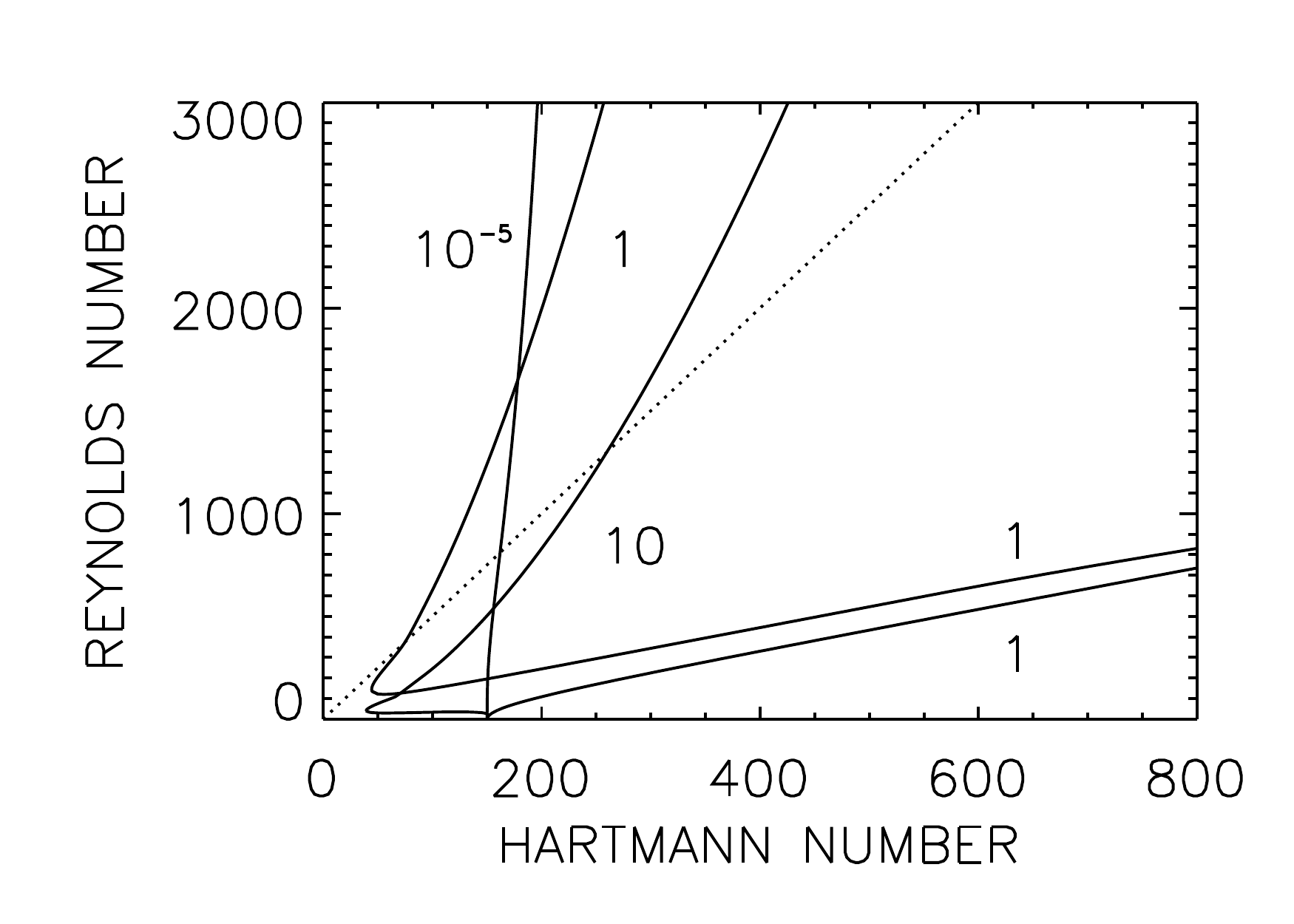}
\includegraphics[width=8cm]{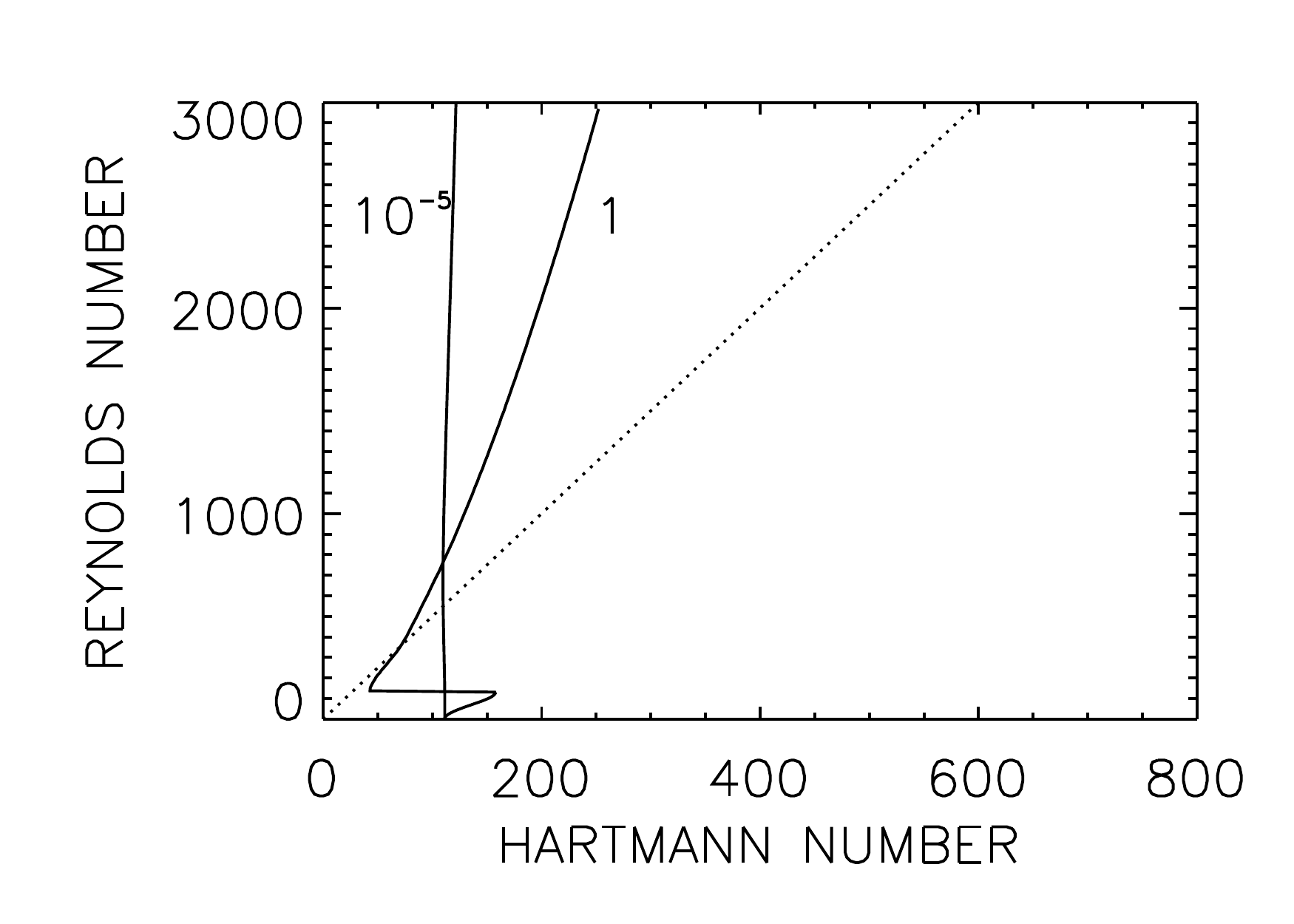}
\caption{Quasi-uniform background flow  for perfectly conducting boundaries (left) and insulating boundaries (right) for various $\Pm$. The lines for $\Pm=10^{-5}$ are valid for all $\Pm\lsim 0.01$.   For $\Pm=1$ the dotted lines  represent $\Mm=5$. $m=1$, $\rin=0.5$, $\mu_B=2\mu_\Om=1$.}
\label{g42}
\end{figure}

Even the simplest model, with approximately uniform flow and field, belongs to the Chandrasekhar class of MHD flows which scale with $\Rey$ and $\Ha$ for $\Pm\to 0$. If $\mu_B=2\mu_\Om=1$, then $U_\phi$ and $B_\phi$ have the same values at both cylinders (for $\rin=0.5$). Background flow $U_\phi$ and background field $B_\phi$ are approximately uniform. The magnetic profile is not current-free between the cylinders. Even without rotation the electric current thus becomes unstable against perturbations with $m>0$ at Hartmann numbers $\Ha_0=109$ for insulating boundaries and $\Ha_0=151$ for perfectly conducting boundaries. These values do not depend on $\Pm$ \cite{RS10}. This Tayler Instability  will be discussed in more detail in Section \ref{TI}. The left panel of Fig.~\ref{g42} also shows an extra instability domain for rapid rotation which has no direct connection to $\Ha_0$. It can thus not be due to the instability of electric current; indeed, the magnetic profile of $\mu_B=1$ also contains the profile $1/R$ which is responsible for AMRI. This AMRI domain (with $\Mm>1$) is easily visible in Fig.~\ref{f26b}, which also shows that for $\Pm\ll1$ the necessary Reynolds numbers for AMRI are too high for Fig.~\ref{g42}. The two instabilities are separated by a stable branch with $\Mm\simeq 1$, where the differential rotation has a stabilizing effect. The extension of the stable branch depends strongly on the boundary conditions. It is very long -- possibly infinitely long -- for perfectly conducting boundaries, but rather short for insulating ones. Even for perfectly conducting cylinders the stable branch disappears for small $\Pm\neq 1$.  Obviously, the narrow stable branches in Fig.~\ref{g42} for $\Pm=1$ reflect the {\em stability} of all ideal MHD flows   fulfilling the Chandrasekhar condition (\ref{chancon}). This the more as for $\Pm=1$  the relation $\Mm=1$ transforms the relation (\ref{chandra2}) to (\ref{chandra1}). The diffusive influences  allow  stability only  in a  rather  narrow strip  close to  the line $\Mm=1$.

Very slow rotation stabilizes the system slightly, but for faster rotation ($\Mm\gsim 1$) and $\Pm\gsim 1$ the instability becomes subcritical, i.e.~it onsets for smaller Hartmann numbers than it does without rotation ($\Ha<\Ha_0$). The phenomenon of subcritical excitation for large $\Pm$ is very characteristic for Chandrasekhar-type flows. It only appears for slow rotation and $\Pm\lsim 1$. The resulting stable branch around the line $\Mm=1$ is also characteristic for this sort of stability map. It separates the region of the TI (for slow rotation) from the region of the AMRI (due to differential rotation). This separation effect does not exist for rigid rotation. As expected for $\Mm\gg 1$ the strong differential rotation suppresses the nonaxisymmetric instability pattern, but again the effect is small for small $\Pm$.

\begin{figure}[htb]
\centering
\includegraphics[width=8cm]{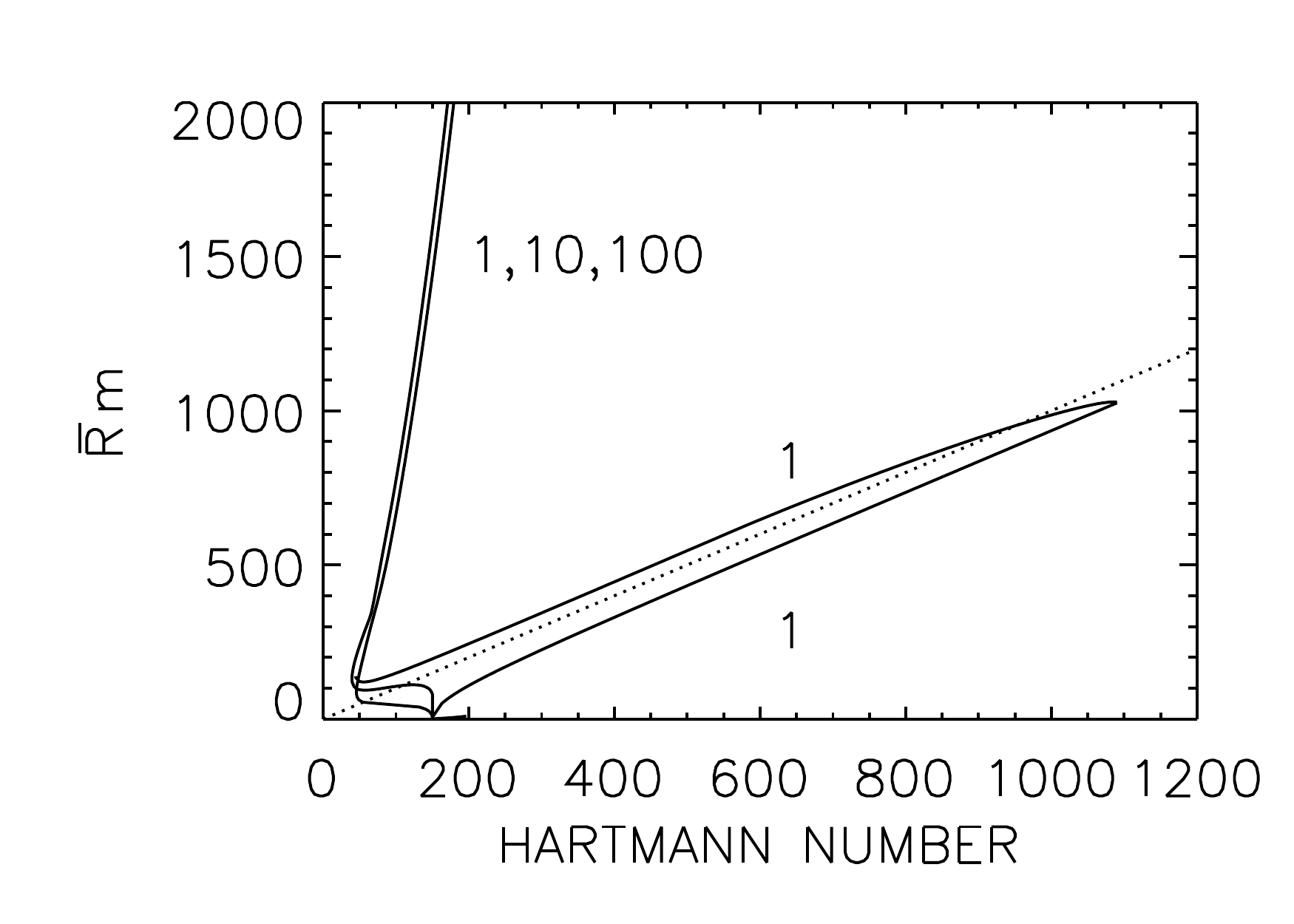}
 \includegraphics[width=8cm]{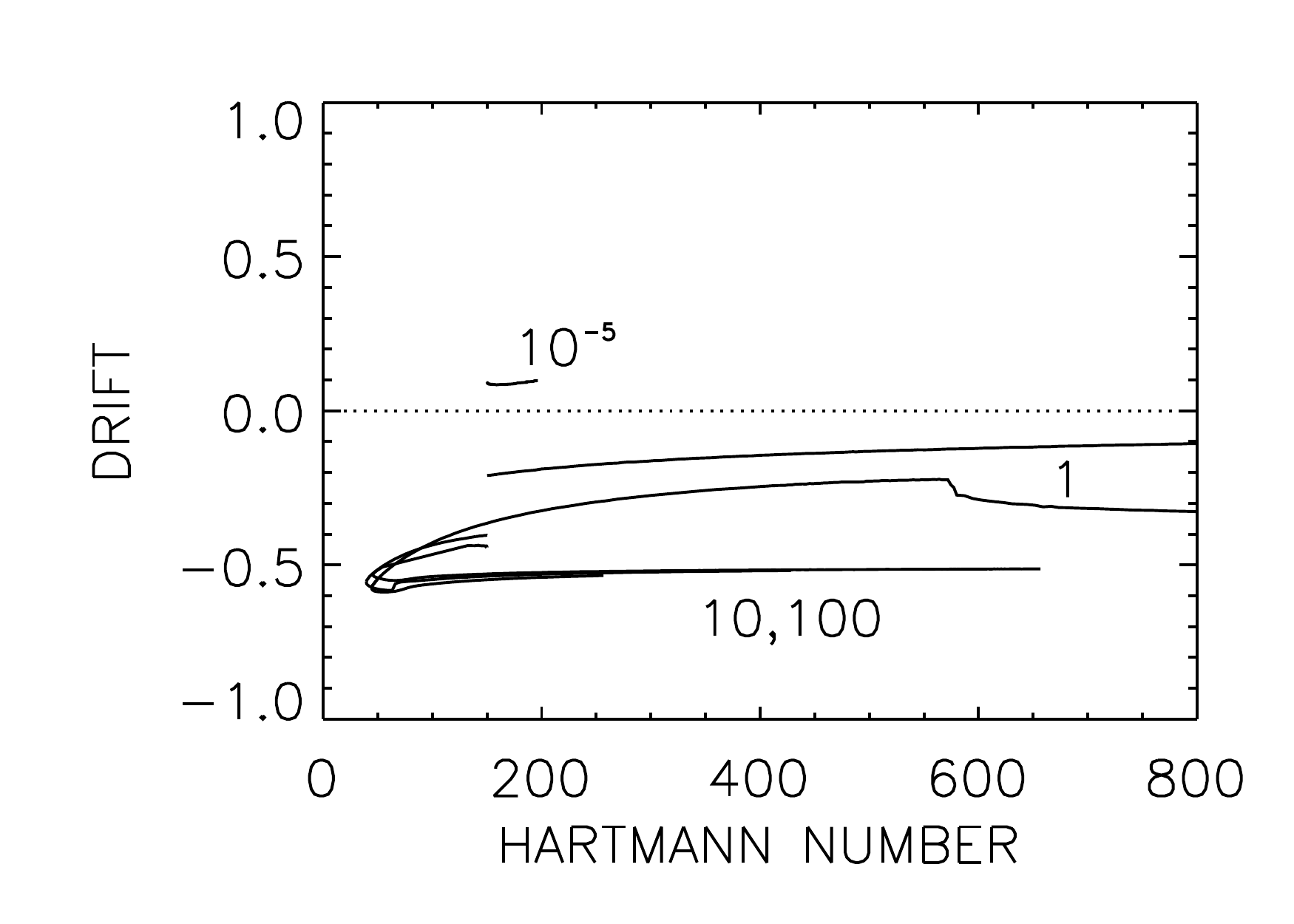}
\caption{Neutral stability curves (left) and drift rates  $\omdr$ (right)  for quasi-uniform field and for large $\Pm$ (marked). For $\Pm>1$ the curves converge in the ($\Ha/\Rmquer$) plane. For $\Pm=1$  the dotted line represents $\Mm=1$. The sign of $\omdr$ differs for small and large $\Pm$, changing for 
$\Pm\simeq 0.1$.    $m=1$, $\rin=0.5$,  $\mu_B=2\mu_\Om=1$.  Perfectly conducting boundaries.}
 \label{g421}
 \end{figure}

It remains to clarify the asymptotic behavior of the stability lines of the $m=1$ mode for large $\Pm$. We shall find a substantial discrepancy between the instability domains for small and large magnetic Prandtl numbers. While for small $\Pm$ the curves converge in the ($\Ha/\Rey$) plane, for large $\Pm$ they converge in the ($\Ha/\Rmquer$) plane (Fig.~\ref{g421}, left). Since $\Mm=\Rmquer/\Ha$, it is obvious that for large $\Pm$ the instability also exists for large magnetic Mach numbers. Rapid rotation does not suppress the instability in this case. For large $\Pm$ combinations of Reynolds and Hartmann numbers with $\Mm>1$ also become unstable, which is not the case for very small $\Pm$. Another consequence is that for a fixed Hartmann number the critical Reynolds numbers behave like $\Rey\propto \Pm^{-1/2}$ for $\Pm\to \infty$, so that the magnetic Reynolds number increases as $\Rm\propto \Pm^{1/2}$ for large $\Pm$. The drift rates also depend on the magnetic Prandtl numbers. Figure \ref{g421} (left) shows these to be negative for $\Pm\geq 1$ and positive for $\Pm\ll 1$.

An exception from this rule, however,  is given by the potential flow with $\mu_B=2\mu_\Om=0.5$ which in Section \ref{potflow} has been discussed as a prominent  application of AMRI. The result  was that the stability lines of the potential  flow converge for $\Pm\to \infty$ in the ($\Ha/\Rm$) plane (Fig.~\ref{f16}) so that the stability curve scales for large $\Pm$ with the magnetic Reynolds number $\Rm$ rather than with the average Reynolds number $\Rmquer$.

The profile $\Om\propto 1/R$ characterizes the rotation of galaxies in their outer parts. If it is further assumed that their azimuthal fields are approximately uniform in this region, then Chandrasekhar states with $\mu_B=2\mu_\Om=1$ may well apply to galaxies. The axial component of the magnetic field is maximally 10\% of the azimuthal field. Also typical for galaxies is the relation $\Mm\simeq 5$, as given in Fig.~\ref{g42} (right) by a dashed line. This line is located almost everywhere to the right of the instability lines for $\Pm\lsim 1$, so that galactic fields together with the rotation according to $U_\phi\simeq$const should develop nonaxisymmetric magnetic perturbations\footnote{Estimates for galaxies are $\Rey\simeq 1000$, $\Ha\simeq 200$ and $\Pm\simeq 1$, the latter due to the interstellar turbulence.}.

Interesting is also the stable branch in Fig. \ref{g421} which for $\Pm=1$ separates the unstable areas close to the line $\Mm=1$. Below this stable  branch one may consider the unstable solutions as due to TI under the influence of differential rotation while above they represent  AMRI solutions under the influence of weak electric currents. For too high Hartmann numbers these currents become too strong so that the stable branch disappears. We know that the combination of quasi-uniform field and quasi-uniform flow (i.e. $\mu_B=2\mu_\Om=1$)  is stable for ideal fluids. The stable `finger' in Fig. \ref{g421} (left) which only appears for $\Pm=1$ can be understood as  a consequence of the stability  theorem for ideal flows as indeed $\Pm=1$  best fits 
magnetohydrodynamics  of ideal media \cite{Sh07}.

We have also computed (not shown)  the stability maps for $\mu_B=2\mu_\Om=1.5$ between the models  with quasi-uniform field ($\mu_B=1$) and the rigidly rotating $z$-pinch ($\mu_B=2$). 
The critical Hartmann number  without rotation is $\Ha_0=57$ for all $\Pm$ and for perfectly conducting boundaries. Again the   curves converge  for small $\Pm$ in the ($\Ha/\Rey$) plane  and they converge  for large  $\Pm$ in the ($\Ha/\Rmquer$) plane  -- as is also true for $\mu_B=2\mu_\Om=1$ (Figs.~\ref{g42} and  \ref{g421}). It is thus  finally clear  that between  the rotation laws $\Om\propto 1/R^2$ and $\Om=$~const the Chandrasekhar-type flows (\ref{chandra2}) indeed scale in the described sense for small and large $\Pm$.
\subsection{Rigidly-rotating $z$-pinch}\label{pinch}
Even rigid-body rotation with $\mu_\Om=1$ can be a prominent example of the Chandrasekhar theorem, provided that the associated magnetic profile also satisfies the condition (\ref{chancon}). This implies a uniform current throughout the entire region $R<\Rout$, known as a $z$-pinch configuration in plasma physics. Any resulting instability is purely current-driven. Such instabilities can occur for $\Rey=0$ but not for $\Ha=0$. A nonrotating pinch is only unstable against nonaxisymmetric perturbations with $m=1$ \cite{T57}. Acheson showed that the {\em necessary} condition for magnetic instability with $m>1$ is not fulfilled for this flow \cite{A78}. This finding remains true for rigid rotation: we found no unstable modes with $m>1$. For rigidly rotating Taylor-Couette flows in a wide gap with $B_\phi\propto R$ global calculations provided stability in the inviscid approximation \cite{F83}. 

\begin{figure}[htb]
\centering
\includegraphics[width=8cm]{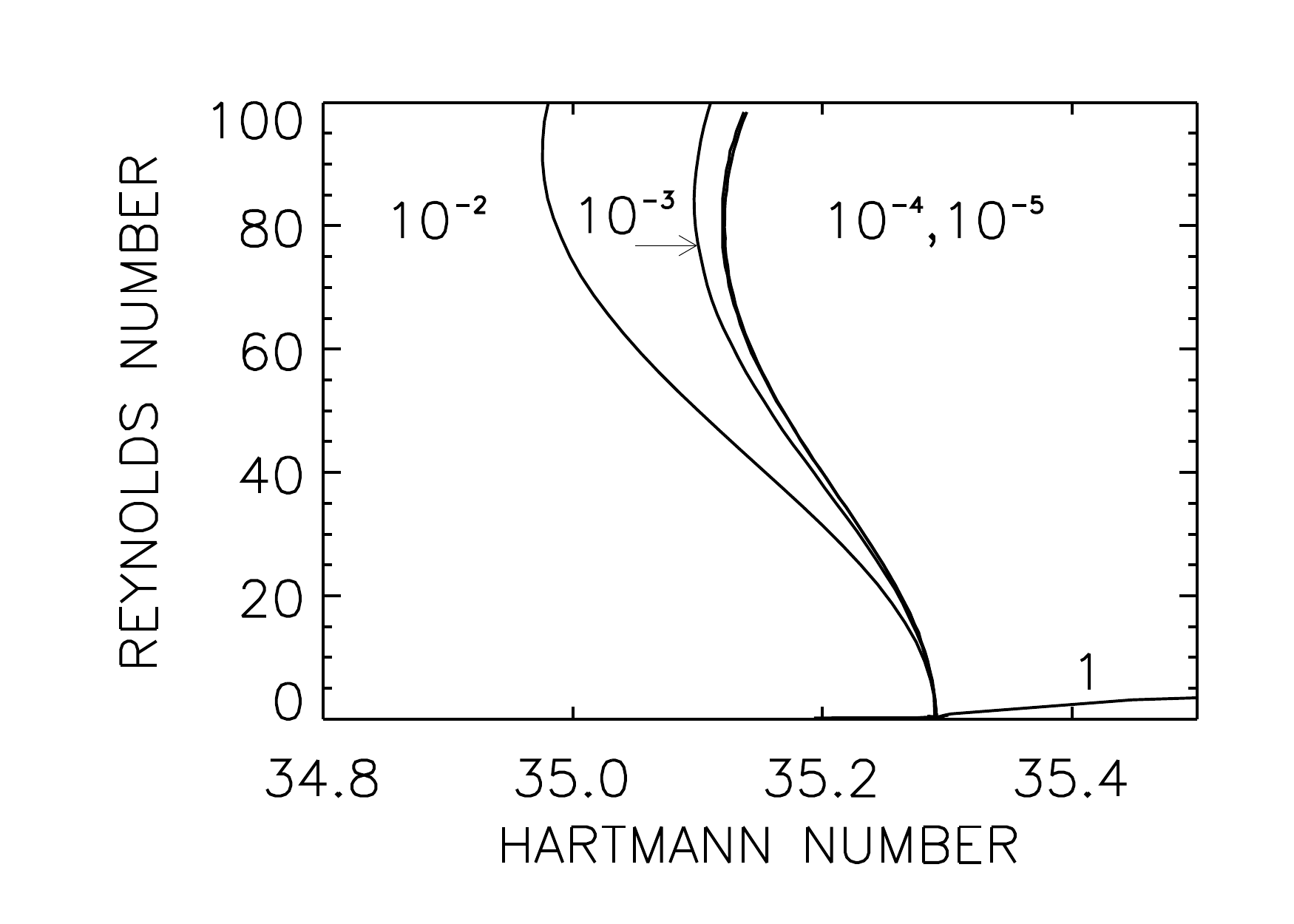}
\includegraphics[width=8cm]{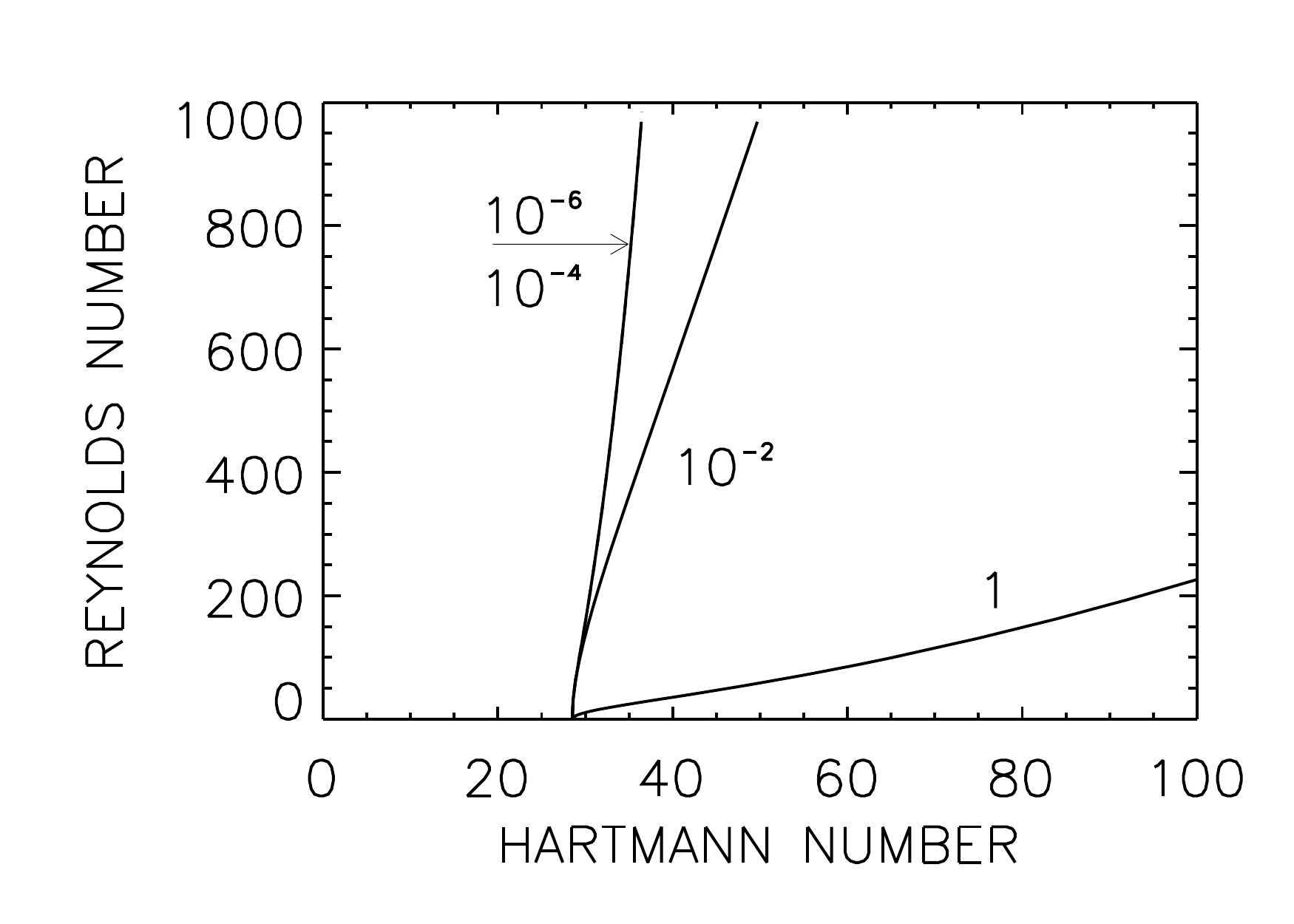}
\caption{Stability maps for the only unstable mode $m=1$ of the rigidly rotating $z$-pinch for perfectly conducting (left) and insulating (right) boundaries. The curves are marked with $\Pm$; they always converge for $\Pm\to 0$. The two (different) values of $\Ha_0$ for $\Rey=0$ do not depend on $\Pm$. $\rin=0.5$, $\mu_B=2\mu_\Om=1/\rin$.}
\label{g5}
\end{figure}

The stability curves for $m=1$ are shown in Fig.~\ref{g5} for conducting and insulating boundary conditions. The curves basically differ from the former examples as the characteristic minima no longer exist. For both boundary conditions  the stabilizing effect of rigid rotation on the Tayler instability is clearly  demonstrated for $\Pm=1$ \cite{PT85}. In this representation the rotational suppression becomes weaker for smaller (and larger, not shown) magnetic Prandtl numbers. In the ($\Ha$/$\Rey$) plane the curves converge for $\Pm\to 0$, hence the eigenvalues also scale with $\Rey$ and $\Ha$. We find that for all models along the Chandrasekhar sequence in the ($\Ha$/$\Rey$) plane the lines of marginal stability for $m=1$ do not depend on $\Pm$ for sufficiently small $\Pm$. The magnetic Mach number 
\beg
\Mm=\sqrt{\Pm}\ \frac{\Rey}{\Ha}
\label{Mm}
\ende
of the solutions for small $\Pm$ remains smaller than unity. A rotating pinch with small $\Pm$ and $\Mm>1$ is always stable.

Figure \ref{g5} also illustrates the influence of the boundary conditions. Perfectly conducting cylinders yield $\Ha_0=35.3$, whereas insulating cylinders yield $\Ha_0=28.5$. For the conducting boundary conditions a subcritical excitation for slow rotation is clearly visible, $\Ha<\Ha_0$, but only if $\Pm< 1$. The solutions for $\Pm=1$  show the  rotational suppression for all Reynolds numbers, while for $\Pm<1$ the suppression only exists for sufficiently rapid rotation.
Without rotation $\omega_{\rm dr}=0$ always holds, and the pattern is stationary in the laboratory system. For the rotating pinch the instabilities drift in the rotation direction for $\Pm\geq 1$, but in the opposite direction for $\Pm<1$. For $\Pm\to \infty$ the lines in the right panel of Fig.~\ref{g5} converge slightly below the line for $\Pm=1$ (see Fig.~\ref{f12} below). We have thus the exceptional situation that both the limits for very small and very large $\Pm$ appear  in one and the same coordinate system. The consequences of this phenomenon are described in Section \ref{diffrot}.
\begin{figure}[h]
\centering
 \includegraphics[width=9cm]{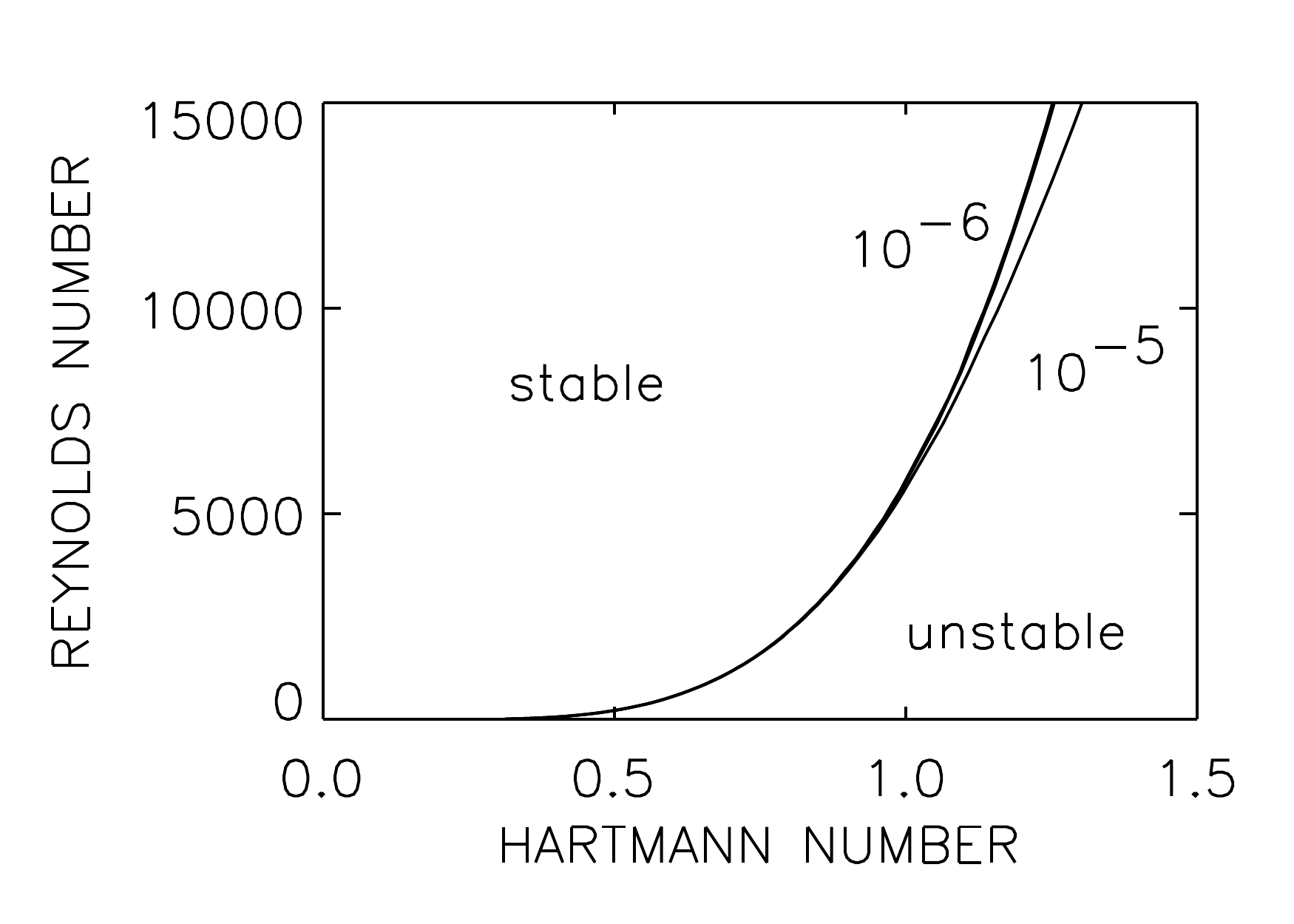}
 \caption{A rigidly rotating $z$-pinch in a wide gap for small $\Pm$ (marked).  The curves in the ($\Ha/\Rey$) plane coincide for small $\Pm$. The Hartmann number (\ref{Hartmannin}) is  formed with the inner magnetic field (see text). $m=1$, $\mu_B=1/\rin$, $\mu_\Om=1$,  $\rin=0.05$. Insulating boundary conditions.}
 \label{f11}
 \end{figure}

One may ask whether the rotational stabilization can also be probed in the laboratory. A rigidly rotating wide pinch with $\rin=0.05$ and $\mu_B=1/\rin$ with insulating cylinders is thus considered for the small magnetic Prandtl numbers of liquid metals. The Chandrasekhar condition (\ref{chancon}) is fulfilled with rigid rotation ($\mu_\Om=1$). Without rotation the inner critical Hartmann number according to (\ref{Hartmannin}) is $\Ha_0= 0.31$ for this container\footnote{With outer values and the   definition (\ref{Hartout}) it is $\Ha_{\rm out}= 28.4$.}, independent of $\Pm$. We find the rotational stabilization is rather weak for not too fast rotation (Fig.~\ref{f11}). The figure also perfectly shows the scaling of the eigenvalues in the ($\Ha/\Rey$) plane which is typical for the Chandrasekhar-type MHD flows. For a Reynolds number  $\Rey\simeq 10^3$ the supercritical magnetic field needed for instability is (only) two times larger than $\Ha_0$. It should thus easily be possible to find the basic effect of the rotational suppression of the pinch-type instability in the laboratory. The constellation analyzed by Fig.~\ref{f11} forms an ideal experimental setup for studies of the instability characteristics of a Chandrasekhar-type MHD flow.
\begin{figure}[htb]
\centering
 \includegraphics[width=7.8cm]{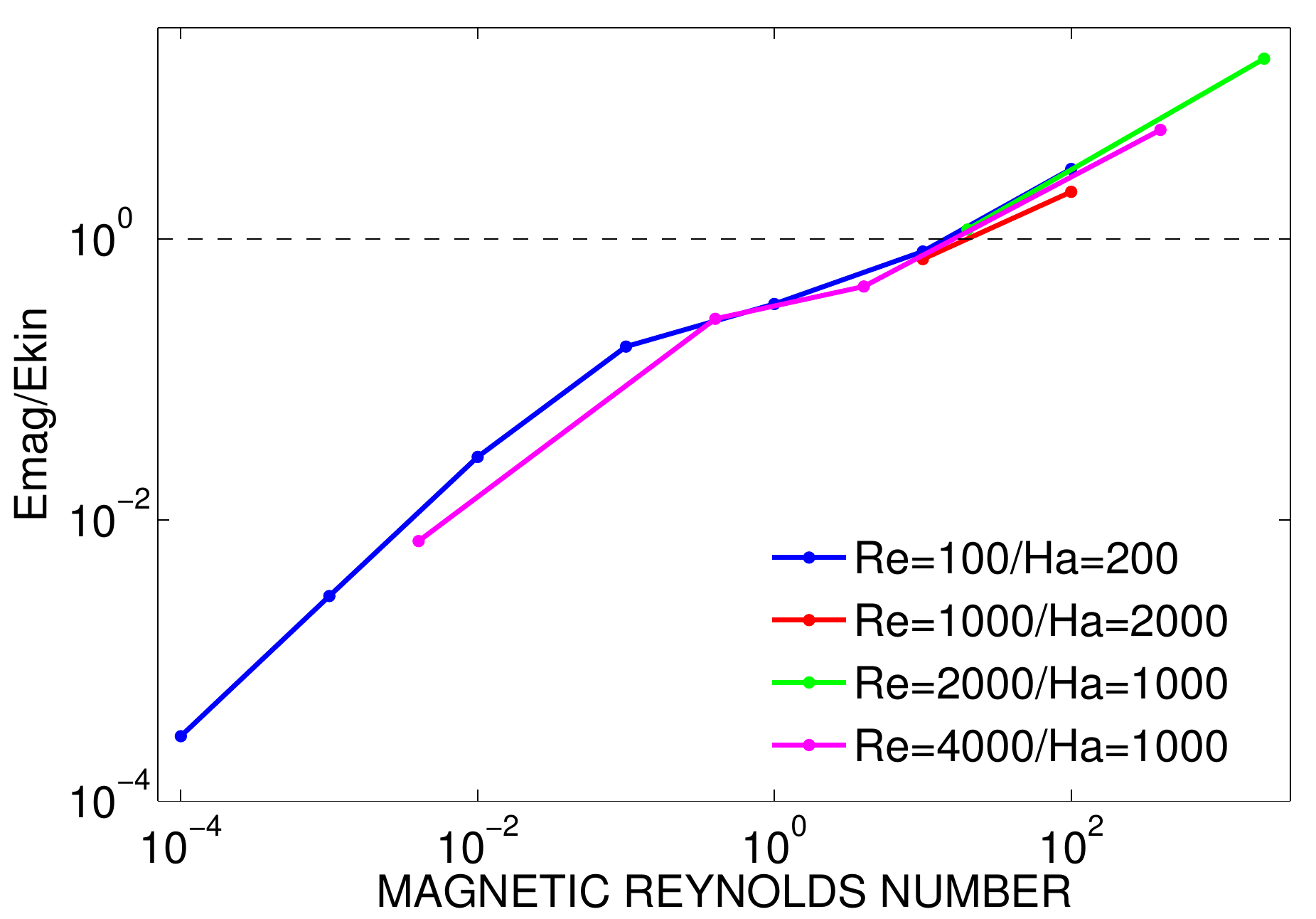}
  \includegraphics[width=8cm]{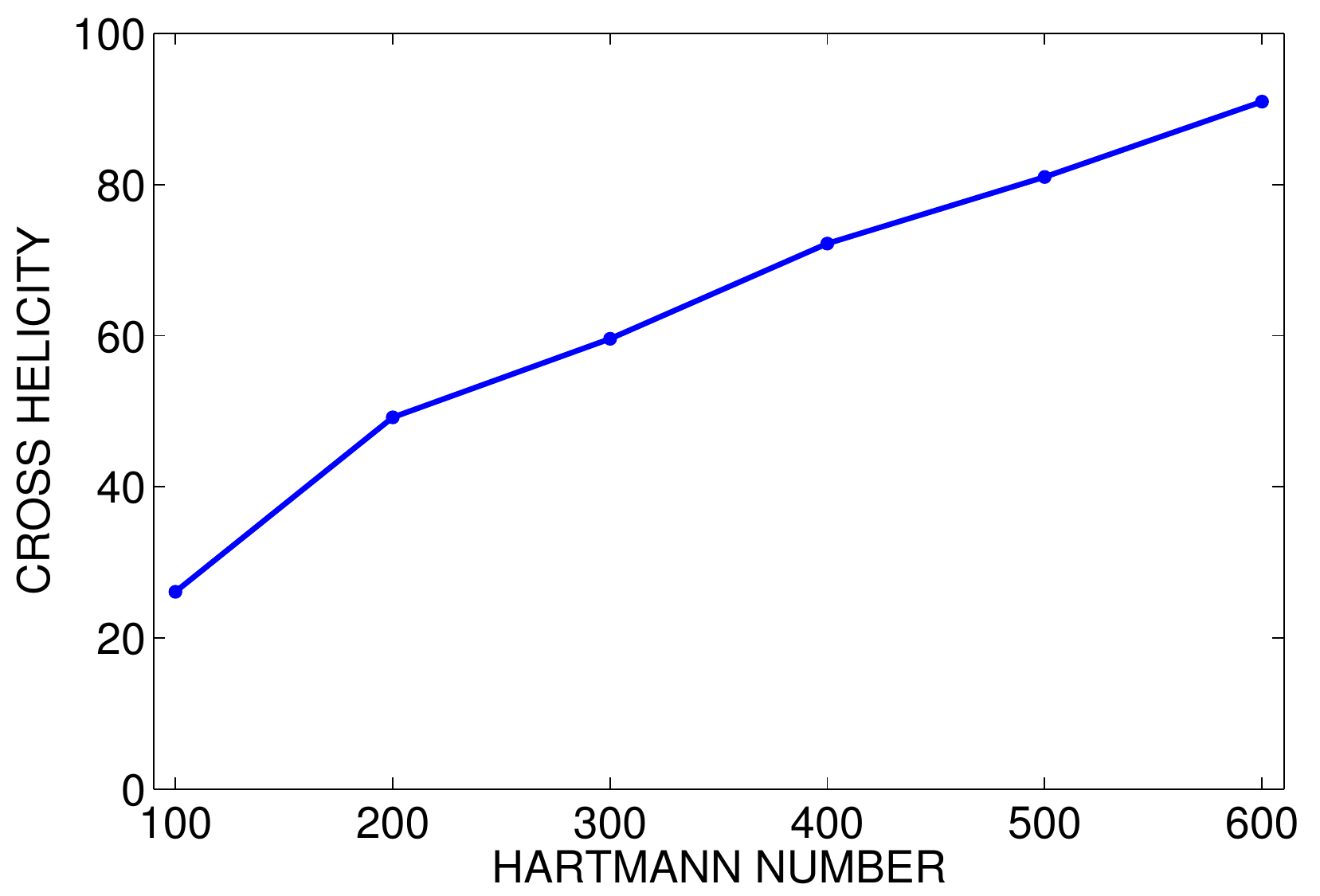}
 \caption{Rigidly rotating $z$-pinch. Left: energy ratio (\ref{ratio}) for many models   as function of the magnetic Reynolds number.  The models show equipartition of the two energies (see the dotted line) only for $\Rm\simeq 20$. Dominating magnetic energy requires higher magnetic Reynolds numbers \cite{GR16}. 
 Right: cross-helicity  measured in units of $\nu B_{\rm in}/R_0$. $\Rey=200$, $\Pm=0.1--1$.
  $\rin=0.5$, $\mu_B=2\mu_\Om=2$. Insulating boundary conditions.}
 \label{g6}
\end{figure}

\subsection{Energies and cross-helicity}
For the energy ratio $\varepsilon$  of magnetic to kinetic energy defined by Eq.~(\ref{ratio}) one finds similar properties as for the potential flow. For the latter it is known that the ratio of the energies is small for small $\Rm$ (Fig.~\ref{f25d}). The same is true for the rigidly rotating $z$-pinch. Figure \ref{g6} demonstrates the result of the  numerical simulations, that $\varepsilon$ of the pinch scales with $\Rm$. Almost independent of $\Pm$, $\varepsilon$ exceeds unity only for $\Rm\gsim 20$, or in other words, if the numerical product of $\Rey$ and $\Pm$ exceeds about 20. The same result also holds for the Chandrasekhar-type flow with quasi-Keplerian rotation \cite{GR16}.

For the rotating pinch the pseudo-scalar $\vec{\Om}\cdot\vec{J}$ should exist, linear in the magnetic field. The question is whether the {\em cross-helicity} $\langle \vec{u}\cdot \vec{b}\rangle $ becomes non-zero in the fluid. For the stationary pinch the cross-helicity must vanish. Indeed, the numerical simulations for the rotating pinch by means of the nonlinear code described in Section \ref{nonsim} provide the surprisingly simple result that for weak fields
\beg
\langle \vec{u}\cdot \vec{b}\rangle = h_{\rm cross} \Ha\ U_{\rm in} B_{\rm in},
\label{cross}
\ende
(with $U_{\rm in}= R_0 \Om_{\rm in}$). This result has been tested for several combinations of low values of $\Rey$ and $\Ha$. One finds from the linear part of the curve in Fig.~\ref{g6} (right) that $h_{\rm cross}\simeq 1.3 \cdot 10^{-3}$, almost independent of the magnetic Prandtl number. The parallel components of the flow and field fluctuations are correlated due to the Coriolis force. Only the global rotation generates such a correlation averaged over the entire container. The expression (\ref{cross}) is symmetric in the dissipation coefficients $\nu$ and $\eta$ via the Hartmann number. Being linear in $\Ha$, the relation  is only valid for not too large $\Ha$. For stronger fields the numerical coefficient $h_{\rm cross}$ is magnetically suppressed. The robustness of the result (\ref{cross}) also shows that the cross-helicity is not a consequence of the initial conditions, which could potentially have been the case as cross-helicity is conserved in ideal fluids \cite{B03}.
\subsection{Azimuthal spectra}
\begin{figure}[h]
\centering
\includegraphics[width=7.5cm,height=5cm]{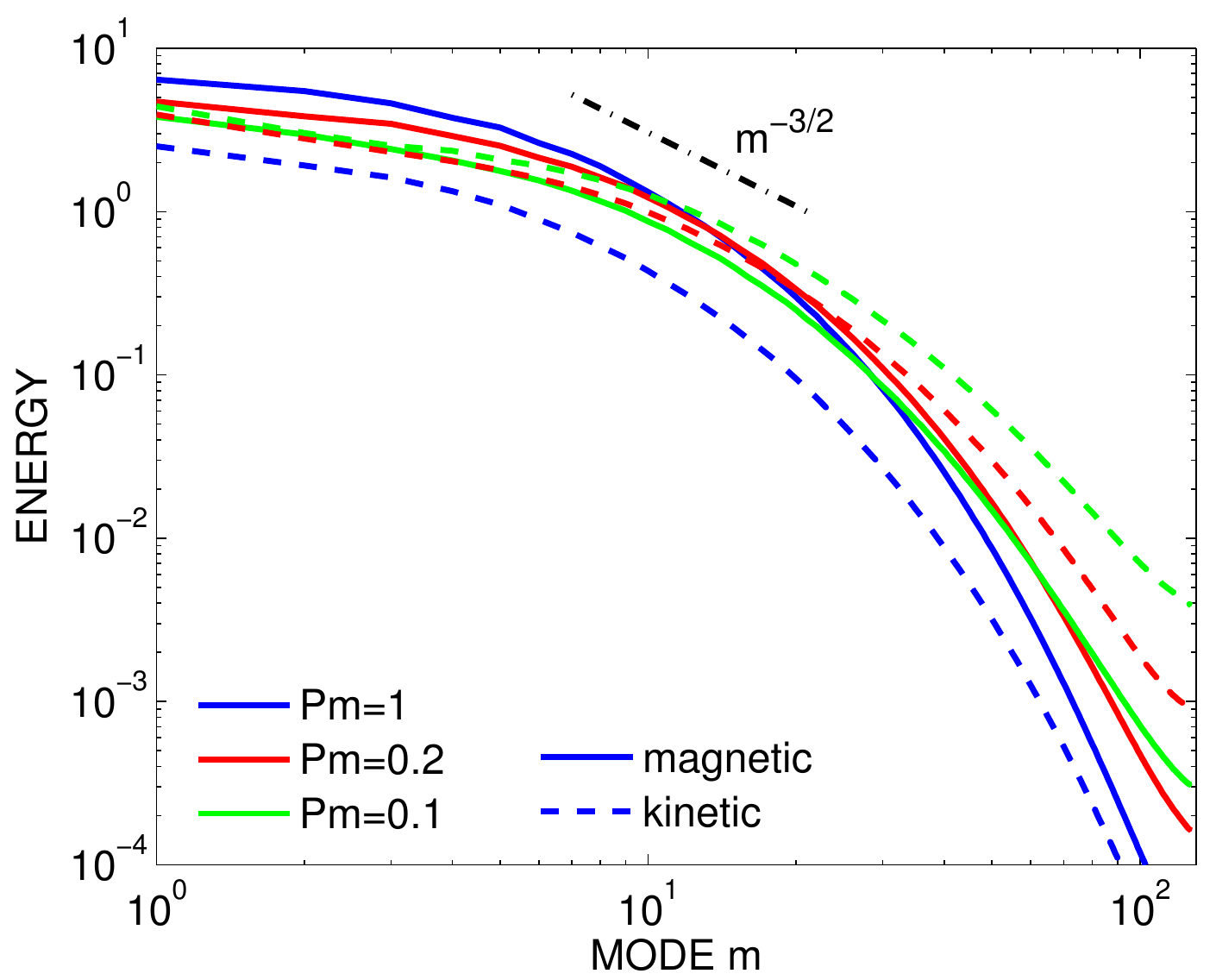}
 \includegraphics[width=7.5cm,height=5cm]{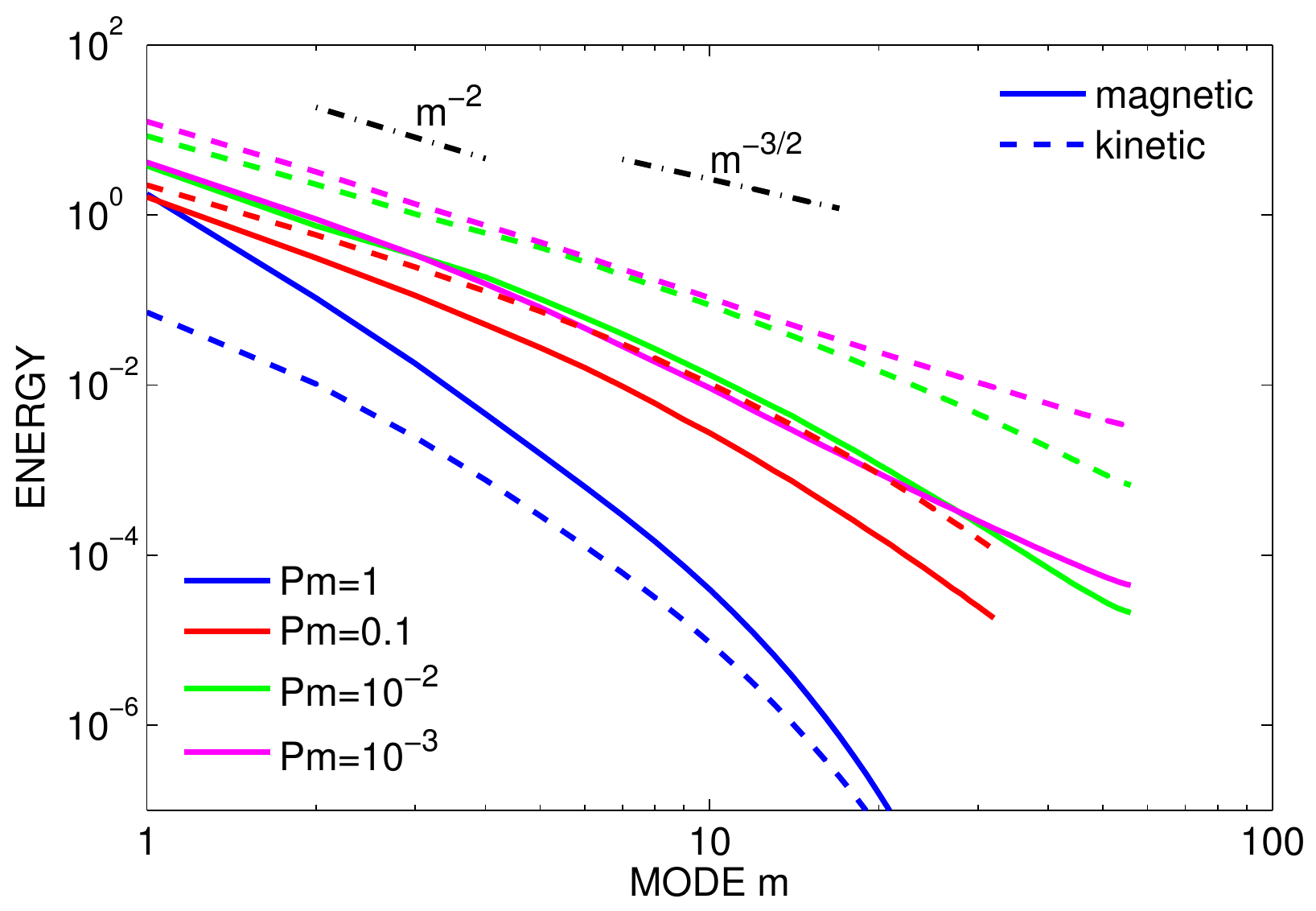}
 \caption{Azimuthal spectra of the magnetic fluctuations (solid lines) and kinetic fluctuations (dashed lines)  for two different  Chandrasekhar-type flows of various magnetic Prandtl numbers.  Left: potential flow ($\mu_B=2\mu_\Om=0.5$) with $\Rey=10,000$ and $\Ha=600$. Right: a rigidly rotating $z$-pinch ($\mu_B=2\mu_\Om=2$) with $\Rey=1000$ and $\Ha=1000$. Insulating boundary conditions.}
 \label{g8}
\end{figure}

For the rigidly rotating pinch only $m=1$ is unstable, but the energy is nonlinearly transferred to modes with higher wave numbers. Figure \ref{g8} shows the resulting power spectra of this model for fixed Reynolds and Hartmann numbers but various magnetic Prandtl numbers. The Mach number varies between $\Mm=0.2$ for $\Pm=0.01$ and $\Mm=2$ for $\Pm=1$. Only the mode $m=1$ provides the energy to initiate the nonlinear cascade so that the spectrum is rather steep. Neither the Iroshnikov-Kraichnan spectrum ($m^{-3/2}$, \cite{B03}) nor the Kolmogorov spectrum ($m^{-5/3}$, \cite{K41}) fit the resulting curves. A scaling $m^{-2}$ that is found in forced turbulence \cite{DT16} comes much closer.

It is typical for the magnetic instability that only the modes with the lowest $m\neq 0$ become unstable for finite $\Ha$ and $\Rey$. The rotating pinch gives an example where only a single linearly unstable mode ($m=1$) injects the energy into the system. For the AMRI with $\mu_B=2\mu_\Om=0.5$ modes with higher $m$ also become unstable. For given $\Ha$ and $\Rey$ the number of unstable modes decreases for decreasing magnetic Prandtl number. This is a consequence of the fact that for AMRI all azimuthal modes scale with $\Rey$ and $\Ha$ for $\Pm\to 0$. Figure \ref{g8} (right) shows the kinetic and magnetic energies for all modes $m$ for a fixed magnetic field with $\Ha=600$ and the high Reynolds number of
$\Rey=10,000$, but for several $\Pm$. The magnetic and kinetic spectra have a similar shape, but they are only close together for large $\Pm$. For small $\Pm$ the magnetic spectrum lies below the kinetic one. For $\Pm$ of order unity the spectrum is rather flat on the low $m$ side, and rather steep for small $\Pm$.

It is also obvious that the spectra for the kinetic and magnetic fluctuations have similar shapes. If a power law is fitted, both would slightly favor the Iroshnikov-Kraichnan spectrum compared with the Kolmogorov spectrum, but the differences are not significant. Although the Iroshnikov-Kraichnan profile is favored for MHD turbulence \cite{ZM04}, Kolmogorov-like spectra are also known from the measurements of turbulence in the solar wind \cite{M03}, as well as the result of 3D MHD simulations \cite{MB00}. Often, however, the direct numerical simulations are done for $\Pm$ of order unity \cite{B14}. A clear preference between Iroshnikov-Kraichnan and Kolmogorov scaling cannot be made.

\section{Helical magnetorotational instability (HMRI)}\label{HMRI}
\begin{figure}[h]
\centering
\includegraphics[width=9cm]{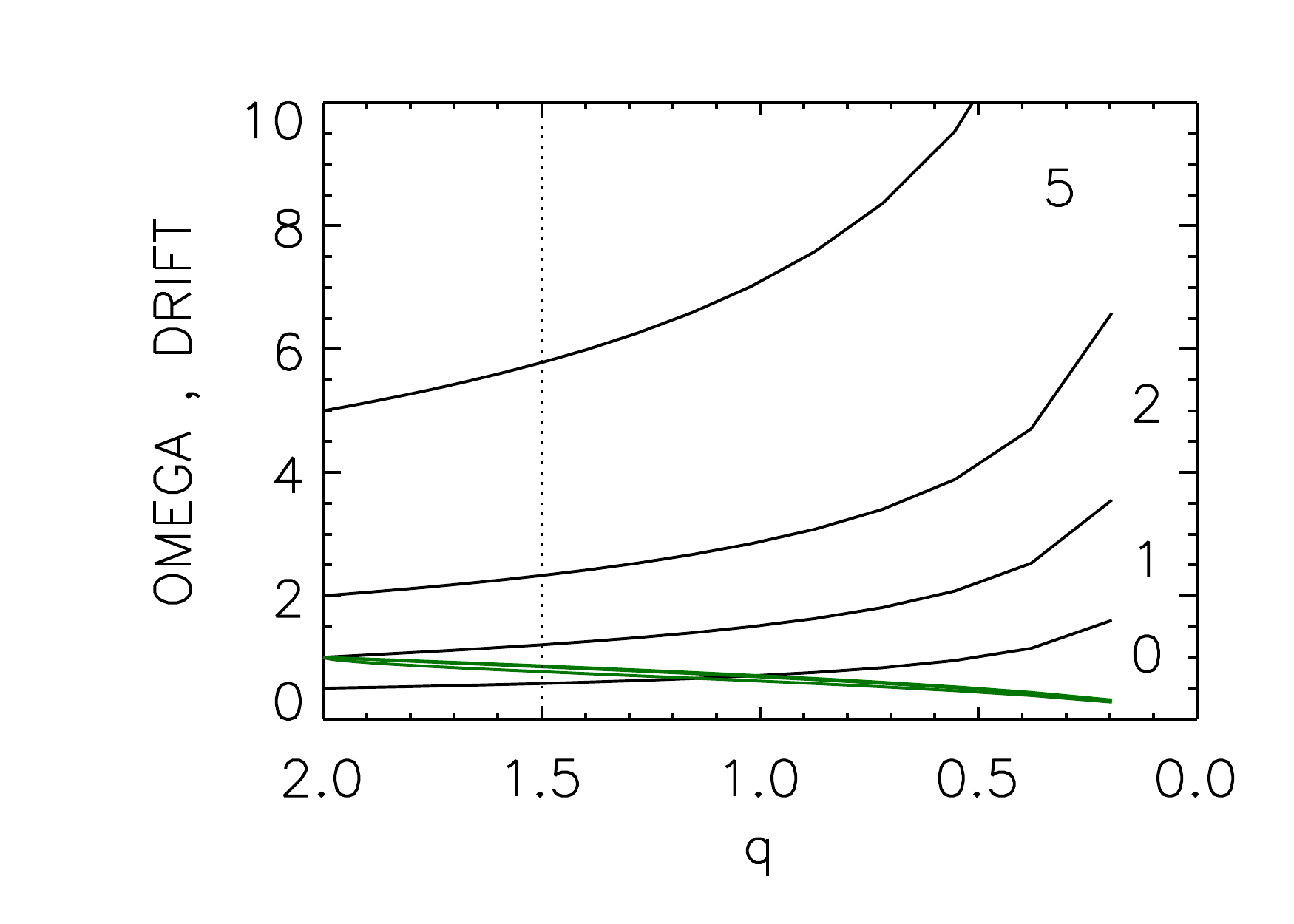}
 \caption{Rotation rate  $\tilde\Om/\Om_{\rm A}$ (black) and  travel frequency  $\Re(\omega)/\Om_{\rm A}$ (green)  for marginal stability of ideal flows. The curves are marked with their value of $\tilde\beta$. Potential flow has $q=2$,  rigid rotation has $q=0$, the shear of the Kepler law is marked by the dotted vertical line. Solutions for negative  $q$ (superrotation)  even for very large $\tilde\beta$ do not exist.}
 \label{liu}
\end{figure}

To the azimuthal  magnetic field  -- current-free between the cylinders --  discussed in Section \ref{AMRI} with respect to its stability a uniform axial magnetic field may be  added resulting in a helical magnetic configuration. After Eq. (\ref{gupta}) with ideal flows such a  system can be unstable  against axisymmetric perturbations  for negative shear (${\rm d}  \Om/{\rm d} R<0$)
but they should be stable for positive shear  (${\rm d}  \Om/{\rm d} R<0$). However, in case of instability the toroidal field basically acts stabilizing with respect to the standard MRI of purely axial fields. The parameter $\beta$ describes the inner value $B_{\rm in}$ of the azimuthal field normalized with the uniform vertical field, i.e.
\begin{equation}
\beta =\frac{B_{\rm in}}{B_0}.
\label{beta}
\end{equation}
The numerical value of $\beta$ gives the  angle between the field line and the axial direction. Almost axial fields possess only small values of $\beta$.
With this  parameter in mind the dispersion relation (\ref{11}) for ideal fluids takes the form
\begin{equation}
(\omega^2-\Om_{\rm A}^2- 2(2-q) \tilde\Om^2)(\omega^2-\Om_{\rm A}^2)-4 (\tilde\Om-\tilde\beta\omega)^2\Om_{\rm A}^2=0
\label{disphmri}
\end{equation}
 with $\tilde\beta=(k_z/k)\beta$ and $\tilde\Om=(k_z/k)\Om$  (see the definitions below Eq. (\ref{11})). The potential flow rotates with  $q=2$, the quasi-Keplerian flow with $q=3/2$ and rigid rotation leads to $q=0$. Negative  $q$ represent superrotation. The solutions of (\ref{disphmri}) for marginal stability (with  $\Im(\omega)=0$) are given in Fig. \ref{liu} for positive values of $\tilde\beta$. Both the 
critical rotation rate $\tilde\Om$ and a travel frequency  $\Re(\omega)$ are given in units of the \A~frequency $\Om_{\rm A}$. The latter only exists for $\beta\neq 0$ but  does not depend on the value of $\beta$.
The figure  demonstrates that without dissipation  finite values of $\beta$ always suppress the standard MRI  which appears  for $\beta=0$ with  $\tilde\Om/\Om_{\rm A}=1/\sqrt{2  q}$.  All curves for $\beta\neq 0$ lie {\em above} this minimum limit.
 For the potential flow  one simply finds   $\tilde\Om\simeq \beta\Om_{\rm A}=\Om_{\rm A,in}$ for all positive $\beta$.

Equation (\ref{disphmri}) does not provide  solutions  with $\Im(\omega)=0$ for  positive shear $q$. The numerical results thus confirm the formulation below Eq. (\ref{gupta}) that dissipationless superrotating flows also in helical fields  are stable against axisymmetric perturbations. Fluids with negative shear, however,  can be unstable but the azimuthal components of the magnetic field always {\em suppress}  the axisymmetric standard MRI with purely axial fields \cite{K96}. All phenomena of subcritical excitation by  additional  azimuthal background fields which we shall describe in the present section are thus of diffusive nature which only exist if at least one of the diffusion coefficients  $\nu$ or $\eta$ have finite values. One can repeat the calculations within the inductionless approximation ($\Pm=0$, see Section \ref{inductionless}) and finds solutions only for $q>1.66$ but also for $q<-9.66$.  Finite values of the magnetic resistivity, therefore, stabilize  flat rotation laws with negative shear  but they even destabilize steep enough rotation laws with positive shear \cite{LG06,LV07,PG07,MS16}, see Section \ref{QKr}. 

\medskip

To study the stability of helical background fields  in the presence of differential rotation is insofar of particular interest as the fundamental (`lowest') modes with axial field are axisymmetric while those with azimuthal current-free fields are nonaxisymmetric. 
The first question concerns the symmetry type of the instability of such helical (or better: twisted) fields with a preferred handedness. It has been shown that possible instabilities of helical background fields can never be stationary so that a possible axisymmetric mode must travel along the rotation axis \cite{K92,K96,BV05,LV07}. The symmetry of the background field  is changed as $z$ and $-z$ are no longer equivalent. One can speculate to utilize this axial drift to observe the instability in a laboratory experiment. To this end it would be important to know the oscillation frequency and its dependence on basic parameters.  The following examples mainly concern the right-handed twisted magnetic field with $\beta=2$ where the axial and the azimuthal field components are of the same order. The Hartmann numbers are now formed with the {\em axial field amplitude} $B_0$ as defined by  (\ref{Hartmann}) -- only for the exceptional case of $\beta=\infty$   the toroidal field $B_{\rm in}$ as in  (\ref{Hartmannin}) is  used.
The geometry of the mixed field instability modes can be described via the relations (\ref{pitch}) and
\beg
\frac{\partial z}{\partial t}\bigg\vert_{\phi}= - \frac{\omega_{\rm dr}}{k},
\ende
which describes the phase velocity in the axial direction of the modes at a fixed azimuth. The wave is traveling upwards if the real part of the eigenfrequency, $\omega_{\rm dr}$, is negative. 

The wave numbers  $k$ and $m$ are both real values, and without loss of generality one of them, e.g. $k$, can be taken as positive. Then $m$ must be allowed to have both signs. The sign of $\beta$ fixes the spiral geometry of the background field with respect to the rotation axis. If the axisymmetric background field possesses positive $B_z$ and $B_\phi$ (as mostly used for the calculations here) then it forms a right-hand spiral.

The introduction of the new parameter $\beta$ makes the situation complex. In the present section we thus only consider azimuthal fields which are current-free in the fluid between the cylinders, i.e.~$\mu_B=0.5$ for $\rin=0.5$. The cylinders always form perfectly conducting boundaries. The only exception is Fig.~\ref{h21}, where for a demonstration of scaling laws for small $\Pm$ an almost uniform azimuthal magnetic field is considered.

We must also question the scaling of the results for small magnetic Prandtl number. From the foregoing sections we know that the MRI scales with $\Rm$ and $\Lu$ for $\Pm\to 0$. The consequence is that the ordinary Reynolds number cannot remain finite for $\Pm\to 0$. The same is true for the AMRI with vanishing axial electric current within the non-potential flows. We should thus expect that the HMRI also scales with $\Rm$ and $\Lu$ for $\Pm\to 0$. However, all models of the Chandrasekhar-type with $\vec{U}=\vec{U}_{\rm A}$ scale with $\Rey$ and $\Ha$ for $\Pm\to 0$. It is thus an open question how the eigenvalues for decreasing $\Pm$ behave for HMRI of the {\em potential flow}. Another prominent example is the Chandrasekhar-type flow with $\mu_B=2\mu_\Om=1$ describing a rotation law with almost uniform azimuthal flow $U_\phi$ and azimuthal field $B_\phi$.
\begin{figure}[htb]
\centering
 \includegraphics[width=8cm]{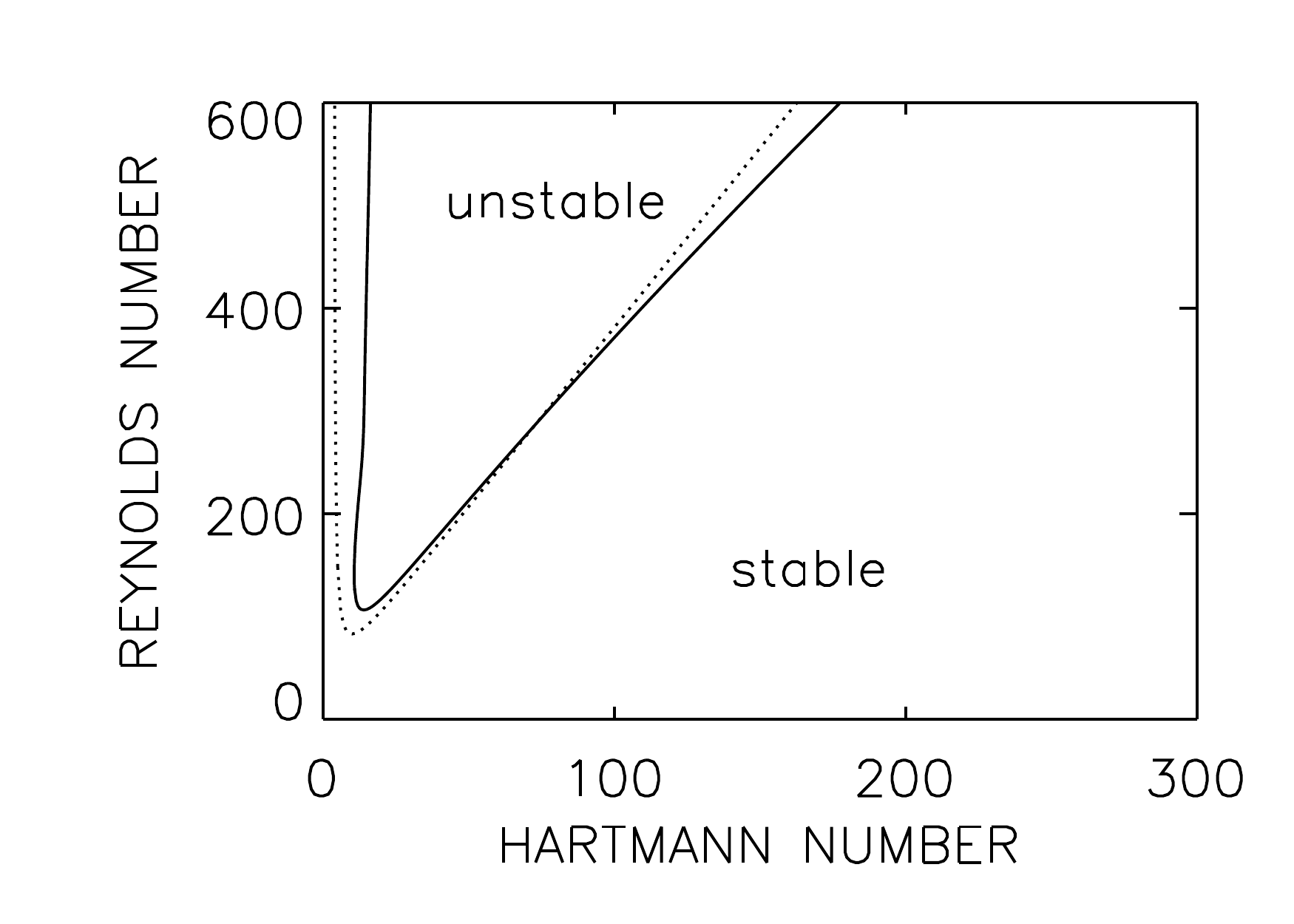}
 \includegraphics[width=8cm]{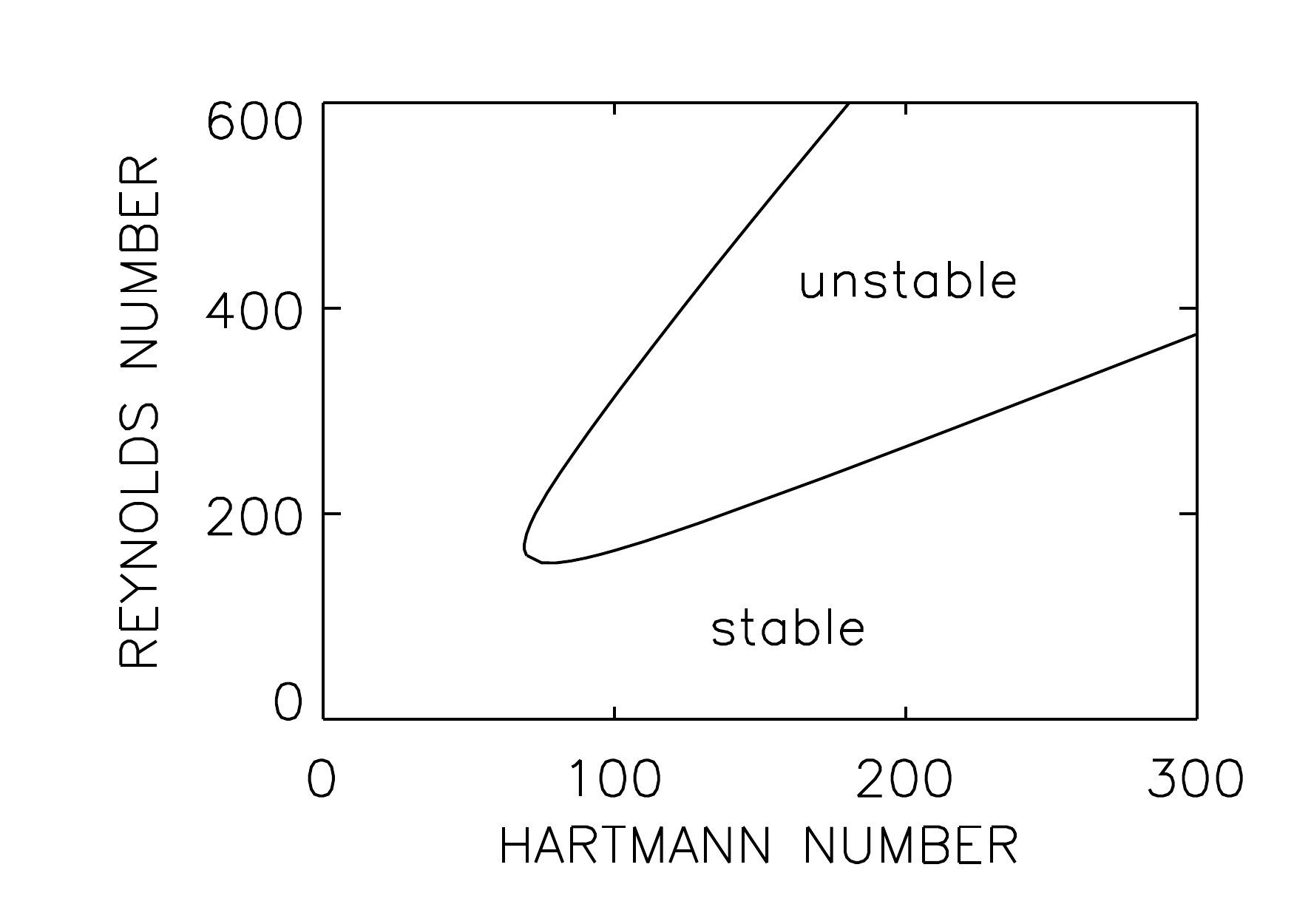}
 \caption{Stability maps  for quasi-uniform flow with uniform axial fields (MRI, left) and  with current-free azimuthal fields (AMRI, $\mu_B=\rin$, right),  . Note the different definitions of the Hartmann numbers: Eq.~(\ref{Hartmann}) for the left panel and Eq.~(\ref{Hartmannin}) for the right panel.  Solid lines: $m=1$, dotted line: $m=0$. The slopes ${\rm d}\Rey/{\rm d}\Ha$ of the solid lines ($m=1$) are always positive but they are not for the dotted line ($m=0$). The modes with the lowest Reynolds numbers are axisymmetric for MRI and nonaxisymmetric for AMRI. $\mu_\Om=\rin=0.5$, $\Pm=1$. Perfectly conducting boundaries.}
 \label{hh1}
\end{figure}
\subsection{From AMRI to HMRI}\label{amrihmri}
We start with the stability of the flow $U_\phi \simeq$~const in the presence of a purely axial field. In this case both axisymmetric and nonaxisymmetric modes may be excited, with the axisymmetric $m=0$ mode being the one with the lowest Reynolds number (Fig.~\ref{hh1}, left). For $\rm Pm=1$ this overall minimum occurs for $\rm Ha\simeq 10$ and $\rm Re\simeq 80$. For larger $\rm Ha$ there is a switch to $m=1$ being the mode with the lowest Reynolds number. The axisymmetric mode only dominates for weak fields, but including also the global minimum $\Rey$ value. It also dominates the weak-field branch of the instability curve. This branch of the axisymmetric instability curve tilts to the left, whereas the strong-field branch tilts to the right. For the nonaxisymmetric mode both branches tilt to the right, forming a characteristic tilted cone. A purely azimuthal field without electric currents between the cylinders and subject to the same rotation law yields an instability for $m=1$ for $\Ha\gsim 80$ and $\Rey \gsim 150$ (Fig.~\ref{hh1}, right). Both the upper and lower branches of the instability curve tilt to the right. For a given Hartmann number, the instability therefore only exists within a finite range of Reynolds numbers.
\begin{figure}[htb]
\centering
 \includegraphics[width=8cm]{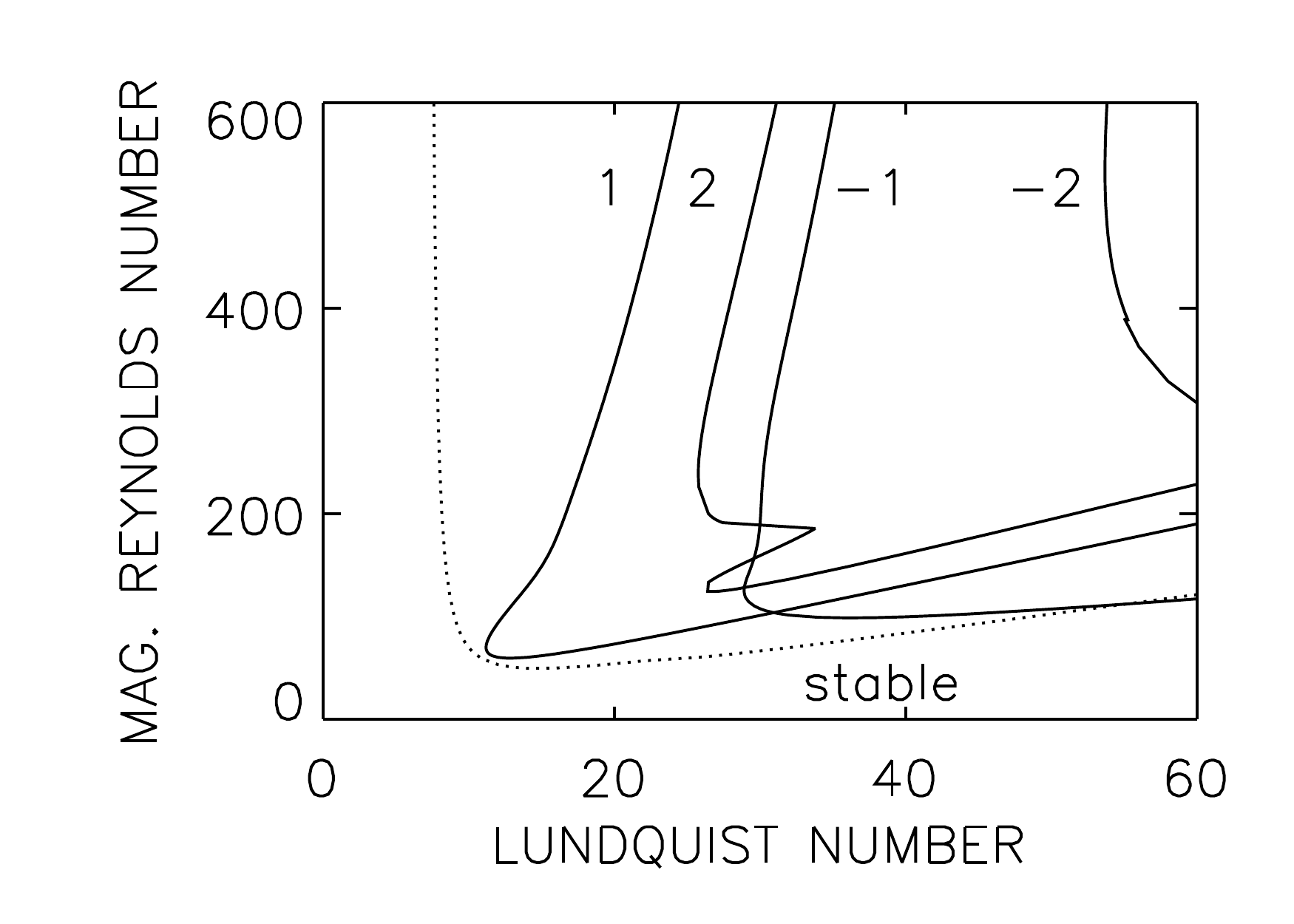}
 \includegraphics[width=8cm]{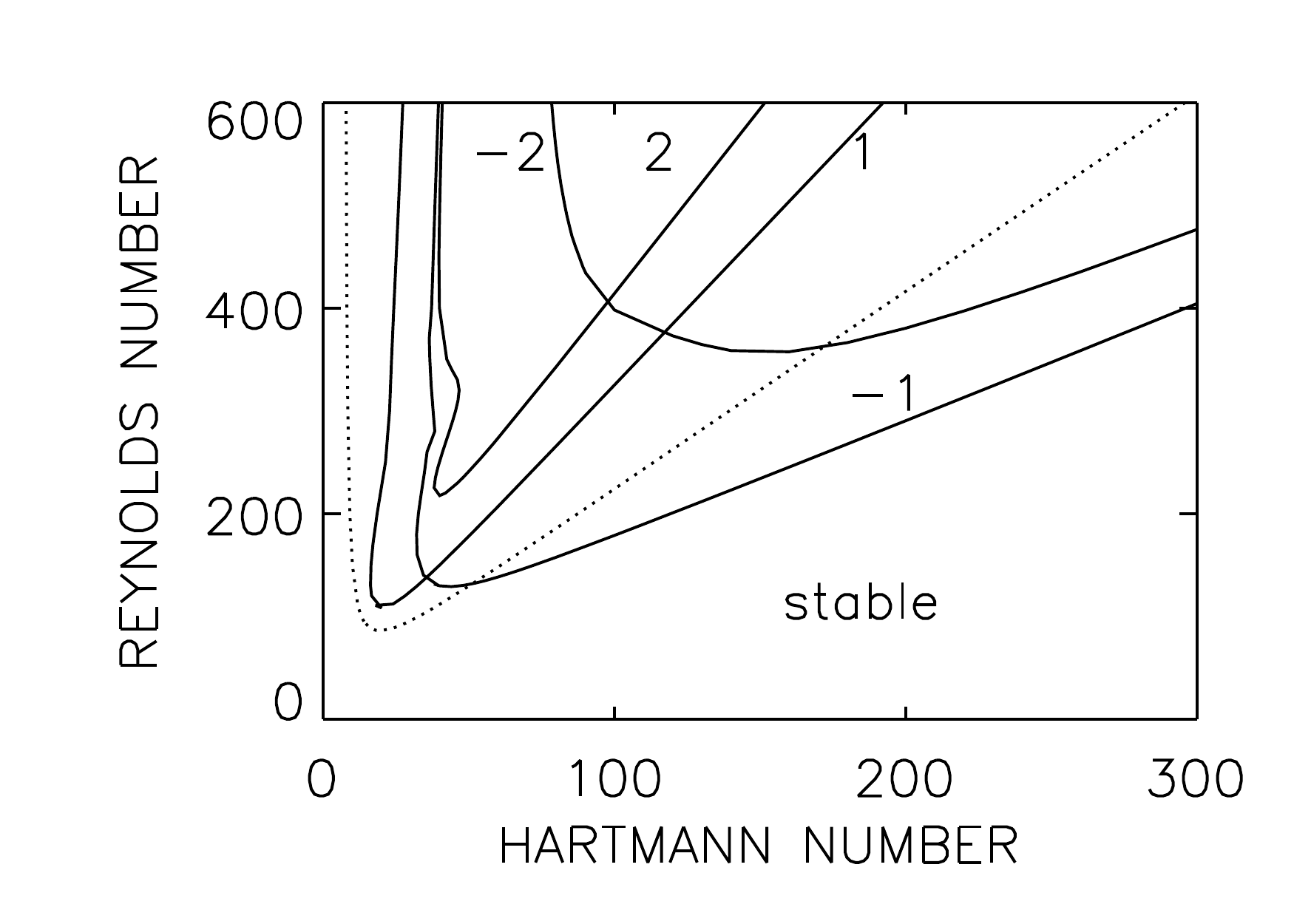}
 \caption{Critical magnetic Reynolds numbers for excitation of HMRI modes with $\beta=2$ for various $m$  (marked). $\Pm=0.01$ (left) and $\Pm=1$ (right). The background field is current-free  and the flow is quasi-uniform. Negative $m$ describe right-hand spirals,  positive $m$ describe left-hand spirals and the dotted lines represent $m=0$. Lundquist and Hartmann numbers are formed as  in (\ref{Hartmann}).  $\mu_\Om=\mu_B=\rin=0.5$. Perfectly conducting cylinders. From \cite{RG10}.}
 \label{h2}
\end{figure}

Figure \ref{h2} (left) shows the results for the combination of azimuthal and axial fields with $\beta=2$. One finds the same general pattern as before: only the weak-field branch of the $m=0$ mode tilts to the left; both branches of all nonaxisymmetric modes tilt to the right. Up to ${\Ha}\approx 50$ the axisymmetric mode is preferred, just as before for the standard MRI. For ${\Ha}>50$ the $m=1$ spiral is preferred. Note also that the minimum Hartmann number for excitation is much smaller than for fields with $B_0=0$.

\begin{figure}
\centering
 \includegraphics[width=8cm]{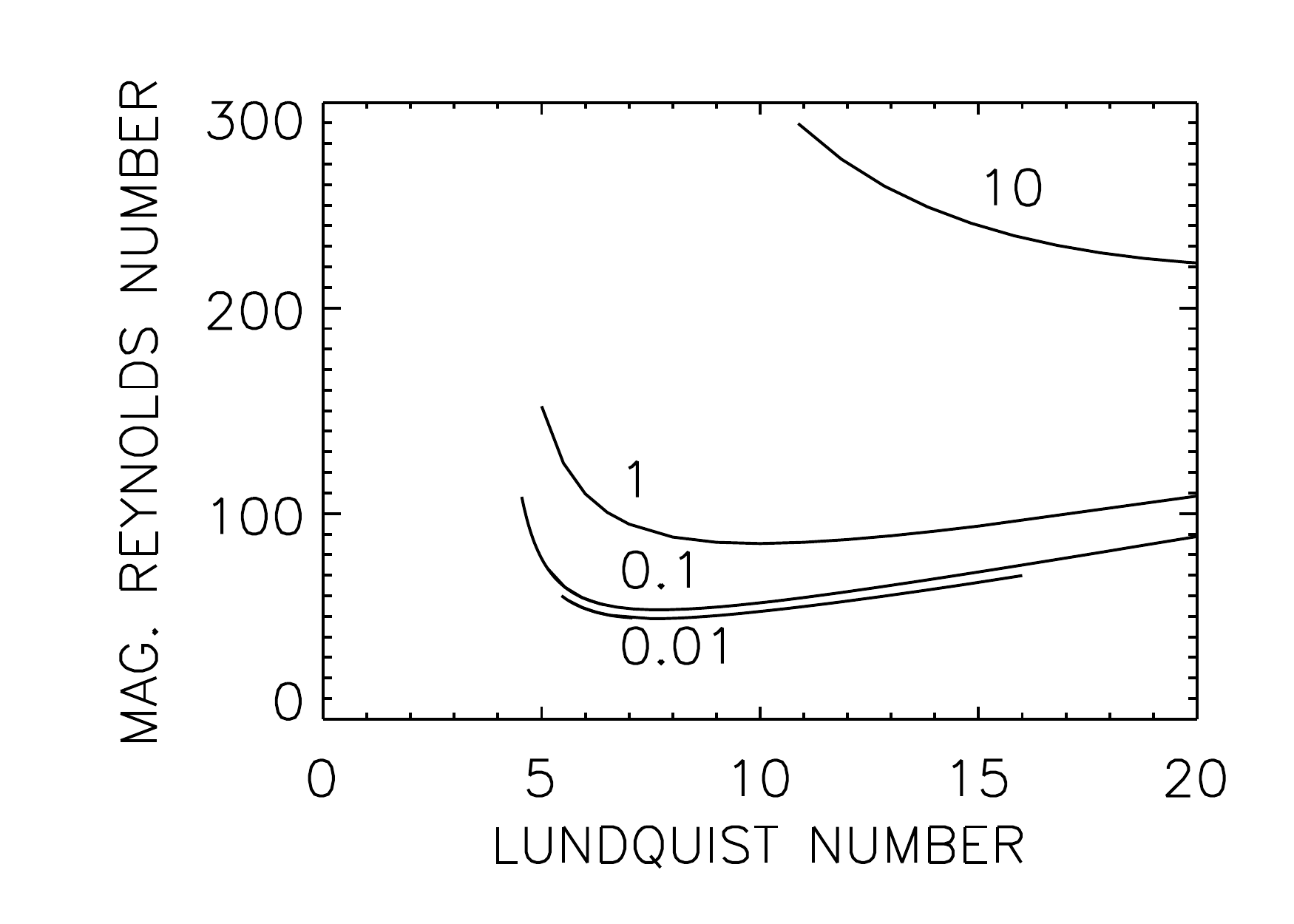}
 \includegraphics[width=8cm]{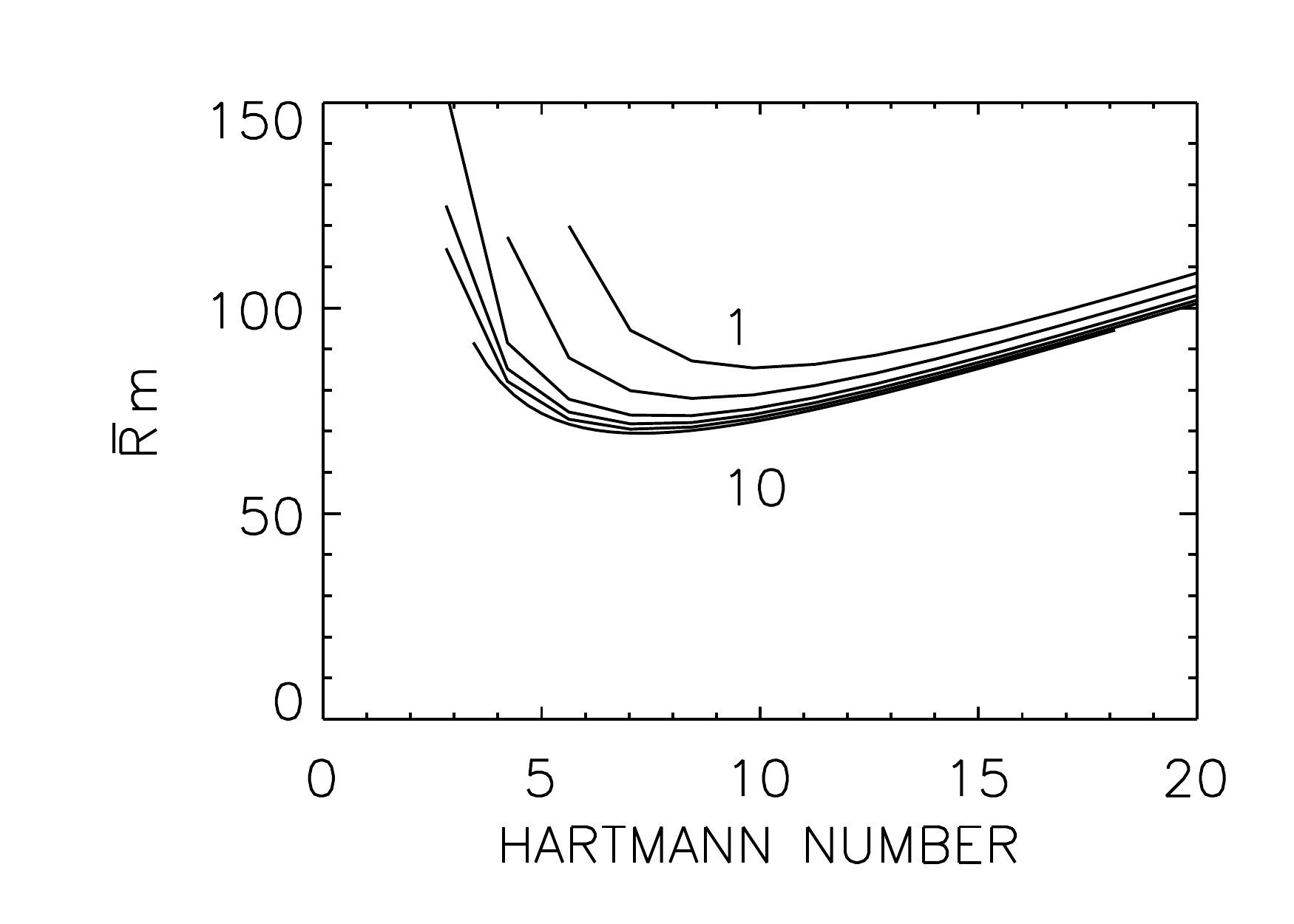}
 \caption{Stability maps   for background fields with $\beta=2$ for small $\Pm$ (left) and for large $\Pm$ (right). The curves are marked with their values of $\Pm$. The azimuthal field and the flow are both quasi-uniform.  One finds the same  scaling laws as for  standard MRI (see Fig.~\ref{f14}).  $\rin=0.5$, $m=0$, $\mu_B=2\mu_\Om=1$. Perfectly conducting boundary conditions.}
 \label{h21}
\end{figure}

Obviously, the (axisymmetric) standard MRI and the (nonaxisymmetric) AMRI are basic elements both influencing the excitation conditions if the background field has a twisted geometry. More specifically, one finds that the weak-field branch of the instability in Fig.~\ref{h2} is very similar to the weak-field branch of the MRI, while the strong-field branch resembles the strong-field branch of AMRI. The absolute minimum values of the Reynolds and Hartmann numbers always belong to the axisymmetric mode. The similarity of the instability maps in Fig.~\ref{h2} for $\Pm=1$ and $\Pm=0.01$ also indicates that the HMRI scales with $\Rm$ and $\Lu$ for $\Pm\to 0$. This finding remains true if the background field satisfies the condition (\ref{chancon}) for Chandrasekhar MHD flows. Figure \ref{h21} demonstrates that the instability lines of this {\em axisymmetric} mode for this magnetic configuration converge in the ($\Lu/\Rm$) plane for $\Pm\to 0$ and in the ($\Ha/\Rey$) plane for $\Pm\to \infty$. This scaling rule of the eigenvalues for $m=0$ in the presence of axial fields is {\em opposite} to the rules of Chandrasekhar-type flows for $m=1$ without any axial field. Obviously, the helical structure of the total background field changes the scaling rules in the sense as they exist for MRI. The fields in Figs.~\ref{h2} and \ref{h21} only differ by the parameter $\mu_B$. In the second case the azimuthal field is of Chandrasekhar-type and in the first case it is not. In both cases, however, the scaling for small $\Pm$ is that of the MRI, which makes experiments with liquid metals so challenging. Consequently, the two branches of each curve in Fig.~\ref{h21} have opposite slopes: the weak-field branch goes to the left while the strong-field branch goes to the right (which is also typical for the axisymmetric modes of MRI rather than for AMRI). Generally, for background fields forming a right-hand spiral ($\beta>0$) the left-hand modes ($m>0$) require a lower Hartmann numbers for their excitation. 

\begin{figure}[htb]
\centering
 \includegraphics[width=9cm]{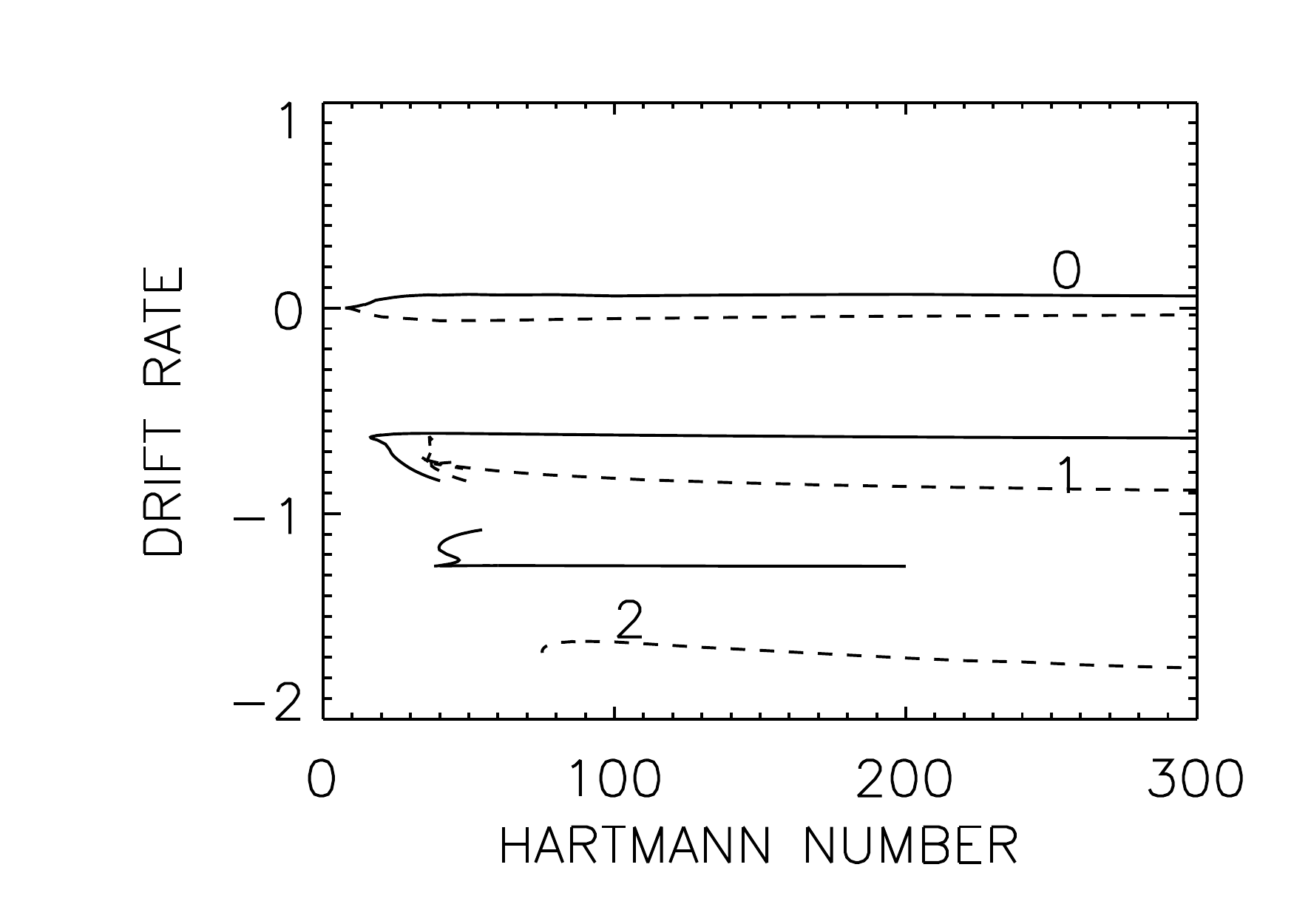}
 \caption{Drift rates $\omega_{\rm dr}$ for  $\beta=2$ (solid) and $\beta=-2$ (dashed) for the models of Fig.~\ref{h2} (right). The curves are marked with the azimuthal wave numbers $m$.  $m=0,1,2$. The axisymmetric modes possess small $\omdr$ of same sign as $\beta$ leading to upward or downward axial pattern migration. The negative $\omdr$ of the nonaxisymmetric modes provide  azimuthal migration in positive $\phi$-direction. $\mu_B=\mu_\Om=\rin=0.5$, $\Pm=1$. Perfectly conducting boundaries.}
 \label{h3}
\end{figure}

Another key phenomenon is the different character of the eigenfrequencies: MRI is stationary, AMRI drifts in azimuthal direction, but the HMRI drifts in $z$ as a necessary consequence of the $\pm z$ symmetry-breaking. The oscillatory nature of the axisymmetric HMRI is reflected by the finite values of the drift frequency $\omega_{\rm dr}$ for $m=0$. They have the same sign as the parameter $\beta$ (Fig.~\ref{h3}). Positive $\beta$ generate positive $\omega_{\rm dr}$ (downwards traveling) and vice versa. Vanishing $\beta$ leads to $\omega_{\rm dr}=0$, i.e.~to stationary axisymmetric instability patterns. The drift rates for $m=0$ (axial migration) are very low while for the nonaxisymmetric modes (azimuthal migration) they are large and negative for $\beta=\pm 2$.

Note that for all $m$ and all $\beta$ the migration frequencies (\ref{dotfi}) of the nonaxisymmetric modes have very similar negative values, which means that all modes approximately corotate with the inner cylinder. They are much higher than the frequency of the axial drift. The negative values demonstrate that all nonaxisymmetric instability patterns migrate in the positive $\phi$ direction. They exceed the value $\mu_\Om=0.5$ (the rotation rate of the outer cylinder in the laboratory system) so that they are always overtaking the outer cylinder. One may assume that the drift rates of the nonaxisymmetric modes are due to the rotation rates while the axial-traveling frequency scales with the viscosity frequency which is here only 1\% of the global rotation.

A direct consequence of the $\pm z$ symmetry-breaking, and the associated axial drift of the axisymmetric HMRI modes, is that the distinction between convective and absolute instabilities becomes important, especially in axially unbounded cylinders. Convective instabilities are disturbances that grow only in a reference frame moving with the perturbation, whereas absolute instabilities grow even at a fixed point in space, as the perturbation drifts past. Absolute instability is thus a more restrictive  condition than convective instability. Correspondingly, the analysis of \cite{PG09,P11}, in which the axial wavenumber $k$ is allowed to be complex, shows that the absolute HMRI exists in a somewhat narrower parameter range than the convective HMRI. The basic scalings and transitions between scalings remain the same though. See also \cite{HSK17}, who computed fully nonlinear solutions in background fields that varied periodically on very long axial wavelengths, and found absolute and convective instabilities to behave similarly even in the nonlinear regime. All nonlinear calculations in axially bounded cylinders are automatically also computing absolute rather than convective instabilities. It is
nevertheless important to take cylinders that are sufficiently long, as modulations may develop on sufficiently long axial length scales \cite{HSK17,Clark2017}.

\subsection{Quasi-potential flow}
Figure \ref{h21} demonstrates that a helical field with $\mu_B=2\mu_\Om=1$ (Chandrasekhar-type) with a uniform axial magnetic component becomes unstable for eigenvalues $\Rm$ and $\Lu$ which are independent of $\Pm$ for small $\Pm$. Standard MRI and AMRI for this flow also scale with $\Rm$ and $\Lu$ for small $\Pm$. As the potential flow in the presence of current-free fields also belongs to the class of Chandrasekhar-type flows with $\mu_B=2\mu_\Om=0.5$, it is thus expected that the combination with a uniform axial field also scales with $\Rm$ and $\Lu$ for small $\Pm$. The calculations, however,  do {\em not} confirm this expectation. The explanation of the low-$\Rey$ and low-$\Ha$ phenomenon for the potential flow in the presence of axial fields is not based on the fact that the current-free azimuthal field together with the potential flow belongs to the class of Chandrasekhar-type  MHD flows.

\begin{figure}[htb]
\centering
\includegraphics[width=5.25cm]{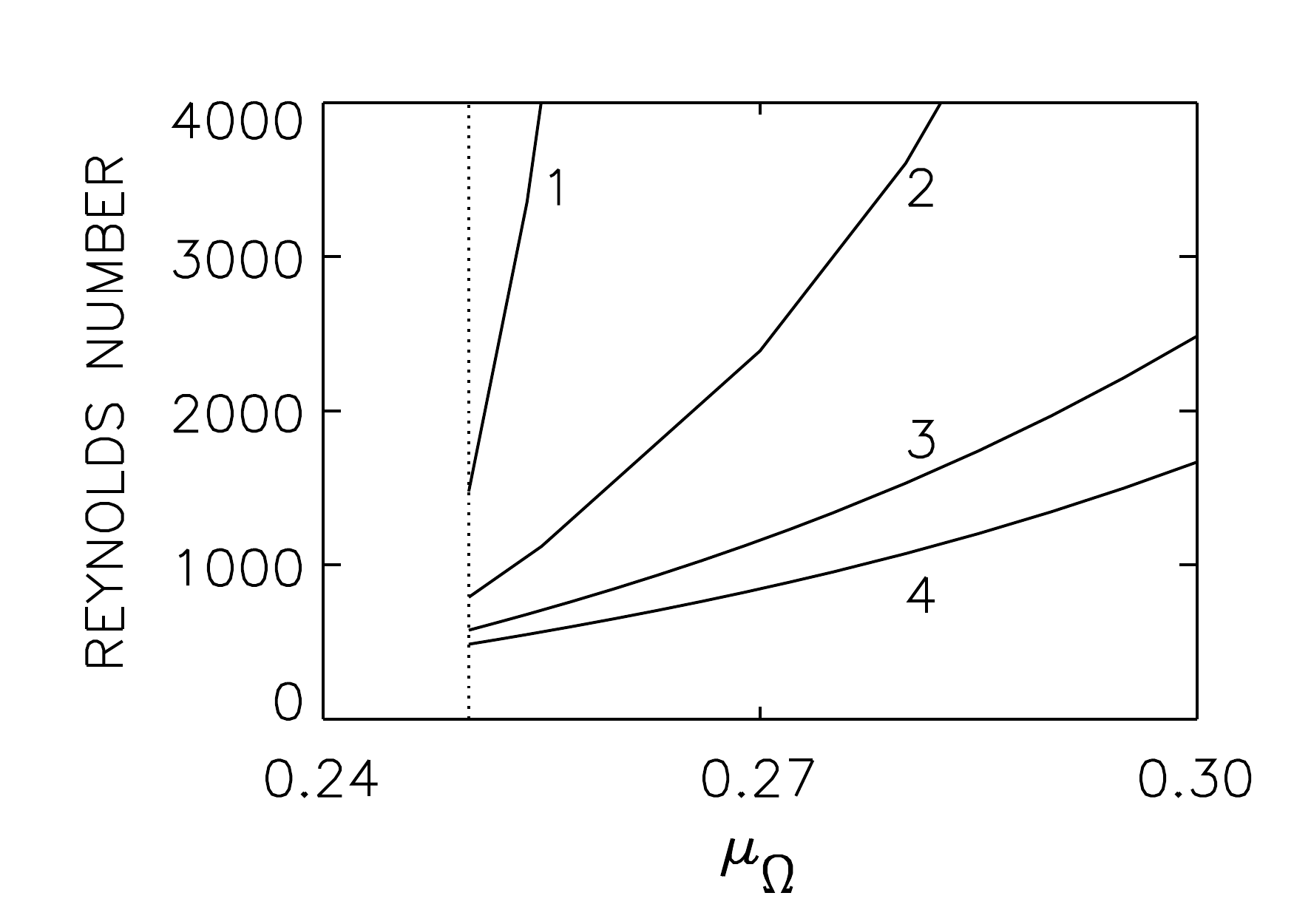} 
\includegraphics[width=5.25cm]{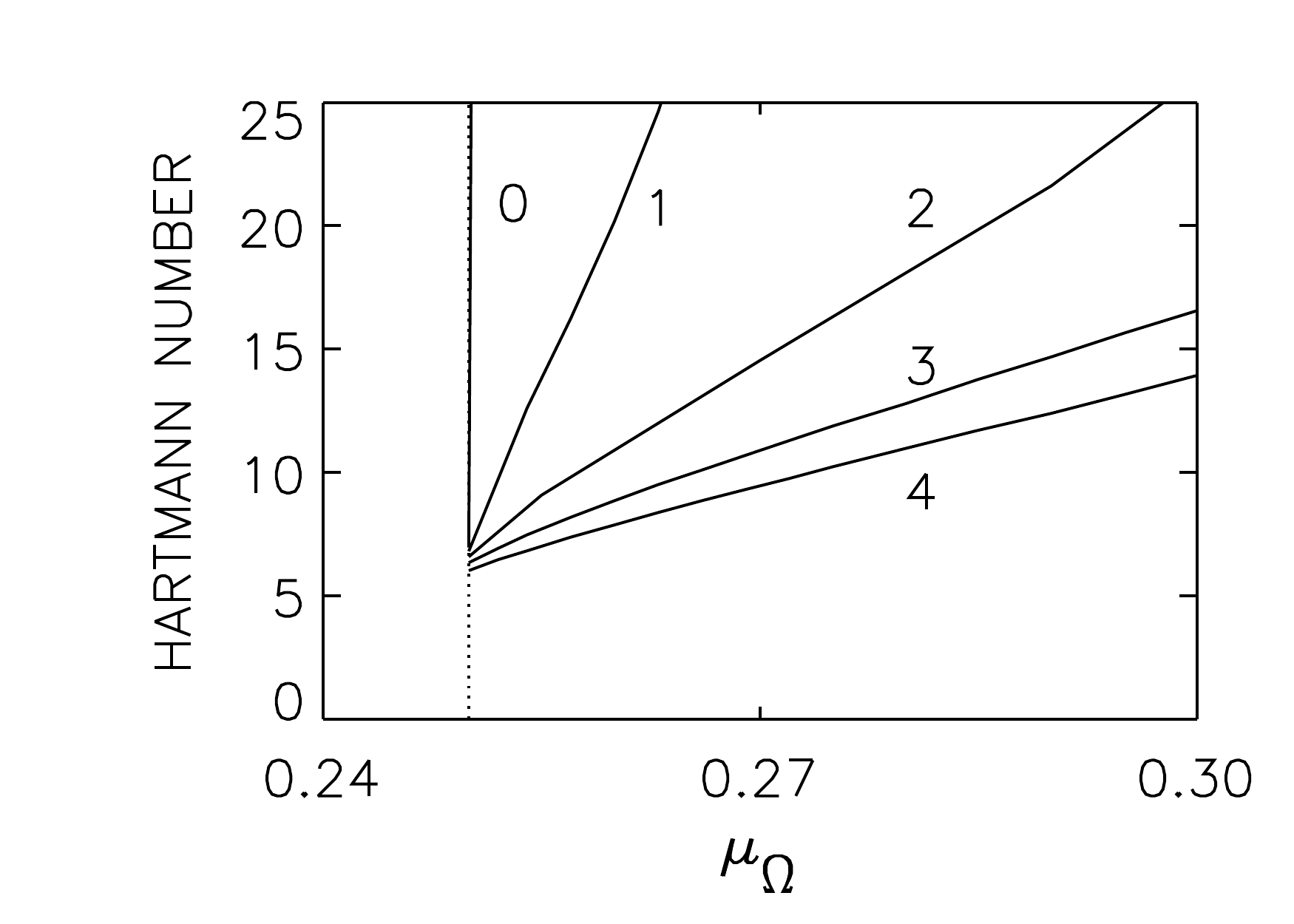} 
\includegraphics[width=5.25cm]{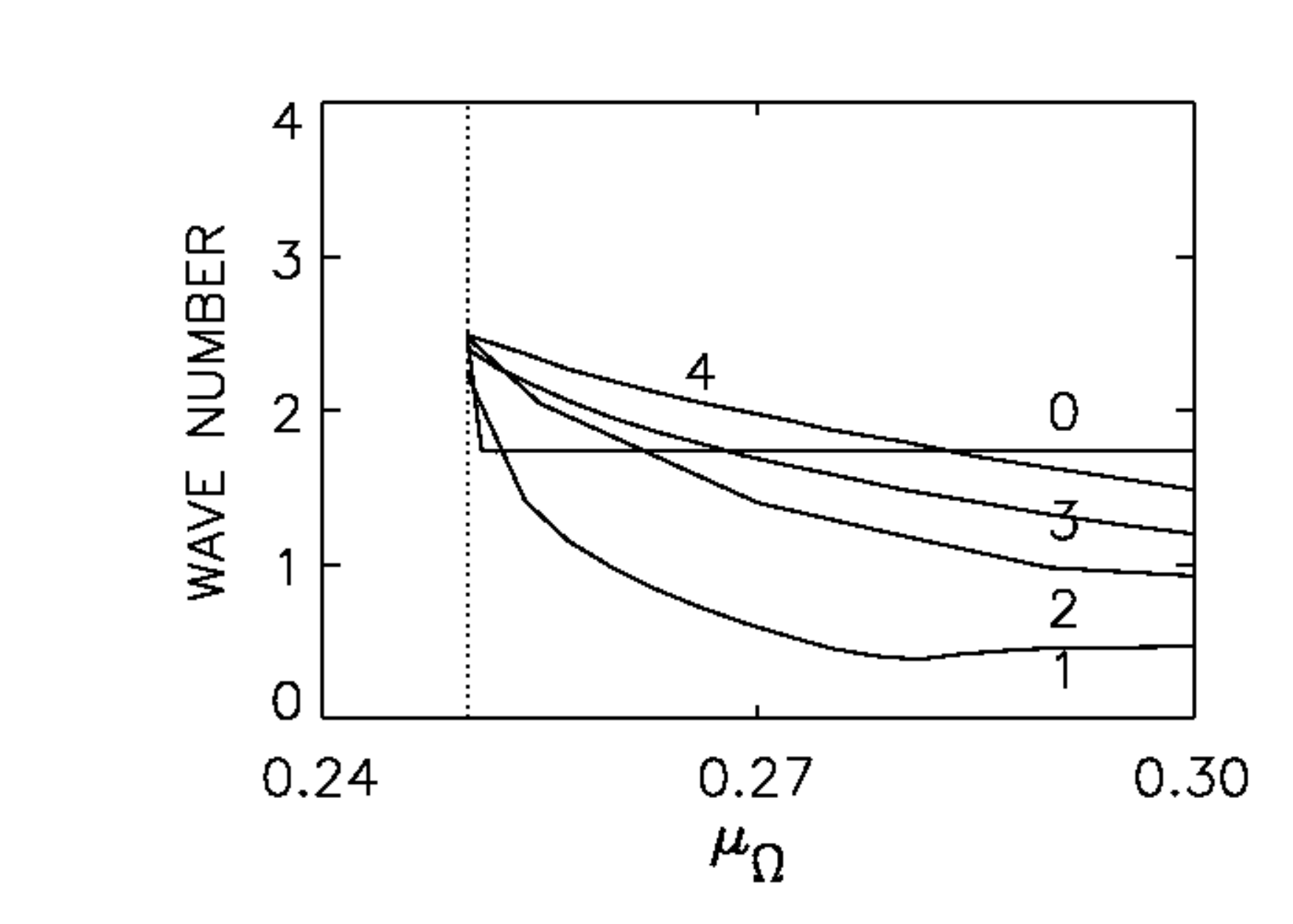} 
\caption{Critical Reynolds numbers (left), critical Hartmann numbers (middle) and the corresponding wave numbers (right) of the axisymmetric modes for various $\beta$ and at and beyond  the Rayleigh line. The cells prove to be always elongated in axial direction.  $\mu_B=\rin=0.5$ (vacuum field), $\rm Pm=10^{-5}$, perfectly conducting cylinders. From \cite{RH05}.}
\label{prom1}
\end{figure}

The transition of HMRI with $\beta\neq 0$ from the potential flow to the quasi-Keplerian flow will now be discussed. The Figs. \ref{prom1} give a detailed insight into how the critical Reynolds number, Hartmann number and wave number behave for small $\Pm$ for the potential flow and beyond. Standard MRI is described by $\beta=0$; immediately beyond the Rayleigh limit its critical Reynolds number jumps to values of $10^6$ (not shown). This is no longer true for finite $\beta$. For $\beta$ of order unity the Reynolds number takes much lower values at and close the Rayleigh line. For $\beta=2$ and for (say) $\mu_\Om = 0.27$ (within the hydrodynamically stable area) low values for $\Rey\simeq$~\ord{10^3} and $\Ha\simeq$~\ord{10} are sufficient to excite the HMRI. Such values can easily be realized in the MHD laboratory by use of sodium or GaInSn as the fluid. 

If, opposite to the standard MRI,  the instability for quasi-potential flow (close to the Rayleigh limit) scales with $\Ha$ and $\Rey$ for small $\Pm$, then as in Section \ref{inductionless} the solution  in the inductionless approximation exists and equals the solution of the full equation system for the limit $\Pm\to 0$. Extensive numerical simulations for axially periodic boundary conditions and perfectly conducting cylinders have thus been done in the quasistationary approximation $\Pm=0$ for axisymmetric perturbations, based on the code developed and described by \cite{YB06}. For infinite cylinders and for $\mu_\Om=0.27$ the flow is always hydrodynamically stable, but with helical magnetic background  field with $\beta=4$ it loses its stability already for the small Reynolds number $\Rey_{\rm crit}=842$. This result well agrees with the value of the linear theory given in the left panel of Fig.~\ref{prom1}.
 \begin{figure}[htb]
\centering
 \includegraphics[width=1.7cm]{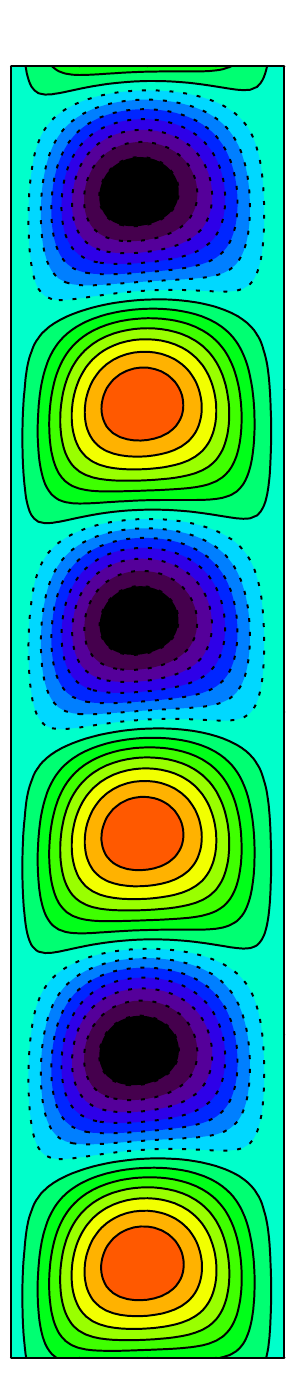}
 \includegraphics[width=1.7cm]{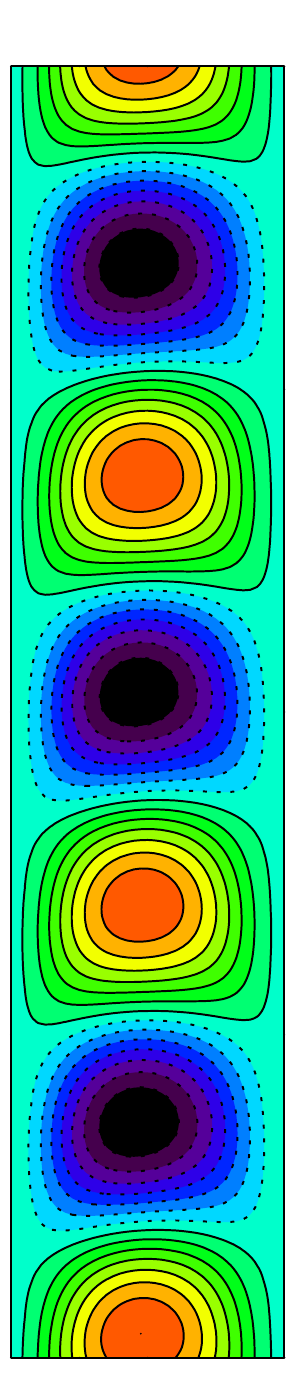}
 \includegraphics[width=1.7cm]{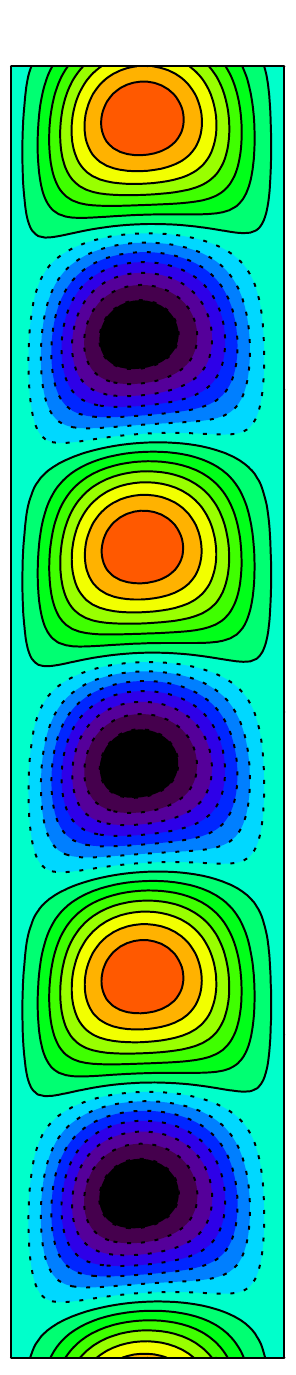}
 \includegraphics[width=1.7cm]{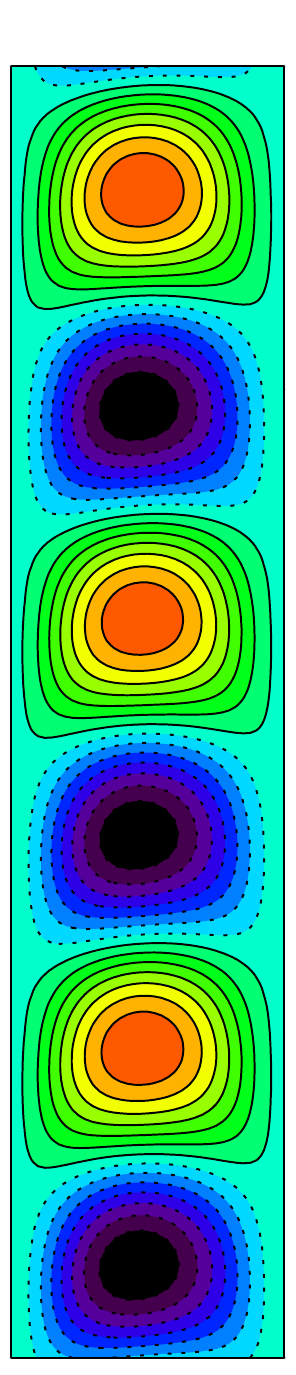}
 \hfill
 \includegraphics[width=1.7cm]{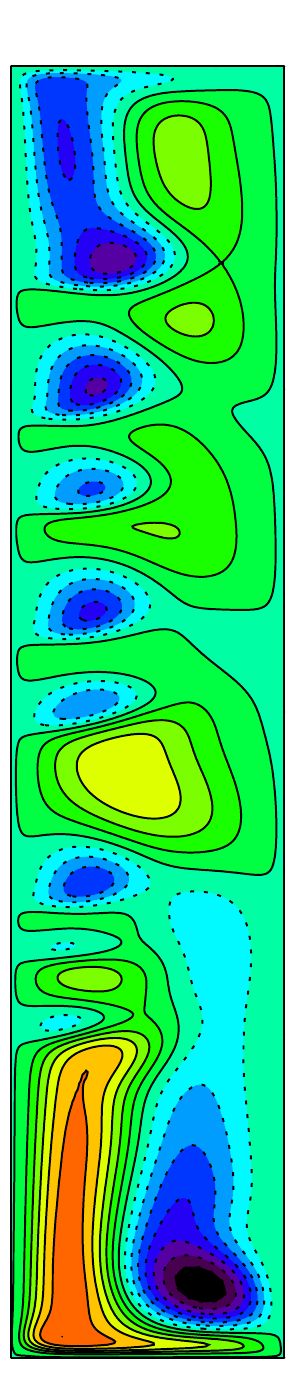}
 \includegraphics[width=1.7cm]{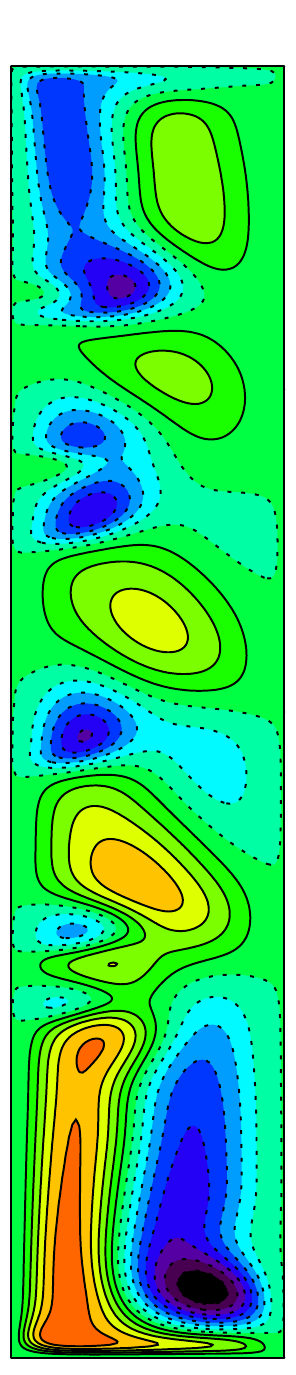}
 \includegraphics[width=1.7cm]{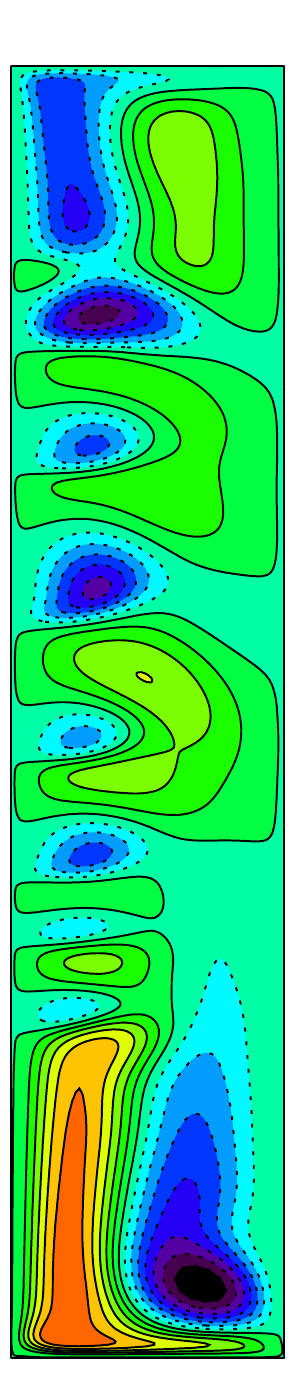}
 \includegraphics[width=1.7cm]{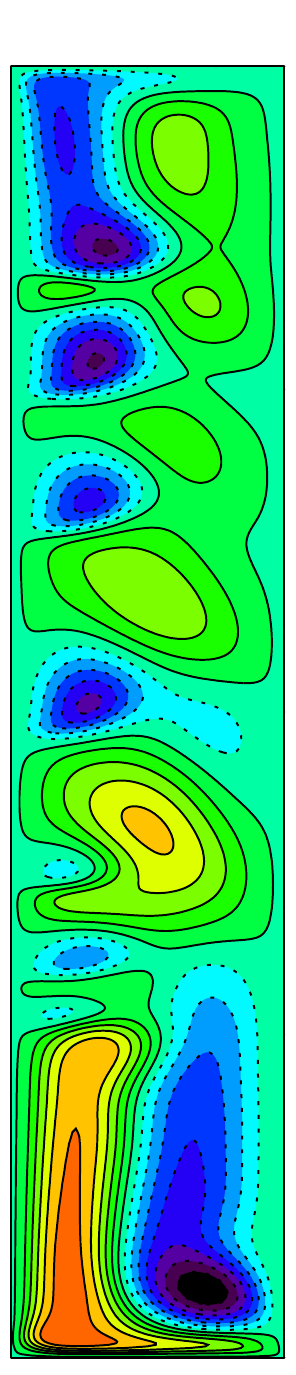}
 \caption{Simulated snapshots (sequenced in time) of numerical simulations representing  downward-traveling wave (for $\beta>0)$ without (left panel) and with endplates (right panel).  The axisymmetric  contourlines of the streamfunction of the flow are shown (solid lines: clockwise, dashed lines: counterclockwise). Left: $ \Rey=900$, $\beta=4$. Right: $ \Rey=1480$, $\beta=6$, $H=10 d$. The upper lid corotates with the outer cylinder while the lower lid is stationary in the laboratory.  It is  $\Ha=9.5$, $\mu_\Om =0.27$, $\Pm=0$. Perfectly conducting cylinders, insulating endplates. From \cite{SR06}.}
 \label{jatzek} 
\end{figure}
Figure \ref{jatzek} shows the downward drift of the streamlines of the HMRI cells without and with (insulating) endplates. The fluid moves along the given contourlines of the streamfunction (positive streamfunction: clockwise, blue color; negative streamfunction: counterclockwise, red color). 
For $\mu_\Om$ between 0.25 and 0.27 the wave travels with $\omdr\simeq 0.13$ in the axially periodic container,  and with $\omdr\simeq 0.12$ in the container with top and bottom endplates. In both cases there is a weak anticorrelation between the values of $\mu_\Om$ and $\omdr$. These values agree with the results of the linear analysis (see the right panel of Fig.~\ref{prom11a}). The axial travel speed of the unbounded model with supercritical $\Rey\simeq 1600$ is about 1 mm/s for gallium \cite{SR06}.
Also  the structure and evolution of  the Ekman-Hartmann layers which develop at the special endplates of the container has been discussed in detail  \cite{SR07}. 

To observe the influence of the boundary conditions in the right panel of Fig.~\ref{jatzek} the lower endplate rests in the laboratory, while the upper endplate corotates with the outer cylinder. Observe that the perturbing influence on the traveling instability pattern is much stronger for the stationary lower lid than it is for the rotating upper lid. These extreme endplates do not prevent the traveling wave though; even the agreement between the linear and the nonlinear results proved to be satisfying. For the maximal axial velocity of the traveling pattern for the models with $\beta=4$ and $\Rey=1500$ the simulations provided 1~mm/s in both cases, close to empirical data for the {\sc Promise} container (see below).

 \begin{figure}[htb]
\centering
 \includegraphics[width=15cm]{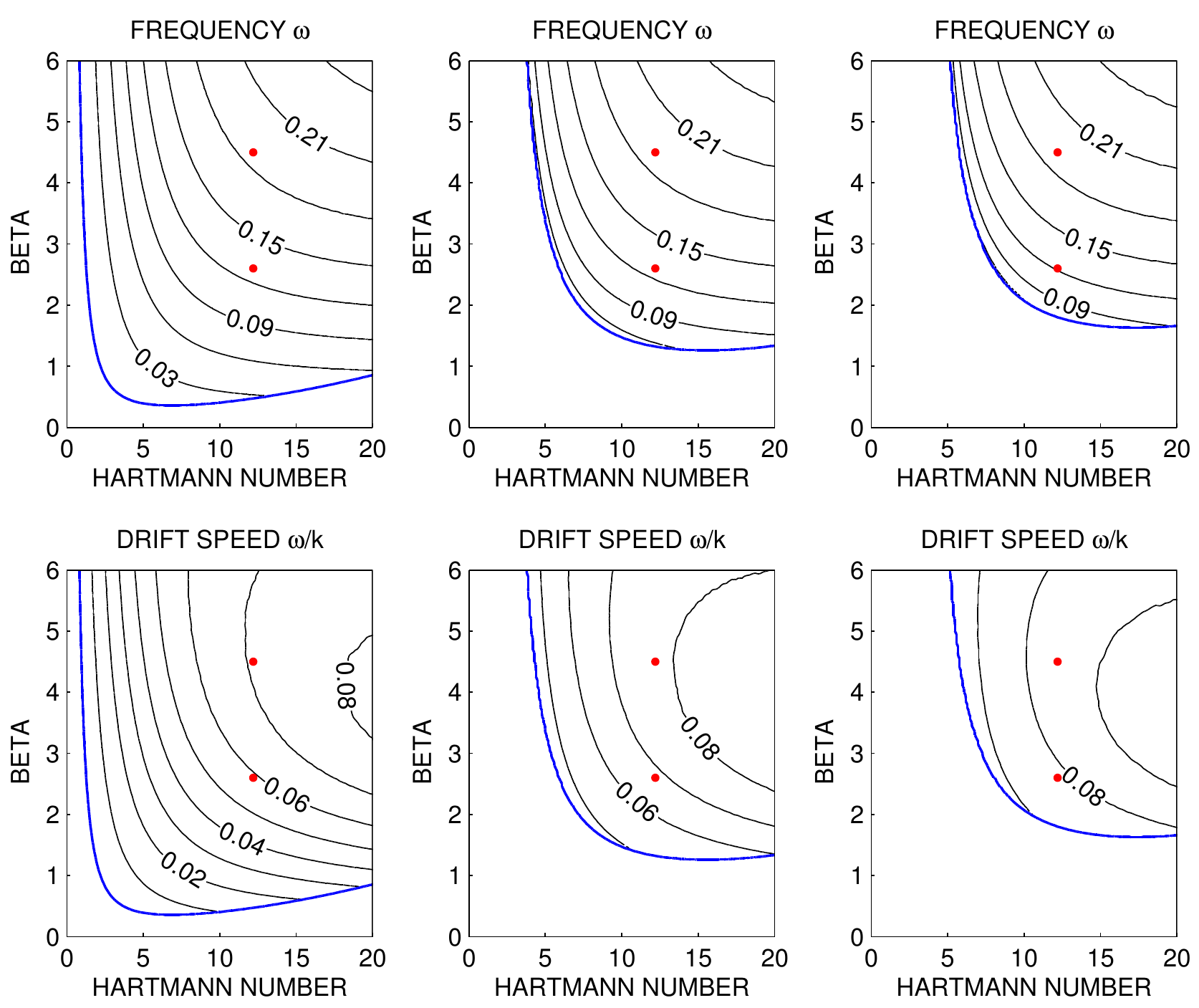}
 \caption{Drift frequency (top) and axial drift speed (bottom) versus Hartmann numbers for marginal stability of flows with $\mu_\Om=0.25$ (left), $\mu_\Om=0.26$ (middle) and $\mu_\Om=0.27$ (right). The maximum Reynolds number is $\Rey=4000$ (blue lines). The red dots concern real experiments discussed below. $\mu_B=\rin=0.5$, $\Pm=10^{-5}$. Perfectly conducting cylinders.}
 \label{prom11a}
\end{figure}

For marginal stability Fig.~\ref{prom11a} provides the corresponding travel frequencies $\omega_{\rm dr}$ and the travel speeds $\omdr/k$ as functions of $\Ha$ and $\beta$ from the linear theory. The travel frequency is the lowest frequency in the system. A typical value for medium $\beta$ is $\omdr\lsim 0.1$. A deeper inspection of the plots in the top row of Fig.~\ref{prom11a} suggests that at least for $\mu_\Om=0.25$, only a very weak dependence of the frequencies on the values of $\beta$ exist. 
Moreover,  the travel frequencies  normalized with the viscosity frequency $\omega_\nu=\nu/R_0^2$ written  as functions of $\beta$ and $\Ha$ show almost no dependence  on the value of $\beta$. 
It has also been shown that only a slight dependence on the magnetic Prandtl number,
 i.e.~$\omega_{\rm dr}\propto {\omega_\nu}{{\rm Pm}^{-1/4}},$
exists, without any influence of $\beta$ \cite{LV07}. For the phase velocity at the Rayleigh line the numerical value $\omega_{\rm dr}/k\simeq 0.01-0.02$ results, almost independent of $\beta$ and scaling linearly with $\Ha$. For the flows slightly beyond the Rayleigh limit the normalized travel velocity also hardly depends on the value of $\beta$ if the Hartmann number is not too large. For the quasi-Keplerian flow the influence of $\beta$ becomes much stronger and even depends on the boundary conditions.


\subsection{Boundary conditions}\label{boundary}
It is also worth comparing the results for perfectly conducting cylinders with those obtained with insulating ones \cite{HR05}. The numerical values explicitly mentioned in this paper are for $\mu_\Om=0.27$ and $\beta=4$, which yield $\Rey=1521$ and $\Ha=16$. For the model with perfectly conducting cylinders we find the smaller values $\Rey=842$ and $\Ha=9.5$. Insulating boundaries thus increase both $\Rey$ and $\Ha$ by almost a factor of two. Hence, an experiment with perfectly conducting boundaries would be  the most promising design for exploring the magnetorotational instability in the laboratory. Note, however, that for TI the situation is different: Reynolds numbers for insulating cylinders are lower than for conducting ones, but the Hartmann number behaves opposite.
 
The material of the endplates for axially bounded containers also plays an important role. It is known that in the transition zone between differentially rotating fluid and rigid endplates an Ekman-Hartmann layer develops in the presence of an axial magnetic field \cite{GB68}. This magnetized shear layer induces electric currents beneath the layer in the bulk of the container. Their radial component, together with the axial background field, provides azimuthal Lorentz forces accelerating or decelerating the global rotation. The rotation is suppressed in the range between the cylinders if the endplates corotate with the outer cylinder. For perfectly conducting endplates the Hartmann current reduces the rotation rate within the gap between the cylinders by almost 50\% (for $\Ha\simeq 10$), but this effect is much weaker for insulating endplates. The material for the endplates, therefore, should ideally be a good insulator \cite{S07diss,SR07}. The dramatic consequences of `wrong' endplates for the rotation profile in the midplane (!) between the endplates are shown in Ref.~\cite{S07} for various amplitudes of the axial magnetic field. This effect is not weak; the profile becomes very steep close to the inner cylinder and rather flat in the gap between the cylinders. Close to the outer cylinder the shear even changes its sign. 

\begin{figure}[h]
\centering
 \includegraphics[width=8cm]{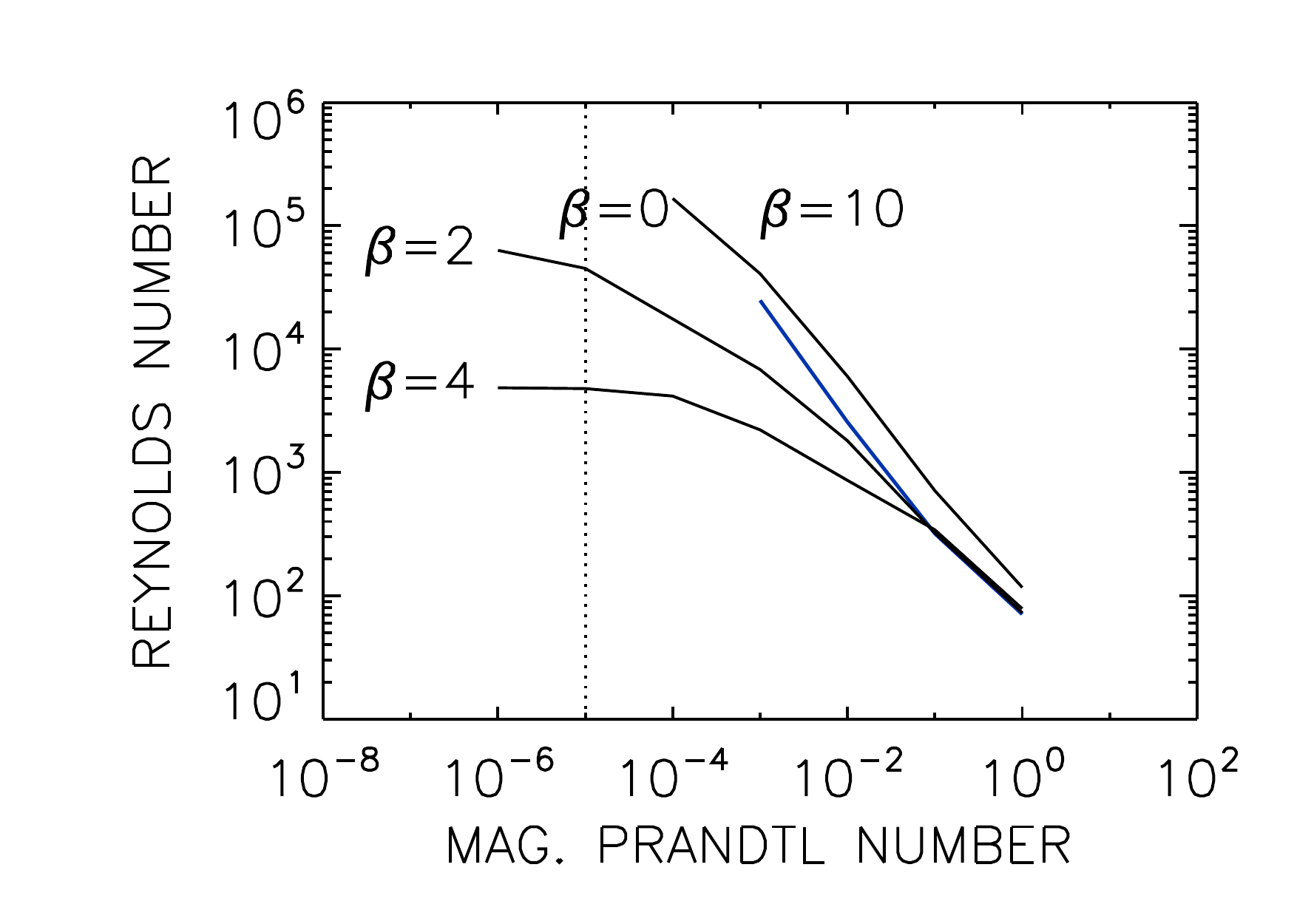}
 \includegraphics[width=8cm]{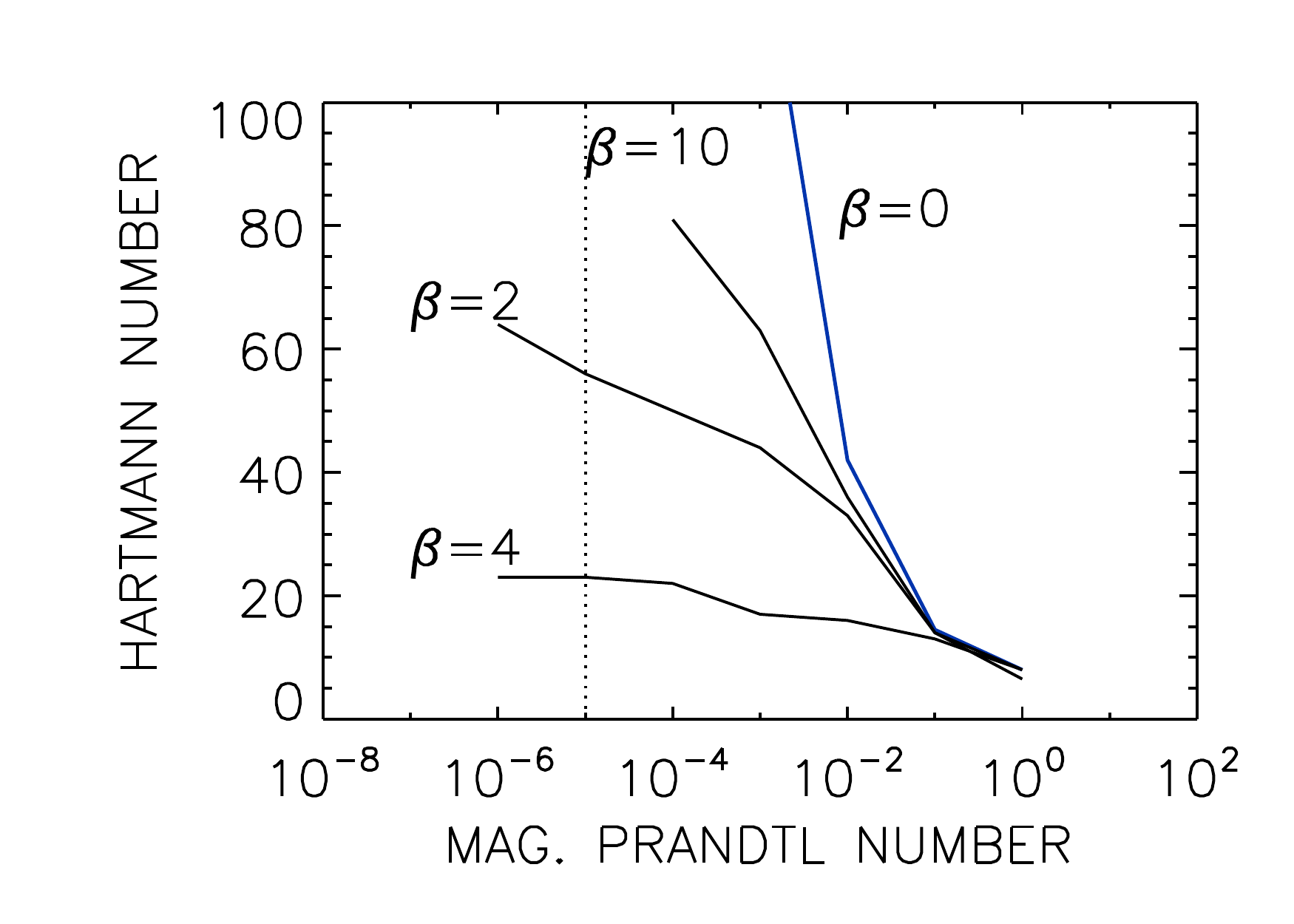}
 \caption{Quasi-Keplerian flow subject to twisted background field with current-free azimuthal component: critical Reynolds numbers (left) and Hartmann numbers (right)  as function of the magnetic Prandtl number for $\beta=0$ (blue lines) and for $\beta=2,4,10$. Hartmann number defined by (\ref{Hartmann}). Observe that  $\beta=0,10$ scale with $\Rm$ and $\beta=2,4$ scale with $\Rey$. Vertical dotted lines mark the magnetic Prandtl number of liquid sodium. $m=0$, $\rin=0.5$, $\mu_B=\rin$, $\mu_\Om=0.35$, perfectly conducting cylinders. }
 \label{h5}
\end{figure}
\subsection{Quasi-Keplerian rotation}\label{QKr}
Another question related to boundary conditions is whether the HMRI for quasi-Keplerian rotation and small $\Pm$ also scales with $\Rey$ and $\Ha$ or with $\Rm$ and $\Lu$. In a local and inviscid approximation  it has been shown that  solutions do not exist for rotation laws with $\mu_\Om> 0.32$, which would suggest the quasi-Keplerian law with $\mu_\Om\simeq 0.35$ to be stable \cite{LG06}.  Calculations for containers with perfectly conducting cylinders and finite $\Pm$ do not confirm this strict result. Figure \ref{h5} shows for quasi-Keplerian rotation that for small $\Pm$ the scaling with $\Rey$ exists, but only for not too large $\beta$. A scaling with ${\Rey}$ only exists for the narrow range of $\beta \simeq 2-4$ but no longer for $\beta\gsim 10$. Very small $\beta$ (MRI) and very large $\beta$ (AMRI) both lead to a scaling with $\Rm$ and $\Lu$ for $\Pm\to 0$, yielding very high Reynolds and Hartmann numbers as eigenvalues for small $\Pm$. For ${\rm Pm}=10^{-5}$ and $\beta=4$, the critical Reynolds number at about 6000 is still rather low in comparison with values \ord{10^6} which are characteristic for MRI. This finding is always true if at least one of the cylinders is perfectly conducting (see below).
\begin{figure}[!h]
\centering
\includegraphics[width=15cm]{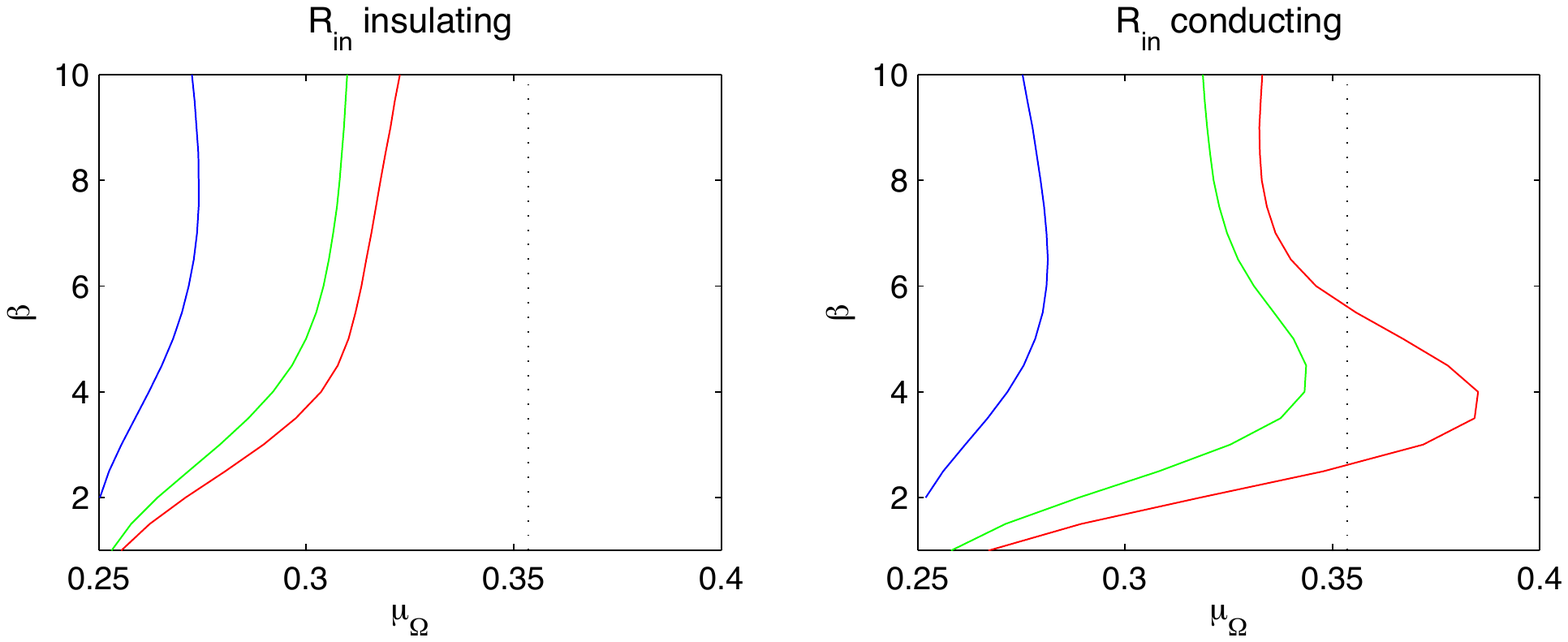}
\vskip-1.5cm
\caption{Isolines of Reynolds numbers as functions of $\mu_\Om$ and $\beta$ for insulating (left) and perfectly conducting (right) inner cylinder. $\Rey=10^3$ (blue), $\Rey=10^4$ (green), $\Rey=10^5$ (red). The dotted vertical line marks the quasi-Keplerian rotation law. $\mu_B=\rin=0.5$, $\Pm=0$.}
\label{keplerholl}
\end{figure}

Figure \ref{h5} also demonstrates  that for $\Pm=1$ the eigenvalues for $\beta>0$ exceed  the eigenvalues for $\beta=0$. When $\Rey=\Rm$ the toroidal field, therefore, basically suppresses the standard MRI. This fact has  been formulated long ago for ideal fluids \cite{K92}. One can indeed argue that  large values of $\Rey=\Rm$ may mimic the case of vanishing $\nu$ and $\eta$ of ideal fluids \cite{Sh07}. For the ideal potential flow also the dispersion relations provide an increase of the critical rotation rate with increasing $\beta$; the {\em reduction} of the critical rotation rate compared with that for $\beta=0$ (as   in Fig. \ref{h5}, left) is a double-diffusive phenomenon. This picture is confirmed by the analytic result that the travel frequency of the instability pattern for the potential flow equals the viscosity frequency $\omega_\nu$  which vanishes for ideal flows \cite{LV07, RS08}.

A similar statement holds for AMRI with quasi-Keplerian rotation as demonstrated in Section \ref{amrikep}. As expected, for $\Pm\to 0$ the eigenvalues for potential flow ($\mu_\Om=0.25$) converge in the ($\Ha/\Rey$) plane, and for the quasi-uniform flow ($\mu_\Om=0.5$) they converge in the ($\Lu/\Rm$) plane. The quasi-Keplerian flow with its shear between the two examples scales with $\Lu$ and $\Rm$, but only for insulating boundary conditions (Fig.~\ref{f26a}, left panel). For conducting boundaries, the eigenvalues behave similarly to those of the potential flow. Obviously, this particular flow forms the transition between the scaling laws for $\Pm\to 0$ of the models with steep and flat radial profiles of the angular velocity.

We know, however,  that for HMRI the scaling for small $\Pm$ switches from $\Rey\simeq$~const close to the Rayleigh line to $\Rm\simeq$~const for more flat rotation profiles (see Fig.~\ref{h55}). Solutions in the inductionless approximation, therefore,  can only exist close to the Rayleigh line. For vanishing magnetic Prandtl number  they { must} thus disappear for a critical $\mu_{\Om,0}$ somewhere beyond this line. The exact value  of this limit depends on the construction  of the model.  Details are given in Fig.~\ref{keplerholl} where for  $\Pm=0$ the isolines of Reynolds number and Hartmann number are plotted as functions of $\beta$ and the shear parameter $\mu_\Om$ for two different inner boundary conditions.  The outer boundary is always taken as insulating. As expected,  one finds for both models maximal values $\mu_{\Om,0}$ which, however, differ strongly for differing boundary conditions  \cite{RH07,P11}. There are only small differences for small $\Rey$ and at the Rayleigh line but drastic differences occur beyond this line. For insulating inner boundary rotation profiles with $\mu_\Om>0.31$ require Reynolds numbers exceeding $10^4$ to become unstable. With conducting inner boundary Reynolds numbers of $10^4$ are sufficient to destabilize flatter profiles up to $\mu_\Om\simeq 0.34$ (with $\beta\simeq 4$). The quasi-Keplerian rotation becomes unstable at $\Rey=10^5$, but only if the inner cylinder is perfectly conducting. Insulating inner cylinders stabilize the quasi-Keplerian flow unless the Reynolds number exceeds a value of $10^6$. The plots for the Hartmann numbers are very similar.

In Ref.~\cite{LG06} with $\mu_{\Om,0}= \rin^{-9.66}$ also an upper limit is given for HMRI stability, i.e.~$\mu_{\Om,0}\simeq 776$ for $\rin=0.5$. If existing, one would interpret this number as a suggestion   that for  strong superrotation (better, for stationary inner cylinder) there is another branch  for HMRI scaling with $\Rey$ and $\Ha$, which exists  in the inductionless approximation. In Section \ref{Ec} we have shown that indeed for superrotation at least with $\beta=0$  eigensolutions appear even  for $\Pm\to 0$.
\subsection{Nonaxisymmetric modes}
\begin{figure}[htb]
\centering
\includegraphics[width=9cm]{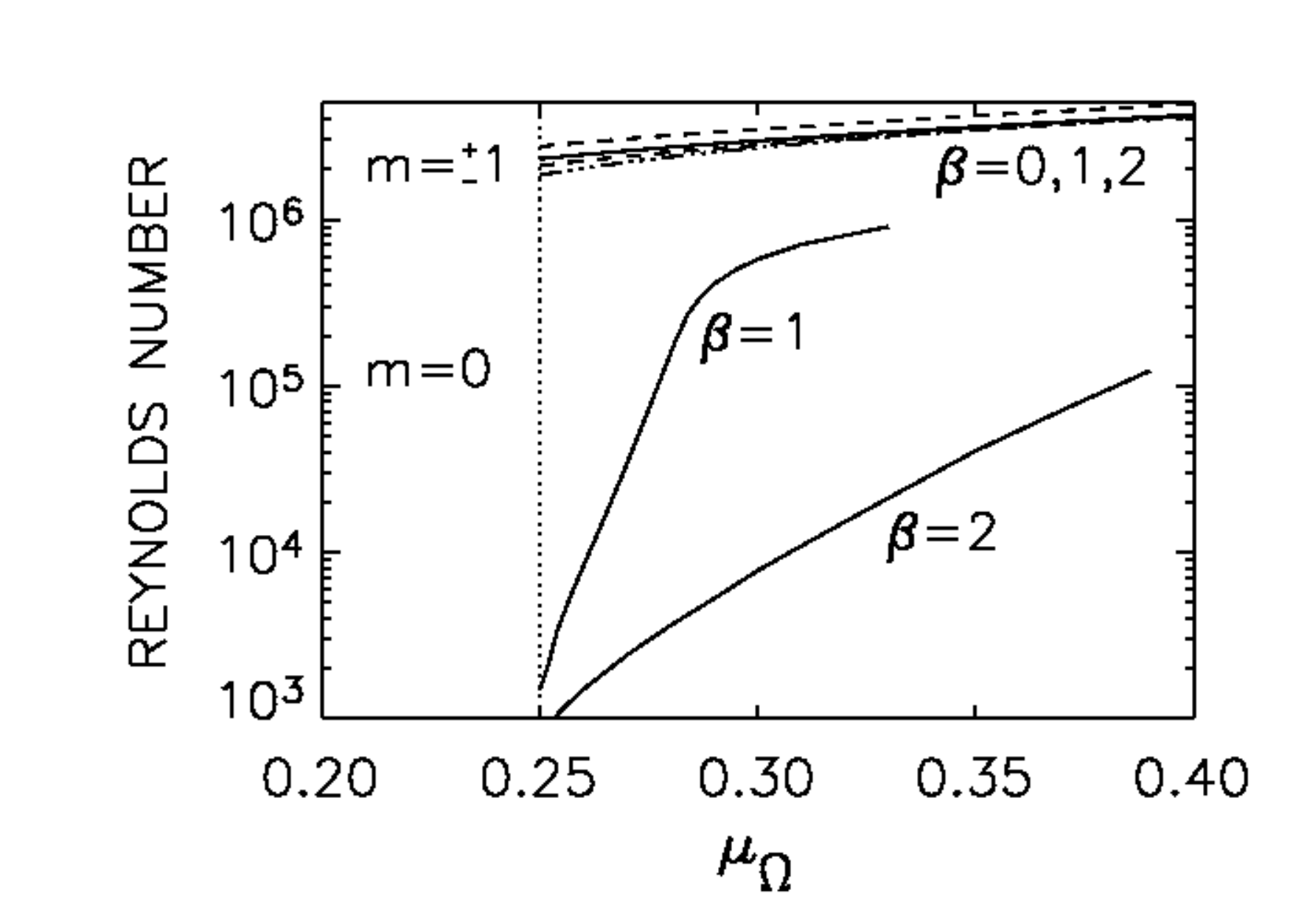}
\caption{Critical Reynolds numbers for excitation of the nonaxisymmetric modes $m= 1$ (dashed lines) in comparison to the axisymmetric modes $m=0$ (solid lines)   for $\beta=0,1,2$. The nonaxisymmetric modes always scale with $\Rm$ for $\Pm\to 0$ (as for the standard MRI) while close to the Rayleigh line the axisymmetric modes with $\beta=1,2$ scale with $\Rey$. $\mu_B=\rin=0.5$. $\Pm=10^{-5}$, perfectly conducting cylinders.}
\label{h55}
\end{figure}

Surprisingly enough, the nonaxisymmetric modes for twisted background fields behave similarly to the axisymmetric mode of MRI if the critical Reynolds and Hartmann numbers are considered as functions of $\Pm$. In Fig.~\ref{h5} for quasi-Keplerian flow the steep blue lines for $m=0$ and $\beta=0$ also represent the nonaxisymmetric mode $m=1$ and $\beta\neq 0$. For $\Pm\to 0$ they all scale with $\Lu$ and $\Rm$ (so that for small $\Pm$ very high Reynolds and Hartmann numbers are needed for excitation \cite{RS08}). This is not only true for quasi-Keplerian rotation but also for {\em all} flows including the potential flow. The behavior of the Reynolds numbers is shown in Fig.~\ref{h55} for $\mu_\Om\geq 0.25$ and $\Pm=10^{-5}$. The solid lines represent the axisymmetric solutions whereas the dashed lines denote the nonaxisymmetric modes with $m=\pm 1$. The critical Reynolds numbers of the modes with $m>0$ hardly change between $\mu_\Om=0.25$ and $\mu_\Om=0.4$; they always exceed $10^6$. There is no actual change of scalings for $\Pm\to 0$ for the nonaxisymmetric modes between the potential flow and beyond. The rather low Reynolds and Hartmann numbers shown for $\Pm\to 0$ in Fig.~\ref{h5} for $\beta\neq 0$ close to the Rayleigh line are thus a basically axisymmetric phenomenon. Nonaxisymmetric modes can hardly be observed along this way.

Figure \ref{h55} also demonstrates how for nonvanishing $\beta$ the Reynolds number for the axisymmetric mode is reduced by orders of magnitude if the rotation law becomes steeper until the Rayleigh line is reached. The potential flow with axial fields and azimuthal fields of the same order thus becomes unstable against axisymmetric perturbations already for Reynolds numbers of \ord{10^3}. This is a consequence of the fact that for azimuthal fields which are current-free in the fluid the potential flow (i.e.~$U_\phi\propto B_\phi\propto 1/R$) belongs to the Chandrasekhar-type of MHD flows and the quasi-uniform flow does not. 

\subsection{Experiment {\sc Promise}}\label{promise}
The simplest idea to realize the MRI in an experiment concerns a Rayleigh-stable flow between differentially rotating cylinders. Such a flow can be destabilized by an externally imposed magnetic field. If the imposed field is purely axial, however, the relevant parameter for the onset of the instability is the magnetic Reynolds number which must exceed about 10 \cite{RZ01,JG01}. The kinetic Reynolds number for excitation of the MRI then becomes $10^6$ or even $10^7$ because of the small magnetic Prandtl numbers of liquid metals. Such large Reynolds numbers are not only difficult to realize in experiments but also end-effects become very important \cite{HF04}.

For a combined axial and azimuthal field the relevant parameter slightly beyond the Rayleigh limit is $\Rey$, which must only be \ord{10^3} for instability (Fig.~\ref{h55}). For decreasing $\beta$ the Reynolds number gradually rises until for $\beta=0$ the necessary Reynolds number is \ord{10^7}, known for MRI with $\Pm\simeq 10^{-6}$. The main difference of the solutions to those for purely axial imposed fields is that the HMRI pattern drifts along the rotation axis of the cylinders. In both cases the modes with the lowest Reynolds numbers are axisymmetric. Provided $B_0>60$ G and $\beta\simeq 3$, Reynolds numbers of only $10^3$ are sufficient to excite the instability waves for conducting cylinders. The threshold numbers for insulating boundaries are higher.
 \begin{figure}[htb]
\centering
 \includegraphics[width=1.6cm]{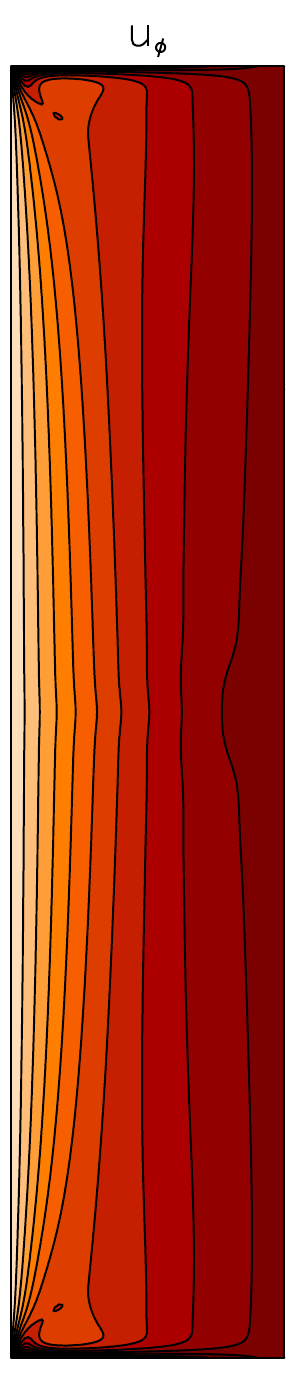}
 \includegraphics[width=1.6cm]{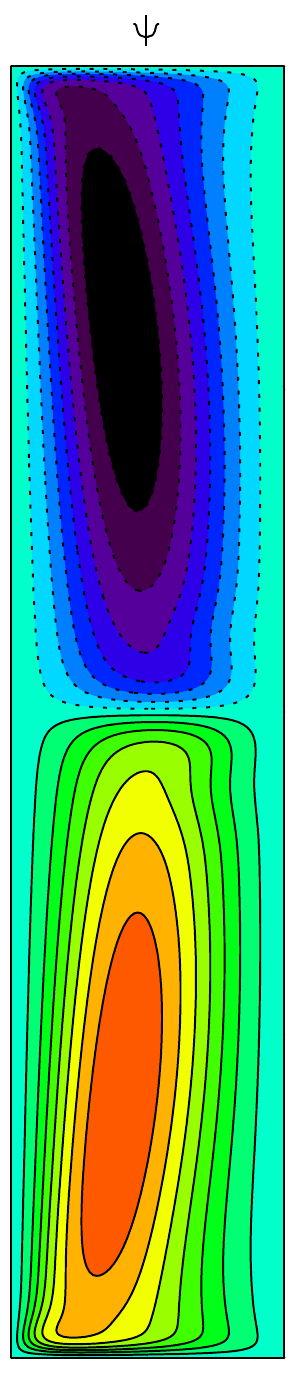}
 \hskip3cm
 \includegraphics[width=1.6cm]{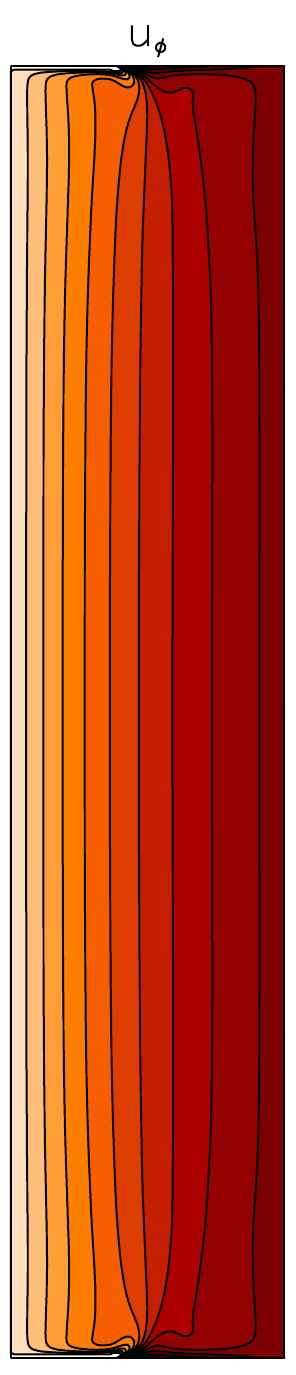}
 \includegraphics[width=1.6cm]{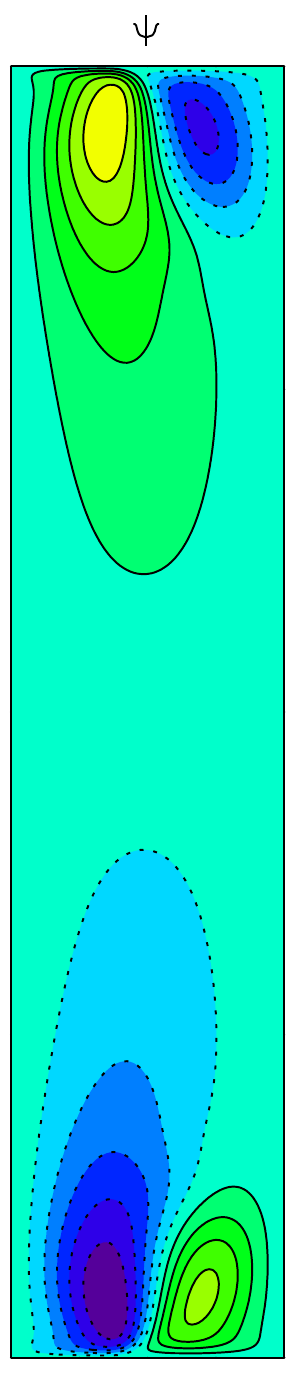}
 \hskip3cm
 \includegraphics[width=1.6cm]{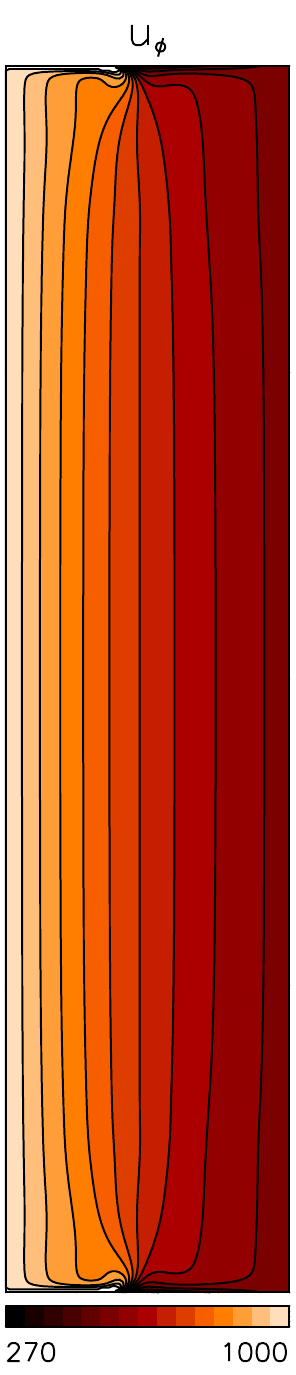}
 \includegraphics[width=1.6cm]{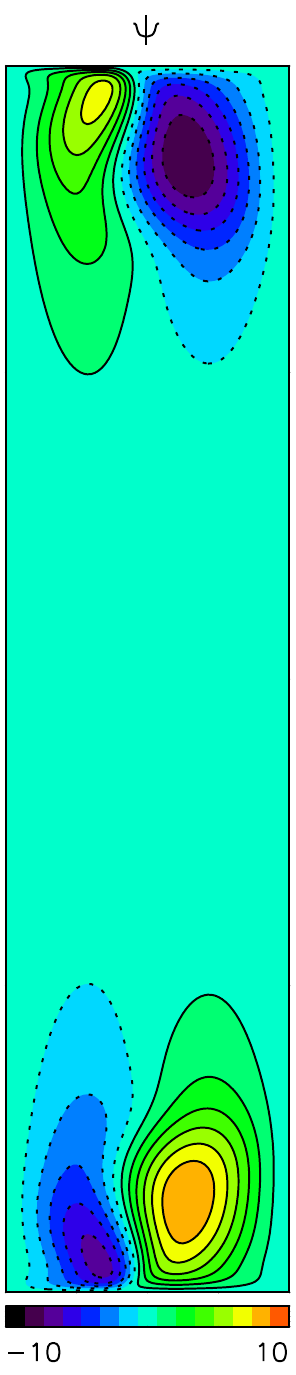}
\caption{Simulated snapshots representing the flow pattern  for rigidly rotating endplates (left) and for endplates formed by two rings attached to the cylinders (middle, right). The isolines of the azimuthal flow and the contourlines of the streamfunctions of the flow are shown. Axial magnetic fields ($\Ha=10$) are  applied only in the simulations for insulating endplates shown by the right panel. 
$\Rey=1000$, $\Pm=0$, $\beta=0$, $\mu_\Om=0.27$. From \cite{S07}.}
 \label{jacek1} 
 \end{figure}

However, the existence of the viscous endplates results in Ekman layers in which the velocity differs from the prescribed rotation law. A global meridional circulation with two Ekman vortices is the immediate consequence. As known for nonrotating endplates a radial inflow close to the boundaries appears and for solid-body rotation a radial outflow appears \cite{SR07}. Figure \ref{jacek1} demonstrates the Ekman layer phenomenon for different sorts of endplates. While for the left panel the rigid endplates rotate with the angular velocity of the outer cylinder the endplates in the middle panel are split at $R_{\rm split}=\Rin+0.4 d$; the inner part is attached to the inner cylinder and the outer part is attached to the outer cylinder. Meridional planes are presented for the variables $U_\phi$ and the streamlines of the meridional flow. We notice for rigid endplates that the mean flow $U_\phi$ in this case significantly depends on $z$ and that two strong Ekman vortices fill the whole container. If the plates are replaced by two rings then $U_\phi$ is almost independent of $z$ in the bulk of the container and the Ekman circulation is strongly suppressed. If the axial magnetic field is applied (for the plots of the right panel with the two rings attached to the two cylinders) it  looks even better as the Ekman vortices are further reduced. The rotation profiles are almost unchanged when compared to the hydrodynamic case. For this result insulating endplates must be used as for perfectly conducting lids the Ekman-Hartmann layer produces basically stronger modifications of the rotation law.

The experiments were done at the {\sc Promise} facility as described in Section \ref{expamri}. This time, however, the coil for the production of an axial field was also used. Since this coil is not cooled, the current in the windings is restricted to values of around 150 A, which corresponds to a Hartmann number of 23.7. The endplates are made of plexiglass which are split into two rings where the inner one is attached to the inner cylinder and the outer one to the outer cylinder. Based on numerical simulations the splitting position is at $R_{\rm split}=56$ mm, minimizing the Ekman pumping of rigid endplates \cite{S07,SG07,SG08}. See also \cite{GGJ2012,Wei2016} for similar calculations
designed to minimize end-effects in the Princeton experiment.
 \begin{figure}[htb]
\centering
 \includegraphics[width=15cm]{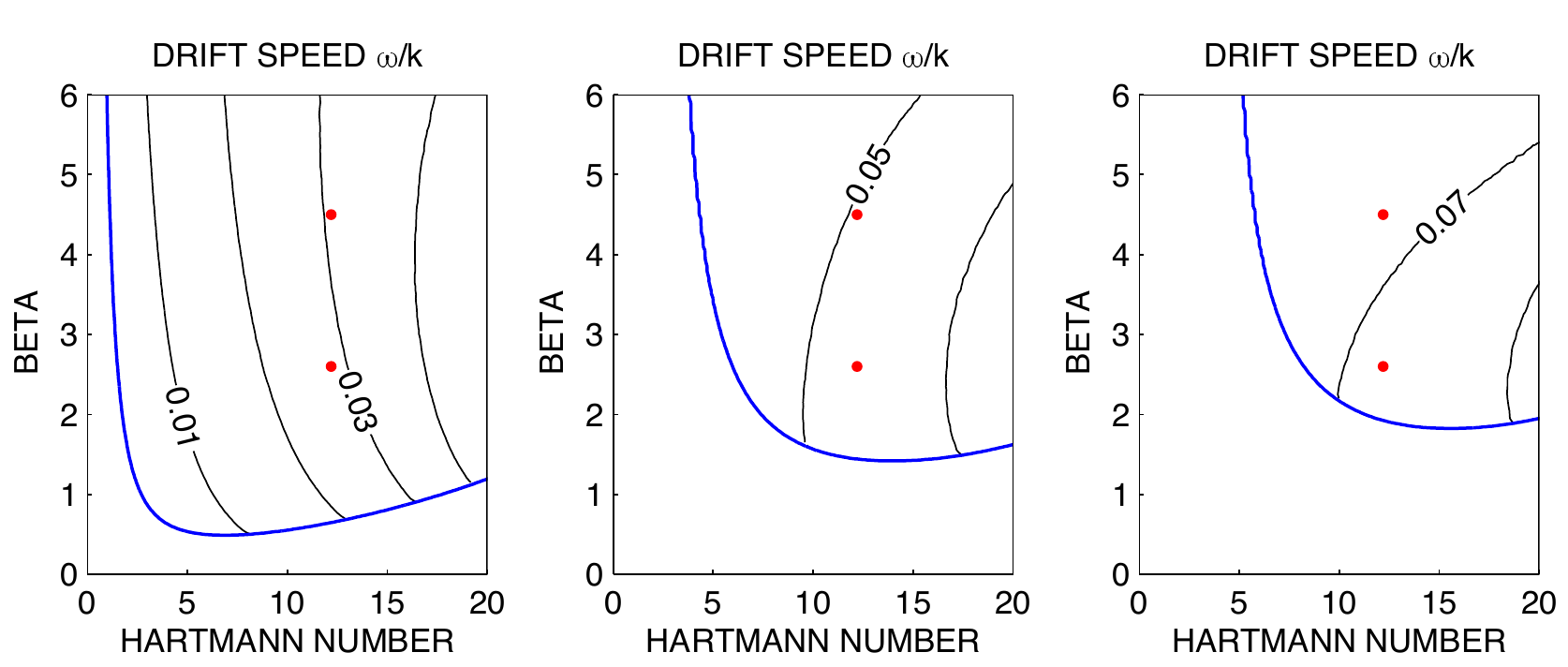}
 \caption{As in Fig.~\ref{prom11a} but for a  fixed Reynolds number of $\Rey=2959$.  $\mu_\Om=0.25$ (left), $\mu_\Om=0.26$ (middle) and $\mu_\Om=0.27$ (right). The solutions are of neutral stability only along the blue lines. The red symbols mark the locations of the values $\Ha$ and $\beta$ used in  real experiments measuring  the travel speeds. Perfectly conducting cylinders.}
 \label{prom5} 
\end{figure}

The azimuthal magnetic field is imposed by a current up to 7 kA through a water-cooled rod along the central axis. The field within the fluid is current-free. The fluid within the vessel is the GaInSn alloy with the material parameters given in Table \ref{t2}. For experiments with this apparatus as an improved version of {\sc Promise}~1 (which worked with rigidly rotating endplates, see \cite{SG06,RH06}) detailed predictions are possible. The main target for the experiments are measurements of the vertical travel velocity $u_z=\omega/k$ by two ultrasonic high-focus transducers mounted on opposite sides of the top endplate.

For the marginal instability with the wave number $k$ leading to the lowest Reynolds number for given $\Ha$ and $\beta$, the resulting normalized drift rates $\omdr$ and axial phase speeds $\omdr/k$ are given in Fig.~\ref{prom11a} for the shear values $\mu_\Om=0.25-0.27$. They are normalized with $\Omin$ and $R_0\Omin$, respectively. The maximum Reynolds number for the calculations is 4000 (the blue lines in the plots). The minima of the lines define the necessary minimal $\beta$ values with which an instability appears. This minimum $\beta$ becomes smaller for greater Reynolds numbers. The limit $\beta\to 0$ would require $\Rey= 55,780$ for $ \mu_\Om=0.25$ (and for galinstan as the fluid conductor) together with $\omdr=0$. The corresponding critical Hartmann number is simply 7, independent of the magnetic Prandtl number (see Section \ref{carlos}). There is obviously a smooth transition from HMRI to standard MRI by this constellation. Along the low-field branches of the blue lines the travel frequency $\omdr$ hardly varies (top row in Fig.~\ref{prom5}). This is in particular true for $\mu_\Om=0.25$, and means that in this case $\Rey\cdot \omdr\simeq$~const, so that it is shown that in this case the drift frequency is determined by the viscosity frequency independently of $\beta$ and $\Rey$. Beyond the Rayleigh line the relations are more complicated.

\begin{figure}[htb]
\centering
 \includegraphics[width=9cm]{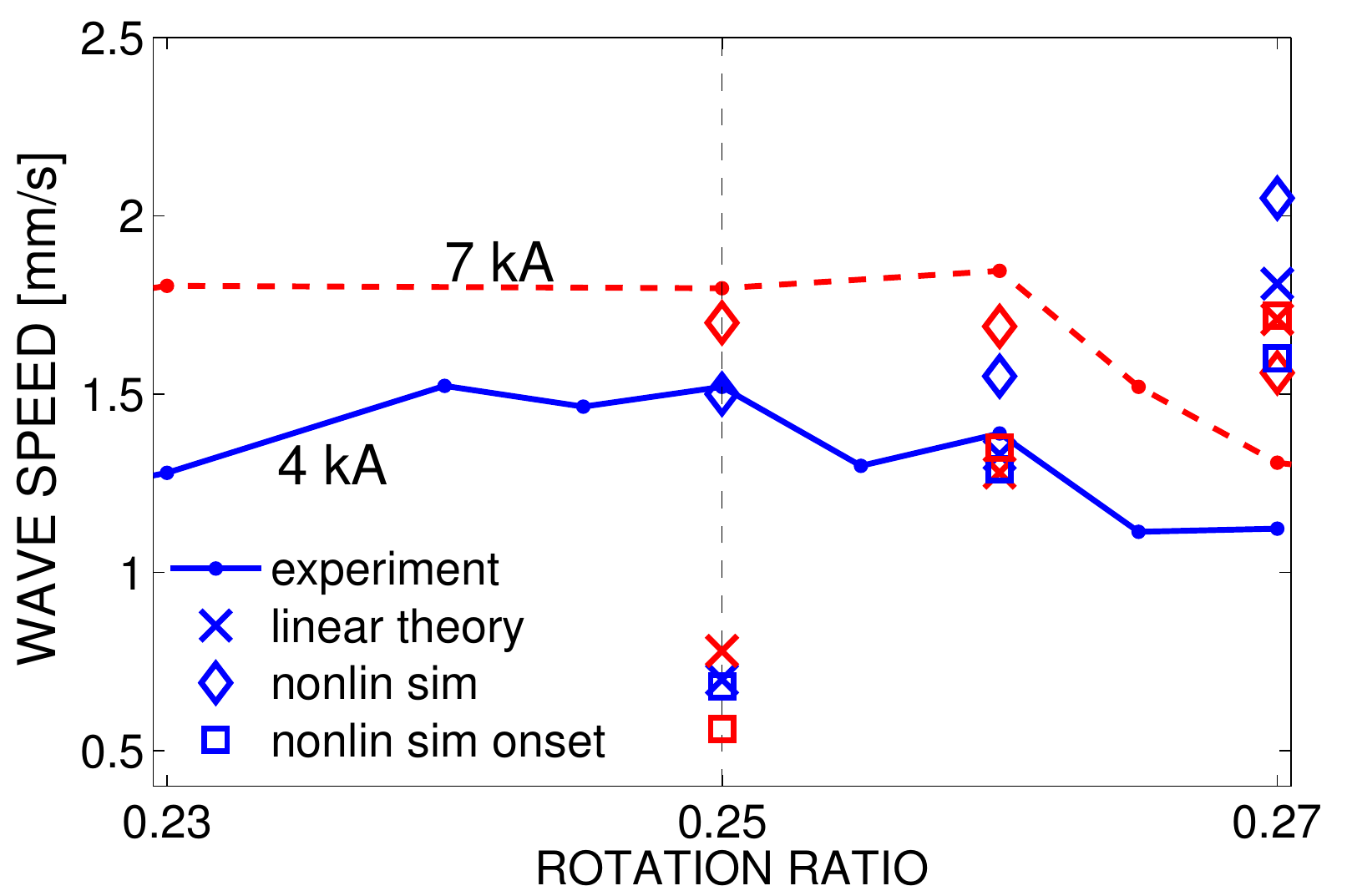}
 \caption{Axial travel speed measured with {\sc Promise} for $\beta=2.6$ (lower line) and $\beta=4.5$ (upper line) as a function of the shear parameter $\mu_\Om$. The values for the crosses are taken from  Fig.~\ref{prom5} while the diamonds are results of nonlinear simulations with $\Pm=10^{-6}$, which also yield the values marked as squares during the onset phase of the instability. The vertical line indicates the Rayleigh limit. The linear speed of the inner cylinder is $u_{\rm in}=2.51$~cm/s.  $\Ha=12$, $\Rey=2959$. Perfectly conducting walls, no endplates. See \cite{SG09}.}
 \label{prom4}
\end{figure}

In Fig.~\ref{prom5}  the linear calculations for  axially unbounded cylinders     shown in Fig.~\ref{prom11a} are repeated but with a fixed and supercritical  Reynolds number ($\Rey=2959$) prescribed for all curves. Again  the red symbols   mark the parameters characterizing the main experiments with {\sc Promise}~2. The  Reynolds number and the  axial magnetic field have  been fixed to $B_0=77.2$~G ($\Ha=12.2$). The azimuthal field is then varied by the application of an axial electric current with 4~kA ($\beta=2.6$) or 7~kA ($\beta=4.5$). The empirical results for the rotation ratios $\mu_\Om=0.23-0.27$ have been described in detail in Ref.~\cite{SG09}. For the Rayleigh limit $\mu_\Om=0.25$ the measured travel speed for the two $\beta$ values varies between 1.5 mm/s and 1.8 mm/s, increasing slightly with $\beta$ (Fig.~\ref{prom4}). With the nonlinear code described in Section \ref{nonsim} (insulating cylinders, no endplates) these measurements can be reproduced exactly (the diamonds in Fig.~\ref{prom4}). The linear approximation with fixed Reynolds number, however, yields values that are too small (the red and blue crosses). Figure \ref{prom5} shows  the axial phase speed $\omega/k$ for the Rayleigh limit and $\Pm=0$. For $\Ha=12.2$ one finds $\omega/k=0.03$ for both $\beta$ values, hence $u_z\simeq 0.75$~mm/s.  Interestingly enough, if the velocities are measured with the code during the linear onset of the instability (the squares in Fig.~\ref{prom4}) then the results perfectly match the data of the linear theory taken from Fig.~\ref{prom5}. 

For $\mu_\Om=0.26$ the agreements are even better; now also the two linear results are close to the measurements. For $\mu_\Om>0.26$, however, all theoretical phase speeds increase for reduced shear while the experimental values decrease. Note that the given theoretical results were obtained for an unbounded container. The wave numbers given in the right panel of Fig.~\ref{prom1} (for the easiest excitation) represent wavelengths of about 15 cm, but the real container has a height of only 40 cm.

\begin{figure}[hbt]
\centering
\includegraphics[width=16cm]{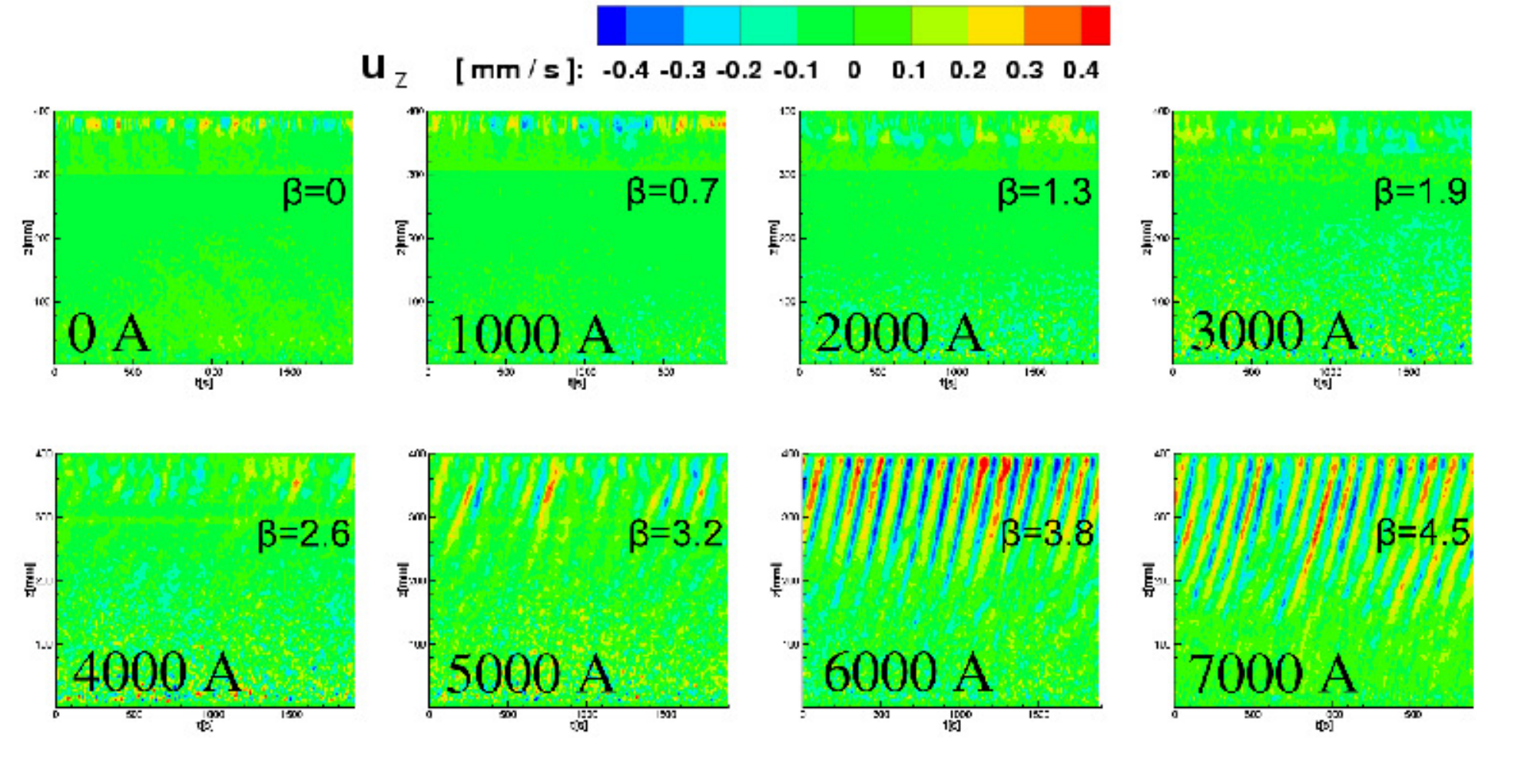}
\caption{Upward traveling HMRI wave of   {\sc Promise} when varying the axial current $I_{\rm rod}$ (opposite  provides downward waves). The height in the cylinders and the time are measured along the axes.  Measured UDV signals in dependence on time and vertical position, for 8 different axial current values. The phase velocity for the experiment with 7 kA is 0.9 mm/s. The different structure of  the endplates (see text) leads to asymmetric wave trains in the axial direction.  $I_{\rm coil}=76$\,A ($\Ha=12$),  $\Rey=1775$, $\rin=0.5,$ $\mu_\Om=0.26$. Conducting cylinders.}
\label{fig:betalinie1}
\end{figure}
\begin{figure}[hbt]
\centering
\includegraphics[width=9cm]{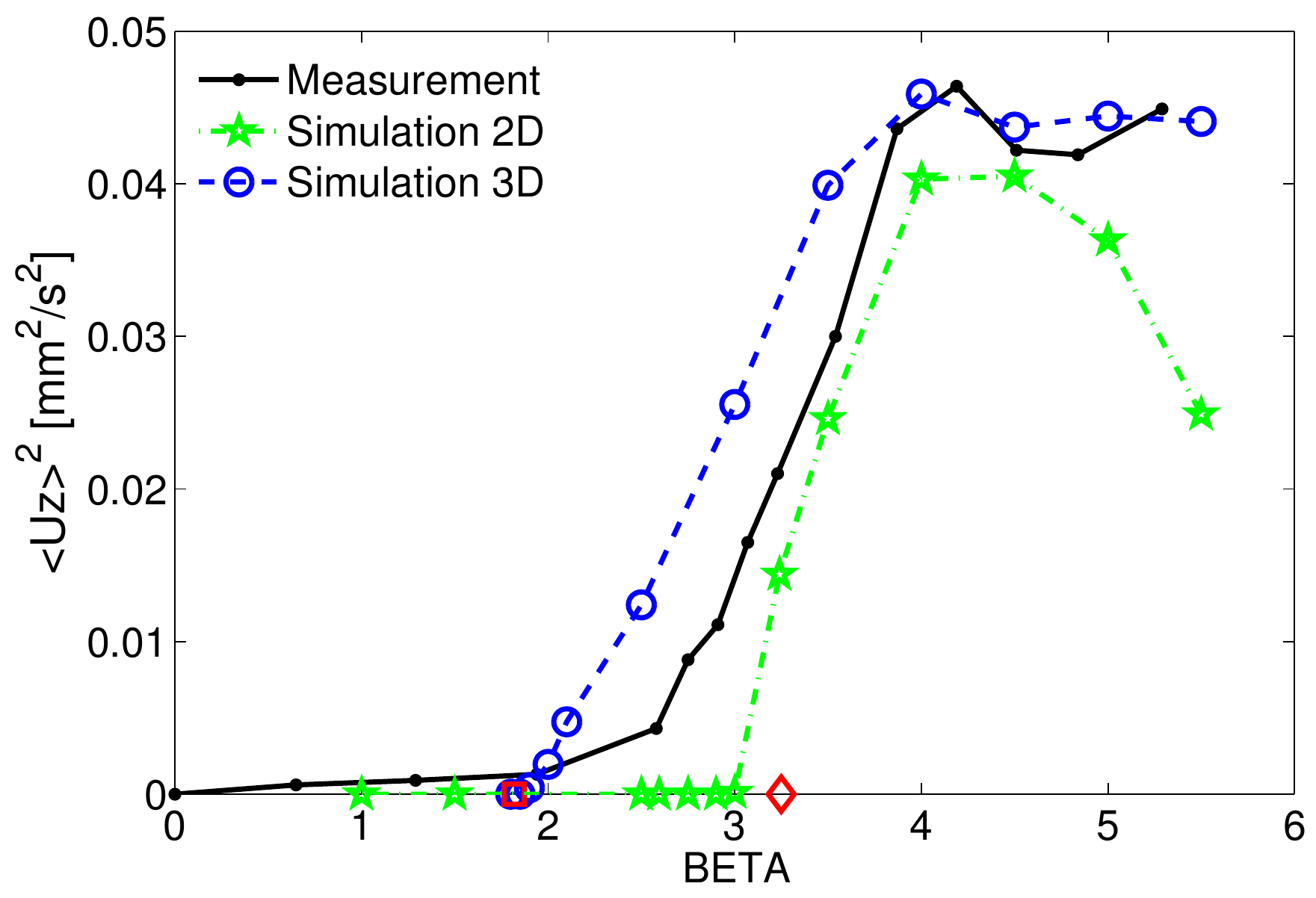}
\caption{The axial  energy $\langle u_z^2\rangle$ at the UDV sensor position in dependence on the normalized toroidal field $\beta$. Same parameters as in Fig.~\ref{fig:betalinie1}. Dashed lines show  numerical results, the full line gives the experimental data. The  2D simulation  for $\Pm=0$ (from  \cite{SG09}) concerned a container with split endcaps  while the 3D simulation for $\Pm=10^{-5}$ used  axial periodicity with $\Gamma=10$. The red marks at the horizontal axis indicate the predictions for the onset of the convective instability (circle) and the absolute instability (diamond). See the remark at the end of Section \ref{amrihmri}.}
\label{promresults} 
\end{figure}
Figure~\ref{fig:betalinie1} illustrates the observed variation of $\beta$ in further detail. Fixing $\Omin=0.38$\,s$^{-1}$, $\mu_\Om=0.26$, $I_{\rm coil}=76$\,A, the axial current is varied between 0 and 7\,kA so that the maximal magnetic field at the inner cylinder  is 350~G. We observe the HMRI wave only above 4\,kA or equivalently, for $\beta\geq 1.9$. The endplates are rather different, resulting in axial differences of the wave trains which become weaker for increasing $\beta$. The upper endplate is insulating and stationary while the lower endplate is conducting and rotates with the outer cylinder (see the right panel of Fig.~\ref{jatzek}).  A comparison of the experimental results with numerical predictions is shown in Fig.~\ref{promresults}. The 3D  simulations have been done  without  endplate effects.  The values are averaged over the whole container including the near-wall domains where the vertical flow in the cells is larger than in the middle of the gap between the cylinders. The results obtained with an inductionless axisymmetric 2D  code for $\Pm=0$  concern the central part between the endplates \cite{SG09}.  The two very different codes provide very similar results for the maximum intensities also in agreement with the measurements. The critical $\beta$ values, however, vary between the   red symbols at the horizontal axis indicating the numerically determined threshold values  for the convective and the absolute instability \cite{PG09}. The experimental data well fit the numerical approaches. The 2D simulations for the bounded container reflect the onset of the {\em absolute} instability while for the unbounded container the onset of the {\em convective} instability is simulated.

By definition of the averaging procedure, the systematic phase velocity $(\omdr/k)$ of the waves is not reproduced in the plots. The instability starts with rather small intensities at the threshold value $\beta_0=1.9$ which is known from the linear theory (see Fig.~\ref{prom1}, left). For slightly larger values the intensity grows like the difference $\beta-\beta_0$. Much stronger saturated intensities are reached for $\beta$ values larger than the theoretical value for the absolute instability. The values are averaged over the axial coordinate $z$, explaining that they are smaller than the amplitude values given in Fig.~\ref{prom4}. The typical value of the axial rms velocity for larger $\beta$ is 0.2 mm/s.

\section{Tayler instability (TI)}\label{TI}
\begin{figure}[htb]
\centering
 \includegraphics[width=13.0cm]{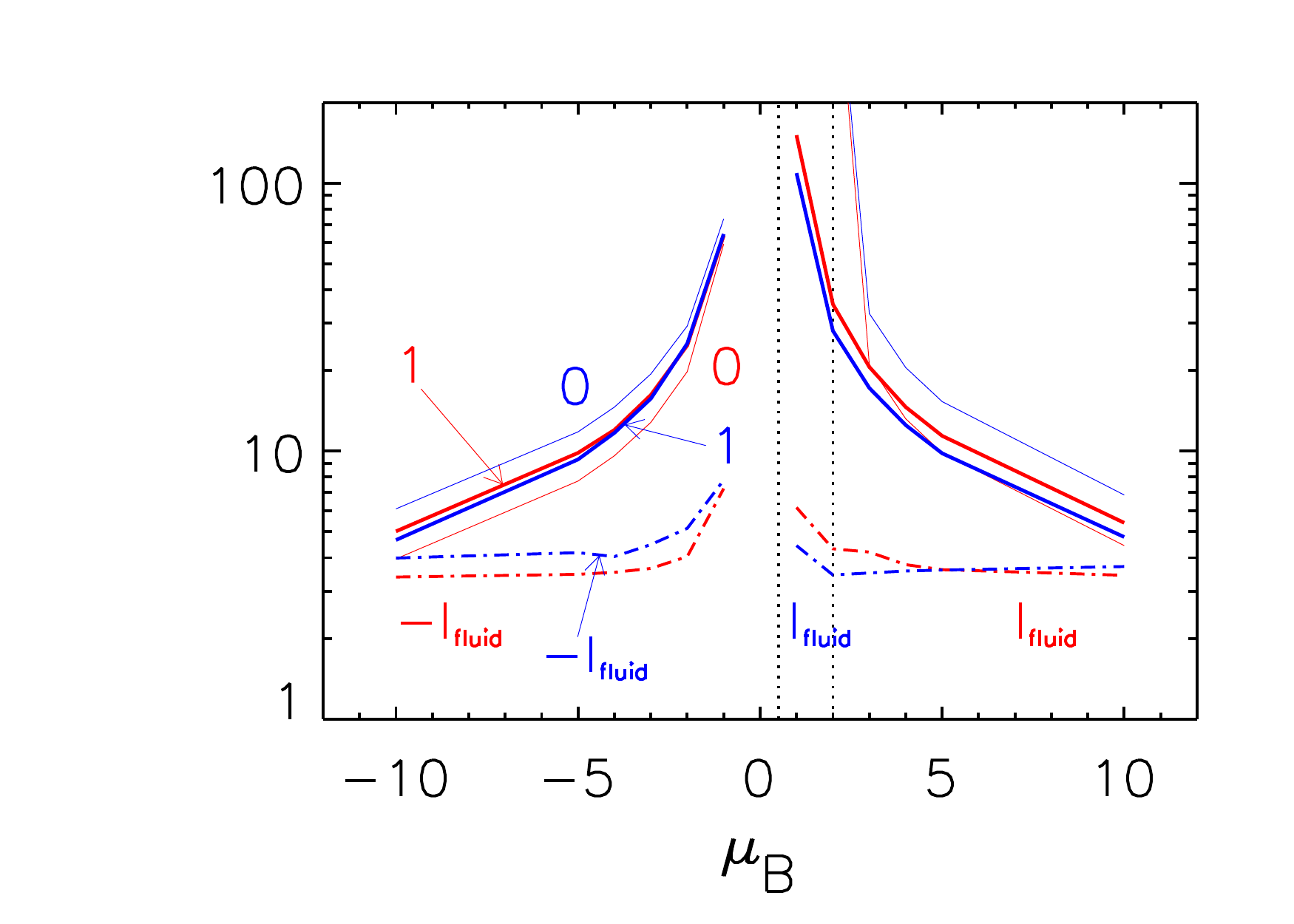}
 \caption{Critical Hartmann numbers $\Ha_0^{(m)}$ for $m=0$ (thin lines) and $m=1$ (thick lines) for perfectly conducting (red) and insulating (blue) boundary conditions for azimuthal fields with various $\mu_B$. The profiles in the central  area around $\mu_B=\rin=0.5$ (left dotted line) are stable against axisymmetric and nonaxisymmetric perturbations.  The right dotted line at $\mu_B=1/\rin=2$
 represents the $z$-pinch  which is stable against axisymmetric perturbations but unstable against nonaxisymmetric perturbations with $m=1$.
 The related electric currents (in kA) are calculated from (\ref{Iin}) for liquid sodium.  Hartmann numbers $\Ha_0$ do not depend $\Pm$.   Data from \cite{RS07}.}
 \label{ti2}
\end{figure}

Almost all applications in the foregoing sections concern toroidal magnetic fields which are current-free in the fluid between the cylinders. In these cases the  instability cannot exist without global differential rotation. In Sections \ref{keplergalactic} and \ref{pinch}, however, models with $\Rey=0$ also proved to be unstable against nonaxisymmetric perturbations for azimuthal fields with $\mu_B> \rin$. Hence, $a_B\neq 0$ in Eq.~(\ref{abB}), so that axial electric currents exist in the fluid. It is also interesting to combine the stability criteria (\ref{bf}) for axisymmetric modes and (\ref{tay}) for nonaxisymmetric modes. As illustrated in Fig.~\ref{ti2} the solution $B_\phi\propto 1/R$ (i.e.~$\mu_B=0.5$) is always stable while the profiles $ B_\phi\simeq$~const  and $B_\phi\propto R$  (i.e.~$\mu_B=1$ and $\mu_B=2$) are  unstable against nonaxisymmetric perturbations. That the $z$-pinch  with uniform electric current between the cylinders  is always stable against $m=0$ follows from the simplified Eq.~(\ref{bphi}), i.e.
\beg
\frac{ {\rm d}^2 b_\phi}{{\rm d} R^2}
+\frac{1}{R}\frac{{\rm d} b_\phi}{{\rm d} R}
-\frac{b_\phi}{R^2} -k^2 b_\phi
- {\textrm{i Pm Re}}\ \omega b_\phi
-R\frac{{\rm d}}{{\rm d} R}\left(\frac{B_\phi}{R}\right) u_R=0,
\label{bphi2}
\ende
which for $B_\phi\propto R$ fully decouples from the hydrodynamics as it also does $b_\phi$ in accordance with (\ref{bR}). All magnetic perturbations, therefore, decay because of missing energy sources.   The last term in (\ref{bphi2}) is only able to destabilize fields with radial profiles {\em steeper} than  $B_\phi\propto R$   against axisymmetric perturbations.

It may be worth to consider the occurrences of instability against axisymmetric perturbations with $m=0$ shown by this line in Fig.~\ref{ti2}. It is no surprise that  in accordance with the condition (\ref{bf}) axisymmetric instabilities exist for $\mu_B>2$ (right vertical dotted line). More interesting is the existence  of an axisymmetric instability   for toroidal fields which change the sign between the boundaries, i.e.~$\mu_B<0$. Note that the electric current $I_{\rm fluid}$ given by
 Eq.~(\ref{Iin}) changes its sign at $\mu_B=\rin$ (left vertical dotted line). Because of 
\beg
\mu_B=\rin \big(1+\frac{I_{\rm fluid }}{I_{\rm axis}}\big)
\label{negmu}
\ende
negative $  \mu_B$ result for electric currents with opposite signs,  $I_{\rm fluid}<-I_{\rm axis}$. Figure \ref{ti2} demonstrates that such fields are unstable against $m=0$ perturbations for $\mu_B<-1$. The reason is that the lefthand side of the relation  (\ref{bf}) changes in sign leading to instability against axisymmetric perturbations which even (for perfectly conducting cylinders) can possess the lowest eigenvalue.

For real fluids in the presence of azimuthal fields the equation system is given in Section \ref{azfi} with the definition  (\ref{Hartmannin}). Without rotation for any value of $\mu_B$ and for a given mode number $m$, the resulting eigenvalue for neutral stability is the Hartmann number $\Ha_0$. One can easily show that for $\Rey=0$ the drift value $\omdr$ vanishes, and $\Ha_0$ does not depend on the magnetic Prandtl number \cite{S06,RS10}.  The critical Hartmann numbers for the excitation of the axisymmetric mode ($m=0$) and the nonaxisymmetric modes with $m=\pm1$ for $-10 \leq \mu_B\leq 10$ are given for $\rin=0.5$ in Fig.~\ref{ti2}. Of particular importance here are the values for $\mu_B=1$ and $\mu_B=2$, describing (approximately) uniform fields and uniform electric currents, respectively. For $\mu_B=1$ the critical Hartmann numbers for excitation of the $m=1$ mode are $\Ha_0=151$ and $109$ for conducting and insulating cylinders, respectively; for $\mu_B=2$ the values are $\Ha_0=35$ and $28$. Uniform currents lead to easier excitations.

Figure \ref{ti2} also reveals the nontrivial influence of the boundary conditions. For perfectly conducting cylinders and negative $\mu_B$ the axisymmetric instability will be excited with the lowest Hartmann number (as in  \cite{G62}) while for insulating cylinders the mode with $m=1$ is the preferred one. For positive $\mu_B$ and insulating cylinders the nonaxisymmetric $m=1$ mode will be excited with the lowest Hartmann number while for perfectly conducting cylinders the numerical value of  $\mu_B$  determines the fundamental mode.

\begin{figure}[htb]
\centering
 \includegraphics[width=8cm]{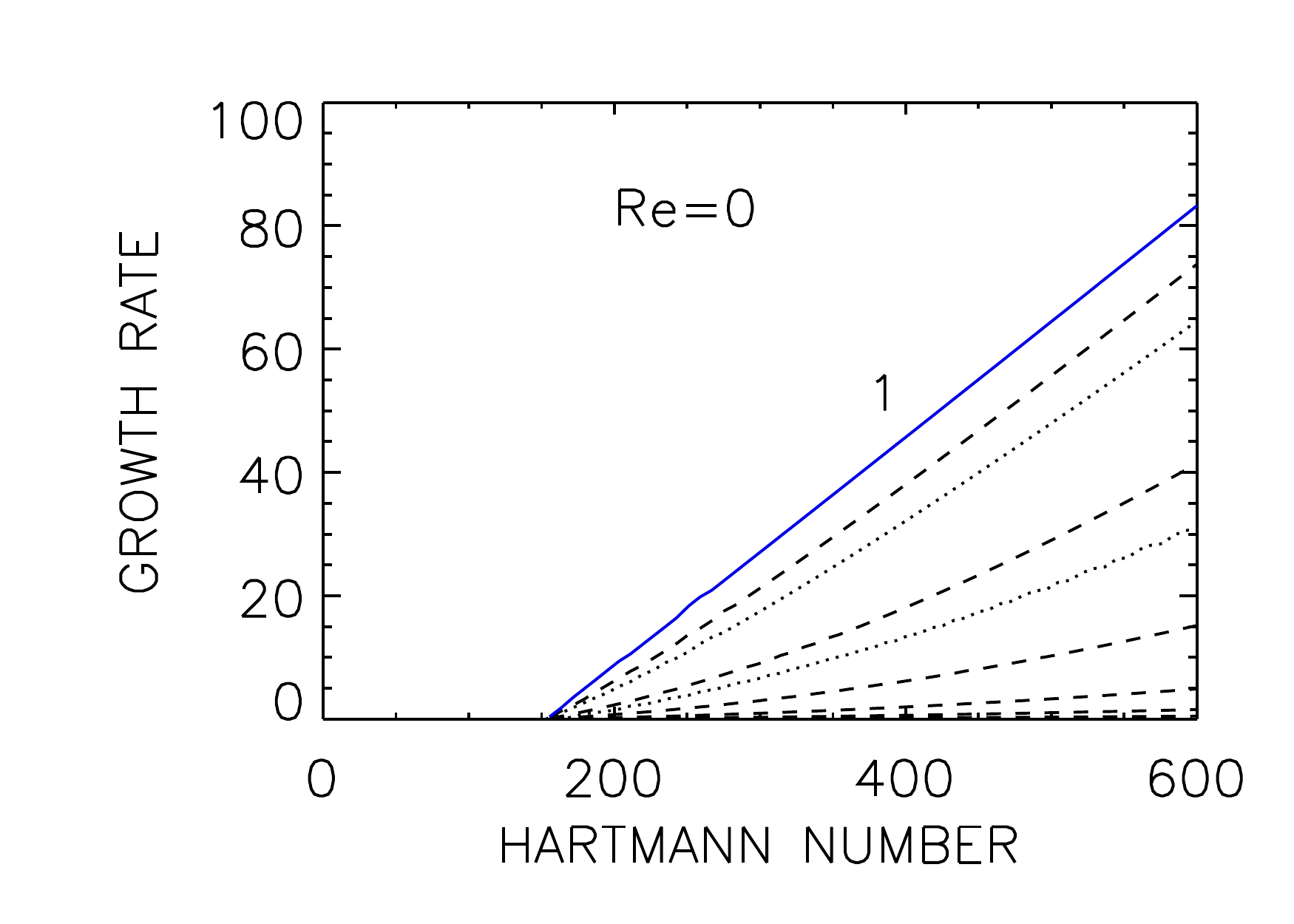}
 \includegraphics[width=8cm]{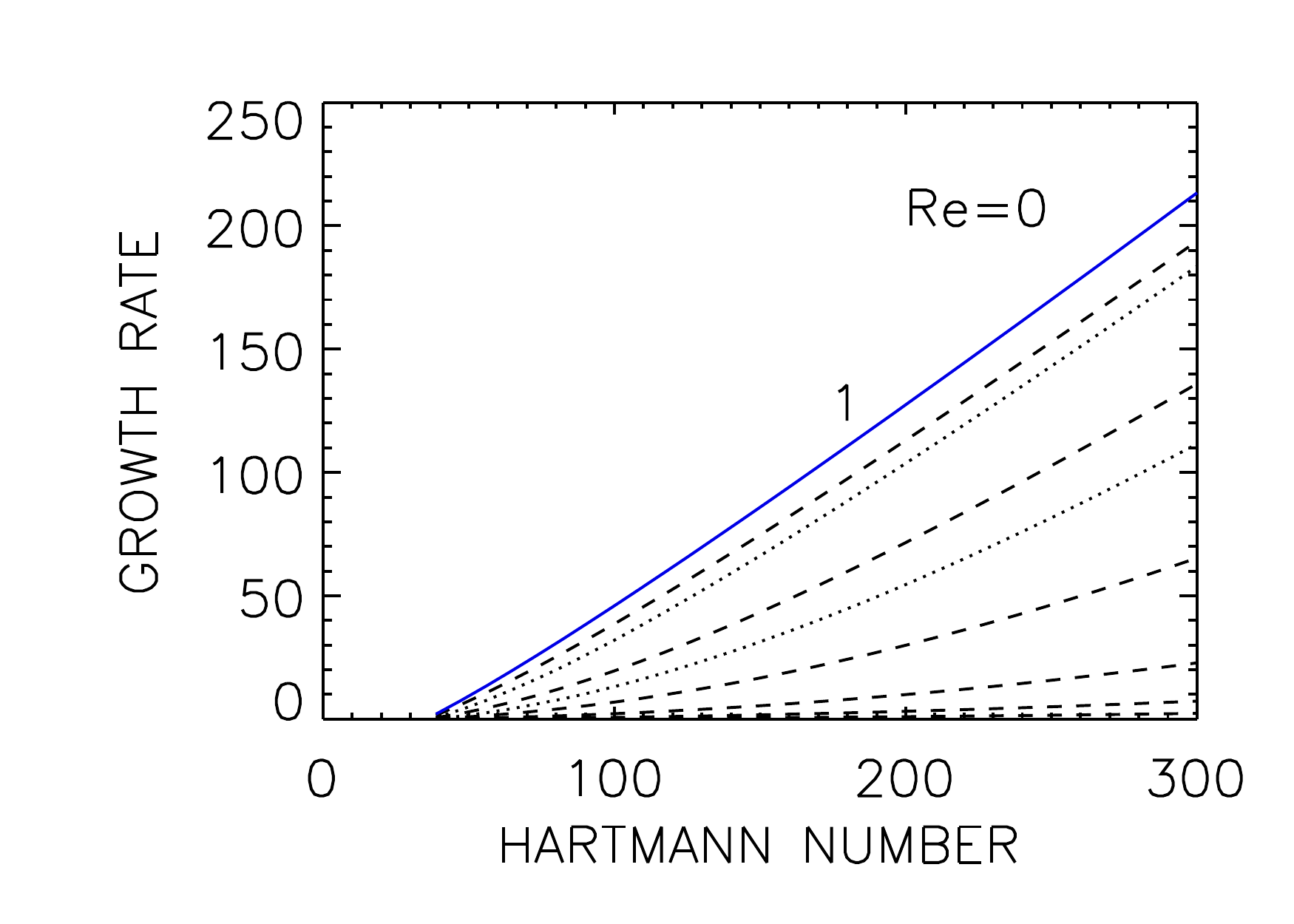}
 \caption{Growth rates of stationary flows normalized with the dissipation frequency  $\overline{\omega}=\sqrt{\omega_\nu \omega_\eta}$ versus supercritical Hartmann numbers (\ref{Hartmannin}) for $\mu_B=1$ (left) and $\mu_B=2$ (right). $\Pm=1$ (blue lines), $\Pm=0.1, 0.01, ...$ (dashed lines) and $\Pm=10, 100, ...$ (dotted lines). The critical values  $H_0$ do  not depend on $\Pm$. The fastest growth belongs to $\Pm=1$. $m=1$, $\Rey=0$,  $\rin=0.5$, perfectly conducting boundaries.}
 \label{ti4}
\end{figure}

For Hartmann numbers exceeding $\Ha_0$ the equation system yields finite growth rates. It is known  that the growth rate of TI grows for growing magnetic fields. The open question is the influence of the magnetic Prandtl number. Figure~\ref{ti4} shows the growth rates for a purely toroidal field with $\rin=0.5$ and $\mu_B=1$ (almost uniform magnetic field) and $\mu_B=2$ (uniform electric current) for various $\Pm$. In this representation they scale almost linearly\footnote{The below discussion of a wide gap flow reveals  a quadratic behavior.} with the Hartmann number, with  a weak dependence on $\Pm$.  Due to the normalization of the growth rates with the averaged frequency $\overline{\omega}=\mdiffquer/R_0^2$, one obtains 
\beg
\omega_{\rm gr} = F(\Pm)\ \Om_{\rm A},
\label{ogr}
\ende
with the \A\ frequency $\Om_{\rm A}= \omquer\ \Ha$ and $F$ as a function of the magnetic Prandtl number. The amplitudes  for $\mu_B=1$ (uniform field) are $F(1)=0.1$ and for $\mu_B=2$ (uniform current) it is $F(1)=1$. In this representation the fastest instability belongs to $\Pm=1$. Such fluids are thus more unstable than those with $\rm Pm\neq 1$. The function $F(\Pm)$ becomes rather small for small and large $\Pm$.
\begin{figure}[h]
\centering
 \includegraphics[width=5cm]{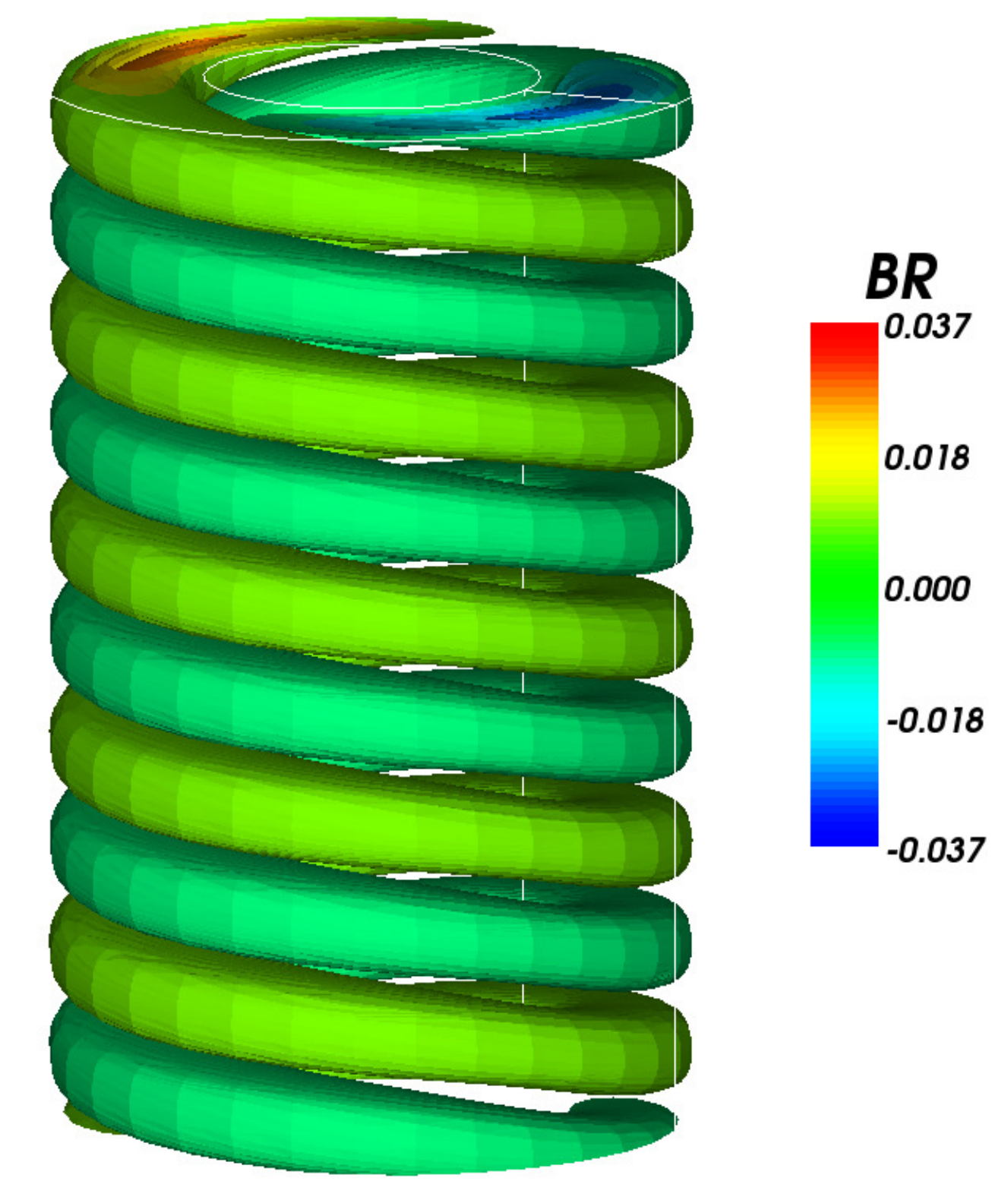}
 \includegraphics[width=5cm]{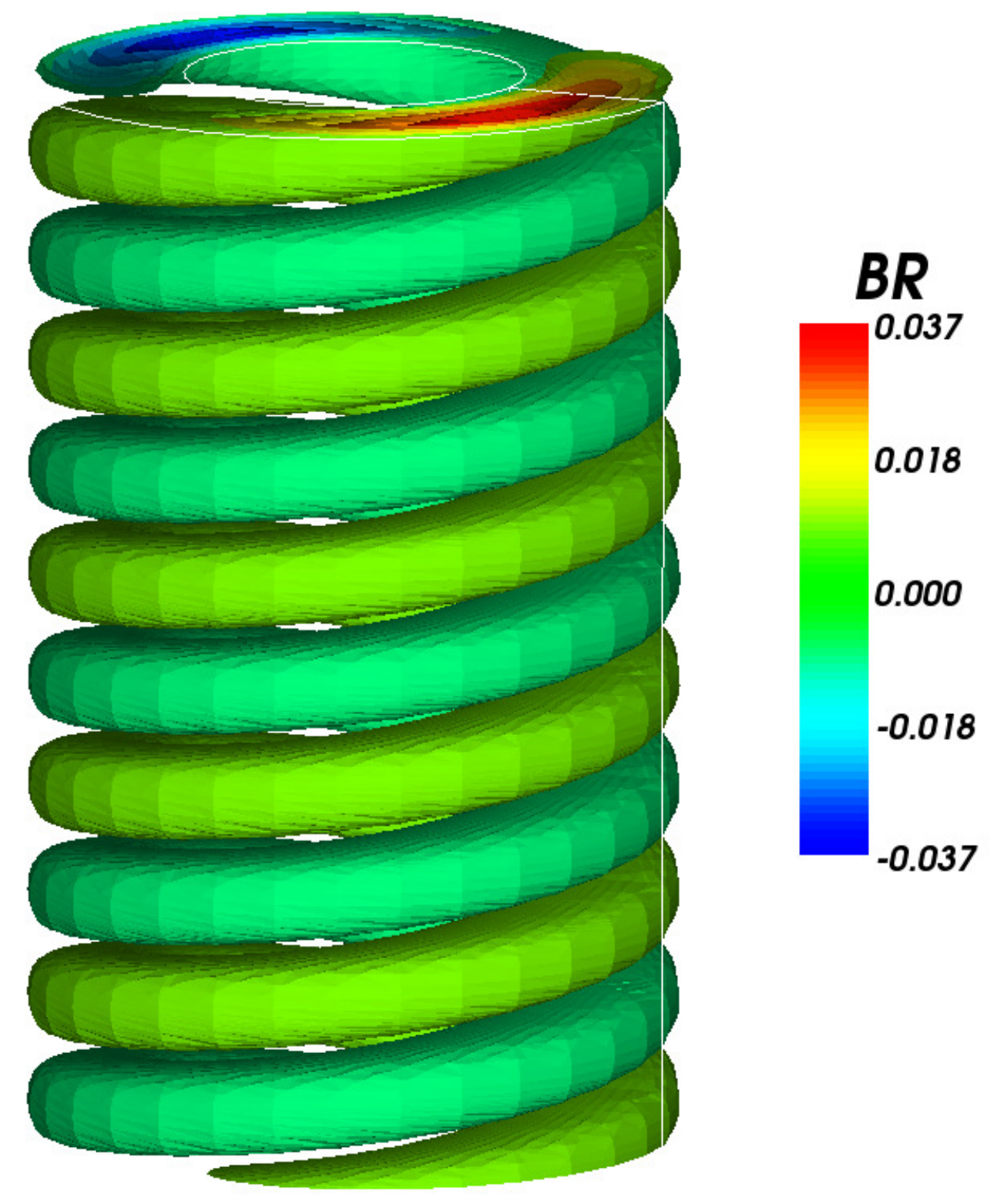}
 \caption{Instability patterns of a purely toroidal quasi-uniform background field  without rotation. The two modes are equivalent: their kinetic helicities are $\pm6.0\cdot 10^{-4}$ and their current helicities are $\pm3.5\cdot 10^{-3}$ (both in units of $\Om^2_{\rm A} R_0$). $\rin=0.5$, $\mu_B=1$, $\Ha=200$, $\Pm=1$, perfectly conducting boundaries. From \cite{GR11}.}
 \label{ti3}
\end{figure}
For a purely toroidal field the azimuthal wave numbers of the modes in Fig.~\ref{ti3} are $m=\pm1$, where the left spiral has $m=1$ and the right spiral $m=-1$. The left-handed and right-handed spirals are degenerate, having exactly the same growth rate. These modes do not drift in the azimuthal direction. Figure \ref{ti3} shows that the nonlinear solutions do not consist of equal mixtures of both modes. Instead, either the left or the right mode suppresses the other. Which mode wins depends on the initial conditions. If the initial condition allows the excitation of both modes, it is the numerical noise that determines the winning mode. Both the kinetic and current helicities of the two possible solutions have the same magnitude but opposite signs. The solution consisting of an equal mixture of both modes proves to be unstable. Other examples of spontaneous parity-breaking bifurcations of this type have been described in Refs.~\cite{HH09,BB12,WG15}.

 \subsection{Wide gaps}
Containers with $\rin=0.05$ may be considered as approaching pipe flows within the outer cylinder. For such models the influence of the inner boundary condition should become negligible. It makes sense for all such cases to work with an {\em outer} Hartmann number according to the rule 
\beg
 {\Ha_{\rm out}} = \frac{B_{\rm out} R_{\rm out} }{\sqrt{\mu_0 \rho \nu \eta}}= \frac{\Ha}{ \sqrt{(1-\rin)\rin^3}}.
 \label{Hartout}
 \ende
 \begin{figure}[htb]
\centering
 \includegraphics[width=9cm]{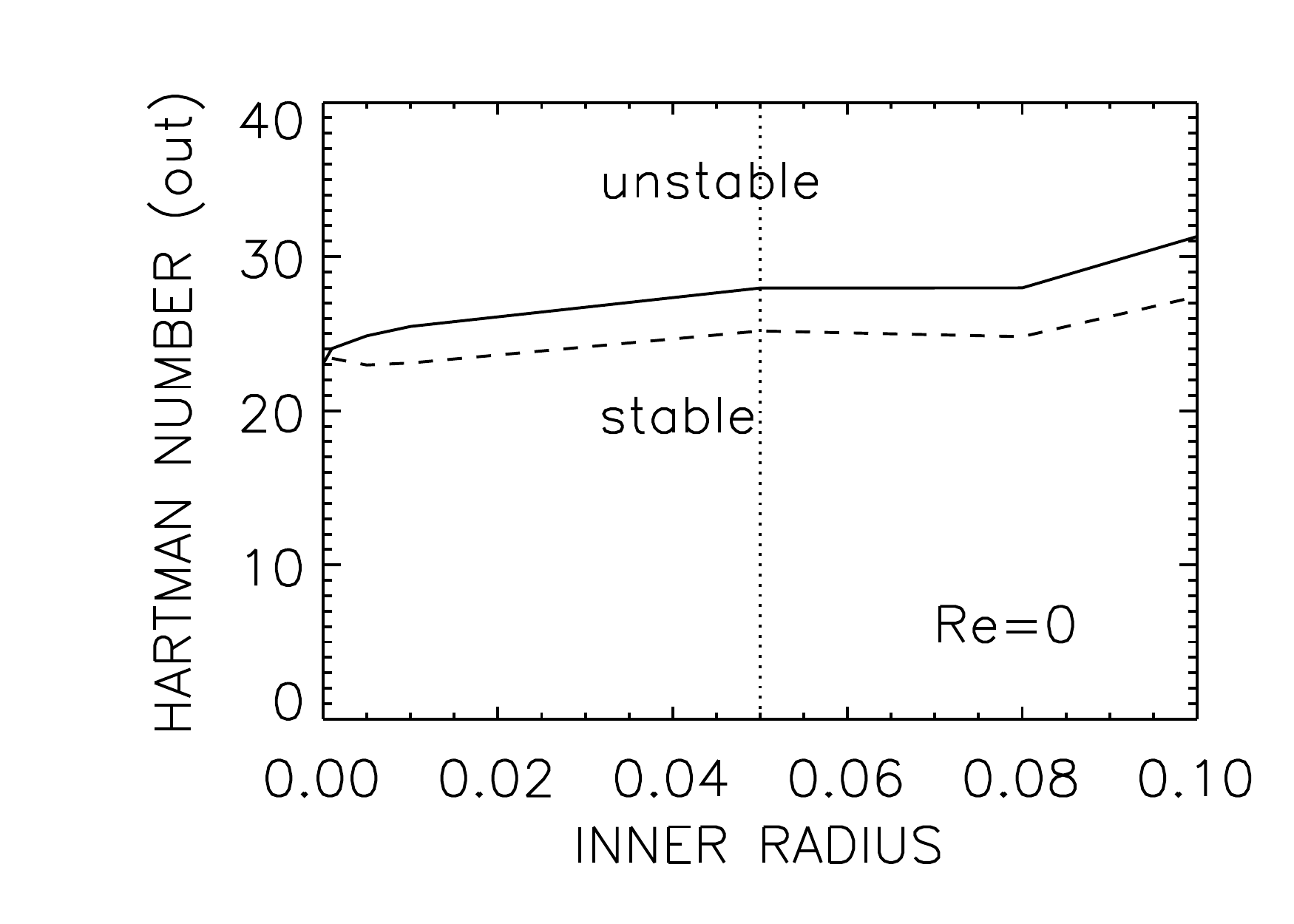}
 \caption{Outer Hartmann number (\ref{Hartout}) for stationary  $z$-pinches   in   wide gaps with $\rin\leq 0.1$. For $\rin\to 0$ the values converge for perfectly conducting (solid line) and insulating (dashed line) cylinders. $\Ha_{\rm out}=28.4$ for $\rin=0.05$ (vertical dotted line) and conducting cylinders. All values  independent of $\Pm$. $m=1$, $\Rey=0$, $\mu_B=1/\rin$. From \cite{RG12}.}
\label{ti6}
\end{figure}
 \begin{figure}[htb]
\centering
 \includegraphics[width=8cm]{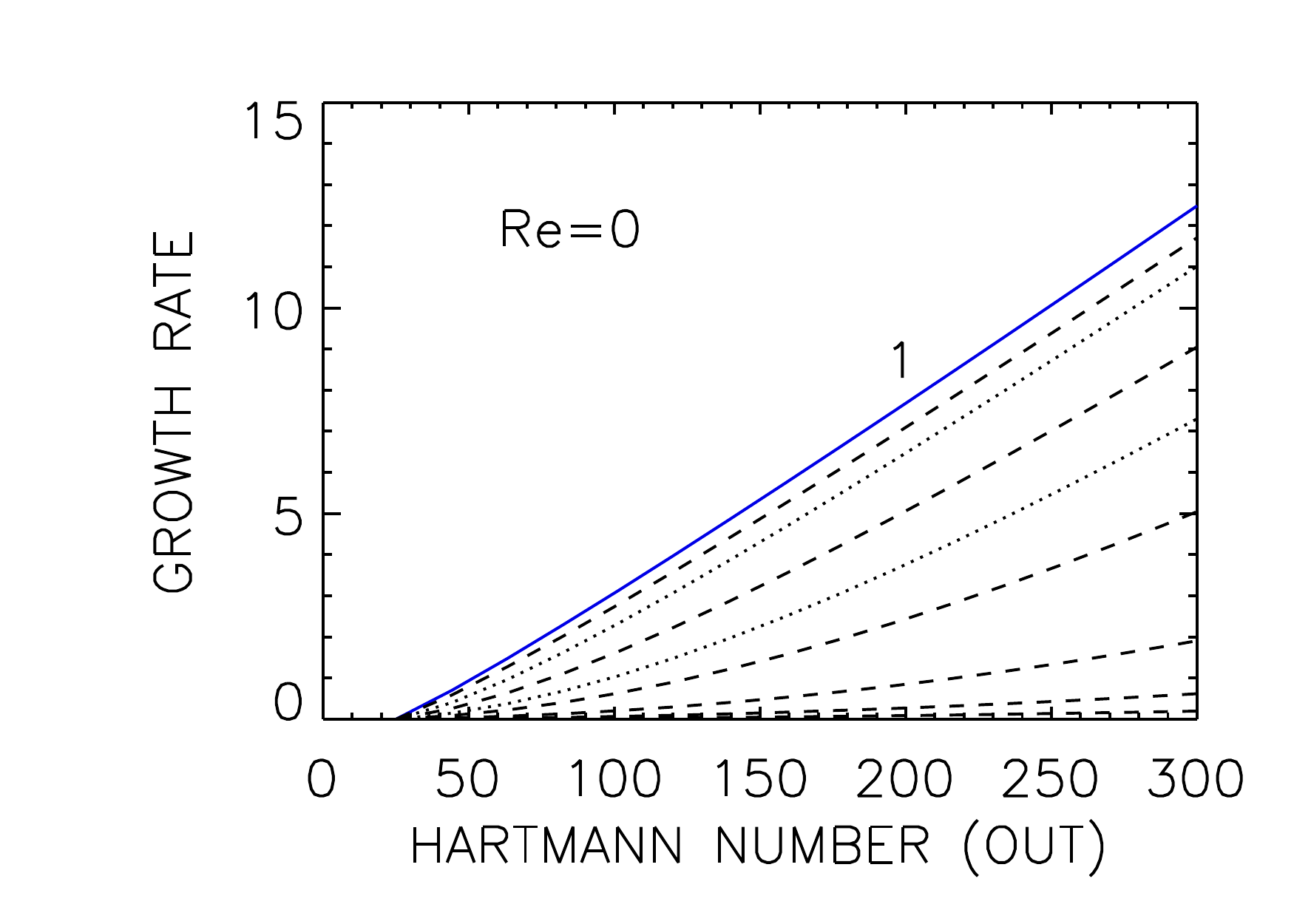}
 \includegraphics[width=8cm]{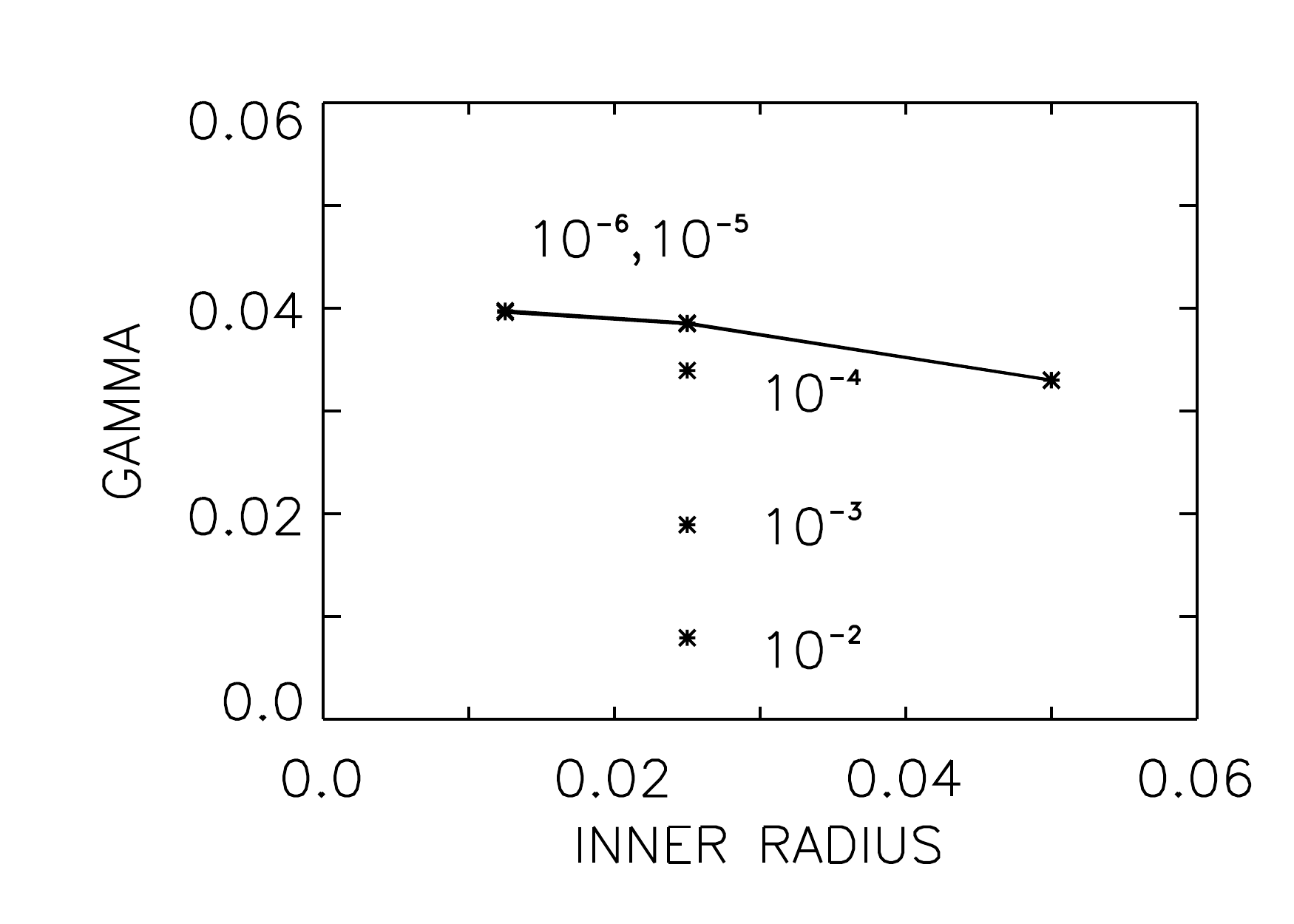}
 \caption{Left: growth rates of the $z$-pinch  normalized with the dissipation frequency $\overline{\omega}=\sqrt{\omega_\nu \omega_\eta}$ versus  outer Hartmann numbers (\ref{Hartout}) for $\rin=0.05$. Right: $\Gamma_{\rm gr}$ (marked with $\Pm$) from Eq.~(\ref{omg}) for various $\rin$.  Observe $\Gamma_{\rm gr}$ for  small $\Pm$  as  almost independent of both  $\rin$ and $\Pm$. Pinch-type field with $m=1$, $\Rey=0$, $\mu_B=1/\rin$, perfectly conducting cylinders. From \cite{RS11}.}
 \label{ti7}
\end{figure}

Figure \ref{ti6} demonstrates the behavior of the outer Hartmann number for $\rin\to 0$ for both types of boundary conditions. For $\rin\to 0$ the two nearly horizontal curves approach (as they should); the dependence on $\rin$ is very weak for $\rin\ll 1$. Note that insulating boundary conditions lead to (slightly) more unstable flows.

The growth rates of the $m=1$ instability of this pinch-type flow for $\Ha>\Ha_0$ are plotted  in Fig.~\ref{ti7} (left) with the same normalization  as used in Fig.~\ref{ti4}. One finds  that the (physical) growth rates in wide gaps behave like
\beg
\omega_{\rm gr} = \Gamma_{\rm gr} \frac{B^2_{\rm out}}{\mu_0\rho\eta},
\label{omg}
\ende
where the coefficient $\Gamma_{\rm gr}$ varies only by a factor of four when the magnetic Prandtl number varies by four orders of magnitude \cite{RS11}. The linear size of the container does not occur in Eq.~(\ref{omg}). It is also surprising that the growth rate is inversely proportional to the diffusion frequency $\omega_\eta=\eta/R_{\rm out}^2$, which means that the growth time  reduces for increasing electric conductivity (in opposition to the diffusion times). For small $\Pm$, $\Gamma_{\rm gr}$ no longer depends on the magnetic Prandtl number (Fig.~\ref{ti7}, right). $\Pm=1$ and $\rin=0.05$ yield $\Gamma_{\rm gr}=0.0009$.
Note also that the growth rates for the wide gap container  are much smaller than those of  the standard gap  displayed by Figs.~\ref{ti4}.

It remains to describe the experimental implication of the critical value $\Ha_0\simeq 30$ for the neutral instability  taken from Fig.~\ref{ti7} (left). The solution of the stationary induction equation inside the outer cylinder in the presence of a uniform electric current $I_{\rm fluid}$ yields $ B_\phi={I_{\rm fluid}}/(5 R_{\rm out})$. With (\ref{Hartout}) it follows 
\beg
I_{\rm fluid}=5 \Ha_{\rm out} \sqrt{\mu_0 \rho \nu \eta}
\label{curr}
\ende
 with $\sqrt{\mu_0 \rho \nu \eta}=8.2$ in cgs units for liquid sodium\footnote{$\sqrt{\mu_0 \rho \nu \eta}=25.8$ in cgs units for liquid gallium.}. Hence, the characteristic value of $\Ha_{\rm out}\simeq 30$ leads to (only) 1.2~kA and/or $B_{\rm out}\simeq 50$~G for (say) $R_{\rm out}=5$~cm.
\subsection{Kinetic and magnetic energy}
\begin{figure}[htb]
\centering
 \includegraphics[width=4.05cm]{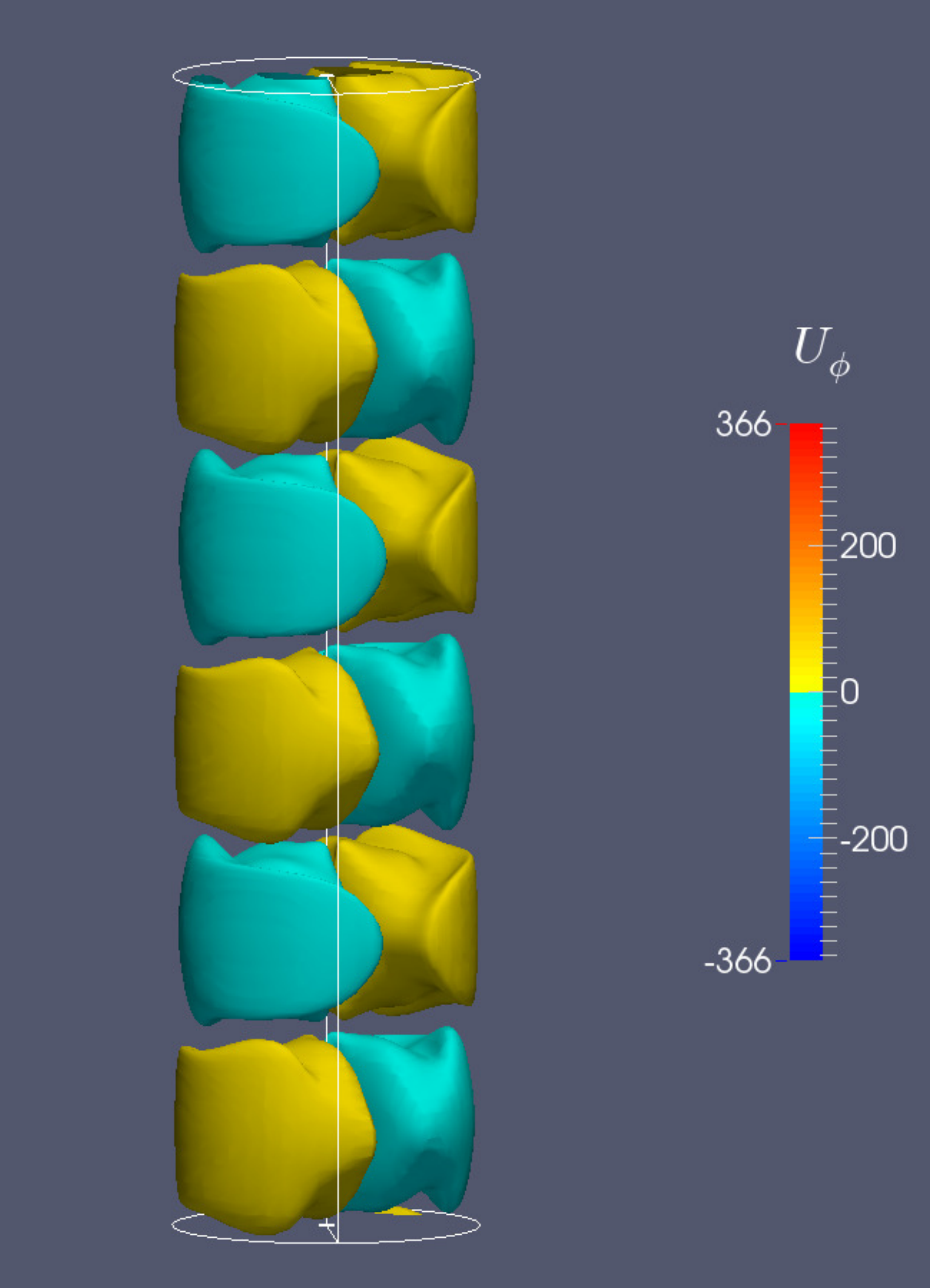}
 \includegraphics[width=4.5cm]{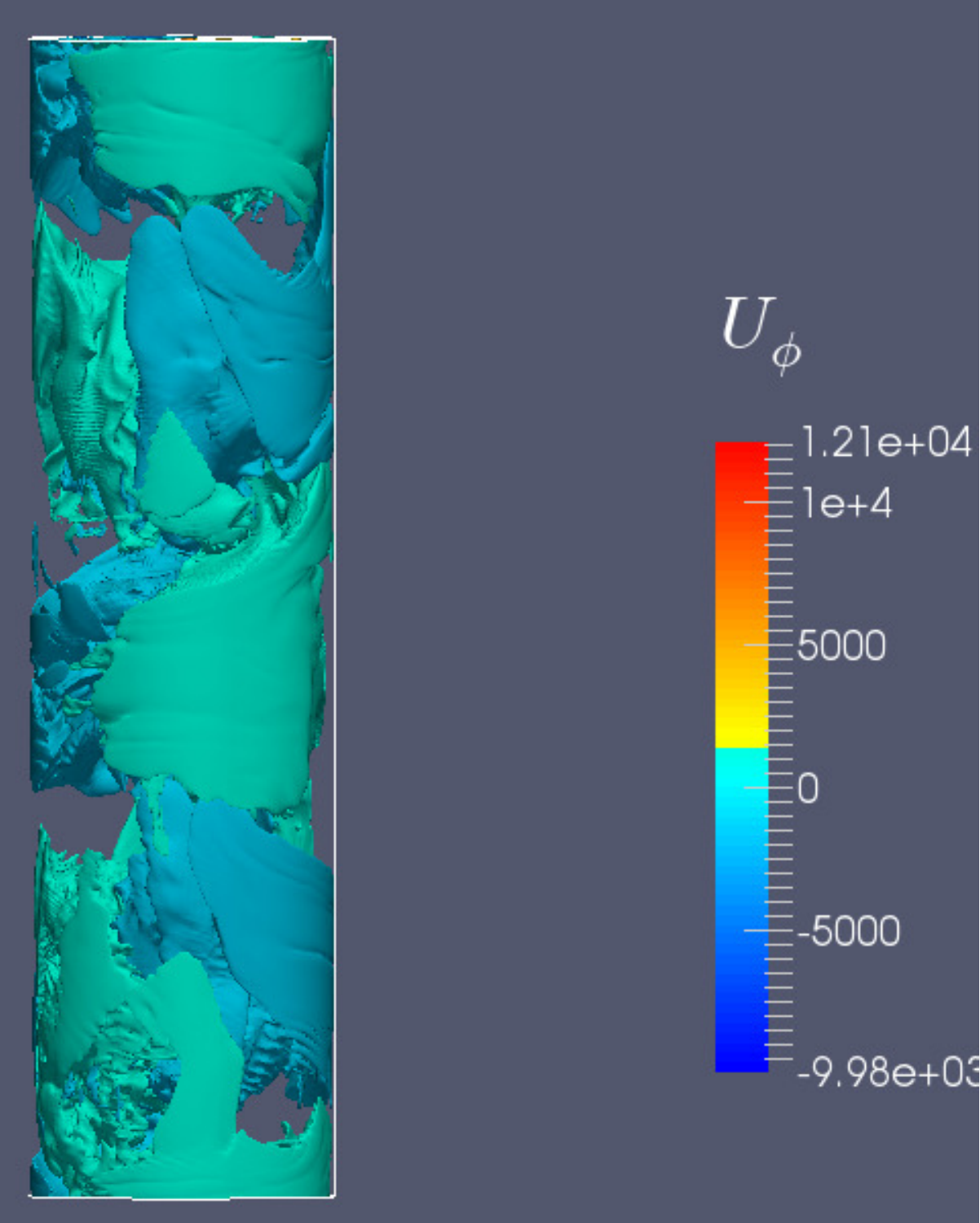}
 \includegraphics[width=4.5cm]{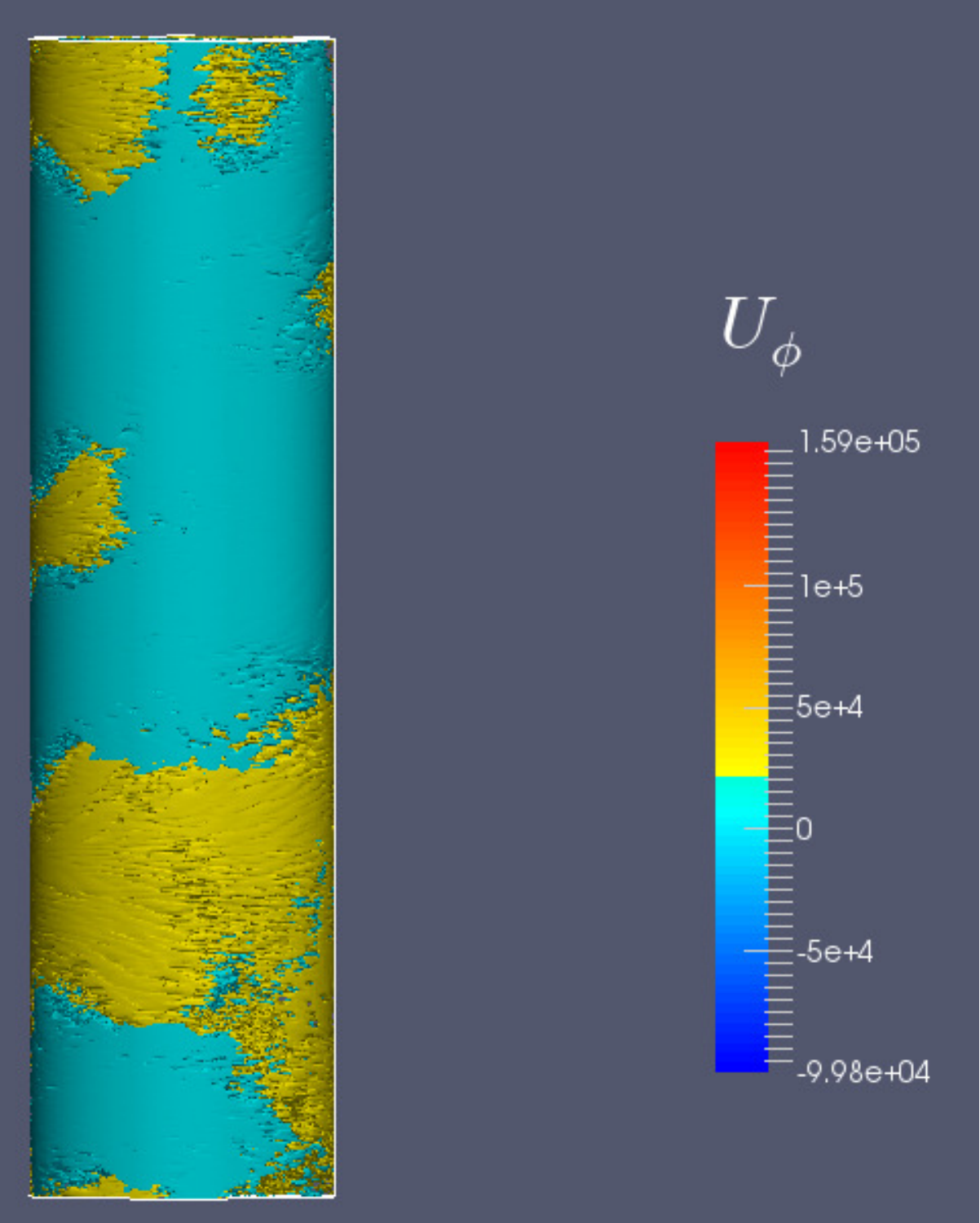}
 \caption{Azimuthal velocity component $u_\phi R_0/\nu$ of the $z$-pinch instability for a wide gap. Left: $\Ha_{\rm out}=40$, middle: $\Ha_{\rm out}=200$, right: $\Ha_{\rm out}=600$. $\rin=0.05$, $\mu_B=1/\rin$, $\Rey=0$,  $\Pm=10^{-5}$. Insulating cylinders.}
 \label{ti72}
\end{figure}

\begin{figure}[htb]
\centering
 \includegraphics[width=3.6cm]{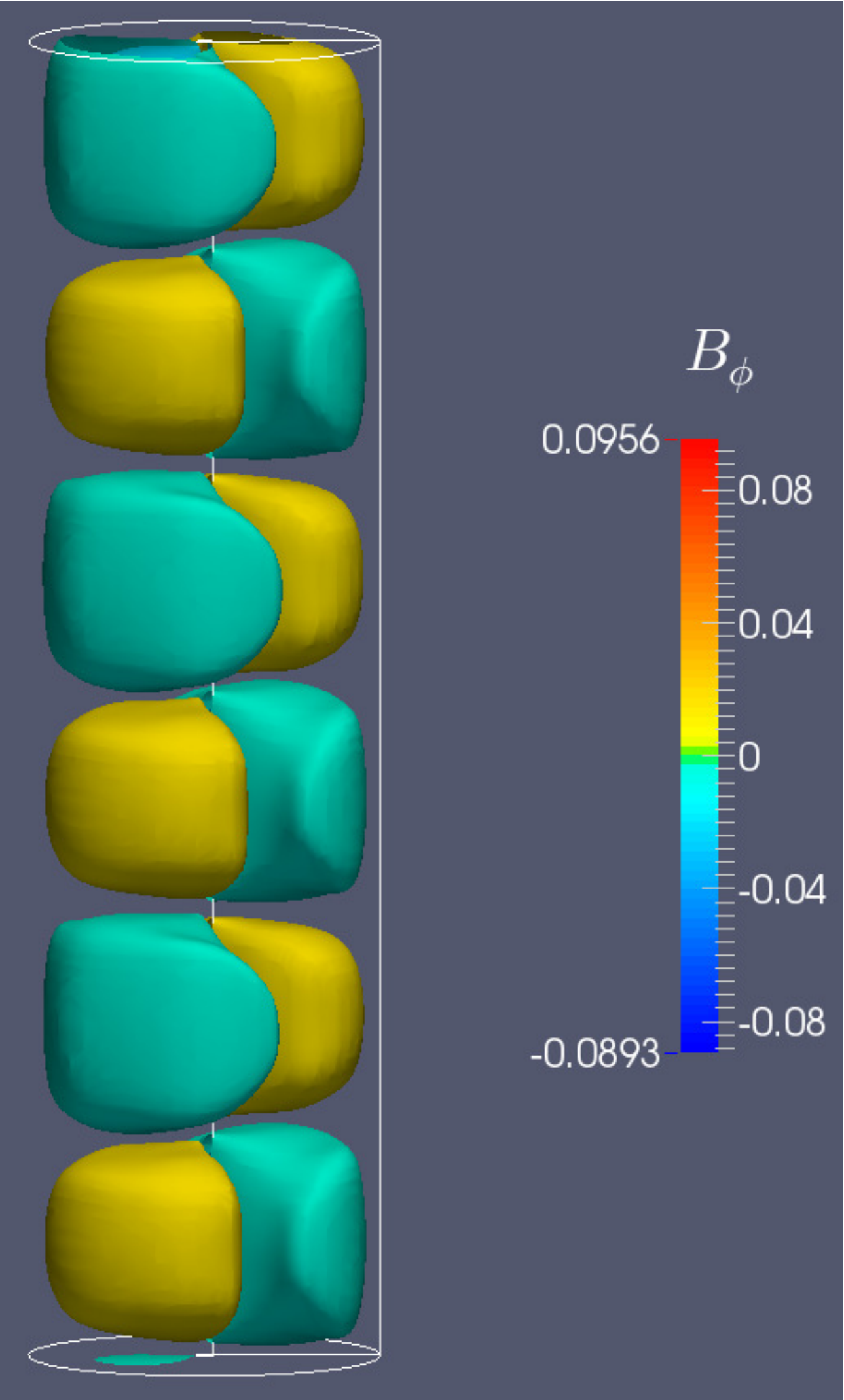}
 \includegraphics[width=4.5cm]{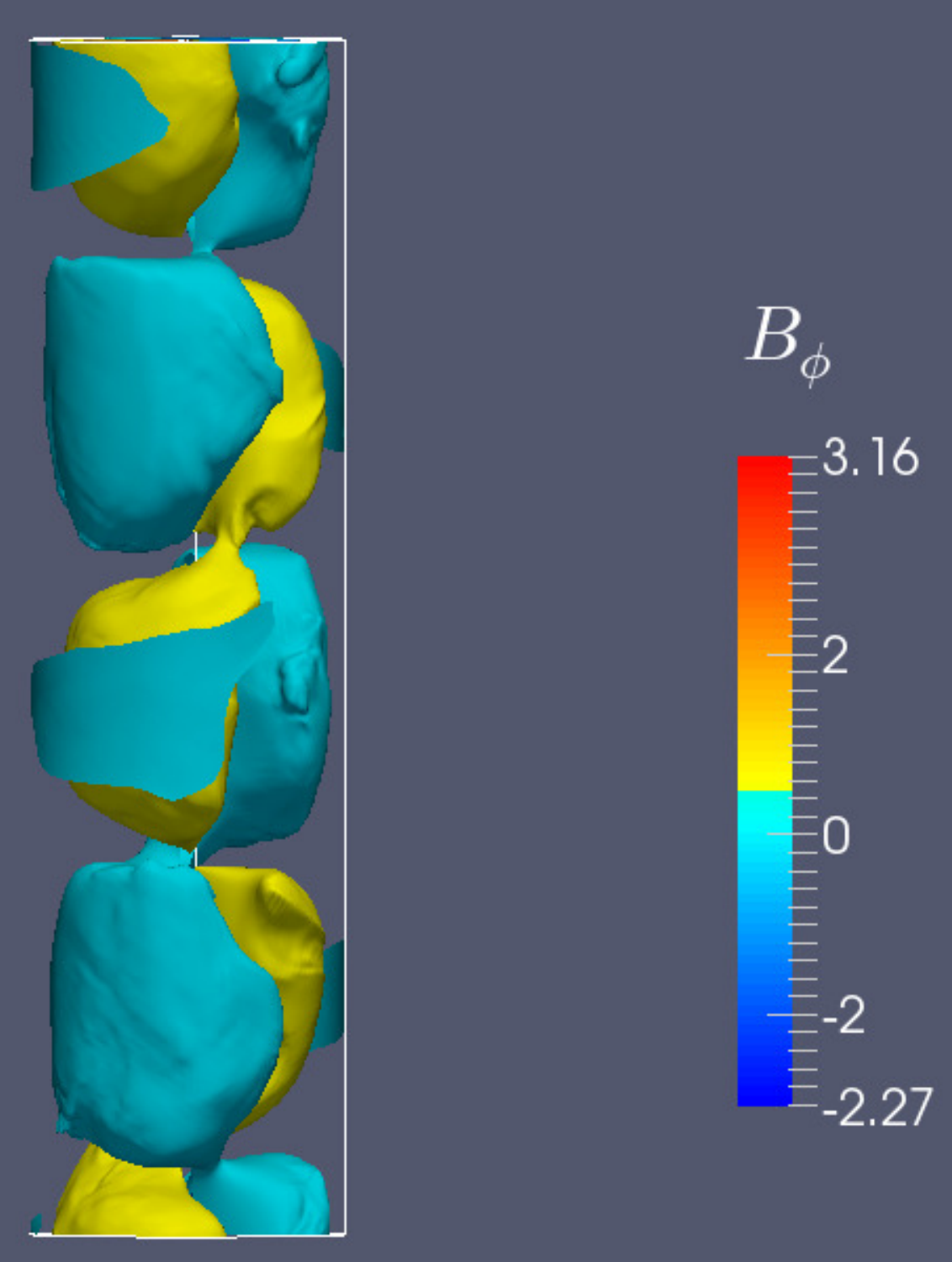}
 \includegraphics[width=4.5cm]{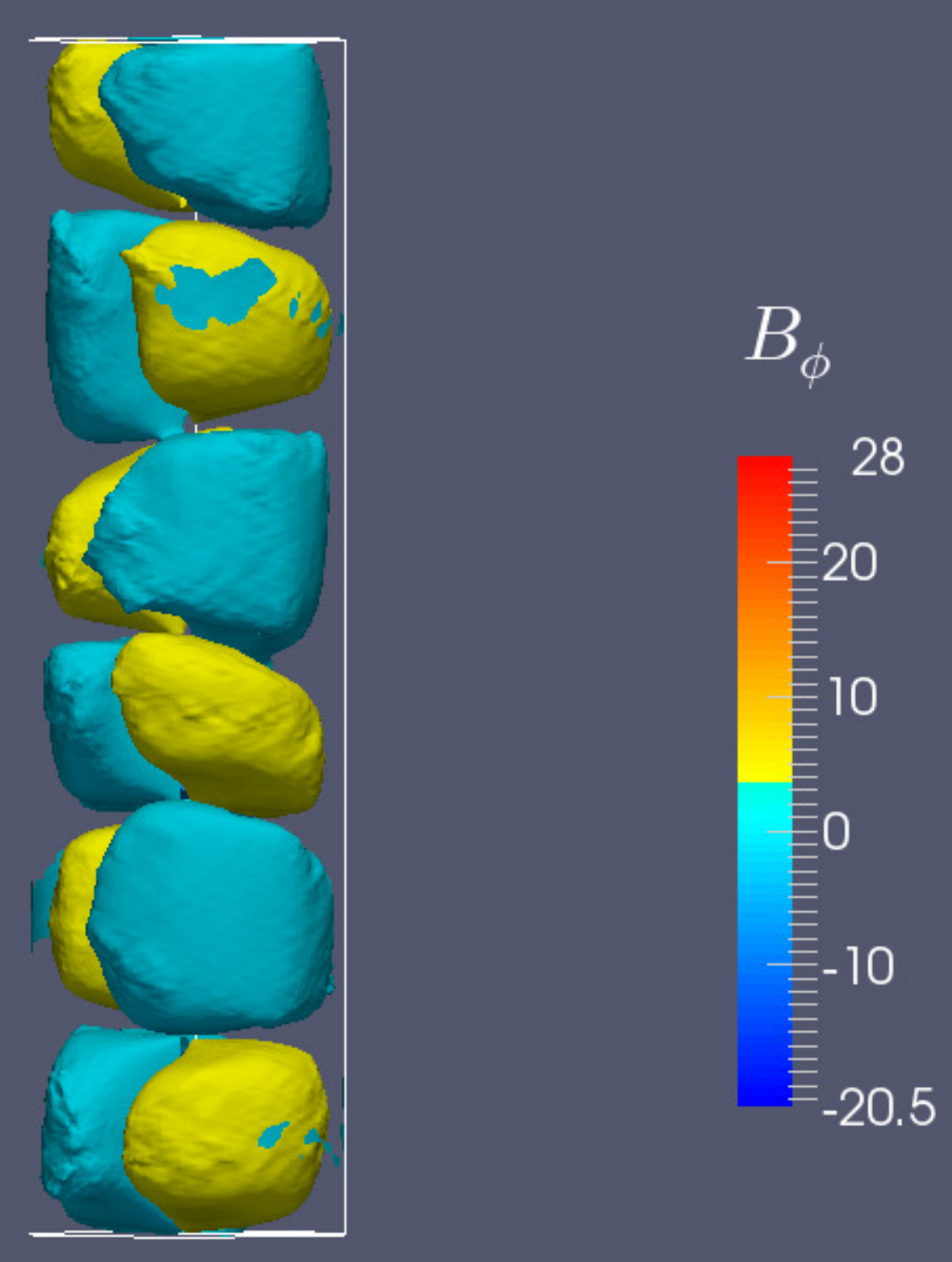}
 \caption{As in Fig.~\ref{ti72} but for the azimuthal magnetic component $b_\phi/B_{\rm in}$.}
 \label{ti73}
\end{figure}

For the very small magnetic Prandtl number $\Pm=10^{-5}$ and for stationary cylinders the TI for increasing electrical currents have been numerically simulated. Figures \ref{ti72} and \ref{ti73} show the azimuthal components of flow and field for $\Ha_{\rm out}=40$ to $\Ha_{\rm out}=600$. While for the weak-field case the expected regular nonaxisymmetric pattern can be observed, stronger fields produce more and more elongated structures and intermittency. Strong currents simultaneously lead to much larger and much smaller axial scales. This effect can be observed at least for the spectrum of the kinetic fluctuations rather than in the spectrum of the magnetic fluctuations. This might be a consequence of the very small magnetic Prandtl number, which leads by the high value of $\eta$ to an effective smoothing of small scales of the magnetic fluctuations. 

\begin{figure}[b]
\centering
 \includegraphics[width=7cm]{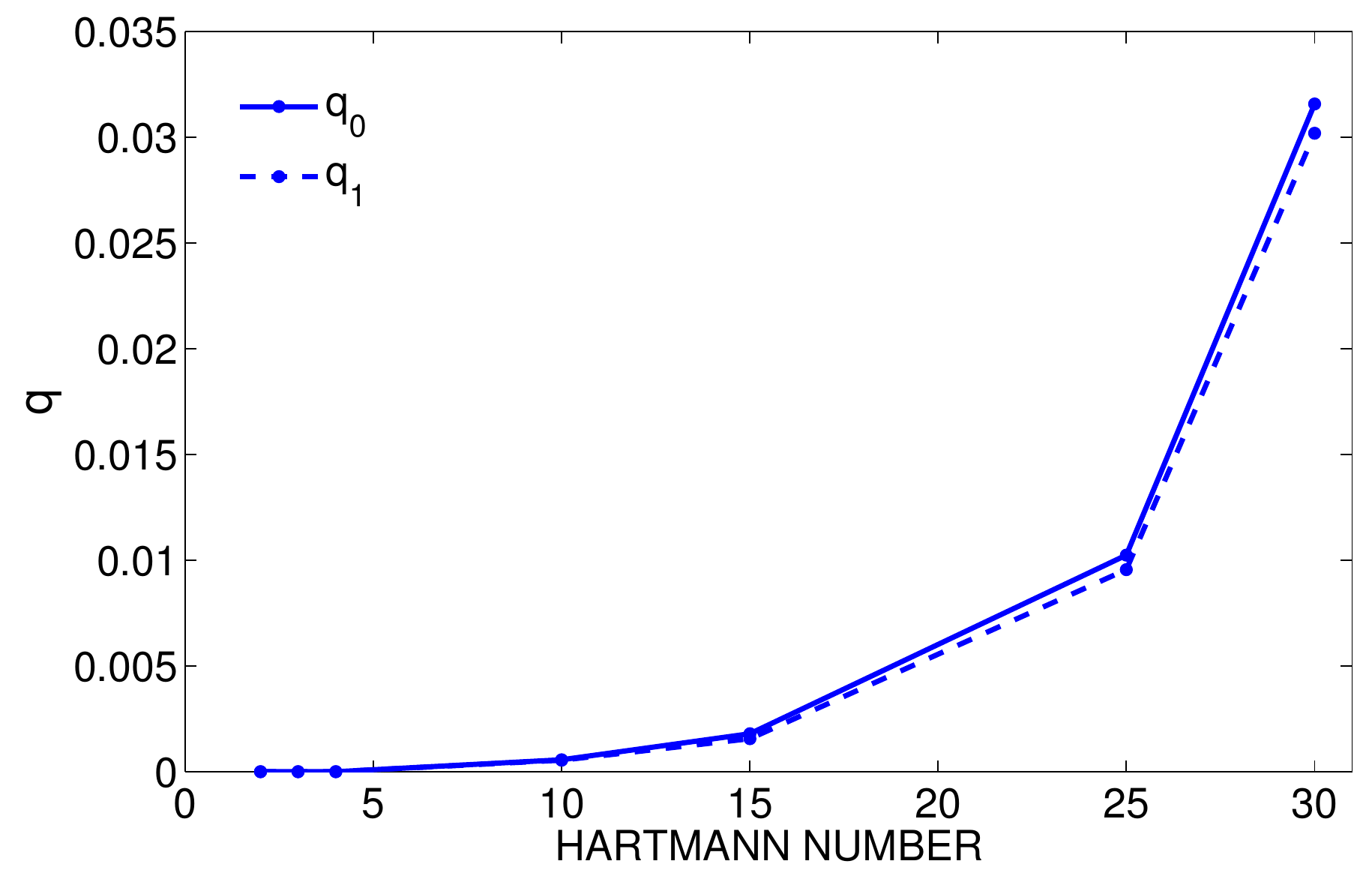}
 \includegraphics[width=7cm]{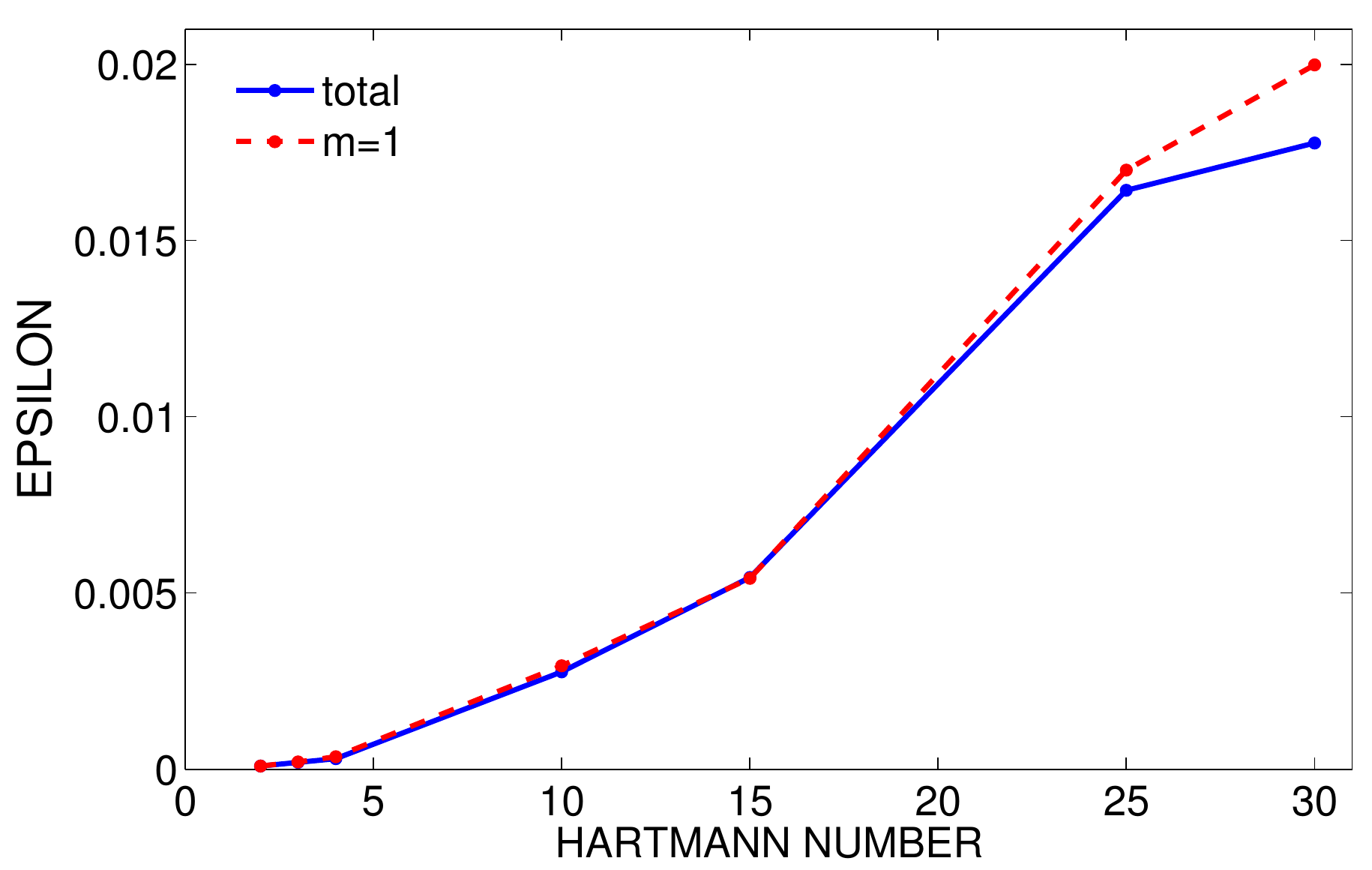}
 \caption{Dependence of $\rm Q$ from (\ref{qu}) (left) and the ratio $\varepsilon$ from (\ref{ratio})  of magnetic to kinetic energies (right) for the stationary $z$-pinch on the Hartmann number. Dased lines are for $m=1$.  These results lead to the relation (\ref{qq}) between the magnetic energy and the microscopic magnetic Reynolds number $\Rm'$.  $\mu_B=1/\rin$, $\rin=0.05$, $\Rey=0$, $\Pm=10^{-5}$, insulating cylinders.}
 \label{ti74}
\end{figure}

The consequences of this situation for the resulting energies may also be discussed. The normalized magnetic energy (\ref{qu}) is plotted in the left panel of Fig.~\ref{ti74} in its dependence on the inner Hartmann number. It is a steep function, ${\rm Q}= Q_0 \Ha^4$, with $Q_0\simeq 4\cdot 10^{-8}$. On the other hand, the energy ratio (\ref{ratio}) only grows linearly with $\Ha$, i.e.~$\varepsilon= E \Ha$ with $E\simeq 0.2$ (right panel). From these expressions it is easy to derive the relation between $\rm Q$ and $\Rm'$ in the form
\beg
{\rm Q}\simeq 92\ \Rm'^{1.6},
\label{qq}
\ende
with $\Rm'=u_{\rm rms} R_0/\eta$. The resulting exponent lies well between the values 1 and 2 for driven turbulence with high and low conductivity \cite{BK74}. It also follows that
\beg
\Rm'\simeq 2.5\ \Lu^{2.5},
\label{qqq}
\ende
if $\Lu=\sqrt{\Pm}\ \Ha$ is used.  The stationary pinch with $\Pm\ll 1$ is not magnetically dominated.
\subsection{The GAllium-Tayler-Experiment ({\sc Gate})}
\begin{figure}[h]
\centering
 \includegraphics[width=0.8\textwidth]{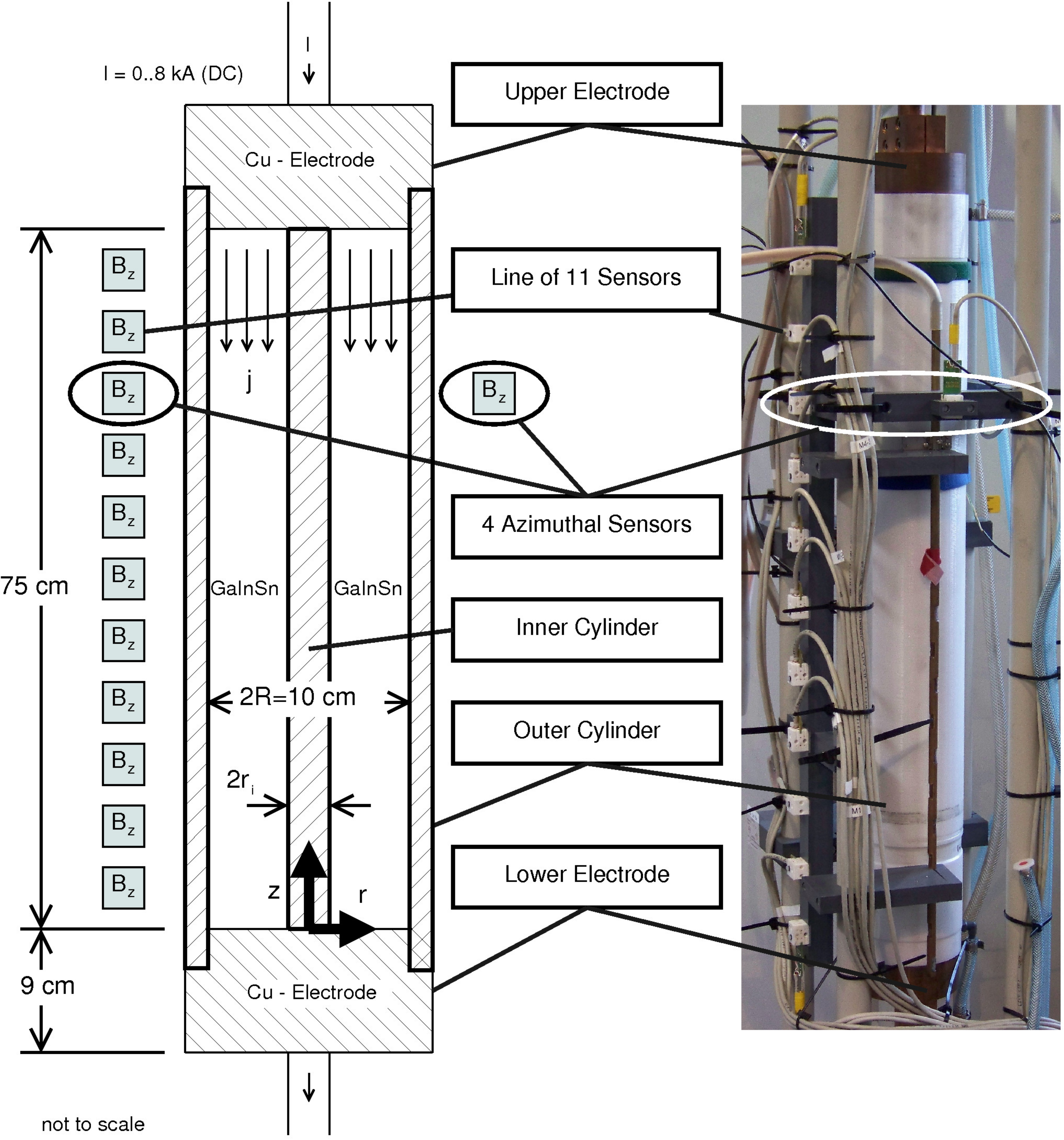}
 \caption{Construction of {\sc Gate} as operated by the Helmholtz-Zentrum Dresden-Rossendorf \cite{SS12}.}
 \label{gate}
\end{figure}
The stationary TI leads to a nondrifting nonaxisymmetric steady-state solution. Because of $\Rey=0$ the eigenvalues $\Ha_0$ do not depend on the magnetic Prandtl number, and can thus be computed for all $\rin$ with a code for (say) $\Pm=1$ (Fig.~\ref{ti6}). On this basis and the calculation of the growth times an experiment can be designed to probe the theoretical predictions for $\Rey=0$ as a first step. The experiment {\sc Gate} consists of an insulating cylinder with a height of 75~cm and a radius $R_{\rm out}=5$~cm which is filled with GaInSn (Fig.~\ref{gate}). The liquid column is in contact with two massive copper electrodes which are connected by water cooled copper tubes to an electric power supply providing up to 8~kA. With 14 fluxgate sensors the modifications of the magnetic fields due to the TI are detected. Eleven of these sensors are positioned along the vertical axis, while the remaining three are positioned along the azimuth in the upper part. Such measurements give the geometry of the field, thus its shape in azimuthal and axial direction as well as the scaling of the growth rates with the applied electric current \cite{SS12}.
\begin{figure}[htb]
\centering
 \includegraphics[width=9cm]{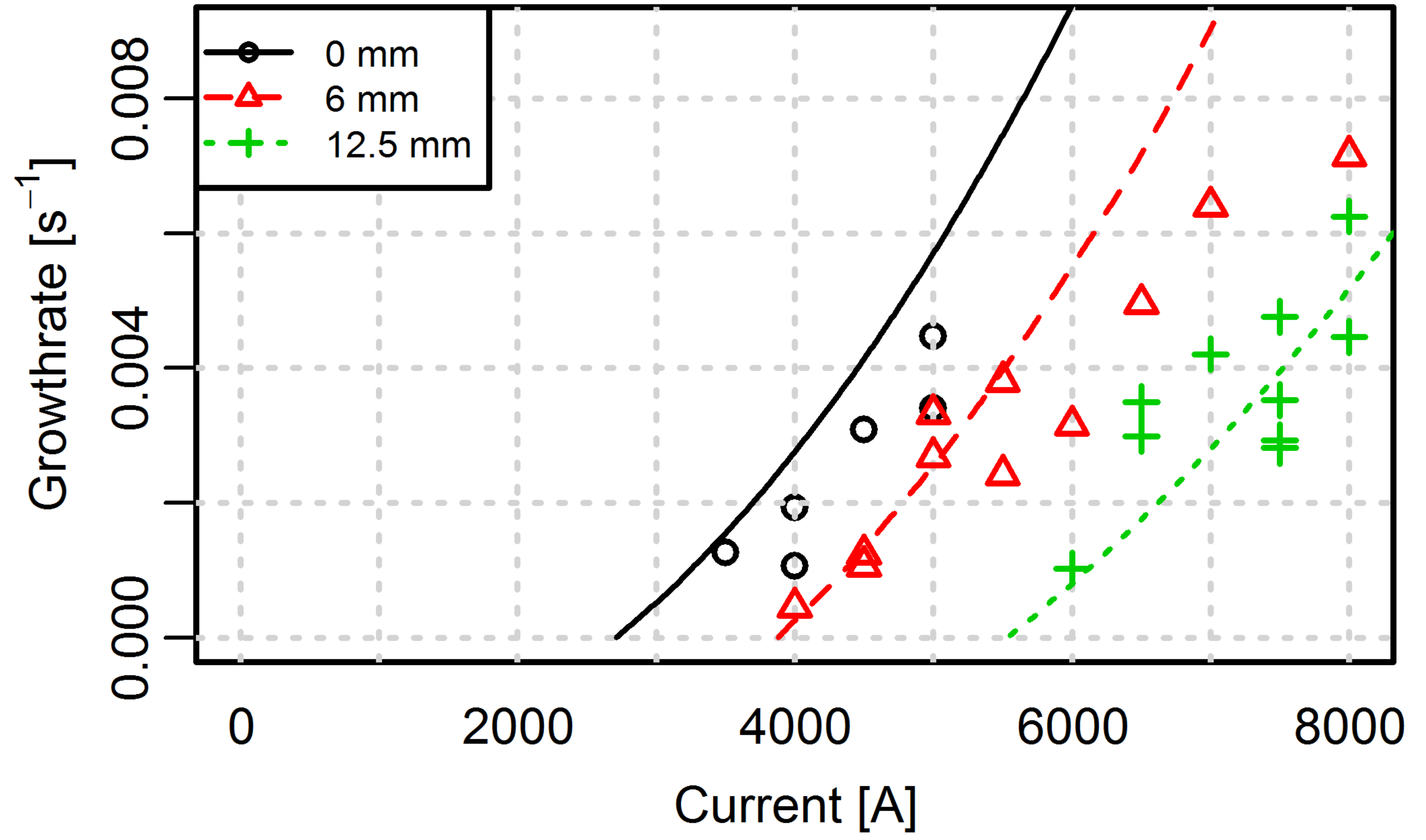}
 \caption{Observed and calculated growth rates  for the stationary $z$-pinch for small $\rin$ (black: $\rin=0$, red: $\rin=0.12$, green: $\rin=0.25$). The measured values for the container without inner cylinder (black circles)  do not exceed the calculated ones, indicating almost perfect agreement between experiment and theory \cite{RG12}.}
 \label{ti8}
\end{figure}

In all cases of instability the observed pattern of the magnetic perturbations is nonaxisymmetric with $m=1$. Figure~\ref{ti8} shows the resulting growth rates and the calculations according to Eq.~(\ref{omg}) for three containers with wide and very wide gaps ($\rin\leq 0.25$). Note that for small $\Pm$ and for very small values of   $\rin<0.1$ the theoretical growth rates are almost independent of $\rin$ (Fig.~\ref{ti7}, right). The predicted threshold value for the electric current is 2.8 kA. For low growth rates the experimental data for all three models fit the theoretical curves very well.  The theory always provides maximal growth rates, optimized over the wave number. The  theoretical values should thus always lie above the observed data, which is indeed the case for the container with $\rin=0$ (black circles).  One finds a relation $\omega_{\rm gr}\simeq \gamma (I^2-I_{\rm crit}^2)$ with $\gamma= 2.7 \cdot 10^{-10}$, so that $\Gamma_{\rm gr}\simeq 0.038$ results. Nearly the same value can be taken from the theoretical results plotted in Fig.~\ref{ti7}. 

Also for the container with the very thin inner cylinder almost all red triangles lie below the theoretical curve. The agreement is less perfect for the container with the relatively wide inner cylinder. However, for any fixed growth rate even for this example all measured values (green crosses) lie at the right-hand side of the red triangles  which conforms with the theoretical result (green dashed line). It is even quite natural that the agreement between observed and calculated becomes less perfect the larger the growth rate and the thicker the inner cylinder.  For the {\em lowest} growth rates in  Fig.~\ref{ti8}, i.e.~for the determination of the amplitudes of the electric currents for the onset of instability,  the agreement between theory and observation is Fig.~\ref{ti8} is almost perfect.
\section{Tayl{\em E}r-Couette flow}\label{Tayler}
In this section the influence of rotation on the Tayler instability will be described. The rotation law $\Om=\Om(R)$  shall have the form (\ref{1.1}) as a stationary solution of the  angular momentum equation varying from $\Om\propto 1/R^2$ (negative shear) via rigid-body rotation to superrotation with positive shear. 
Also the radial profile of the azimuthal magnetic field is a function of two free parameters in accordance to  (\ref{basic}).  
Among many other possibilities the examples of quasi-uniform fields ($\mu_B=1$) and the $z$-pinch due to a uniform electric field ($\mu_B=2$) will be discussed in detail. 
\subsection{Rigid rotation}\label{riro}
\begin{figure}[htb]
\centering
 \includegraphics[width=5.4cm]{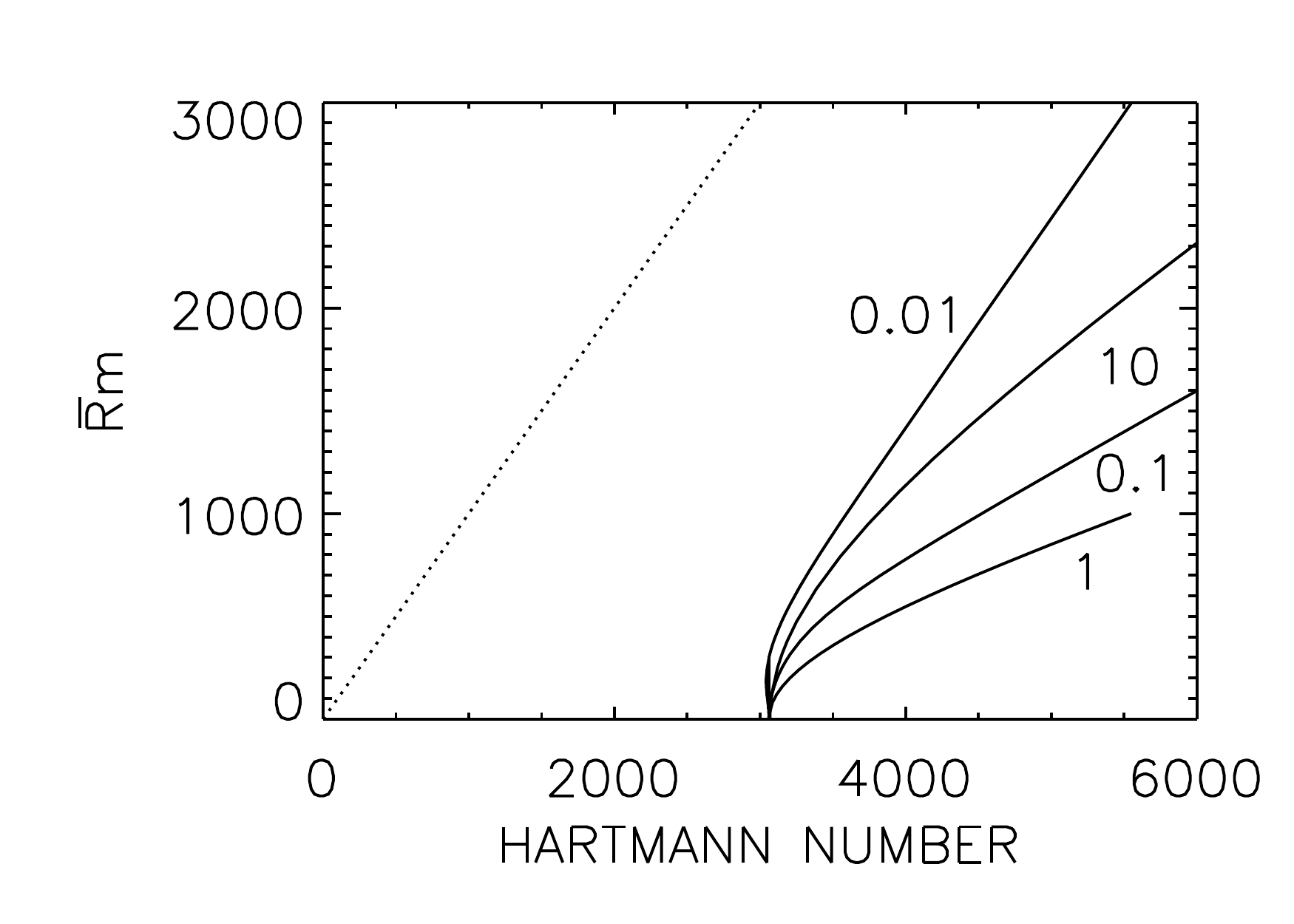}
 \includegraphics[width=5.4cm]{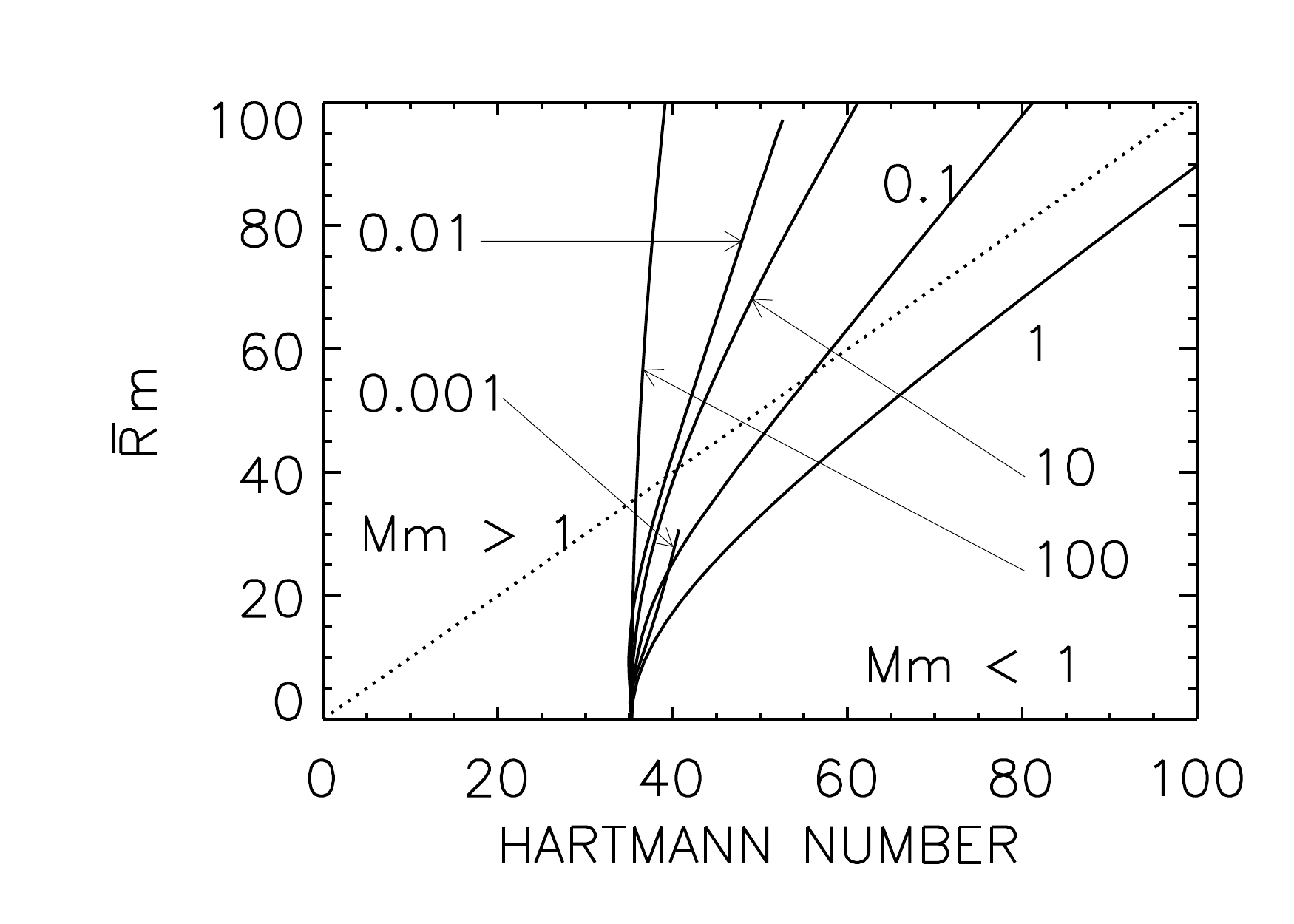}
  \includegraphics[width=5.4cm]{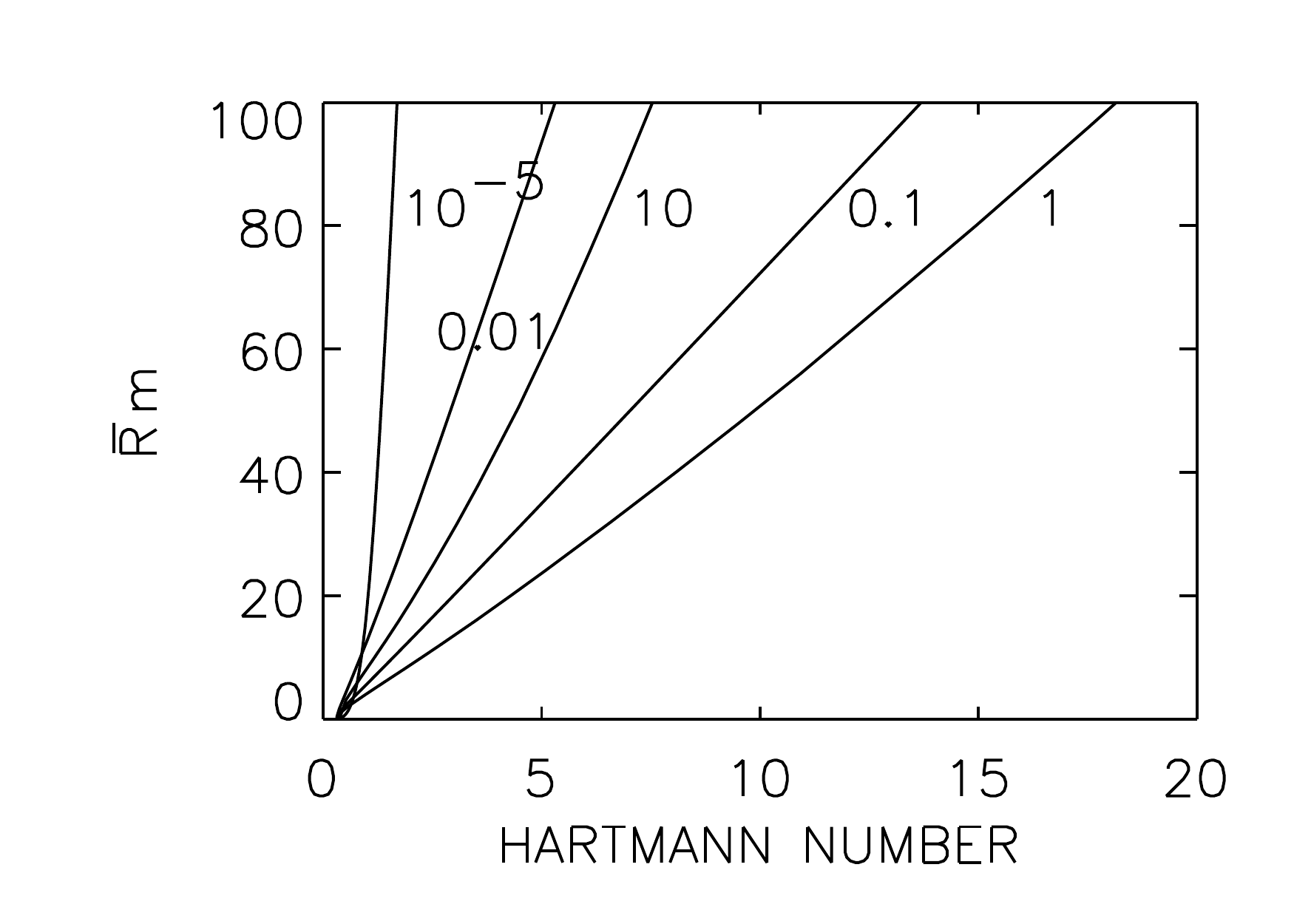}
\caption{Suppression of TI 
by rigid rotation in $z$-pinches within a  narrow gap (left, $\rin=0.95$), a medium  gap (middle, $\rin=0.5$) and a wide gap (right, $\rin=0.05$) plotted in the ($\Ha/\Rmquer$) coordinates. The curves are marked with their magnetic Prandtl numbers. $\Pm=1$ always  plays an exceptional role. For $\Pm\neq 1$ the  Mach numbers are larger than for $\Pm=1$. $m=1$,  $\mu_B=1/\rin$, $\mu_\Om=1$. Perfectly conducting cylinders \cite{RS10}.}
 \label{tirigid1}
 \end{figure}

The most prominent example of this class is formed by a rigidly rotating $z$-pinch due to an uniform axial electric current. It belongs to the Chandrasekhar-type flows, and has $m=1$ as the only unstable mode. The eigenvalues $\Rey$ and $\Ha$ for small $\Pm$ do not depend on $\Pm$. Figure \ref{tirigid1} shows the influence of the magnetic Prandtl number on the suppression of the instability by rigid rotation for three values of the gap width. It makes sense to interpret the results by means of the averaged  Reynolds number $\Rmquer$, because of the convenient possibility to define the magnetic Mach number as the ratio of $\Rmquer$ and $\Ha$  where the quantities on both axes are symmetric in $\nu$ and $\eta$. The dashed lines in the plots correspond to $\Mm=1$. The rotational suppression of TI in this representation is strongest for $\Pm=1$. For very small and very large $\Pm$ there is almost no rotational suppression of TI. In this sense the magnetic Prandtl number $\Pm=1$ plays an exceptional role.   Depending on the magnetic Prandtl number a fluid with  identical  Reynolds numbers and Hartmann numbers can be stable or unstable. One needs basically stronger fields to destabilize rigid rotation for $\Pm=1$ rather than for $\Pm\neq 1$. Clearly, for the magnetic Prandtl numbers used in the middle panel of Fig.~\ref{tirigid1} the Tayler instability for rigid rotation is a sub-Alfv\'enic phenomenon only for  $\Pm=1$. For smaller and larger $\Pm$ the lines of marginal stability cross the   dashed line for $\Mm=1$ and the instability becomes super-Alfv\'enic.

\begin{figure}[h]
\centering
\includegraphics[width=8cm]{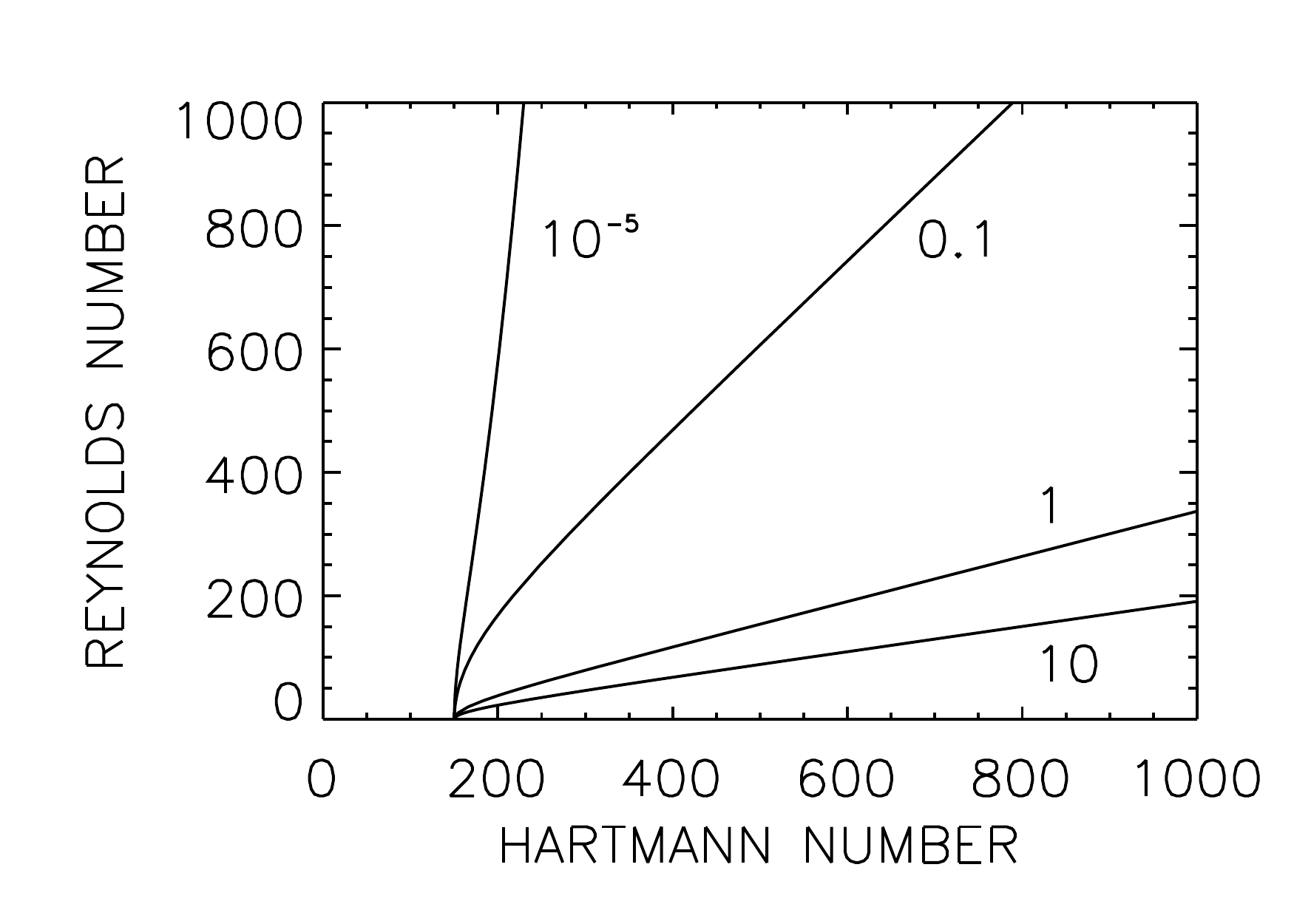}
 \includegraphics[width=8cm]{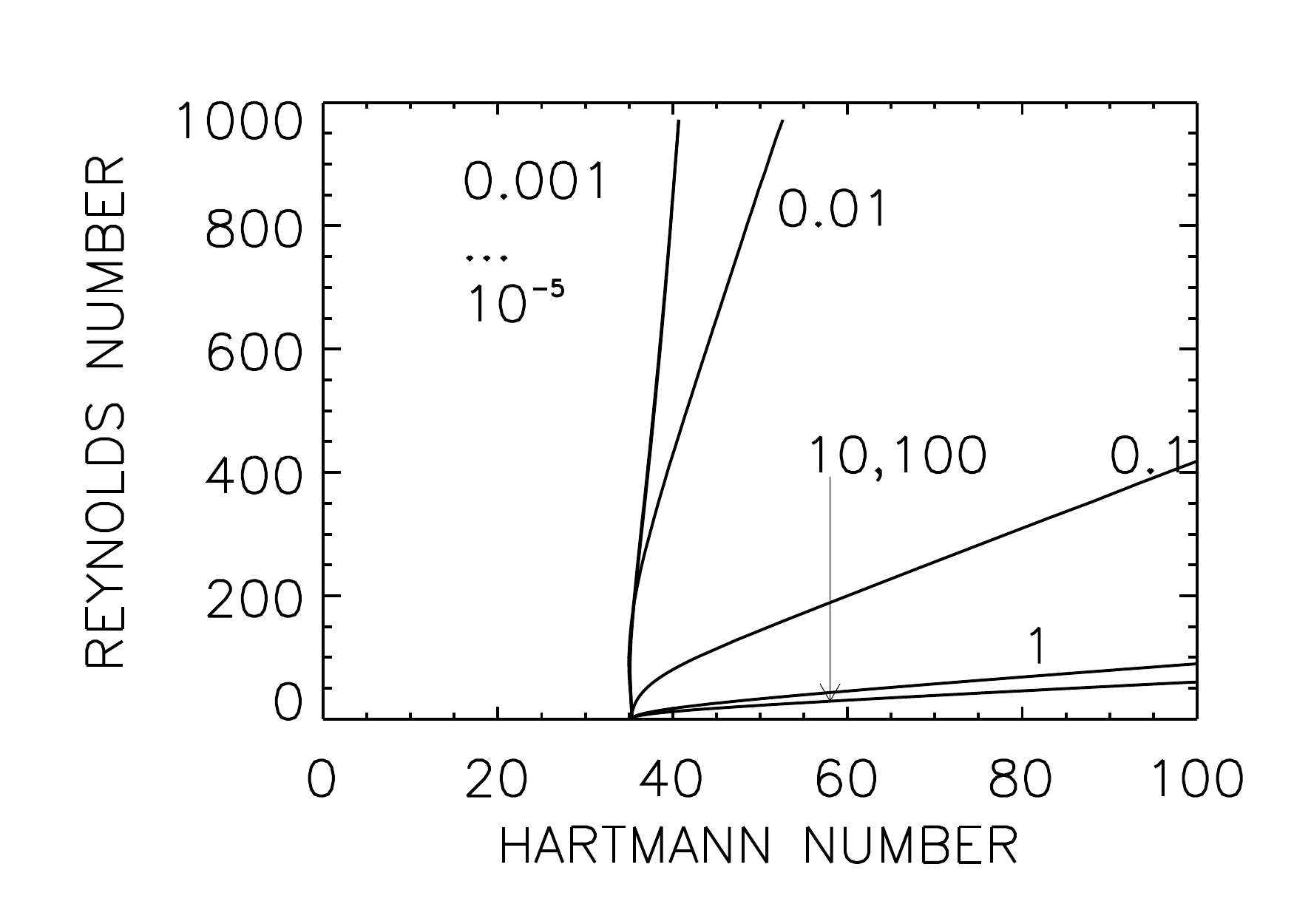}
 \caption{Stabilization  by rigid rotation of quasi-uniform magnetic field ($ \mu_B=1$, left) and of fields due to uniform current ($ \mu_B=2$, right)    for various $\Pm$ (marked) plotted in the $(\Ha/\Rey)$ plane. The parameters used for the right panel satisfy the Chandrasekhar condition (\ref{chancon}). The curves converge in the ($\Ha/\Rey$) plane for both limits $\Pm\to 0$ and $\Pm\to \infty$ (not true in the left panel). $m=1$, $\rin=0.5$, $\mu_\Om=1$, perfectly conducting boundaries.}
 \label{f12}
 \end{figure}

This, however, is not the whole truth. Figure \ref{f12} reflects  the influence of the magnetic Prandtl number on the strength of the rotational suppression for two different radial field profiles in the standard gap ($\rin=0.5$). The left panel concerns the pinch-type field with uniform electric current  and right panel concerns  the almost uniform field. The magnetic Prandtl number varies over many orders of magnitude. The ordinary Reynolds number and the Hartmann numbers are  used. The two standard values $\Ha_0=35$ and $\Ha_0=150$ (for $\rin=0.5$ and  perfectly conducting cylinders) appear. There are  differences between the panels, but the common feature is that the rotational suppression becomes very weak for very small $\Pm$. Note that in the left panel the stability curves for $\Pm\to 0$ converge, unlike the curves in the right panel. For this field profile  the magnetic Mach number $\Mm=\sqrt{\Pm} \Rey/\Ha$ shifts to zero for $\Pm\to 0$, as for $B_\phi\propto R$ and rigid rotation $\Rey$ and $\Ha$ are independent of $\Pm$ for small $\Pm$. Indeed, for $\mu_B=2\mu_\Om=2$ the condition (\ref{chancon}) of Chandrasekhar-type flows is satisfied, so that the convergence of the stability lines for small $\Pm$ in the ($\Ha/\Rey$) plane is not surprising. In summary, for very small magnetic Prandtl numbers  $z$-pinches for $\Pm\to 0$ are unstable for $\Mm\to 0$; all stability curves move more and more below the dashed lines in Figs.~\ref{tirigid1}. This  is not true for the alternative field profile with $\mu_B=1$ (right panel of Fig.~\ref{f12}) which can thus easily reach super-Alfv\'enic values.
 
 Figure \ref{f12} (left) provides another surprise. The plot demonstrates that the curves in the ($\Ha/\Rey$) plane not only converge for $\Pm\to 0$ but also for $\Pm\to \infty$. The rigidly rotating $z$-pinch with perfectly conducting cylinders, therefore,
  is  stable for $\Rey>  \gamma \Ha$ where $\gamma=G$ is a large number for $\Pm\ll1$ and a small number $\gamma=g$ (of order unity) for $\Pm \gg 1$. Hence, the pinch is stable if $\Mm>G \sqrt{\Pm}$ for $\Pm\ll 1$ and  $\Mm>g \sqrt{\Pm}$ for $\Pm\gg1$. 
 The rigidly rotating $z$-pinch is thus easier to keep stable for very small $\Pm$ while for very large $\Pm$  the stabilization requires very rapid rotation.

 Considered in the ($\Ha/\Rmquer$) coordinate system the fluids with $\Pm\neq 1$ are less suppressed by rigid rotation than those with $\Pm=1$. Numerical simulations with $\Pm=1$ may thus be stable although the stability is lost for $\Pm\neq 1$. Instability for $\Pm=1$ requires fields with $\Om_{\rm A }\geq \Omin$ while  much weaker fields become unstable for $\Pm\neq 1$. 
\subsection{Differential rotation}\label{diffrot}
\begin{figure}[h]
\centering
\includegraphics[width=8cm]{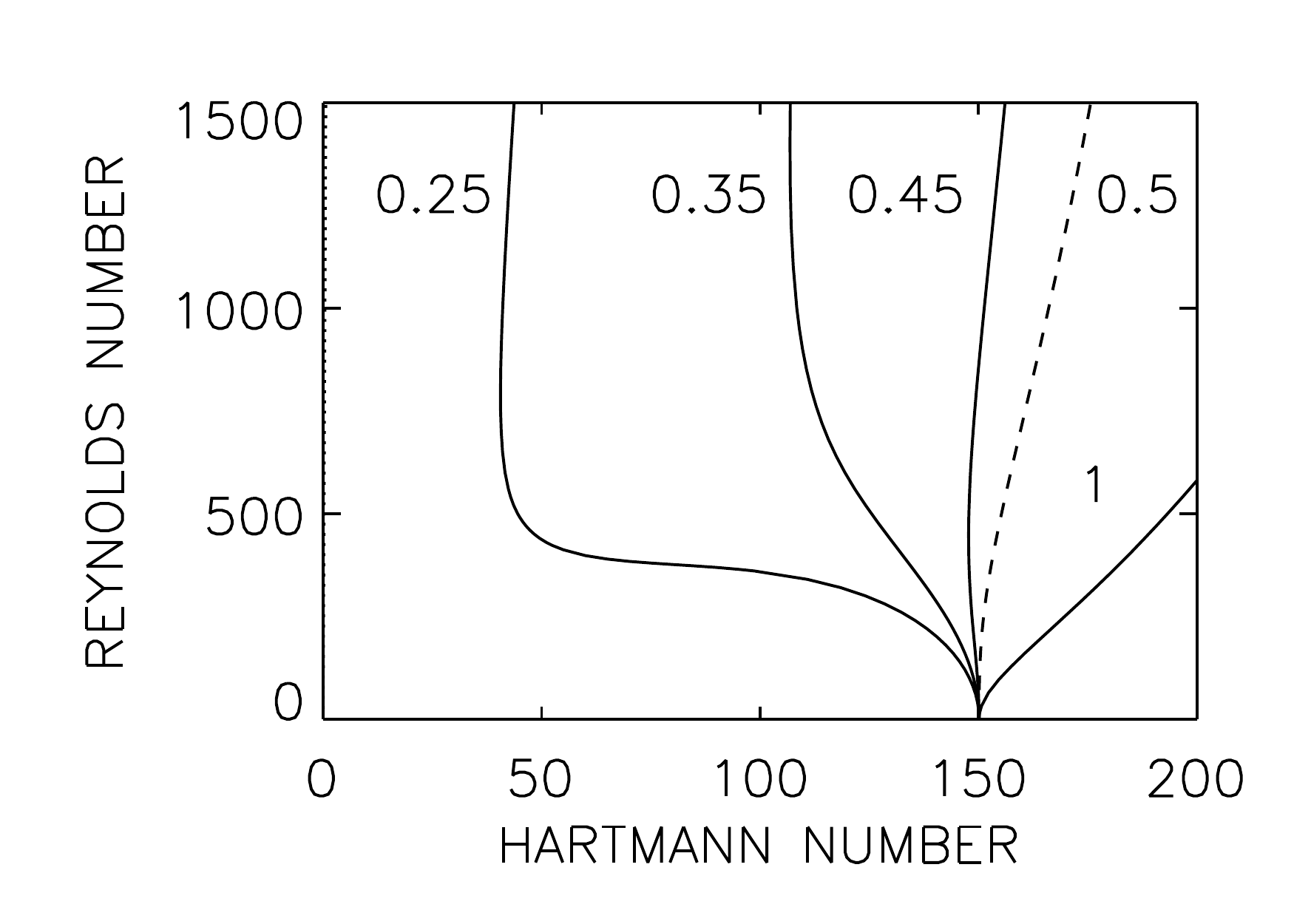}
\includegraphics[width=8cm]{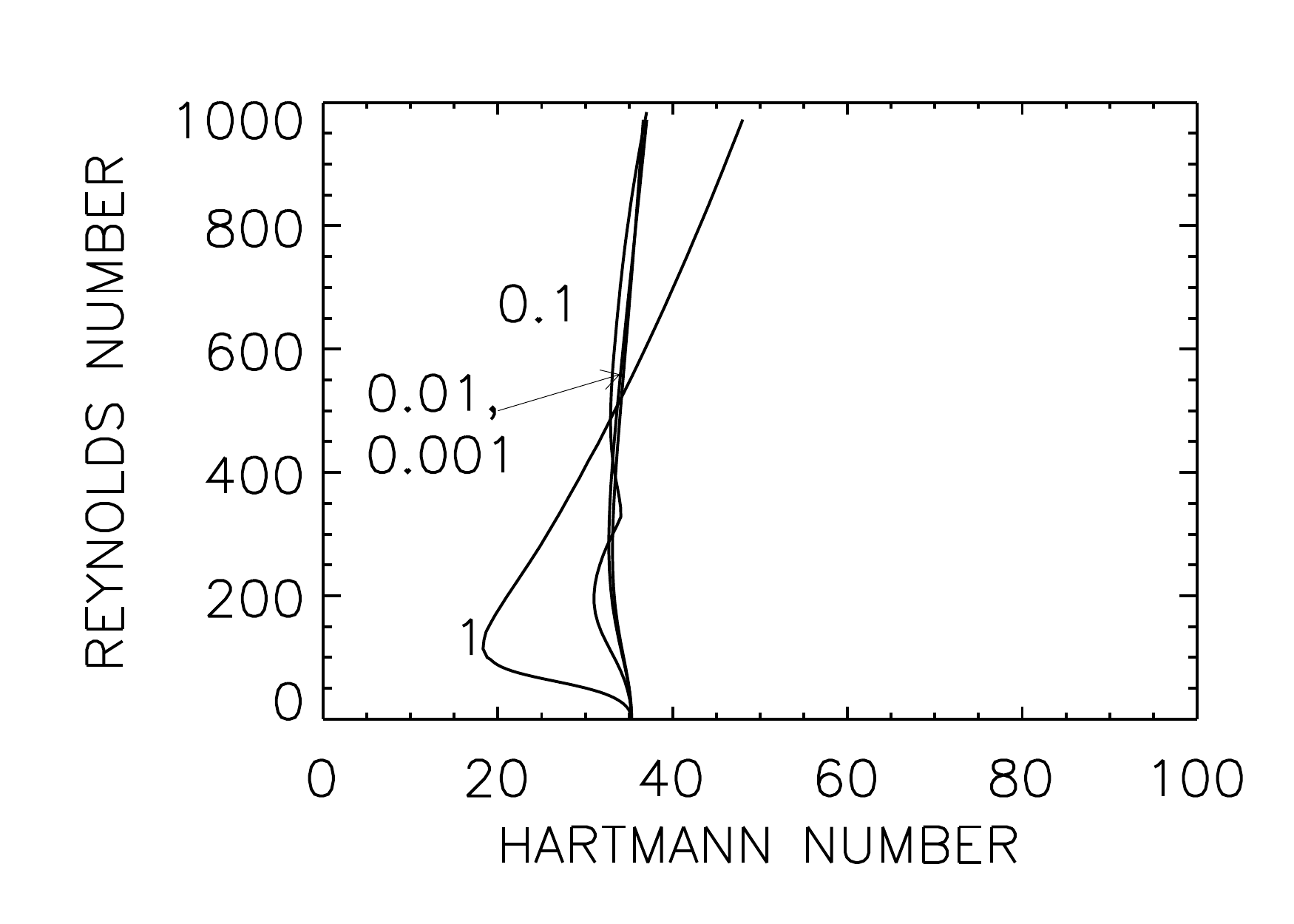}
\caption{Left: Stability lines  for the quasi-uniform magnetic field influenced by  differential rotation for small $\Pm$. The curves are marked with their values of $\mu_\Om$. For slow rotation the rotation profile with $\mu_\Om=\rin$  (uniform flow, dashed)   separates   amplification and suppression. As (only)  this choice of $\mu_\Om$ fulfills the Chandrasekhar condition (\ref{chancon}), the dashed line is valid for all $\Pm\ll1$.  $\mu_B=1$,   $\Pm=10^{-5}$. 
Right: Stability map  of the $z$-pinch for quasi-Keplerian flow. The curves are marked with their values of $\Pm$. The subcritical excitation with $\Ha<\Ha_0$ due to differential rotation almost disappears for $\Pm<1$ \cite{PG16}.   
 $\mu_B=2$, $\mu_\Om=0.35$. 
$m=1$, $\rin=0.5$, perfectly conducting boundaries.}
\label{tidiffrot0}
\end{figure}

For the normalizations used in  Fig.~\ref{ti4} the growth rates for the $m=1$ instability of the stationary pinch are maximal  for $\Pm=1$, but for the rigidly rotating pinch $\Pm=1$ leads to a maximal stabilization (Fig.~\ref{tirigid1}). We shall find that also the combination of the almost uniform magnetic fields $\mu_B=1$ with differential rotation lead to a basic role of the magnetic Prandtl number. For perfectly conducting boundary conditions the critical Hartmann number for $\Rey=0$ is $\Ha_0=150$ (see Fig.~\ref{g42}, left). The form $\Om=\Om(R)$ of the rotational profile plays an important role for the destabilization  of the toroidal fields. For small magnetic Prandtl number Fig.~\ref{tidiffrot0} demonstrates various possibilities for modest Reynolds numbers and for quasi-uniform magnetic field. Rigid rotation basically stabilizes the field as instability requires increasing Hartmann numbers for increasing Reynolds numbers. This stabilization  is much weaker for  rotation laws with negative shear. For fixed $\Rey$ the field amplitudes which become  unstable are much weaker for subrotation than for rigid rotation, hence steep rotation laws effectively destabilize the field. Even a   subcritical excitation with $\Ha<\Ha_0$ exists, but only for  $\mu_\Om<\rin$. Note that  $\mu_\Om=0.5$ and $\mu_B=1$ in Fig.~\ref{tidiffrot0} according to relation  (\ref{chancon}) belongs to the class of Chandrasekhar-type flows which for $ \Pm\to 0$ scale with $\Rey$ and $\Ha$.   The nearly vertical dashed line in  Fig.~\ref{tidiffrot0} is valid for all  $\Pm<1$. It is the  rotation law for quasi-uniform linear velocity, $U_\phi\simeq$~const. The calculations show that for this particular rotation profile the rotational support or suppression of TI is minimized for moderate Reynolds numbers.   For very large Reynolds numbers, however,  {\em all curves}  turn to the right, describing an effective stabilization of the magnetic fields by differential rotation (see the right panel of 
Fig.~\ref{tidiffrot1a}, below). 

In order to reveal the $\Pm$ dependence of the effectiveness of subcritical excitation the right panel of Fig.~\ref{tidiffrot0} presents the stability map for the pinch-type field with quasi-Keplerian rotation in the standard gap with $\rin=0.5$, for various magnetic Prandtl numbers $\Pm\leq 1$. The critical Hartmann number for $\Rey=0$ does not depend on the magnetic Prandtl number, but surprisingly the instability curves for all $\Pm< 1$ also hardly differ if $\Rey<1000$. As in Fig.~\ref{g42} (left) the rotational suppression almost disappears for $\Pm<1$. For slow rotation and for $\Pm\simeq 1$ the instability even becomes subcritical ($\Ha<\Ha_0$), and the stabilization switches  to  destabilization. For faster rotation the subcritical excitation disappears, but the rotational suppression is weaker than for rigid rotation. For the phenomenon of subcritical excitation the influence of the magnetic Prandtl number is obviously  strong. Note that $\Pm=0.1$ already belongs to the small $\Pm$ regime where the subcritical excitations are very weak. Here also fluids with $\Pm=1$ behave exceptionally.

\begin{figure}[h]
\centering
 \includegraphics[width=8cm]{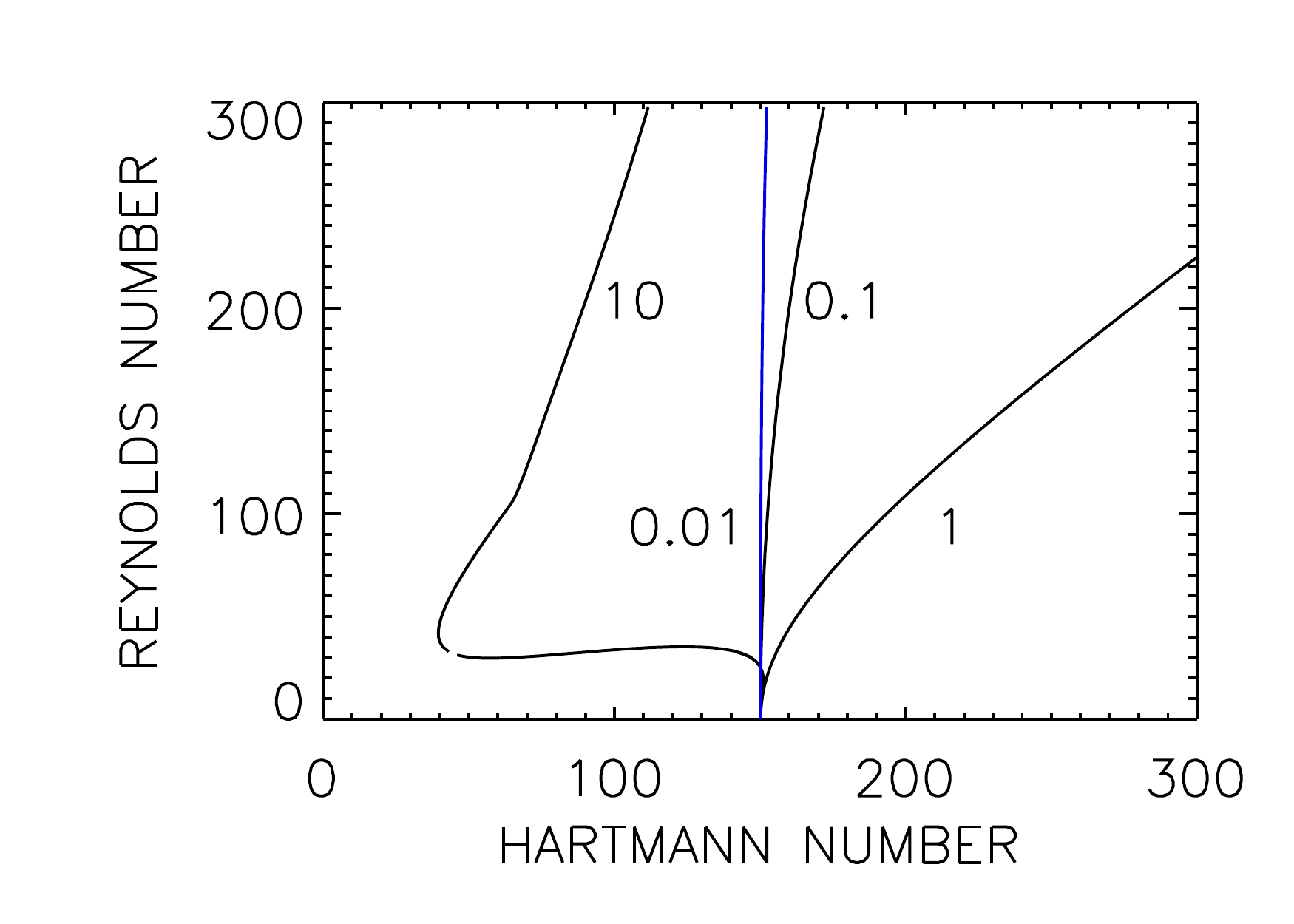}
 \includegraphics[width=8cm]{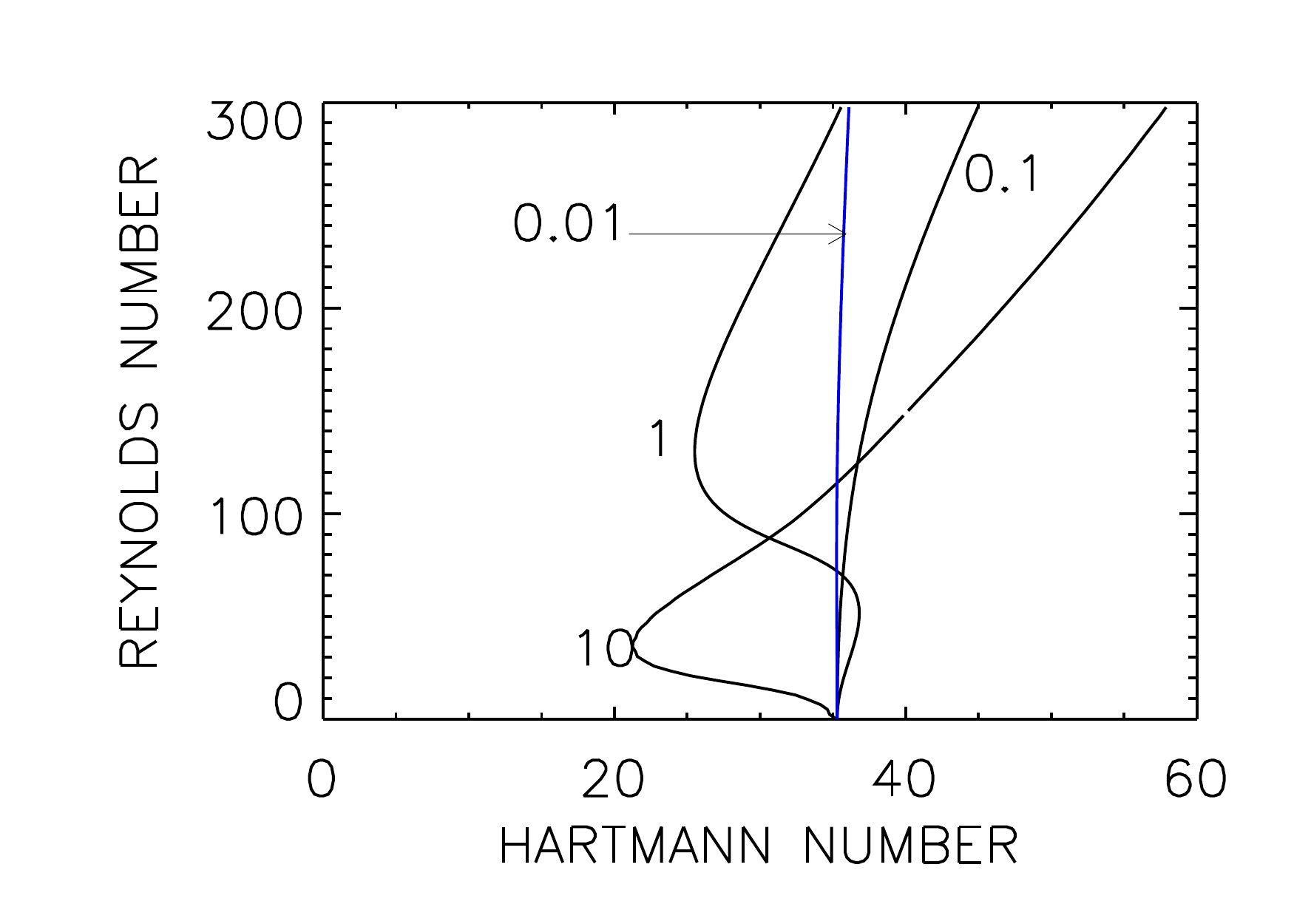}
 \caption{Stabilization  by uniform  flow with $\mu_\Om=\rin$ of    the quasi-uniform magnetic field ($ \mu_B=1$, left) and of fields due to uniform current ($ \mu_B=2$, right)  for various $\Pm$ (marked). The parameters used for the left panel satisfy the Chandrasekhar condition (\ref{chancon}). For small magnetic Prandtl number the lines of marginal stability become almost  perpendicular to the abscissa ($\Ha\simeq\Ha_0$, blue lines) so that the rotational suppression of TI disappears. The patterns of these modes are stationary in the laboratory system.  $m=1$, $\rin=0.5$.  Perfectly conducting boundaries.}
 \label{uniformflow}
 \end{figure}

\begin{figure}
 \centering
 \includegraphics[width=8cm]{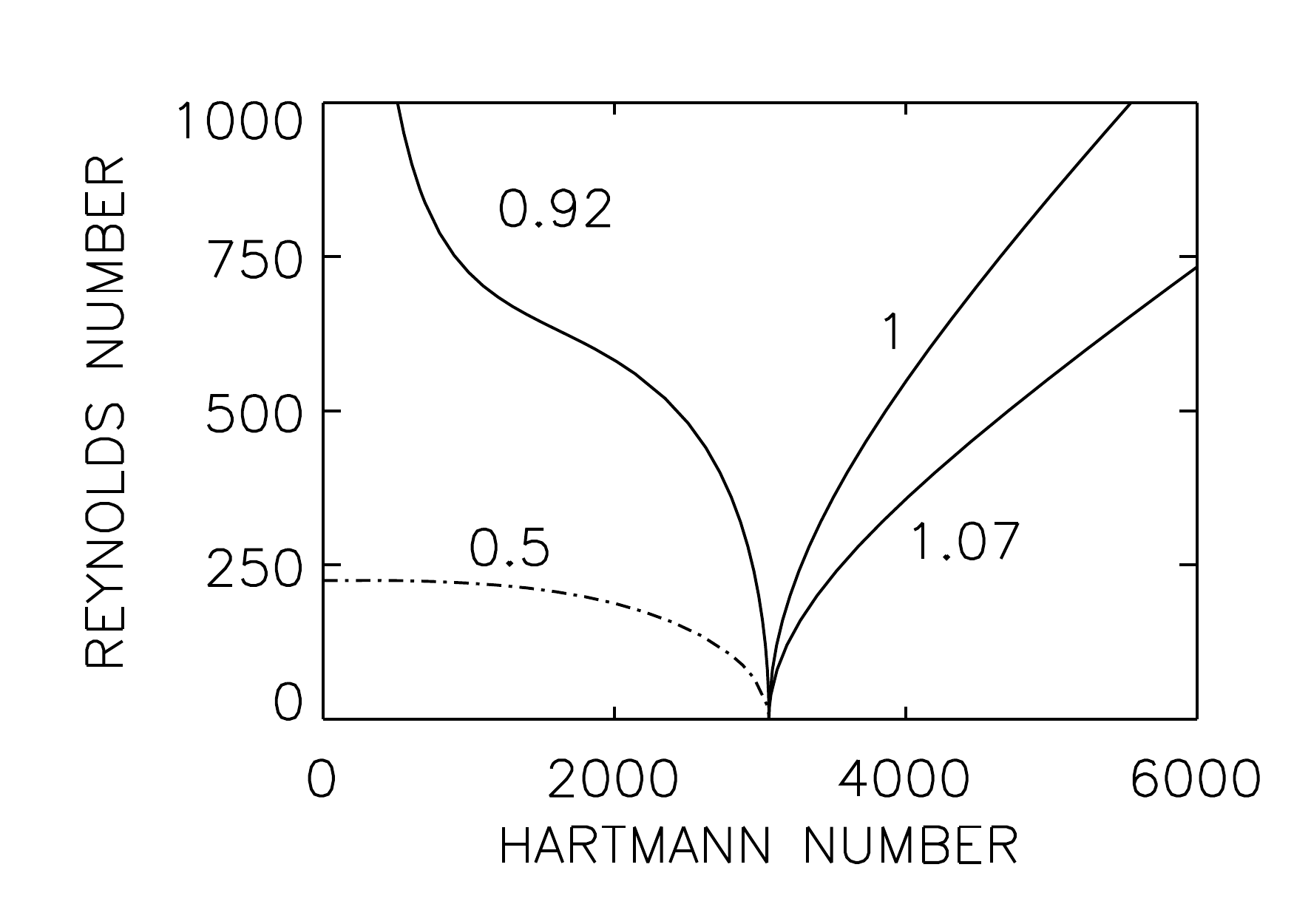}
 \includegraphics[width=8cm]{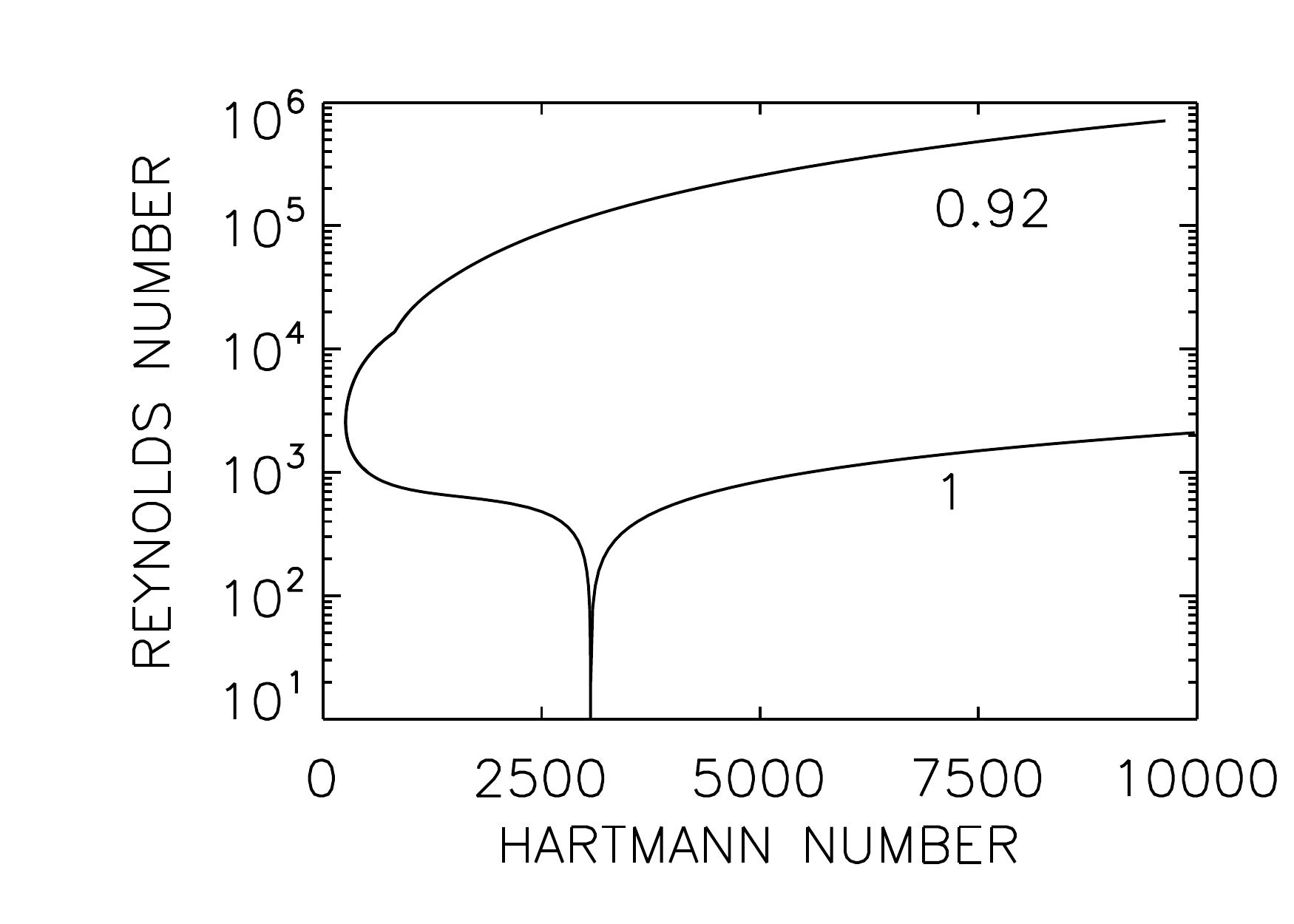}
 \caption{Stability maps for $z$-pinches  in a narrow gap with differential rotation; the curves are marked by the rotation rate ratio $\mu_\Om$. Left: slow rotation ($\mu_\Om=0.5$ is hydrodynamically stable). $\mu_\Om=0.92$ represents quasi-Keplerian rotation.  Right: for higher Reynolds numbers stronger fields are needed for instability. For large Reynolds numbers  rotational stabilization exists for all types of differential rotation.   $m=1$,  $\Pm=1$, $\rin=0.95$,  $\mu_B= 1/\rin$. Perfectly conducting boundaries, \cite{RS10}.}
 \label{tidiffrot1a}
\end{figure}

The described  effect of missing  rotational suppression is demonstrated in more detail by Figs.~\ref{uniformflow}   for various magnetic field profiles and magnetic Prandtl numbers. Indeed, for small magnetic Prandtl number the lines of marginal stability for both examples  become more and more perpendicular, fulfilling the condition $\Ha\simeq \Ha_0$. The entire domain with $\Ha>\Ha_0$ is unstable, independent of the Reynolds number as long as the rotation is slow  enough.
 
As another example for the complex character of the rotational  stabilization/destabilization, Fig.~\ref{tidiffrot1a} shows the results of varying rotation profiles on the $z$-pinch with uniform axial electric current ($\mu_B= 1/\rin$) in a narrow gap. Rotation profiles with both negative and positive shear are considered. In this container the profile $\mu_\Om=0.5$ is centrifugally unstable even without magnetic fields, but note that the magnetic field destabilizes the flow. The other rotation laws are stable in the hydrodynamic regime. Rigid rotation and also superrotation stabilize the magnetic field. For subrotation the behavior is opposite. While for rigid rotation and superrotation the critical Hartmann numbers grow for growing Reynolds number with $\Ha>\Ha_0$, for (slow) subrotation the associated Hartmann numbers represent subcritical excitation, i.e.~$\Ha<\Ha_0$. Again, the characteristic rotation parameter $\mu_\Om\simeq \rin$ seems to separate the two regimes.
One should think that strong rotational shear of any sign tends to suppress nonaxisymmetric patterns. We therefore expect that for sufficiently large $\Rey$ the subrotation curves would eventually also turn over toward larger Hartmann numbers. As seen in Fig.~\ref{tidiffrot1a} (right panel), this is indeed the case, for both rigid rotation {\em and} subrotation. Rigid rotation seems to be more effective in stabilizing the TI. The critical Hartmann numbers required for instability are much higher for rigid rotation than for differential rotation. There is thus an extra destabilization effect by differential rotation suppressing the nonaxisymmetric field perturbations. The magnetic Mach number for strongest fields remain smaller than unity for rigid rotation, in opposition to subrotation where the given parameters yield $\Mm\gg 1$. However, the curves suggest that in all cases $\Mm<1$ for $\Ha\to \infty$.

\begin{figure}[htb]
\centering
\includegraphics[width=4cm,height=8cm]{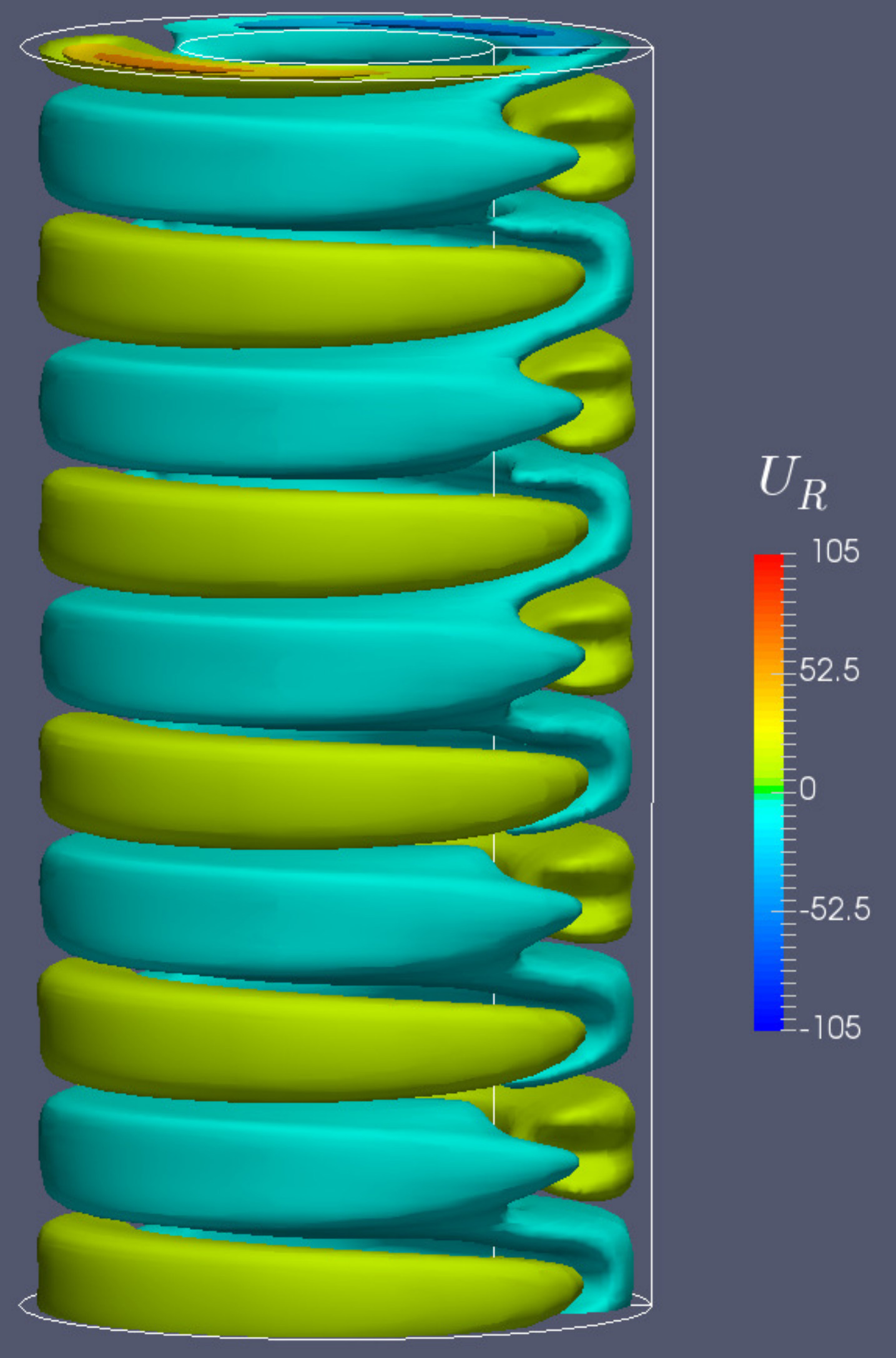}
\includegraphics[width=4cm,height=8cm]{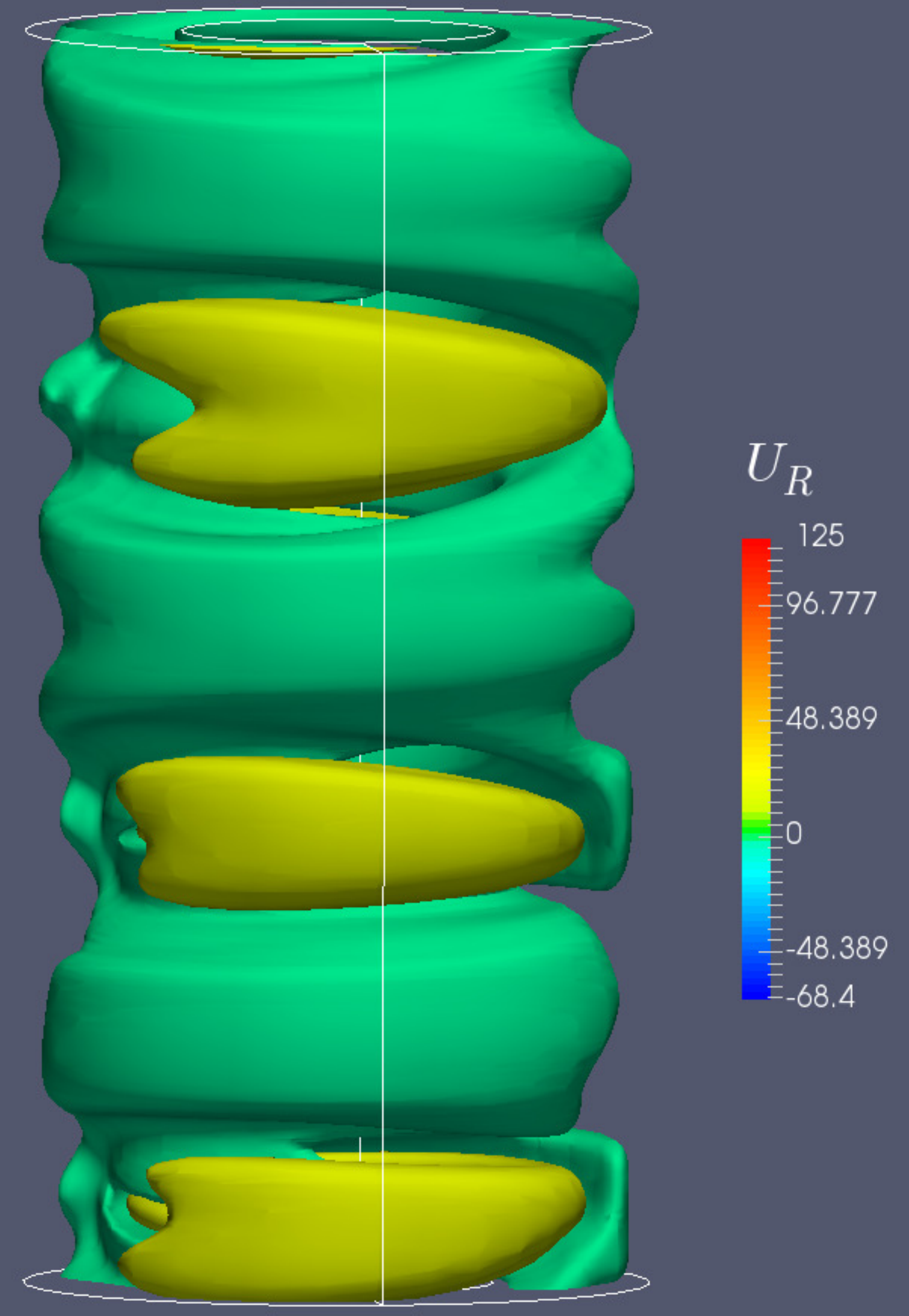}
\includegraphics[width=4cm,height=8cm]{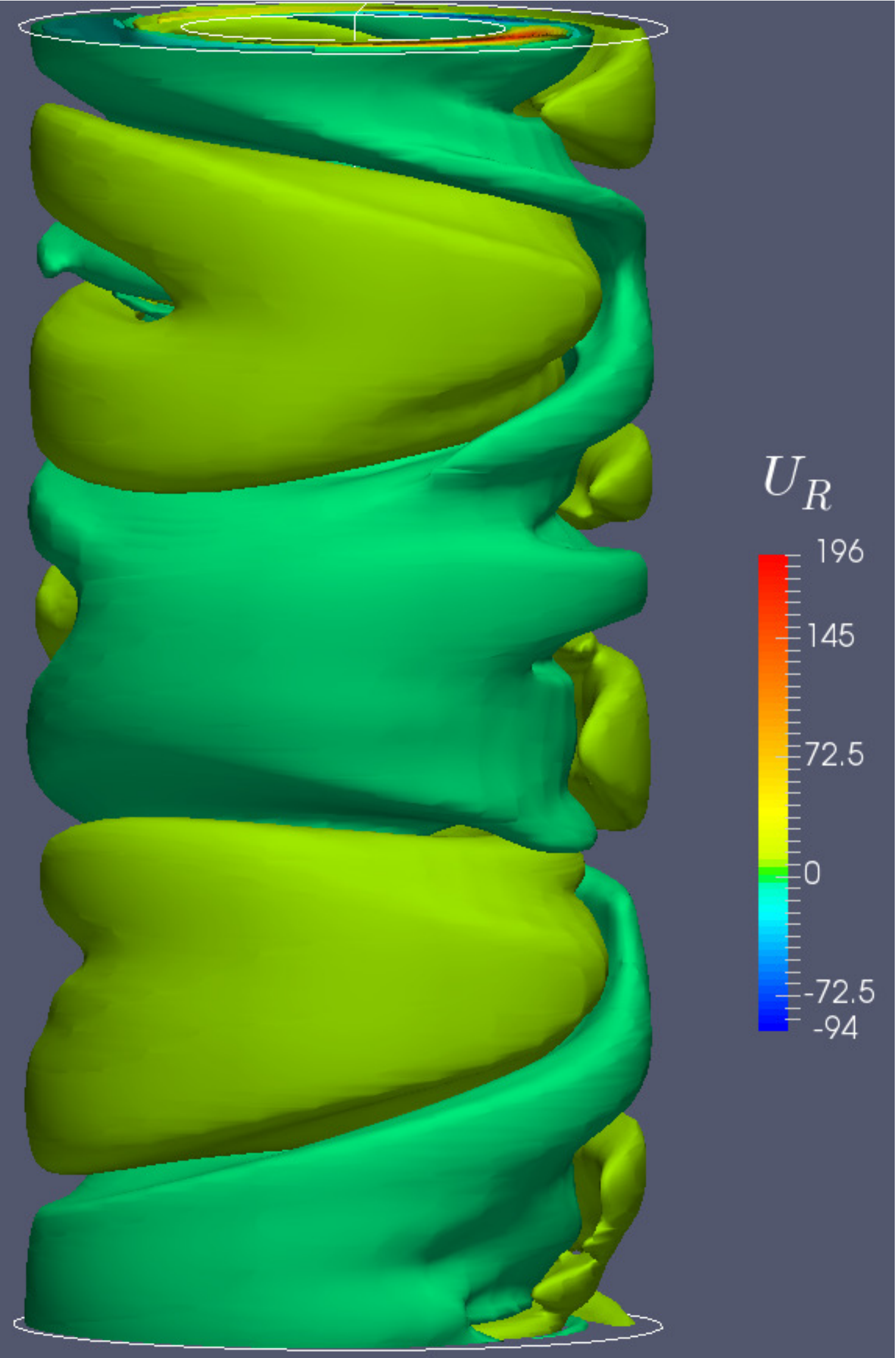}
\caption{$z$-pinch and quasi-Keplerian rotation: isolines of $u_R$ given as Reynolds numbers $u_R R_0/\nu$ for $\Rey=500$ (left), $\Rey=1000$ (middle) and $\Rey=1500$ (right). $\Ha=80$. $\rin=0.5$, $\mu_B=1/\rin=2$, $\mu_\Om=0.35$, $\Pm=0.1$. Perfectly conducting cylinders.}
\label{tidiffrot3}
\end{figure}

The flow pattern for a supercritical $z$-pinch under the influence of quasi-Keplerian rotation is shown in Fig.~\ref{tidiffrot3}  for three Reynolds numbers. The Hartmann number is fixed ($\Ha=80$). The magnetic Prandtl number $\Pm=0.1$ is small enough, according to the classification suggested by  Fig.~\ref{tidiffrot0} (right). The instability is clearly nonaxisymmetric and  the velocity amplitude increases linearly with the Reynolds number.
 Within the same interval the axial wavelength  grows for faster rotation. Surprisingly,  detailed numerical simulations lead to the result that  the rms value of the radial  velocity   remains constant  for faster rotation.  Simultaneously, the fluctuations become more and more asymmetric so  that, e.g. the rms values of the axial flow perturbations  grow for growing Reynolds numbers \cite{PG16}.
\subsection{Superrotation}
\begin{figure}[htb]
 \centering
 \includegraphics[width=5.3cm]{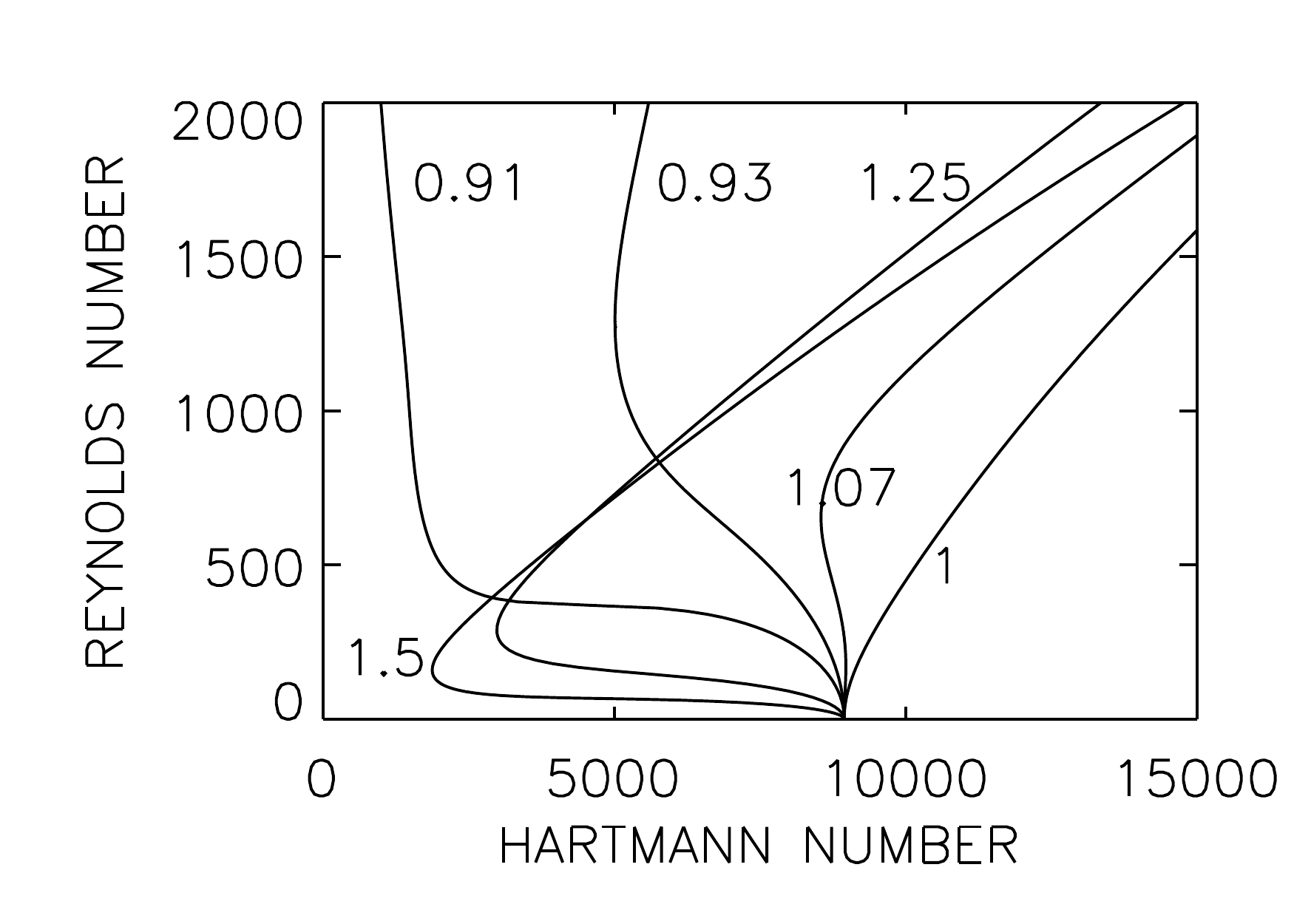}
 \includegraphics[width=5.3cm]{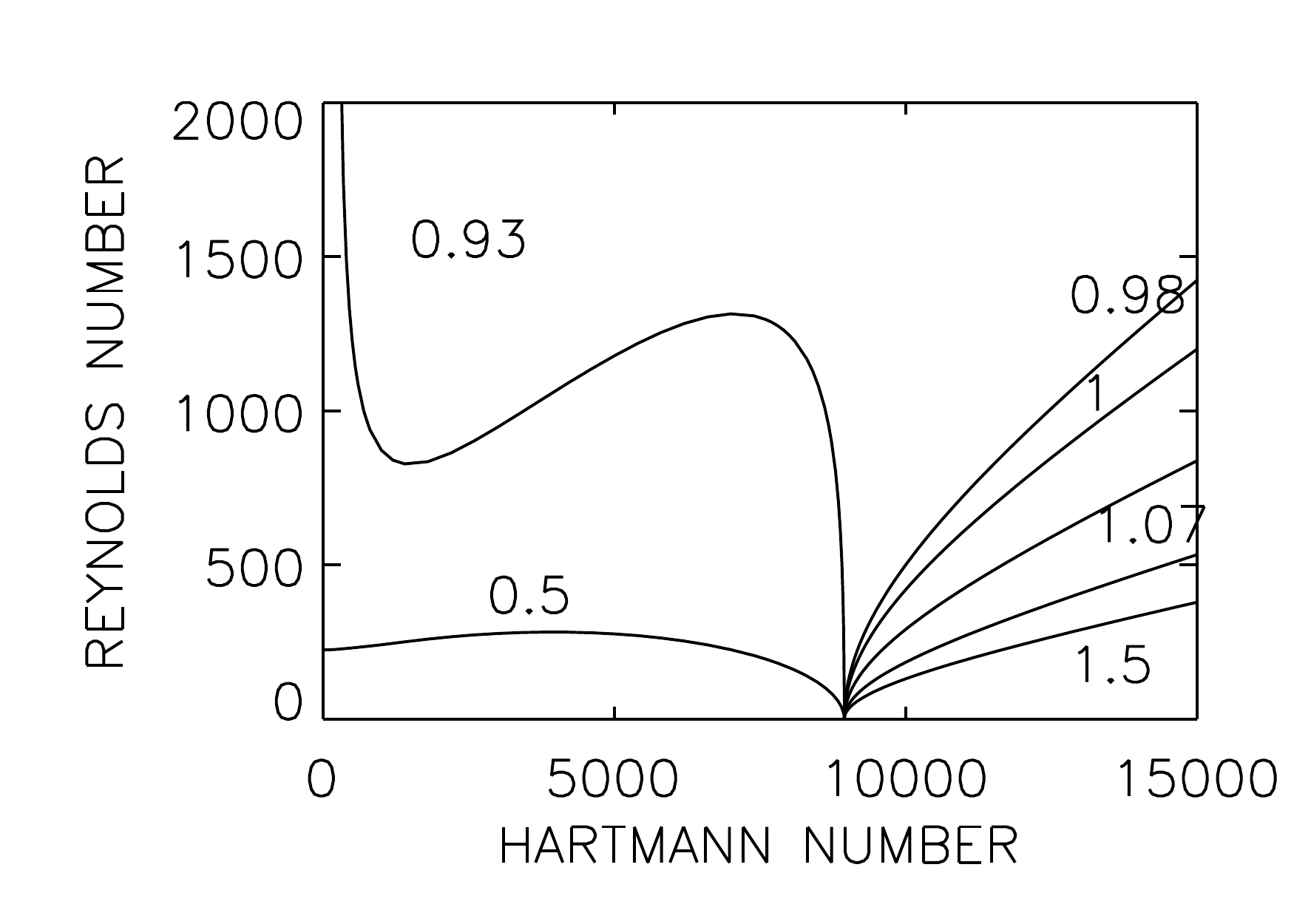}
 \includegraphics[width=5.3cm]{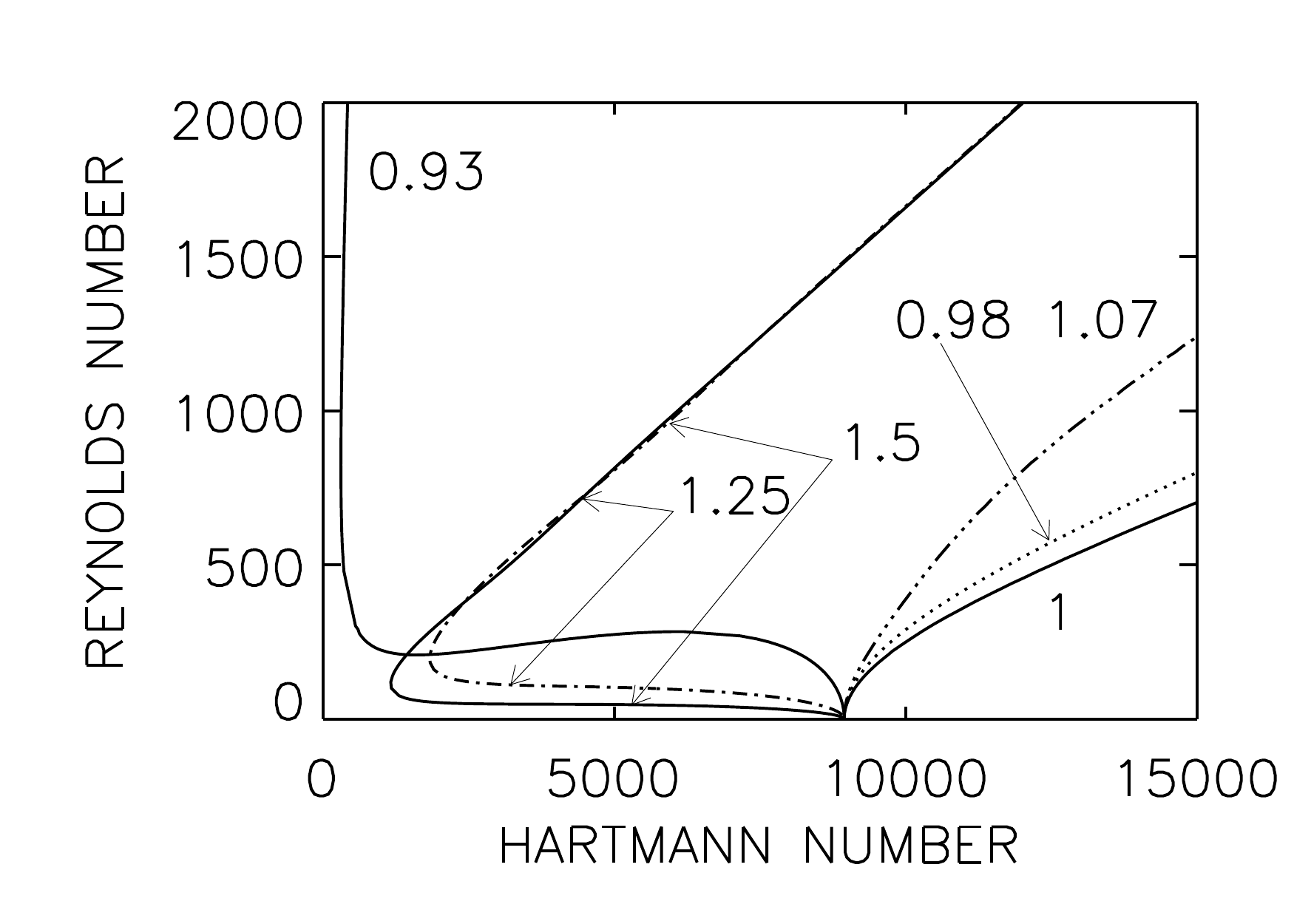} 
 \caption{Quasi-uniform field and differential rotation of negative and positive shear in a narrow gap for $\Pm=0.1$ (left), $\Pm=1$ (middle) and $\Pm=10$ (right). The curves are marked with  $\mu_\Om$. $\rin=0.95$, $\mu_B=1$, $m=\pm 1$, perfectly conducting boundaries. See the corresponding plots for the $z$-pinch with $\mu_B=1/\rin$ in Ref.~\cite{RS16}. }
 \label{TIdiffrot4}
 \end{figure}

The stability of the almost uniform field $\mu_B=1$ for rotation profiles with negative and positive shear is studied next, in dependence on the magnetic Prandtl number. For the narrow gap with $\rin=0.95$ Fig.~\ref{TIdiffrot4} gives maps for $\Pm=0.1$, $\Pm=1$ and $\Pm=10$. The Hartmann number without rotation is $\Ha_0\simeq 8945$. For $\Pm=1$ rigid-body rotation and superrotation are always stabilizing ($\Ha>\Ha_0$), opposite to subrotation with $\mu_\Om<\rin$. Sufficiently strong subrotation leads to subcritical excitation with $\Ha<\Ha_0$. Rotation laws with negative shear (here $\mu_\Om=0.93$) are strongly destabilizing. For small $\Pm$ the domain of stability in Fig.~\ref{TIdiffrot4} is larger. For sufficiently rapid rotation, however, the lines for subrotation must also turn to the right, stabilizing the system, since strong shear of either sign always suppresses nonaxisymmetric patterns. The lines of marginal stability for rigid rotation and for superrotation lie below $\Mm=1$.

Note, however, that for both $\Pm> 1$ (right panel) and $\Pm< 1$ (left panel) and for slow rotation, the rotation profiles with positive shear lead to subcritical excitations, but not for $\Pm=1$. Superrotation can only provide subcritical excitation for $\nu\neq \eta$ (double-diffusive instability). The magnetic Mach number for the subcritical excitation and for both magnetic Prandtl numbers is only $\Mm\simeq 0.05$. The curves for rigid rotation and for superrotation are again always located below the line $\Mm=1$. To exist for $\Mm>1$ the TI needs the action of a differential rotation law with strong negative shear.

\begin{figure}[htb]
 \centering
 \includegraphics[width=8cm]{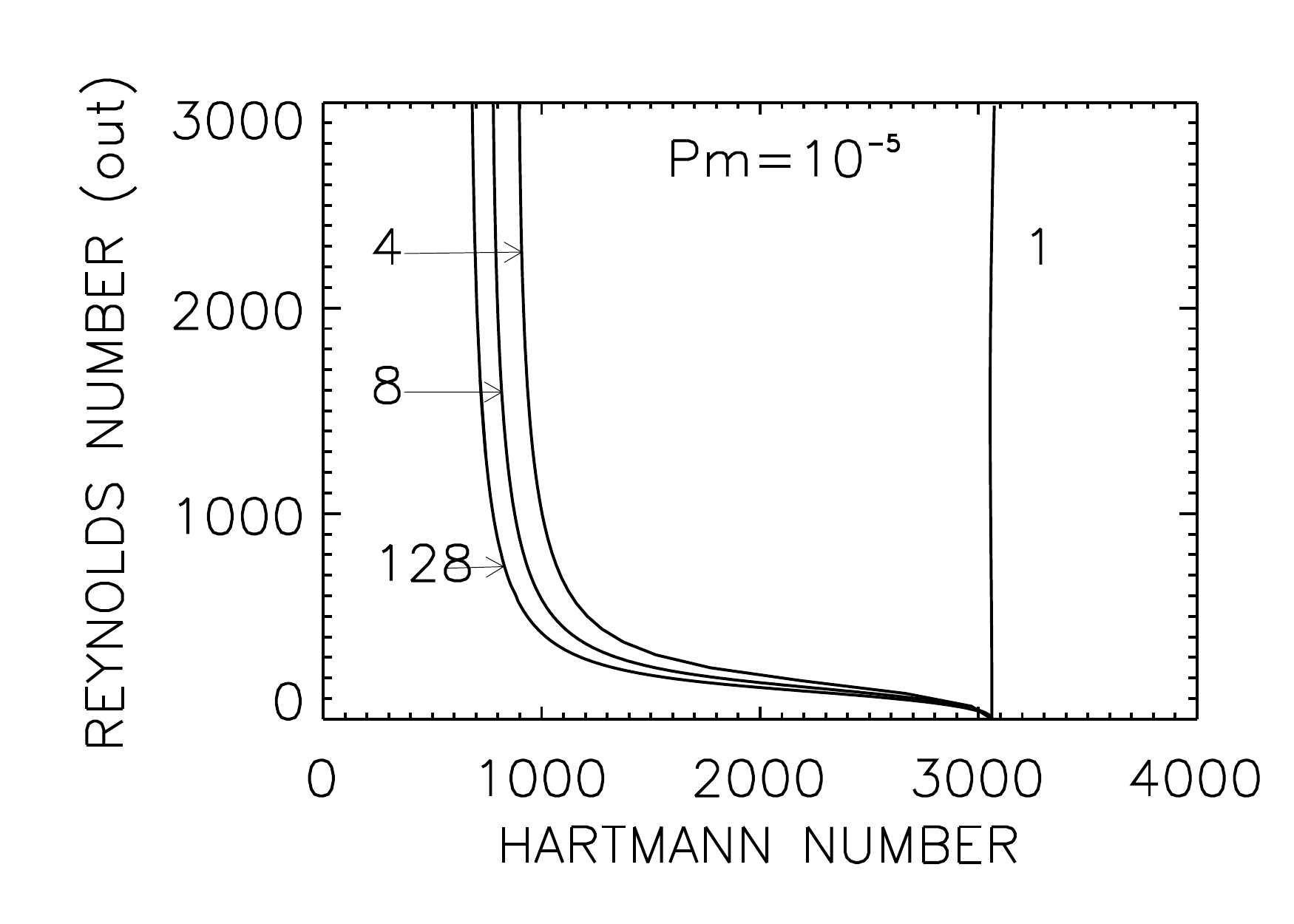}
 \includegraphics[width=8cm]{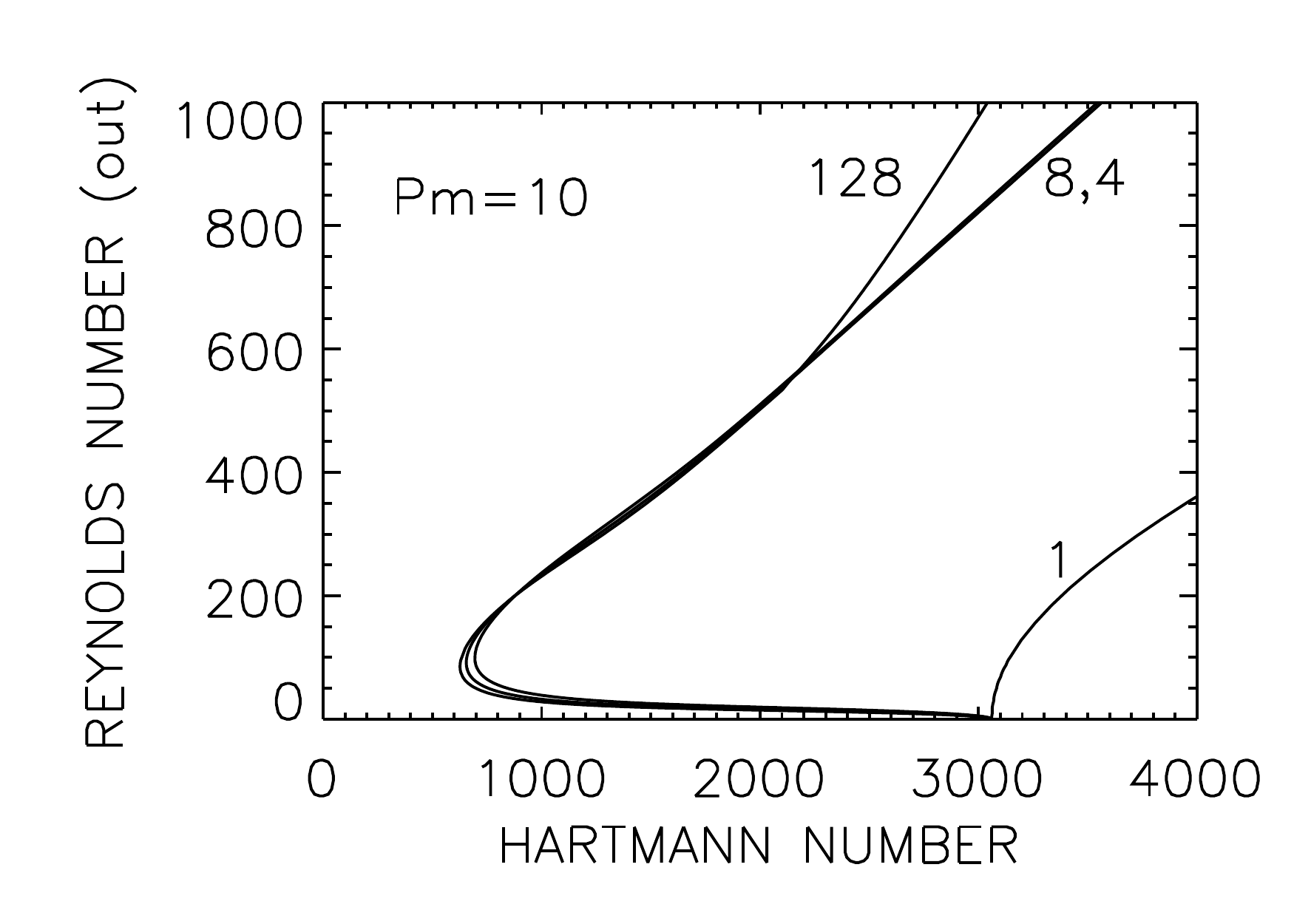}
 \caption{Stability maps  for  the $z$-pinch in a narrow gap subject to superrotation for small and large $\Pm$ (left: $\rm Pm=10^{-5}$, right: $\Pm=10$); the lines are marked with their values of $\mu_\Om$. Note that the Reynolds numbers are defined with the {\em outer} rotation rate. $m=1$, $\mu_B=1/\rin$, $\rin=0.95$, $\mu_\Om= 1 - 128$. Perfectly conducting cylinders.}
 \label{TIdiffrot5}
 \end{figure}

Superrotation at high Reynolds numbers is stabilizing for all $\Pm$, with the effect greater for small $\Pm$ than for large $\Pm$. For $\Pm=10$ the stabilization by superrotation is even weaker than that of rigid rotation. Here also large $\Pm$ destabilize nonuniform rotation, while small $\Pm$ stabilize them. The question arises about the possible existence of a minimum Hartmann number for steeper and steeper superrotation laws. The existence of such a limit is suggested by the suppression of nonaxisymmetric magnetic field perturbations by differential rotation whose effectiveness grows with increasing shear. The line of marginal stability can never cross the vertical axis, since nonmagnetically superrotation is always stable. Figure \ref{TIdiffrot5} for a $z$-pinch with uniform electric current shows converging lines up to $\mu_\Om\to 128$, so that a minimum Hartmann number $\Ha_{\rm min}$ exists and can be estimated as smaller by a factor of three compared with $\Ha_0=3060$. For the small magnetic Prandtl number used for Fig.~\ref{TIdiffrot5} (left) the ratio $\Ha_{\rm min}/\Ha_0$ is surprisingly small. For large $\Pm$ (right panel of Fig.~\ref{TIdiffrot5}, $\Pm=10$) the subcritical excitation also occurs with similar values. For larger Reynolds numbers almost all curves (except the curve for rigid rotation) are nearly identical; they only weakly depend on the numerical values of shear and electric current. Compared with the curves for small $\Pm$, however, the curves have a different form.

Another striking feature results from the comparison of Figs.~\ref{samri1} and Figs.~\ref{TIdiffrot5}, both for $\Pm\neq 1$ and for narrow gaps. The radial profiles of the magnetic fields completely differ: AMRI (no background current between the cylinders) for Fig.~\ref{samri1} and TI (background current between the cylinders) for Fig.~\ref{TIdiffrot5}. For the rapid-rotation branches the dependence of the Reynolds number on the Hartmann number is extremely weak. These plots show, however,  that  the dependence of the eigenvalues on the radial profile of $B_\phi(R)$ is extremely weak. The lines of neutral stability of the flow with and without axial current for rapid rotation almost coincide. For positive shear and rapid rotation the presence of the electric current becomes irrelevant for the occurrence of an instability. One can show that all possible radial magnetic profiles between $B_\phi\propto 1/R$ and $B_\phi\propto R$ provide more or less the same instability curves, revealing that any differential rotation for $\Pm\neq 1$ is able to deliver the entire energy for the maintenance of the instability patterns. The magnetic field only acts as a catalyst.

\subsection{Influence of the boundary conditions}
\begin{figure}[htb]
 \centering
 \includegraphics[width=8cm]{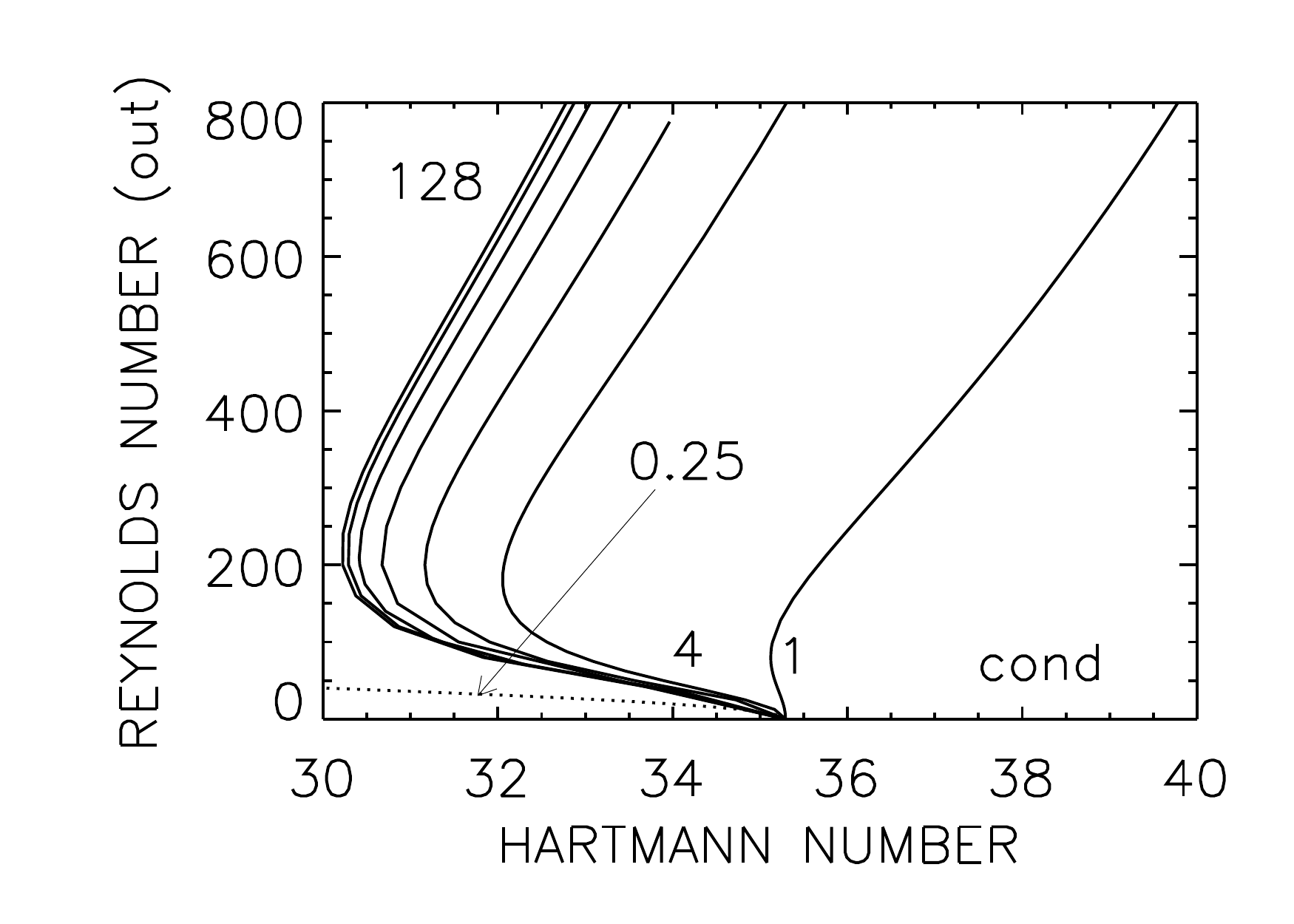}
 \includegraphics[width=8cm]{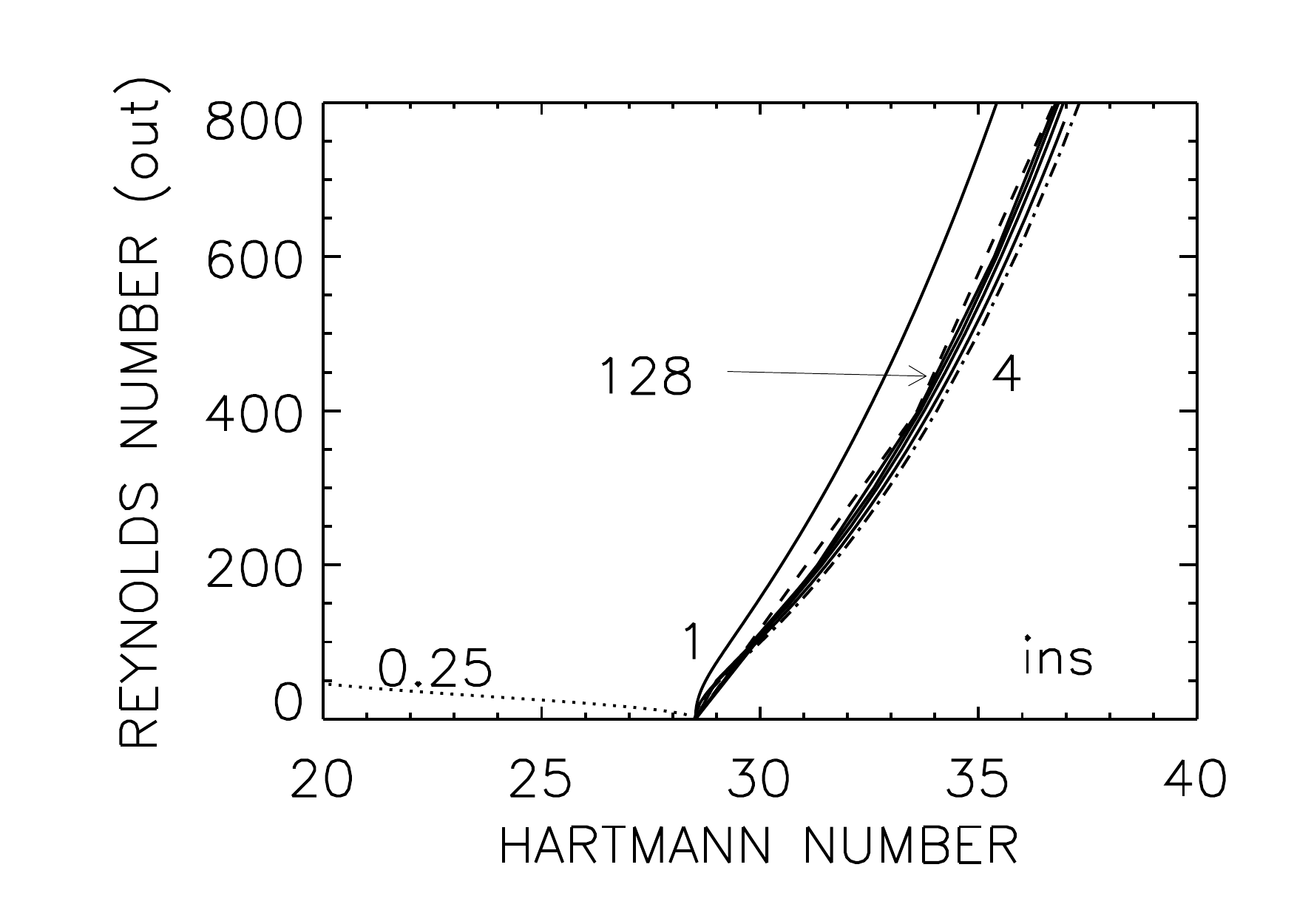}
 \caption{Stability maps of the $z$-pinch in a wide gap subject to superrotation for small $\Pm$ and for perfectly conducting (left) and insulating (right) cylinders. The Reynolds numbers are defined with the outer rotation rate, the curves are marked with their values of $\mu_\Om$. The solutions for rigid rotation ($\mu_\Om=1$) and  the potential flow ($\mu_\Om=0.25$) are given for comparison. Subcritical excitation ($\Ha<\Ha_0$) for superrotation  does {\em not } exist for insulating boundary conditions. For subrotation ($\mu_\Om=0.25$, dotted lines) it exists independent of the boundary conditions. 
$m=1$,   $\mu_B=1/\rin$,  $\rin=0.5$, $\Pm=10^{-5}$, $\mu_\Om= 1 - 128$. Perfectly conducting cylinders.}
 \label{TIdiffrot6}
 \end{figure}

For wide gaps there is a surprisingly strong influence of the boundary conditions, similar to the combination of AMRI and superrotation (see Section \ref{AMRIBC}). Figure \ref{TIdiffrot6} shows the stability maps for the standard container with $\rin=0.5$ for perfectly conducting and insulating cylinders. Note that the values for $\Ha_0$ for stationary  perfectly conducting  cylinders are larger than for stationary insulating cylinders. On the other hand, the critical Hartmann numbers for faster  rotation (say, $\Rey\gsim 150$) are nearly equal for both boundary conditions.  The subcritical excitation which can be observed in the left panel of Fig.~\ref{TIdiffrot6} (where  the superrotation laws are characterized by $1 < \mu_\Om \leq 128$) is a simple consequence of the strong influence of the boundary conditions for slow rotation (TI) and the weak influence of the boundary conditions for fast  rotation (AMRI). For comparison also the rotation law with $\mu_\Om=0.25$ is used, which shows the instability-supporting behavior (i.e.~subcritical excitation, $\Ha<\Ha_0$),  for b
oth boundary conditions occurring for all rotation laws with negative shear. For positive shear, however, this behavior only exists for cylinders made from perfectly conducting material. With insulating boundary conditions the superrotation laws {\em stabilize} the pinch with the uniform electric current ($\Ha>\Ha_0=35$). Observe   the convergence of the eigensolutions for $\mu_\Om\to \infty$. For large $\mu_\Om$ the Reynolds numbers taken for the outer cylinder also hold for the case of { stationary} inner cylinder.

For fast rotation (large Reynolds number) all nonaxisymmetric magnetic instabilities are suppressed, reducing the strong influence of the boundary conditions. Note, however, how easily a $z$-pinch can be stabilized by means of a slowly rotating outer cylinder made from insulating material. This effect  vanishes for high-conductivity material, large magnetic Prandtl number and fast rotation.

 \section{Twisted fields}\label{Twisted}
For combined axial and azimuthal background fields the nonaxisymmetric modes with $m$ and $-m$ (corresponding to left and right spirals) differ by the excitation conditions. If the imposed field has both axial and azimuthal components, the system no longer exhibits $\pm z$ symmetry \cite{K96}. We shall see that for nonaxisymmetric modes, therefore, the $\pm z$ asymmetry of the background field breaks the $\pm m$ symmetry of the instabilities. Spiraling either in the same or the opposite sense of the twisted field geometry is possible. This azimuthal-symmetry breaking by helical background fields forms a characteristic difference between rapid rotation (AMRI) and slow rotation (TI). For rapid rotation the most unstable mode spirals opposite to the imposed field; for slow rotation it spirals in the same sense (see Fig.~\ref{ti10} below as an illustrative example).

We are interested in the linear stability of the background field $\vec{B}= (0, B_\phi(R), B_0)$ with $B_z=B_0=$~const. For the current helicity of the background field one finds ${\rm curl}\vec{B}\cdot \vec{B} =2 a_B B_0$, which may be either positive or negative. Both signs yield the same instability curves with left and right spirals interchanged. This current helicity vanishes for fields which are current-free between the cylinders ($a_B=0$).  In accordance with Eq.~(\ref{Hartmannin}) the Hartmann number will be defined with the azimuthal field value $B_{\rm in}$, contrary to the definition (\ref{Hartmann}) used in Sections \ref{standardMRI} and \ref{HMRI}.

 \subsection{Quasi-uniform azimuthal field}
Following Refs.~\cite{RG10,BU11,RS11} we start to consider quasi-uniform azimuthal fields $B_\phi$ with  $ \mu_B=1$ and turn later to the pinch-type fields   due to  homogeneous electric currents with $\mu_B=2$. As in (\ref{beta}) the inner field amplitude $B_{\rm in}$ will be normalized with the uniform axial field $B_0$. 
Then the current helicity of the background field is
\begin{equation}
{\rm curl}\vec{B}\cdot \vec{B} =\frac{2\beta}{3} \frac{B_0^2}{R_{\rm in}}
\label{bet}
\end{equation}
with $\beta=B_{\rm in}/B_0$. The sign of $\beta$ determines the sign of the helicity of the background field. Interchanging $\pm\beta$ simply interchanges left and right spirals. As an exception, for almost uniform azimuthal background fields with $\mu_B=1$ both the Hartmann number and the ratio $\beta$ can also be imagined to be formed with the outer field amplitudes. If the axisymmetric background field possesses positive $B_\phi$ and $B_z$ then its current helicity is positive forming a right-hand spiral.

The phase relation (\ref{pitch}) gives the angle between the components of the perturbation field patterns. If the axial wave number $k$ is defined as a positive number (as we shall always do) then $m$ must be allowed to have both signs. Negative $m$ describe right-hand spirals, and positive $m$ describe left-hand spirals.

\begin{figure}[htb]
\centering
 \includegraphics[width=8cm]{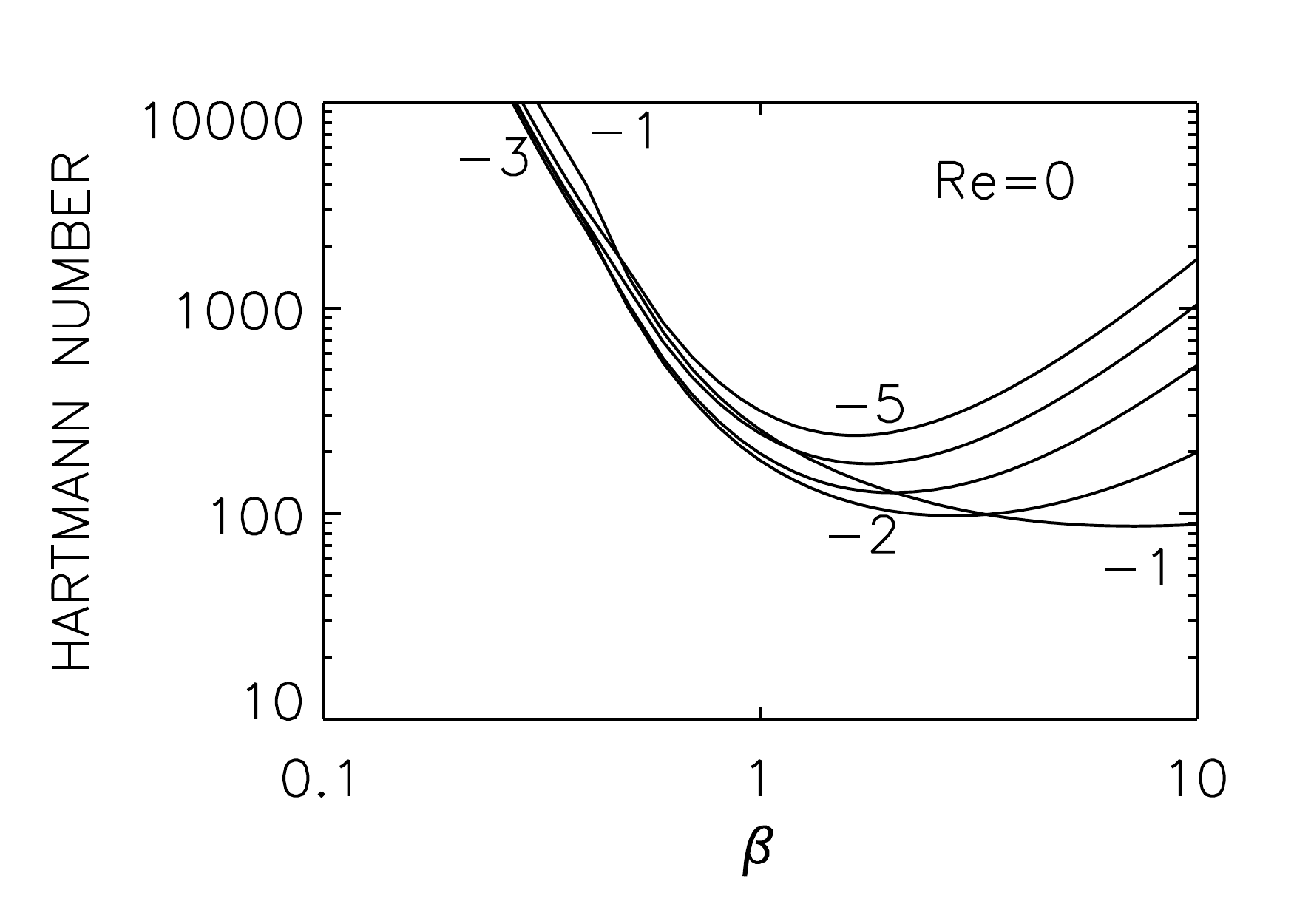}
 \includegraphics[width=8cm]{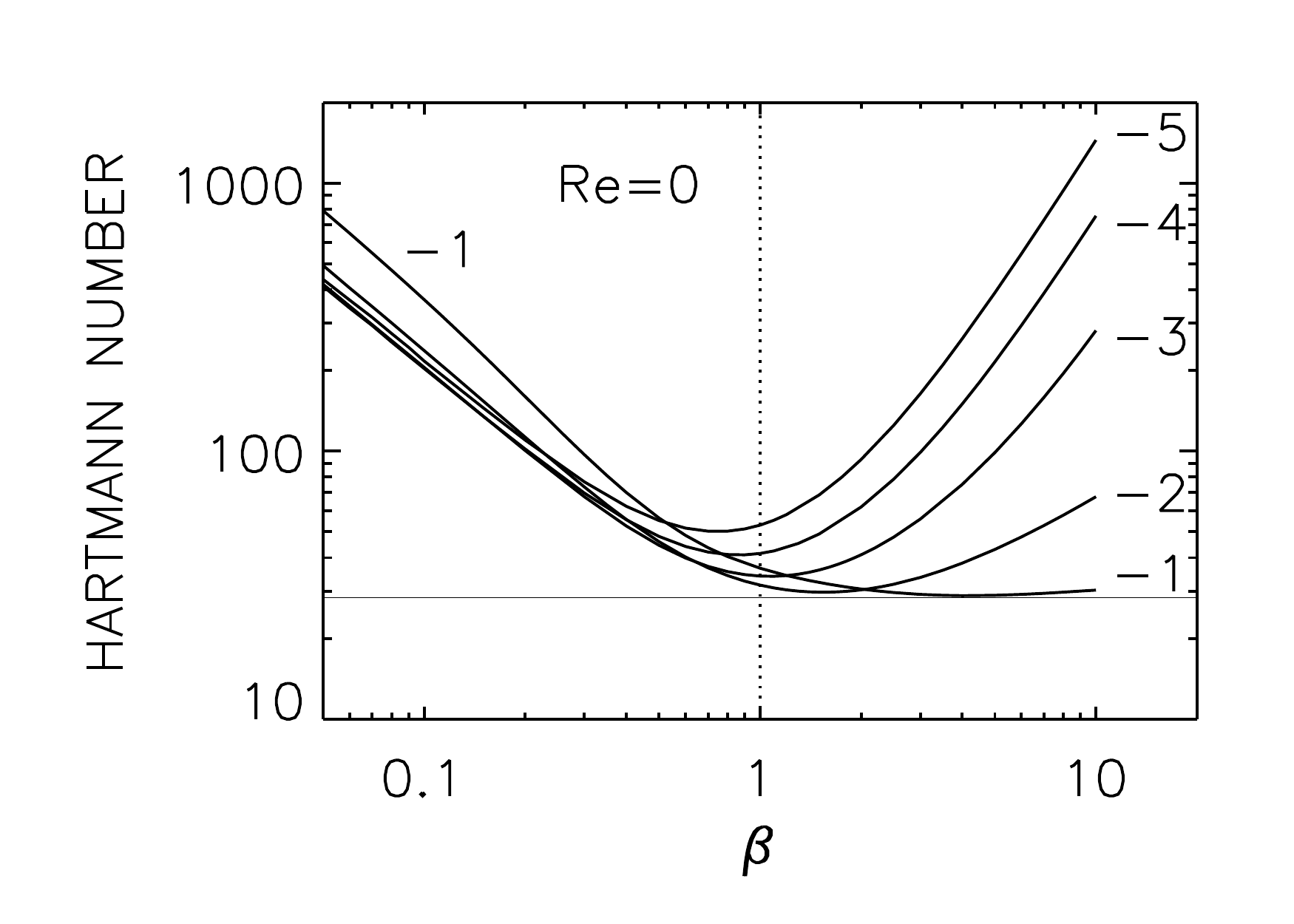}
 \caption{Twisted background fields:  Hartmann numbers $\Ha_0$ versus the  ratio $\beta>0$ for stationary cylinders. The curves are marked with their value of $m$, they are invariant against the simultaneous transformation $m\to -m$, $\beta\to -\beta$. Twisted fields are more unstable than non-twisted fields. Strong axial magnetic field components have a stabilizing influence. $\mu_B=1$ (left), $\mu_B=2$ (right). $\rin=0.5$, $\Rey=0$, all $\Pm$. Perfectly conducting cylinders. From \cite{RG10}.}
 \label{ti88}
\end{figure}

The critical Hartmann numbers $\Ha_0$ for nonrotating containers do not depend on $\Pm$. Hence, the results for $\Om=0$ in Fig.~\ref{ti88} for the modes with $m =-1,\dots, -5$ are valid for large magnetic Prandtl numbers and also for the small magnetic Prandtl numbers of liquid metals.
For $\mu_B=1$, $m=1$ and for perfectly conducting boundaries we have $\Ha_0=150$ for purely toroidal fields, i.e.~$\beta\to\infty$. For decreasing $\beta$ the critical Hartmann number is reduced to about 100. The most unstable mode is $m=-1$ for $\beta \gsim 8$. For $\beta$ of order unity the mode with $m=-2$ yields the lowest Hartmann number. For $\beta \lsim 0.4 $ the mode with $m=-3$ starts to be preferred. Even higher $m$ occur for smaller $\beta$ but an increase of the axial field component ($\beta \ll 1$) is strongly stabilizing, more so as the normalized differences of the critical Hartmann numbers for various $m$ become smaller and smaller. The energy needed to excite the nonaxisymmetric unstable modes  grows strongly with decreasing $\beta$. If the axial field for $\beta < 2$ starts to dominate the azimuthal field then the system becomes more and more stable. In the limit $\beta \to 0$ there is no unstable mode remaining. These results do not change if formulated with the Hartmann number of the {\em axial} field rather than with the Hartmann number of the toroidal field. For positive $\beta$ the twist of the background field is right-handed as is the twist of the most unstable modes. While without rotation for $B_z=0$ no preferred helicity exists for the instability pattern, with axial field the resulting twist is the same as that of the background field. 

\begin{figure}[htb]
\centering
 \includegraphics[width=5.2cm]{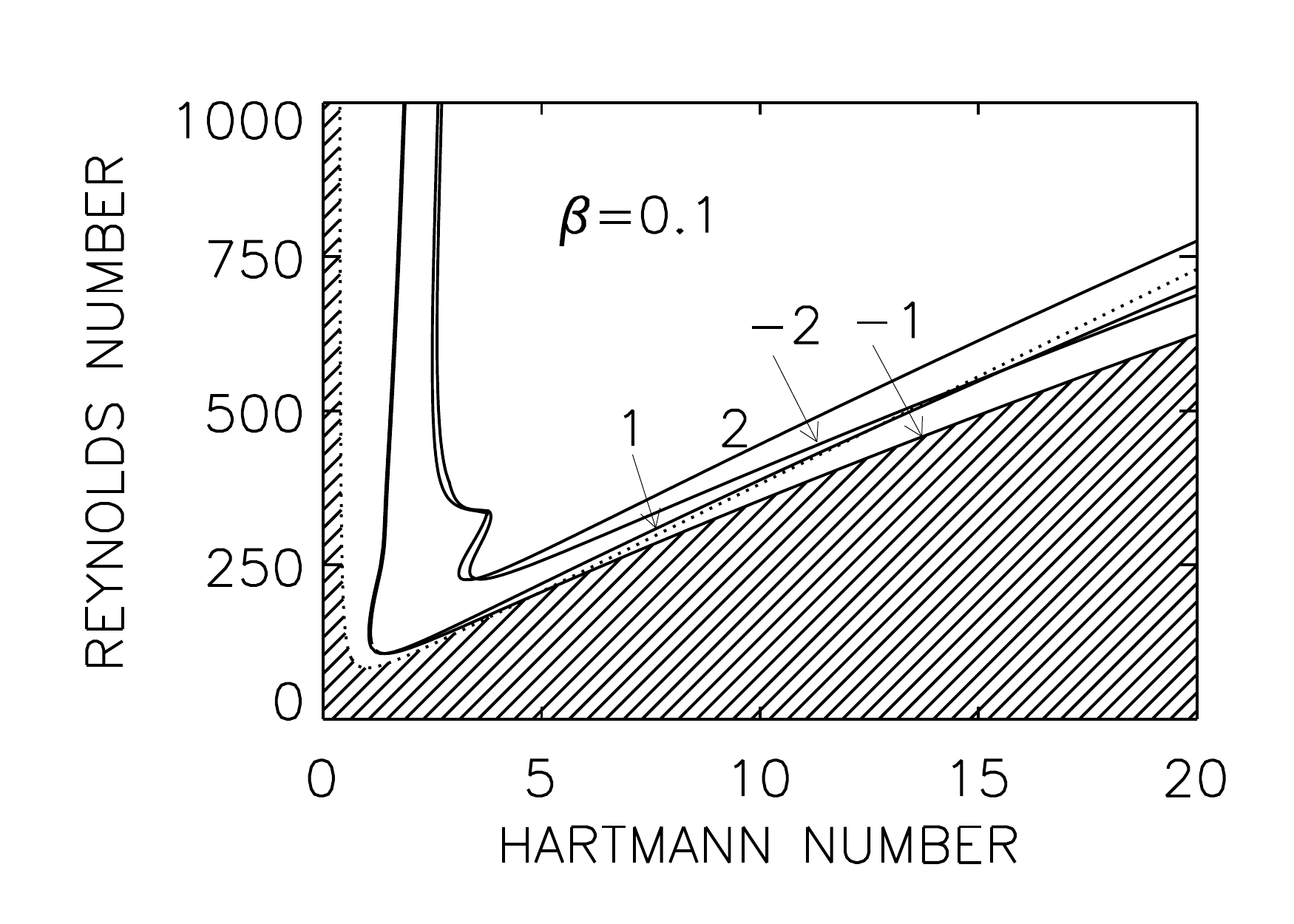}
 \includegraphics[width=5.2cm]{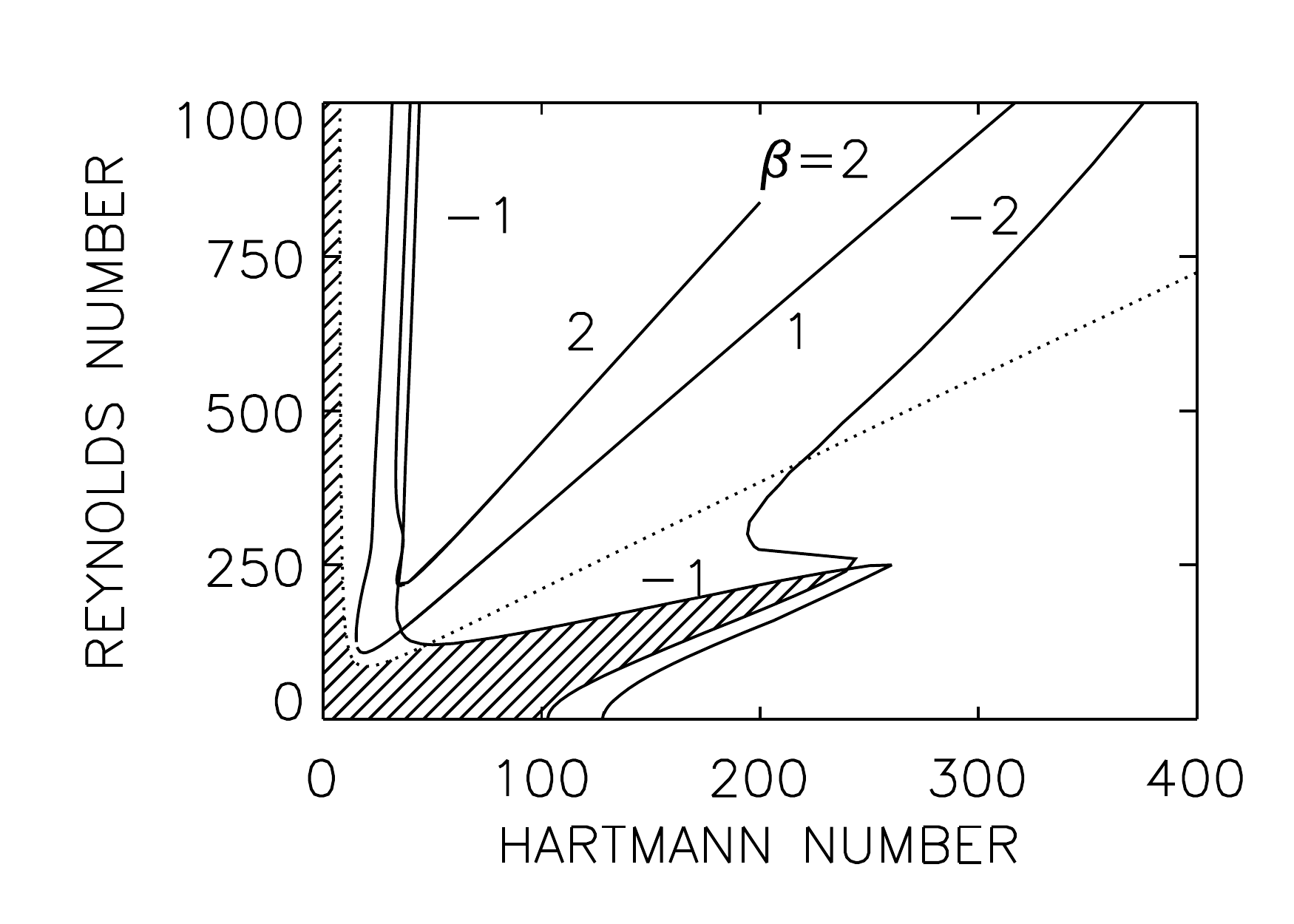}
 \includegraphics[width=5.2cm]{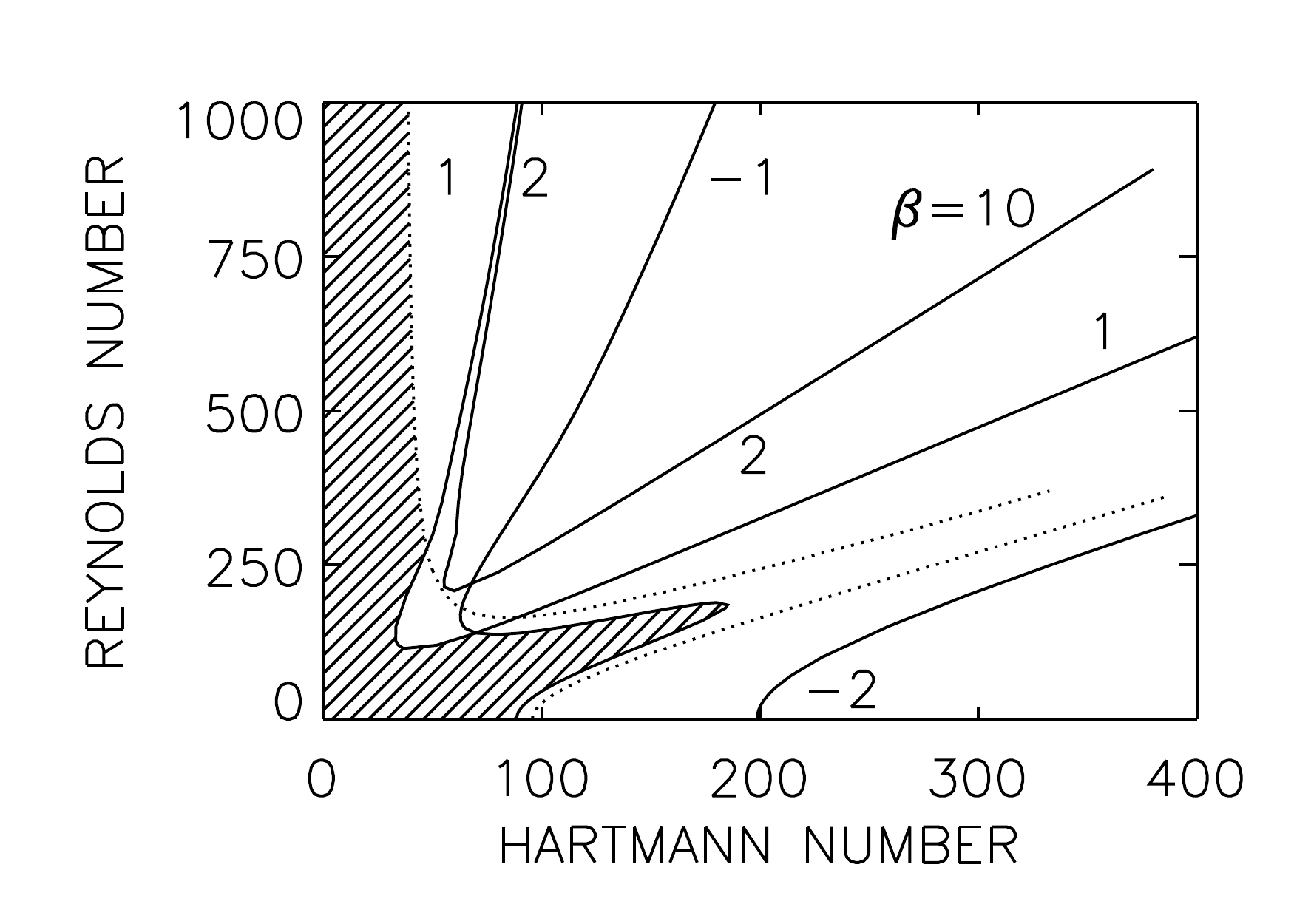}
 \caption{Stable (hatched) and unstable domains in the ($\Ha/\Rey$) plane for quasi-uniform azimuthal field, quasi-uniform flow and for various $\beta$. Left: $\beta=0.1$, middle: $\beta=2$, right: $\beta=10$. The curves are marked with their values of  $m$, for $m=0$ the lines are dotted. Negative $m$ stand for right-hand spirals and positive $m$ stand for left-hand spirals. $\rin=0.5$,  $\mu_B=2\mu_\Om=1$, $\Pm=1$, perfectly conducting cylinders.}
 \label{ti9}
\end{figure}

If the nearly homogeneous field with $\mu_B=1$ is subject to differential rotation with $\Om\propto 1/R$ (quasi-uniform rotation velocity), then the field and the flow belong to the Chandrasekhar-type considered in Section \ref{Chandra}. For $\Pm\to 0$ the corresponding eigenvalues $\Rey$ and $\Ha$ lose their dependence on  $ \Pm$. For $\Pm=1$ the instability curves for $\mu_B=2\mu_\Om=1$ are given in Fig.~\ref{ti9}. For $\beta\lsim 1$ the $m=0$ mode (dotted line) yields the instability with the lowest Reynolds number. As also  in Fig.~\ref{g42} for $\beta$ of order unity a stability branch develops along the line for $\Mm=1$. We find for the AMRI domain ($\Mm>1$) that for large $\beta$ the lowest Reynolds number belongs to nonaxisymmetric modes. The transition from nonaxisymmetry to axisymmetry can be accomplished simply by increasing the axial component of the background field. It is thus clear that there is a smooth transition from AMRI to the standard MRI.

The middle plot of Fig.~\ref{ti9} for $\beta$ of order unity shows as Fig.~\ref{ti88} that the mode $m=-2$ indeed possesses lower $\Ha_0$ than $m=-1$ (see \cite{BU08}). Small shear and larger $\beta$, however, bring $m=-1$ back to the leading mode with the lowest critical Hartmann number.

Note also that the slopes of the lines in Fig.~\ref{ti9} change from positive for nonaxisymmetric modes to negative for axisymmetric modes.  If the preferred modes with the lowest Reynolds numbers are nonaxisymmetric (for large $\beta$) then the spirals are always left-handed in the AMRI domain ($\Mm>1$) and right-handed in the TI domain ($\Mm<1$). 
\begin{figure}[htb]
\centering
 \includegraphics[width=4cm,height=6cm]{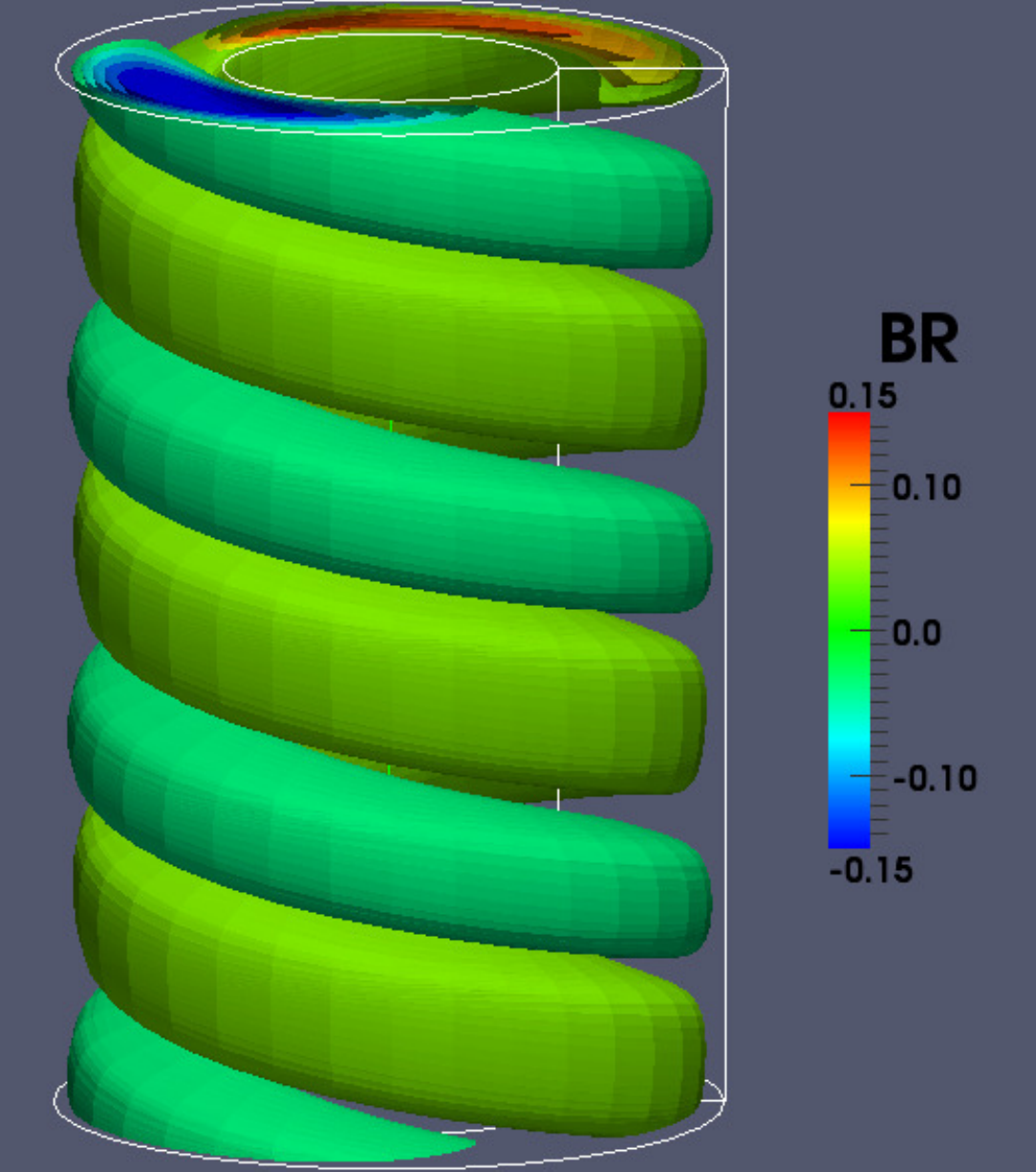}
 \includegraphics[width=4cm,height=6cm]{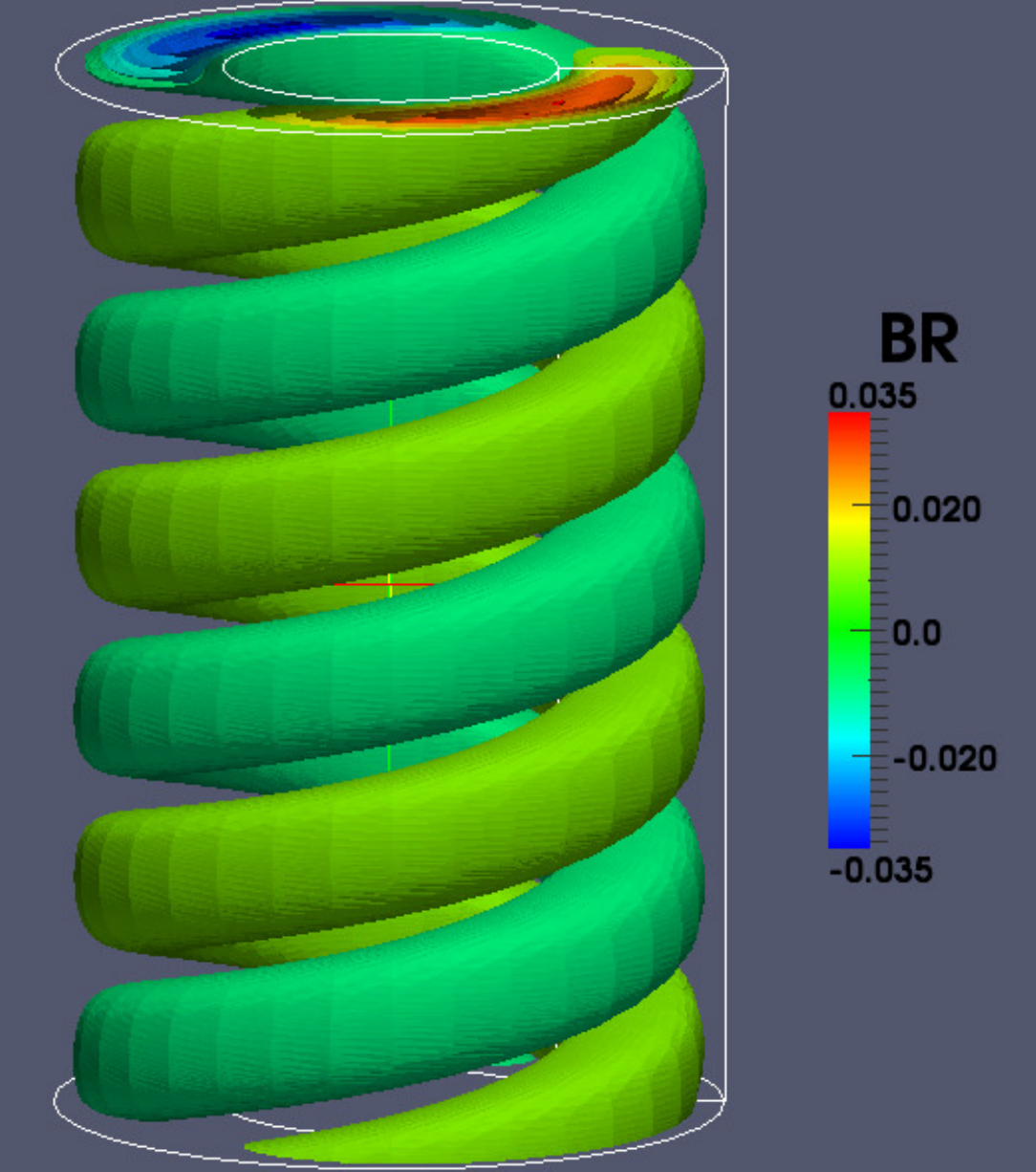}
 \caption{The radial component of the magnetic pattern for models taken from the right panel of Fig.~\ref{ti9}. Left: fast rotation with $\Mm=2.5$ ($\Rey=200$, $\Ha=80$, AMRI-type). Right: slow rotation with   $\Mm= 0.23$ ($\Rey=30$, $\Ha=130$, TI-type). The fields are normalized with $B_{\rm in}$. $\mu_B=2\mu_\Om=1$, $ \beta=10$, $\Pm=1$, perfectly conducting cylinders. Adapted from \cite{RG10}.}
 \label{ti10}
\end{figure}

For nonlinear simulations we begin by noting that transforming $\beta\to -\beta$ has the expected result that positive/negative $\beta$ do indeed yield right/left spirals. The helical structure of all solutions is clearly visible, dominated by low Fourier modes $m=1$ and/or $m=2$ in agreement with the linear analysis. The solutions are stationary, except for a drift in the azimuthal direction. Figure \ref{ti10} (left) concerns the AMRI domain for fixed $\beta=10$. One finds the expected $m=1$ left-hand spirals in agreement with the linear results in Fig.~\ref{ti9}. The nonaxisymmetric modes in the AMRI domain for large $\beta$ have the signature $m=1$. For the TI domain the most unstable mode is $m=-1$ for $\beta \gsim 10$ and $m=-2$ for $\beta\simeq 1$. Obviously, the unstable modes which characterize AMRI and TI according to the magnetic Mach number have different helicities. No mode mixture exists.

The simulations also provide the amplitudes of the kinetic helicity $\langle\vec{u}\cdot {\rm curl}\ \vec{u}\rangle$ of the perturbations (averaged over $\phi$). The two models of Fig.~\ref{ti10} with positive $\beta=10$ provide negative values of order of $\langle\vec{u}^2\rangle / R_0$. The signs of the kinetic helicity and the current helicity of the background field are opposite. According to Fig.~\ref{ti10} AMRI with $\Mm>1$ produces instability patterns with higher field strengths than TI with $\Mm<1$ does. This effect may be due to the action of the differential rotation. Indeed, the magnetic energy $\rm Q$ (normalized with $B^2_{\rm in}$) calculated with the amplitudes of both examples differs by a factor of almost 20, which is just of the order of the ratio of the two magnetic Mach numbers. 

 \subsection{Uniform axial current}
We turn next to the pinch-type field due to a uniform electric current, Ref.~\cite{RS11}. Without rotation the critical Hartmann number $\Ha_0$ does not depend on $\Pm$ and the azimuthal drift of the nonaxisymmetric instability pattern vanishes. We also know that for the nonaxisymmetric mode with $|m|=1$ for very large $\beta$ we have $\Ha_0=35$ for $\mu_B=2$. 
This value is reduced if a small uniform axial field is added to the system. The axial field supports the pinch-type instability of the toroidal field. The critical Hartmann number reduces to $\Ha_0\simeq 30$. However, for $|m|>1$ the destabilization of the toroidal field by axial fields is much stronger, so that for $\beta$ of order unity all modes with different mode numbers $m$ possess the same critical Hartmann number. We thus find a destabilizing effect by axial fields components compared to fields of purely toroidal fields. If $B_\phi$ and $B_z$ are of the same order then the field is more unstable than it is for $B_z=0$ or $B_\phi=0$.

For $\beta=4$ we find ${\Ha_0}=29$ as the absolute minimum of the stability curve for $m=-1$. For stronger axial fields, the critical Hartmann number increases strongly to reach values of about 500 for $\beta \simeq 0.1$. Again, for strong axial fields the modes with $m<-1$ possess {\em lower} critical Hartmann numbers than those with $m=-1$. For the smallest $\beta$ in Fig.~\ref{ti88} (left) the $m=-4$ mode possesses the lowest critical Hartmann number. However, the pinch-type instability of toroidal fields in the presence of a uniform axial magnetic field without rotation is strongly suppressed by strong axial fields. The maximal stabilization happens for $m=-1$. With a sufficiently strong axial field rather strong toroidal fields can be stabilized.

The growth rates in units of the diffusion frequency $\omega_\eta=\eta/R_0^2$ for fixed $\beta=1$ and $\Pm=1$ are plotted in the left panel of Fig.~\ref{ti13}. The plot clearly demonstrates that the growth rates scale with the Hartmann number. For stronger fields, differences for the growth rates of various $m$ appear. One finds the maximum growth rates belonging to azimuthal wave numbers $|m|>1$.

\begin{figure}[htb]
\centering
 \includegraphics[width=4cm,height=6cm]{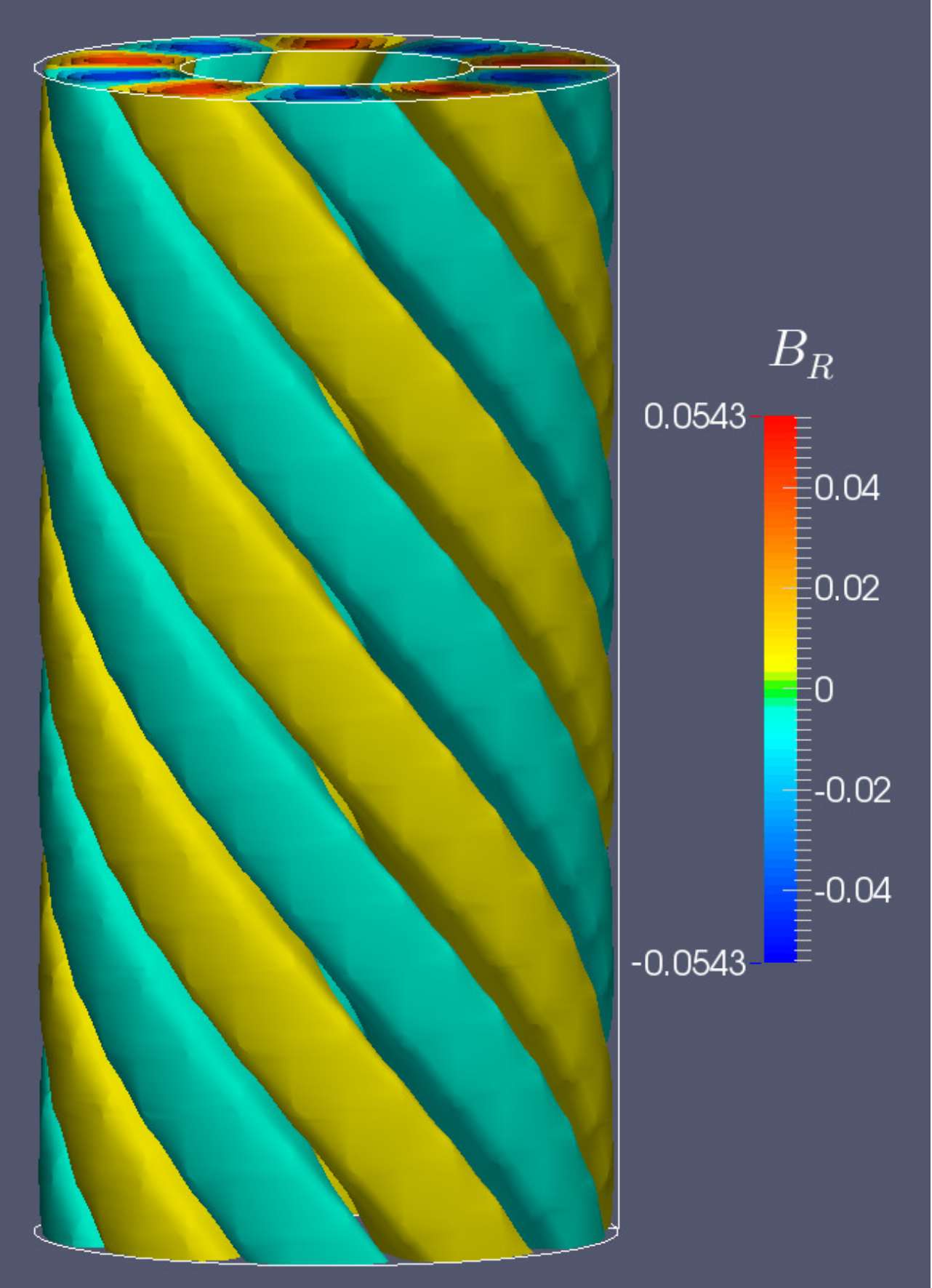}
 \includegraphics[width=4cm,height=6cm]{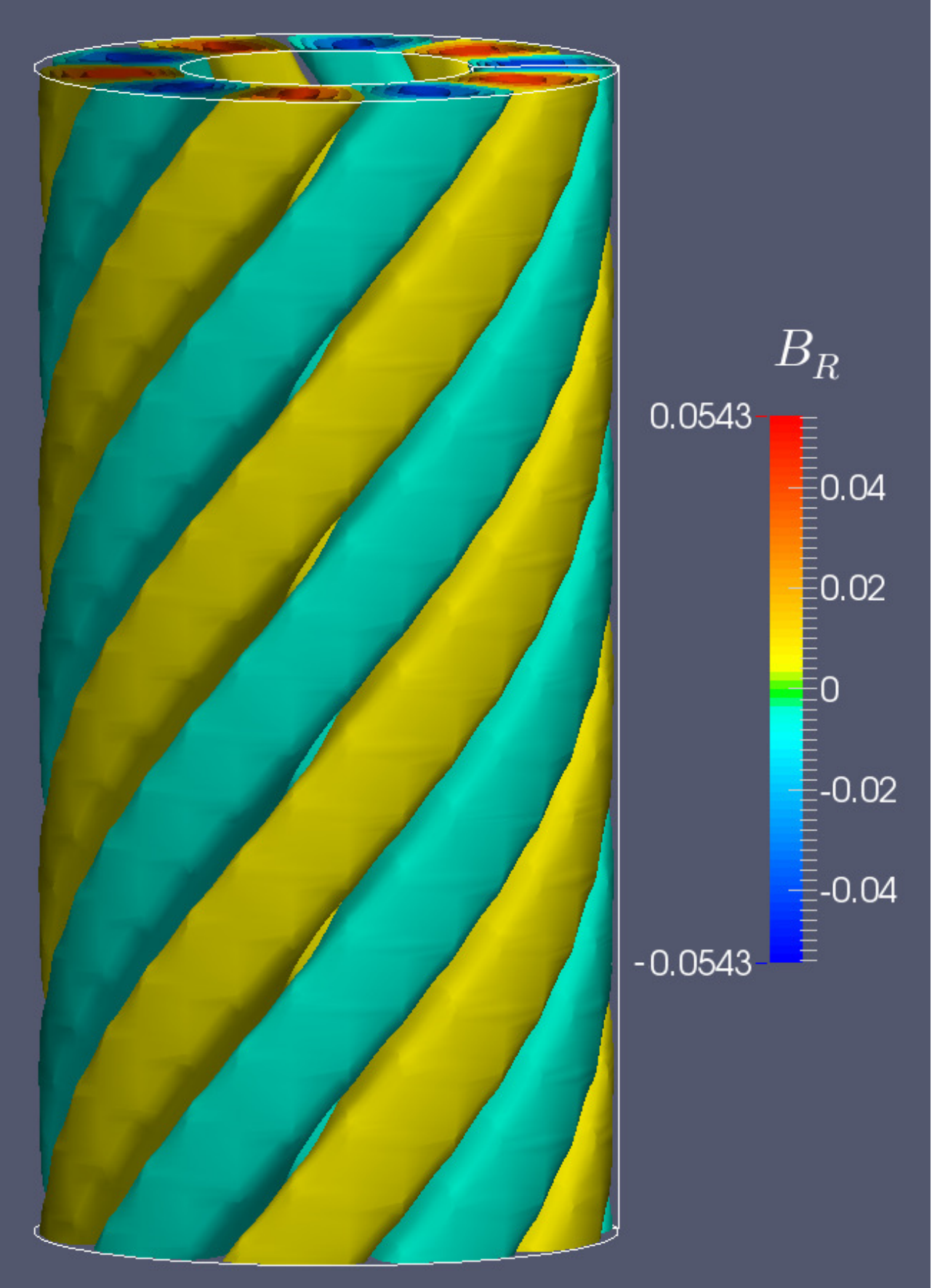}
 \caption{Numerical simulation of the instability patterns of a twisted background field of pinch-type without rotation with $\beta=-0.5$ (left) and $\beta=0.5$ (right). The signs of the helicities of background field and instability always coincide. $\rin=0.5$, $\mu_B=1/\rin$, $\Rey=0$, $\Ha=150$,  $\Pm=0.1$.}
 \label{ti12}
\end{figure}

\begin{figure}[htb]
\centering
  \includegraphics[width=5.25cm]{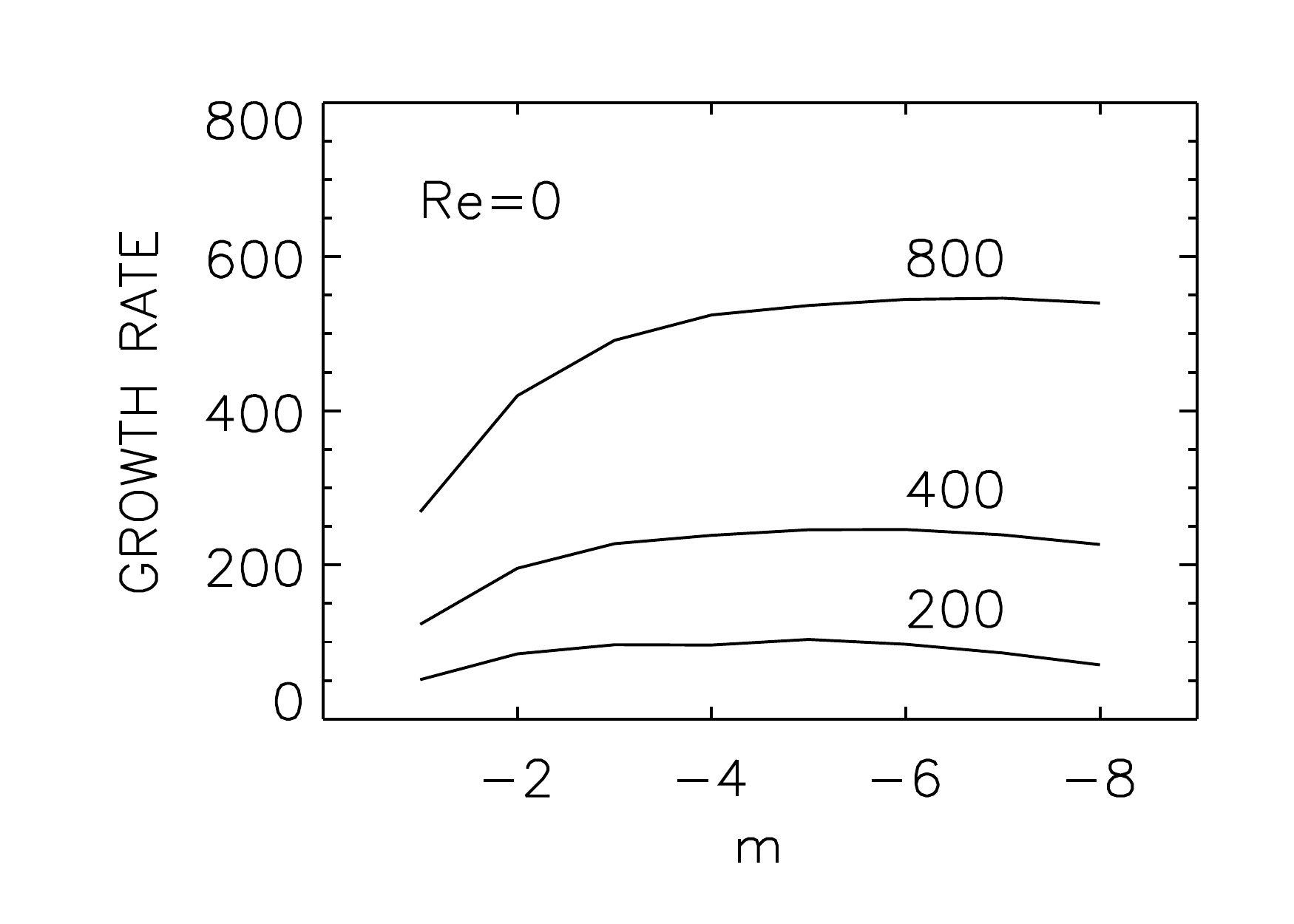}
 \includegraphics[width=5.25cm]{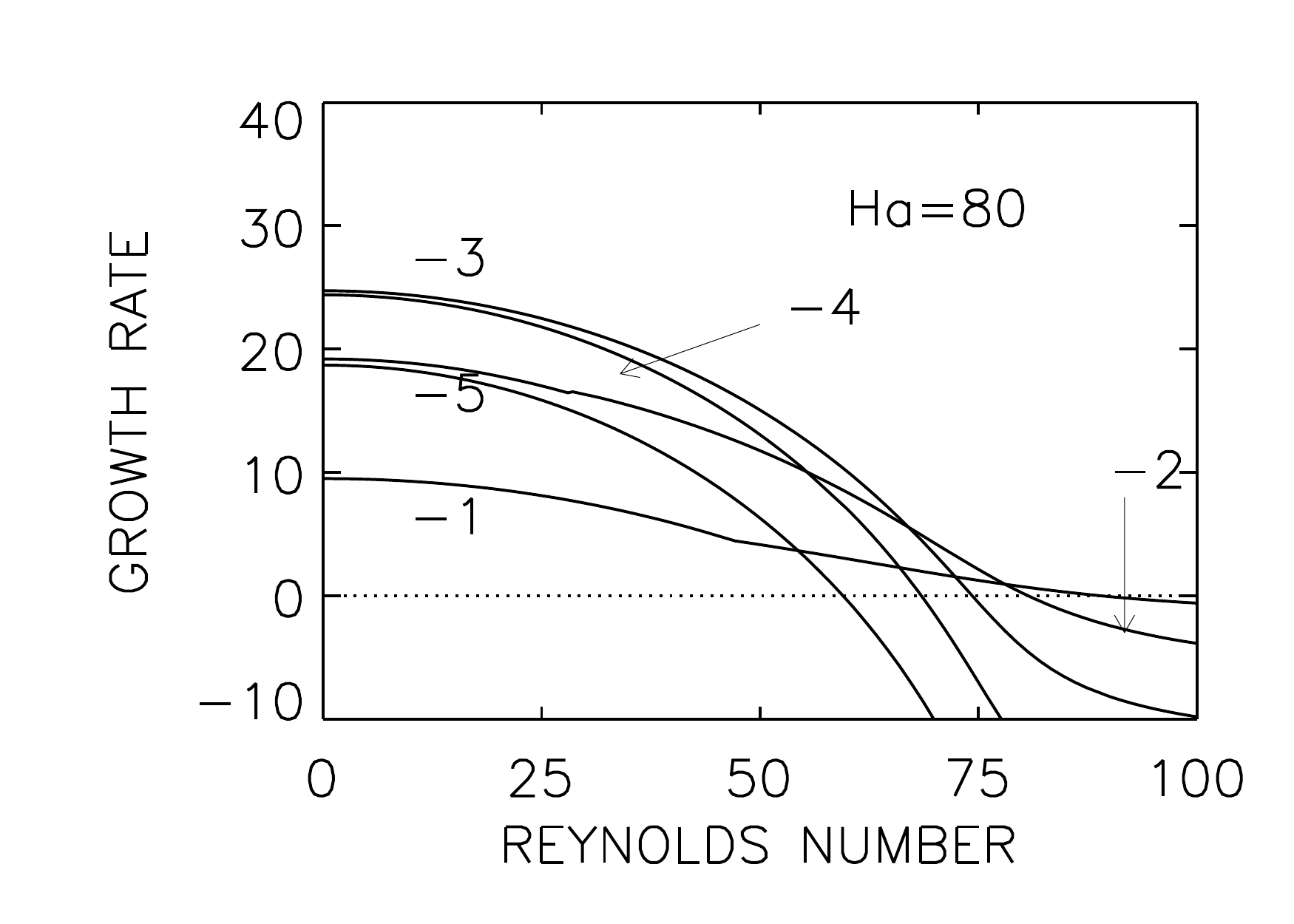}
 \includegraphics[width=5.25cm]{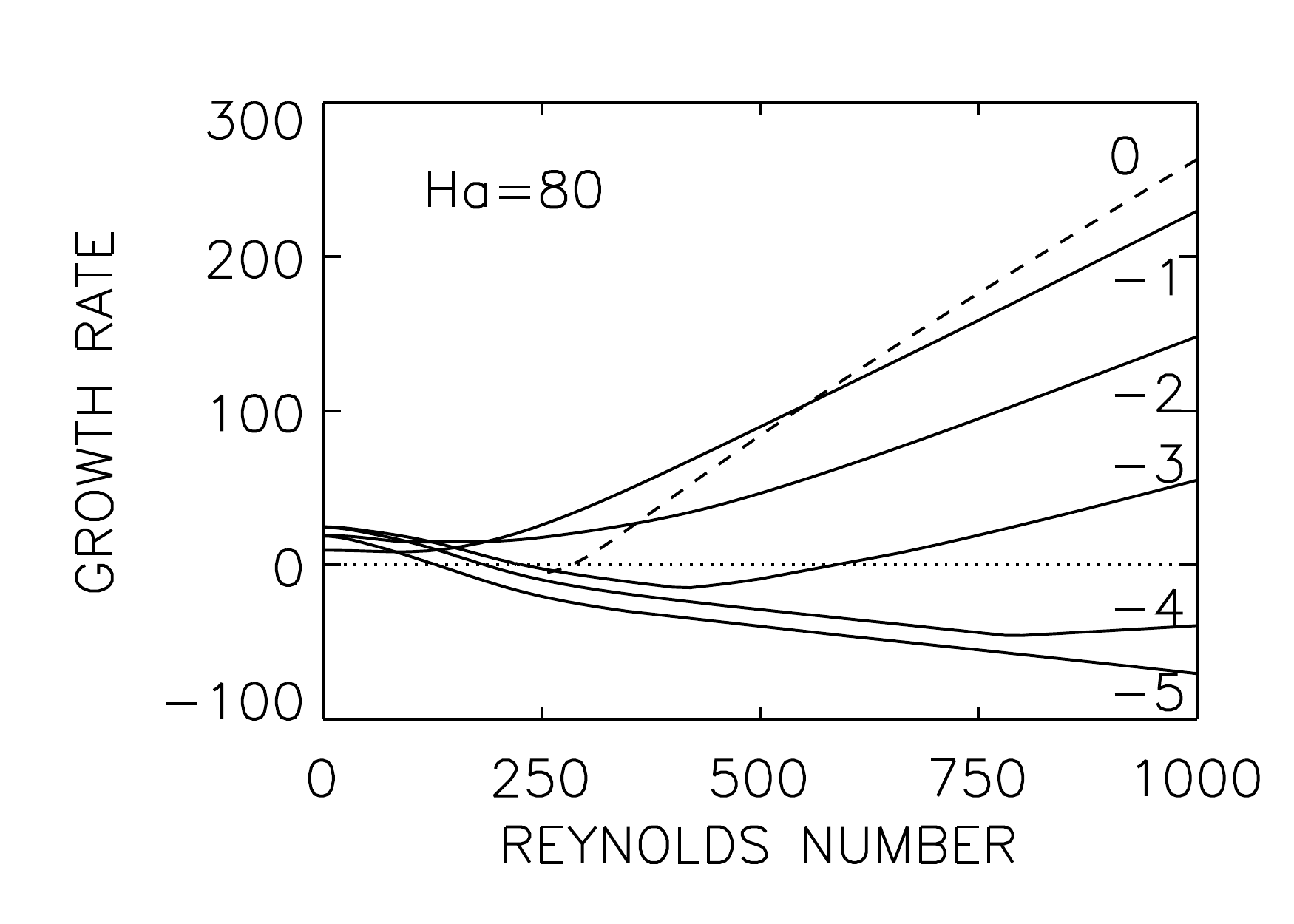}
 \caption{Growth rates normalized with the resistivity frequency $\omega_\eta$ without rotation (left) and under the influence of rigid rotation ($\mu_\Om=1$, middle) and differential rotation ($\mu_\Om=0.5$, right). The dashed line represents the growth rates for the axisymmetric MRI mode $m=0$. The curves  are invariant against the simultaneous transformation $m\to -m$, $\beta\to -\beta$. $\rin=0.5$, $\Ha=80$, $\mu_B=1/\rin$, $\beta=1$, $\Pm=1$. See \cite{RS11}.}
 \label{ti13}
\end{figure}

These findings are confirmed by numerical simulations of the instability. They show the dependence of the handedness of the patterns on the sign of the helicity of the background field, i.e.~the sign of $\beta$ (Fig.~\ref{ti12}). There is no clear dependence of the results on the magnetic Prandtl number. The main result concerns the azimuthal wave number $m$. The nonlinear simulations indeed show the prevalence of the higher modes $m=-3$ or $m=-4$ within the instability patterns. While the instability pattern of the nonrotating pinch with $B_z=0$ is dominated by the mode with $|m|=1$, the addition of a uniform axial field leads to the excitation of much more complex instability patterns.

Rigid rotation stabilizes the magnetic perturbations while differential rotation supports the instability. For the $z$-pinch with $\mu_B=2$ the growth rates were calculated for supercritical $\Ha$ and for $\beta=1$ (Fig.~\ref{ti13}, middle). The critical $\Ha_0$ for $\beta=1$ is $\sim 35$. For $\Ha=80$ one finds positive growth rates for slow rotation while for rapid rotation there is stability. For $\Rey=0$ the mode with $m=-3$ grows fastest. The instability does not survive for $\Mm>1$. The mode with $m=-1$ withstands the rotational suppression best. The modes with the highest $m$ are already suppressed by lower Reynolds numbers. The dominance of the modes with $|m|>1$ disappears for rigid rotation. Note that the pinch with $\mu_B=2\mu_\Om=2$ is a Chandrasekhar-type flow which for $\Pm\to0$ scale with $\Ha$ and $\Rey$. The only unstable mode is $|m|=1$.

Another situation holds for nonuniform rotation. The growth rates for ${\Ha}=80$ and the rotation law with $\mu_B=0.5$ are given in the right panel of Fig.~\ref{ti13}. The slow rotation curve is almost identical to the rigid rotation curve. The modes are rotationally stabilized. Only $|m|=1$ has positive growth rate for all rotation rates. Modes with $|m|>1$ do not contribute to the instability for high Reynolds numbers because they are damped by strong differential rotation. For $\Mm \gsim 1$, however, the magnetic instability is re-animated, but at most for the lower modes including the axisymmetric MRI mode with $m=0$. Finally the $m=0$ mode becomes dominant because its growth rate becomes higher and higher, finally scaling with the rotation frequency. For $\Mm>1$ the growth rate increases with increasing $\Om$ rather than with $\Om_{\rm A}$.

\begin{figure}[htb]
\centering
 \includegraphics[width=9cm]{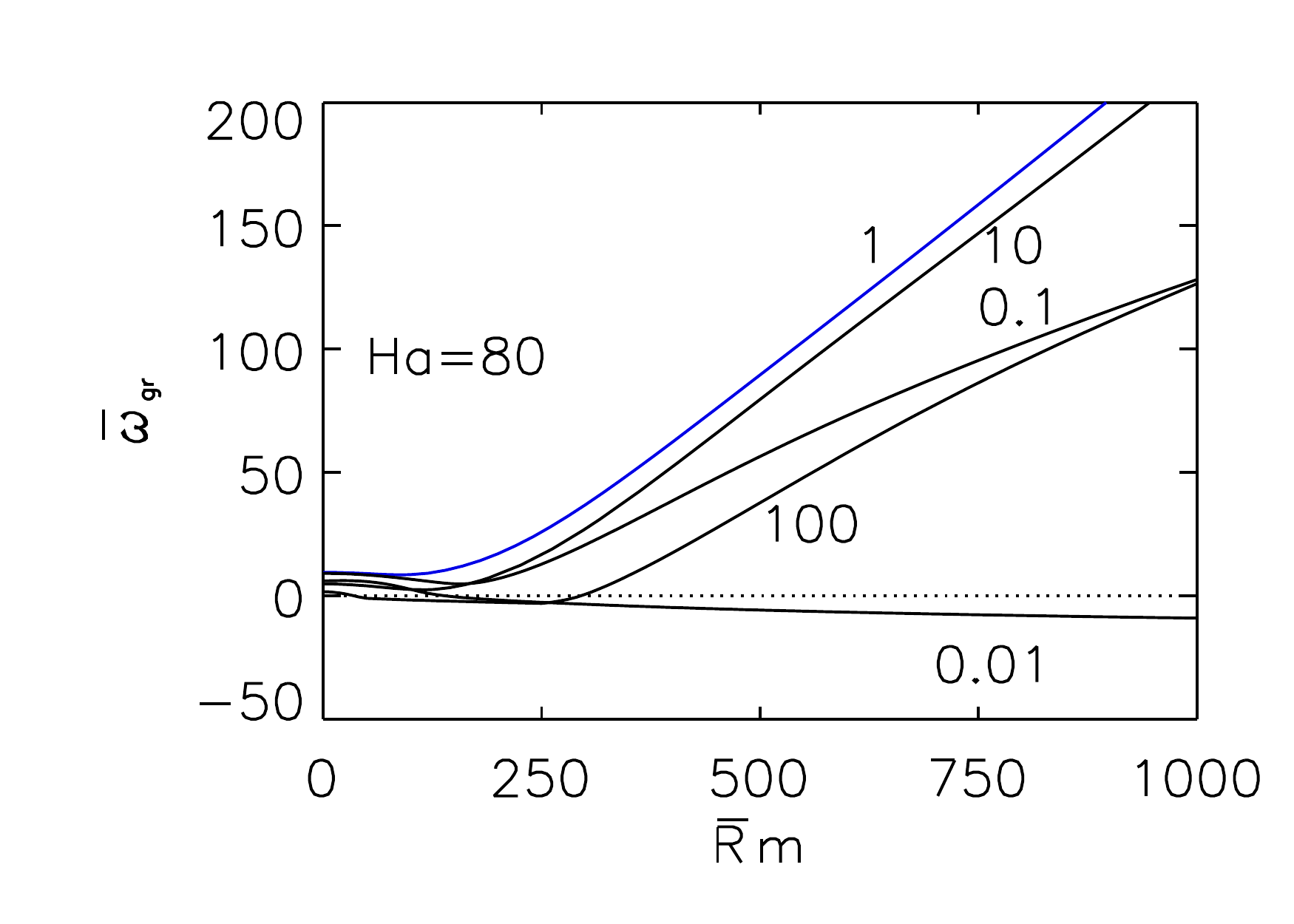}
 \caption{Growth rates  for  $m=-1$  normalized with  the dissipation frequency $\omquer=\sqrt{\omega_\nu \omega_\eta}$ in  dependence on the averaged Reynolds number $\Rmquer$ and the magnetic Prandtl number $\Pm$ (marked) for  quasi-uniform flow,.  The fastest growth of the perturbations for all $\Rmquer$
 ($\Om=0$ included) belongs to  $\Pm=1$. $m=1$, $\Ha=80$, $\beta=1$, $\rin=0.5$, $\mu_B=1/\rin$,  $\mu_\Om=0.5$. Perfectly conducting cylinders.  }
 \label{ti14}
\end{figure}

The dependence of the growth rates on the magnetic Prandtl number is a puzzling problem which can be demonstrated by use of new variables. Figure \ref{ti14}  demonstrates  that the mode with $|m|=1$ grows fastest for $\Pm=1$ if the growth rate is normalized with the geometrical average $\overline{\eta}=\sqrt{\nu\eta}$ of both diffusivities. This is also valid for stationary cylinders. A flow can thus be unstable for $\Pm=1$ whereas it is stable for $\Pm\neq 1$.
For differential rotation with $\mu_\Om=0.5$ (almost uniform $U_\phi$) the  Fig.~\ref{ti14} gives the growth rates for a supercritical Hartmann number. The growth rates of the mode $m=-1$ and the global rotation rate are again normalized with $\mdiffquer$. One again finds that with  the special normalization models with $\Pm=1$ always have maximum growth rates for slow and fast rotation. Both small or large magnetic Prandtl numbers lead to slower growth of the instability. This effect is so strong that the considered field pattern can even be stabilized for too small or too large $\Pm$. This is a remarkable  restriction for numerical simulations with $\Pm=1$. Magnetic instability strongly depends on the magnetic Prandtl number of the fluid. For a fixed Hartmann number two regimes for the rotational influence on the growth rates exist. There is only a weak influence of small $\Rmquer$ on the growth rate. For large averaged Reynolds numbers $\Rmquer$ one finds linearly increasing growth rates. High values of $\Rmquer$ strongly accelerate the instability in accordance with $\omquer\propto \Rmquer$, which leads to $\omgr\propto \Omin$. In this case the physical growth rate results in $0.2 \Omin$, so that the growth time is shortened by the rotation to approximately only one rotation time.

\section{Transport coefficients by the pinch-type instability}\label{Transport}
A  consistent model for magnetoturbulence might easily  be originated by the nonaxisymmetric pinch-type instability which appears in an axially unbounded Taylor-Couette flow of an electrically conducting fluid between the stationary or rotating cylinders. The fluid is permeated by a homogeneous and axial current which produces a radius-dependent azimuthal magnetic field. The simplest case of this configuration with stationary cylinders forming a wide gap  has   been realized in a laboratory experiment {\sc Gate} (Section \ref{TI}).  
The resulting flow and magnetic field fluctuations are able to transport magnetic flux, angular momentum,  or passive scalars like concentration  of chemicals or temperature.

The Hartmann number is defined by Eq.~(\ref{Hartmannin}). First estimates of the perturbation velocity and the cell size for the unstable pinch are simple and can be taken from the information given by   Fig.~\ref{tidiffrot3}. The vertical cell size roughly equals the gap width, and for the Reynolds number of the fluctuation, $\Rey'=\urms d/\nu$, the value 10 is provided.  One obtains, therefore, $10\nu$ for the product of flow speed and cell size, which is often used as a first orientation for the viscosity or the resistivity of a turbulent fluid. Taking into account the standard correction factor of order 0.1, then the value for the instability-generated diffusivities is only $\etaT\sim \nu$, which  is certainly a rather small value. We shall probe the relation of the eddy diffusivities to the pinch parameters in the following with more sophisticated methods. For the numerical simulations the nonlinear code described in Section \ref{azifield} is used.  Up to $M=16$ Fourier modes are used, and the order of the polynomials is varied between $N=8$ and $N=12$.  The cylinder material is perfectly conducting and the  container is axially periodic.

 \subsection{Electromotive force}
By definition, the eddy diffusivity connects the turbulence-induced electromotive force with the axial electric current in the pinch, i.e.
\beg
\langle \vec{u}\times \vec{b}\rangle = - \etaT \rot\, \vec{B},
\label{EMF}
\ende
where  for simplicity all possible anisotropies due to rotation and magnetic field are ignored, see  \cite{KP94,RK03,BS09}. The axial component of the electromotive force is ${\cal E}_z= \langle u_Rb_\phi- u_\phi b_R\rangle $, hence 
 \beg
\etaT=\Rin\frac{\langle u_\phi b_R-u_Rb_\phi\rangle}{2B_{\rm in}} \ \frac{1-\rin^2}{\mu_B\rin-\rin^2}.
\label{eta}
\ende
As the angular momentum transport is also due to the Maxwell stress of the fluctuations, the turbulent viscosity should always exceed the molecular viscosity. The question is whether this is also the case for the instability-induced magnetic resistivity. First nonlinear simulations for not too small magnetic Prandtl numbers revealed the axial component of (\ref{EMF}) as negative in the entire container, which ensures the expression (\ref{eta}) as positive definite \cite{GR09a}. The resulting eddy resistivity $\etaT$ did not  depend strongly on the magnetic Prandtl number and the Reynolds number of rotation. The  simulations presented below in particular focus on the influence of the strength of the background field. The applied magnetic field must be due to an axial current; calculations for pure AMRI are thus not possible.
\subsubsection{Stationary pinch}
\begin{figure}[h]
\centering
\includegraphics[width=3cm,height=6cm]{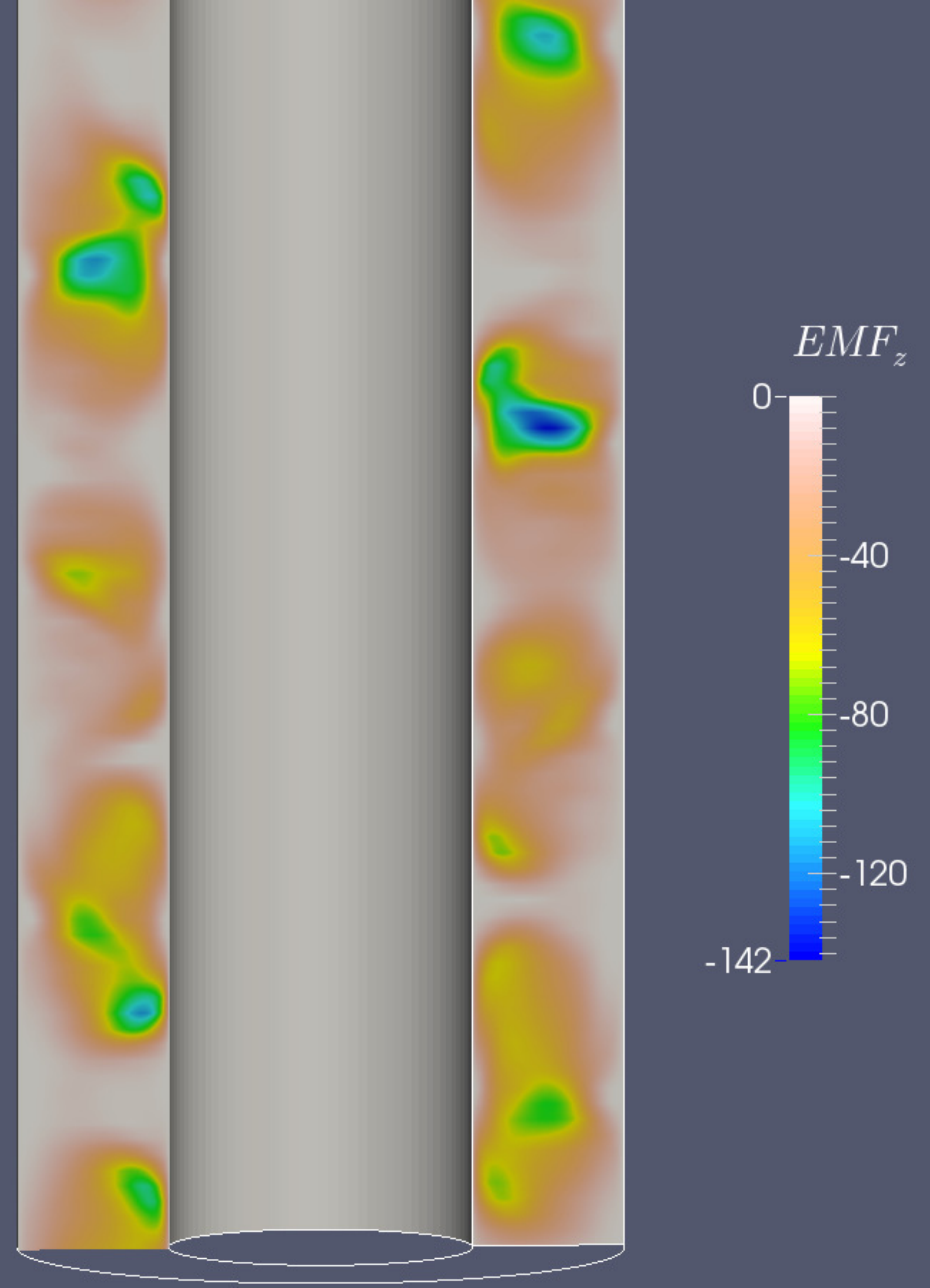} \ 
\includegraphics[width=3cm,height=6cm]{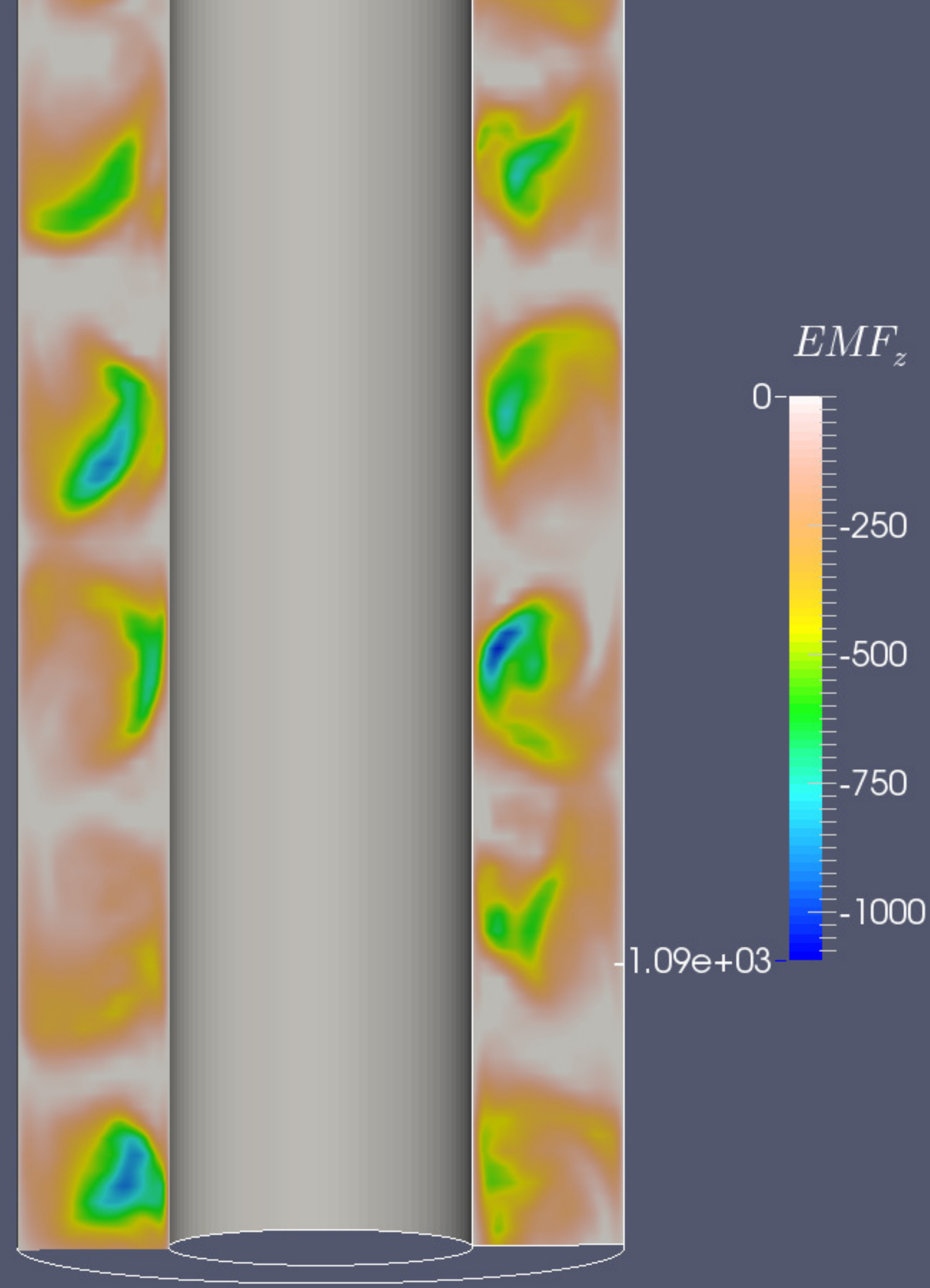}
\includegraphics[height=6.5cm]{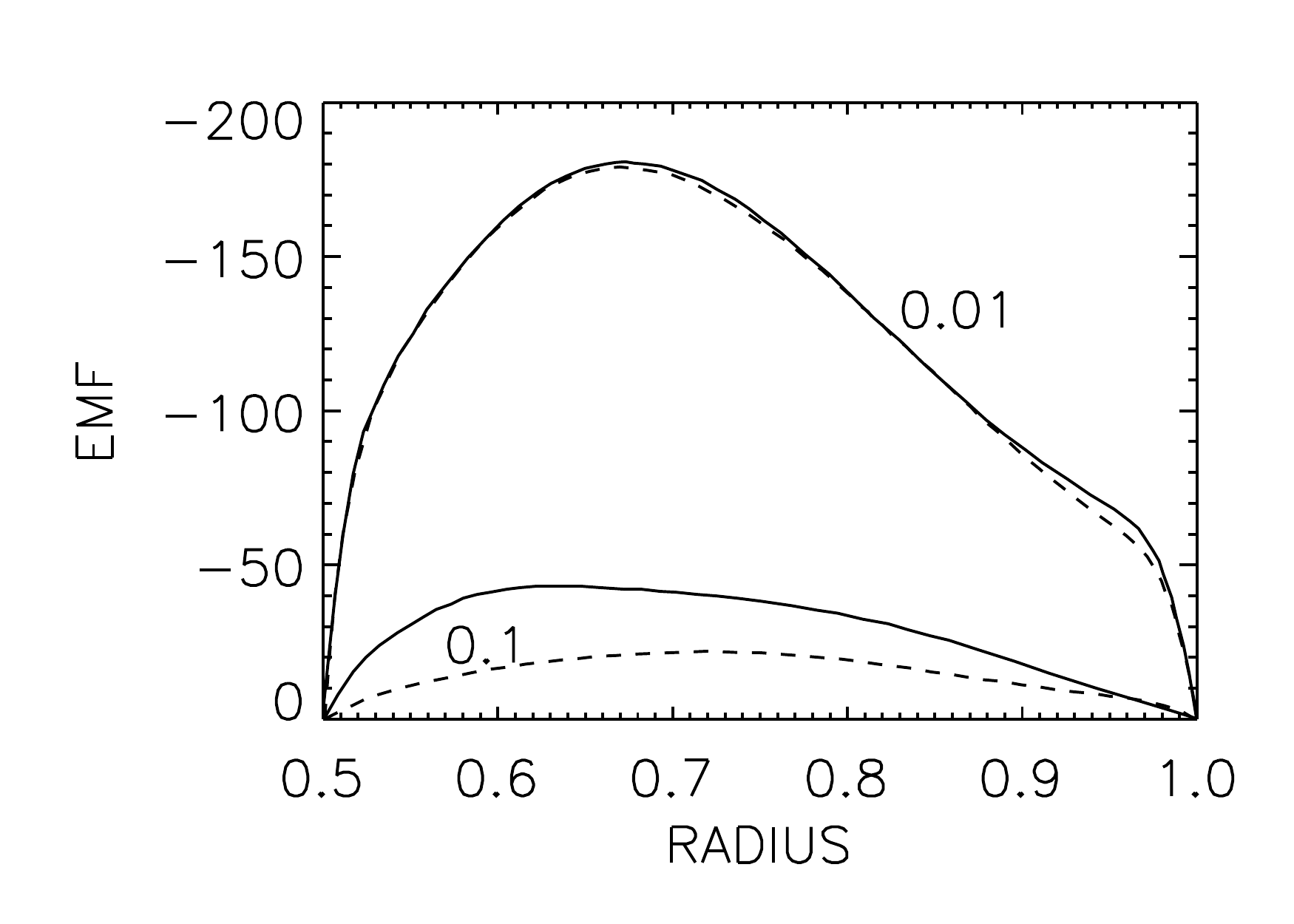}\ \ \ \ \ \ 
\caption{Left: Axial component $u_Rb_\phi- u_\phi b_R$ of the electromotive force for the stationary $z$-pinch with $\Pm=0.1$ (left) and $\Pm=0.01$ (right). Right: The  $z$-component of the electromotive force after averaging time and azimuth for $\Rey=0$ (solid lines)  and $\Rey=500$ (dashed lines) for quasi-Keplerian flow is always negative. Upper curves: $\Pm=0.01$, lower curves:  $\Pm=0.1$.  $\Ha=100$,  $\rin=0.5$, $\mu_B=1/\rin$.  Insulating boundary conditions.}
\label{figEMFa} 
\end{figure}

We start with the stationary pinch with $\mu_B=1/\rin$ -- which strongly simplifies (\ref{eta}) -- for various magnetic field amplitudes. Figure \ref{figEMFa} (left) presents snapshots of the non-averaged axial EMF for two models with one and the same $\Ha$ but different magnetic Prandtl numbers. 
The fluctuating axial EMF values are always negative. The dependence on the magnetic Prandtl number seems to be strong in the sense that the values scale with $1/\Pm$. This is also shown by the radial profiles of the axial EMF after averaging over all snapshots and the meridional planes. As shown by  Fig.~\ref{figEMFa} (right)  the  influence of the basic rotation is very weak, but the maxima of the curves are anticorrelated with the magnetic Prandtl number.

\begin{figure}[h]
\centering
\includegraphics[width=8cm]{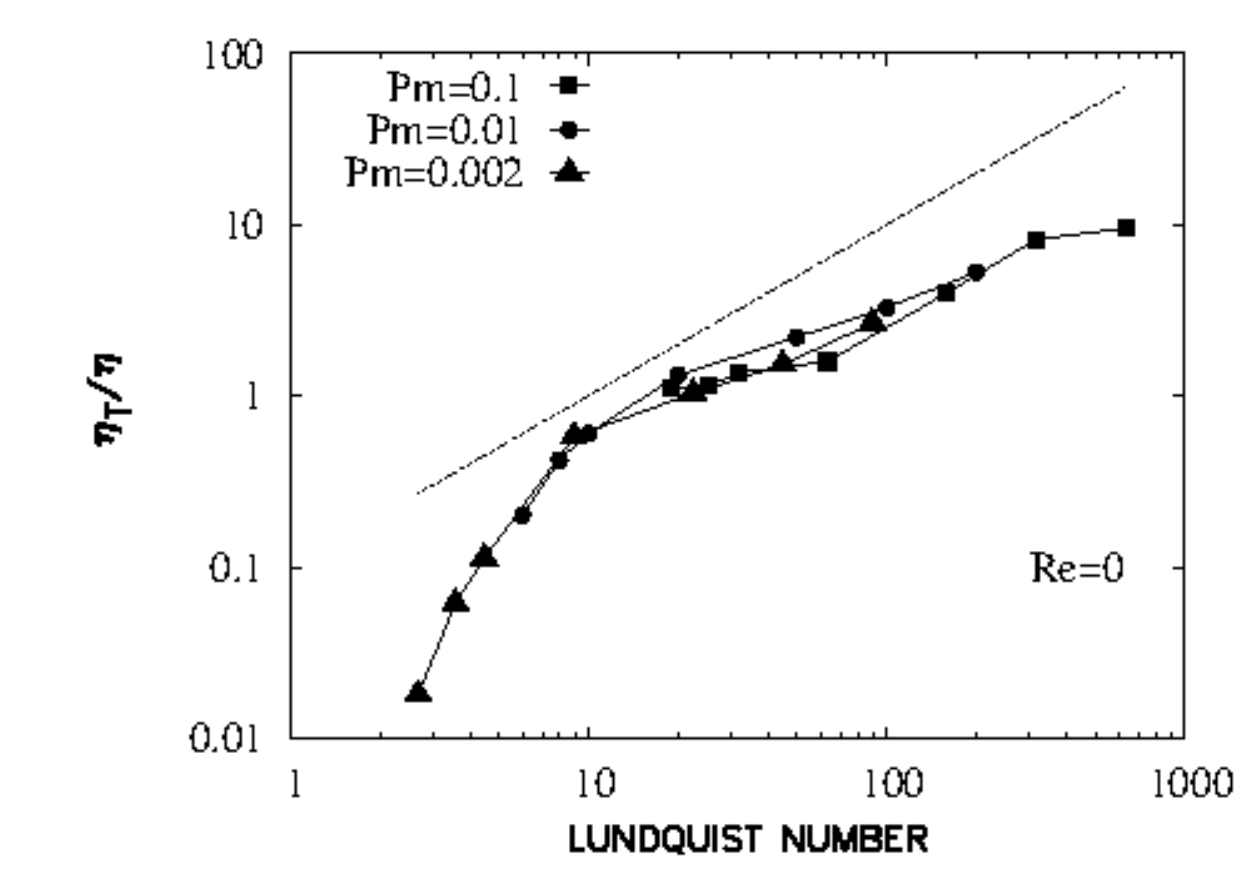}
\includegraphics[width=8cm]{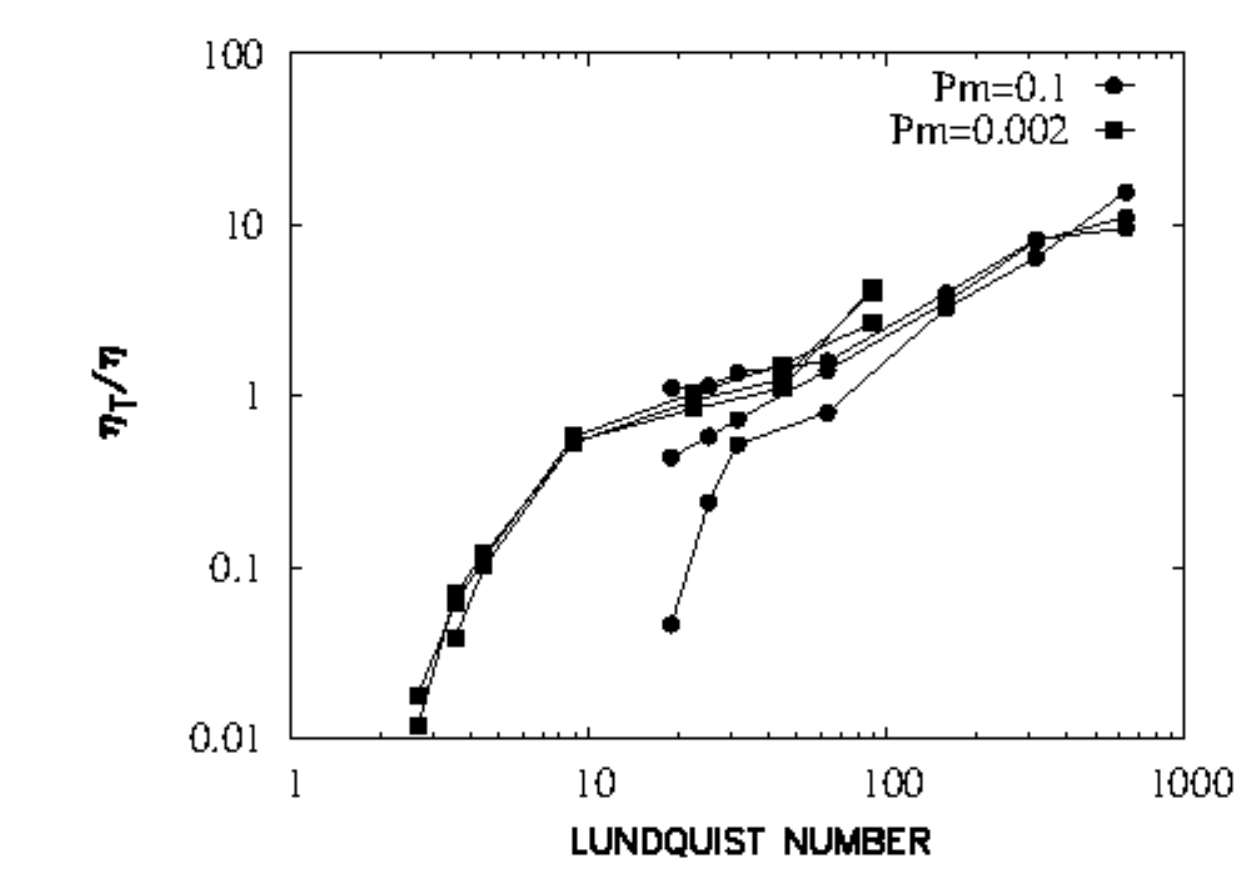}
\caption{Instability-induced resistivity $\etaT/\eta$ for the stationary  $z$-pinch (left) and for quasi-Keplerian rotation with $\Rey=500$ and $\Rey=1000$  (right, $\Rey=0$ also included)  as  function of the Lundquist number for various $\Pm$.  The dotted line gives $\Lu/10$. $\rin=0.5$,  $\mu_B=1/\rin$.}
\label{TIdiff1} 
\end{figure}

Figure \ref{TIdiff1} gives the final  eddy resistivity values  after averaging over the radius and normalized with the molecular resistivity for various  magnetic Prandtl numbers as a function of the Lundquist number $\Lu$ for stationary  cylinders. In this representation the influence of the magnetic Prandtl number almost vanishes, though $\Pm$ varies over two orders of  magnitude. This is because of the fact that the normalization used in Fig.~\ref{figEMFa} also scales with $1/\Pm$.

For small Lundquist number $\Lu$ the curve  scales with $\Lu^4$, while a much flatter linear dependence on $\Lu$ appears for models with $\Lu>10$. The form of the curve even suggests a saturation of $\etaT/\eta$ for  large $\Lu$. For those central parts of the curve where $\etaT/\eta\propto \Lu$, the eddy resistivity loses its dependence on the molecular diffusivity, so that simply $\etaT\propto \Om_{\rm A}R_0^2$ with the \A\ frequency $ \Om_{\rm A}=B_{\rm in}/\sqrt{\mu_0\rho R_0^2}$. On the other hand, if by Eq.~(\ref{qqq}) $\Lu\propto \Rm'^{0.4}$, then for the  central  parts of the curves  one finds 
\beg
\frac{\eta_{\rm T}}{\eta}\simeq 0.7\ \Rm'^{0.4}.
\label{qqqq}
\ende
For the steep weak-field part of the profile  with $\Lu^4$ this yields $\etaT/\eta\propto \Rm'^{1.6}$. Very similar relations are empirically known  from liquid-metal experiments \cite{DN08}.

The values of $\etaT/\eta$ for the stationary pinch are numerically small. For the smallest $\Lu$ the eddy resistivity is $\etaT\ll \eta$, while even the much larger fields with $\Lu\simeq 1000$ only yield $\etaT/\eta\simeq 10$. The transition $\etaT\simeq \eta$ happens at $\Lu\simeq 30$, this value only depending on the value of $\rin$.  A rough description of the results  is given by $\etaT/\eta\simeq 0.1 \Lu$, again leading to $\etaT\simeq 0.1 \Om_{\rm A} R_0^2$. As the typical Lundquist number for experiments with liquid metals does not exceed unity, one can hardly expect to find values of $\etaT>\eta$ \cite{RB01,MB07,DN08,FD08,FN10,ND12}.
\subsubsection{Quasi-Keplerian flow}
It is no problem to extend the calculations to the presence of  (differential)  rotation. The right panel of Fig.~\ref{TIdiff1} gives  the results of many simulations for slow and modest quasi-Keplerian rotation. The three lines in the plot belong to stationary cylinders, $\Rey=500$ and $\Rey=1000$. Compared with the findings  for $\Rey=0$ for   small  $\Lu$ the lines are almost identical. Hence, a rotational influence of the small magnetic Reynolds numbers $\Rm\lsim 2$ can not be observed.
The curves for the larger Reynolds numbers, however, are located below the one for the stationary pinch. A rotational suppression is  clearly  visible but only   for  larger Lundquist numbers $\Lu\gsim 10$. On the other hand, for even larger $\Lu$  there is no rotational suppression for the Reynolds numbers shown. The conclusion is that the rotational suppression scales with the magnetic Mach number $\Mm=\Rm/\Lu$. As a rough description of the data  for $\Lu\gsim 10$ the relation $\etaT/\eta\propto \Lu/(1+\Mm^2)$ can be considered.
Hence, for small $\Pm$ the molecular viscosity does not influence  the results for the instability-induced resistivity and its rotational quenching.

\subsection{Angular momentum transport}
The same simulations can serve for calculations of  the eddy viscosity by use of  relation (\ref{T1}). It is sufficient to compute the angular momentum transport (\ref{Trfi}) for various $\mu_B$. The angular momentum transport should change its sign for uniform rotation; it indeed is positive for negative shear profiles, and negative for positive shear profiles \cite{GR08}.

\begin{figure}[h]
\centering
\includegraphics[width=9cm]{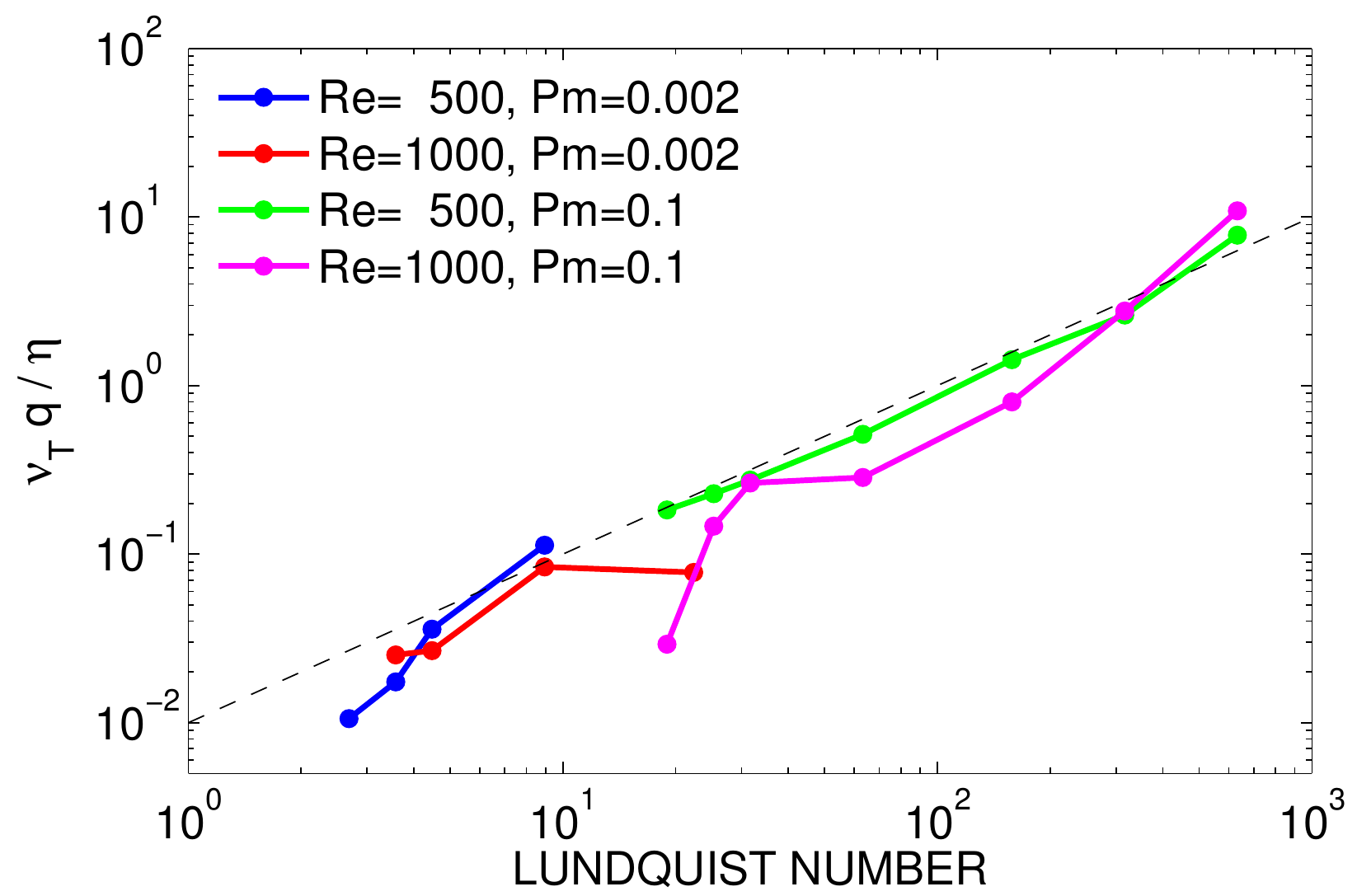}
\caption{The quantity $q\nuT/\eta$ for quasi-Keplerian rotation as a function of $\Lu$, similar to Fig.~\ref{TIdiff1}.  The dotted line gives $\Lu/100$. $\rin=0.5$, $\mu_B=1/\rin$, $\mu_\Om=0.35$. Perfectly conducting boundaries.}
\label{TIdiff3}
\end{figure}

The model already discussed in Fig.~\ref{TIdiff1} for the $z$-pinch in the presence of quasi-Keplerian rotation will now be applied to the calculation of the eddy viscosity. In dimensionless form it is
\beg
\frac{\nuT}{\nu} =\frac{1} {q\Rey} \langle u_R u_\phi - \frac{\Ha^2}{\Pm} b_R b_\phi\rangle
\label{nu2}
\ende
as the sum of Reynolds stress and Maxwell stress, with the radial function $q$ 
\beg
q = -\frac{R}{\Om_{\rm in}} \frac{{\rm d}\Om}{{\rm d}R}.
\label{qq1}
\ende
The dimensionless  flow components are here measured in the form of  Reynolds numbers, and the field components are normalized with $B_{\rm in}$. For the quasi-Keplerian flow a simple  approximation is $q = 1.5 {\Om}/{\Omin}$,
which for $\rin=0.5$ is of order unity. Figure \ref{TIdiff3} leads to the estimate $q \nuT/\eta\simeq 0.01\ \Lu$ or simply to $\nuT\simeq 0.01 \Om_{\rm A} R_0^2$.

For the instability-induced magnetic Prandtl number one immediately finds $\nuT/\etaT\simeq 0.1/q$, independent of $\Pm$. The magnetic Prandtl number due to the pinch-type instability falls below unity. This is unexpected as for driven turbulence the eddy viscosity is formed by both the Reynolds stress and the Maxwell stress, while only the kinetic energy contributes to the eddy diffusivity \cite{VK83}. Hence, one should expect the turbulent Prandtl number to be larger than unity, which is not the case though for these magnetically-induced instabilities under the influence of differential rotation.
\subsection{Mixing of a passive scalar}
The transport of a passive scalar is governed by the diffusion equation
\beg
{\partial \rho C \over \partial t} + {\rm div} \left(\rho C \vec{U} - \rho
D \vec{\nabla} C\right) = 0,
\label{pip2}
\ende
where $C$ is  the fluctuating concentration, $\vec{U}$ the fluctuating flow field and $D$ the microscopic diffusion coefficient. For   $D\to\infty$ all possible fluctuations are immediately smoothed out and any mean-field transport decays. The adiabatic approximation requires $D=0$. In the sense of the anelastic approximation we shall always apply the source-free condition of the mass flux, 
${\rm div}\ \rho \vec{u} = 0$.
As usual, the  concentration is split into a mean and a fluctuating part, $C={\bar C}+c$, so that the diffusion equation in the presence of turbulence becomes
\beg
{\partial \rho {\bar C} \over \partial t} + {\rm div} \left(\rho
\langle c\vec{u}\rangle - \rho D \vec{\nabla}
{\bar C}\right) = 0
\label{pip5}
\ende
with $\vec{u}$ as the fluctuations of  the flow and  $\bar C$ as the large-scale part of the concentration field. The bars denoting the average procedure are dropped in the following. The ensemble-average will also be replaced by averaging over the azimuthal coordinate and time. The influence of a possible large-scale circulation $\bar{\vec{U}}$ is neglected here. We have thus to compute the turbulent concentration flux vector $\langle c\vec{u}\rangle$, which is necessary to formulate the mean-field diffusion equation. In the sense of Boussinesq the concentration-flux vector may be written as an anisotropic diffusion in terms of the mean concentration gradient, i.e.
\beg
\langle c u_i\rangle = - D_{ij} \ {\partial {C} \over
\partial x_j}
\label{pip16}
\ende
\cite{B97}. This is only reasonable, of course, if scales are considered which exceed the correlation scales (in space and time). The framework of magnetically-induced instabilities may well serve as a tool to study the various approximations of the diffusion theory in detail.

A basic anisotropy results if the turbulence is subject to a global rotation. Then the structure of the diffusion tensor is
\beg
D_{ij} = D_{\rm T}\ (d_1 \delta_{ij}
+ d_2 \Om_i \Om_j),
\label{pip19}
\ende
which describes an extra diffusion in the $z$-direction as a consequence of the Taylor-Proudman theorem. Terms linear in $\Om$ do not exist.

Note that the material mixing occurs only by the action of the kinetic part of the momentum transport tensor rather than by its magnetic part, so that the diffusion tensor can be approximated by
\beg
D_{ij}\simeq \frac{1}{2} \tau_{\rm corr}\ \langle u_i(\vec{x},t)u_j(\vec{x},t)\rangle.
\label{pip20}
\ende
There is no magnetic influence on the diffusion coefficient except the magnetic suppression of the correlation tensor of the fluctuations. A nonlinear code must be used to compute the eddy diffusion of a passive scalar in the radial direction. The time-dependent dimensionless transport equation 
\beg
\frac{\partial C}{\partial t} + \div (C\vec{U} ) =\frac{1}{\rm Sc} \Delta C \label{TEPS} 
\ende
for the fluctuating concentration  values  $C$  is added to the equation system. The microscopic Schmidt number 
\beg
{\rm Sc}=\frac{\nu}{D} 
\label{Sc}
\ende
is used in Eq.~(\ref{TEPS}) with $D$ as the molecular diffusivity of the fluid. No concentration fluxes are allowed through  the walls of the container.

Almost all of the existing simulations work with $\rm Sc=1$, e.g. \cite{BK04}. The Schmidt number for gases is of order unity, while for fluids it is $O(100)$. In the present section the molecular Schmidt number is varied from $\rm Sc=0.1$ to $\rm Sc=2$. The effective diffusivity can be modeled by the sum
$
D_{\rm eff} = D + \DT,
$
where $\DT$ is due only to the magnetically-induced instability.
To find $\DT/D$ a numerical simulation is performed until the instability is fully developed. Then the transport equation (\ref{TEPS}) is switched on and several models with different $\rm Sc$ numbers are simulated for quasi-Keplerian rotation and with $\Pm=0.1$. The simulations indicate that $\DT/D$ always scales linearly with $\rm Sc$, so that the striking relation $\DT\propto \nu$ results for $\rm Sc>0.1$. As it must, the effect vanishes for $\rm Sc\to 0$. Schatzman suggested such a relation $\DT=\Rey^*\ \nu$ with $\Rey^*=$\ord{100} in order to explain diffusion processes in radiative zones of stars \cite{S69,S77,Z92,BH08}. Obviously, the factor $\Rey^*$, which in a wider sense can be considered as a microscopic Reynolds number (of a pattern cell) can be computed as due to magnetic instability with the presented models. For slow rotation, $\Rey^*$ scales linearly with $\Mm$, then reaches a maximum at $\Mm\simeq 2$, and finally decreases rapidly for larger $\Mm$, saturating around $\Rey^{*}=1$. The magnetic Mach number $\Mm$ 
represents the global rotation in relation to the magnetic field strength. 
\begin{figure}[h]
\centering
\includegraphics[width=10cm]{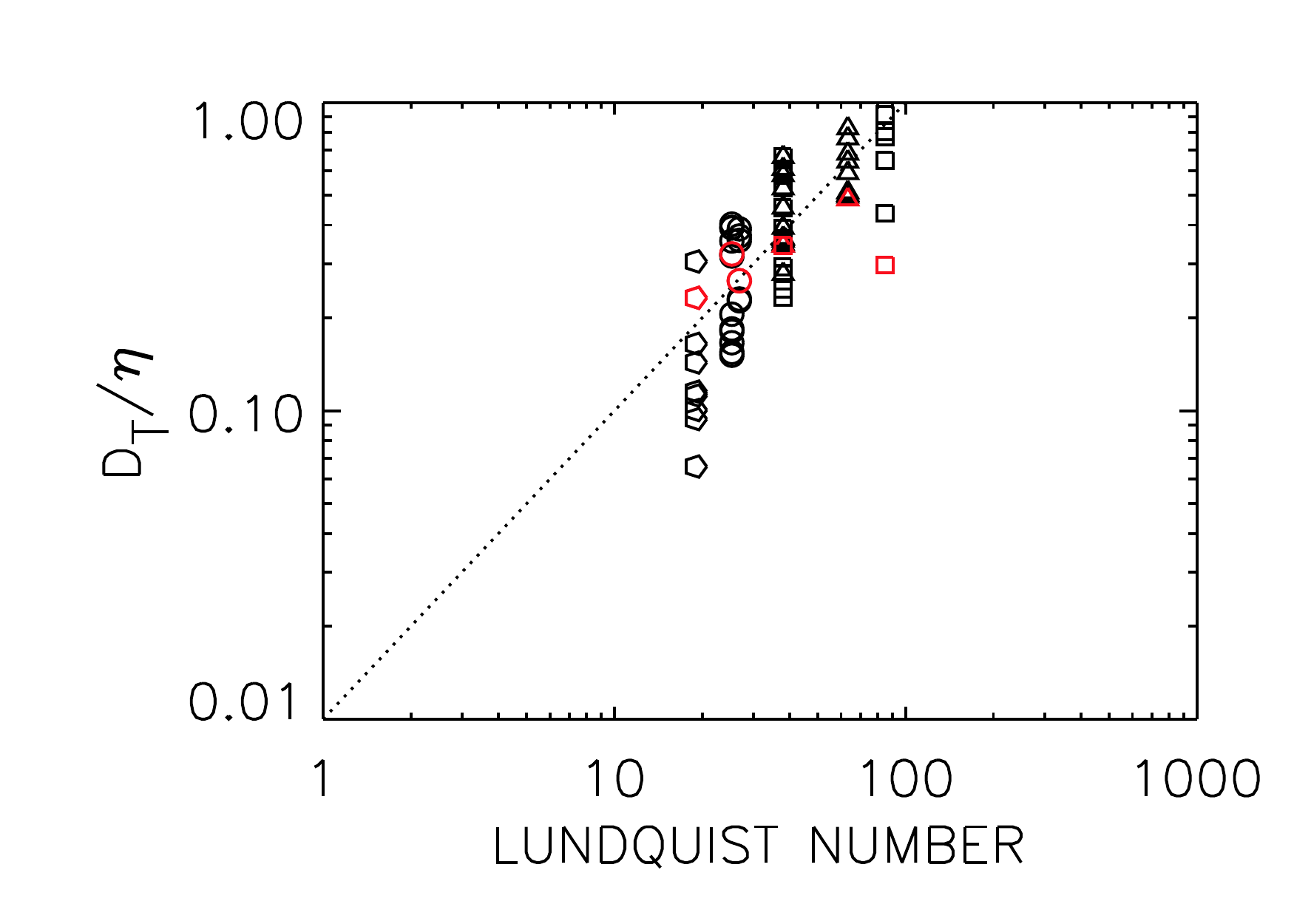}
\caption{$\DT/\eta$ versus Lundquist number $\Lu$ of all computed diffusion models. Red symbols mark nonrotating models. The dotted line represents a linear dependence, $\etaT\propto S$. $\rin=0.5$, $\mu_B=1/\rin$, $\mu_\Om=0.35$. Perfectly conducting boundaries, \cite{PG16}.}
\label{TIdiff5} 
\end{figure}

In the right panel of Fig.~\ref{TIdiff5} all models simulated in Ref.~\cite{PG16} are summarized, showing $D_{\rm T}/\eta$ as functions of the Lundquist number $\Lu$ defined by Eq.~(\ref{Lund2}). Nonrotating models are marked in red, and are located among the other models. The dotted line represents the relation $D_{\rm T}/\eta= 0.01 \Lu$ or $\DT\simeq 0.01 \Om_A R_0^2$, similar to Fig.~\ref{TIdiff3} for the instability-induced viscosity. The corresponding Schmidt number resulting from the simulations of differentially rotating $z$-pinches is thus also of order unity.

\section{Helicities, alpha effect}\label{Helicities}
Numerical simulations suggest that the MRI alone should be sufficient for the operation of the accretion disk dynamo \cite{BN95,HG96,GusevaPRL}. For low $\Pm$ it remained uncertain though whether the MRI~dynamo has physical or numerical origin \cite{FP07}. Another possibility was discussed by Spruit who suggested that differential rotation and TI can jointly drive a dynamo \cite{S99,S02}. Radial displacements converting toroidal into poloidal field are necessary for any dynamo. A dynamo effect, however,  is not guaranteed by the joint action of differential rotation and a magnetic instability converting toroidal field into poloidal. There are doubts especially concerning the TI which, in contrast to MRI, develops at the expense of magnetic energy. Estimations of dynamo parameters are necessary to probe the dynamo effectivity of a magnetic instability. The ability of turbulence to produce a mean electromotive force (EMF) along the background magnetic field plays a basic role in turbulent dynamos, i.e.
 \begin{equation}
 \langle {\vec u}\times {\vec b}\rangle = \alpha {\vec{B}} - \dots, 
 \label{rze62}
 \end{equation}
where the term on the right side of this relation is called the $\alpha$ effect which, by definition, must be odd in the magnetic field. The $\alpha$ value (or better, the $\alpha$ tensor)  and also the kinetic and/or current helicity represent pseudo-scalars (or better, pseudo-tensors), and must be even in $\vec{B}$. In rotating, radially stratified cosmic bodies the pseudo-scalar $\vec{g}\cdot \vec{\Om}$ (with $\vec{g}$ as the vector of stratification) always  exists. In Taylor-Couette flows unbounded in the axial direction no stratification vector parallel to the rotation axis exists. The only possible pseudo-scalar even in $\vec B$ is of magnetic nature: $\vec{B}\cdot \rot\vec{B}$ exists when the field geometry allows electric currents parallel to field components. For such fields the instabilities may produce finite values for the helicities and the $\alpha$ effect. Sign and amplitude of these quantities will now be discussed for twisted fields where axial currents are indeed parallel to an axial field component.

Helicity and $\alpha$ effect only exist in rotating turbulent {and} stratified media. In order to obtain finite values of these pseudo-scalars after the averaging procedure the density and/or the turbulence intensity must be nonuniform. The latter always happens close to the boundaries. A typical example of helicity formation due to the boundary effect without density stratification showed that in a geodynamo model the helicity mainly appears along the tangential cylinder of the inner spherical core similar to a boundary layer effect \cite{R06}. Another situation exists in cases with any axial stratification. The stability of a system with axial magnetic fields and an {\em axial} gradient of the angular velocity has been considered in Ref.~\cite{B06}. A pseudo-scalar $B_iB_j\Om_{i,j}$ exists in this system, yielding finite values of the helicities and the $\alpha$ effect (see below). 
\subsection{Tayler instability}
We proceed by evaluating the kinetic and current helicities and the $\alpha$ effect for the TI. Solving the linear stability problem may serve to estimate the sign and latitudinal profile of the kinetic and current helicities
 \begin{equation}
 {\cal H}^\mathrm{kin} = \langle {\vec u}\cdot
 {\rm curl}\,{\vec u}\rangle , \ \ \ \ \ \ \ \ \ \ \ \ \ \ \ \ \ \ \ \ \ \ \ \ \ \ \ \ {\cal H}^\mathrm{curr} = \langle {\vec b}\cdot
 {\rm curl}\,{\vec b}\rangle,
 \label{rze63}
 \end{equation}
which for driven turbulence both contribute to the $\alpha$ effect \cite{PF76,BS05}. The averaging in Eqs.~(\ref{rze62}) and (\ref{rze63}) is over the azimuth. If only the toroidal background field is present it follows that the helicity has opposite signs for positive and negative azimuthal wave number $m$, i.e.~${\cal H}^\mathrm{kin}(m = 1) = - {\cal H}^\mathrm{kin}(m=-1)$, see \cite{RK13}. For any unstable mode with finite helicity, there is thus another unstable mode with the same growth and drift rates but opposite helicity (see Fig.~\ref{ti3}). If all modes are excited, and there is no symmetry-breaking bifurcation in the nonlinear regime, the instability of purely toroidal field cannot produce finite kinetic helicity, as the resulting net helicity vanishes \cite{GR11}. The same argument leads to the same conclusion for the current helicity ${\cal H}^\mathrm{curr}$. The EMF also reverses when the sign of $m$ is changed hence the $\alpha$~effect also vanishes. The same is true for a possible ${\vec\Om}\times{\vec J}$-term which may appear in the expression (\ref{rze62}) for the EMF as a consequence of a rotationally-induced anisotropy of the diffusivity tensor. The ${\vec\Om}\times{\vec J}$~effect due to the Tayler instability of toroidal fields also does not exist. 
\subsection{Twisted background fields}
The Hartmann number is defined by  Eq.~(\ref{Hartmannin}). For twisted background fields with  finite azimuthal ($B_\phi$) and axial ($B_0$) components, we present two series of solutions with helical fields. The definition (\ref{beta}) is used for the ratio $\beta$ of the azimuthal and axial field components. The profiles of the azimuthal flow and field components form a system of the Chandrasekhar-type with $\mu_B=2\mu_\Om=1$. We present  $\Ha=100$, $\Rey=200$ for the first series, and $\Ha=200$, $\Rey=20$ for the second, with $\Pm=1$ for both. The first series is rotationally dominated ($\Mm>1$) in contrast to the second one ($\Mm<1$).
\begin{figure}
\centering
\includegraphics[width=8cm]{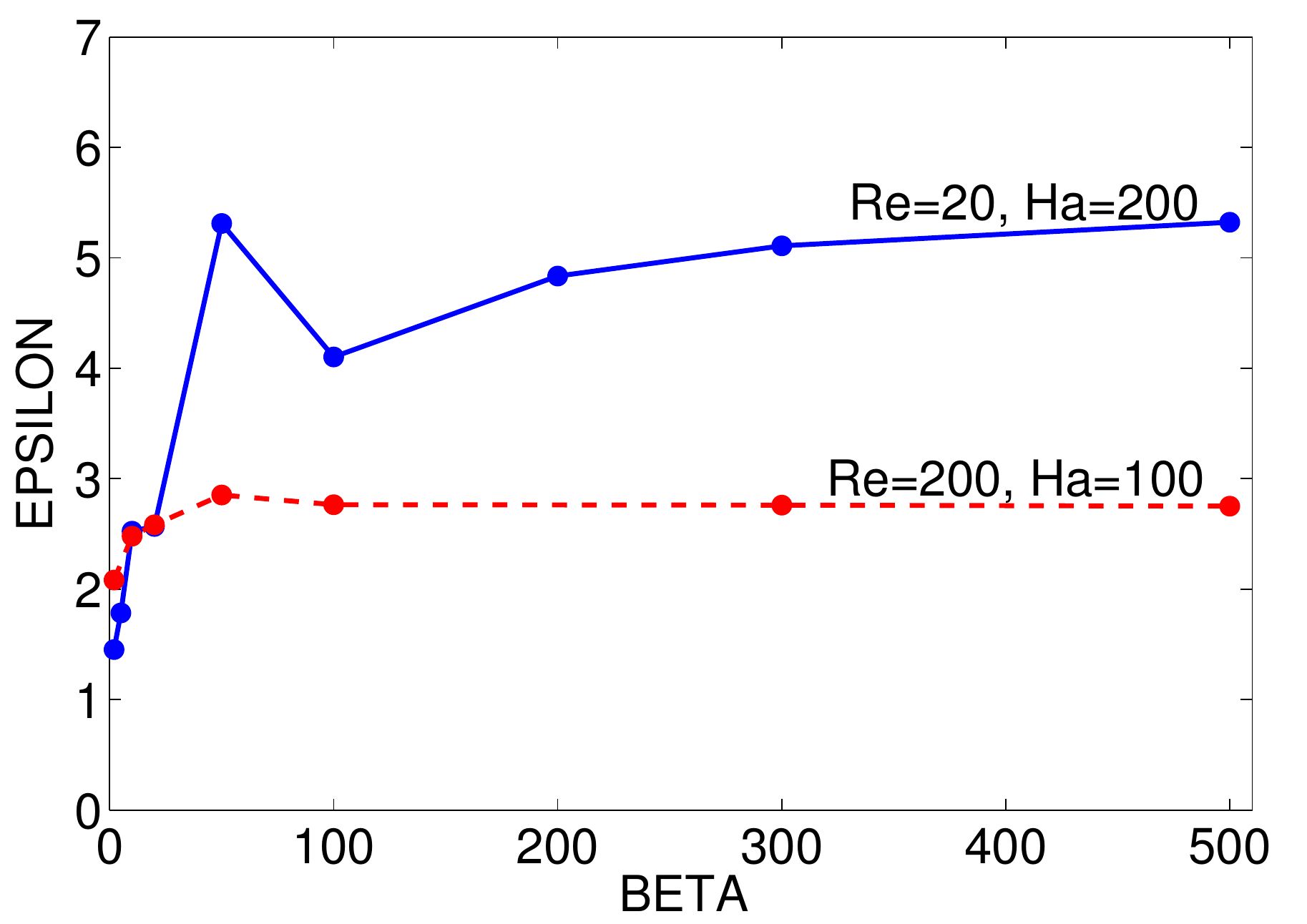}
\caption{Ratio (\ref{ratio}) of  magnetic to kinetic energy for  models with $\Mm<1$ (blue line) and $\Mm>1$ (red line) and  large magnetic Reynolds numbers. See  Fig.~\ref{ti10} for the special case $\beta=10$. $\Ha$ is defined with the azimuthal field $B_ {\rm in}$.  $\rin=0.5$, $\mu_B=2\mu_\Om=1$, $\Pm=1$. Perfectly conducting cylinders.}
\label{energy}
\end{figure}
For the two  parameter sets  Fig.~\ref{energy} shows the ratio $\varepsilon$ of the magnetic and kinetic energies. For sufficiently large $\beta$ the axial magnetic component is too weak to have any significant influence. Generally, $\varepsilon>1$ for all parameters, which for large $\beta$ is consistent with the results plotted in Figs.~\ref{f20} and \ref{g6}. This is not true for small $\beta$, where the axial field starts to dominate. For $\beta<1$ the instability is so strongly stabilized that the resulting energies of the perturbations are reduced. 

\begin{figure}[h]
\centering
\includegraphics[width=7cm]{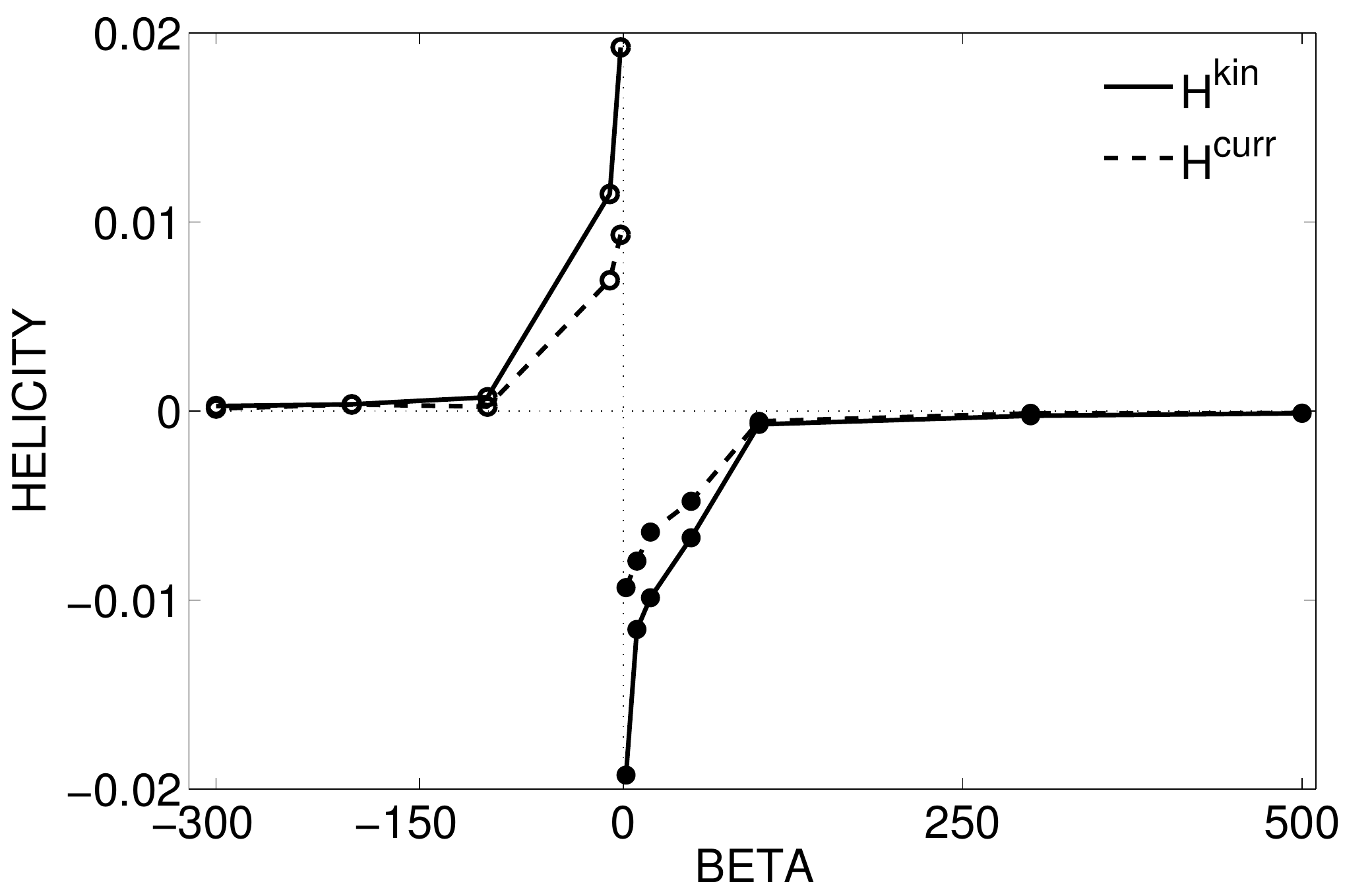}
\includegraphics[width=7cm]{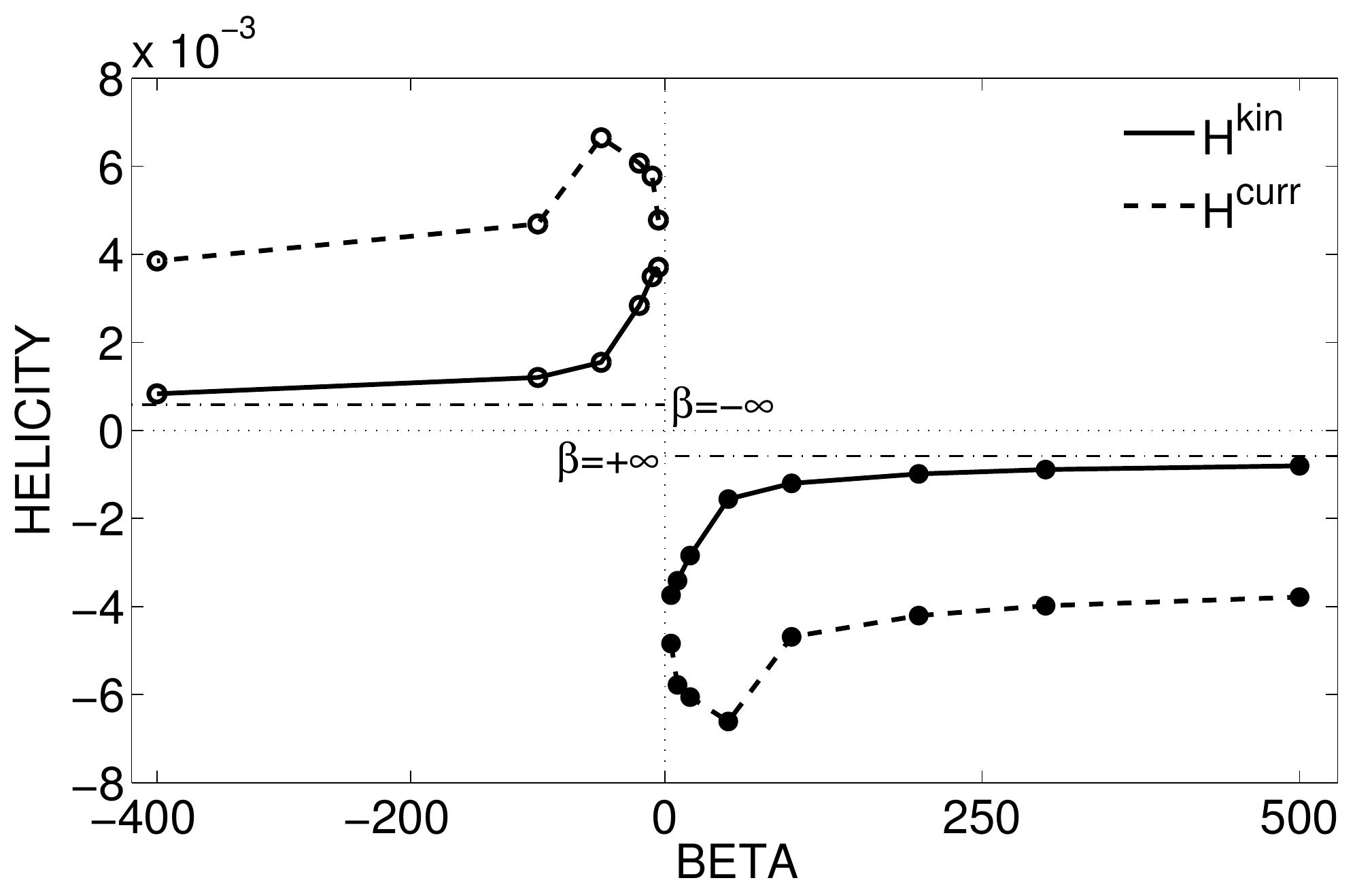}
\caption{
Kinetic helicity (solid lines) and current  helicity (dashed lines) as defined by (\ref{rze63}) as functions of $\beta$ for  models taken from   Fig.~\ref{energy} for small and large magnetic Mach numbers. The dashed lines indicate the limits $\pm 6\cdot 10^{-4}$ of the kinetic helicity of the modes in Fig.~\ref{ti3}. Helicities and $\beta$ are anticorrelated.}
\label{helk}
\end{figure}

The helicities for the two runs are presented in Fig.~\ref{helk}. For both series, both helicities have the same sign but  opposite to the sign of $\beta$. For $\beta$ of order unity the background field has a strong twist that forces the instabilities to have a  parity of opposite sign. If one then gradually increases $\beta$, each time using the previous solution as the new initial condition, this parity of the instabilities is preserved all the way to $\beta\to\infty$, where the basic state no longer has a twist, and both left and right instabilities could exist equally well as in Fig.~\ref{ti3}.
For large $\beta$ the basic state makes sufficiently little distinction between left and right modes that both could exist, but because of the way we have reached the model with $\beta=500$, we consistently obtain the right mode although a left mode would also be possible. This feature that both left and right modes are allowed for sufficiently large $\beta$ but not for smaller $\beta$ is analogous to an imperfect pitchfork bifurcation. It is thus important to specify carefully the nature of the initial conditions used in each run.

\begin{figure}[h]
\centering
\includegraphics[width=9cm]{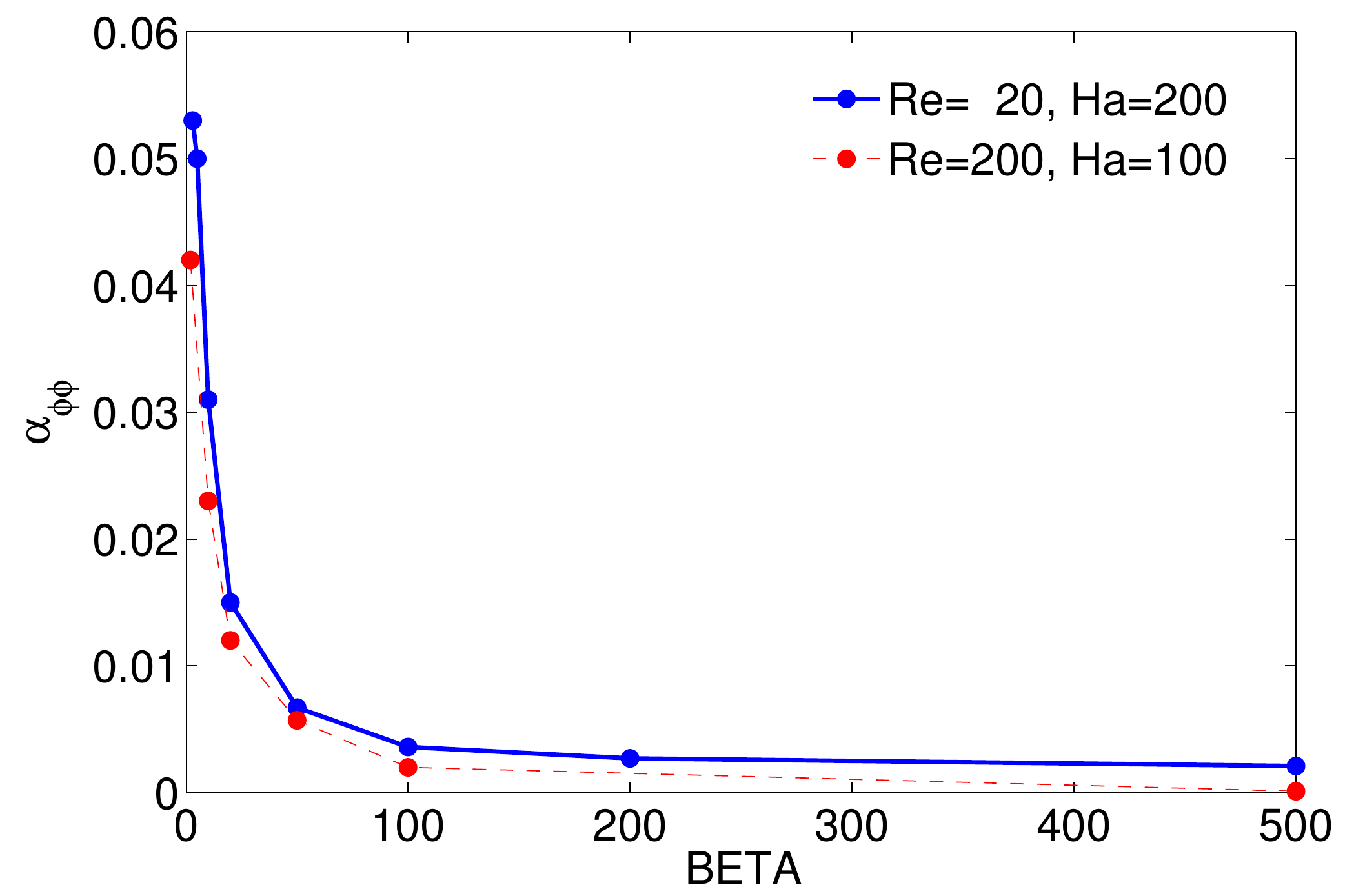}
\caption{\label{alpha}
Dimensionless $C_\alpha$ according to (\ref{Calf}) of $\alpha_{\phi\phi}$ as a function of $\beta$ for slow and rapid rotation. One finds  
 $\alpha$ effect and background helicity as positively correlated. Note that for $\Mm>1$ (red line)  $\alpha_{\phi\phi}\to 0$ for $\beta\to \infty$, as it must. The blue line for $\Mm<1$ reflects  for very large $|\beta|$ the effect of spontaneous parity breaking in accordance to Fig.~\ref{helk} (right). $\rin=0.5$,  $\mu_B=2\mu_\Om=1$, $\Pm=1$. Perfectly conducting cylinders.}
\end{figure}

We are also interested in the signs and amplitudes of the $\alpha$ effect, in both azimuthal and axial directions. According to the general rule that the azimuthal $\alpha$ effect is anticorrelated with the (kinetic) helicity, we expect the azimuthal $\alpha$ effect to be positive for $\beta>0$. The expected sign of the axial 
$\alpha$ effect is not clear. There are theories and simulations leading to $\alpha_{\phi\phi}$ and $\alpha_{zz}$ with opposite signs (\cite{RH04} for an overview).
Figure~\ref{alpha} gives the numerical results for slow and rapid rotation with the  dimensionless $\alpha$ in the form 
\beg
C_\alpha= \frac{\alpha R_0}{\eta}
\label{Calf}
\ende
 only  for $\alpha_{\phi\phi}$. It yields $C_\alpha>0$ almost everywhere in the meridional plane. The influence of rotation on $\alpha$ is not strong, but $C_\alpha$ is smaller for rapid rotation than for slow rotation (Fig.~\ref{alpha}). This surprising result is opposite to the well-known behavior of the $\alpha$ effect for rotating and stratified convection. The signs of $\alpha_{\phi\phi}$ and $\beta$ coincide. The plot mainly shows how the amplitudes of $\alpha_{\phi\phi}$ vary with $\beta$, being roughly inversely proportional in both cases. For large values of $\beta$ the $\alpha$ effect scales as $c/\beta$ with $c\simeq 0.05$, so that for $B_{\rm in}\gg B_0$ we have $C_\alpha\ll 1$.

Obviously, $C_\alpha$ is much too small for the operation of an $\alpha^2$ dynamo. On the other hand, an $\alpha\Om$ dynamo always leads to large $\beta$, which leads to small $C_\alpha$. Too small $C_\alpha$ requires stronger differential rotation to maintain the dynamo action. Stronger differential rotation, however, leads to higher $\beta$, and so on. The formal argument is as follows: Dynamo waves of $\alpha\Om$ type require for self-excitation that $C_\alpha C_\Om \geq 1$, with the magnetic Reynolds number of the differential rotation $C_\Om= - {R_0^3}/{\eta}\ {{\rm d} \Om}/{{\rm d} R}$. The amplitudes of the field components $B_\phi$ and $B_R$ can be estimated by
\begin{equation}
\frac{|B_\phi|}{|B_R|} \simeq \sqrt{\frac{C_\Om}{C_\alpha}}
\label{bfr}
\end{equation}
so that dynamo excitation requires
\begin{equation}
\frac{|B_\phi|}{|B_R|} C_\alpha\geq 1.
\label{bfr1}
\end{equation}
With $C_\alpha \simeq c/\beta$ follows
\begin{equation}
c> \frac{|B_R|}{|B_z|}
\label{C}
\end{equation}
for dynamo action by differential rotation and current-driven $\alpha$ effect. For disk dynamos $B_R$ dominates $B_z$, and for spherical dynamos $B_R$ is comparable to $B_z$. The condition for self-excitation, therefore, becomes $c>1$, which according to Fig.~\ref{alpha} cannot be fulfilled. On the basis of the numerical results given in Fig.~\ref{alpha} an $\alpha\Om$ dynamo cannot operate for this particular choice of the magnetic Prandtl number \cite{ZB07,GR11}.

\subsection{Axial shear}\label{axshear}
\begin{figure}[h]
\centering
\includegraphics[width=8cm]{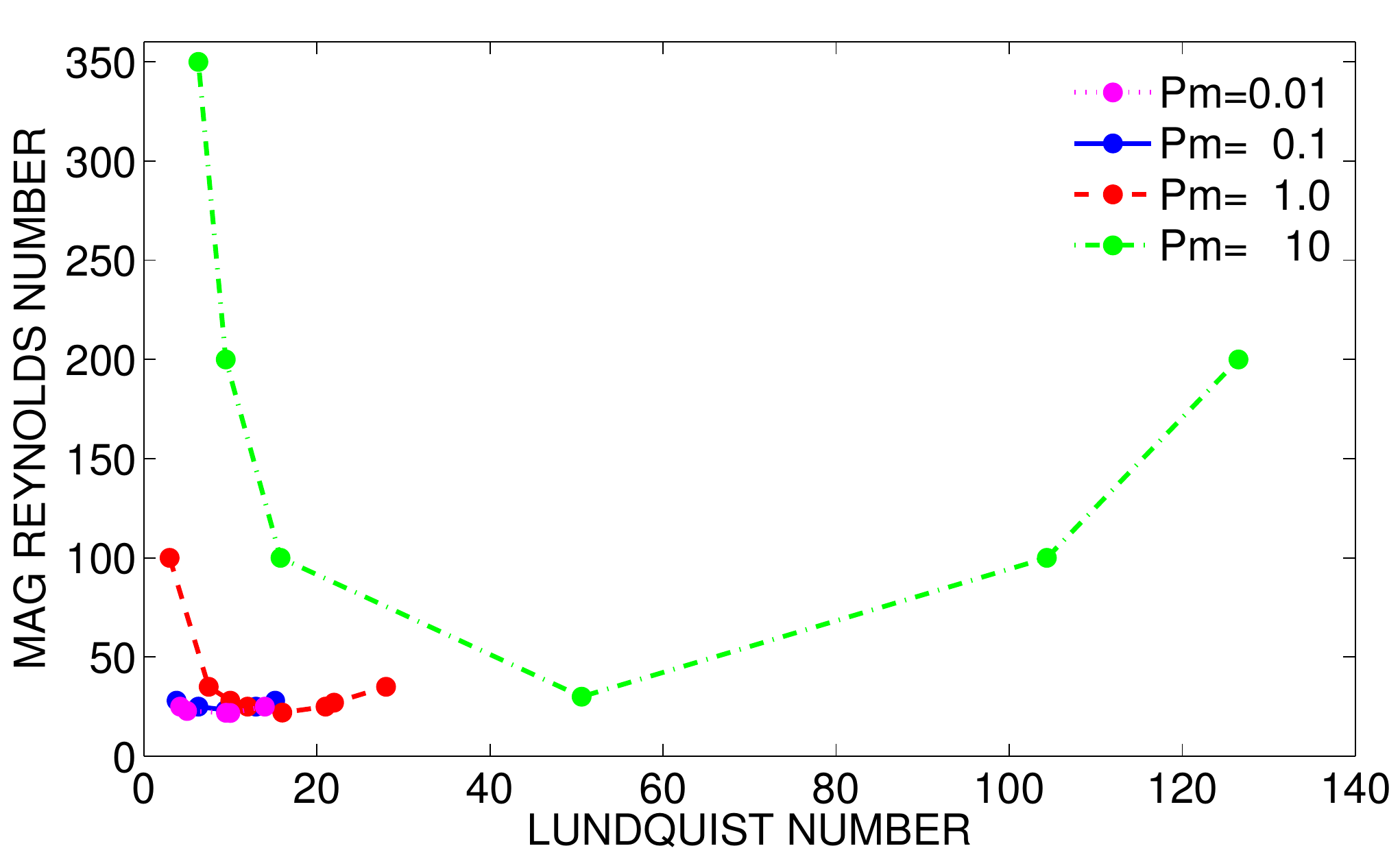}
\includegraphics[width=8cm]{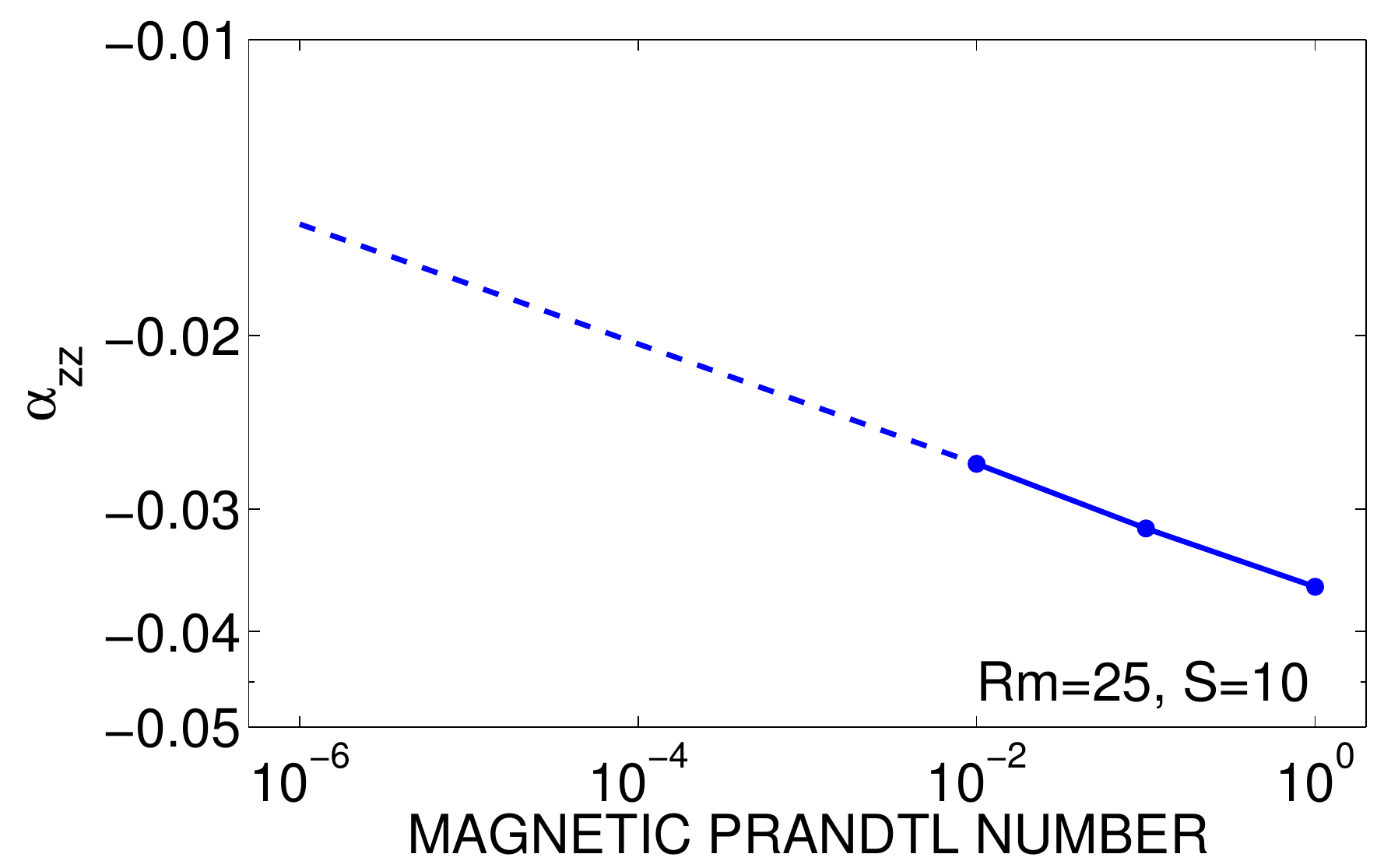}
\caption{Axial shear $\Om\propto z$ and uniform axial field $B_0$.  Left:  lines of neutral stability for small $\Pm$; for small $\Pm$ the curves scale with $\Lu$ and $ \Rm$. Right: $\Pm$-dependence of the $\alpha_{zz}$  in units of the maximal velocity of the driving endplate. The dashed line suggests a possible extrapolation to smaller $\Pm$ (not yet confirmed). $\Rm=25$, $\Lu=10$. Perfectly conducting boundaries.}
\label{TG1}
\end{figure}

In the majority of the models the radial profile of the toroidal field was prescribed. The simplest way to obtain a natural radial profile is to consider the result of an axial shear ${\rm d}\Om/{\rm d}z$ acting on a given uniform axial field $B_0$ \cite{B06}. If the induced toroidal field $B_{\phi}$ becomes strong enough a Tayler instability can be observed, leading to a growing nonaxisymmetric field. Not only must the magnitude of $B_{\phi}$ be strong enough, but also a certain limit $B_{\phi}/B_0$ must be exceeded, as an additional poloidal field component suppresses the instability. As an illustration we mention that for $\mathrm{Pm}=15$ the instability sets in at a Lundquist number of the axial field of order 20 \cite{GRE08}. The excitation condition for lower magnetic Prandtl numbers are given in Fig.~\ref{TG1} (left). The curves for small $\rm Pm$ seem to converge. The minimum values of ${\rm Rm}$ are about 20, while the corresponding $\Lu$ defined by  (\ref{Lund1}) are of order 10. These numbers correspond to the values for MRI given in Table \ref{MRI} which for small $\Pm$ indeed lead to Reynolds numbers exceeding 10$^6$.

A new nonlinear relation between the $\alpha$ effect and the external field and differential rotation is 
\begin{equation}
{\cal \alpha} \propto B_i \, B_{j}\, \Om_{i,j}, 
\label{h1}
\end{equation}
where $B_{i}$ means the axial external field and $\Om_{i,j}$ the axial shear of the basic rotation. The sign of this pseudo-scalar does not depend on the sign of the magnetic field, but does depend on the sign of the shear. The relation (\ref{h1}) requires a quadratic law, $\alpha \propto B_0^2$, which is indeed confirmed by the simulations. 

The data of the right panel of Fig.~\ref{TG1} have been obtained for a model with piecewise constant $\Om$ and a jump between the two cylinder parts. Only one of the endplates must be forced to rotate so that $\Rm_{\rm out}=$\ord{10} should be possible. The axial component $\alpha_{zz}$ is directly computed for positive and negative shear via the $z$-component of the electromotive force (\ref{rze62}). One finds the same signs for $\alpha_{zz}$ and ${\rm d}\Om/{\rm d}z$. The rotating endplate with a radius $R_{\rm out}$ may rotate with an angular velocity of $\Om_{\rm out}$. Then Fig.~\ref{TG1} yields $\alpha_{zz}\simeq 0.02 R_{\rm out}\Om_{\rm out}$. In the sense of an order-of-magnitude estimate the normalized $\alpha$ thus becomes
\begin{equation}
{\cal C_\alpha} =\frac{|\alpha_{zz}| R_{\rm out}}{\eta} \simeq 0.02\ \Rm_{\rm out} 
\label{h2a}
\end{equation}
with 
\beg
\Rm_{\rm out}=\frac{R^2_{\rm out} \Om_{\rm out}}{\eta}.
\label{rmout}
\ende
The $\Pm$-dependence is very weak. With $\Rm_{\rm out}\simeq 10$ the $C_\alpha$ is maximally of order 0.2.

In order to probe the observability of this effect we start with the potential difference for axial shear between the endplates as  
\beg
\Delta \Phi= 10^{-8} \alpha_{zz} B_0 H,
\label{Volt}
\ende
measured in Volt, Gauss and cm/s and with $H$ the height of the container. With (\ref{h2a}) follows
 \beg
\Delta \Phi= 10^{-8} C_\alpha \eta B_0 \Gamma.
\label{Volt1}
\ende
Hence, with $ \eta\simeq 10^3$ cm$^2$/s for sodium or gallium, 1 kG for the axial field and $\Gamma=H/R_{\rm out}=10$ a potential difference of $\Delta \Phi\simeq 0.1 C_\alpha$ (in V) is generated, which according to (\ref{h2a}) leads to about 10 mV. For longer containers the potential difference grows linearly. The container filled with a liquid metal acts as a generator of an observable potential difference between its endplates -- if the above mentioned instability conditions can be fulfilled. 

The $\alpha$ experiment in Riga worked with $B\simeq 1$ kG and velocities of the order of m/s, so that  $\Delta {\Phi}$ exceeded 10 mV \cite{SK67}. This experiment, however,  used a prescribed helical geometry to mimic the symmetry-breaking between left and right helicities. It has not been demonstrated so far that a rotating fluid with a non-prescribed helicity leads to an observable $\alpha$ effect. By nonlinear numerical simulations with the  {\sc Pencil} code the mean electromotive force in plane Couette flows of a nonrotating conducting fluid under the influence of a large-scale magnetic field on driven turbulence has been calculated. A vertical stratification of the turbulence intensity results in an observable $\alpha$ effect owing to the presence of horizontal shear \cite{RB13}.

\section{Influence of the Hall effect}\label{Influence}
Fluids with Hall effect can be described as conductors with conductivity tensors with off-diagonal elements. Under these conditions a feedback of toroidal to poloidal field exists which in combination with differential rotation -- inducing toroidal fields from poloidal ones -- makes the magnetic field unstable even for an axisymmetric geometry. The instability, however, can only exist if the timescale of the Hall effect is shorter than the diffusion time and longer than the shear time. Otherwise the diffusion or the Hall effect would dominate, destroying any instability. We shall first show that the growth time of such an instability is determined by the rotation time, so that the instability is basically fast. 
 
\subsection{The Shear-Hall Instability (SHI)}\label{SectSHI}
It is known that a stable rotational shear in a  fluid with Hall effect can destabilize a magnetic background field \cite{W99,BT01,RS04}. This `shear-Hall instability' is a basic property of only the induction equation without any contribution by the momentum equation. The mechanism is reminiscent of global dynamo models where the differential rotation transforms poloidal field components to toroidal field components and the meridional flow generates the poloidal fields from the toroidal fields \cite{DJ89}. According to Cowling's theorem such a mechanism can only maintain nonaxisymmetric fields against the magnetic resistivity losses. It is indeed possible to imagine a replacement of  the meridional flow by the Hall term which itself is also able to produce poloidal fields from toroidal ones. As even axisymmetric field configurations can be destabilized by this process it is immediately clear that SHI is by no means a dynamo mechanism. One needs nondecaying background fields to feed the entire system.

The induction equation with Hall effect included is
\begin{equation}
\frac{\partial \vec{B}}{\partial t} = {\rm curl}\, (\vec{U} \times
\vec{B}) + \eta \Delta \vec{B} - \beta_{\rm Hall}\ {\rm curl}\ ({\rm curl}\, \vec{B} 
\times \vec{B})
\label{hall21}
\end{equation}
where the Hall parameter $\beta_{\rm Hall}$ does not depend on the magnetic field. One can show that the Hall effect exactly conserves the magnetic energy. The only source of energy is due to the shear, so that the Hall term alone is unable to feed an instability \cite{HR02,WH09}. The sign of the Hall parameter $\beta_{\rm Hall}$ depends on the definition of the elementary charge; we shall only use it as a positive number.

The linearized version of Eq.~(\ref{hall21}) for a current-free background field becomes
\begin{equation}
\frac{\partial \vec{b}}{\partial t} = {\rm curl}\, (\vec{u} \times
\vec{B}) + {\rm curl}\, (\vec{U} \times
\vec{b}) + \eta \Delta \vec{b} - \beta_{\rm Hall}\ {\rm curl}\, ({\rm curl}\ \vec{b} 
\times \vec{B}).
\label{hall2}
\end{equation}
If the Hall term exceeds the first term on the right side of this equation then the induction equation decouples from the Navier-Stokes equation, and one may ask whether the remaining equation can have its own solution. This is indeed the case. For plane short waves subject to a global rotation with $\Om\propto R^{-q}$ a dispersion relation
\begin{equation}
({\tilde \omgr}+1)^2 +\Rb(\Rb-q {\tilde \Rm})=0
\label{hall22}
\end{equation}
results where $\tilde \omgr$ is the growth rate normalized with the resistivity frequency $\omega_\eta=\eta k^2$ (with $\vec{k}$ as the wave number). $\tilde \Rm$ is also formed with $\eta k^2$. Note that for neutron stars the ratio 
\begin{equation}
{\rm Rb}=\frac{\tau_{\rm diff}}{\tau_{\rm Hall}}=\frac{\beta_{\rm Hall} B_0}{\eta}
\label{Rb}
\end{equation}
lies between 1 and 100 for magnetic fields of order $10^{12}$ G. In dependence on the orientation of the field this parameter can have both signs. An instability can indeed exist if for positive or negative $\Rb$ we have $|\Rb|<|q|{\tilde \Rm}$ and $q$ and $\Rb$ are of the same sign. There must also be a lower bound of $\Rb$, as the Hall effect can also be too weak for an instability. It can also be too strong as only the shear produces the needed energy rather than the Hall effect. For a rotation profile depending only on the radial coordinate  the necessary condition for shear-Hall instability is
\begin{equation}
	(\vec{k}\cdot{\vec{B}})\ k_{z}
\frac{\partial\Om}{\partial R}<0
\label{hall4}
\end{equation}
 \cite{UR05}. For axial $\vec{B}$ this condition simplifies to $q B_{z}>0$, hence there is instability if the signs of the two factors are equal. SHI is thus even able to destabilize flows with positive shear ${\rm d}\Om/{\rm d}R$ if the magnetic field is antiparallel with the rotation axis. That the sign of the magnetic field here plays an important role is a direct consequence of the nonlinear  Eq.~(\ref{hall21}).

Introducing a  dimensionless quantity which only includes material parameters we write
\beg
\beta_0 = \frac{{\rm Rb}}{\Lu}
\label{beta2}
\ende
with the Lundquist number $\Lu=\sqrt{\Pm}\Ha$ and $\Ha$ defined by Eq.~(\ref{Hartmann}). The parameter $\beta_0$ may also have both signs, depending on the orientation of the magnetic field relative to the rotation axis. The amplitude of $\beta_0$ can be imagined as smaller than \ord{1}. 
\begin{figure}[htb]
\centering
\includegraphics[width=9cm]{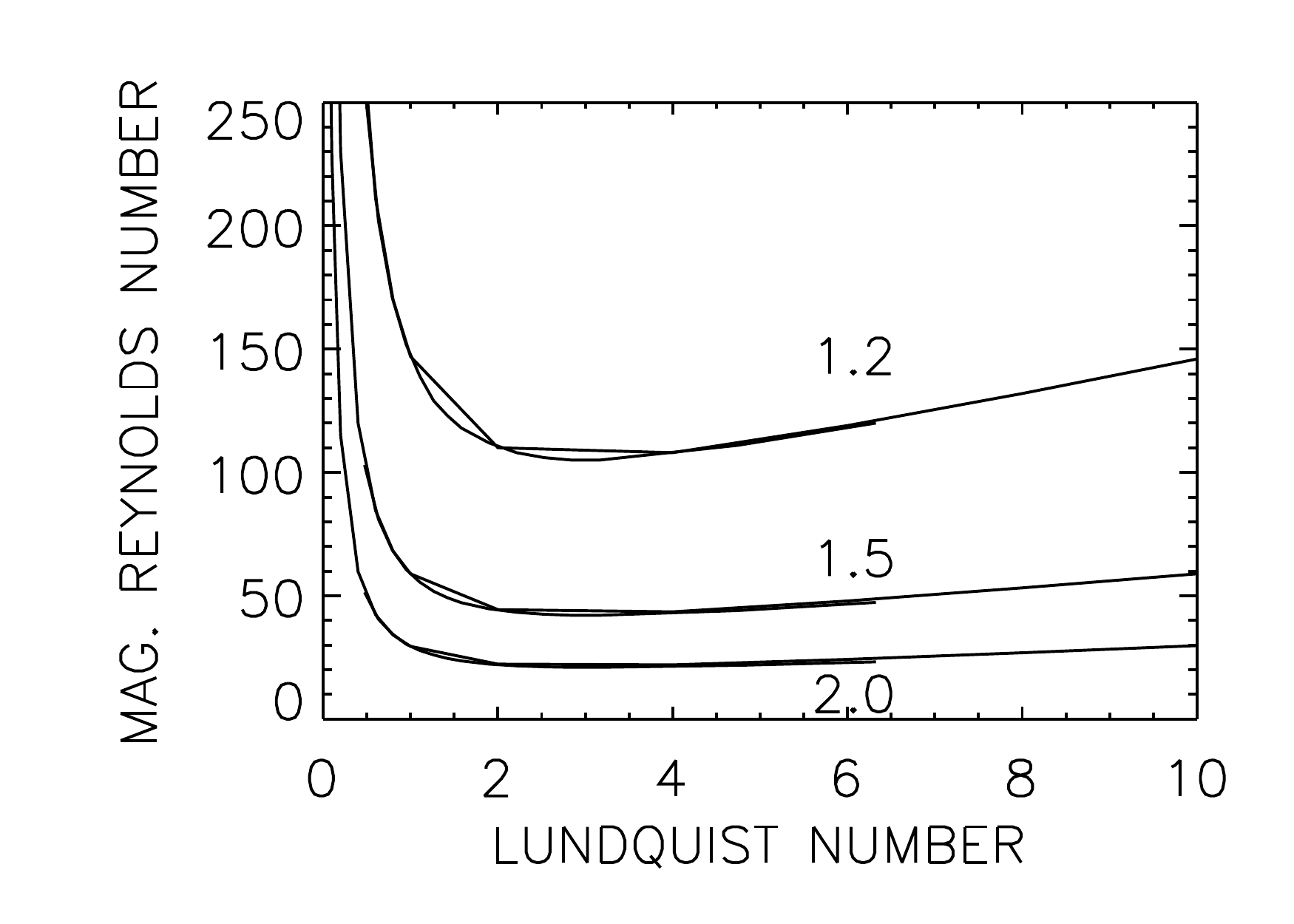}
\caption{Shear-Hall instability  with axial magnetic fields  for positive shear.  As must be the case, $\Pm=10^{-5}$ and $\Pm=1$ provide identical curves which have been  calculated for $\mu_\Om>1$ as dedicated. Their minima possess large magnetic Mach numbers. Hall parameter $\beta_0=-1$. $m=0$, $\rin=0.5$, insulating boundary conditions.} 
 \label{SH1}
\end{figure}

Figure~\ref{SH1} illustrates SHI of superrotating fluids for insulating boundary conditions. The boundary conditions for the hydromagnetic Taylor-Couette flow are not influenced by the Hall effect. The calculation of the Hartmann number uses the axial field strength as prescribed by the definition (\ref{Hartmann}). The flow is unstable only for $\beta_0<0$ when rotation axis and magnetic field are antiparallel. Equation (\ref{hall2}) has been solved for three rotation profiles with positive shear. The rather strong differences of the resulting characteristic Reynolds numbers for fixed Hartmann number demonstrate how important the differential rotation for the instability mechanism is. For weaker shear one needs faster rotation to excite the instability. In Fig.~\ref{SH1} for Prandtl numbers differing by five orders of magnitude the differences of the curves number are very small. Hence, the eigenvalues scale for $\Pm\to 0$ with the Lundquist number $\Lu$ and the magnetic Reynolds number $\Rm$. It is then easy to calculate the magnetic Mach number as $\Mm=\Rm/\Lu$. In all cases with $\Pm\lsim 1$ the instability exists for rapid rotation with $\Mm>1$. 

With spherical models it has been shown that the growth rates of SHI scale with the rotation rate and not with the rather long Hall time \cite{KR11,KR12}. The dispersion relation (\ref{hall22}) leads to the same conclusion. The maximum of the first bracket as a function of $\Rb$ is taken for $\Rb=q \tilde\Rm/2$ so that ${\tilde \omega}_{\rm gr,max}+1=q{\tilde\Rm}/2$. Dropping the tildes indeed leads to $\omega_{\rm gr,max}\propto \Om$ for the growth rates of SHI.

\subsection{Hall-MRI}\label{HallMRI}
\begin{figure}[htb]
\centering
\includegraphics[width=5.3cm, height=5cm]{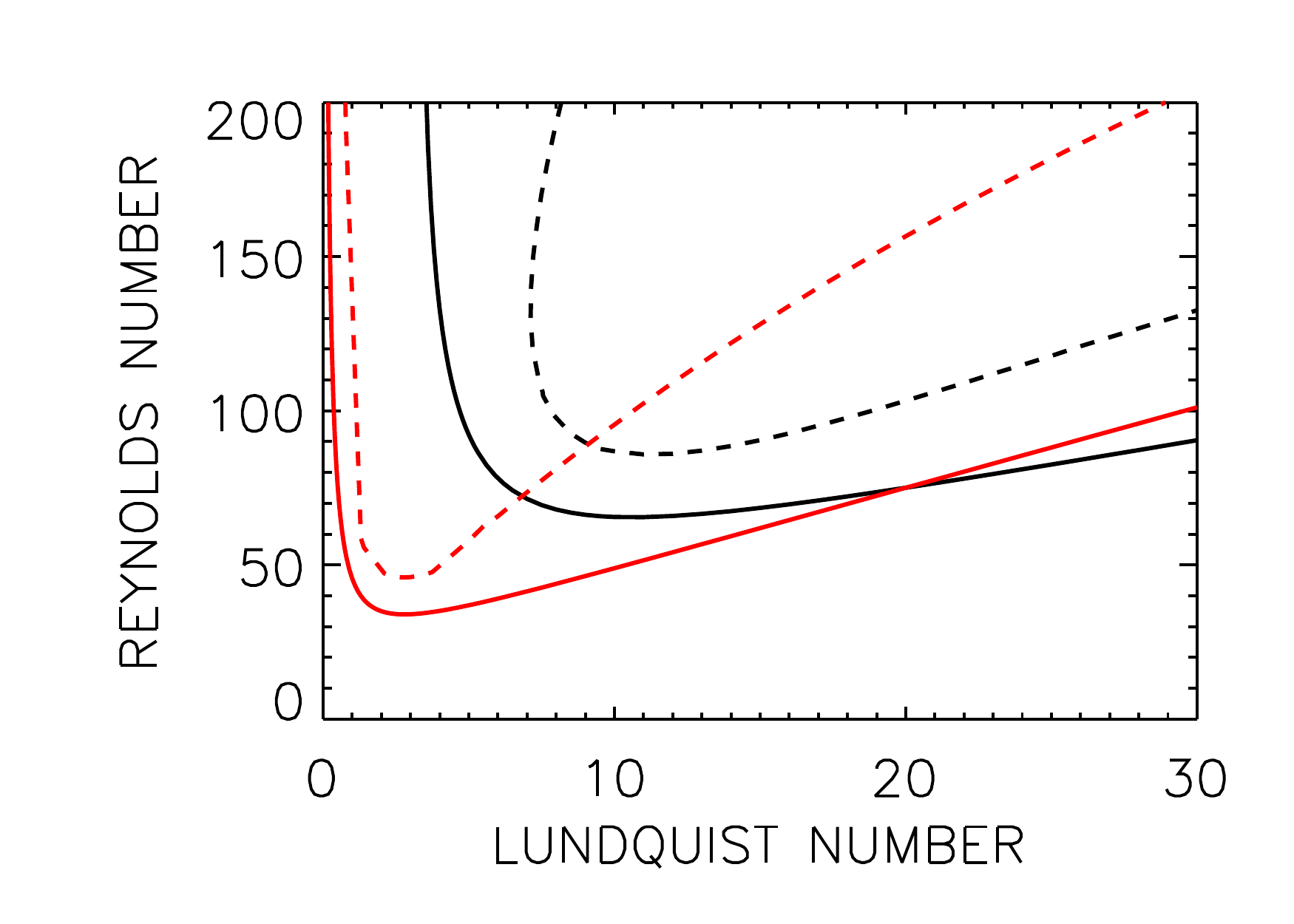}
\includegraphics[width=5.3cm,height=5cm]{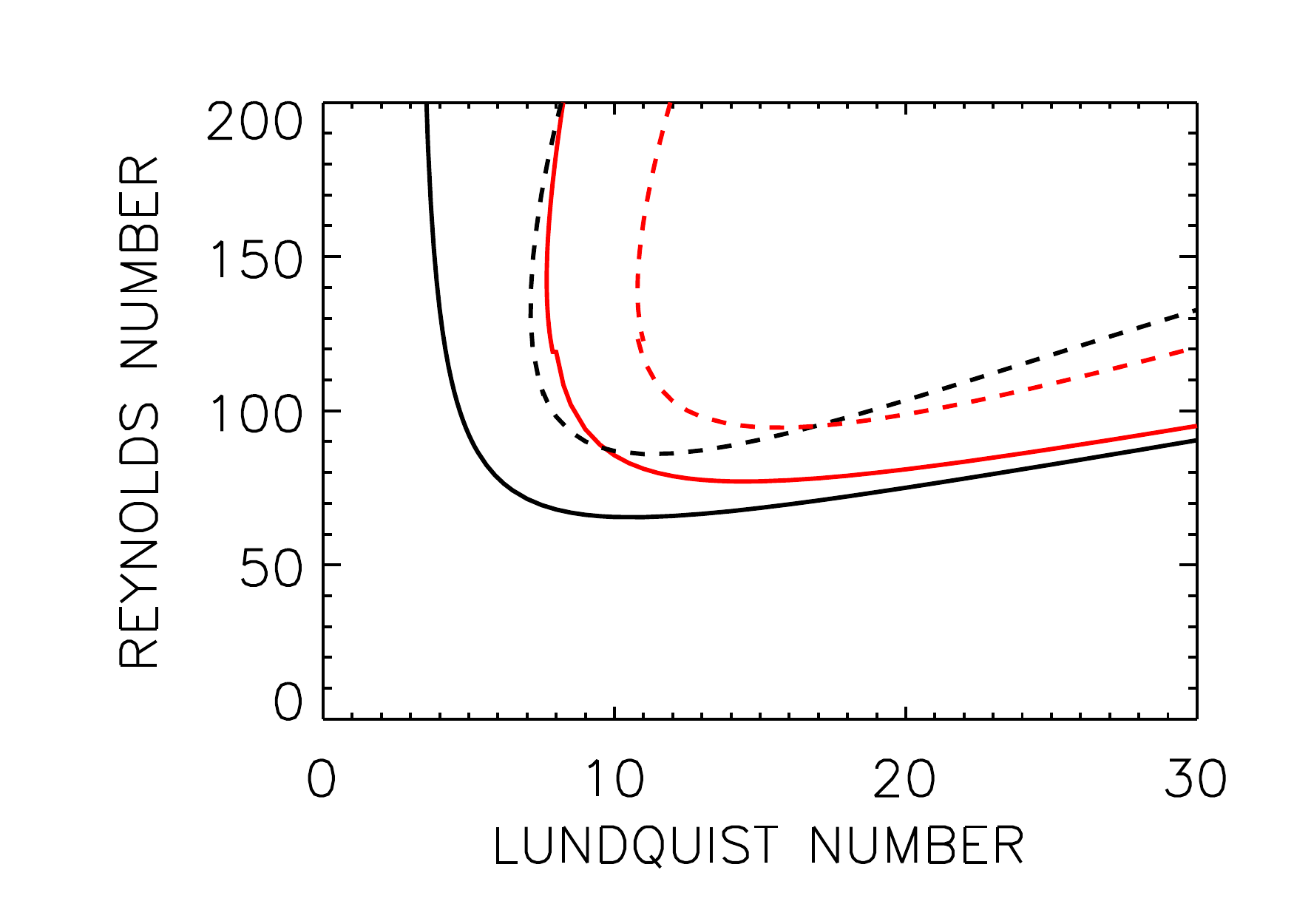}
 \includegraphics[width=5.3cm,height=5cm]{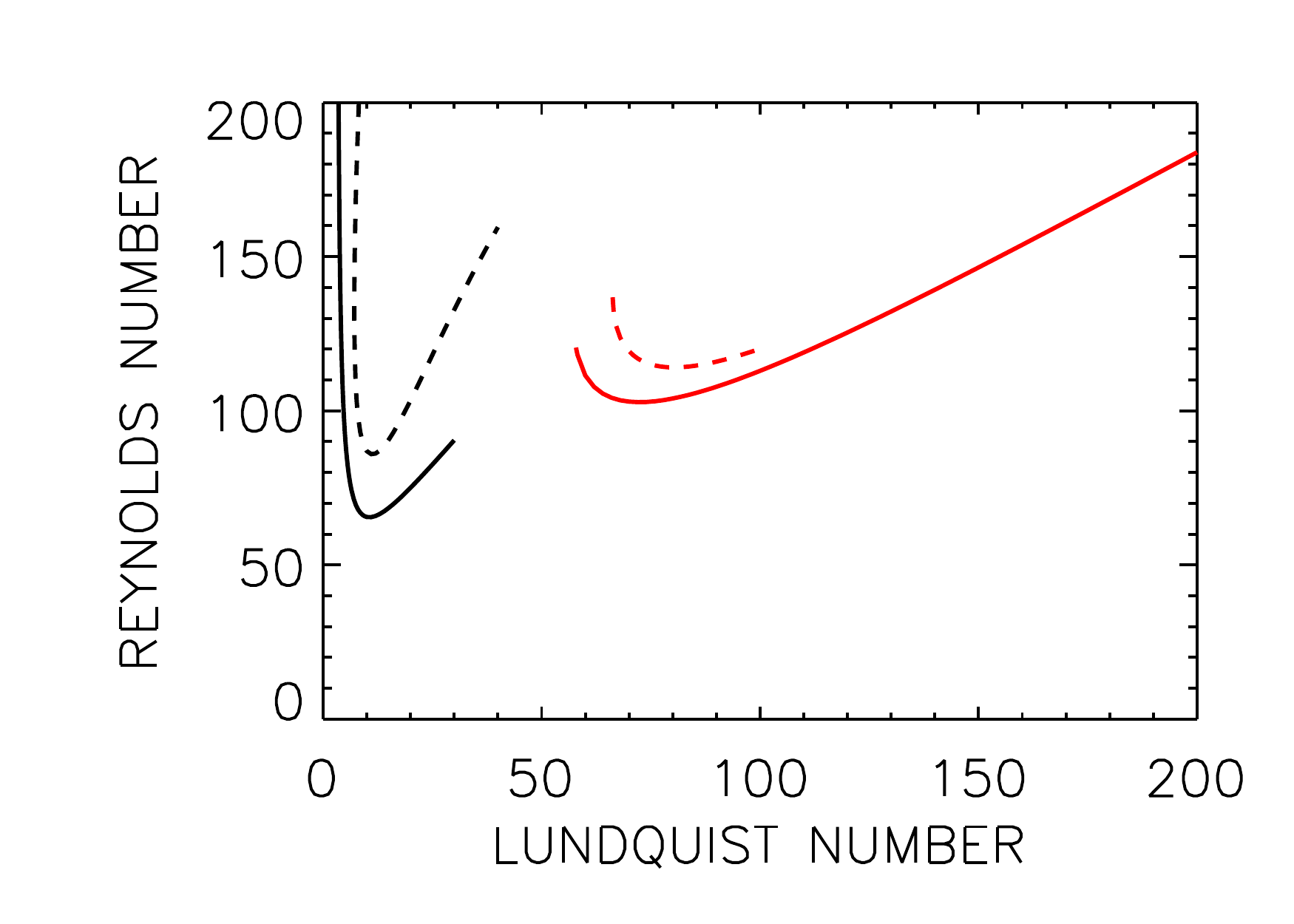}
\caption{Stability maps for quasi-Keplerian flow subject to axial fields with (red) and without (black)  Hall effect for $m=0$ (dashed lines) and $m=1$ (solid lines). Red lines: $\beta_0=1$ ({left}), $\beta_0=-0.1$ ({middle}) and $\beta_0=-1$ (right). Quasi-Keplerian  rotation is maximally destabilized by positive Hall effect (field parallel to rotation axis).  $\rin=0.5$, $\mu_\Om=0.35$, $\rm Pm=1$. Insulating boundary conditions.} 
\label{SH2}
\end{figure}

Flows with {negative} shear (positive  $q$) are much more complicated, as they can even be unstable in the presence of magnetic fields without Hall effect. We expect, however, that for positive $\Rb$ the shear-Hall instability supports the MRI but the strength of the support must be calculated. The dark lines in Fig.~\ref{SH2} indicate marginal stability without Hall effect for axisymmetric and nonaxisymmetric MRI modes. As usual the instability domain for the nonaxisymmetric mode is much smaller than for the axisymmetric mode. It is increased, however, for fluids with Hall effect if the magnetic axis and the rotation axis have the same orientation so that $B_{z}{{\rm d}\Om}/{{\rm d}R}<0$. This case is realized in the left panel of Fig.\ \ref{SH2} for strong Hall effect with $\beta_0=1$. The minimum Lundquist number for instability is now smaller than for MRI, but for large magnetic fields the critical Reynolds number with Hall effect becomes greater than without Hall effect. Hence, for positive $\beta_0$ the Hall effect destabilizes for weak fields and stabilizes for strong fields. The red dashed lines representing the $m=1$ mode with Hall effect are also shifted so that nonaxisymmetric modes become more (less) unstable for weak (strong) fields. Note also that the dashed red line for the mode $m=1 $ in the left panel of Fig.~\ref{SH2} no longer has the positive slope of both branches as it appears for nonaxisymmetric MRI modes without Hall effect. In the Hall regime the nonaxisymmetric mode has the same open geometry as the axisymmetric mode of MRI, so that it is not suppressed for rapid rotation. The different geometry of the neutral stability curves of the nonaxisymmetric modes with and without Hall effect seems to be the most striking consequence of the Hall-MRI. The Hall-MRI thus produces much more complex patterns than the axisymmetric rings which are mainly excited by standard MRI without Hall effect. The minimum  magnetic Reynolds number is also reduced by the Hall effect, but this reduction remains small. One only finds a reduction by a factor of $\lsim 2$, as realized in Fig.~\ref{SH2} (left) for strong positive Hall effect.

For the opposite case of negative Hall effect, for fields antiparallel to the rotation axis, the MRI is suppressed for weak fields and enhanced for strong fields. In Fig.~\ref{SH2} the plots for negative $\beta_0$ (middle and right panels) show the red lines for Hall-MRI as shifted towards large $\Lu$. Even the small value $\beta_0=-0.1$ gives a drastic stabilization of the axisymmetric and nonaxisymmetric modes. Moreover, both branches of the $m=1$ mode have the typical positive slopes, so that there is also a maximum Reynolds number beyond which the nonaxisymmetric modes are suppressed. The characteristic Lundquist numbers for instability at the global minimum Reynolds number also strongly increase for $\beta_0=-1$. It seems that the stabilization by negative $\beta_0$ appears to be much more effective than the destabilization by positive $\beta_0$.

\subsection{Hall-TI}
\begin{figure}[h]
\centering
\includegraphics[width=8cm]{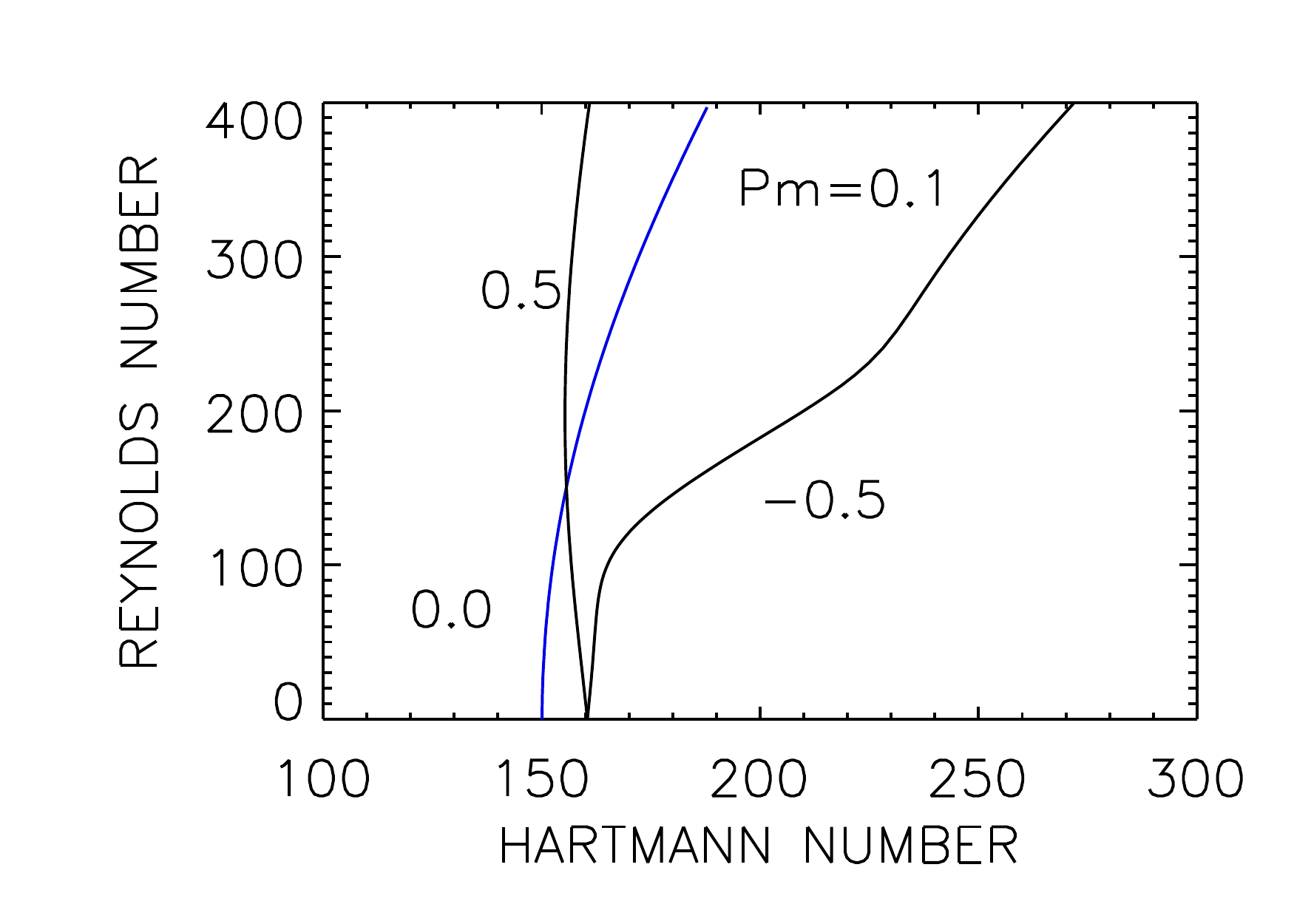}
\includegraphics[width=8cm]{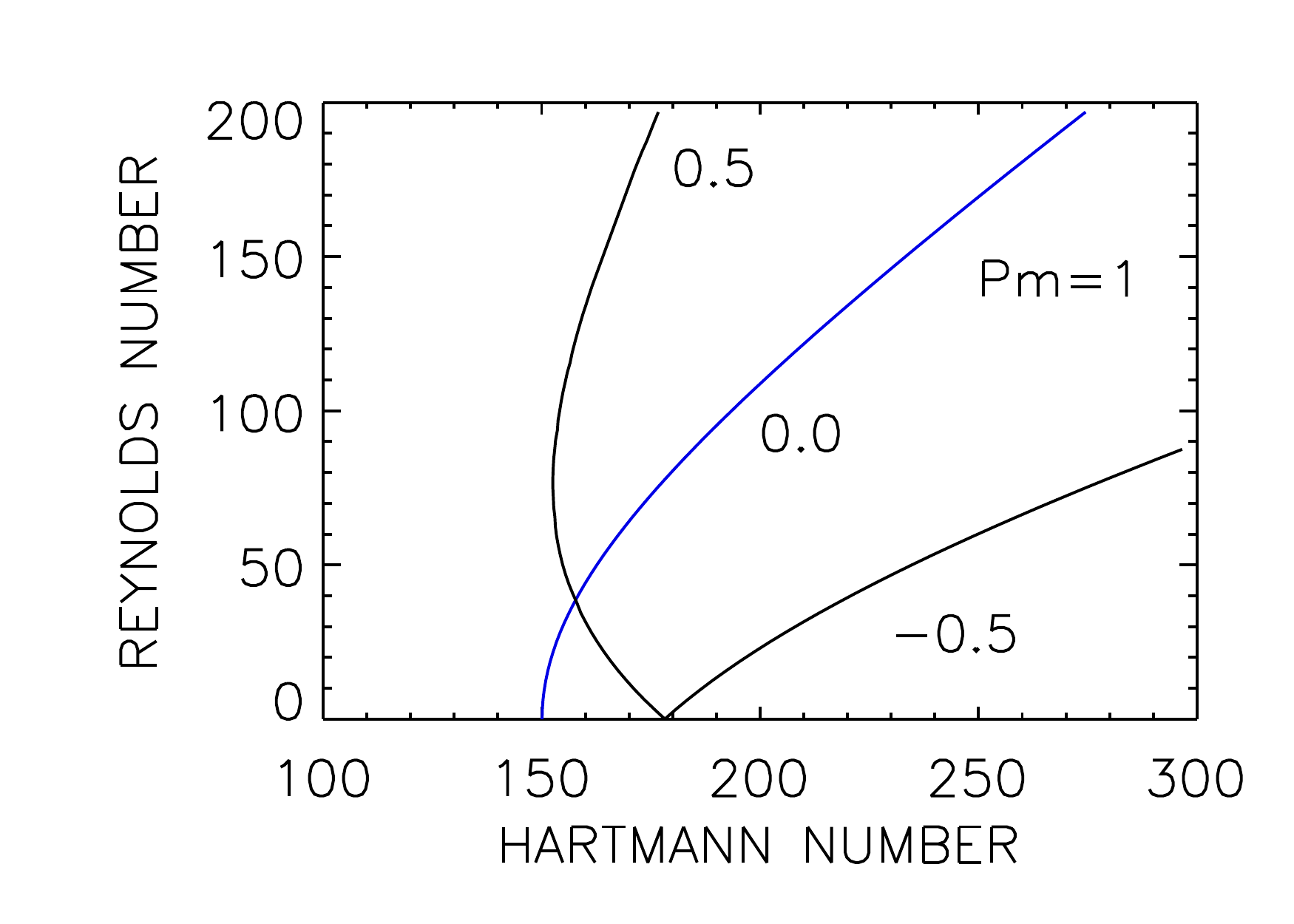}
\caption{Hall-TI of quasi-uniform azimuthal fields and quasi-uniform flows with two magnetic Prandtl numbers. The curves are labeled by the Hall parameter $\beta_0$. $\rm Pm =0.1$ (left), $\Pm=1$ (right). Note the increase of $\Ha_0$ for both signs of $\beta_0$ (blue line $\beta_0=0$). The plots for $\beta_0\to -\beta_0$ and simultaneously $m\to -m$ are identical. $m=1$, $\rin=0.5$, $\mu_B=2\mu_\Om=1$. Perfectly conducting boundaries.}
\label{mainplots}
\end{figure}

Another situation holds if the magnetic background field contains electric currents. The influence of the electric current is twofold. It enters the expression of the Hall effect in Eq.~(\ref{hall21}) and produces an own pinch-type instability, so that strong modifications of the TI must be expected \cite{RSS09}. We start with the Chandrasekhar-type flow with almost uniform toroidal field, i.e.~$\mu_B=2\mu_\Om=1$, for which we know that for $\Pm\to 0$ it scales with Reynolds number and Hartmann number. For azimuthal fields the Hartmann number is defined by Eq.~(\ref{Hartmannin}). The numerical results are given in Fig.~\ref{mainplots} for Hall parameters $\beta_0$ between $-0.5$ and $0.5$. The value $\beta_0$ and the magnetic Prandtl number $\Pm$ are the free parameters of the system for a prescribed hydromagnetic Taylor-Couette flow. We again define $\Ha$ as positive and use both signs of $\beta_0$ corresponding to opposite magnetic field orientations.

The Hall-free curves start at $\Ha_0=150$ for $\Rey=0$. For $\beta_0\neq 0$, $\Ha_0>150$. The increase does not depend on the sign of the Hall term. For either sign, in stationary containers the Hall effect  stabilizes the azimuthal field. For $\beta_0<0$ the rotation together with the Hall effect has a strongly stabilizing influence. To become unstable the magnetic field must be much stronger under the influence of rotation than without rotation. On the other hand, for $\beta_0>0$ the Hall effect has a destabilizing influence ($\Ha<\Ha_0$) if the rotation is not too rapid. The destabilization is thus similar to that of the shear-Hall instability with axial fields. Axial fields with positive $\beta_0$ also destabilize flows with negative shear. 
\begin{figure}[h]
\centering
\includegraphics[width=8cm]{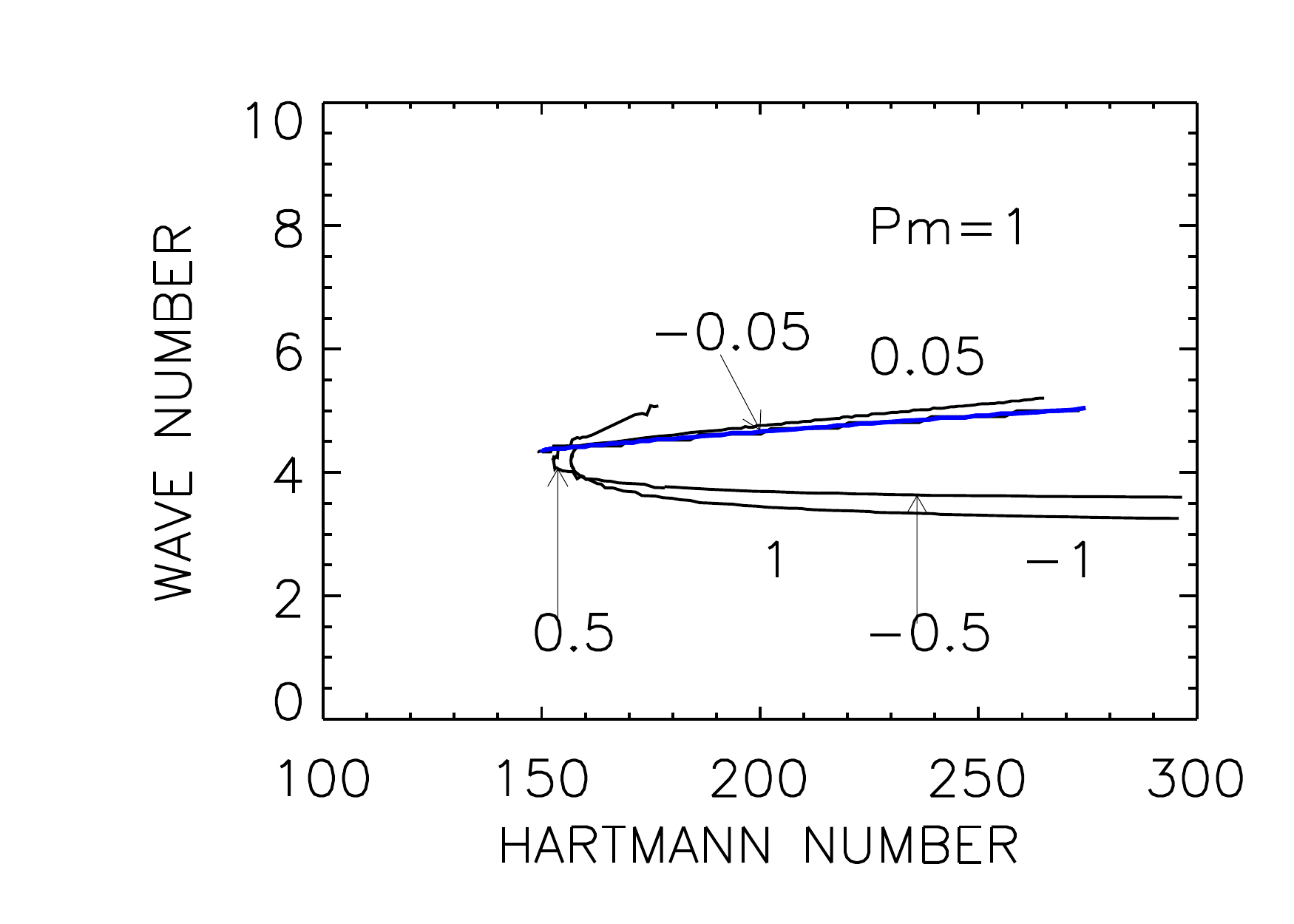}
\includegraphics[width=8cm]{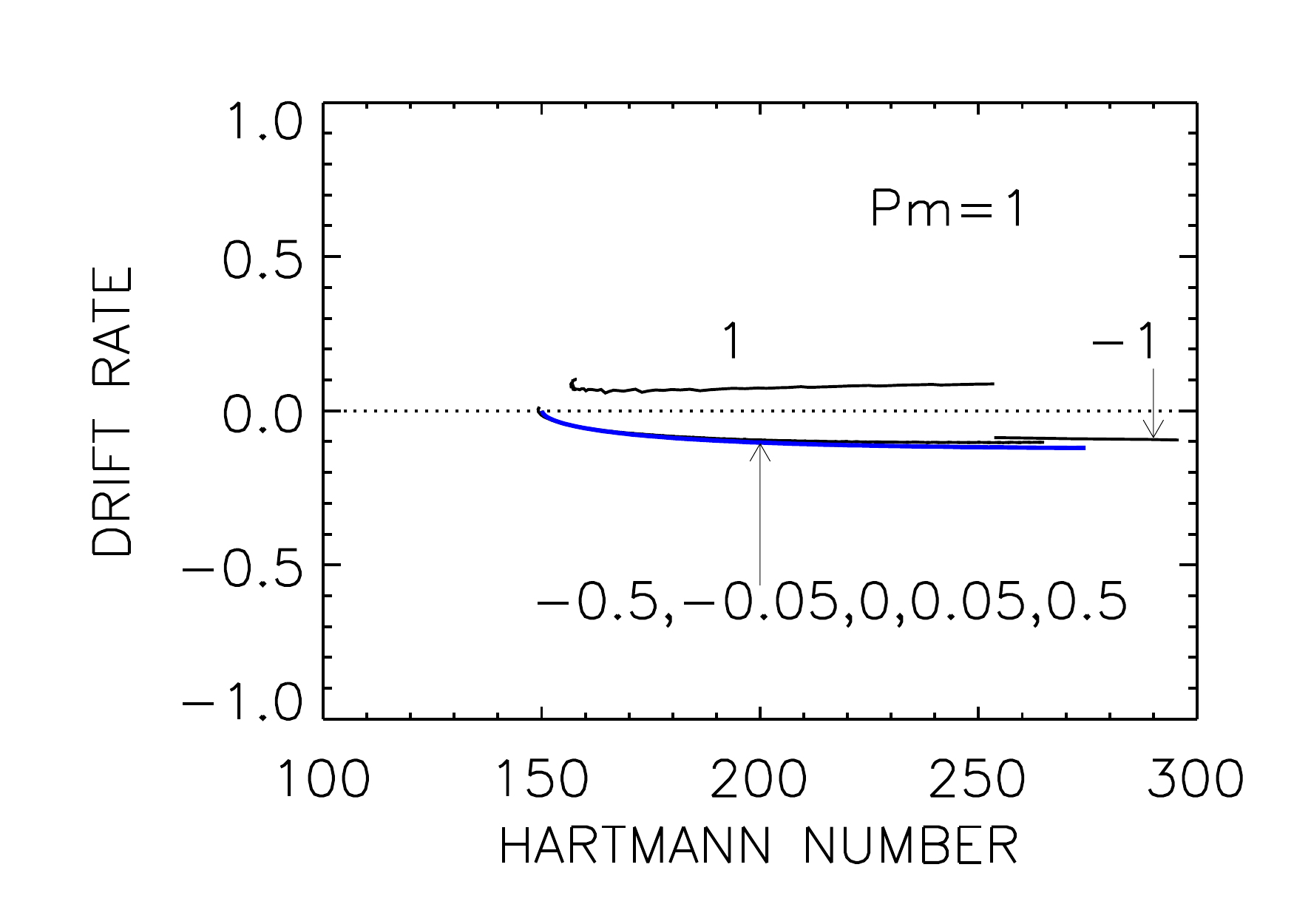}
\caption{Same as in Fig.~\ref{mainplots} (right) but for  wave numbers (left) and drift rates (right) along the neutral lines of the Hall-TI. The lines are marked with $\beta_0$; the blue line denotes $\beta_0=0$. The influence of the Hall effect is only weak. $m=1$, $\rin=0.5$, $\mu_B=2\mu_\Om=1$, $\Pm=1$. Perfectly conducting boundaries.}
\label{hallti}
\end{figure}

In accordance with the relation (\ref{hall4}), for azimuthal fields the positive Hall effect also destabilizes flows with negative shear. For positive $\beta_0$ the stability domain is reduced, and for negative $\beta_0$ it is increased. The stabilization (destabilization) of negative (positive) Hall $\beta_0$ is a common phenomenon of all models. In other words, for positive $q$,  positive $B_\phi$ (i.e.~$\beta_0>0$) lead to smaller critical field amplitudes than negative $B_\phi$ (i.e.~$\beta_0<0$). If the nonaxisymmetric Tayler instability would limit the strength of the toroidal fields $B_\phi$, then the resulting amplitudes are different for different signs of $B_\phi$ due to the action of the Hall effect. The effects, however, are not substantial. Wave number and drift rates are influenced even less by the Hall effect (Fig.~\ref{hallti}). Generally, positive (negative) Hall effect decreases (increases) the axial size of the instability cells. If the Hall effect destabilizes then it acts against the Taylor-Proudman theorem. On the other hand, a stabilizing Hall term elongates the cells in the axial direction.

\begin{figure}[htb]
\centering
\includegraphics[width=8cm]{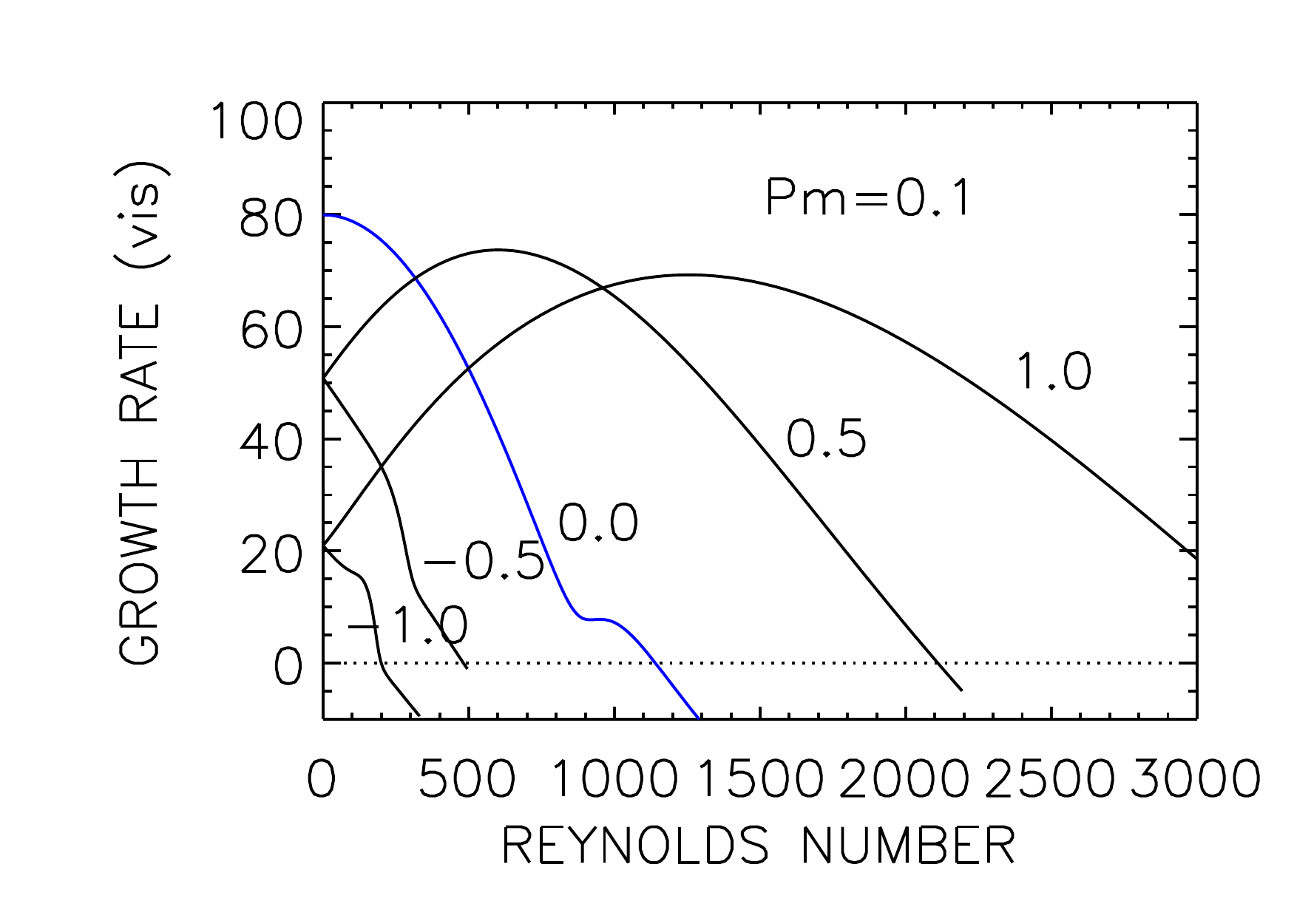}
\includegraphics[width=8cm]{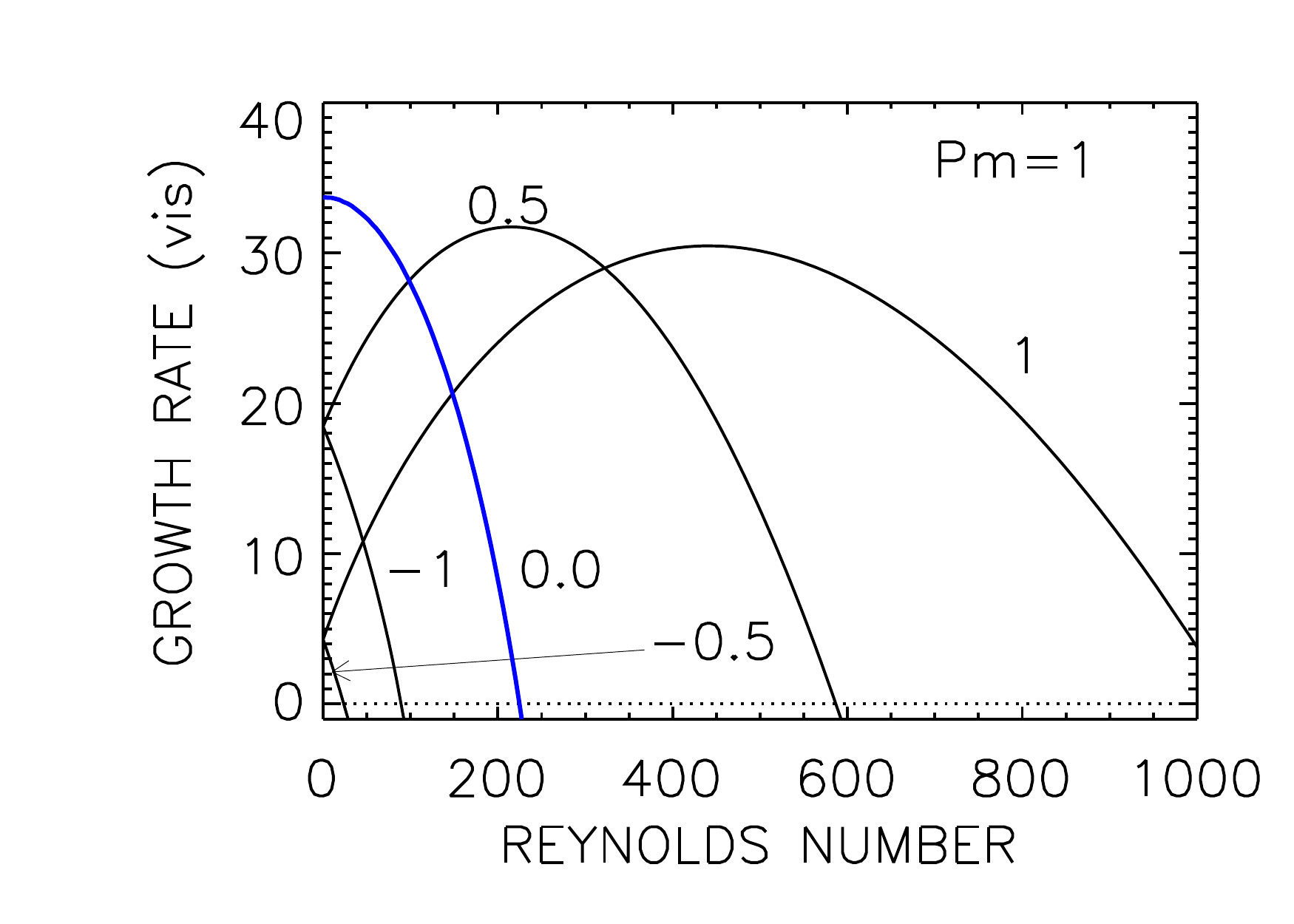}
\caption{Growth rates for Hall-TI normalized with $\omega_\nu$  with and without Hall effect for $\Ha=300$.  $\Pm =0.1$ (left) and $\Pm=1$ (right). The curves are marked by their Hall parameter $\beta_0$; the blue lines give the TI without Hall effect. $m=1$, $\rin=0.5$, $\mu_B=2\mu_\Om=1$. Perfectly conducting boundaries.}
\label{growthhall}
\end{figure}

The question still remains how the Hall effect modifies the growth times of the TI. Figure \ref{growthhall} shows the growth rates of the Hall-TI for various parameters. They are computed for $\rm Ha=300$ and for increasing rotation rates. The growth rates -- normalized here with the viscosity frequency $\omega_\nu=\nu/R_0^2$ -- vanish at the stability lines. One finds that for the considered parameters the Hall effect strongly influences the growth rates of TI. For negative $\beta_0$ the rotational stabilization of TI is  amplified. For positive $\beta_0$, however, the rotational suppression without Hall effect {\it is compensated} by the Hall effect. Hence, the {\em maximal} growth rate is always given by the value for $\Om=0$, and this quantity in the normalization used scales with $\Ha/\sqrt{\Pm}$, so that the physical growth rate for $\Om=0$ scales with the \A\ frequency $\Om_{\rm A}$. The positive Hall effect (almost) reproduces this value at a certain magnetic Reynolds number where the Hall-influenced growth rate has its maximum. This surprising phenomenon only occurs for positive Hall effect and under the presence of differential rotation. For positive Hall effect the TI grows much faster than for negative Hall effect. These findings strongly resemble the consequences of the SHI effect for axial fields described in Section \ref{SectSHI}. Indeed, the relation (\ref{hall4}) does not exclude the existence of an azimuthal SHI. Contrary to the solutions with positive $m$, however, for $m<0$ (if $k_z$ is assumed as positive definite) the destabilization occurs for negative $\beta_0$ rather than for positive values. For destabilization the product of $m \beta_0$ must be positive, while for stabilization it must be negative.

The situation at the vertical axis, $\Rey=0$, is also of interest. The Hall effect considerably reduces the growth rates of TI without rotation, and this reduction is the same for both signs of $\beta_0$. On the other hand, the growth rates vanish for characteristic upper Reynolds numbers beyond which the TI with $m=1$ decays. These Reynolds numbers $\Rey_{\rm max}$ also reflect the suppressing action of negative $\beta_0$, and the enhancing action of positive $\beta_0$. These actions, however, are asymmetric: the increase of $\Rey_{\rm max}$ for positive $\beta_0$ compared with $\Rey_{\rm max}$ for $\beta_0=0$ (blue lines) is much larger than the same difference for negative $\beta_0$.

One also finds that even a weak Hall effect -- resulting in a long Hall time of order of the diffusion time -- does not generally prolong the growth time of the Tayler instability, which also in this case scales with the \A\ time. In this sense the Hall effect is only a modification of another instability and does not impose its own timescale on the instability. 
\begin{figure}[htb]
\centering
\includegraphics[width=8cm]{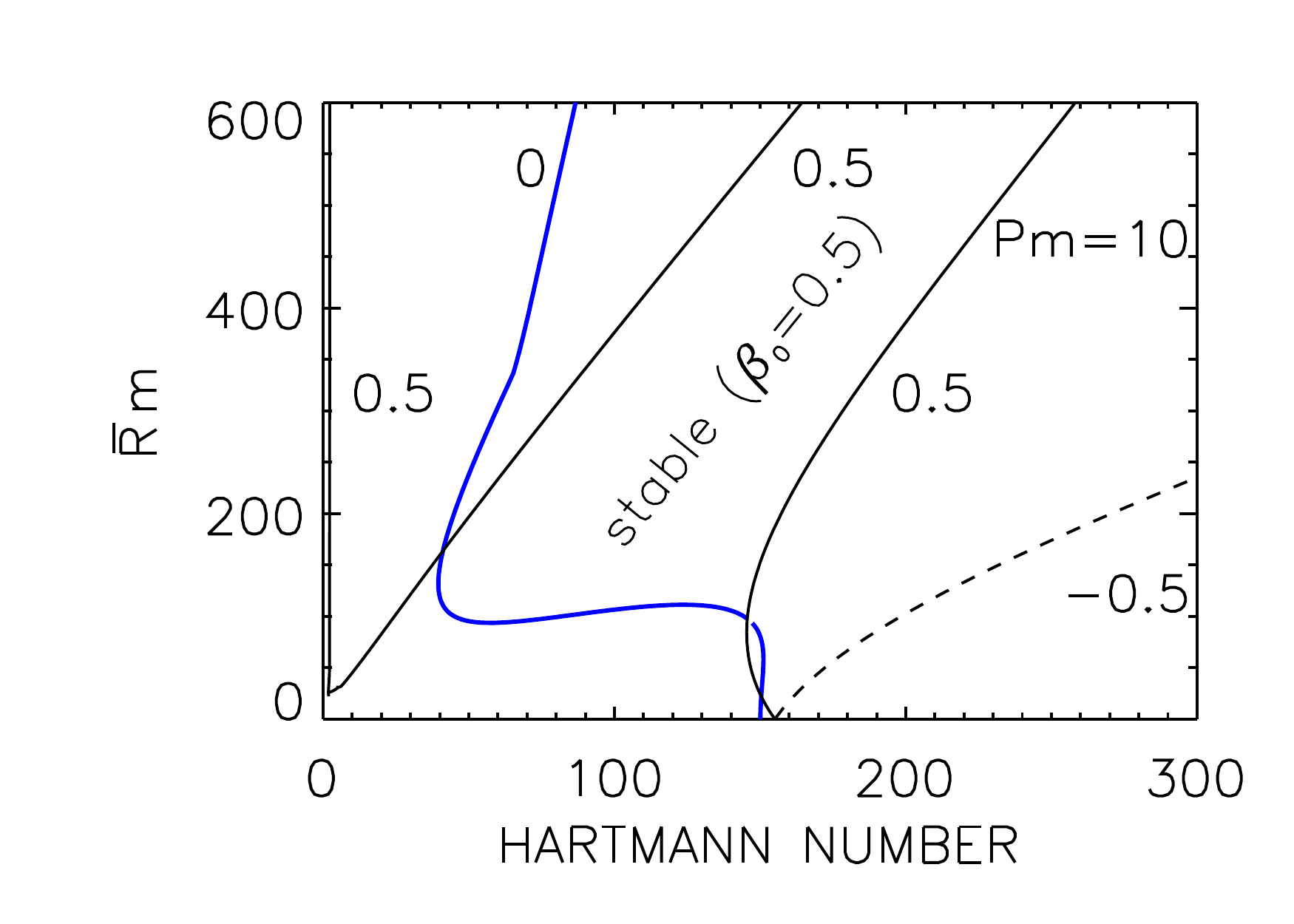}
\includegraphics[width=8cm]{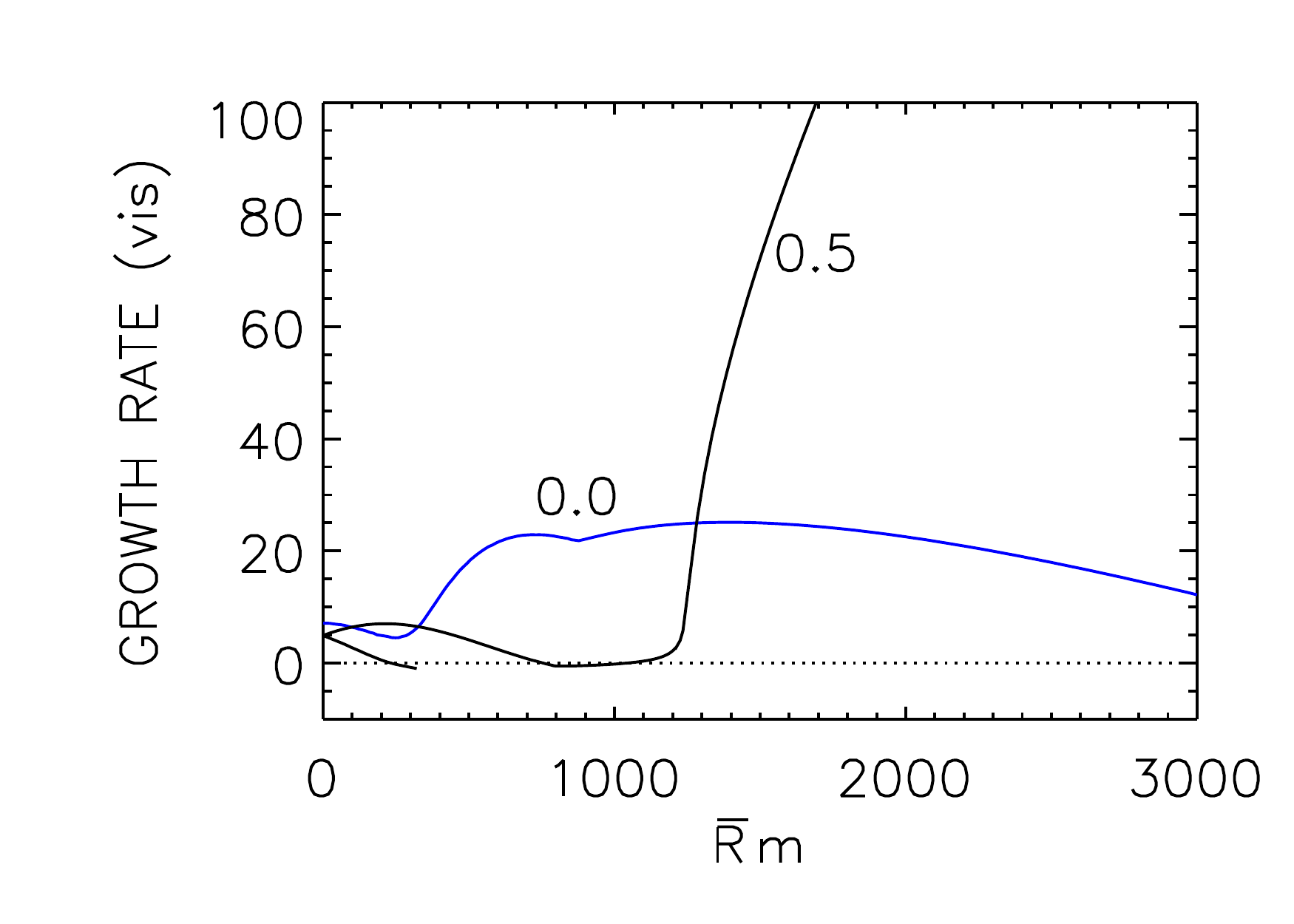}
\caption{Stability map (left) and growth rate in units of $\omega_\nu$ (right, $\Ha=300$) for Hall-TI with large $\Pm$. The curves are marked by their Hall parameter $\beta_0$; the blue lines are for $\beta_0=0$. $m=1$, $\Pm =10$, $\rin=0.5$, $\mu_B=2\mu_\Om=1$. Perfectly conducting boundaries.}
\label{hallpm10}
\end{figure}

The results for large $\Pm$ are also interesting. Without Hall effect the differential rotation strongly supports the TI as long as the rotation is slow and the magnetic Prandtl number is large. The blue line in the left panel of Fig.~\ref{hallpm10} reflects a distinct subcritical excitation with $\Ha<\Ha_0$ for the lower Reynolds numbers. Because of the action of the differential rotation on nonaxisymmetric modes this phenomenon is finally compensated, leading to $\Ha>\Ha_0$ for sufficiently large Reynolds numbers. We take from Fig.~\ref{hallpm10} that the negative Hall effect destroys the subcritical excitation for slow rotation. The positive Hall effect produces a stable branch in the ($\Ha/\Rmquer$) plane which separates two unstable branches. The large-field branch shows no subcritical excitation behavior but the weak-field branch introduces extremely weak fields for which the system becomes unstable. This instability domain has no relation to the TI but it is due to the SHI for negative shear. According to the condition (\ref{hall4}) it will disappear for $m=-1$.
\section{Prospect of future experiments}

We have systematically assessed the
instabilities that arise in Taylor-Couette flows under the 
influence of magnetic fields with diverse geometries.
Starting with the classical problems of 
flows with stationary 
outer cylinder and the standard MRI for 
quasi-Keplerian rotation, the focus has 
moved to various types of diffusive 
instabilities, in particular AMRI, HMRI and TI, 
which survive also at low magnetic Prandtl numbers. One finds for their lines of neutral stability  convergence in the ($\Ha/\Rey$) coordinate plane for decreasing magnetic Prandtl number   $\Pm\to 0$, which 
can also be obtained with  the  inductionless approximation  of the MHD equations for $\Pm=0$. Both issues are typical for the special class of MHD flows introduced by Chandrasekhar \cite{C60}  which is defined by identical radial profiles   of the background field and  flow. The potential flow under the influence of a current-free azimuthal field  as well as  the rigidly rotating $z$-pinch are  examples of such Chandrasekhar-type flows which have  been discussed in detail. 
It is this very feature which  
makes them suitable for being investigated in
liquid metal experiments 
with comparably low effort.
We have discussed in detail the experiments on 
HMRI and AMRI carried out at the  {\sc Promise} facility, and 
the TI experiment at the {\sc Gate} facility.

A new large-scale liquid sodium 
Taylor-Couette experiment,
which is presently under construction in the framework
of the {\sc Dresdyn} project \cite{SE12}, will 
achieve significantly higher magnetic Reynolds numbers 
($\Rm \approx 40$) and Lundquist numbers ($\Lu \approx 10$) 
than the {\sc Promise} experiment.  The same split-lid technique
as successfully used in the {\sc Promise} experiment will be applied,
with the fallback option of developing a more complicated
multi-ring lid system with planetary gears.
The first and foremost 
aim of this new
experiment will be to approach standard
MRI supposed to start at $\Rm \simeq 20$ and $\Lu \simeq 4$ (see Table \ref{MRI})
by setting out from  the  regime of 
HMRI and increasing
$\Rm$ and $\Lu$, while simultaneously decreasing the ratio 
of azimuthal to axial field. 
 This is a legitimate 
procedure, since HMRI and standard MRI are 
indeed connected in a continuous and monotonic manner, although 
this connection is quite subtle as the standard MRI does not exist for $\Pm=0$.

The second aim of the large Taylor-Couette experiment 
will be to study various combinations of MRI and TI 
by adding, to the axial current along the central  axis, some 
parallel current in the liquid sodium.  
A specific goal here is to prove the extension
of HMRI and AMRI to Keplerian rotation 
just by slightly flattening the
radial profile of the azimuthal magnetic field.
The corresponding 
quasi-Keplerian Chandrasekhar-type flow will be destabilized
for ${\Rey} \approx 10^5$ and 
${\Ha} \approx 300$ (from Fig.~\ref{g2}) which is  achievable 
with the new experimental setting.

The technical feasibility of another liquid sodium 
experiment devoted to Super-AMRI is presently under scrutiny.
The principle of such an experiment,
and the necessity for using a narrow-gap setup has been demonstrated in 
Section \ref{Super}, where also the very strong influence of the boundary conditions 
on the critical values of the Hartmann and Reynolds numbers (for small $\Pm$)   is demonstrated. 
Perfectly conducting boundaries
lead to a reduction of the values needed for the onset of the instability by 
a factor 3 compared with insulating boundaries. The use of 
copper walls, having a conductivity $\sim5$ times higher than liquid sodium,
should indeed make such an experiment possible.

Finally, it is noteworthy how much work remains
to be done to explore the fully nonlinear regime for all of the instabilities
presented in this review. There are so many parameters in the problem --
$\Om_{\rm in},\ \Om_{\rm out},\ B_{\rm in},\ B_{\rm out},\ B_z,\ \Pm$ -- that
simply mapping out how the linear onset depends on all of them has been a major
undertaking, as presented in this review. Understanding the nonlinear equilibration,
both computationally and analytically (e.g.\ \cite{Julien}) is the next task, and
will undoubtedly reveal new results, such as subcritical instabilities
(e.g.\ \cite{GusevaPRL}), or connections between instability modes that are
completely separate in the linear regime. Nonlinear theories and computations,
including in the fully turbulent regime, will allow the more thorough
calculation of turbulent transport coefficients, such as eddy viscosity and
effective diffusivity, as well as the possible occurrence of helicities and the
corresponding $\alpha$ effect. The implications of these findings for basic
astrophysical problems, such as angular momentum transport or nonlinear 
dynamo action in accretion disks and  the radiative interior of massive stars, 
are also a matter of ongoing research.

\newpage
\section{References}
\bibliographystyle{elsarticle-num}
\bibliography{article.bib}
\section{Acknowledgments}
This work was partly supported by the German Leibniz Gemeinschaft  within the SAW program, by the  Deutsche Forschungsgemeinschaft in
frame of the SPP 1488 (PlanetMag), and by the Helmholtz-Gemeinschaft Deutscher Forschungszentren  in frame of the Helmholtz
alliance LIMTECH.
\end{document}